\documentclass[rmp,aps,twocolumn,amsfonts,showpacs,superscriptaddress]
{revtex4-1}

\setcitestyle{numbers,square,sort}
\usepackage{amsmath,amssymb,bm} 
\usepackage{graphicx}
\usepackage{color}
\usepackage{subfigure}    
\usepackage[colorlinks=true]{hyperref} 
\hypersetup{
    bookmarks=true,         
    unicode=false,          
    pdftoolbar=true,        
    pdfmenubar=true,        
    pdffitwindow=false,     
    pdfstartview={FitH},    
    pdftitle={Holographic quantum matter},    
    pdfauthor={Sean A. Hartnoll, Andrew Lucas, Subir Sachdev},     
    pdfsubject={},   
    pdfcreator={},   
    pdfproducer={}, 
    pdfkeywords={} {} {}, 
    pdfnewwindow=true,      
    colorlinks=true,       
    linkcolor=magenta, 
    citecolor=blue,        
    filecolor=magenta,      
    urlcolor=blue           
} 
\usepackage{amsfonts}
\usepackage{upgreek}
\usepackage{slashed}
\usepackage{latexsym}

\usepackage{todonotes}

\newcommand{\beq}{\begin{equation}}
\newcommand{\eeq}{\end{equation}}

\newcommand{\nn}{\nonumber \\}

\usepackage{euscript}
\usepackage{amssymb}
\usepackage{amsfonts}
\usepackage{amsmath}
\usepackage{amsbsy}
\usepackage{epsfig}
\usepackage{amsthm}
\usepackage{amscd}
\usepackage{amstext}
\usepackage{verbatim}
\usepackage{amsmath}
\usepackage{cancel}
\usepackage{capt-of}
\usepackage{empheq}
\usepackage{slashed}
\usepackage{xcolor}
\usepackage{setspace}

\usepackage{tikz}
\usepackage{tikz-cd}
\usetikzlibrary{arrows}
\usetikzlibrary{intersections}
\usetikzlibrary{shapes.geometric}
\usetikzlibrary{decorations.pathmorphing, patterns,shapes}

\allowdisplaybreaks

\def\ben{\begin{equation}}
\def\een{\end{equation}}
\def\half{{\textstyle{1\over2}}}

\let\a=\alpha \let\b=\beta \let\g=\gamma \let\d=\delta 
   \let\k=\kappa
\let\l=\lambda     \let\r=\rho
\let\s=\sigma \let\t=\tau

\let\w=\omega \let\G=\Gamma

  \let\re=\ref
  
\def\nn{\nonumber}

\let\pa=\partial
\def\be{\begin{equation}}
\def\ee{\end{equation}}
\def\beq{\begin{equation}}
\def\eeq{\end{equation}}
\def\ba{\begin{array}}
\def\ea{\end{array}}

\def\dalemb#1#2{{\vbox{\hrule height .#2pt
       \hbox{\vrule width.#2pt height#1pt \kern#1pt
               \vrule width.#2pt}
       \hrule height.#2pt}}}

\newcommand{\bea}{\begin{eqnarray}}
\newcommand{\eea}{\end{eqnarray}}
\newcommand{\tr}{{\rm tr} }

\def\ep{{\epsilon}}
\def\vep{{\varepsilon}}

\def\R{{{\Bbb R}}}

\def\Z{{{\Bbb Z}}}

\def\Lag{{\mathcal{L}}}

\def\ocal{{\mathcal{O}}}

\newcommand{\specialcell}[2][c]{%
  \begin{tabular}[#1]{@{}c@{}}#2\end{tabular}}
  
  \usepackage{titlesec}

\renewcommand{\thesection}{\arabic{section}}
\renewcommand{\thesubsection}{\thesection.\arabic{subsection}}
\renewcommand{\thesubsubsection}{\thesubsection.\arabic{subsubsection}}
\renewcommand{\theparagraph}{\thesubsubsection.\arabic{paragraph}}

\makeatletter

\renewcommand{\p@subsection}{}
\renewcommand{\p@subsubsection}{}
\renewcommand{\p@paragraph}{}

\renewcommand{\@dotsep}{10000}

\DeclareRobustCommand\bfseriesitshape{%
  \not@math@alphabet\itshapebfseries\relax
  \fontseries\bfdefault
  \fontshape\itdefault
  \selectfont
}

\titleformat{\paragraph}
{\normalfont\normalsize\itshape}{\theparagraph)}{1em}{}
\titlespacing*{\paragraph}
{0pt}{3.25ex plus 1ex minus .2ex}{1.5ex plus .2ex}
\def\toclevel@paragraph{4}
\def\l@paragraph{\@dottedtocline{4}{7em}{3.5em}}
\makeatother

\begin{document}

\title{Holographic quantum matter}

\author{Sean A. Hartnoll}
\email{hartnoll@stanford.edu}
\affiliation{Department of Physics, Stanford University, 
Stanford, CA 94305, USA}

\author{Andrew Lucas}
\email{ajlucas@stanford.edu}
\affiliation{Department of Physics, Stanford University, 
Stanford, CA 94305, USA}

\author{Subir Sachdev}
\email{sachdev@g.harvard.edu}
\affiliation{Department of Physics, Harvard University, Cambridge, 
MA 02138, USA}
\affiliation{Perimeter Institute for Theoretical Physics, Waterloo, Ontario, Canada N2L 2Y5}

\date{\today}

\begin{abstract}%
We present a review of theories of states of quantum matter without
quasiparticle excitations. Solvable examples of such states are provided through a holographic duality with 
gravitational theories in an emergent spatial dimension.  We review  the duality  between gravitational backgrounds and the various states of quantum matter which live on the boundary.   We then describe quantum matter at a fixed commensurate density (often described by conformal field theories), and also compressible quantum matter with variable density, providing an
extensive discussion of transport in both cases. We present a unified discussion of the holographic theory of transport with memory matrix and hydrodynamic methods, allowing a direct connection to experimentally realized quantum matter.  We also explore other important challenges in non-quasiparticle physics, including symmetry broken phases such as superconductors and non-equilibrium dynamics.  
\end{abstract}


\maketitle

\tableofcontents

\pagebreak


\section{The holographic correspondence}

\subsection{Introducing `AdS/CMT'}

This review is about a particular interface between condensed matter physics, gravitational physics and string and quantum field theory. The defining feature of this interface is that it is made possible by holographic duality, to be introduced briefly in this first section. This interface has been called `AdS/CMT' -- a pun on the name of the best understood version of holography, the AdS/CFT correspondence. Let us be clear from the outset, however, that neither AdS nor CFT are essential features of the framework that we will develop.

Holographic condensed matter physics encompasses a body of work with a wide range of motivations and objectives. Here are three different informal and zeroth order descriptions of the remit of this review:

\begin{enumerate}

\item For the condensed matter physicist: {\it AdS/CMT is the study of condensed matter systems without quasiparticles. In particular, AdS/CMT provides a class of models without an underlying quasiparticle description in which controlled computations can nonetheless be performed.}

\item For the relativist: {\it AdS/CMT is the study of event horizons in spacetimes whose asymptopia acts as a confining box. Asymptotically AdS spacetime is the simplest such box. Of particular interest are charged horizons, as well as various flavors of novel `hairy' black holes.}

\item For the string or field theorist: {\it AdS/CMT is the study of certain properties of large N field theories and their corresponding holographically dual spacetimes. Of particular interest are dissipative properties and the phases of the theory as a function of temperature and chemical potentials.}

\end{enumerate}

Of course our reader should feel no need to place herself in a particular category! The point is that under the umbrella of `AdS/CMT' one will find, for example, discussions of experimental features of non-quasiparticle transport, ways around no-hair theorems for black holes, and discussions of objects from string theory such as D-branes. We aim to address all these perspectives in the following. While we have tried to make our discussion intelligible to readers from different backgrounds, for basic introductory material we will refer the reader to the literature. We also note two recent books \cite{ammon2015gauge,zaanen2015holographic} which cover related topics on applications of holography to condensed matter physics.

\subsection{Historical context I: quantum matter without quasiparticles}
\label{sec:qmwqp}

A ubiquitous property of the higher temperature superconductors (found in the cuprates, the pnictides, and related compounds)
is an anomalous `strange metal' state found at temperatures, $T$, above the superconducting critical temperature, $T_{\mathrm{c}}$ \cite{Sachdev:2011cs}.
Its transport properties deviate strongly from those of conventional metals described by Fermi liquid theory, and spectral probes
display the absence of long-lived quasiparticle excitations. The value of $T_{\mathrm{c}}$ is determined by a balance between the free
energies of the superconductor and the strange metal, and so a theory for the strange metal is a pre-requisite for any theory of 
$T_{\mathrm{c}}$. This is one of the reasons for the great interest in developing a better understanding of strange metals.

Speaking more broadly, we may consider the strange metal as an important realization of a state of quantum matter without
quasiparticle excitations. We can roughly define a quasiparticle as a long-lived, low energy elementary excitation of a many-body state
with an `additive' property: we can combine quasiparticles to create an exponentially large number of composite excitations. 
In Landau theory \cite{pines1994theory}, the energy of a multi-quasiparticle excitation is given 
by sum of the energies of the quasiparticles and an additional mean-field interaction energy, dependent upon 
the quasiparticle density. It is important to note that the quasiparticles need not have the same quantum numbers
as a single electron (or whatever happens to be the lattice degree of freedom in the model). In quantum states with long-range
quantum entanglement, such as the fractional quantum Hall states, the quasiparticles are highly non-trivial combinations of the 
underlying electrons, and carry exotic quantum numbers such as fractional charge, and obey fractional statistics. Nevertheless, Landau's
quasiparticle framework can be used to build multi-particle excitations even in such cases.

In the states of quantum matter of interest to us here, no quasiparticle description exists. However, such a definition is unsatisfactory because it makes it essentially impossible to definitively establish whether a particular state is in a non-quasiparticle category.
Given a state and a Hamiltonian, we can propose some set of possible quasiparticles, and rule them out as valid excitations.
However, it is difficult to rule out all possible quasiparticles: short of an exact solution, we cannot be sure that
there does not exist a `dual' description of the model in which an unanticipated quasiparticle description can emerge.

A more satisfactory picture emerges when we view the issue from a dynamic perspective. We focus on one of the defining
properties of a quasiparticle, that it is `long-lived'. Imagine perturbing the system of interest a little bit from a state at thermal
equilibrium at a temperature $T$. If quasiparticles exist, they will be created by the perturbation; eventually they will collide
with each other and establish local thermal equilibrium. 
We denote the time required to establish local thermal equilibrium by $\tau_\varphi$;
alternatively, we can also consider $\tau_\varphi$ to be the time over which the local quantum phase coherence is lost after the external perturbation.
The upshot of this discussion is that states with quasiparticles are expected to have a long $\tau_\varphi$, while those without 
quasiparticles have a short $\tau_\varphi$. Can we make this boundary more precise? In a Fermi liquid, a standard Fermi's golden
rule computation shows that $\tau_\varphi$ diverges as $T \rightarrow 0$ as $1/T^2$. In systems with an energy gap, this time
is even longer, given the diluteness of the thermal quasiparticle excitations, and we 
have $\tau_\varphi \sim \mathrm{e}^{\Delta/T}$ \cite{Damle98} , 
where $\Delta$ is the energy gap. But how short can we make $\tau_\varphi$ in a system without quasiparticles? 
By examining a variety of models with
and without quasiparticles, and also by arguments based upon analytic continuation of equivalent statistical mechanics models
in imaginary time, it was proposed \cite{ssbook} that there is a lower bound
\beq
\tau_\varphi \geq C \frac{\hbar}{k_{\mathrm{B}} T} \,, \label{SSbound}
\eeq
which is obeyed by all infinite many-body quantum systems as $T \rightarrow 0$. Here $C$ is a dimensionless number of order unity
which is independent of $T$. Note that even without specifying $C$,  
(\ref{SSbound}) is a strong constraint {\em e.g.\/} it rules out systems in which
$\tau_\varphi \sim 1/\sqrt{T}$, and no such systems have been found. 
So now, we can give a more assertive criterion for a system without quasiparticles \cite{ssbook,ZaanenNature04}: it is a system which saturates the lower bound
in (\ref{SSbound}) as $T \rightarrow 0$ {\em i.e.\/} $\tau_{\varphi} \sim 1/T$.

In recent work, using inspiration from gravitational analogies, Maldacena, Shenker, and Stanford \cite{Maldacena:2015waa}
have taken a quantum
chaos perspective on related issues. 
They focused attention on a Lyapunov time, $\tau_{\mathrm{L}}$, defined as a measure of the time for a many-body system to lose
memory of its initial state \cite{Larkin1969}. 
This can be measured by considering the magnitude-squared of the commutator of two observables a time $t$
apart: 
the growth of the commutator with $t$ is then a measure of how the quantum state at the later
time has been perturbed the initial observation. With a suitable choice of observables, the growth is initially exponential, $\sim \exp (t/\tau_{\mathrm{L}})$, and this defines $\tau_{\mathrm{L}}$.
Modulo some reasonable physical assumptions, they established a lower bound for $\tau_{\mathrm{L}}$
\beq
\tau_{\mathrm{L}} \geq \frac{1}{2 \pi} \frac{\hbar}{k_{\mathrm{B}} T}. \label{MSSbound}
\eeq
As we will discuss further in this article, there are two important systems 
which precisely saturate the lower bound in (\ref{MSSbound}):
({\em i\/}) a mean-field model of a strange metal called the Sachdev-Ye-Kitaev (SYK) model \cite{SY92,kitaev2015talk} (see \S \ref{sec:SYK}),
and ({\em ii\/}) the holographic duals of black holes
in most theories of gravity. More generally, our interest here is in strongly-coupled states without quasiparticles obeying
$\tau_{\mathrm{L}} \sim \hbar/(k_{\mathrm{B}} T)$, which reach quantum chaos in the shortest possible time.

\subsection{Historical context II: horizons are dissipative}
\label{sec:horizons}

The most interesting results in holographic condensed matter physics are made possible by the fact that the \emph{classical} dynamics of black hole horizons is thermodynamic and dissipative. This is the single most powerful and unique aspect of the holographic approach. In this subsection we briefly outline these basic intrinsic properties of event horizons, independently of any holographic interpretation. Later, via the holographic correspondence, the classical dynamics of event horizons will be reinterpreted as statements about the deconfined phase of gauge theories in the 't Hooft matrix large $N$ limit. The 't Hooft limit will be the subject of the following \S \ref{sec:thooft}.

To emphasize the central points, we discuss stationary, neutral and non-rotating black holes. The important quantities characterizing the black hole are then its area $A$, mass $M$ and surface gravity $\kappa$. The surface gravity is the force required at infinity to hold a unit mass in place just above the horizon. The horizon is the surface of no return -- this irreversibility inherent in classical gravity underpins the connection with thermodynamics. Bardeen, Carter and Hawking showed that the classical Einstein equations imply \cite{Bardeen:1973gs}:
\begin{enumerate}
\item The surface gravity $\kappa$ is constant over the horizon.
\item Adiabatic changes to the mass and size of the black hole are related by
\be\label{eq:firstlaw}
\delta M = \frac{\kappa}{8 \pi G_{\mathrm{N}}} \, \delta A \,.
\ee
Here $G_{\mathrm{N}}$ is Newton's constant.
\item The horizon area always increases: $\delta A \geq 0$.
\end{enumerate}
These are clearly analogous to the zeroth through second laws of thermodynamics. Hawking subsequently showed that black holes semiclassically radiate thermal quanta, and thereby obtained the temperature and entropy \cite{Hawking:1974sw}
\be\label{eq:BH}
T = \frac{\hbar \, \kappa}{2 \pi} \,, \qquad S = \frac{1}{4} \frac{A}{\hbar \, G_{\mathrm{N}}} = \frac{1}{4} \frac{A}{L_{\mathrm{p}}^2} \,.
\ee
Note the explicit factors of Planck's constant $\hbar$. This shows that while Hawking radiation is a quantum mechanical effect, the entropy of the black hole is so large (the area is measured in Planck units $L_{\mathrm{p}}$) that the thermodynamic nature of black holes is imprinted on the classical equation (\ref{eq:firstlaw}) -- the factors of $\hbar$ in the temperature and entropy have cancelled. This is an important lesson for our purposes: black holes are able to geometrize thermodynamics because they describe systems with a huge number of degrees of freedom.

Granted that black holes in equilibrium behave as a thermodynamic system, the next question is whether perturbations of stationary black holes are described by dissipative linear response theory. Early work in the 70s by Hawking, Hartle, Damour and Znajek showed that perturbed black holes exhibited dissipative processes such as Joule heating. These observations were formalized as the `membrane paradigm' of black holes \cite{Price:1986yy, Thorne:1986iy, Parikh:1997ma}. For example, near an event horizon electromagnetic radiation must be `infalling' (that is, directed into rather than out of the black hole). Maxwell's equations then relate the electric and magnetic fields parallel to the horizon as
\be\label{eq:firstin}
E_\parallel = \hat n \times B_\parallel \,, 
\ee
where $\hat n$ is a spacelike outward pointing unit normal to the horizon. Because the horizon acts as a boundary of the spacetime, properly satisfying the variational principle leading to the Maxwell equations requires the association of charge and current densities to the horizon. The charge density is found to be proportional to the electric flux through the horizon while the current density $j$ along the horizon is related to the magnetic field at the horizon via
\be
B_\parallel = g^2_{\mathrm{H}} \, j \times \hat n \,,
\ee
where $g^2_{\mathrm{H}}$ is the electromagnetic coupling evaluated at the horizon (allowing in general for a space-dependent `dilaton' coupling).
Putting the previous two equations together gives Ohm's law on the horizon
\be
E_\parallel = g^2_{\mathrm{H}} \, j \,.
\ee
Hence the two dimensional spatial slice of the horizon has a universal dimensionless resistivity of $g^2_{\mathrm{H}}$, much like a quantum critical medium in two spatial dimensions. Solving the Einstein equations at quadratic order in the electromagnetic fields one finds that the area of the horizon increases in precisely the way required to describe the entropy generated by this current and resistivity.

The membrane description of black hole dissipation in asymptotically flat spacetime was complicated by the fact that such black holes have an unstable underlying thermodynamics (in particular, a negative specific heat) as well as finite size horizons. As we shall see, the holographic correspondence gives a clean framework for making sense of -- and putting to use -- the dissipative dynamics of horizons. The membrane paradigm was directly incorporated into holography in \cite{Kovtun:2003wp, Iqbal:2008by, Donos:2015gia}, to be discussed below.

The membrane paradigm describes perturbations of the event horizon by long-lived, hydrodynamic modes of currents and conserved charges. More general perturbations of a black hole relax quickly to equilibrium \cite{Vishveshwara}. For the black holes appearing in holographic scenarios, the thermalization timescale characterizing this decay is found to be $\tau_\varphi \sim \hbar/(k_{\mathrm{B}} T)$, as first emphasized in \cite{Horowitz:1999jd}. More recently,
it was recognized \cite{Shenker:2013pqa} that black holes are maximally chaotic, and their dynamics implies a saturation of the Lyapunov time $\tau_{\mathrm{L}}$ in (\ref{MSSbound}). The saturation of the timescale (\ref{SSbound}) for local equilibration indicates that the degrees of freedom underlying black hole thermodynamics must have strongly interacting non-quasiparticle dynamics.
This leads us to the following \S\ref{sec:thooft}.

\subsection{Historical context III: the 't Hooft matrix large $N$ limit}
\label{sec:thooft}

The essential fact about useful 
large $N$ limits of quantum fields theories is that there exists a set of operators $\{\ocal_i\}$ with no fluctuations to leading order at large $N$ \cite{Witten:1979pi, Coleman:1980nk}:
\be\label{eq:factorize}
\langle \ocal_{i_1} \ocal_{i_2} \cdots \ocal_{i_n} \rangle = \langle \ocal_{i_1} \rangle \langle \ocal_{i_2} \rangle \cdots \langle \ocal_{i_n} \rangle \,.
\ee
Operators that obey this large $N$ factorization are {\it classical} in the large $N$ limit. One might generally hope that, for a theory with many degrees of freedom `per site', each configuration in the path integral has a large action, and so the overall partition function is well approximated by a saddle point computation. However, the many local fields appearing in the path integral will each be fluctuating wildly. To exhibit the classical nature of the large $N$ limit one must therefore identify a set of `collective' operators $\{\ocal_i\}$ that do not fluctuate, but instead behave classically according to (\ref{eq:factorize}). Even if one can identify the classical collective operators, it will typically be challenging to find the effective action for these operators whose saddle point determines their expectation values. The complexity and richness of different large $N$ limits depends on the complexity of this effective action for the collective operators.

A simple set of examples are given by vector large $N$ limits. One has, for instance, a bosonic field $\vec \Phi$ with $N$ components and, crucially, interactions are restricted to be $\mathrm{O}(N)$ symmetric. This means the interaction terms must be functions of
\be\label{eq:sigma0}
\sigma \equiv \vec \Phi \cdot \vec \Phi \,.
\ee
(Slightly more generally, they could also be functions of $\vec \Phi \cdot \partial_{\mu_1} \cdots \partial_{\mu_2} \vec \Phi$.)  While the individual vector components of $\vec\Phi$ fluctuate about zero, $\sigma$ is a sum of the square of $N$ such fluctuating fields.  We thus expect that $\sigma$ behaves classically up to fluctuations which are subleading in $N$, analogous to the central limit theorem.  Indeed, by means of a Hubbard-Stratonovich transformation that introduces $\sigma$ as an auxiliary field, the action can be represented as a quadratric function of $\vec \Phi$ coupled to $\sigma$. Performing the path integral over $\vec \Phi$ exactly, one then obtains an effective action for $\sigma$ alone with an order $N$ overall prefactor (from the $N$ functional determinants arising upon integrating out $\vec \Phi$). This is the desired effective action for the collective field $\sigma$ that can now be treated in a saddle point approximation in the large $N$ limit, see e.g. \cite{ZinnJustin:2002ru}.

For our purposes the vector large $N$ limits are too simple. The fact that one can obtain the effective action for $\sigma$ from a few functional determinants translates into the fact that the theory is essentially a weakly interacting theory in the large $N$ limit.  As we will emphasize repeatedly throughout this review, weak interactions means long lived quasiparticles, which in turn mean essentially conventional phases of matter and conventional Boltzmann descriptions of transport. The point of holographic condensed matter is precisely to realize inherently strongly interacting phases of matter and transport that are not built around quasiparticles. Thus, while vector large $N$ limits do admit interesting holographic duals \cite{Giombi:2012ms}, they will not be further discussed here. As we will explain below, strongly interacting large $N$ theories are necessary to connect directly with the classical dissipative dynamics of event horizons discussed in the previous \S \ref{sec:horizons}.

The 't Hooft matrix large $N$ limit \cite{'tHooft:1973jz}, in contrast to the vector large $N$ limit, has the virtue of admitting a strongly interacting saddle point description. The fields in the theory now transform in the adjoint rather than the vector representation of some large group such as $\mathrm{U}(N)$. Thus the fields are large $N \times N$ matrices $\Phi_I$ and interactions are functions of traces of these fields
\be\label{eq:ocal}
\ocal_i = \tr \left(\Phi_{I_1} \Phi_{I_2} \cdots \Phi_{I_m} \right) \,.
\ee
These `single trace' operators are straightforwardly seen to factorize and hence become classical in the large $N$ limit \cite{Witten:1979pi, Coleman:1980nk}. There are vastly more classical operators of the form (\ref{eq:ocal}) than there were in the vector large $N$ limit case (\ref{eq:sigma0}). Furthermore there is no obvious prescription for obtaining the effective action whose classical equations of motion will determine the values of these operators. For certain special theories in the 't Hooft limit, this is precisely what the holographic correspondence achieves. The fact that was unanticipated in the 70s is that the effective classical description involves fields propagating on a higher dimensional curved spacetime.

In the simplest cases, the strength of interactions in the matrix large $N$ limit is controlled by the 't Hooft coupling $\lambda$ \cite{'tHooft:1973jz}. When the Lagrangian is itself a single trace operator then this coupling appears in the overall normalization as, schematically,
\be\label{eq:Lag}
{\mathcal L} = \frac{N}{\lambda} \tr \Big(\pa^\mu \Phi \, \pa_\mu \Phi +  \cdots \Big) \,.
\ee
When $\lambda$ is small, the large $N$ limit remains weakly coupled and can be treated perturbatively. In this limit the theory is similar to the vector large $N$ limit. When $\lambda$ is large, however, solving the theory would require summing a very large class of so-called planar diagrams. Through the holographic examples to be considered below, we will see that the large $\lambda$ limit is indeed strongly coupled when the theory is gapless: 
there are no quasiparticles and the single trace operators, while classical, acquire significant anomalous dimensions. This review is a study of the phenomenology of these strongly interacting large $N$ theories

A separate large $N$ limit has been the focus of much recent attention. This is intermediate 
between the vector and matrix large $N$ limits described above \cite{StanfordStrings}, and is realized in 
the Sachdev-Ye-Kitaev models. This limit does not have quasiparticles, and will be discussed in \S \ref{sec:SYK}.

\subsection{Maldacena's argument and the canonical examples}
\label{sec:malda}

We will now outline Maldacena's argument that connects the physics of event horizons described in \S \ref{sec:horizons} with the 't Hooft matrix large $N$ limit of \S \ref{sec:thooft} \cite{Maldacena:1997re, Maldacena:2011ut}. This argument involves concepts from string theory. Experimentalists have our permission to skip this section on a first reading. However, we will only attempt to get across the essential logic of the argument, which we hope will be broadly accessible. This argument is the source of the best understood, canonical examples of holographic duality.

\begin{figure}
\centering
\includegraphics[height = 0.2\textheight]{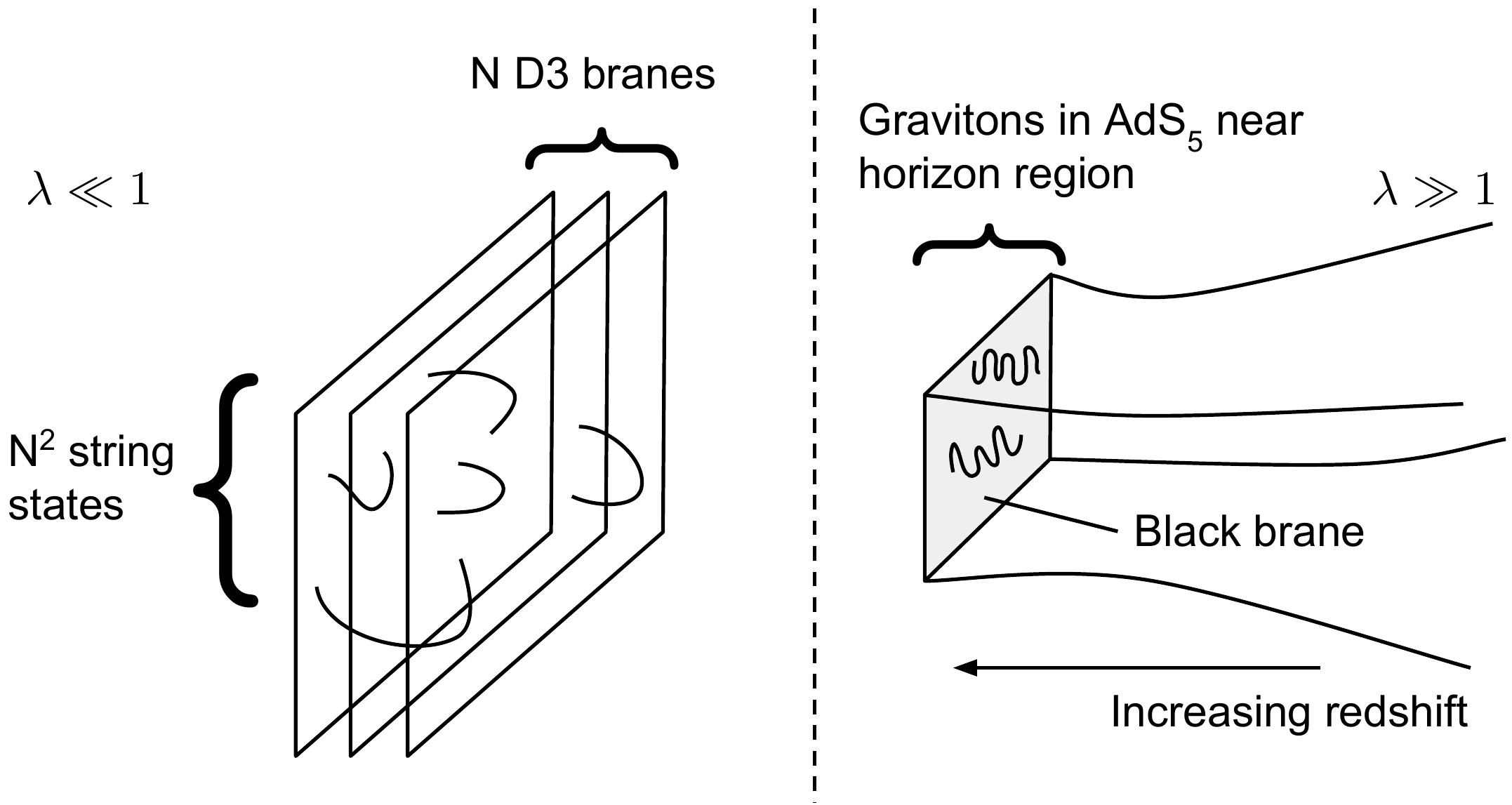}
\caption{\label{fig:branes} {\bf Low energy excitations} of a stack of D3 branes at weak and strong 't Hooft coupling. Weak coupling: $N^2$ light string states connecting the $N$ D3 branes. Strong coupling: classical gravitational excitations that are strongly redshifted by an event horizon.}
\end{figure}

The central idea is to describe a certain physical system in two different limits. Specifically, we consider a large number $N$ of `D$p$ branes' stacked on top of each other, in an otherwise empty spacetime. D$p$ branes are objects in string theory with $p$ spatial dimensions. Thus a D0 brane is a point particle, a D1 brane is an extended string, a D2 brane is a membrane, and so on. As a physical system, the coincident D$p$ branes have certain low energy excitations. We proceed to describe these excitations in two limits. These limits are illustrated in figure \ref{fig:branes} and explained in more detail in the following.

The simplest case of all is for D3 branes, so we will start with these. D3 branes are objects with three spatial dimensions, extended within the nine spatial dimensions of (type IIB) string theory.  Like all objects, the D3 branes gravitate. How strongly they gravitate depends on their tension and the strength of the gravitational force. The gravitational backreaction turns out to be determined by the number $N$ of superimposed D3 branes and the dimensionless string coupling constant $g_{\mathrm{s}}$. It is found that when
\be\label{eq:lam}
\lambda \equiv 4 \pi g_{\mathrm{s}} N \ll 1 \,,
\ee
then the gravitational effects of the D3 branes are negligible. This is because the tension of the $N$ D3 branes goes like $N/g_{\mathrm{s}}$, whereas the strength of gravity goes like $g_{\mathrm{s}}^2$. In this limit we can just imagine the D3 branes sitting in (nine dimensional) flat space. The low energy degrees of freedom in this case are known to be strings stretching between pairs of D3 branes. The lightest string states are massless because the D3 branes are coincident in space. There are $N^2$ such strings, because these strings can begin and end on any of the $N$ branes. At the lowest energy scales, it is known that these $N^2$ massless strings are described by a field theory called ${\mathcal N} = 4$ super Yang-Mills theory. This theory describes excitations that propagate on the three spatial dimensions of the D3 branes and, schematically, has Lagrangian density
\be\label{eq:Nis4}
\Lag \sim \frac{N}{\lambda} \tr \left( F^2 + \left( \nabla \Phi \right)^2 + \mathrm{i} {\bar \Psi} \slashed{D} \Psi + \mathrm{i} {\bar \Psi} [\Phi,\Psi] - [\Phi,\Phi]^2 \right) \,.
\ee
This expression describes a theory with an $\mathrm{SU}(N)$ field strength $F$ coupled to bosonic fields $\Phi$ and fermionic fields $\Psi$, all in the adjoint representation. In ${\mathcal N} = 4$ super Yang-Mills there are six bosonic fields and four fermionic fields. All of these fields are $N \times N$ matrices (matrix indices give the branes on which strings begin/end). The above Lagrangian therefore takes the form we discussed previously in (\ref{eq:Lag}), with $\lambda$ in (\ref{eq:lam}) now revealed as the 't Hooft coupling! Therefore, we have found that the low energy excitations of the D3 branes in the regime (\ref{eq:lam}) are described by a weakly interacting matrix large $N$ theory.

In the opposite regime
\be\label{eq:strong}
\lambda \equiv 4 \pi g_{\mathrm{s}} N \gg 1 \,,
\ee
the D3 branes gravitate very strongly. Under the strong effects of gravity the branes collapse upon themselves and form what is known as a black brane. These black branes are described by an appropriate analogue of well-known black hole solutions of general relativity. The horizon of the black brane is, by definition, a surface of infinite gravitational redshift. Therefore, for a far away observer, the low energy excitations of the system are any excitations that occur very close to the horizon. This region is called the near horizon geometry. This (ten dimensional) geometry is known to be a solution called $\mathrm{AdS}_5 \times \mathrm{S}^5$, with spacetime metric
\be\label{eq:ads5s5}
\mathrm{d}s^2 = L^2 \left( \frac{-\mathrm{d}t^2 + \mathrm{d}\vec x^2_3 + \mathrm{d}r^2}{r^2} + \mathrm{d}\Omega^2_5 \right) \,.
\ee
The coordinates $\{t,\vec x\}$ are those along the D3 brane worldvolume, the `radial' coordinate $r$ is orthogonal to the D3 branes, while the five sphere with round metric $d\Omega^2_5$ surrounds the D3 branes. The horizon itself is at $r \to \infty$ in these coordinates. The scale $L$ is called the AdS radius and can be related (see e.g. \cite{Maldacena:2011ut}) to the two fundamental scales in string theory, the string length $L_{\mathrm{s}}$ and the Planck length $L_{\mathrm{p}}$
\be\label{eq:stringy}
L = \lambda^{1/4} L_{\mathrm{s}} = (4 \pi N)^{1/4} L_{\mathrm{p}} \,.
\ee
In particular, in the strong 't Hooft coupling regime (\ref{eq:strong}) and at large $N \gg 1$, the radius of curvature $L$ of the $\mathrm{AdS}_5 \times \mathrm{S}^5$ geometry is very large compared to both the string and the Planck lengths. It follows that for many purposes we can neglect effects due to highly excited string states (controlled by $L_{\mathrm{s}}$) as well as quantum gravity effects (controlled by $L_{\mathrm{p}}$). The excitations of the near horizon region will simply be described by classical gravitational perturbations of the background (\ref{eq:ads5s5}).

The upshot of the previous two paragraphs is that the low energy excitations of $N$ D3 branes look very different in two different limits. At weak 't Hooft coupling they are described by a weakly interacting matrix large $N$ field theory. At strong 't Hooft coupling they are described by gravitational perturbations about $\mathrm{AdS}_5 \times \mathrm{S}^5$ spacetime (\ref{eq:ads5s5}). The original version of the holographic correspondence \cite{Maldacena:1997re} was the conjecture that there existed a decoupled set of degrees of freedom that interpolated between these weak and strong coupling descriptions. In particular, this implies that the {\it classical} gravitational dynamics of $\mathrm{AdS}_5 \times \mathrm{S}^5$ {\it is precisely} the strong coupling description of large $N$ ${\mathcal N} = 4$ super Yang-Mills theory. This is the long-sought effective classical description of a matrix large $N$ limit. Remarkably, the effective classical fields are gravitating and live in a higher number of spacetime dimensions! We now gain our first glimpse of a microscopic understanding of the dissipative dynamics of horizons that we discussed in \S \ref{sec:horizons}: gravity is in fact the dynamics of a strongly interacting matrix large $N$ quantum field theory in disguise.

The logic of the above argument can be applied to a very large number of configurations of branes in string theory. Many `dual pairs' of quantum field theories and classical gravitational theories are obtained in this way. In Table \ref{tab:holog} 
\begin{table*}[htp]
\caption{\label{tab:holog} {\bf The canonical examples of holographic duality for field theories in 3+1, 2+1 and 1+1 spacetime dimensions.} The first column shows the string theory setup of which one should consider the low energy excitations. In the limit of no gravitational backreaction the low energy degrees of freedom are described by a quantum field theory, given in the second column. In the limit of strong gravitational back reaction, the low energy degrees of freedom are described by classical gravitational dynamics about the backgrounds shown in the third column.}
\begin{center}
\begin{tabular}{|c|c|c|}
\hline
String theory system & Quantum field theory & Gravitational theory \\
\hline
\hline
\specialcell{$N$ D3 branes in \\ IIB string theory on $\R^{1,9}$}
 &  ${\mathcal N} = 4$ SYM theory (3+1) & \specialcell{IIB supergravity on \\ $\mathrm{AdS}_5 \times \mathrm{S}^5$}
\\
\hline
\specialcell{$N$ M2 branes in \\ M theory on $\R^{1,2} \times \R^{8}/\Z_k$}
 &  ABJM theory (2+1) & \specialcell{11d supergravity on \\ $\mathrm{AdS}_4 \times \mathrm{S}^7/\Z_k$}
\\
\hline
\specialcell{$Q_1$ D1 branes and $Q_5$ D5 branes in \\ IIB string theory on $\R^{1,5} \times M^4$}
 &  ${\mathcal N} = (4,4)$ SCFT (1+1) & \specialcell{IIB supergravity on \\ $\mathrm{AdS}_3 \times \mathrm{S}^3 \times M^4$}
\\
\hline
\end{tabular}
\end{center}
\label{default}
\end{table*}%
we list what might be considered the canonical, best understood examples of holographic duality for field theories in 3+1, 2+1 and 1+1 spacetime dimensions. We will not explain the terminology appearing in this table. Entry points to the literature include the original papers \cite{Maldacena:1997re, Aharony:2008ug} and the review \cite{Aharony:1999ti}. Our main objective here is to make clear that explicit examples of the duality are known in various dimensions and that they are found by using string theory as a bridge between quantum field theory and gravity.

\subsection{The essential dictionary}
\label{sec:essential}

\subsubsection{The GKPW formula}

While string theory is useful in furnishing explicit examples of holographic duality, much of the machinery of the duality is quite general and can be described using only concepts from quantum field theory (QFT) and gravity.

Basic observables that characterize the QFT are the multi-point functions of operators in the QFT. In particular, at large $N$, the basic observables are multi-point functions of the single trace operators $\ocal_i$, described above in (\ref{eq:ocal}). As examples of such operators, we can keep in mind charge densities $\rho = j^t$ and current densities $\vec \jmath$, associated with symmetries of the theory. The multi-point functions can be obtained if we know the generating functional
\be\label{eq:ZQFT}
Z_\text{QFT}[\left\{h_i(x)\right\}] \equiv \left\langle \mathrm{e}^{\mathrm{i} \sum_i \int \mathrm{d}x \, h_i(x) \ocal_i(x)} \right\rangle_\text{QFT} \,.
\ee

Observables in gravitating systems can be difficult to characterize, because the spacetime itself is dynamical. In the case where the spacetime has a boundary, however, observables can be defined on the boundary of the spacetime. We can, for instance, consider a Dirichlet problem in which the values of all the `bulk' fields (i.e. the dynamical fields in the theory of gravity) are fixed on the boundary. The boundary itself is not dynamical, giving the observer a `place to stand'. 
We can then construct the partition function of the theory as a function of the boundary values $\left\{h_i(x)\right\}$ of all the bulk fields $\{\phi_i\}$
\be\label{eq:gravity}
Z_\text{Grav.}[\left\{h_i(x)\right\}] \equiv \int^{\phi_i \to h_i} \left(\prod_i {\mathcal D}\phi_i \right) \mathrm{e}^{\mathrm{i} S[\left\{\phi_i\right\}]} \,.
\ee
It is a nontrivial mathematical problem to show that the Dirichlet problem for gravity is in fact well-posed, even in the classical limit (see e.g. \cite{Marolf:2012dr} for a recent discussion). However, we will see below how in practice the boundary data determines the bulk spacetime.

\begin{figure}
\centering
\includegraphics[height = 0.15\textheight]{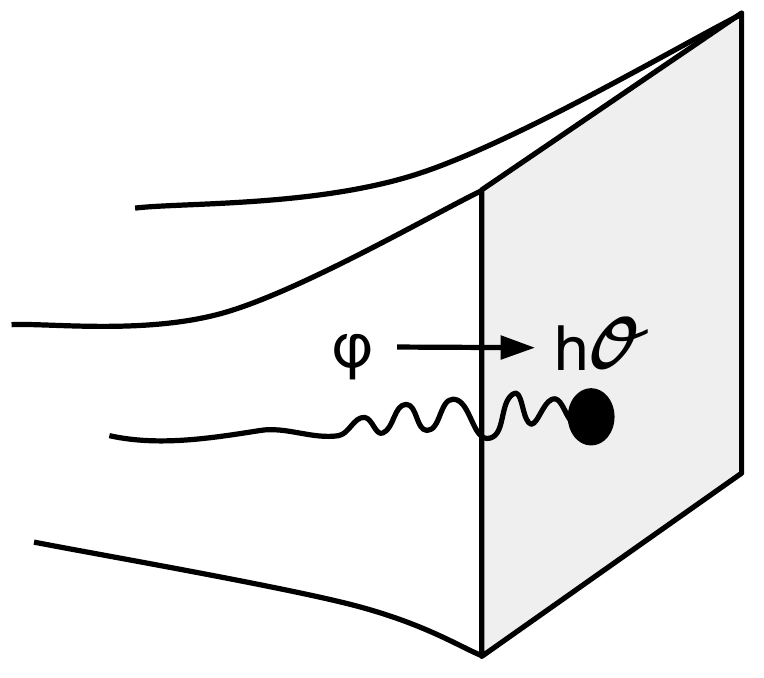}
\caption{\label{fig:source} {\bf Essential dictionary:} The boundary value $h$ of a bulk field $\phi$ is a source for an operator $\ocal$ in the dual QFT.}
\end{figure}

Suppose that a given QFT and theory of gravity are holographically dual. The essential fact relating observables in the two dual descriptions (of the same theory) is that there must be a one-to-one correspondence between single trace operators $\ocal_i$ in the QFT, and dynamical fields $\phi_i$ of the bulk theory. For example, a given scalar operator such as $\tr \left( \Phi^2 \right)$ in the QFT with schematic Lagrangian (\ref{eq:Nis4}) will be dual to a particular scalar field $\phi$ in the bulk theory. The scalar $\phi$ will have its own bulk dynamics given by the action $S$ in (\ref{eq:gravity}). We will be more explicit about the bulk action shortly. Having matched up bulk operators and boundary fields in this way, we can write the essential entry in the holographic dictionary, as first formulated by Gubser-Klebanov-Polyakov and Witten (GKPW) \cite{Witten:1998qj, Gubser:1998bc}:
\be\label{eq:dictionary}
Z_\text{QFT}[\left\{h_i(x)\right\}] = Z_\text{Grav.}[\left\{h_i(x)\right\}] \,.
\ee
That is, the generating functional of the QFT with source $h_i$ for the single trace operator $\ocal_i$ is equal to the bulk partition function with the bulk field $\phi_i$ corresponding to $\ocal_i$ taking boundary value $h_i$. This relationship is illustrated in Figure \ref{fig:source}.

The reader may have many immediate questions: How do we know which bulk field corresponds to which operator? There are a very large number of single trace operators, and so won't the bulk description involve a very large number of fields and hence be unwieldy? Before addressing these questions, we will illustrate the dictionary (\ref{eq:dictionary}) at work.

\subsubsection{Fields in AdS spacetime}

The large $N$ limit is supposed to be a classical limit, so let us evaluate the bulk partition function semiclassically
\be\label{eq:classicalsource}
Z_\text{Grav.}[\left\{h_i(x)\right\}] = \mathrm{e}^{\mathrm{i} S\left[\left\{\phi^\star_i \to h_i \right\}\right]} \,.
\ee
Here $S[\left\{\phi^\star_i \to h_i \right\}]$ is the bulk action evaluated on a saddle point, i.e. a solution to the bulk equations of motion, $\left\{\phi^\star_i\right\}$, subject to the boundary condition that $\phi^\star_i \to h_i$.

We need to specify the bulk action. The most important bulk field is the metric $g_{ab}$. The QFT operator corresponding to this bulk field will be the energy momentum tensor $T^{\mu\nu}$. This means that the boundary value of the bulk metric (more precisely, the induced metric on the boundary) is a source for the energy momentum tensor in the dual QFT, which seems natural. In writing down an action for the bulk metric, we will consider an expansion in derivatives, as one usually does in effective field theory. Later in this section we describe the circumstances in which this is a correct approach. The general bulk action for the metric involving terms that are at most quadratic in derivatives of the metric is
\be\label{eq:EH}
S[g] = \frac{1}{2 \kappa^2} \int \mathrm{d}^{d+2}x \sqrt{-g} \left(R + \frac{d(d+1)}{L^2} \right) \,.
\ee
The first term is the Einstein-Hilbert action, with $R$ the Ricci scalar and $\kappa^2 = 8\pi G_{\mathrm{N}} \propto L_{\mathrm{p}}^{d}$ defines the Planck length in the bulk $d+2$ dimensions (i.e. the boundary QFT has $d$ spatial dimensions\footnote{In this review, a $d$-dimensional system \emph{always} refers to a boundary theory in $d$ spatial dimensions and one time dimension. This is the convention of condensed matter physics.  We warn readers that much of the (early) holographic literature uses $d$ for the number of spacetime dimensions.}). The second term is a negative cosmological constant characterized by a lengthscale $L$ (holography with a positive cosmological constant remains poorly understood \cite{Anninos:2012qw}). The equations of motion following from this action are
\be\label{eq:vacE}
R_{ab} = - \frac{d+1}{L^2} g_{ab} \,.
\ee
The simplest solution to these equations is Anti-de Sitter spacetime, or $\mathrm{AdS}_{d+2}$:
\be\label{eq:adssol}
\mathrm{d}s^2 = L^2 \left( \frac{-\mathrm{d}t^2 + \mathrm{d}\vec x_{d}^2 + \mathrm{d}r^2}{r^2} \right) \,.
\ee
This spacetime has multiple symmetries. The easiest to see is the dilatation symmetry: $\{r,t,\vec x\} \to \lambda \{r,t,\vec x\}$. In fact there is an $\mathrm{SO}(d+1,2)$ isometry group \cite{Aharony:1999ti}. This is precisely the conformal group in $d+1$ spacetime dimensions. The solution (\ref{eq:adssol}) describes the vacuum of the dual QFT. If we now perturb the solution by turning on sources $\{h_i\}$, these perturbations will need to transform under the $\mathrm{SO}(d+1,2)$ symmetry group of the background. This suggests that the $d+2$ dimensional metric (\ref{eq:adssol}) dually describes a `boundary' conformal field theory (CFT) in $d+1$ spacetime dimensions. Let us see this explicitly.

Perturbing the metric itself is a little complicated, so consider instead an additional field $\phi$ in the bulk with an illustrative simple action
\be\label{eq:phi}
S[\phi] = -\int \mathrm{d}^{d+2}x \sqrt{-g}\left(\frac{1}{2} \left(\nabla \phi \right)^2 + \frac{m^2}{2} \phi^2 \right) \,.
\ee
As noted above, $\phi$ will correspond to some particular scalar operator in the dual quantum field theory. For concreteness we can have in the back of our mind an operator like $\tr\left(\Phi^2\right)$. To see the effects of adding a source $h$ for this operator, we must solve the classical bulk equations of motion
\be\label{eq:scalarwave}
\nabla^2 \phi = m^2 \phi \,,
\ee
subject to the boundary condition $\phi \to h$, and then evaluate the action on the solution as specified in (\ref{eq:classicalsource}). We have already noted that the horizon of $\mathrm{AdS}_{d+2}$ is the surface $r \to \infty$ in (\ref{eq:adssol}) -- this is properly called the Poincar\'e horizon. This is the surface of infinite redshift where $g_{tt} \to 0$. The `conformal boundary' at which we impose $\phi \to h$ is the opposite limit, $r \to 0$. In the metric (\ref{eq:adssol}) this corresponds to an asymptotic region where the volume of constant $r$ slices of the spacetime are becoming arbitrarily large. This should be thought of as imposing boundary conditions `at infinity'.

The equation of motion (\ref{eq:scalarwave}) is a wave equation in the $\mathrm{AdS}_{d+2}$ background (\ref{eq:adssol}). We can solve this equation by decomposing the field into plane waves in the $t$ and $x$ directions
\be
\phi = \phi(r) \mathrm{e}^{- \mathrm{i} \omega t + \mathrm{i} k \cdot x} \,.
\ee
The equation for the radial dependence then becomes
\be\label{eq:radial}
\phi'' - \frac{d}{r} \phi' + \left(\omega^2 - k^2 - \frac{(m L)^2}{r^2} \right) \phi = 0 \,.
\ee
To understand the asymptotic boundary conditions, we should solve this equation in a series expansion as $r \to 0$. This is easily seen to take the form
\be\label{eq:boundaryexpand}
\phi(r\rightarrow0) = \frac{\phi_{(0)}}{L^{d/2}}r^{d+1-\Delta}  + \cdots + \frac{\phi_{(1)}}{L^{d/2}}r^\Delta  + \cdots 
\ee
where
\be\label{eq:delta}
\Delta (\Delta - d-1) = \left(m L\right)^2 \,.
\ee
There are two constants of integration $\phi_{(0)}$ and $\phi_{(1)}$. The first of these is what we previously called the boundary value $h$ of $\phi$.  We shall understand the meaning of the remaining constant $\phi_{(1)}$ shortly. In (\ref{eq:boundaryexpand}) we see that in order to extract the boundary value $h$ of the field, one must first strip away some powers of the radial direction $r$. These powers of $r$ will now be seen to have an immediate physical content. We noted below (\ref{eq:adssol}) that rescaling $\{r,t,\vec x\} \to \lambda \{r,t,\vec x\}$ is a symmetry of the background $\mathrm{AdS}_{d+2}$ bulk spacetime. Like the metric, the scalar field $\phi$ itself must remain invariant under this transformation (the symmetry is a property of a particular solution, not of the fields themselves). Therefore the source must scale as $\phi_{(0)} = h \to \lambda^{\Delta - d-1} h$ in order for (\ref{eq:boundaryexpand}) to be invariant. Recalling that the source $h$ couples to the dual operator $\ocal$ as in (\ref{eq:ZQFT}), we see the dual operator must rescale as
\be\label{eq:BigDelta}
\ocal(x) \to \lambda^{-\Delta} \ocal(\lambda x) \,.
\ee
Therefore $\Delta$ is nothing other than the scaling dimension of the operator $\ocal$. Equation (\ref{eq:delta}) is a very important result that relates the mass of a bulk scalar field to the scaling dimension of the dual operator in a CFT. We can see that for relevant perturbations ($\Delta < d+1$), the $\phi_{(0)}$ term in (\ref{eq:boundaryexpand}) goes to zero at the boundary $r \to 0$, while for irrelevant perturbations ($\Delta > d+1$), this term grows towards the boundary. This will fit perfectly with our interpretation in \S \ref{sec:wilson} below of the radial direction as capturing the renormalization group of the dual quantum field theory, with $r \to 0$ describing the UV.

For a given bulk mass, equation (\ref{eq:delta}) has two solutions for $\Delta$. For most masses we must take the larger solution $\Delta_+$ in order for $\phi_{(0)}$ to be interpreted as the boundary value of the field in (\ref{eq:boundaryexpand}). Exceptions to this statement will be discussed later. Note that $d + 1 - \Delta_\pm = \Delta_\mp$.

\subsubsection{Simplification in the limit of strong QFT coupling}

We can now address the concern above that a generic matrix large $N$ quantum field theory has a very large number of single trace operators. For instance, with only two matrix valued fields one can construct operators of the schematic form $\tr \left(\Phi_1^{n_1} \Phi_2^{n_2} \Phi_1^{n_3}  \Phi_2^{n_4} \cdots \right)$. These each correspond to a bulk field. The bulk action should therefore be extremely complicated. In theories with tractable gravity duals an unexpected simplification takes place: in the limit of large 't Hooft coupling, $\lambda \to \infty$, all except a small number of operators acquire parametrically large anomalous dimensions. According to the relation (\ref{eq:delta}) this implies that all but a small number of the bulk fields have very large masses.\footnote{We are ignoring the presence of so-called bulk Kaluza-Klein modes in this discussion. These modes can lead to many light bulk fields even in the strong 't Hooft coupling limit $\lambda \to \infty$. See \S \ref{sec:consistent}.} The heavy fields can be neglected for many purposes. This leads to a tractable bulk theory.

In the concrete string theory realizations of holographic duality discussed above in \S \ref{sec:malda}, the hierarchy in the bulk can be easily understood. The excitations of the theory are string states. From the relation quoted in (\ref{eq:stringy}) we see that large 't Hooft coupling $\lambda \gg 1$ implies that $L/L_{\mathrm{s}} \gg 1$. The mass of a typical excited string state is $m \sim 1/L_{\mathrm{s}}$, and therefore $L m \gg 1$ for these states. Only a handful of low energy string states survive. This perspective also justifies a derivative expansion of the bulk action (and hence the neglect of higher derivative terms in (\ref{eq:EH}) and (\ref{eq:phi})). Higher derivative terms are suppressed by powers of $L_{\mathrm{s}}/L$, as they arise due to integrating out heavy string states.\footnote{A handful of higher derivative terms with unsuppressed couplings are in principle allowed, and can be traded for additional fields. This often leads to ghosts (fields with wrong-sign kinetic terms) or other pathologies \cite{Camanho:2014apa}. A finite number of unsuppressed and well-behaved higher derivative terms would correspond to fine tuning the bulk theory. We have already encountered an instance of fine tuning: to obtain the basic $\mathrm{AdS}_{d+2}$ solution (\ref{eq:adssol}), the cosmological constant term in the action balances the Ricci scalar term, which has more derivatives. This is only possible because the cosmological constant has been tuned to be very small in Planck units ($L/L_{\mathrm{p}} \gg 1$). In the dual QFT, this tuning is just the fact that we have taken $N$ to be large! The minimal statement
for a bulk theory dual to a QFT with an operator gap is that there cannot be an infinite number of unsuppressed higher derivative terms in the bulk, which would lead to a non-local bulk action \cite{Heemskerk:2009pn}.}

Beyond specific string theory realizations, the lesson of the previous paragraph is that a holographic approach is useful for QFTs in which {\it two} simplifications occur. Firstly,
\begin{align}
\text{Large $N$ limit} \qquad \Leftrightarrow \qquad \text{Classical bulk theory} \,,
\end{align}
and secondly
\begin{align}
&\text{Gap in single trace operator spectrum} \notag \\
& \text{(e.g. large $\lambda$ limit)} \notag \\
& \qquad \qquad \Leftrightarrow \notag \\
& \text{Derivative expansion in bulk,} \notag \\
& \text{small number of bulk fields}  \,. \label{eq:gap}
\end{align}
While the second condition may appear restrictive, string theory constructions show that these simplifications do indeed arise in concrete models. The most important point is that this is a way (a large $N$ limit with a gap in the operator spectrum) to simplify the description of QFTs without going to a weakly interacting quasiparticle regime. We will repeatedly see below how this fact allows holographic theories to capture generic behavior of, for instance, transport in strongly interacting systems that is not accessible otherwise. In contrast, weakly interacting large $N$ limits, such as the vector large $N$ limit, cannot have such a gap in the operator spectrum. This is because to leading order in large $N$ the dimensions of operators add, and in particular operators such as  $\vec \Phi \cdot \partial_{\mu_1} \cdots \partial_{\mu_2} \vec \Phi$ give an evenly spaced spectrum of conserved currents with no gap. Beyond the holographic approach discussed here, the development of purely quantum field theoretic methods exploiting the existence of an operator gap has recently been initiated \cite{Fitzpatrick:2010zm, ElShowk:2011ag, Fitzpatrick:2012cg, Hartman:2014oaa}.

\subsubsection{Expectation values and Green's functions}

A second result that can be obtained from the asymptotic expansion (\ref{eq:boundaryexpand}) is a formula for the expectation value of an operator. From (\ref{eq:ZQFT}), (\ref{eq:dictionary}) and (\ref{eq:classicalsource}) above
\be\label{eq:vevfull}
\langle \ocal_i(x) \rangle = \frac{1}{Z_\text{QFT}} \frac{\delta Z_\text{QFT}}{\delta h_i(x)} =
\mathrm{i} \frac{\delta}{\delta h_i(x)} S\left[\left\{\phi^\star_i \to h_i \right\}\right] \,.
\ee
The scalar action (\ref{eq:phi}) becomes a boundary term when evaluated on a solution. Because the volume of the boundary is diverging as $r \to 0$, we take a cutoff boundary at $r= \epsilon$. Then \begin{widetext}
\begin{align}
\langle \ocal(x) \rangle  &=  - \frac{\mathrm{i}}{2} \frac{\delta}{\delta \phi_{(0)}(x)} \int_{r=\epsilon} \mathrm{d}^{d+1}x^\prime \sqrt{-\gamma} \, \phi(x^\prime) \, n^a \nabla_a \phi  \notag \\
&=  \frac{\mathrm{i}}{2} \frac{\delta}{\delta \phi_{(0)}(x)} \int\limits_{r=\epsilon} \mathrm{d}^{d+1}x^\prime \left(\frac{L}{\epsilon}\right)^d \left((d+1-\Delta) \frac{\epsilon^{2d+1-2\Delta}}{L^d} \phi_{(0)}(x^\prime)^2 + (d+1) \, \frac{\phi_{(0)}(x^\prime) \phi_{(1)}(x^\prime)}{L^d} + \cdots \right). \label{eq:diverge}  
\end{align}
\end{widetext}
In the first line $\gamma$ is the induced metric on the boundary (i.e. put $r= \epsilon$ in (\ref{eq:adssol})) and $n$ is an outward pointing unit normal (i.e. $n^r = - \epsilon/L$). In the second line we have used the boundary expansion (\ref{eq:boundaryexpand}). Upon taking $\epsilon \to 0$, the first term diverges, the second is finite, and the remaining terms all go to zero. The divergent term is an uninteresting contact term, to be dealt with more carefully in the following section. We ignore it here. In evaluating the second term, note that because the bulk field $\phi$ satisfies a linear equation of motion, then $\phi_{(1)x}=(\delta \phi_{(1)}/\delta \phi_{(0)})_{xy} \phi_{(0)y}$ holds as a matrix equation. Thus we obtain
\be\label{eq:vev}
\langle \ocal(x) \rangle \propto \phi_{(1)}(x) \,.
\ee
We have not given the constant of proportionality here, as obtaining the correct answer requires a more careful computation, which will appear in \S \ref{sec:wilson}. The important point is that we have discovered the following basic relations:
\begin{subequations}
\bea
\lefteqn{\text{Field theory source $h$}} \nonumber \\
 & & \Leftrightarrow \qquad \text{Leading behavior $\phi_{(0)}$ of bulk field} \,. \\
\lefteqn{\text{Field theory expectation value $\langle \ocal \rangle$}} \nonumber \\
 &  & \Leftrightarrow \qquad \text{Subleading behavior $\phi_{(1)}$ of bulk field}  \,. \label{eq:subl}
\eea
\end{subequations}
This connection will turn out to be completely general. For instance, the source could be the chemical potential and the expectation value the charge density. Or, the source could be an electric field and the expectation value could be the electric current.

For a given set of sources $\{h_i\}$ at $r \to 0$, regularity of the fields in the interior of the spacetime (e.g. at the horizon $r \to \infty$) will fix the bulk solution completely. In particular, the subleading behavior near the boundary will be fixed. Therefore, by solving the bulk equations of motion subject to specified sources and regularity in the interior, we will be able to solve for the expectation values $\langle \ocal_i \rangle$. We can illustrate this with the simple case of the scalar field. The radial equation of motion (\ref{eq:radial}) is solved by Bessel functions. The boundary condition at the horizon is that the modes of the bulk wave equation must be `infalling', that is, energy flux must fall into rather than come out of the horizon. We will discuss infalling boundary conditions in detail below. For now, we quote the fact that the boundary conditions at the horizon and near the asymptotic boundary pick out the solution (let us emphasize that $t$ here is real time, we will also be discussing imaginary time later)
\be
\phi \propto h \, r^{(d+1)/2} \mathrm{K}_{\Delta-\frac{d+1}{2}}\left(r \sqrt{k^2 - \omega^2} \right) \mathrm{e}^{- \mathrm{i} \omega t + \mathrm{i} k \cdot x} \,.
\ee
The overall normalization is easily computed but unnecessary and not illuminating.
Here $\mathrm{K}$ is a modified Bessel function and $\Delta$ is the larger of the two solutions to (\ref{eq:delta}).
By expanding this solution near the boundary $r \to 0$, we can extract the expectation value (\ref{eq:vev}). In momentum space, the retarded Green's function of the boundary theory is given by the ratio of the expectation value by the source:
\be\label{eq:easyG}
G^{\mathrm{R}}_{\ocal\ocal}(\omega,k) = \frac{\langle \ocal \rangle}{h} \propto \frac{\phi_{(1)}}{\phi_{(0)}} \propto \left(k^2 - \omega^2 \right)^{\Delta - (d+1)/2} \,.
\ee
This is precisely the retarded Green's function of an operator $\ocal$ with dimension $\Delta$ in a CFT. This confirms our expectation that perturbations about the $\mathrm{AdS}_{d+2}$ background (\ref{eq:adssol}) will dually describe the excitations of a CFT.

\subsubsection{Bulk gauge symmetries are global symmetries of the dual QFT}
\label{sec:global}

The computation of the Green's function (\ref{eq:easyG}) illustrates how knowing the bulk action and the bulk background allows us to obtain correlators of operators in the dual strongly interacting theory. For this to be useful, we would like to know which bulk fields correspond to which QFT operators. This can be complicated, even when the dual pair of theories are known explicitly. Symmetry is an important guide. In particular, gauge symmetries in the bulk are described by bulk gauge fields. These include Maxwell fields $A_a$, the metric $g_{ab}$ and also nonabelian gauge bosons. The theory must be invariant under gauge transformations of these fields, including `large' gauge transformations, where the generator of the transformation remains constant on the asymptotic boundary. We can illustrate this point easily with a bulk gauge field $A_a$, which transforms to $A_a \rightarrow A_a + \nabla_a \chi$ for a scalar function $\chi$. If $\chi$ is nonzero on the boundary, then the boundary coupling in the action transforms to 
\bea
\lefteqn{\int \mathrm{d}^{d+1}x \sqrt{-\g} (A_\mu + \nabla_\mu \chi) J^\mu } \notag \\
&& = \int \mathrm{d}^{d+1}x \sqrt{-\g} (A_\mu J^\mu - \chi\nabla_\mu J^\mu ), 
\eea
where we integrated by parts in the last step.  Invariance under the bulk gauge transformation requires that $\nabla_\mu J^\mu = 0$, so that the current is conserved (in all correlation functions).
Therefore:
\bea
&& \text{Bulk gauge field (e.g. $A_a, g_{ab})$} \nonumber \\*
&& \qquad \qquad \Leftrightarrow \notag \\
&& \text{Conserved current of global symmetry in QFT} \notag \\
&& \text{(e.g. $J^\mu, T^{\mu\nu}$)} \,. \label{eq:JA}
\eea
Similarly, fields that are charged under a bulk gauge field will be dual to operators in the QFT that carry the corresponding global charge. Thus quantities such as electric charge and spin must directly match up in the two descriptions.

Matching up operators beyond their symmetries is often not possible in practice, and indeed is not really the right question to ask. The bulk is a self-contained description of the strongly coupled theory. The spectrum and interactions of bulk fields define the correlators and all other properties of a set of dual operators. Reference to a weakly interacting QFT Lagrangian description is not necessary and potentially misleading. Nonetheless, it can be comforting to have in the back of our minds some familiar operators. Thus a low mass scalar field $\phi$ in the bulk might be dual to QFT operators such as $\tr F^2, \tr \, \Phi^2$ or $\tr \, \overline \Psi \Psi$. Here we are using the notation of the schematic QFT action (\ref{eq:Nis4}). A low mass fermion $\psi$ in the bulk will be dual to an operator such as $\tr \, \Phi \Psi$.\footnote{
Single trace operators in QFT are dual to fields in the bulk. What about multi-trace operators? The insertion of multi-trace operators in the action in the sense of (\ref{eq:ZQFT}) will be discussed in the following section on holographic renormalization. A simpler aspect of multi-trace operators relates to the operator-state correspondence for CFTs. Here the single trace operator $\ocal$ corresponds to the state created by a quantum of the dual field $\phi$. The double-trace operator $\ocal^2$ corresponds to creating two quanta of the dual field. In the bulk, quantum effects are suppressed at large $N$ and so these different quanta are simply superimposed solutions of free bulk wave equations. For this reason, single trace operators in large $N$ theories (whose dimension does not scale with $N$) are sometimes referred to as generalized free fields, e.g. \cite{ElShowk:2011ag}.}

The general condensed matter systems we wish to study are not CFTs (although CFTs do comprise an interesting subset). Instead there will be multiple scales of interest: temperature, chemical potential, and scales generated by renormalization group flow of relevant operators. These are all understood within the framework of holographic renormalization, that we turn to next.

\subsection{The emergent dimension I: Wilsonian holographic renormalization}
\label{sec:wilson}

\begin{figure}
\centering
\includegraphics[height = 0.25\textheight]{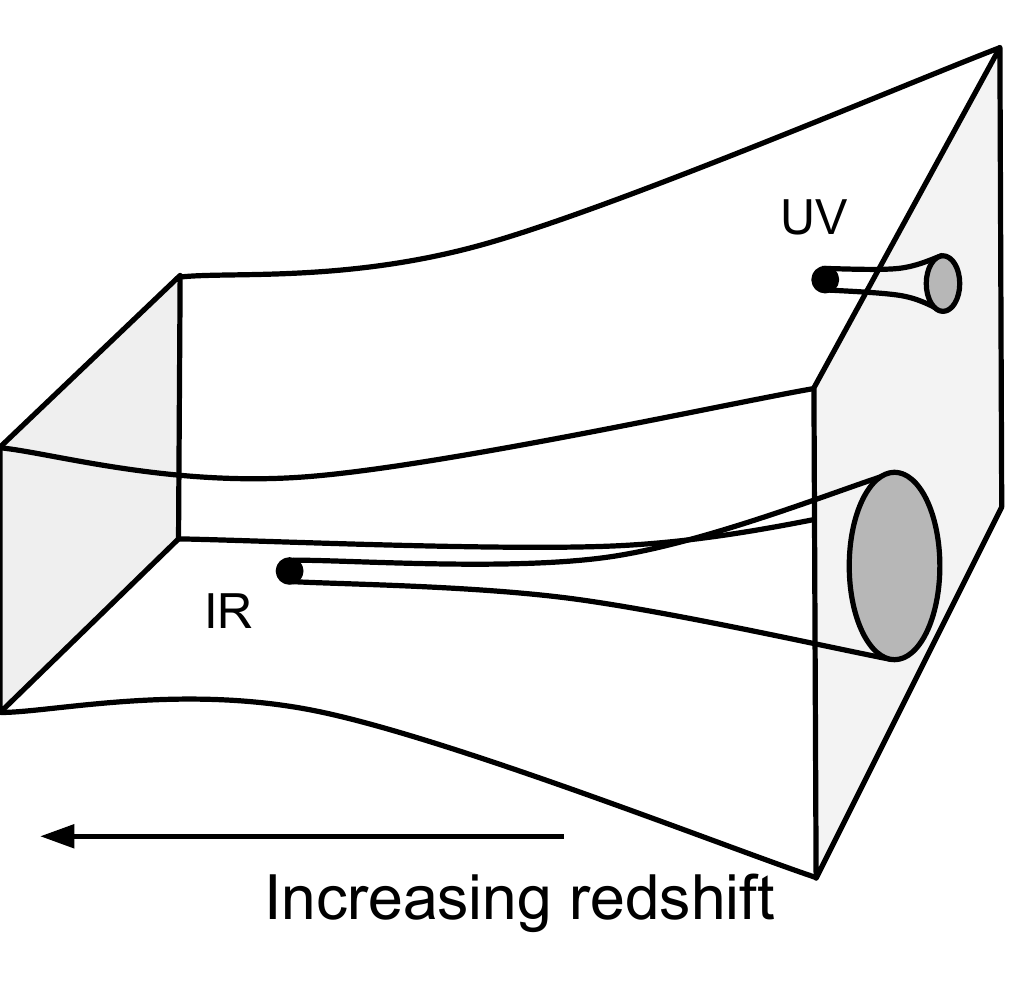}
\caption{\label{fig:RG} {\bf The radial direction:} Events in the interior of the bulk capture long distance, low energy dynamics of the dual field theory. Events near the boundary of the bulk describe short distance, UV dynamics in the field theory dual. The simplest way to think about this is that the interior events are increasingly redshifted relative to the boundary energy scales (a similar logic was used in the decoupling argument of \S \ref{sec:malda} to derive holographic duality, here we are discussing redshifts within the near horizon AdS region itself).}
\end{figure}

An essential aspect of the duality map (\ref{eq:dictionary}) is that the gravitating `bulk' spacetime has an extra spatial dimension relative to the dual `boundary' CFT. In this section and the following we will gain some intuition for the meaning of this extra `radial' dimension. The short answer is that the radial dimension geometrizes the renormalization group: processes close to the boundary of the bulk correspond to high energy physics in the dual QFT while dynamics deep in the interior of the bulk describes low energy physics in the QFT. This will be one of the conceptual pillars of holographic condensed matter physics, another being the already mentioned fact that horizons geometrize dissipation. The relation is illustrated in the Figure \ref{fig:RG}.

The radial or `holographic' dimension captures all of the standard renormalization physics: (\emph{i}) isolating universal low energy dynamics while parameterizing our ignorance about short distance physics, (\emph{ii}) computing beta functions for running couplings and (\emph{iii}) determining the structure of short distance divergences as a UV regulator is removed. In this  section we work through the illustrative case of a scalar field in a fixed background geometry, as in the previous section. We will develop the concepts further as we need them throughout the review.

\subsubsection{Bulk volume divergences and boundary counterterms}

We have already encountered a divergence when we tried to evaluate the on-shell action in (\ref{eq:diverge}). This divergence
is due to the infinite volume of the bulk spacetime, that is integrated over in evaluating the action. In (\ref{eq:diverge}) we regulated the divergence by cutting off the spacetime close to the boundary at $r = \epsilon$. We will see momentarily that this cutoff appears in the dual field theory as a short distance regulator of UV divergences. This connection between infinitely large scales in the bulk and infinitesimally short scales in the field theory is sometimes called the UV/IR correspondence \cite{Susskind:1998dq}. Taking our cue from perturbative renormalization in field theory, we can regulate the bulk action by adding a `counterterm' boundary action. For the case of the scalar, let
\be\label{eq:counter}
S \to S + S_\text{ct.} = S + \frac{a}{2 L} \int_{r=\epsilon} \mathrm{d}^{d+1}x \sqrt{-\gamma} \phi^2 \,.
\ee
Choosing $a=\Delta-d-1$, the counterterm precisely cancels the divergent term in (\ref{eq:diverge}). Because it is a boundary term defined in terms of purely intrinsic boundary data, it neither changes the bulk equations of motion nor the boundary conditions of the bulk field. The counterterm action furthermore contributes to the finite term in the expectation value of the dual operator: We can now write (\ref{eq:vev}) more precisely as
\be\label{eq:nicevev}
\langle \ocal(x) \rangle =  (2 \Delta - d-1) \phi_{(1)} \,.
\ee
The counterterm boundary action is a crucial part of the theory if we wish the dual field theory to be well-defined in the `continuum' limit $\epsilon \to 0$. A bulk action that is finite as $\epsilon \to 0$ allows us to think of the dual field theory as being defined by starting from a UV fixed point. This is the first of several instances we will encounter in holography where boundary terms in the action must be considered in more detail than one might be accustomed to.

The counterterm action (\ref{eq:counter}) is a local functional of boundary data. It is a nontrivial fact that all divergences that arise in evaluating on shell actions in holography can be regularized with counterterms that are local functionals of the boundary data. The terms that arise precisely mimic the structure of UV divergences in quantum field theory in one lower dimension, even though the bulk divergences are entirely classical. This identification of the counterterm action is called holographic renormalization. Entry points into the vast literature include \cite{deHaro:2000xn, Skenderis:2002wp}. An important conceptual fact is that in order for the bulk volume divergences to be local in boundary data, the growth of the volume towards the boundary must be sufficiently fast. Spacetimes that asymptote to AdS spacetime are the best understood case where such UV locality holds.

\subsubsection{Wilsonian renormalization as the Hamilton-Jacobi equation}

In this review we will frequently be interested in universal emergent low energy dynamics. For such questions, as in conventional quantum field theory, the precise short distance regularization or completion of the theory is unimportant. To understand the running of couplings and emergent dynamics we require a Wilsonian perspective on holographic theories. We outline how this works following \cite{Heemskerk:2010hk, Faulkner:2010jy}, which built on earlier work \cite{deBoer:1999tgo}.

Recall that a QFT is defined below some UV cutoff $\Lambda$ by the path integral
\be\label{eq:lambda}
Z = \int_\Lambda {\mathcal D}\Phi \mathrm{e}^{\mathrm{i} \left(I_0[\Phi] + I_\Lambda[\Phi] \right)} \,. 
\ee
Here $I_0[\Phi]$ is the microscopic action and $I_\Lambda[\Phi]$ are the terms obtained by integrating out all degrees of freedom at energy scales above the cutoff scale $\Lambda$. By requiring that the partition function $Z$ is independent of $\Lambda$ we obtain the renormalization group equations for $I_\Lambda[\Phi]$. By taking the cutoff to low energies we can obtain (in principle) the emergent low energy dynamics. By seeing how this structure emerges in holography we can gain some intuition for the meaning of the radial direction.

In the presence of a UV cutoff, there is a generalization of the fundamental holographic dictionary (\ref{eq:dictionary}). Consider a radial cutoff in the bulk at $r=\epsilon$. The cutoff gravitational partition function is now expressed in terms of the values of fields $\phi_i^\epsilon \equiv \phi_i(\epsilon)$ at the cutoff. Considering a single bulk field $\phi$ for notational simplicity, the gravity partition function (\ref{eq:gravity}) becomes
\be\label{eq:gravityep}
Z^\epsilon[\phi^\epsilon] \equiv \int^{\phi \to \phi_\epsilon} {\mathcal D}\phi \, \mathrm{e}^{\mathrm{i} S[\phi]} \,.
\ee
This quantity is naturally related to the partition function of the dual QFT integrated only over modes below some UV cutoff $\Lambda$. In particular, it is natural to generalize (\ref{eq:dictionary}) to
\be\label{eq:dic}
Z^\epsilon[\phi^\epsilon] = \int_\Lambda {\mathcal D}\Phi \, \mathrm{e}^{\mathrm{i} I_0[\Phi] + \mathrm{i} \int \mathrm{d}x \, \phi^\epsilon(x) \ocal(x)} \,.
\ee
Here $\int_\Lambda {\mathcal D}\Phi$ is the regulated field theory path integral as in (\ref{eq:lambda}).
Thus, the values of the bulk fields at the cutoff boundary are now interpreted as sources in an effective QFT valid at scales below a field theoretic cutoff $\Lambda$. While we schematically write $\Lambda \sim 1/\epsilon$, the precise nature of the field theory cutoff $\Lambda$ corresponding to the bulk cutoff $\epsilon$ is not known. For instance, the regulator preserves any gauge invariance of the boundary field theory and is therefore not a hard momentum cutoff. Determining precisely the field theory renormalization group scheme defined by a cutoff in the bulk would probably amount to a proof of holography. The perspective we will take is that the radial cutoff is a natural UV regulator in theories with holographic gravity duals, and we wish to check now that it leads to sensible consequences. In particular, we proceed to derive the renormalization group equations from the bulk. The following few paragraphs are a little technical, but allow us to show how the second order bulk equations of motion are related to first order renormalization group flow equations.

\begin{figure}
\centering
\includegraphics[height = 0.25\textheight]{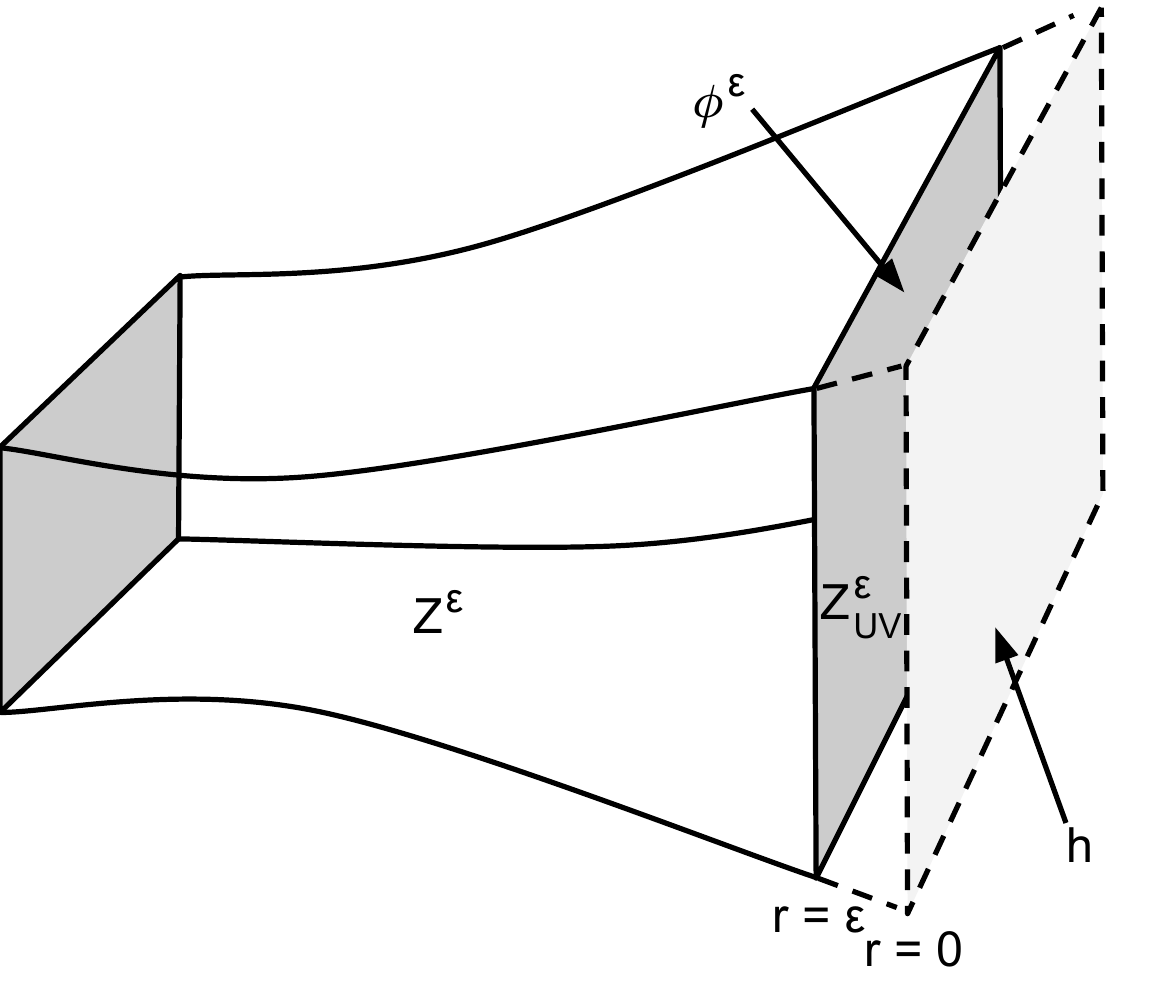}
\caption{\label{fig:glue} {\bf The Wilsonian cutoff:} The cutoff is at $r=\epsilon$, while the UV fixed point theory is at $r=0$. The partition function $Z^\epsilon[\phi^\epsilon]$ is over all modes at $r > \epsilon$, subject to the boundary condition $\phi(\epsilon) = \phi^\epsilon$. The partition function $Z^\epsilon_\text{UV}[\phi^\epsilon,h]$ is over all modes between $r=\epsilon$ and $r=0$, with boundary conditions at both ends.}
\end{figure}

The full bulk partition function (\ref{eq:gravity}) with source $h$ at the boundary is related to the truncated partition function (\ref{eq:gravityep}) by gluing path integrals together in the usual way (see Figure \ref{fig:glue}):
\begin{align}
Z[h] &=  \int {\mathcal D \phi^\epsilon} Z^\epsilon[\phi^\epsilon] Z^\epsilon_\text{UV}[\phi^\epsilon,h] \notag \\
&\equiv \int {\mathcal D \phi^\epsilon} \left[ Z^\epsilon[\phi^\epsilon] \int_{\phi \to \phi_\epsilon}^{\phi \to h} {\mathcal D \phi} \, \mathrm{e}^{\mathrm{i} S[\phi]} \right] \,.
\end{align}
Here $Z^\epsilon_\text{UV}[\phi^\epsilon,h]$ is the bulk path integral over the UV modes between the cutoff $\epsilon$ and the boundary at $r=0$, with boundary conditions at both ends.
Putting this previous equation together with (\ref{eq:dic}) and (\ref{eq:lambda}) we obtain
\be\label{eq:integraltransform}
e^{i I_\Lambda[\ocal]} = \int {\mathcal D \phi^\epsilon} \, \mathrm{e}^{\mathrm{i} \int \mathrm{d}x \, \phi^\epsilon(x) \ocal(x)} Z^\epsilon_\text{UV}[\phi^\epsilon,h] \,.
\ee
This equation gives a holographic expression for the effective action for the QFT operator $\ocal$ (a single trace operator built from the field theory degrees of freedom $\Phi$) obtained by integrating out high energy modes above the cutoff $\Lambda$. It is given by an integral transform of the bulk partition function where we only integrate over modes living on the bulk spacetime between the boundary at $r=0$ and $r=\epsilon$, subject to boundary conditions at both ends.

The semiclassical limit of (\ref{eq:integraltransform}) is the Legendre transform relation
\begin{align}
I_\Lambda[\ocal] &= \int \mathrm{d}^{d+1}x \, \phi_\star^\epsilon(x) \ocal(x) + S[\phi^\epsilon_\star \gets \phi_\star \to h] \notag \\
&\ocal = - \frac{\delta S_\star}{\delta \phi^\epsilon_\star} \,.\label{eq:Leg}
\end{align}
Here $S_\star \equiv S[\phi^\epsilon_\star \gets \phi_\star \to h]$ is the action evaluated on a solution $\phi_\star$ subject to the boundary conditions indicated. The action is the integral of a Lagrangian density $S = \int \mathrm{d}^{d+2}x \sqrt{-g} \Lag$, together with any boundary counterterms of the type we discussed above. We see that $\ocal$ in (\ref{eq:Leg}) is precisely the momentum conjugate to $\phi$ under radial evolution. A functional differential equation for $I_\Lambda[\ocal]$ can now be obtained using manipulations familiar from Hamilton-Jacobi theory.

To be concrete, consider the previous bulk action for the scalar (\ref{eq:phi}) generalized to arbitrary potential
\be
\Lag = - \frac{1}{2} \left(\nabla \phi \right)^2 - V(\phi) \,.
\ee
Using standard manipulations of Hamiltonian mechanics, with the radial direction playing the role of time \cite{Heemskerk:2010hk, Faulkner:2010jy}, from (\ref{eq:Leg}) one can
obtain the (Legendre transformed) Hamilton-Jacobi functional equation
\bea\label{eq:functional}
\lefteqn{\sqrt{g^{rr}}\pa_\ep I_\Lambda[\ocal] =} \\
&& \;\; \int \mathrm{d}^{d+1}x \sqrt{-\g}\left(\frac{1}{2 \gamma} \ocal^2 + \frac{1}{2} \left(\nabla_{(d+1)} \phi_\star^\epsilon \right)^2 + V(\phi_\star^\epsilon) \right) \,, \notag
\eea
where
\be
\phi_\star^\epsilon = \frac{\delta I_\Lambda[\ocal]}{\delta \ocal}\,.
\ee
In the above we assumed for simplicity that the metric takes the form $\mathrm{d}s^2 = g_{rr} \mathrm{d}r^2 + \gamma_{\mu\nu}(r) \mathrm{d}x^\mu \mathrm{d}x^\nu$, with no cross terms between the radial and boundary directions. $\nabla_{(d+1)}$ denotes the derivative along the boundary directions only. The second equation in (\ref{eq:functional}) follows from (\ref{eq:Leg}).

The functional equation (\ref{eq:functional}) is to be solved for $I_\Lambda[\ocal]$, and gives the evolution of the effective field theory action as a function of the cutoff $\epsilon$. If we expand the action in powers of $\ocal$ then we can write
\be\label{eq:effectiveL}
I_\Lambda[\ocal] = \sum_n \int \mathrm{d}^{d+1}x \sqrt{-\g} \,\lambda_n(x,\Lambda) \ocal(x)^n \,.
\ee
Plugging this expansion in the flow equation (\ref{eq:functional}), we obtain the beta functions for all of the couplings $\lambda_n$. Therefore, we have shown that varying the bulk cutoff $\epsilon$ leads to explicit first order flow equations of exactly the kind we anticipate for couplings in quantum field theories as function of some Wilsonian cutoff $\Lambda$. As we noted below (\ref{eq:dic}) above, the exact nature of the QFT cutoff $\Lambda$ is not known.

\subsubsection{Multi-trace operators}
\label{sec:multitrace}

It is useful to note that the Wilsonian effective action (\ref{eq:effectiveL}) includes multi-trace operators $\ocal^n$. Even -- as is often the case -- we start with a single trace action in the UV, these terms are generated under renormalization. Explicit holographic computations typically only make use of the bulk field $\phi$, which is dual to the single trace operator $\ocal$. However, $\phi$ satisfies a second order equation in the bulk. What has happened is that the first order beta function equations for all the couplings (including the multi-trace couplings) have been repacked into a second order equation for a single bulk field $\phi$. This approach, which is useful at large $N$, has been called the quantum renormalization group in \cite{Lee:2013dln}.

Multi-trace operators can also be inserted directly into the UV action. It can be shown that these correspond to changing the boundary conditions of the field as $r \to 0$ \cite{Witten:2001ua, Marolf:2006nd}. To deform the field theory action by
\be
\Delta S = W[\ocal] \,,
\ee
where $W$ is some functional of the single trace operator $\ocal(x)$, we must impose a specific relation between the coefficients $\phi_{(0)}$ and $\phi_{(1)}$ appearing in the near boundary expansion (\ref{eq:boundaryexpand}). This boundary condition is
\be\label{eq:genboundary}
\phi_{(0)} = \left. \frac{\delta W[\ocal]}{\delta \ocal}\right|_{\ocal = \frac{2 \Delta - d-1}{L} \phi_{(1)}} \,.
\ee
This general prescription includes the case of deformations by single trace operators we had considered previously in (\ref{eq:ZQFT}). That case corresponds to the simple linear relation $W[\ocal] = \int \mathrm{d}^{d+1}x \phi_{(0)}(x) \ocal(x)$.

An important special case is the double trace deformation $W[\ocal] \propto \int \mathrm{d}^{d+1}x \ocal(x)^2$, in the case when the operator dimension (\ref{eq:delta}) of $\ocal$ is in the range $\frac{d+1}{2} \leq \Delta_+ \leq \frac{d+1}{2} + 1$. It is a general fact about large $N$ conformal field theories that such an irrelevant double trace deformation induces a flow leading to a new UV fixed point theory \cite{Gubser:2002vv}. Under this flow, the dimension of the operator $\ocal$ changes from the larger root $\Delta_+$ of (\ref{eq:delta}) to the smaller root $\Delta_-$, while no other single trace operators receive any change in dimension to leading order at large $N$. This new theory therefore corresponds precisely to what is known as the alternate quantization of the bulk 
\cite{Klebanov:1999tb}. That is, the role of $\Delta_+$ and $\Delta_-$ are exchanged -- one can also think of this as the roles of $\phi_{(0)}$ and $\phi_{(1)}$ being exchanged, keeping $\Delta$ fixed, although we will prefer not to. The alternate quantization of the bulk is possible when the mass of the scalar field (\ref{eq:scalarwave}) is in the range $-\frac{(d+1)^2}{4} \leq m^2 \leq -\frac{(d+1)^2}{4} + 1$ (note that slightly negative values of the mass squared are stable in anti-de Sitter spacetime, so long as $\Delta$ is real -- we will discuss this in more detail later). The important fact about this range of masses is that both modes in (\ref{eq:boundaryexpand}) fall off sufficiently quickly that either one can be taken to be the `normalizable' mode, and the other the source. While the standard quantization required the boundary counterterm action (\ref{eq:counter}), the alternate quantization requires instead
\be
S_\text{ct.}^\text{(alt)} = \int_{r=\epsilon} \mathrm{d}^{d+1}x \sqrt{-\gamma} \left(\phi \, n^a \nabla_a \phi + \frac{\Delta}{2 L} \phi^2 \right) \,.
\ee
Here $\Delta = \Delta_-$ is now the smaller of the two exponents. Unlike the previous counterterm (\ref{eq:counter}), this expression includes a derivative normal to the boundary. This is the term responsible for changing the boundary conditions so that the $\Delta_+$ rather than $\Delta_-$ term becomes the source. The expression (\ref{eq:nicevev}) for the expectation value still holds.

\subsubsection{Geometrized versus non-geometrized low energy degrees of freedom}
\label{sec:nongeom}

From the Wilsonian holographic renormalization picture outlined above it is clear, in principle, how to zoom in on the low energy universal physics. We need to consider the dynamics of the fields in the far interior of the bulk spacetime. We will also need to understand how to match this interior dynamics to the sources that appear as boundary conditions on the asymptotic spacetime. We shall do this explicitly several times later in this review.

There is an interesting caveat to the previous paragraph. Holography geometrizes the renormalization of the order $N^2$ matrix degrees of freedom in the dual large $N$ field theory. Sometimes, however, there can be order $1 \ll N^2$ light degrees of freedom that are missed by this procedure. These degrees of freedom are extended throughout the bulk spacetime and not localized in the far interior. Effectively, they are not fully geometrized. An example are Goldstone modes of spontaneously broken symmetries. A framework, known as semi-holography, has been developed to characterize these cases, e.g. \cite{Faulkner:2010tq, Nickel:2010pr, Faulkner:2010jy}. We shall describe these methods later, as the various modes manifest themselves as poles in Green's functions that we will be computing. The contribution of these (low energy, non geometrized) modes running in loops in the bulk can lead to non-analyticities in low energy observables at subleading order in the $1/N$ expansion, causing subtle breakdowns of the large $N$ expansion. Examples are quantum oscillations due to bulk Fermi surfaces (\S \ref{sec:qo}), destruction of long-range order in low dimensions by fluctuating Goldstone modes (\S \ref{sec:coleman}) and long time tails in hydrodynamics (\S \ref{sec:quasinormal}).

\subsection{The emergent dimension II: Entanglement entropy}\label{sec:EE}

The previous section showed how the radial direction of the bulk spacetime geometrized the energy scale of the dual quantum field theory. The energy scale was understood within a functional integral, Wilsonian perspective on the QFT. An alternative way to conceive the bulk radial direction is as a certain geometric re-organization of the QFT Hilbert space.

\begin{figure}
\centering
\includegraphics[height = 0.17\textheight]{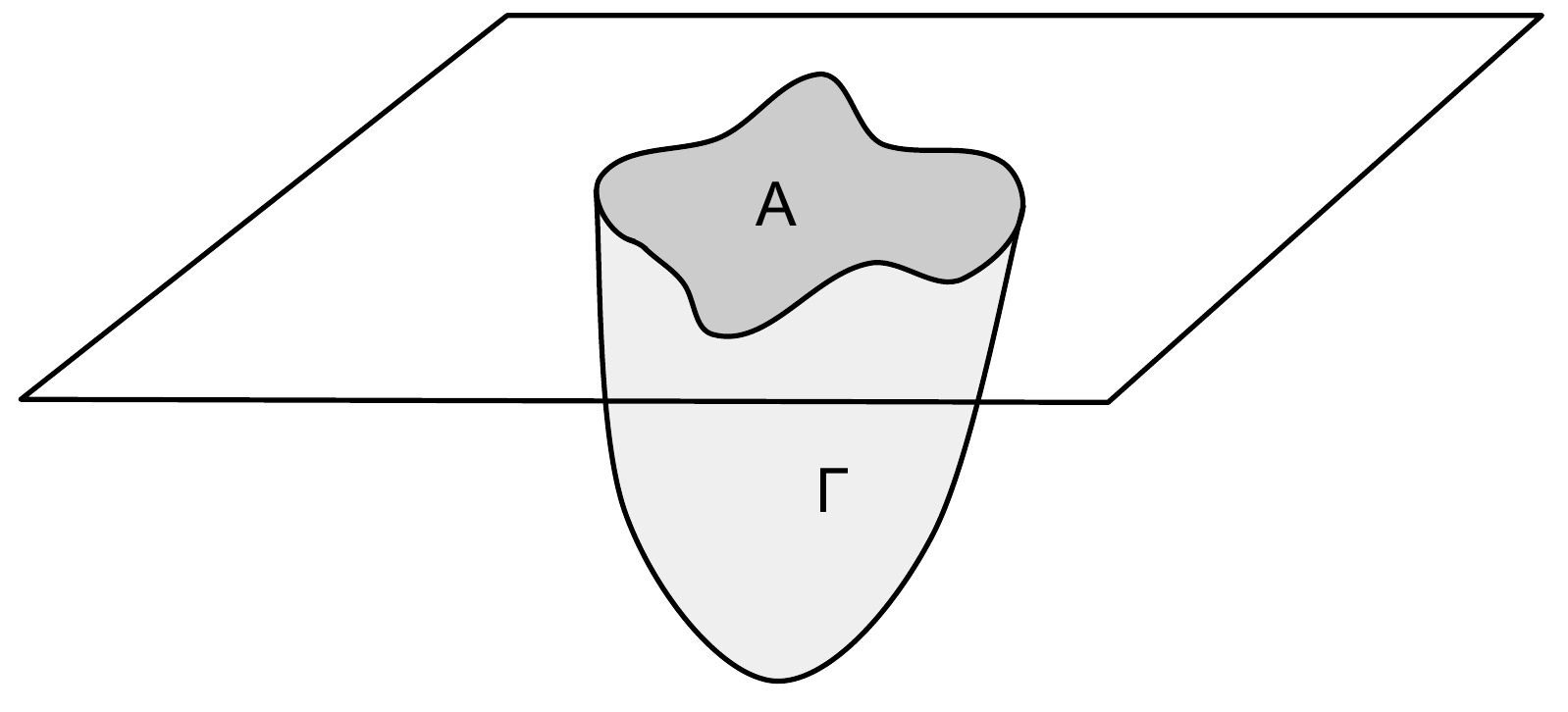}
\caption{\label{fig:RT} {\bf Minimal surface} $\Gamma$ in the bulk, whose area gives the entanglement entropy of the region $A$ on the boundary.}
\end{figure}

Shortly after the discovery of the holographic correspondence, it was realized that the bulk can be thought of as being made up of order $N^2$ QFT degrees of freedom per AdS radius \cite{Susskind:1998dq}. The recent discovery \cite{Ryu:2006bv, Ryu:2006ef, Nishioka:2009un} and proof \cite{Lewkowycz:2013nqa} of the Ryu-Takayangi formula for entanglement entropy in theories with holographic duals suggests that a more refined understanding of how the bulk reorganizes the QFT degrees of freedom is within reach. In the simplest case, in which the bulk is described by classical Einstein gravity coupled to matter, the Ryu-Takayanagi formula states that the entanglement entropy of a region $A$ in the QFT is given by
\be\label{eq:rt}
S_{\mathrm{E}} = \frac{A_\Gamma}{4 G_{\mathrm{N}}} \,,
\ee
where $G_{\mathrm{N}}$ is the bulk Newton's constant and $A_\Gamma$ is the area of a minimal surface (i.e. a soap bubble) $\Gamma$ in the bulk that ends on the boundary of the region $A$. The region $A$ itself is at the boundary of the bulk spacetime. We illustrate this formula in figure \ref{fig:RT}.
The formula (\ref{eq:rt}) generalizes the Bekenstein-Hawking entropy formula (\ref{eq:BH}) for black holes. This last statement is especially clear for black holes describing thermal equilibrium states, which correspond to entangling the system with a thermofield double \cite{Maldacena:2001kr}.

Let us now illustrate the Ryu-Takayanagi formula, and in particular how it reveals the physics of the bulk radial direction, by computing the entanglement entropy of a spherical region of radius $R$ in a CFT, following \cite{Ryu:2006ef}.    By spherical symmetry, we know that the minimal surface will be of the form $r(\rho)$, where $\rho$ is the radial coordinate on the boundary (so that $\mathrm{d}\vec x^2_d = \mathrm{d}\rho^2 + \rho^2 \mathrm{d}\Omega^2_{d-1}$). The induced metric on this surface is given by putting $r = r(\rho)$ and $t=0$ in the $\mathrm{AdS}_{d+2}$ spacetime (\ref{eq:adssol}) to obtain: \begin{equation}
\mathrm{d}s^2 = \frac{L^2\mathrm{d}\rho^2}{r^2}\left[1+\left(\frac{\mathrm{d}r}{\mathrm{d}\rho}\right)^2\right] + \frac{L^2\rho^2}{r^2}\mathrm{d}\Omega_{d-1}^2 \,.
\end{equation}
From the determinant of this induced metric, the area of the surface is given by:
\begin{equation}
A_\Gamma = L^d \, \Omega_{d-1} \int \frac{\mathrm{d}\rho}{r^d}\rho^{d-1}\sqrt{1+\left(\frac{\mathrm{d}r}{\mathrm{d}\rho}\right)^2}.  \label{eq:Agamma}
\end{equation}
$\Omega_{d-1}$ denotes the volume of the unit sphere $\mathrm{S}^{d-1}$.  It is simple to check that \begin{equation}
r(\rho) = \sqrt{R^2-\rho^2} \,, \label{eq:Agammasol}
\end{equation}
solves the Euler-Lagrange equations of motion found by minimizing the area (\ref{eq:Agamma}). This is therefore the Ryu-Takayanagi surface. Note that $\mathrm{d}r/\mathrm{d}\rho = -\rho/r$.

The entanglement entropy (\ref{eq:rt}) is given by evaluating (\ref{eq:Agamma}) on the solution (\ref{eq:Agammasol}).  For dimensions $d>1$,
\begin{equation}
A_\Gamma = L^d\Omega_{d-1} \int\limits_0^{R} \mathrm{d}\rho \; \frac{R\rho^{d-1}}{(R^2-\rho^2)^{(d+1)/2}}.
\end{equation}
This integral is dominated near $r=0$. Switching to the variable $y = R-\rho$, we obtain that
\begin{equation}
A_\Gamma \approx L^d\Omega_{d-1} \int\limits_\delta  \mathrm{d}y \frac{R^d}{(2Ry)^{(d+1)/2}} \sim L^d \left(\frac{R}{\delta}\right)^{(d-1)/2} \,,
\end{equation}
where $\delta$ is a cutoff on the bulk radial dimension near the boundary that, as in the previous Wilsonian section, we will want to interpret as a short distance cutoff in the dual field theory. This result becomes more transparent in terms of the equivalent cutoff on the bulk coordinate $r=\epsilon$; using (\ref{eq:Agammasol}) one has that  $\epsilon = \sqrt{2R\delta}$.  Hence, the entanglement entropy is
\begin{equation}
S_{\mathrm{E}} \sim \frac{L^d}{G_{\mathrm{N}}} \left(\frac{R}{\epsilon}\right)^{d-1} + \text{subleading}.  \label{eq:arealaw}
\end{equation}
The first term is proportional to the area of the boundary sphere $R$, and hence (\ref{eq:arealaw}) is called the area law of entanglement \cite{Srednicki:1993im}.   It can be intuitively understood as follows -- entanglement in the vacuum is due to excitations which reside partially inside the sphere, and partially outside. In a local QFT, we expect that the number of these degrees of freedom is proportional to the area of the sphere in units of the short distance cutoff, $(R/\epsilon)^{d-1}$.  This is true regardless of whether or not the state is gapped. Equation (\ref{eq:arealaw}) therefore gives another perspective on the identification of a near-boundary cutoff on the bulk geometry with a short-distance cutoff in the dual QFT. In the holographic result (\ref{eq:arealaw}), the proportionality coefficient $L^d/G_{\mathrm{N}} \sim (L/L_{\mathrm{p}})^d \gg 1$. This is consistent with the general expectation that there must be a large number of degrees of freedom ``per lattice site" in microscopic QFTs with classical gravitational duals.  While the coefficient of the area law is sensitive to the UV cutoff $\epsilon$ (since we have to count the number of degrees of freedom residing near the surface),  some subleading coefficients are universal \cite{Ryu:2006ef}.

In $d=1$, the story is a bit different.   Now, \begin{align}
A_\Gamma &= 2L\int\limits_0^{R} \frac{\mathrm{d}\rho \; R}{R^2-\rho^2} \approx 2L\int\limits_\delta^R \frac{\mathrm{d}\rho}{2\rho} = L\log\frac{R}{\delta} + \cdots \notag \\
&= 2L\log\frac{R}{\epsilon} +\cdots \,.
\end{align}
The prefactor of 2 is related to the fact that the semicircular minimal surface has two sides.   Using the result (beyond the scope of this review) that the central charge $c$ of the dual CFT2, which counts the effective degrees of freedom, is given by \cite{Brown:1986nw} \begin{equation}
c = \frac{3L}{2G_{\mathrm{N}}},
\end{equation}
we recover the universal CFT2 result \cite{Calabrese:2004eu}
\begin{equation}
S = \frac{c}{3}\log \frac{R}{\epsilon} + \cdots \,.
\label{eq:CFT2EE}
\end{equation}
Note that the area law, which would have predicted $S=\text{constant}$, has logarithmic violations in a CFT2.  We will see an intuitive argument for this shortly, as shown in Figure \ref{fig:MERA}.

\subsubsection{Analogy with tensor networks}

It was pointed out in \cite{Swingle:2009bg} that the Ryu-Takayangi formula resembles the computation of the entanglement entropy  in quantum states described by tensor networks. We will now describe this connection, following the exposition of \cite{Hartman:2013qma}. While, at the time of writing, these ideas remain to be fleshed out in any technical detail, they offer a useful way to think about the emergent spatial dimension in holography. For a succinct introduction to the importance of entanglement in real space renormalization, see \cite{vidal}.

Consider a system with degrees of freedom living on lattice sites labelled by $i$, taking possible values $s_i$. For instance each $s_i$ could be the value of a spin. Tensor network states (see e.g. \cite{eisert}) are certain wavefunctions $\psi(\{s_i\})$. The simplest example of a tensor network state is a Matrix Product State (MPS) for a one dimensional system with $L$ sites. For every value that the degrees of freedom $s_i$ can take, construct a $D \times D$ matrix $T_{s_i}$. Here $D$ is called the bond dimension. The physical wavefunction is now given by
\be\label{eq:MPS}
\psi(s_1, \cdots, s_L) = \tr \left( T_{s_1} T_{s_2} \cdots T_{s_L}  \right) \,.
\ee
For the case that each $s_i$ describes a spin half, these Matrix Product States parametrize a $2 L D^2$ dimensional subspace of the full $2^L$ dimensional Hilbert space. The importance of these states is that they capture the entanglement structure of the ground state of gapped systems, as we now recall.

To every site $i$ in the lattice, associate two auxiliary vector spaces of dimension $D$. Let these have basis vectors $|n\rangle_{i1}$ and $|n\rangle_{i2}$ respectively, with $n=1, \cdots , D$. Now form a state, in this auxiliary space, made of maximally entangled pairs between neighboring sites
\be\label{eq:bigpsi}
|\Psi\rangle = \prod_{i=1}^L \left(\sum_{n=1}^D |n\rangle_{i2}|n\rangle_{(i+1)1} \right) \,.
\ee
For periodic boundary conditions we can set $|n\rangle_{(L+1)1} = |n\rangle_{11}$. It is clear that if we trace over some number of adjacent sites, the entanglement entropy of the reduced density matrix will be $2 \log D$, corresponding to maximally entangled pairs at each end of the interval we have traced over. To obtain a state in the physical Hilbert space, we now need to project at each site. That is, let $P_i: \R^D \times \R^D \to \R^2$ be a projection, then the physical state
\be
|\psi\rangle = \prod_{i=1}^L P_i | \Psi\rangle \,.
\ee
If we write the projection operators as $P_i = \sum_{m,n,s_i} T^{mn}_{s_i} \, |s_i \rangle \, {}_{i1}\langle m | {}_{i2}\langle n |$, then we recover precisely the MPS state (\ref{eq:MPS}). The advantage of this perspective is that we now see that if we trace out a number of adjacent sites in this physical state $\psi$, the entanglement entropy of the reduced density matrix is bounded above by $S_{\mathrm{E}} \leq 2 \log D$.\footnote{This follows from the fact that, focusing on a single cut between two sites for simplicity, the state (\ref{eq:bigpsi}) in the auxiliary space can be written $|\Psi\rangle = \sum_{p=1}^D |p\rangle_{\mathrm{L}} |p\rangle_{\mathrm{R}}$ for some set of states $|p\rangle_{\mathrm{L}}$ and $|p\rangle_{\mathrm{R}}$ to the left and right of the cut, respectively. The projectors act independently on the left and right states, and necessarily map each $D$ dimensional subspace onto a vector space of dimension $D$ or lower. Schmidt decomposing the projected state, we have $|\psi\rangle = \sum_{p=1}^D c_p \widetilde{|p\rangle}_{\mathrm{L}} \widetilde{|p\rangle}_{\mathrm{R}}$, for coefficients $c_p$ and states $\widetilde{|p\rangle}_{\mathrm{L/R}}$ in the physical Hilbert space on either side of the cut. Hence the entanglement entropy is bounded above by $\log D$.} In gapped systems, the entanglement entropy of the ground state grows with the correlation length. At a fixed correlation length, the entanglement structure of the state can therefore be well approximated by a MPS state with some fixed bond dimension $D$ (if the inequality $S_{\mathrm{E}} \leq 2 \log D$ is approximately saturated).

\begin{figure}
\centering
\includegraphics[height = 0.2\textheight]{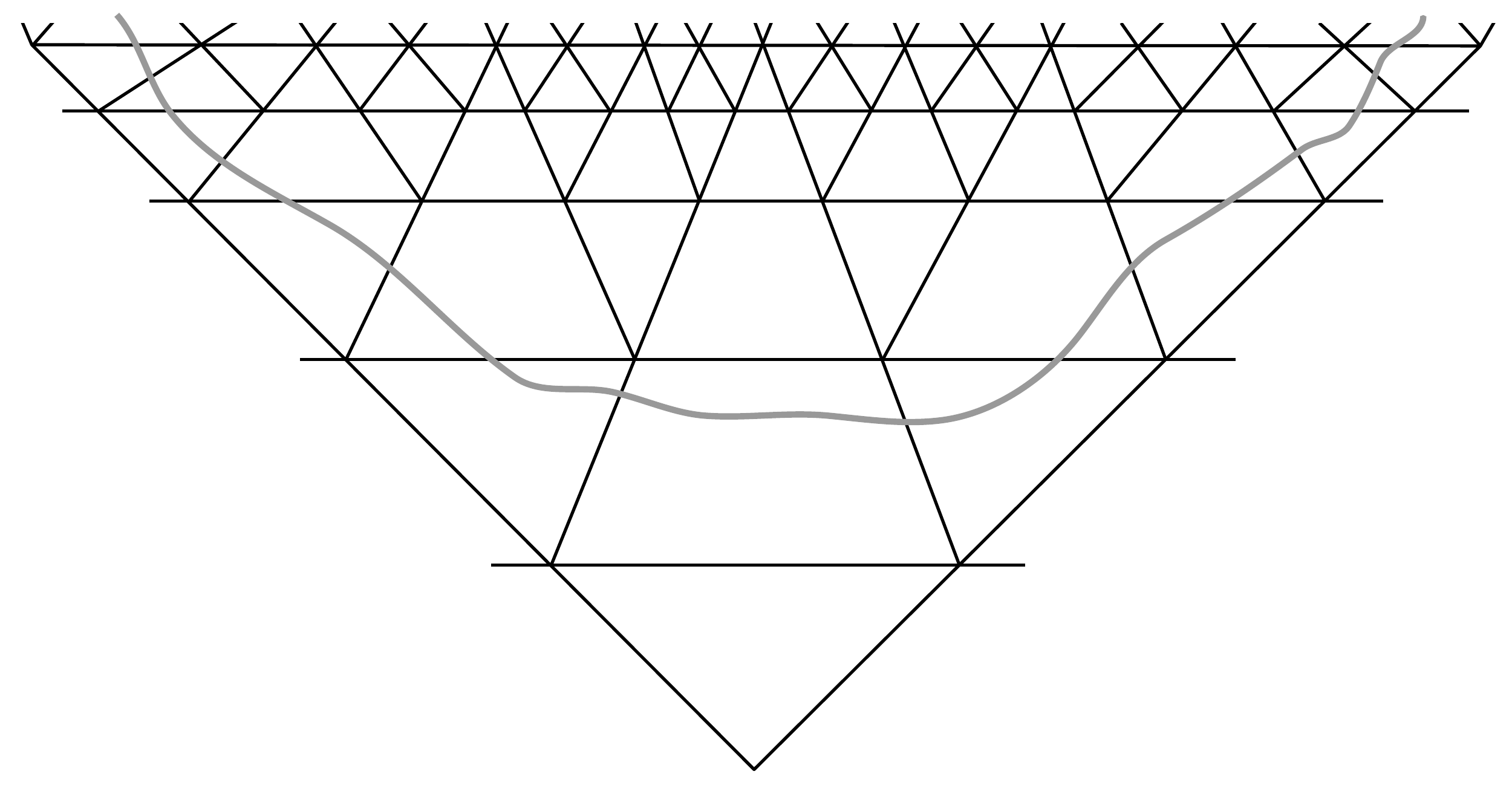}
\caption{\label{fig:MERA} {\bf MERA network} and a network geodesic that cuts a minimal number of links allowing it to enclose a region of the physical lattice at the top of the network. Each line between two points can be thought of as a maximally entangled pair in the auxiliary Hilbert space, while the points themselves correspond to projections that glue the pairs together. At the top of the network, the projectors map the auxiliary Hilbert space into the physical Hilbert space. The entanglement entropy of the physical region is bounded by the number of links cut by the geodesic: $S_{\mathrm{E}} \leq \ell \log D$.}
\end{figure}

For gapless systems such as CFTs, a more complicated structure of tensor contractions is necessary to capture the large amount of entanglement. A structure that realizes a discrete subgroup of the conformal group is the so-called MERA network, illustrated in figure \ref{fig:MERA}.
Viewing the network as obtained by projecting maximally entangled pairs, the entanglement bound discussed above for MPS states generalizes to the statement that the entanglement entropy of a sequence of adjacent sites is bounded above $\ell \log D$, where $\ell$ is the number of links in the network that must be cut to separate the entangled sites from the rest of the network. The strongest bound is found from the minimal number of such links that can be cut. This effectively defines a `minimal surface' within the network. As with the minimal surface appearing in the Ryu-Takayanagi formula, the entanglement entropy is proportional (if the bound is saturated) to the length of the surface. Such a surface in a MERA network is illustrated in figure \ref{fig:MERA}.

The similarity between the tensor network computation of the entanglement entropy and the Ryu-Takayangi formula suggests that the extended tensor network needed to capture the entanglement of gapless states can be thought of as an emergent geometry, analogous to that arising in holography \cite{Swingle:2009bg}. From this perspective, the emergent radial direction is a consequence of the large amount of entanglement in the QFT ground state, and the tensor network describing the highly entangled state is the skeleton of the bulk geometry.  Further discussion may be found in \cite{Swingle:2009bg, Swingle:2012wq, Pastawski:2015qua}. A key challenge for future work is to show how the large $N$ limit can flesh out this skeleton to provide a local geometry at scales below the AdS radius.

\subsection{Microscopics: Kaluza-Klein modes and consistent truncations}
\label{sec:consistent}

This section may be skipped by readers of a more pragmatic bent. The following discussion is however conceptually and practically important for understanding the simplest explicit holographic dual pairs, such as those in Table \ref{tab:holog} above.

In our discussion of the gap in the operator spectrum in \S\ref{sec:essential} we glossed over an important issue. In many explicit examples of holographic duality, e.g. those in Table \ref{tab:holog}, the bulk spacetime has additional spatial dimensions relative to the dual QFT, beyond the `energy scale' dimension discussed in previous sections. These dimensions are referred to as the `internal space'. For instance in the $\mathrm{AdS}_5 \times \mathrm{S}^5$ spacetime (\ref{eq:ads5s5}), the five sphere is the internal space. We can see that near the boundary $r \to 0$, the five sphere remains of constant size while the $\mathrm{AdS}_5$ part of the metric becomes large. This means that the sphere is not part of the boundary dimensions. Here is the complication: while the statement in (\ref{eq:gap}) that there is a small number of bulk fields as $\lambda \to \infty$ remains true, these fields can have an arbitrary dependence on the internal dimensions. Thus, for every field $\phi$ in the full spacetime we obtain an infinite tower of fields in $\mathrm{AdS}_5$, one for each `spherical harmonic' on the internal space. These are called Kaluza-Klein modes and their dimensions do not show a gap (because $L M_\text{KK} \sim L/L \sim 1$). That is, while the large $\lambda$ limit does get rid of many operators (those corresponding to excited string states in the bulk), it does not remove the infinitely many Kaluza-Klein modes. These modes are mostly a nuisance from the point of view of condensed matter physics. While emergent locality in the radial direction -- explored in the previous two sections -- usefully geometrized the renormalization group, emergent locality in the internal space leads to the QFT growing undesired extra spatial dimensions at low energy scales. In this section we discuss these modes and how to avoid them.

The most compelling strategy would be to get rid of the internal space completely. This is difficult because the best understood consistent bulk theories of quantum gravity require ten or eleven spacetime dimensions. However, `non-critical' string theories -- defined directly in a lower number of spacetime dimensions -- do exist. In fact, non-critical string theory lead to early intuition about the existence of holographic duality \cite{Polyakov:1998ju}. Holographic backgrounds of non-critical string theory have been considered in, for instance \cite{Kuperstein:2004yk, Klebanov:2004ya, Bigazzi:2005md}. The technical difficulty with these solutions is that while Kaluza-Klein modes can be eliminated or reduced, the string scale is typically comparable to the AdS radius (i.e. the 't Hooft coupling $\lambda \sim 1$) and therefore the infinite tower of excited string states must be considered -- undoing our motivation for considering these models in the first place.

The second best way to get rid of the Kaluza-Klein modes would be to find backgrounds in which the internal space is parametrically smaller than the AdS radius (unlike, e.g. the $\mathrm{AdS}_5 \times \mathrm{S}^5$ spacetime (\ref{eq:ads5s5}), in which the $\mathrm{AdS}_5$ and the internal space have the same radius of curvature). This way, all the Kaluza-Klein modes acquire a large bulk mass and can be mostly ignored (i.e. $L M_\text{KK} \sim L/L_\text{internal} \gg 1$). The necessary tools to find backgrounds with a hierarchy of lengthscales have been developed in the context of the `string landscape' \cite{Denef:2007pq}. A first application of these methods to holography can be found in \cite{Polchinski:2009ch}. In our opinion this direction of research is underdeveloped relative to its importance.

The most developed method for getting rid of the Kaluza-Klein modes is called consistent truncation. Here, one shows that it is possible to set all but a finite number of harmonics on the internal space to zero and that these harmonics are not sourced by the handful of modes that are kept. Because the equations of motion in the bulk are very nonlinear, and typically all modes are coupled, it can be technically difficult to identify a consistent set of modes that may be retained. While symmetries can certainly help, a general method does not exist -- finding consistent truncations is something of an art form. While less satisfactory than the other approaches mentioned above, consistent truncation gives a manageable bulk action with only one extra spatial dimension relative to the dual QFT. Furthermore, the dynamics of this action consistently captures the dynamics of a subset of operators in the full theory. An important drawback of consistent truncations is that they can miss instabilities in modes that have been thrown out in the truncation. They may, therefore, mis-identify the ground state.

Classical instabilities arising from Kaluza-Klein modes can potentially be avoided by in constructions built using `baryonic' branes \cite{Herzog:2009gd}. More generally, string theoretic realizations of holographic backgrounds may be subject to additional quantum instabilities. It has been suggested that all non-supersymmetric holographic flux vacua are unstable \cite{Ooguri:2016pdq}.

We proceed to give a list of simple examples of bulk actions that can be obtained by consistent truncation, referring to the literature for details. These examples have been chosen (from a large literature) to illustrate and motivate the kind of bulk dynamics we will encounter later in this review.  These constructions lead to so-called ``top-down holography" and are in contrast to ``bottom-up" holography, wherein one postulates a reasonable bulk action with convenient properties. The objective of bottom-up holography is to explore the parameter space of possible strong coupling dynamics. It is useful to have a flavor for what descends from string theory and supergravity to build a sensible bottom-up model.

\begin{widetext}\begin{enumerate}

\item {\bf Einstein-Maxwell theory:} The four dimensional bulk action
\be\label{eq:S4}
S_{\text{EM4}} = \int \mathrm{d}^4x \sqrt{-g} \left[ \frac{1}{2 \kappa^2} \left(R + \frac{6}{L^2} \right) - \frac{1}{4 e^2} F^2 \right] \,,
\ee
is a consistent truncation of $M$ theory compactified on any Sasaki-Einstein seven-manifold, see e.g.
\cite{Herzog:2007ij, Gauntlett:2007ma, Denef:2009tp}. Therefore this action captures the dynamics of a sector of a large class of dual $2+1$ dimensional QFTs. The simplest Sasaki-Einstein seven-manifold is $\mathrm{S}^7$. The bulk Maxwell field, $F = \mathrm{d}A$, is dual to a particular $\mathrm{U}(1)$ symmetry current operator $J^\mu$, called the R-symmetry, in the dual superconformal field theory. Similarly, the five dimensional action
\be\label{eq:S5}
S_{\text{EM5}} = \int \mathrm{d}^5x \sqrt{-g} \left[ \frac{1}{2 \kappa^2} \left(R + \frac{12}{L^2} \right) - \frac{1}{4 e^2} F^2 \right] 
+ \frac{\kappa}{\sqrt{6} \, e^3} \frac{1}{3!} \int \mathrm{d}^5x \epsilon^{abcde} A_a F_{bc} F_{de} \,,
\ee
is a consistent truncation of IIB supergravity compactified on any Sasaki-Einstein five-manifold, e.g. \cite{Gauntlett:2007ma}. This theory describes a sector of a large number of dual $3+1$ dimensional QFTs. The simplest Sasaki-Einstein five-manifold is $\mathrm{S}^5$, and so these theories include ${\mathcal N} = 4$ SYM theory. The additional Chern-Simons term in the above action is common in five bulk spacetime dimensions and dually describes an anomaly in the global $\mathrm{U}(1)$ R-symmetry of the QFT \cite{Witten:1998qj}.

Einstein-Maxwell theory is the workhorse for much of holographic condensed matter physics.

\item {\bf Einstein-Maxwell theory with charged scalars:} The above consistent truncations to Einstein-Maxwell theory by compactification on Sasaki-Einstein manifolds may be extended to include additional bulk fields. In particular the five dimensional action (\ref{eq:S5}) can be extended to \cite{Gubser:2009qm, Gubser:2009gp}
\be
S = S_\text{EM5} - \frac{1}{2 \k^2} \int \mathrm{d}^5x \sqrt{-g} \left[ \frac{1}{2} \left(\nabla \eta \right)^2 + V(\eta) + f(\eta) \left(\nabla \theta - \frac{\sqrt{6} \k}{e L} A \right)^2 \right] \,.
\ee
Here $\eta$ and $\theta$ are the magnitude and phase of a complex scalar field that is charged under the Maxwell field $A$. The two functions appearing in the action are
\be
V(\eta) = \frac{3}{L^2} \sinh^2 \frac{\eta}{2}\left(\cosh \eta - 3 \right) \,, \qquad f(\eta) = \frac{1}{2} \sinh^2\eta \,.
\ee
The expression for $V$ is slightly different compared to that in \cite{Gubser:2009qm, Gubser:2009gp} because we have extracted the cosmological constant term in (\ref{eq:S5}). The complex scalar field is dual to a particular charged scalar operator $\ocal$ in the dual QFT \cite{Gubser:2009qm}. Similarly the four dimensional action (\ref{eq:S4}) can be extended to \cite{Denef:2009tp, Gauntlett:2009dn, Gauntlett:2009bh, Donos:2012yu}
\bea
\lefteqn{S = \frac{1}{2 \k^2} \int \mathrm{d}^4x \sqrt{-g} \Bigg[ R - \frac{1}{2} \left(\nabla \sigma \right)^2 - \frac{\tau(\sigma)}{4 e^2} F^2} \nonumber \\
&& - \frac{1}{2} \left(\nabla \eta \right)^2 - V(\sigma,\eta) - f(\eta) \left(\nabla \theta - \frac{1}{2eL} A \right)^2 + \frac{1}{8 e^2}  \vartheta(\sigma) \epsilon^{abcd} F_{ab} F_{cd}\Bigg] \,. \label{eq:4dsc}
\eea
Actually this action is not quite an extension of (\ref{eq:S4}) -- the Sasaki-Einstein compactification needs to be `skew-whiffed', as described in the above references.
This theory contains a neutral pseudoscalar field $\sigma$ in addition to the charged scalar $\{\eta,\theta\}$. The functions in the action are now
\be
\tau(\sigma) = \frac{1}{\cosh \sqrt{3} \sigma} \,, \qquad \vartheta(\sigma) = \tanh \sqrt{3} \sigma \,, \qquad
f(\eta) = 2 \sinh^2 \eta \,,
\ee
and
\be
V(\sigma,\eta) = - \frac{6}{L^2} \cosh \frac{\sigma}{\sqrt{3}} \cosh^4 \frac{\eta}{2} \left[1 - \frac{4}{3} \tanh^2\frac{\eta}{2} \cosh^2\frac{\sigma}{\sqrt{3}} \right] \,.
\ee
Note that in a background with both electric and magnetic charges, the pseudoscalar is sourced by the final term in the action and hence cannot be set to zero in general. On the other hand we can set $\eta = 0$ and remove the complex scalar.

Charged scalars in the bulk will play a central role in the theory of holographic superconductivity.

\item {\bf Einstein-Maxwell theory with neutral scalars:} Both of the next two examples are taken from \cite{Cvetic:1999xp, Gubser:2009qt}. The five dimensional bulk action
\be
S = \frac{1}{2 \kappa^2} \int \mathrm{d}^5x \sqrt{-g} \left[  R  - \frac{e^{4 \alpha}}{4} F^2 - 12 \left(\nabla \alpha \right)^2 + \frac{1}{L^2} \left(8 e^{2 \alpha} + 4 e^{- 4 \alpha} \right) \right] \,,
\ee
is a consistent truncation of IIB supergravity compactified on $\mathrm{S}^5$, and therefore describes a sector of ${\mathcal N}=4$ SYM theory. Here $\alpha$ is a neutral scalar field. Similarly the four dimensional action
\be
S = \frac{1}{2 \kappa^2} \int \mathrm{d}^4x \sqrt{-g} \left[  R  - \frac{e^{\alpha}}{4} F^2 - \frac{3}{2} \left(\nabla \alpha \right)^2 + \frac{6}{L^2} \cosh \alpha \right] \,,
\ee
is a consistent truncation of M-theory compactified on $\mathrm{S}^7$, provided $\epsilon^{abcd} F_{ab} F_{cd} = 0$. If this last condition does not hold, an additional pseudoscalar must be included -- similarly to in (\ref{eq:4dsc}) above -- and the action becomes more complicated \cite{Cvetic:1999xp}.

Exponential functions of scalars multiplying the Maxwell action are ubiquitous in consistent truncations and will underpin  our discussions of Lifshitz geometries and hyperscaling violation.

\end{enumerate}
\end{widetext}

The interested reader will find many more instances of consistent truncations by looking through the references in and citations to the papers mentioned above. It is also possible to find consistent truncations in which the four and five bulk dimensional actions above (and others) are coupled to fermionic fields. The resulting actions are a little complicated and we will not give them here, see \cite{Gauntlett:2011mf, Belliard:2011qq, Gauntlett:2011wm, DeWolfe:2011aa, DeWolfe:2012uv, DeWolfe:2014ifa}. Bulk fermion fields will be important in our discussion below of holographic Fermi surfaces.

\section{Zero density matter}
\label{sec:zerodensity}

The simplest examples of systems without quasiparticle excitations are realized in quantum matter at special densities which are commensurate with an underlying lattice \cite{ssbook}. These are usually modeled by quantum field theories which are particle-hole symmetric, and have a vanishing value of a conserved U(1) charge linked to the particle number of the lattice model --- hence `zero' density. Even in the lattice systems, it is conventional to measure the particle
density in terms of deviations from such a commensurate density.

In this section we will first give some explicit examples of non-quasiparticle zero density field theories arising at quantum critical points in condensed matter systems. This will include an example with emergent gauge fields. We will then discuss how such quantum critical systems can arise holographically. Our discussion in this section will cover quantum critical scaling symmetries as well as deformations away from scaling by temperature and relevant operators. Charge dynamics in these zero density quantum critical theories will be the subject of \S \ref{sec:qctransport}.

\subsection{Condensed matter systems}
\label{sec:sit}

The simplest example of a system modeled by a zero density quantum field theory is graphene. This QFT is furthermore relativistic at low energies.\footnote{We will see shortly that Lorentz invariance is by no means essential for holography.} Graphene consists of a honeycomb lattice of carbon atoms, and the important electronic  excitations all reside on a single $\pi$-orbital
on each atom. `Zero' density corresponds here to a density of one electron per atom. A simple computation of the band structure of free
electrons in such a configuration yields an electronic dispersion identical to massless 2-component Dirac fermions, $\Psi$, which have a `flavor' index extending
over 4 values. The Dirac `spin' index is actually a sublattice index, while the flavor indices correspond to the physical $S=1/2$ spin and a `valley' index
corresponding to 2 Dirac cones at different points in the Brillouin zone.
The density of electrons can be varied by applying a chemical potential, $\mu$, by external gates, so that the dispersion
is $\epsilon ({\bf k}) = v_{\mathrm{F}} |{\bf k}| - \mu$, where $v_{\mathrm{F}}$ is the Fermi velocity \cite{netormp}. 
Electronic states with $\epsilon ({\bf k}) < 0$ are occupied at $T=0$,
and zero density corresponds to $\mu=0$. 

At first glance, it would appear that graphene is not a good candidate for holographic study. The interactions between the electrons 
are instantaneous, non-local Coulomb interactions, because the velocity of light $c \gg v_{\mathrm{F}}$. These interactions 
are seen to be formally irrelevant at low energies under the renormalization group, so that the electronic quasiparticles are well-defined. However, as we will briefly note later,
the interactions become weak only logarithmically slowly \cite{Guinea94,Fritz08}, and there is a regime with $T \gtrsim \mu$ where graphene can be modeled as a dissipative
quantum plasma.  Thus, insight from holographic models can prove quite useful.   Experimental evidence for this quantum plasma has emerged recently \cite{Crossno1058}, as we will discuss further in this review. More general electronic models
with short-range interactions on the honeycomb lattice provide examples of quantum critical points without quasiparticle excitations even at $T=0$,
and we will discuss these shortly below.

\begin{figure}
\centering
\includegraphics[height = 0.23\textheight]{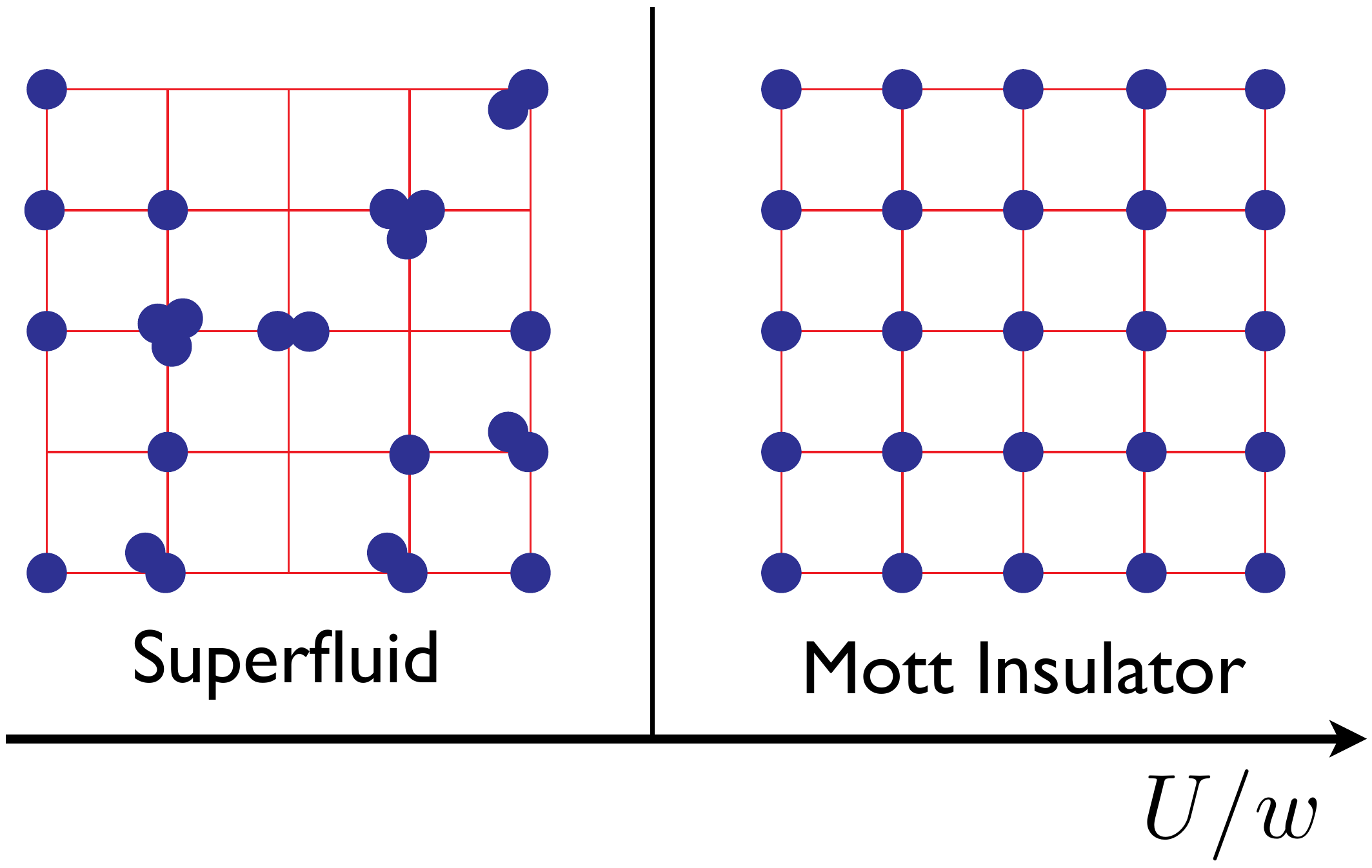}
\caption{\label{fig:bosehubbard} \textbf{Ground states of bosons on a square lattice with tunneling amplitude between the sites $w$, 
and on-site repulsive interactions $U$.} Bosons are condensed in the zero momentum state in the superfluid, and so there are large number
flucutations in a typical component of the wavefunction shown above. The Mott insulator is dominated by a configuration with exactly one particle
on each site.}
\end{figure}

For a more clear-cut realization of quantum matter without quasiparticles we turn to a system of ultracold atoms in optical lattices. 
Consider a collection of bosonic atoms (such as $^{87}$Rb) placed in a periodic potential, with a density such that there is precisely 
one atom for each minimum of the periodic potential. The dynamics of the atoms can be described by the boson Hubbard model:
this is a lattice model describing the interplay of boson tunneling between the minima of the potential (with amplitude $w$),
and the repulsion between two bosons on the same site (of energy $U$). For small $U/w$, we can treat the bosons as nearly free, 
and so at low temperatures $T$ they condense into a superfluid with macroscopic occupation of the zero momentum single-particle state, as shown in Figure \ref{fig:bosehubbard}.
In the opposite limit of large $U/w$, the repulsion between the bosons localizes them into a Mott insulator: this is a state adiabatically
connected to the product state with one boson in each potential minimum. We are interested here in the quantum phase transition between these
states which occurs at a critical value of $U/w$. Remarkably, it turns out that the low energy physics in the vicinity of this critical point
is described by a quantum field theory with an emergent Lorentz invariant structure, but with a velocity of `light', $c$, given by the velocity
of second sound in the superfluid \cite{ssbook,FWGF}. In terms of the long-wavelength boson annihilation operator $\psi$, the Euclidean time ($\tau$) 
action for the field theory is 
\be
S = \int \mathrm{d}^2 x \, \mathrm{d}\tau \left[c^2 |\nabla_x \psi |^2 +  |\partial_\tau \psi |^2 + s |\psi|^2 + u |\psi|^4 \right] \,. \label{Swf}
\ee
Here $s$ is the coupling which tunes the system across the quantum phase transition at some $s=s_{\mathrm{c}}$. For $s > s_{\mathrm{c}}$, the field theory
has a mass gap and no symmetry is broken: this corresponds to the Mott insulator. The gapped particle and anti-particle states associated with the 
field operator $\psi$ correspond to the `particle' and `hole' excitations of the Mott insulator. These can be used as a starting point
for a quasiparticle theory of the dynamics of the Mott insulator via the Boltzmann equation. 
The other phase with $s<s_{\mathrm{c}}$ corresponds to the superfluid: here the global U(1) symmetry of $S$ is broken, 
and the massless Goldstone modes correspond to the second sound excitations. Again, these Goldstone excitations are well-defined quasiparticles,
and a quasiclassical theory of the superfluid phase is possible (and is discussed in classic condensed matter texts). 

Our primary interest here is in the special critical point $s=s_{\mathrm{c}}$, which provides the first example of a many body quantum state without quasiparticle
excitations. And as we shall discuss momentarily, while this is only an isolated point at $T=0$, the influence of the non-quasiparticle description widens to a finite range of couplings at non-zero temperatures. Renormalization group analyses show that the $s=s_{\mathrm{c}}$ critical point is described
by a fixed point where $u \rightarrow u^\ast$, known as the Wilson-Fisher fixed point \cite{ssbook,WilsonFisher}. 
The value of $u^\ast$ is small in a vector large $M$ expansion in which $\psi$ is generalized to an $M$-component vector, or for small $\epsilon$ in a field theory generalized to $d=3-\epsilon$ spatial dimensions. Given only the assumption of scale invariance at such a fixed point, the two-point $\psi$ Green's function obeys
\be
G^{\mathrm{R}}_{\psi\psi} (\omega, k) = k^{-2 + \eta} F\left(\frac{\omega}{k^z} \right) \,, \label{Gscal}
\ee
at the quantum critical point. Here $\eta$ is defined to be the anomalous dimension of the field $\psi$ which has scaling dimension
\be
[ \psi ] = (d+z-2+\eta)/2 \,,
\ee
in $d$ spatial dimensions. The exponent $z$ determines the relative scalings of time and space, and is known as the dynamic critical exponent: $[t] = z[x]$. 
The function $F$ is a scaling function which depends upon the particular critical point under study. In the present situation, we know from the structure
of the underlying field theory in Eq.~(\ref{Swf}) that $G^{\mathrm{R}}_{\psi\psi}$ has to be Lorentz invariant ($\psi$ is a Lorentz scalar), and so we must have $z=1$
and a function $F$ consistent with
\be
G^{\mathrm{R}}_{\psi\psi} (\omega, k) \sim \frac{1}{(c^2k^2 - \omega^2)^{(2 -\eta)/2}} \,. \label{Geta}
\ee
Indeed, the Wilson-Fisher fixed point is not only Lorentz and scale invariant, it is also conformally invariant and realizes a conformal field theory (CFT),
a property we will exploit below.

A crucial feature of $G^{\mathrm{R}}$ in Eq.~(\ref{Geta}) is that for $\eta \neq 0$ 
its imaginary part does not have any poles in the $\omega$ complex plane, only branch cuts at
$\omega = \pm ck$. This is an indication of the absence of quasiparticles at the quantum critical point. However, it is not a proof of absence, because
there could be quasiparticles which have zero overlap with the state created by the $\psi$ operator, and which appear only in suitably defined observables.
For the case of the Wilson-Fisher fixed point, no such observable has ever been found, and it seems unlikely that such an `integrable' structure
is present in this strongly-coupled theory. Certainly, the absence of quasiparticles can be established at all orders in the $\epsilon$ or vector $1/M$ expansions. 
In the remaining discussion we will assume that no quasiparticle excitations are present, and develop a framework to describe the physical properties of such
systems. We also note in passing that CFTs in 1+1 spacetime dimensions ({\it i.e.\/} CFT2s) are examples of theories in which
there are in fact quasiparticles present, but whose presence is not evident in the correlators of most field variables \cite{Herzog:2007ij}. This is a consequence of  integrability properties in CFT2s which do not generalize to higher dimensions.

We can also use a scaling framework to address the ground state properties away from the quantum critical point. In the field theory $S$ we tune
away from the quantum critical point by varying the value of $s$ away from $s_{\mathrm{c}}$. This makes $s-s_{\mathrm{c}}$ a relevant perturbation at the fixed point, 
and its scaling dimension is denoted as
\be
[s-s_{\mathrm{c}}] = 1/\nu.
\ee
In other words, the influence of the deviation away from the criticality becomes manifest at distances 
larger than a correlation length, $\xi$,  which diverges as
\be
\xi \sim |s-s_{\mathrm{c}}|^{-\nu}. 
\ee
We can also express these scaling results in terms of the dimension of the operator conjugate to $s-s_{\mathrm{c}}$ in the action
\be
\left[|\psi|^2 \right] = d+z - \frac{1}{\nu}. \label{thermalop}
\ee
For $s>s_{\mathrm{c}}$, the influence of the relevant perturbation to the quantum critical point is particularly simple: the theory acquires a mass gap $\sim c/\xi$
and so
\be
G^{\mathrm{R}}_{\psi\psi} (\omega, k) \sim \frac{1}{(c^2 (k^2 + \xi^{-2}) - \omega^2)} \,, \label{Gxi}
\ee
at frequencies just above the mass gap. Note now that a pole has emerged in $G^{\mathrm{R}}$, confirming that quasiparticles are present in the $s>s_{\mathrm{c}}$ Mott insulator.

\begin{figure}[h]
\centering
\includegraphics[height = 0.27\textheight]{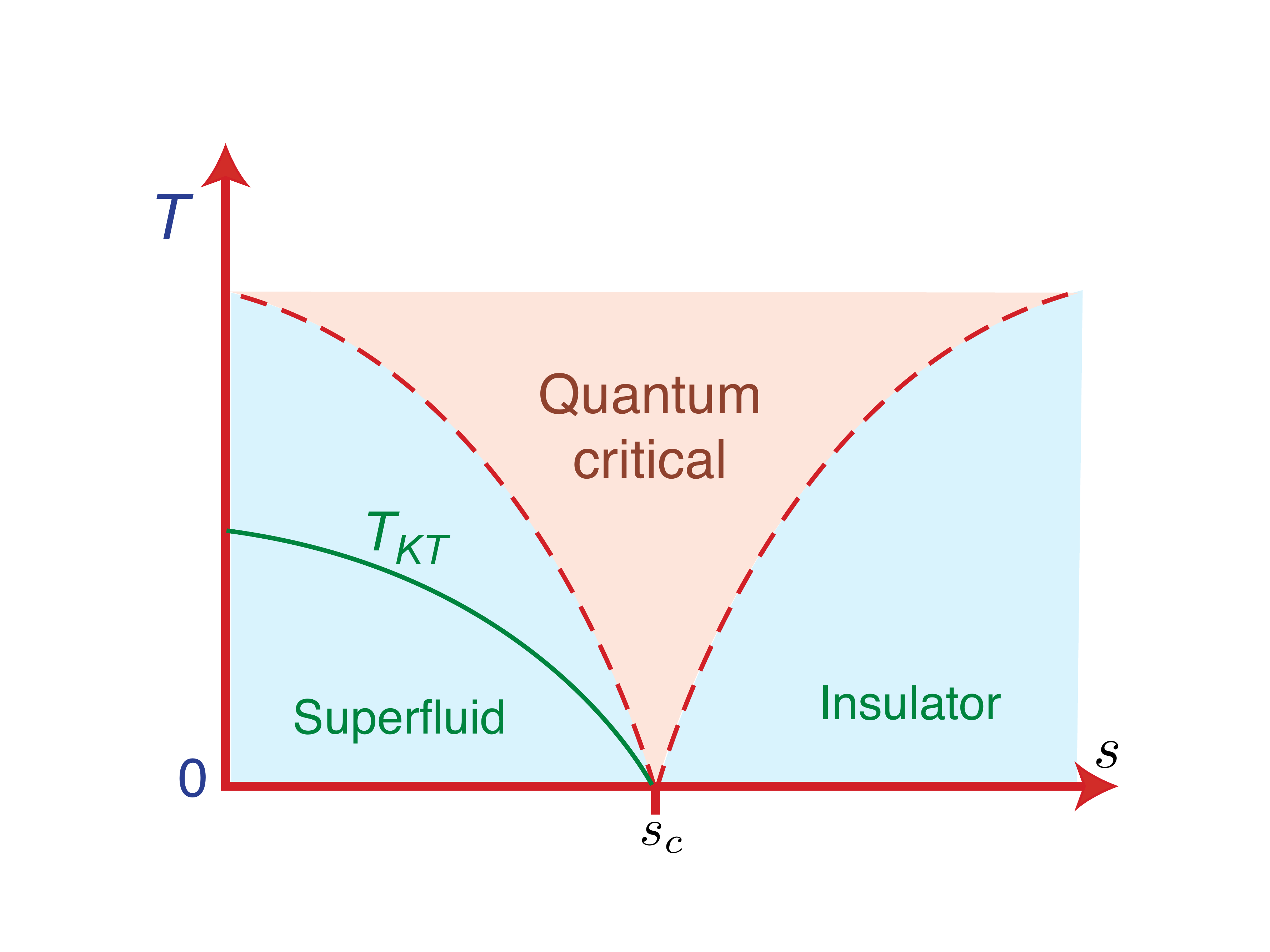}
\caption{\label{fig:qc} \textbf{Phase diagram of $S$ as a function of $s$ and $T$.} The quantum critical region is bounded by crossovers at $T \sim |s-s_{\mathrm{c}}|^{z \nu}$ indicated by the dashed lines. Conventional quasiparticle or classical-wave dynamics applies in the non-quantum-critical regimes 
(colored blue) including a Kosterlitz-Thouless phase transition above which the superfluid density is zero. One of our aims in this review is to develop a theory
of the non-quasiparticle dynamics within the quantum critical region.
}
\end{figure}

We now also mention the influence of a non-zero temperature, $T$, on the quantum critical point, and we will have much more to say about this later.
In the imaginary time path integral for the field theory, $T$ appears only in boundary conditions for the temporal direction, $\tau$: bosonic (fermionic) fields
are periodic (anti-periodic) with period $\hbar/(k_{\mathrm{B}} T)$. At the fixed point, this immediately implies that $T$ should scale as a frequency. So we can generalize 
the scaling form (\ref{Gscal}) to include a finite $\xi$ and a non-zero $T$ by
\be
G^{\mathrm{R}}_{\psi\psi} (\omega, k) = \frac{1}{T^{(2-\eta)/z}} \mathcal{F}\left( \frac{k}{T^{1/z}}, \frac{\xi^{-1}}{T^{1/z}}, \frac{\hbar \omega}{k_{\mathrm{B}} T} \right) .
\label{Gscal2}
\ee
We have included fundamental constants in the last argument of the scaling function $\mathcal{F}$
because we expect the dependence on $\hbar \omega/(k_{\mathrm{B}} T)$ to be fully 
determined by the fixed-point theory, with no arbitrary scale factors. Note also that for $T > \xi^{-z} \sim |g-g_c|^{z \nu}$, we can safely set
the second argument of $\mathcal{F}$ to zero. So in this ``quantum critical'' regime, the finite temperature dynamics is described by the
non-quasiparticle dynamics of the fixed point CFT3: see Figure \ref{fig:qc}. 
In contrast, for $T < \xi^{-z} \sim |g-g_c|^{z \nu}$ we have the 
traditional quasiparticle dynamics associated with excitations similar to those described by (\ref{Gxi}). Our discussion in the next few sections will focus on the quantum-critical region of Figure \ref{fig:qc}. 

In the remainder of this subsection, we note  extensions to other examples of zero density systems without quasiparticles. 

\subsubsection{Antiferromagnetism on the honeycomb lattice}
\label{sec:honeycomb}

\begin{figure}
\centering
\includegraphics[height = 0.32\textheight]{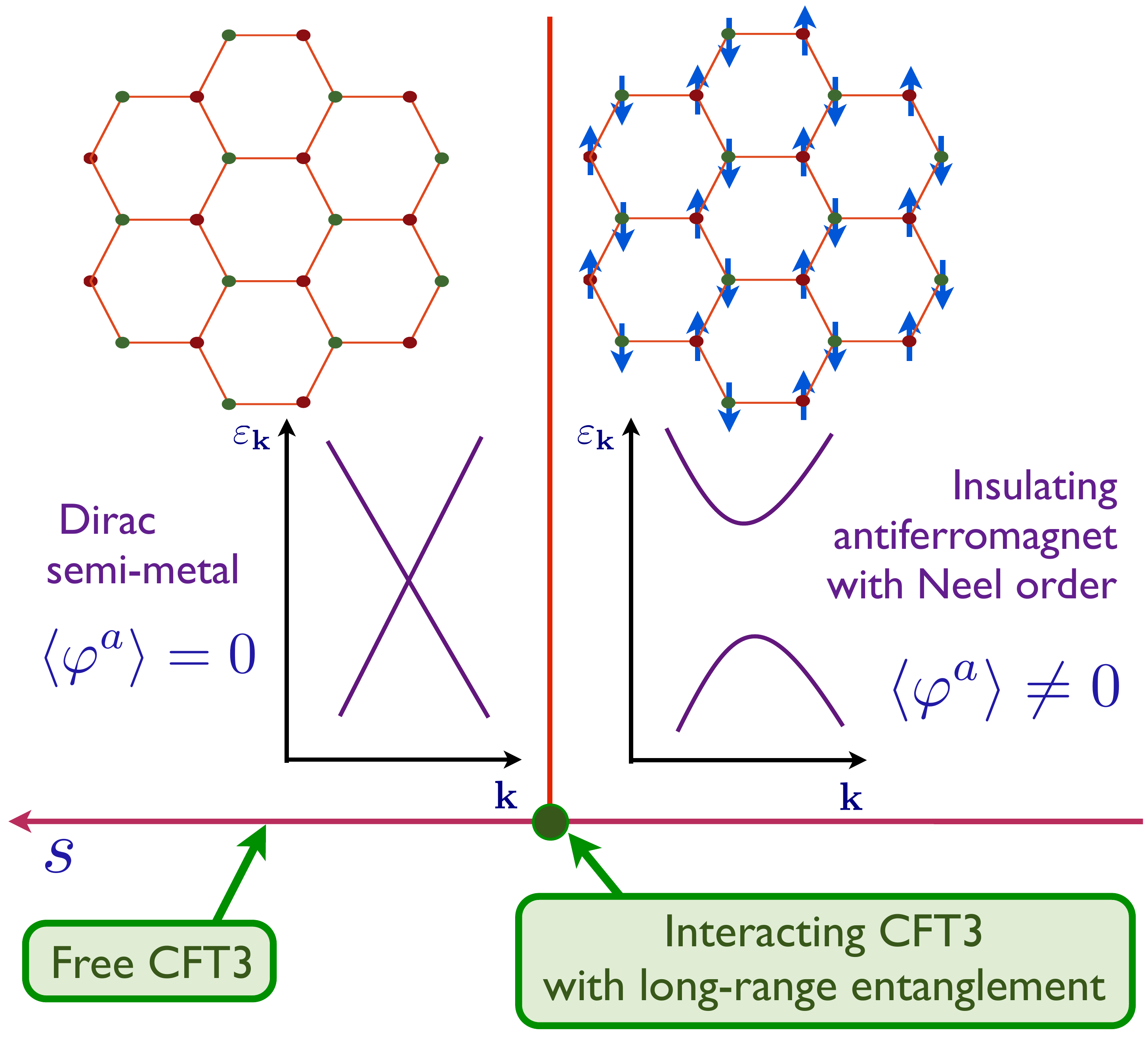}
\caption{\label{fig:honeycomb} \textbf{Phase diagram of the Hubbard model for spin $S=1/2$ fermions on the honeycomb lattice
at a density of one fermion per site.} The large $U/w$ ($s < s_{\mathrm{c}}$) state breaks spin rotation symmetry, and is an insulating
antiferromagnet with a gap to all charged excitations. The small $U/w$ ($s> s_{\mathrm{c}}$) state is described at low energies by a 
CFT3 of free Dirac fermions.}
\end{figure}

We return to electronic models defined on the honeycomb lattice, relevant for systems like graphene, mentioned at the
beginning of \S \ref{sec:sit}. In particular, we consider a Hubbard model for spin $S=1/2$ fermions (`electrons') on the honeycomb lattice at a density of one fermion per site. As for the boson Hubbard model, the two energy scales are the repulsion
energy, $U$, between two electrons on the same site (which necessarily have opposite spin), and the electron hopping amplitude $w$ between nearest-neighbor sites. For small $U/w$, the electronic spectrum is that of 4 flavors of massless 
2-component Dirac fermions, $\Psi$, and the short-range interactions only weakly modify the free fermion spectrum.
In contrast, for large $U/w$, the ground state is an insulating antiferromagnet, as illustrated in Figure \ref{fig:honeycomb}.
This state is best visualized as one with fermions localized on the sites, with opposite spin orientations on the two
sublattices. The excitations of the insulating antiferromagnet come in two varieties, but both are amenable to a quasiparticle
description: ({\em i\/}) there are the bosonic Goldstone modes associated with the breaking of spin rotation symmeter, and 
({\em ii\/}) there are fermionic quasiparticle excitations with the same quantum numbers as the electron above an energy gap.

The quantum field theory for the phase transition \cite{honeycomb15} between the large $U/w$ and small $U/w$ phases is known in the particle-theory
literature as the Gross-Neveu model (where it is a model for chiral symmetry breaking). Its degrees of a freedom are a relativistic
scalar $\varphi^a$, $a=1,2,3$ representing the antiferromagnetic order parameter, and the Dirac fermions $\Psi$. The action
for $\varphi^a$ is similar to that the superfluid order parameter $\psi$ in (\ref{Swf}), and this is coupled to
the Dirac fermions via a `Yukawa' coupling, leading to the field theory
\begin{align}
S_\varphi &= \int \mathrm{d}^2 x \, \mathrm{d}\tau \left[ c^2(\partial_x \varphi^a )^2 + (\partial_\tau \varphi^a )^2 + s (\varphi^a)^2 \right. \notag \\
&+ \left. u \left((\varphi^a)^2\right)^2 + \mathrm{i} \overline{\Psi} \gamma^\mu \partial_\mu \Psi + \lambda \overline{\Psi} \Gamma^a \Psi \, \varphi^a \right] \,. \label{Sgn}
\end{align}
Here $\gamma^\mu$ are suitable $2 \times 2$ Dirac matrices, while the $\Gamma^a$ are `flavor matrices' which act
on a combination of the valley and sublattice indices of the fermions. A Dirac fermion
`mass' term is forbidden by the symmetry of the honeycomb lattice.
We have chosen units to that the Fermi velocity, $v_{\mathrm{F}} = 1$, and then the
boson velocity, $c$, cannot be adjusted.
The $\gamma^\mu$ and the $\Gamma^a$ turn
out to commute with each other, and so when the velocity $c=v_{\mathrm{F}}=1$, the action $S_\varphi$ becomes invariant under Lorentz transformations. The equal velocities are dynamically generated under the renormalization group flow, and so we find an emergent Lorentz invariance near the quantum critical point,
similar to that found above for the superfluid-insulator transition for bosons. The tuning parameter across the quantum critical point, $s$, is now a function of $U/w$. The phase with broken spin-rotation symmetry ($s < s_{\mathrm{c}}$) has $\left\langle \varphi^a \right\rangle \neq 0$, we then see that the Yukawa term, $\lambda$, endows the Dirac fermions with a `mass' {\em i.e.\/} there is a gap
to fermionic excitations; this `mass' is analogous to the Higgs boson endowing the fermions with a mass in the Weinberg-Salam model.  The quantum critical point at $s=s_{\mathrm{c}}$ can be analyzed by renormalization group and other
methods, as for the boson-only theory $S$ in Eq.~(\ref{Swf}). In this manner, we find that the 
quantum critical theory is a CFT, described by a renormalization group fixed at $u=u^\ast$ and $\lambda = \lambda^\ast$. 
This is a state without quasiparticle excitations, with Green's functions for $\varphi^a$ and $\Psi$ similar to those
of the other CFT3s noted above.

\subsubsection{Quadratic band-touching and $z \neq 1$}
\label{sec:qbt}

This subsection briefly presents an example of a system at commensurate density with dynamic
critical exponent $z \neq 1$; all other models described in the current \S \ref{sec:sit} have an emergent
relativistic structure and $z=1$. 
We consider a semiconductor in which the parabolic conduction and valence bands touch
quadratically at a particular momentum in the Brillouin zone: such band-touching
can arise generically in materials with strong spin-orbit coupling \cite{PhysRev.102.1030}. 
Without doping, the electron chemical potential is precisely at the touching point, and so the free
fermion Hamiltonian has a scale invariant structure with $z=2$. In $d=3$, the long-range Coulomb
interaction is a relevant perturbation to the free fermion fixed point \cite{Abrikosov1,Abrikosov2}, and its effects can
be described by the action \cite{2013PhRvL.111t6401M}
\begin{align}
S_{\mathrm{qbt}} &= \int \mathrm{d}^3 x \mathrm{d} \tau \Bigl[ \psi^\dagger \left( \partial_\tau - \mathrm{i} e \varphi + 
\mathcal{H}_0 (-\mathrm{i} \vec{\nabla}) \right) \psi  \notag \\
& \;\;\;\; + c_0 \left( \vec{\nabla}_i \varphi \right)^2 \Bigr],
\end{align}
where $\psi$ is the fermion operator for both the conduction and valence bands, $\mathcal{H}_0 (-\mathrm{i} \vec{\nabla})$ contains the quadratic dispersion of both bands, and the scalar field mediates the 
Coulomb interaction. A renormalization group analysis \cite{2013PhRvL.111t6401M} shows that the
coupling $e$ flows to a non-trivial fixed point with $z$ smaller than 2. A recent experiment \cite{2015NatCo...610042K} on 
the pyrochlore iridate Pr$_2$Ir$_2$O$_7$ displays evidence for this non-trivial critical behavior.

\subsubsection{Emergent gauge fields}
\label{sec:emerge1}

\begin{figure}
\centering
\includegraphics[height = 0.16\textheight]{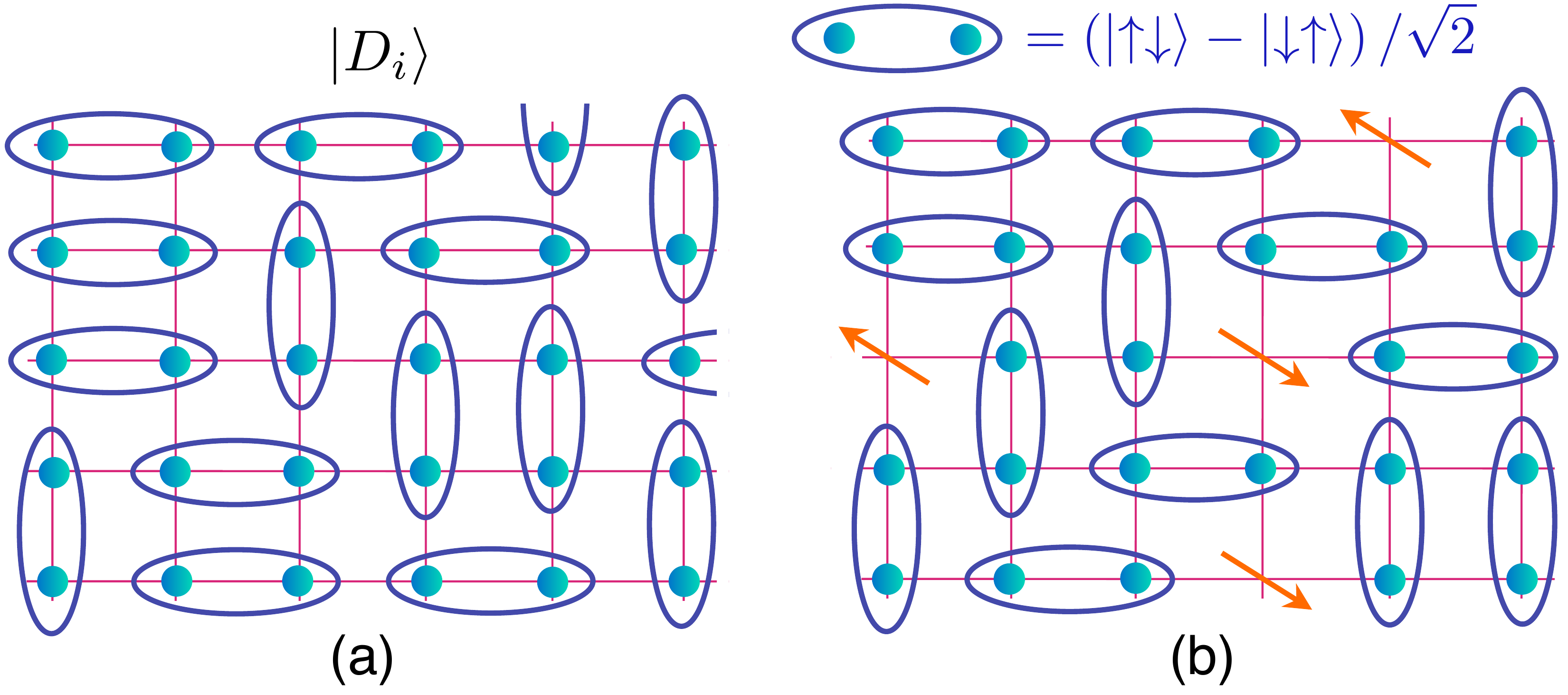}
\caption{\label{fig:rvb} \textbf{RVB states and excitations.} (a) Sketch of a component of a resonating valence bond wavefunction on the square lattice.
(b) Excited state with neutral excitations carrying spin $S=1/2$. In the field theory $\mathcal{S}_z$ each excitation is a quantum of the $z_\alpha$
particle, while the gauge field $A$ represents the fluctuations of the valence bonds (see Fig.~\ref{fig:qed}).}
\end{figure}

\begin{figure}
\centering
\includegraphics[height = 0.2\textheight]{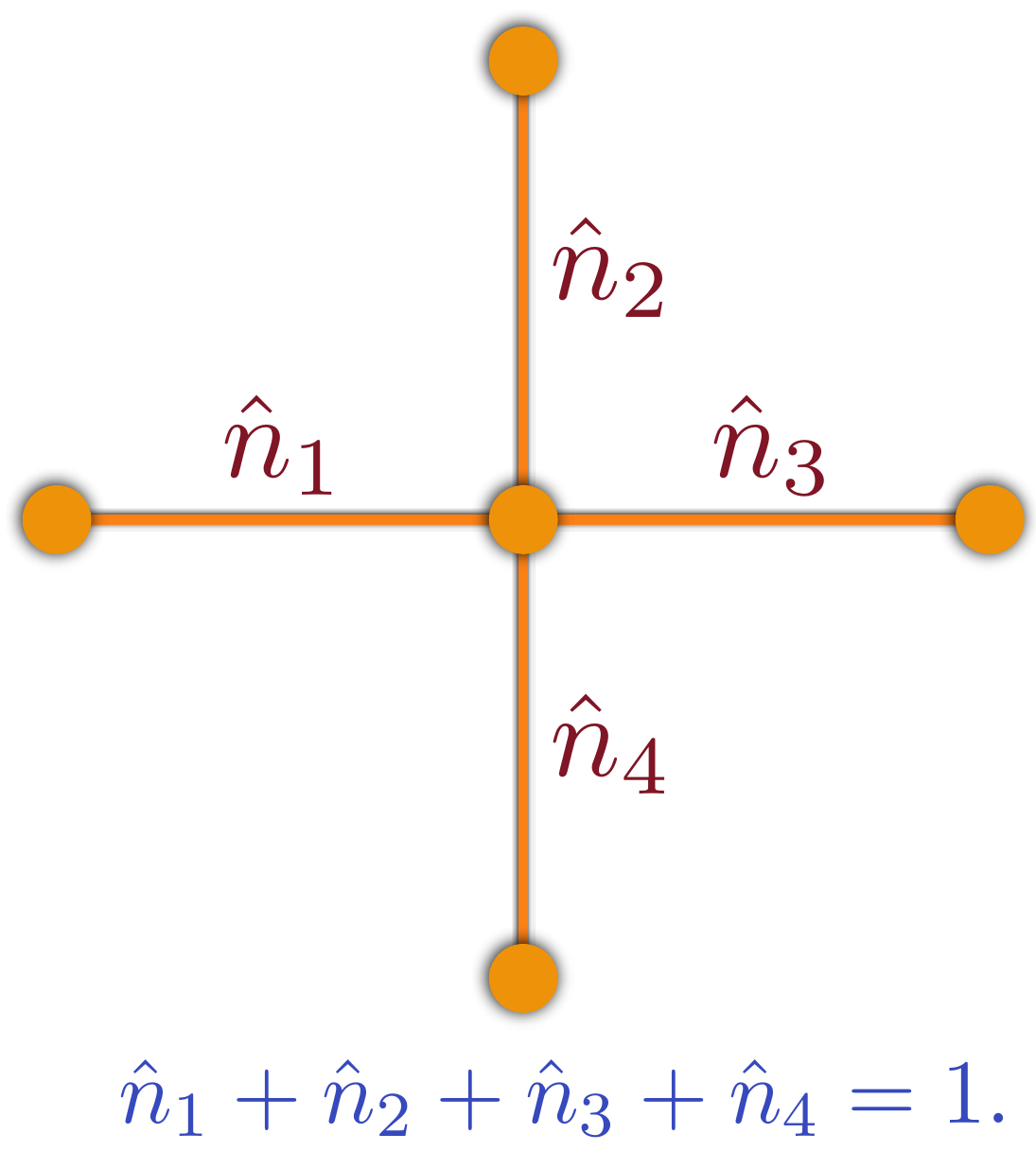}
\caption{\label{fig:qed} \textbf{Numbers operators}, $\hat{n}$, counting the number of singlet valence bonds on each link of the square lattice; here we have numbered links by
integers, but in the text we label them by the sites they connect.
Modulo a phase factor, these operators realize the electric field operator of a compact U(1) lattice gauge theory.}
\end{figure}

Condensed matter models also lead to field theories which have fluctuating dynamic gauge fields, in contrast to those in 
Eq.~(\ref{Swf}) and (\ref{Sgn}). These arise most frequently in theories of quantum antiferromagnets, and can be illustrated
most directly using the `resonating valence bond' (RVB) model \cite{Pauling49,Anderson73}. 
Fig.~\ref{fig:rvb}a shows a component, $\left| D_i \right\rangle$, of a wavefunction
of an antiferromagnet on the square lattice, in which nearby electrons pair up in to singlet valence bonds---the complete wavefunction is a superposition of numerous components with different choices of pairing between the electrons:
\beq
\left| \Psi \right\rangle = \sum_i c_i \left| D_i \right\rangle \,,
\eeq
where the $c_i$ are unspecified complex numbers.
In a modern language, such
a resonating valence bond wavefunction was the first to realize topological order and 
long-range quantum entanglement \cite{Melko15}. In more practical terms, the
topological order is reflected in the fact that an emergent gauge fields is required to describe the low energy 
dynamics of such a state \cite{GBPWA88,EFSK90}. 
A brief argument illustrating the origin of emergent gauge fields is illustrated in Fig.~\ref{fig:qed}.
We label each configuration by a set of integers $\{ \hat{n} \}$ representing the number of singlet valence bonds 
on each link. As each electron can form only one singlet bond, we have a local constraint on each site, as shown in Fig.~\ref{fig:qed}. Now, introducing an `electric field' operator on each link defined by $\hat{E}_{i \alpha} = 
(-1)^{i_x + i_y} \hat{n}_{i \alpha}$ (here $i \equiv (i_x, i_y)$ labels sites of the square lattice, and $\alpha = x, y$ labels the two directions; $\hat{E}_{i \alpha}$ and $n_{i \alpha}$ are on the 
link connecting $i$ to $i+ \hat{\alpha}$), this
local constraint can be written in the very suggestive form 
\beq
\Delta_\alpha \hat{E}_{i \alpha} = \rho_i , \label{eq:gauss}
\eeq
where $\Delta_\alpha$ is a discrete lattice derivative, and $\rho_i \equiv (-1)^{i_x + i_y}$ is a background `charge' density. 
Eq.~(\ref{eq:gauss}) is analogous to Gauss's Law in electrodynamics, and a key indication that the physics of resonating valence
bonds is described by an emergent compact U(1) gauge theory. 

\begin{figure}
\centering
\includegraphics[height = 0.30\textheight]{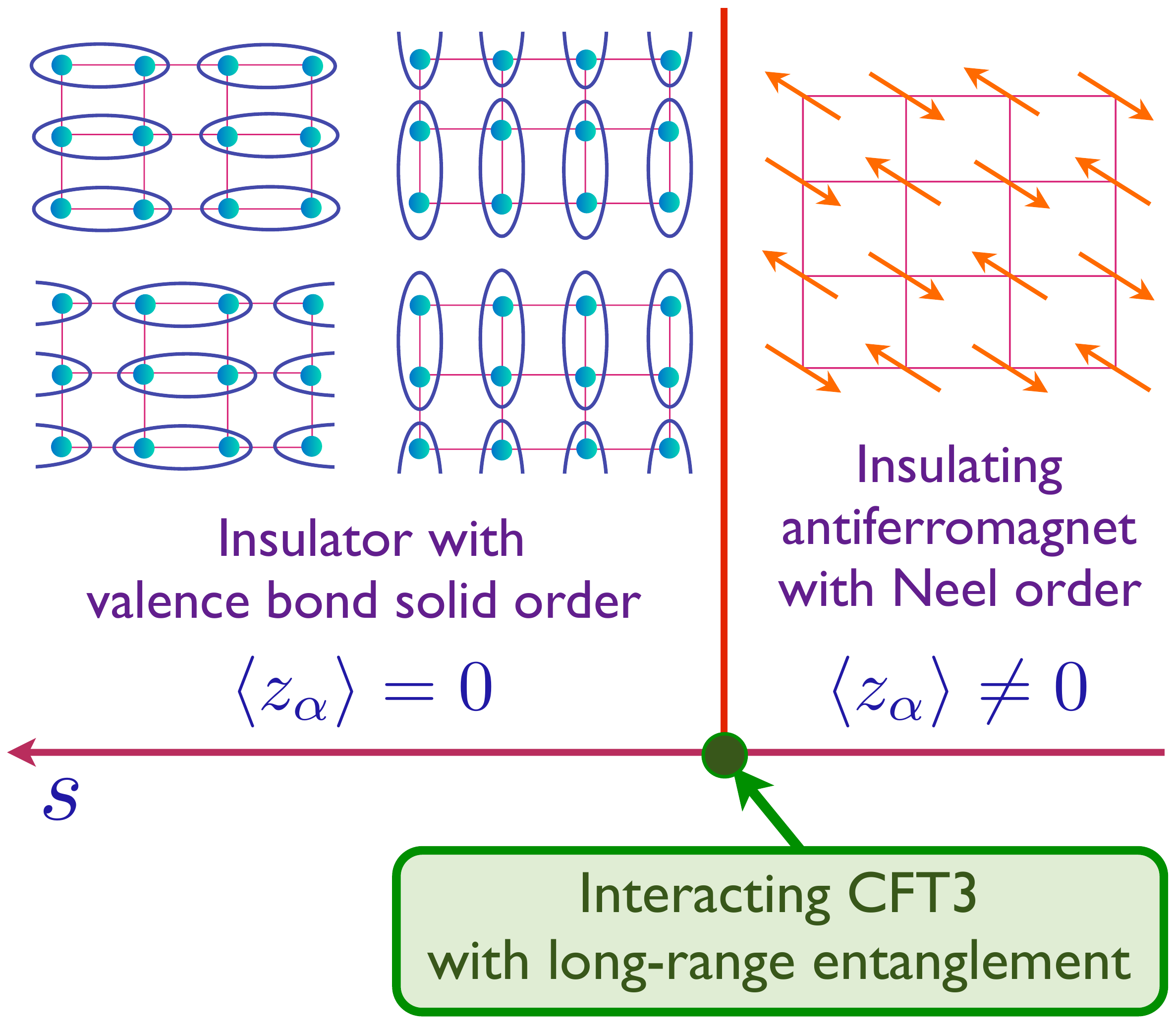}
\caption{\label{fig:square_afm} \textbf{Phases of a square lattice antiferromagnet.} The vicinity of the critical point is described by the theory $S_z$ in Eq.~(\ref{Sz})
(compare Fig.~\ref{fig:honeycomb}). The phase with N\'eel order is the Higgs phase of the gauge theory, while the Coulomb phase of $S_z$ is destabilized by monopoles, leading
to valence bond solid order.}
\end{figure}

CFTs with emergent gauge fields appear when we consider a quantum phase transition out of the resonating valence bond state into an ordered antiferromagnet with broken spin rotation symmetry. To obtain antiferromagnetic order, we need to consider a state in which some on the electrons are unpaired, as illustrated in Fig.~\ref{fig:rvb}b. For an appropriate spin liquid, these excitations behave like relativistic bosons, $z_\alpha$, at low energy; here $\alpha = \uparrow, \downarrow$ is the `spin' index, but note here that $\alpha$
does not correspond to spacetime spin, and behaves instead like a `flavor' index associated with the global SU(2) symmetry of the underlying lattice antiferromagnet. The quantum transition from the resonating valence bond state to the antiferromagnet
is described by the condensation of the $z_\alpha$ bosons \cite{NRSS89,NRSS90}, and the quantum critical point is proposed to be a CFT3 \cite{SVBSF}; see Fig.~\ref{fig:square_afm}.
Such a transition is best established for a set of quantum antiferromagnetics with a global SU($M$) symmetry, as reviewed recently by 
Kaul and Block \cite{2015arXiv150205128K}. These models are expressed in terms of SU($M$) spins, $\vec{S}$, residing on the sites of various bipartite lattices, with 
bilinear and biquadratic 
exchange interactions between nearest neighbor sites. The low energy physics in the vicinity of the critical point is described by a field theory with an emergent U(1) gauge field $A$ and the relativistic scalar $z_\alpha$ (now $\alpha = 1\ldots M$) 
\begin{align}
S_z &= \int \mathrm{d}^2 x \, \mathrm{d}\tau \left[ c^2 |(\nabla_x - \mathrm{i} A_x) z_\alpha |^2 +  |(\partial_\tau - \mathrm{i} A_\tau) z_\alpha |^2 \right. \notag \\
& \left. \;\;\;\; + s |z_\alpha|^2 + u \left(|z_\alpha|^2\right)^2 + \frac{1}{2 e^2} F^2 \right], \label{Sz}
\end{align}
where $F = \mathrm{d}A$. The order parameter for the broken symmetry in the antiferromagnet
involves a gauge-invariant bilinear of the $z_\alpha$. In the present case, it turns out that the resonating valence bond phase where the $z_\alpha$ are gapped is unstable to the proliferation of monopoles in the U(1) gauge field, which leads ultimately to the appearance of a broken lattice symmetry with `valence bond solid' (VBS) order (see Fig.~\ref{fig:square_afm}) \cite{NRSS89,NRSS90}.
However, the monopoles are suppressed at the quantum critical point at $s=s_{\mathrm{c}}$, and then
 $S_z$ realizes a conformal gauge theory \cite{SVBSF}: numerical studies of the lattice Hamiltonian are consistent with many
features of the $1/N$ expansion of the critical properties of the gauge theories. 

Our interest here is primarily on the non-quasiparticle dynamics of the 
$T>0$ quantum critical region near $s=s_{\mathrm{c}}$. Here, the scaling structure is very similar to our discussion of the Wilson-Fisher critical theory,
with the main difference being that the associated CFT is now also a gauge theory.

\subsection{Scale invariant geometries}
\label{sec:scaling}

We have already seen in \S\ref{sec:essential} above that the simplest solution to the most minimal bulk theory (\ref{eq:EH}) is the $\mathrm{AdS}_{d+2}$ spacetime
\be\label{eq:ads2}
\mathrm{d}s^2 = L^2 \left( \frac{- \mathrm{d}t^2 + \mathrm{d}\vec x^2_{d} + \mathrm{d}r^2}{r^2} \right) \,.
\ee
We noted that this geometry has the $\mathrm{SO}(d+1,2)$ symmetry of a $d+1$ (spacetime) dimensional CFT. We verified that a scalar field $\phi$ in this spacetime was dual to an operator $\ocal$ with a certain scaling dimension $\Delta$ determined by the mass of the scalar field via (\ref{eq:delta}). We furthermore verified that the retarded Green's function (\ref{eq:easyG}) of $\ocal$, as computed holographically, was indeed that of an operator of dimension $\Delta$ in a CFT. More generally, the dynamics of perturbations about the bulk background (\ref{eq:ads2}) gives a holographic description of zero temperature correlators in a dual CFT.

\subsubsection{Dynamical critical exponent $z>1$}

A CFT has dynamic critical exponent $z=1$. This section explains the holographic description of general quantum critical systems with $z$ not necessarily equal to $1$. The first requirement is a background that geometrically realizes the more general scaling symmetry: $\{t, \vec x \} \to \{\lambda^z t , \lambda \vec x \}$. The geometry
\be\label{eq:Lif}
\mathrm{d}s^2 = L^2 \left(- \frac{\mathrm{d}t^2}{r^{2z}} + \frac{\mathrm{d}\vec x^2_{d} + \mathrm{d}r^2}{r^2} \right) \,,
\ee
does the trick \cite{Kachru:2008yh}. These are called `Lifshitz' geometries because the case $z=2$ is dual to a quantum critical theory with the same symmetries as the multicritical Lifshitz theory. These backgrounds do not have additional continuous symmetries beyond the scaling symmetry, spacetime translations and spatial rotations. They are invariant under time reversal ($t \to -t$). We will shortly see that correlation functions computed from these backgrounds indeed have the expected scaling form (\ref{Gscal}). Firstly, though, we describe how the spacetime (\ref{eq:Lif}) can arise in a gravitational theory.

All Lifshitz metrics have constant curvature invariants. It can be proven that the only effects of quantum corrections in the bulk is to renormalize the values of $z$ and $L$ \cite{Adams:2008zk}. In this sense the scaling symmetry is robust beyond the large $N$ limit. However, for $z \neq 1$ these metrics suffer from so-called pp singularities at the `horizon' as $r \to \infty$. Infalling observers experience infinite tidal forces. See \cite{Hartnoll:2009sz} for more discussion. Such null singularities are likely to be acceptable within a string theory framework, see e.g. \cite{Horowitz:1997uc, Bao:2012yt}. So far, no pathologies associated with these singularities has arisen in classical gravity computations of correlators and thermodynamics.

Unlike Anti-de Sitter spacetime, the Lifshitz geometries are not solutions to pure gravity. Therefore, a slightly more complicated bulk theory is needed to describe them. Two simple theories that have Lifshitz geometries (\ref{eq:Lif}) as solutions are Einstein-Maxwell-dilaton theory \cite{Taylor:2008tg, Goldstein:2009cv} and Einstein-Proca theory \cite{Taylor:2008tg}. The Einstein-Proca theory couples gravity to a massive vector field:
\begin{equation}
S[g,A] = \frac{1}{2 \k^2} \int \mathrm{d}^{d+2}x \sqrt{-g} \left( R + \Lambda  - \frac{1}{4} F^2 -  \frac{m^2_A}{2} A^2 \right)
\end{equation}with 
\begin{subequations}\begin{align}
\Lambda &=  \frac{(d+1)^2 + (d+1)(z-2) + (z-1)^2}{L^2},  \\
m^2_A &= \frac{z \, d}{L^2}.
\end{align}\end{subequations}
For convenience, we have parametrized $\Lambda$ and $m_A^2$ in terms of $z$ and $L$, which will appear as parameters in simple scaling backgrounds of this theory, as will be seen shortly. In a microscopic bulk construction it would be $\Lambda$ and $m_A^2$ that are given, from which $z$ and $L$ are derived by finding the scaling solutions. It is easily verified that the equations of motion following from this action have (\ref{eq:Lif}) as a solution, together with the massive vector
\be
A = \sqrt{\frac{2(z-1)}{z}} \frac{L \, \mathrm{d}t}{r^z} \,.
\ee
This is only a real solution for $z \geq 1$. We will discuss physically allowed values of $z$ later.

Alternatively, the Einstein-Maxwell-dilaton theory
\begin{align}
S[g,A,\phi] &=\frac{1}{2 \k^2} \int \mathrm{d}^{d+2}x \sqrt{-g} \left( R + \Lambda \right. \notag \\
&\left.\;\;\;\;- \frac{1}{4} \mathrm{e}^{2 \alpha \phi} F^2 - \frac{1}{2} \left( \nabla \phi \right)^2 \right)
\end{align}
with \begin{equation}
\Lambda =  \frac{(z+d)(z+d-1)}{L^2}, \quad \alpha= \sqrt{\frac{d}{2(z-1)}},
\end{equation} 
has the Lifshitz metric (\ref{eq:Lif}) as a solution, together with the dilaton and Maxwell fields
\be
A = \sqrt{\frac{2(z-1)}{z+d}} \frac{L \, \mathrm{d}t}{r^{z+d}} \,, \qquad \mathrm{e}^{2 \alpha \phi} = r^{2 d} \,. 
\ee
Again, these solutions make sense for $z > 1$. In these solutions the scalar field $\phi$ is not scale invariant. We will see below that these kinds of running scalars can lead to interesting anomalous scaling dimensions. Because the Einstein-Mawell-dilaton theory involves a nonzero conserved flux for a bulk gauge field (dual to a charge density in the QFT), it is most naturally interpreted as a finite density solution, and will re-appear in \S\ref{sec:compressible} below. There is a large literature constructing Lifshitz spacetimes from consistent truncations of string theory, starting with \cite{Balasubramanian:2010uk, Donos:2010tu}. In our later discussion of compressible matter, we will describe some further ways to obtain the scaling geometry (\ref{eq:Lif}).

Repeating the analysis we did for the CFT case in \S\ref{sec:essential}, we can consider the two point function of a scalar operator in these theories. As before, consider a scalar field $\phi$ (not the dilaton above) that satisfies the Klein-Gordon equation (\ref{eq:scalarwave}), but now in the Lifshitz geometry (\ref{eq:Lif}). Once again writing $\phi = \phi(r) \mathrm{e}^{- \mathrm{i} \omega t + \mathrm{i} k \cdot x}$, the  radially dependent part must now satisfy
\be\label{eq:lifeq}
\phi'' - \frac{z+d-1}{r} \phi' + \left(\omega^2 r^{2z-2} - k^2 - \frac{(mL)^2}{r^2} \right)\phi = 0 \,.
\ee
This equation cannot be solved explicitly in general. Expanding into the UV region, as $r \to 0$, we find that the asymptotic behavior of (\ref{eq:boundaryexpand}) generalizes to
\be\label{eq:boundaryL}
\phi = \phi_{(0)} \left(\frac{r}{L}\right)^{D_\text{eff.}-\Delta} + \cdots + \phi_{(1)}\left(\frac{r}{L}\right)^{\Delta} + \cdots \qquad (\text{as} \quad r \to 0) \,.
\ee
We introduced the effective number of spacetime dimensions
\be\label{eq:deff}
D_\text{eff.} = z + d \,,
\ee
because these theories can be thought of as having $z$ rather than $1$ time dimensions,
and the (momentum) scaling dimension $\Delta$ now satisfies
\be\label{eq:deltaL}
\Delta (\Delta - D_\text{eff.}) = \left(m L\right)^2 \,.
\ee
 As in our discussion of AdS in \S\ref{sec:essential}, $\phi_{(0)}$ is dual to a source for the dual operator $\ocal$, whereas
$\phi_{(1)}$ is the expectation value. By rescaling the radial coordinate of the equation (\ref{eq:lifeq}), setting $\rho = r \omega^{1/z}$, and then using the asymptotic form (\ref{eq:boundaryL}), we can conclude that the retarded Green's function has the scaling form required for an operator of dimension $\Delta$ in a scale invariant theory with dynamical critical exponent $z$:
\be\label{eq:scalingformwq}
G^{\mathrm{R}}_{\ocal\ocal}(\omega,k) = \frac{2 \Delta - D_\text{eff.}}{L} \frac{\phi_{(1)}}{\phi_{(0)}} = \omega^{(2 \Delta- D_\text{eff.})/z} F\left(\frac{\omega}{k^z}\right) \,.
\ee
To obtain the correct normalization of the second expression, the discussion of holographic renormalization in \S \ref{sec:wilson} must be adapted to Lifshitz spacetimes \cite{Taylor:2008tg}.

The scaling function $F(x)$ appearing in (\ref{eq:scalingformwq}) will depend on the theory; it is not constrained by symmetry when $z \neq 1$. Remarkably, a certain asymptotic property of this function has been shown to hold both in strongly coupled holographic models and also in general weakly interacting theories. Specifically, consider the spectral weight (the imaginary part of the retarded Green's function) at low energies ($\omega \to 0$), but with the momentum $k$ held fixed. We expect this quantity to be vanishingly small, because the scaling of on-shell states as $\omega \sim k^z$ implies that there are no low energy excitations with finite $k$ \cite{Hartnoll:2012rj, Hartnoll:2012wm}. The precise holographic result is obtained from a straightforward WKB solution to the differential equation (\ref{eq:lifeq}), in which $k^z/\omega$ is the large WKB parameter. For the interested reader, we will go through some explicit WKB derivations of holographic Green's functions (very similar to the computations that pertain here) in later parts of the review. The essential point is that if the wave equation (\ref{eq:lifeq}) is written in the form of a Schr\"odinger equation, then the imaginary part of the dual QFT Green's function is given by the probability for the field to tunnel from the boundary of the spacetime through to the horizon. Here we can simply quote the result \cite{Keeler:2015afa}
\be
\text{Im} \left( G^{\mathrm{R}}_{\ocal\ocal}(\omega,k)\right) \sim \mathrm{e}^{ - A \left(k^{z}/\omega \right)^{1/(z-1)}} \,, \qquad (\omega \to 0) \,. \label{eq:imglif}
\ee
The positive coefficient $A$ is a certain ratio of gamma functions. Precisely the same relation (\ref{eq:imglif}) has been shown to hold in weakly interacting theories \cite{Keeler:2015afa}. At weak coupling the constant $A$ goes like the logarithm of the coupling. The weak coupling results follows from considering the spectrum of on-shell excitations that are accessible at some given order in perturbation theory. Taking the holographic and weak coupling results together suggests that the limiting behavior (\ref{eq:imglif}) may be a robust feature of the momentum dependence of general quantum critical response functions, beyond the existence of quasiparticles.

We should note also that when $z=2$, the full Green's function can be found explicitly in terms of gamma functions \cite{Kachru:2008yh,Taylor:2008tg}, which is typically a sign of a hidden $\mathrm{SL}(2,\R)$ symmetry.

\subsubsection{Hyperscaling violation}
\label{sec:hsv1}

A more general class of scaling metrics than (\ref{eq:Lif}) take the form
\be\label{eq:hsv}
\mathrm{d}s^2 = L^2 \left(\frac{r}{R}\right)^{2\theta/d} \left(- \frac{\mathrm{d}t^2}{r^{2z}} + \frac{\mathrm{d}\vec x^2_{d} + \mathrm{d}r^2}{r^2} \right) \,.
\ee
Here $L$ is determined by the bulk theory while $R$ is a constant of integration in the solution.
On the face of it, these metrics do not appear to be scale invariant. Under the rescaling
$\{t, \vec x, r \} \to \{\lambda^z t , \lambda \vec x, \lambda r \}$, the metric transforms as $\mathrm{d}s^2 \to \lambda^{2 \theta/d} \mathrm{d}s^2$.
However, this covariant transformation of the metric amounts to the fact that the energy density operator (as well as other operators) in the dual field theory
has acquired an anomalous dimension. Recall from \S\ref{sec:essential} that the bulk metric is dual to the energy-momentum tensor in the QFT. From the essential holographic dictionary (\ref{eq:dictionary}), the boundary value $\delta \gamma_{\mu\nu}$ of a
bulk metric perturbation sources the dual energy density $\vep$ via the field theory
coupling
\be\label{eq:Rcouple}
\frac{1}{R^\theta} \int \mathrm{d}^{d+1}x \, (\delta \gamma)_t{}^t  \, \vep \,.
\ee
The factors of $R$ appear because the boundary metric is obtained, analogously to (\ref{eq:boundaryexpand}), from (\ref{eq:hsv}) by stripping off the powers of $r$ and $L$ (but not $R$, which is a property of the solution) from each component of the induced boundary metric. As in our discussion around (\ref{eq:BigDelta}) above, the scaling action $\{t, \vec x, r \} \to \{\lambda^z t , \lambda \vec x, \lambda r \}$ must be combined with a scaling of parameters in the solution that leave the field (the metric in this case) invariant. Therefore, under this scaling we have $\mathrm{d}^{d+1}x \to \lambda^{z+d} \mathrm{d}^{d+1}x$,
$(\delta \gamma)_t{}^t$ is invariant (because one index is up and the other down), $\vep(x) \to \lambda^{-\Delta_\vep} \vep(\lambda x)$
and $R \to  \lambda R$. Scale invariance of the coupling (\ref{eq:Rcouple}) then requires
\be\label{eq:Edim}
\Delta_\vep = z + d  - \theta = \Delta_{\vep,0} - \theta \,,
\ee
where $\Delta_{\vep,0}$ is the canonical dimension of the energy density operator. The existence of such an anomalous dimension is known as hyperscaling violation, and $\theta$ is called the hyperscaling violation exponent \cite{Huijse:2011ef}. We will see later that this anomalous dimension indeed controls thermodynamic quantities and correlation functions of the energy-mometum tensor in the expected way. It is important to note, however, that because of the dimensionful parameter $R$ in the background, typically not all correlators of all operators will be scale covariant. An example is given by scalar fields with a large mass in the bulk \cite{Dong:2012se}. In particular, when $z=1$, theories with $\theta \neq 0$ are not CFTs. Hyperscaling violation arises when a UV scale does not decouple from certain IR quantities.

Hyperscaling violating metrics (\ref{eq:hsv}) will be important below in the context of compressible phases of matter. However, they can also describe zero density fixed points and therefore fit into this section of our review. For instance, hyperscaling violating metrics of the form (\ref{eq:hsv}) with $z=1$ arise as solutions to the Einstein-scalar action \cite{Chamblin:1999ya, Charmousis:2010zz}:
\begin{equation}\label{eq:einscal}
S= \frac{1}{2 \k^2} \int \mathrm{d}^{d+2}x \sqrt{-g} \left( R + V_0  \mathrm{e}^{\beta \phi} - \frac{1}{2} \left( \nabla \phi \right)^2 \right)
\end{equation}
with \begin{equation}
V_0 = \frac{(d+1-\theta)(d-\theta)}{L^2} \,, \qquad  \beta^2 = \frac{2 \theta}{d \, (\theta-d)}\,,
\end{equation}
where on the solution
\be\label{eq:scal}
\mathrm{e}^{\beta \phi} = r^{-2\theta/d} \,.
\ee
Thus we find Lorentzian hyperscaling-violating theories. One sees in the above that for $\beta$ to be real one must have $\theta < 0$ or $\theta > d$. In fact, this is a special case of a more general requirement that we now discuss.

A spacetime satisfies the null energy condition if $G_{ab}n^a n^b \geq 0$ for all null vectors $n$. Here $G_{ab}$ is the Einstein tensor. The null energy condition is a constraint on matter fields sourcing the spacetime (because Einstein's equations are $G_{ab} = 8\pi G_{\mathrm{N}} T_{ab}$). In a holographic context, the null energy condition is especially natural because it guarantees that holographic renormalization makes sense \cite{Freedman:1999gp}. In particular, it ensures that the spacetime shrinks sufficiently rapidly as one moves into the bulk, cf. \cite{Heemskerk:2010hk}. For the hyperscaling violating spacetimes (\ref{eq:hsv}), the null energy condition requires the following two inequalities to hold
\begin{subequations}\label{eq:null}
\begin{align}
(d-\theta)[d \, (z-1) - \theta] \geq 0 \,, \\
(z-1)(d+z - \theta) \geq 0 \,.
\end{align}\end{subequations}
With $\theta = 0$, these inequalities require that $z \geq 1$, consistent with our observations above. While perhaps plausible, even this $\theta = 0$ condition on the dynamic critical exponent remains to be understood from a field theory perspective.

Unlike the Lifshitz geometries, hyperscaling-violating spacetimes do not have constant curvature invariants, and these invariants generically (with one potentially interesting exception \cite{Shaghoulian:2011aa, Lei:2013apa}) diverge at either large or small $r$. These are more severe singularities than the pp-singularities of the Lifshitz geometries, and likely indicate the hyperscaling-violating spacetime only describes an intermediate energy range of the field theory, with new IR physics needed to resolve the singularity. For instance, microscopic examples of theories with zero density hyperscaling-violating intermediate energy regimes are non-conformal D-brane theories, such as super Yang-Mills theory in 2+1 dimensions \cite{Dong:2012se}.

\subsubsection{Galilean-invariant `non-relativistic CFTs'}\label{sec:gal}

None of the scaling geometries we have discussed so far (except for AdS itself) have symmetries beyond dilatations, spacetime translation and rotations. In \cite{Son:2008ye, Balasubramanian:2008dm} a gravitational geometry was proposed as a dual to non-relativistic CFTs. These are theories in which the scaling symmetry is enhanced by a Galilean boost symmetry and a particle number symmetry. For the case $z=2$, further enhancement by a special conformal symmetry is possible. These are the symmetries of the Schr\"odinger equation for a free particle, and for this reason these geometries are known as `Schr\"odinger' spacetimes. The various algebras are discussed in \cite{Hartnoll:2009sz}. The primary motivation of \cite{Son:2008ye, Balasubramanian:2008dm} was to model strongly interacting critical regimes of cold atomic gases. This body of work is somewhat tangential to the main thrust of our review, so we limit ourselves to some brief comments and pointers to the literature. 

The vacuum poses a challenge for gravity duals of Galilean-invariant systems. While vacua of e.g. relativistic field theories are highly nontrivial, the state with zero particle number in a Galilean-invariant system has no dynamics whatsoever. It seems puzzling, therefore, that it could be dually described by an emergent spacetime, which is certainly not trivial. This difficulty manifests itself in the fact that the geometries proposed in \cite{Son:2008ye, Balasubramanian:2008dm} involve a compactified null circle, which is well-known to be a delicate thing (see e.g. \cite{Maldacena:2008wh}). It is natural, therefore, to consider states of the theory with a large particle number \cite{Maldacena:2008wh, Herzog:2008wg, Adams:2008wt, Kovtun:2008qy}. While the resulting geometries are better behaved, the underlying compactified null circle in the construction directly results in thermodynamic properties (as a function of temperature and particle number) that are quite unlike those of more conventional Galilean-invariant systems \cite{Maldacena:2008wh}. It is important to see if these difficulties can be overcome by more general holographic constructions \cite{Balasubramanian:2010uw, Janiszewski:2012nb}.

\subsection{Nonzero temperature}
\label{sec:temp}

The most universal way to introduce a single scale into the scale invariant theories discussed above is to heat the system up.
The thermal partition function is given by the path integral of the theory in Euclidean time with a periodic imaginary time direction, $\tau \sim \tau + 1/T$:
\be\label{eq:ZT}
Z_\text{QFT}(T) = \int_{\mathrm{S}^1 \times \R^{d}} \mathcal{D} \Phi \, \mathrm{e}^{-I_{\mathrm{E}}[\Phi]} \,.
\ee
Here $I_{\mathrm{E}}[\Phi]$ is the Euclidean action of the quantum field theory.

The essential holographic dictionary (\ref{eq:dictionary}) identifies the space in which in the field theory lives as the asymptotic conformal boundary of the bulk spacetime. In particular, once we have placed the theory in a nontrivial background geometry ($\mathrm{S}^1 \times \R^d$ in this case), then the bulk geometry must asymptote to this form. In order to compute the thermal partition function (\ref{eq:ZT}) holographically, in the semiclassical limit (\ref{eq:classicalsource}), we must therefore find a solution to the Euclidean bulk equations of motion with these boundary conditions.

\subsubsection{Thermodynamics}
\label{sec:thermoBH}

The required solutions are Euclidean black hole geometries. Allowing for general dynamic critical exponent $z$ and hyperscaling violation, the solutions can be written in the form
\be\label{eq:BH2}
\mathrm{d}s^2 = L^2 \left(\frac{r}{R}\right)^{2\theta/d} \left(\frac{f(r) \mathrm{d}\tau^2}{r^{2z}} + \frac{\mathrm{d}r^2}{f(r) r^2}+ \frac{\mathrm{d}\vec x^2_{d} }{r^2} \right) \,.
\ee
The new aspect of the above metric relative to the scaling geometry (\ref{eq:hsv}) is the presence of the function $f(r)$. There is some choice in how the metric is written, because we are free to perform a coordinate transformation: $r \to \rho(r)$.
The above form is convenient, however, because often the function $f(r)$ is simple in this case. We will see some examples of $f(r)$ shortly, but will first note some general features.

Asymptotically, as $r \to 0$, the metric must approach the scaling form (\ref{eq:hsv}), and therefore $f(0) = 1$. Recall from the discussion in \S\ref{sec:wilson} that the near-boundary metric corresponds to the high energy UV physics of the dual field theory. Therefore, this boundary condition on $f(r)$ is the familiar statement that turning on a nonzero temperature does not affect the short distance, high energy properties of the theory. The temperature will, however, have a strong effect on the low energy IR dynamics. In fact, we expect the temperature to act as an IR cutoff on the theory, with long wavelength processes being Debye screened. It should not come as a surprise, therefore, that in the interior of the spacetime we find that at a certain radius $r_+$ we have $f(r_+) = 0$. At this radius the thermal circle $\tau$ shrinks to zero size and the spacetime ends. This is the Euclidean version of the black hole event horizon. The radial coordinate extends only over the range $0 \leq r \leq r_+$. 

\begin{figure}
\centering
\includegraphics[height = 0.15\textheight]{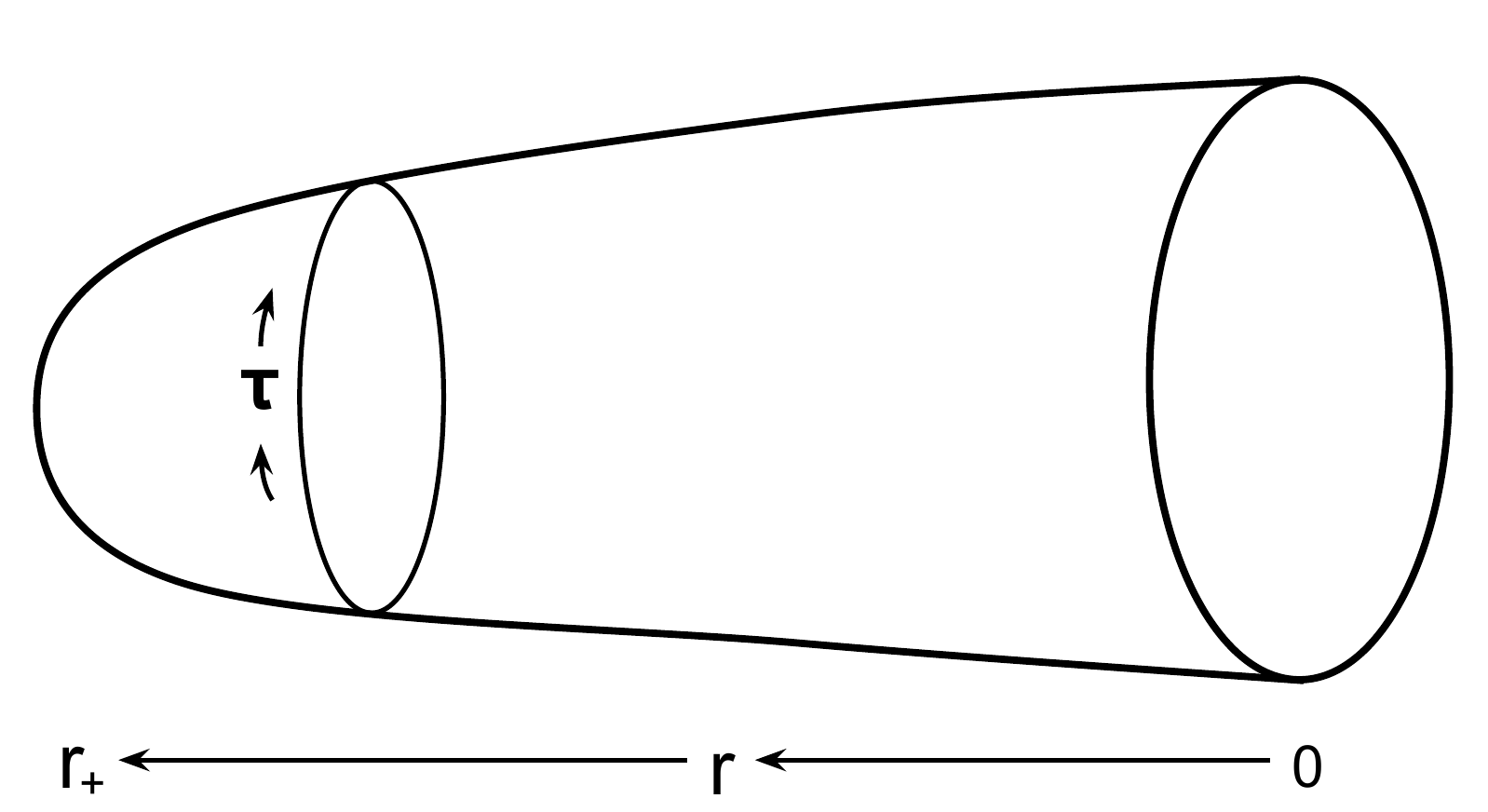}
\caption{\label{fig:cigar} {\bf Cigar geometry.} The $r$ coordinate runs from $0$ at the boundary to $r_+$ at the horizon, where the Euclidean time circle shrinks to zero.}
\end{figure}

The radius $r_+$ is related to the temperature by imposing that the spacetime be regular at the Euclidean horizon. If we expand the metric near the horizon we find
\begin{align}\label{eq:BHnearh}
\mathrm{d}s^2 &=  A_+ \left( \frac{|f'(r_+)| (r_+ - r) \mathrm{d}\tau^2}{r_+^{2z}} \right. \notag \\
&\left.\;\;\;\;+ \frac{\mathrm{d}r^2}{|f'(r_+)| (r_+ - r) r_+^2} + \frac{\mathrm{d}\vec x^2_{d} }{r_+^2} \right) + \cdots \,.
\end{align}
Performing the change of coordinates
\be\label{eq:coordchange}
r  = r_+ - \frac{r_+^2 |f'(r_+)|}{4 \, A_+} \rho^2 \,, \qquad \tau = \frac{2 \, r_+^{z-1}}{|f'(r_+)|} \, \varphi \,,
\ee
the near-horizon geometry becomes
\be
\mathrm{d}s^2 = \rho^2 \mathrm{d}\varphi^2 + \mathrm{d}\rho^2 + \frac{\mathrm{d}\vec x^2_{d} }{r_+^2} + \cdots \,.
\ee
We immediately recognize this metric as flat space in cylindrical polar coordinates. In order to avoid a conical singularity at $\rho = 0$, we must have the identification $\varphi \sim \varphi + 2 \pi$. But $\varphi$ is defined in terms of $\tau$ in (\ref{eq:coordchange}), and $\tau$ already has the periodicity $\tau \sim \tau + 1/T$. Therefore
\be\label{eq:T}
T = \frac{|f'(r_+)|}{4 \pi r_+^{z-1}} \,.
\ee
The (topological) plane parametrized by the $\tau$ and $r$ coordinates is called the cigar geometry and is illustrated in Figure \ref{fig:cigar}.

The Einstein-scalar theories with action (\ref{eq:einscal}) are an example of cases in which the function $f(r)$ can be found analytically. The equations of motion following from this action are found to have a black hole solution given by the metric (\ref{eq:BH2}), with $z=1$, with the scalar remaining unchanged from the zero temperature solution (\ref{eq:scal}) and
\be
f(r) = 1 - \left( \frac{r}{r_+}\right)^{d+1-\theta} \,.   \label{eq:emblackfactor}
\ee
It follows from (\ref{eq:T}) that for these black holes
\be\label{eq:Tgo}
T = \frac{d+1-\theta}{4 \pi r_+} \,.
\ee
In particular, when $\theta = 0$ this gives the temperature of the (planar) AdS-Schwarzschild black hole in pure Einstein gravity. In other cases, such as the Einstein-Proca theory considered in the previous section, it will not be possible to find a closed form expression for $f(r)$. In these cases, the Einstein equations will give ODEs for $f(r)$, which will typically be coupled to other unknown functions appearing in the solution. These ODEs must then be solved numerically subject to the boundary conditions at $r=0$ and $r=r_+$ that we gave in the previous paragraph. These numerics are straightforward and are commonly performed either with shooting or with spectral methods. In performing the numerics, because the constant $r_+$ is the only scale, it can be scaled out of the equations by writing $r = r_+ \hat r$. This scaling symmetry together with the formula (\ref{eq:T}) for the temperature implies that in generality
\be\label{eq:Trp}
T \sim r_+^{-z} \,.
\ee

With the temperature at hand, we can now compute the entropy as a function of temperature. It is a general result in semiclassical gravity that the entropy is given by the area of the horizon divided by $4 G_{\mathrm{N}}$ \cite{Gibbons:1976ue, Gibbons:1979xm}. In our conventions, Newton's constant $G_{\mathrm{N}} = \k^2/(8 \pi)$. Therefore, for the black hole spacetime (\ref{eq:BH2}) we have the entropy density
\be\label{eq:s}
s = \frac{S}{V_{d}} = \frac{2 \pi L^{d}}{\k^2} \frac{r^{\theta-d}_+}{R^\theta} \sim \frac{L^{d}}{L_{\mathrm{p}}^{d}} \frac{T^{(d - \theta)/z}}{R^\theta} \,.
\ee
For the last relation we used (\ref{eq:Trp}) and also re-expressed $\kappa$ in terms of the $d+2$ dimensional Planck length $\kappa^2 \sim L_{\mathrm{p}}^{d}$. For a given theory with an explicit relation between temperature and horizon radius, such as (\ref{eq:Tgo}), we easily get the exact numerical prefactor of the entropy density. Two observations should be made about the result (\ref{eq:s}) for the entropy. Firstly, the temperature scaling $s \sim T^{(d-\theta)/z}$ is exactly what is expected from (\ref{eq:Edim}) for a scale-invariant theory with exponents $z$ and $\theta$. Secondly, in the semiclassical limit, the radius of curvature $L$ of the spacetime is much larger than the Planck length $L_{\mathrm{p}}$. Therefore, the entropy density $s \sim (L/L_{\mathrm{p}})^{d} \gg1$ in this limit. In this way we explicitly realize the intuition from \S \ref{sec:horizons}, \S \ref{sec:thooft} and \S \ref{sec:essential} above that the thermodynamics captured by classical gravity is that of a dual large $N$ theory. For instance, in the canonical duality involving ${\mathcal N} = 4$ SYM theory, we have $(L/L_{\mathrm{p}})^{3} \sim N^2 \gg 1$ (this expression is compatible with (\ref{eq:stringy}) above, because in (\ref{eq:stringy}) $L_{\mathrm{p}}$ refers to the ten dimensional Planck length; the five dimensional Planck length appearing in (\ref{eq:s}) is then $L_{\mathrm{p},5}^3 = L_{\mathrm{p},10}^8/L^5$, upon dimension reduction on the $\mathrm{S}^5$).

The temperature and entropy are especially nice observables in holography because they are determined entirely by horizon data. The free energy is a more complicated quantity because it depends on the whole bulk solution. In particular
\be\label{eq:onshell}
F = - T \log Z_\text{QFT} = - T \log Z_\text{Grav.} = T S[g_\star] \,.
\ee
Here we used the equivalence of bulk and QFT partition functions (\ref{eq:dictionary}) and also the semiclassical limit in the bulk (\ref{eq:classicalsource}). In the final expression, $S[g_\star]$ is the bulk Euclidean action evaluated on the black hole solution (\ref{eq:BH2}). The action involves an integral over the whole bulk spacetime and diverges due to the infinite volume near the boundary as $r \to 0$. The divergences are precisely of the form of the expected short distance divergences in the free energy of a quantum field theory. This is an instance of the general association of the near boundary region of the bulk with the UV of the dual QFT, illustrated in Figure \ref{fig:RG} above. If we believe that the scale invariant theory we are studying is a fundamental theory, then it makes sense to holographically renormalize the theory by adding appropriate boundary counterterms as described in \S\ref{sec:wilson} above. However, more typically, we expect in condensed matter physics that the scaling theory is an emergent low energy description with some latticized short distance completion. The divergences we encounter in evaluating the action are then physical and simply remind us that the free energy is sensitive to (and generically dominated by) non-universal short distance degrees of freedom. For this reason we will refrain from explicitly evaluating the on shell action (\ref{eq:onshell}) for the moment, and note that the entropy (\ref{eq:s}) is a better characterization of the universal emergent scale-invariant degrees of freedom at temperatures that are low compared to some UV cutoff.

\subsubsection{Thermal screening}

Beyond equilibrium thermodynamic quantities, in \S\ref{sec:qctransport} the Lorentzian signature version of the black hole backgrounds (\ref{eq:BH2}) will be the starting point for computations of linear response functions such as conductivities. At this point we can show how black holes cause thermal screening of spatial correlation. Correlators with $\omega = 0$ are the same in Lorentzian and Euclidean signature. To exhibit thermal screening, consider the wave equation of a massive scalar field in the black hole background (\ref{eq:BH2}). For this computation, let us not consider the effects of hyperscaling violation and so set $\theta = 0$. The wave equation (\ref{eq:scalarwave}), at Euclidean frequency $\omega_n$, becomes
\begin{align}
\phi^{\prime\prime} &+ \left(\frac{f^\prime}{f} - \frac{z+d-1}{r} \right) \phi^\prime \notag \\
&- \frac{1}{f}\left(k^2 + \frac{(mL)^2}{r^2} + \frac{r^{2z} \omega_n^2}{r^2 f}\right)\phi = 0 
\label{eq:thermalscreening}
\end{align}
This equation cannot be solved analytically in general. By rescaling the $r$ coordinate as described above equation (\ref{eq:scalingformwq}) above, and with the assumption that $r_+$ is the only scale in the bulk solution, so that $f$ is a function of $r/r_+$, with $r_+$ related to $T$ via (\ref{eq:Trp}), we can again conclude that the Green's function will have the expected scaling form
\be\label{eq:scalingformwqT}
G^{\mathrm{R}}_{\ocal\ocal}(\omega,k) = \omega^{(2 \Delta- D_\text{eff.})/z} F\left(\frac{\omega}{k^z}, \frac{T}{k^z} \right) \,.
\ee

Useful intuition for how thermal screening arises is obtained by solving the above equation in the WKB limit and with $\omega_n=0$. This approximation holds for large $m L \gg 1$. The WKB analysis is especially simple for equation (\ref{eq:thermalscreening}) as there are no classical turning points. Imposing an appropriate boundary condition at the horizon (which will be discussed in generality in \S\ref{sec:qctransport}), one finds that the $\omega_n= 0$ correlator is
\bea\label{eq:kWKB}
\lefteqn{\left\langle \ocal \ocal \right\rangle(k) = } \\*
&& r_+^{-2 m L} \exp\left\{- 2 { \int_0^{r_+}  \frac{\mathrm{d}r}{r} \left(\sqrt{\frac{(mL)^2 + (k r)^2}{f}} - m L\right)}\right\} \,. \notag
\eea
To see thermal screening, consider the real space correlator at large separation. This is found by taking the Fourier transform of (\ref{eq:kWKB}) and evaluating using a stationary phase approximation for the $k$ integral as $x \to \infty$ to obtain:
\be\label{eq:oox}
\left\langle \ocal \ocal \right\rangle(x) = \int_{-\infty}^\infty \frac{\mathrm{d}^{d}k}{(2\pi)^{d}}  \left\langle \ocal \ocal \right\rangle(k) \, \mathrm{e}^{- \mathrm{i} k \cdot x} \sim \mathrm{e}^{- \mathrm{i} k_\star x}\,.
\ee
It is simple to determine $k_\star$. From (\ref{eq:kWKB}) and (\ref{eq:oox}), $k_\star$ must satisfy
\be
2  \int_0^{r_+} \frac{k_\star r \mathrm{d}r}{\sqrt{f}\sqrt{(mL)^2 + (k_\star r)^2}} + \mathrm{i} x = 0 \,.
\ee
In this expression we see that at large $x$, $k_\star$ will be independent of $x$ to leading order. Specifically: to balance the $\mathrm{i} x$ term, the integral must become large. This occurs if the integrand is close to developing a pole at $r = r_+$ (which would give a logarithmically divergent integral). For this to happen it must be that $(mL)^2 + (k_\star r)^2$ is close to vanishing at $r=r_+$ (where $f$ already vanishes). To leading order, then, we have $k_\star = \pm \mathrm{i} m L/r_+$. Choosing the physical sign, it follows that at large $x$
\be\label{eq:decayspace}
\left\langle \ocal \ocal \right\rangle(x) \sim \mathrm{e}^{- m L \, x/r_+} \,, \qquad \text{as} \qquad x \to \infty \,.
\ee
We have therefore shown that the thermal screening length is
\be\label{eq:screenT}
L_\text{thermal} = \frac{r_+}{m L} \sim \frac{1}{T^{1/z}} \,,
\ee
as we might have anticipated.

\begin{figure}
\centering
\includegraphics[height = 0.15\textheight]{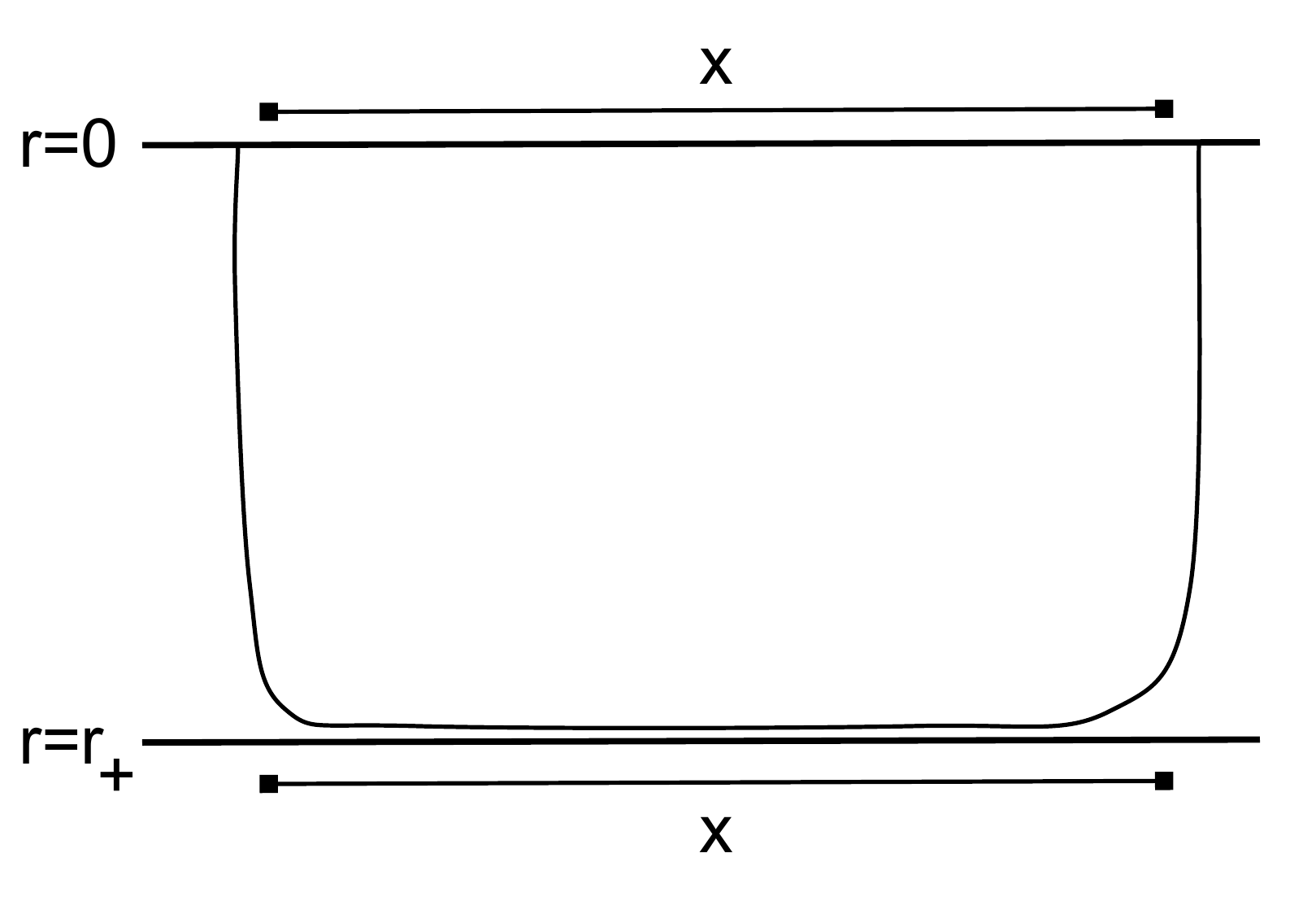}
\caption{\label{fig:screen} {\bf Thermal screening}. The geodesic runs along the horizon over a distance $x$. This contribution to its length dominates the correlation function and leads to an exponentially decaying correlation (\ref{eq:decayspace}) in space, with scale set by the horizon radius $r_+$.}
\end{figure}

The screening length (\ref{eq:screenT}) arose because the momentum integral in (\ref{eq:oox}) was dominated by momenta at the horizon scale. This can be seen even more directly by recalling that the WKB limit of the Klein-Gordon equation in a geometry describes the motion of very massive particles along geodesics. The correlation function at a separation $x$ is then given by the length of a geodesic in the bulk that connects two points on the boundary separated by distance $x$. The simplest quantity to consider here is the equal time correlation function. For points separated by much more than the horizon length $r_+$, the geodesic essentially falls straight down to the horizon, and runs along the horizon for a distance $x$, as shown in Figure \ref{fig:screen}. Thereby one recovers (\ref{eq:decayspace}) from
\be\label{eq:geod}
\left\langle \ocal \ocal \right\rangle(x) \sim \mathrm{e}^{- m \times \text{(Geodesic length)}} \sim \mathrm{e}^{- m L \, x/r_+} \,.
\ee
Thus we see a very geometric connection between the presence of a horizon and thermal screening.

\subsection{Theories with a mass gap}
\label{sec:gapped}

The geometries we have discussed so far give the holographic duals to quantum critical theories at zero and nonzero temperature. The simplest way to move away from criticality is to deform the theory by a relevant operator. The RG flow triggered by the relevant deformation may lead to another fixed point in the IR, or may gap out the theory.

The essential holographic dictionary discussed in \S\ref{sec:essential} and \S\ref{sec:scaling} tells us how to deform a critical theory by a relevant operator $\ocal$ of dimension $\Delta < D_\text{eff.}$. Namely, one introduces a field $\Phi$ in the bulk dual to $\ocal$ and requires that near the boundary ($r \to 0$), the field behave as $\Phi \sim \Phi_{(0)} r^{D_\text{eff.}-\Delta}$. Recall from \S\ref{sec:sit} that $\Phi_{(0)} \equiv s-s_{\mathrm{c}}$ is the dual field theory coupling of the operator $\ocal$. This boundary condition, together with $\Delta < D_\text{eff.}$, implies that the field $\Phi$ grows as one moves into the bulk away from $r=0$. This growth corresponds to the growth of the relevant coupling $\Phi_{(0)}$ into the IR as described in \S\ref{sec:wilson} above.

As the field $\Phi$ grows as a function of the radial coordinate, it will eventually backreact on the metric (as well as any other fields that were present in the scaling solution with $\Phi = 0$). One must therefore solve the coupled equations of motion for the field and the metric. For simplicity, we will focus on the Lorentz invariant case of $z=1$ and with hyperscaling holding in the original undeformed theory. Then, we expect to find a (Euclidean) bulk solution that can be written in the general form 
\be\label{eq:AB}
\mathrm{d}s^2 = \mathrm{e}^{2 A(r)} \left(\mathrm{d}\tau^2 + \mathrm{d}r^2 + \mathrm{d}\vec x^2_{d} \right) \,, \qquad \Phi = \Phi(r) \,.
\ee
This is a general form for a metric that preserves Poincar\'e invariance along the RG flow. On this ansatz, the equations of motion will become coupled ODEs for $\{A, \Phi\}$ as functions of $r$. These must be solved subject to the UV boundary condition that $A \to - \log (r/L)$ and $\Phi \to \Phi_{(0)} r^{d+1 - \Delta}$ as $r \to 0$.

Two qualitatively different types of IR behavior are possible. One is that the solution extends to $r \to \infty$ and that the function in the metric behaves as $A \sim (\frac{\theta_\text{IR}}{d} - 1) \log r$. This signals the emergence of a new far IR fixed point, possibly with hyperscaling violation. In this case the solutions that interpolate between scaling geometries in the UV and IR are called `domain wall' spacetimes. For a discussion (without hyperscaling violation) see \cite{Skenderis:1999mm}. We will be greatly interested in the emergence of IR scaling in slightly more general, nonzero charge density, setups in \S\ref{sec:compressible} below. The reference \cite{Skenderis:1999mm} also explains how the equations of motion for gravity coupled to a scalar field can be substantially simplified when the potential for the scalar field can be written in terms of a `superpotential'. Such a rewriting is also useful to construct explicit examples of the gapped physics that we will only outline below. 

The other option is that the IR is gapped. In a gapped theory correlators decay exponentially at late Euclidean times. This is similar to our earlier discussion of exponential decay at large spatial separation for thermal screening. In fact, the discussion around (\ref{eq:geod}) above tells us what needs to occur for the geometry (\ref{eq:AB}) to give exponential decay at late Euclidean time.
We will need the solution to terminate at some $r = r_{\mathrm{o}}$, and furthermore $A(r_{\mathrm{o}})$ should be finite. This latter condition ensures that the tension of timelike geodesics running along $r = r_{\mathrm{o}}$ is finite. This is important, because such geodesics determine late time correlations, analogously to how the geodesic in (\ref{eq:geod}) determines large separation correlation. In order for the geometry to terminate at $r=r_{\mathrm{o}}$ with $A(r_{\mathrm{o}})$ remaining finite, one or both of $A'(r)$ and $\Phi(r)$ should diverge as $r \to r_{\mathrm{o}}$. This indicates a singular solution, and so care is required to determine if the singularity is physical or not. See, for example, the discussion in \cite{Gubser:2000nd}.

In the above scenario, the mass gap will be set by the geodesic tension at $r=r_0$. As the only dimensionful input (from the boundary QFT perspective) into the problem is the UV coupling $\Phi_{(0)} \equiv s-s_{\mathrm{c}}$, it must be this scale that determines the mass gap
\be
M \sim \mathrm{e}^{A(r_{\mathrm{o}})} \sim (s-s_{\mathrm{c}})^\nu\,,
\ee
with $\nu = 1/(d + 1 - \Delta)$. As with thermal screening, the termination of the geometry at some finite radial distance is a very natural geometrization of a mass gap. In our discussion of Wilsonian renormalization in \S\ref{sec:wilson} we noted that the increasing redshift as one moves into the bulk correspondeds to the decreasing energy scale of the dual QFT. If the bulk terminates at some finite redshift, as we have just described, then there simply do not exist holographic degrees of freedom with energies below this lowest redshift scale. That is, the spectrum is gapped.

In fact, the best understood examples of holographic flows to gapped phases are not quite of the form just described. Instead, the best understood solutions involve Kaluza-Klein modes (see \S\ref{sec:consistent} above) in an essential way, e.g. \cite{Witten:1998zw,Klebanov:2000hb,Maldacena:2000yy}. This is because one or several of the internal dimensions can smoothly collapse to zero size at some finite value of the radial coordinate $r \sim r_{\mathrm{o}}$ (in an entirely analogous way to how the thermal circle collapses to zero at the tip of the Euclidean cigar geometry of figure \ref{fig:cigar}). This description has the advantage of being nonsingular. One disadvantage of this approach is that typically the gap scale will be of order the Kaluza-Klein scale. This makes it difficult to decouple the physics of the gap from microscopic details that pertain to a higher dimensional auxiliary theory.

Most discussion of gaps in holography has been related to the physics of confinement, as the existence of horizons is closely tied to the existence of deconfined phases of large $N$ gauge theories \cite{Witten:1998zw}. Some of this discussion may be directly relevant for the description of deconfined quantum critical points. Beyond this connection, the use of holography to study crossovers between quantum critical scaling and gapped regimes appears to be relatively underexplored. Some caution is called for. The bulk dynamics of fields about a gapped geometry is literally that of fields in a box. The power of holography to describe dissipation classically via event horizons is lost, as the gapped geometry has no regions of infinite redshift. Instead, in the semiclassical bulk limit all excitations are perfectly stable and do not dissipate. This is an artifact of the large $N$ expansion in gapped phases (cf. \cite{'tHooft:1973jz}).

We have characterized the gapped phase by exponential decay of correlators in Euclidean time. One expects the absence of low energy degrees of freedom to manifest itself in the absence of a power law specific heat at low temperatures. To compute the temperature dependence of the specific heat, a solution describing a black hole must be found. For instance, in the framework described above these will be solutions of the form
\be\label{eq:ABBH}
\mathrm{d}s^2 = \mathrm{e}^{2 A(r)} \left(f(r) \mathrm{d}\tau^2 + \frac{\mathrm{d}r^2}{f(r)} + \mathrm{d}\vec x^2_{d} \right) \,, \qquad \Phi = \Phi(r) \,.
\ee
The functions $\{A,\Phi\}$ will change relative to the zero temperature solution (\ref{eq:AB}), but must still satisfy the same boundary conditions in the UV, namely, that the relevant operator is sourced. It is typically found that a black hole solution satisfying the appropriate boundary conditions either does not exist or is not the dominant saddle point of the path integral at temperatures below some critical temperature $T_{\mathrm{c}} \sim M$. For temperatures $T < T_{\mathrm{c}}$ the dominant saddle is just the zero temperature spacetime (\ref{eq:AB}), but now with the Euclidean thermal circle identified. The absence of a finite size horizon for $T < T_{\mathrm{c}}$ then means that the entropy density, and hence specific heat, will be zero to leading order at large $N$ for $T < T_{\mathrm{c}}$. Thus we see the gap in the low energy degrees of freedom very vividly: at large $N$ the specific heat $c$ has a step function discontinuity at $T=T_c$, so that $c \sim \Theta(T-T_c)$. For $T > T_{\mathrm{c}}$, the dominant saddle will be a black hole spacetime of the form (\ref{eq:ABBH}). For an example of such a transition see \cite{Witten:1998zw, Mateos:2007ay}. These types of transitions between geometries with and without horizons are referred to as Hawking-Page transitions. The original and simplest case of such a transition occurs if the boundary is considered on a spatial sphere rather than a plane \cite{Hawking:1982dh, Witten:1998qj, Witten:1998zw, Aharony:2003sx}.

\section{Quantum critical transport}
\label{sec:qctransport}

\subsection{Condensed matter systems and questions}
\label{sec:cmtransport}

Let us consider the simplest case of transport of a conserved U(1) ``charge'' density $\rho (x, t)$
(we will explicitly write out the time coordinate, $t$, separately from the spatial coordinate $x$ in this
section). Its conservation implies that there is a current $J_i (x,t)$ (where $i =1\ldots d$) such that
\be
\partial_t \rho + \partial_i J_i = 0 \,.
\ee
We are interested here in the consequences of this conservation law at $T>0$ for correlators of $\rho$ and $J_i$ in quantum systems
without quasiparticle excitations.

We begin with a simple system in $d=1$: a narrow wire of electrons which realize a Tomonaga-Luttinger liquid \cite{giamarchi03}.
For simplicity, we ignore the spin of the electron in the present discussion. In field-theoretic terms, such a liquid is a CFT2.
In the CFT literature it is conventional to describe its correlators in holomorphic (and anti-holomorphic) variables constructed out of space and Euclidean time, $\tau$. 
The spacetime coordinate is expressed in terms of a complex number $z=x+\mathrm{i}\tau$, and the density and the current combine to yield the holomorphic current $\mathcal{J}$.
A well-known property of all CFT2s with a conserved $\mathcal{J}$ is the correlator
\be
\left\langle \mathcal{J} (z) \mathcal{J} (0) \right\rangle = \frac{\mathcal{K}}{z^2}, \label{km}
\ee
where $\mathcal{K}$ is the `level' of the conserved current. Here we wish to rephrase this correlator in more conventional condensed matter
variables using momenta $k$ and Euclidean frequency $\omega$. Fourier transforming (\ref{km}) we obtain
\be
\left \langle \left| \rho (k, \omega) \right|^2 \right \rangle = \mathcal{K} \frac{ c^2 k^2}{c^2 k^2 + \omega^2} \,, \label{nn}
\ee
where $c$ is the velocity of `light' of the CFT2. A curious property of CFT2s is that (\ref{nn}) holds also at $T>0$. This is a consequence of the conformal
mapping between the $T=0$ planar spacetime geometry and the $T>0$ cylindrical geometry; as shown in \cite{Herzog:2007ij}, 
this mapping leads to {\em no\/}
change in the Euclidean time correlator in (\ref{nn}) apart from the restriction that $\omega$ is an integer multiple of $2 \pi T$ ({\it i.e.\/} it is a
Masturbara frequency). We can also analytically continue from the Euclidean correlation in (\ref{nn}), via $i \omega \rightarrow \omega + i \epsilon$, 
to obtain the retarded two-point correlator of the density 
\be
G_{\rho\rho}^R = \mathcal{K} \frac{ c^2 k^2}{c^2 k^2 - (\omega + \mathrm{i} \epsilon)^2} \, , \qquad d=1 \,, \label{nnR}
\ee
where $\epsilon$ is a positive infinitesimal. We emphasize that (\ref{nnR}) holds for all CFT2s with a global U(1) symmetry at any temperature $T$.

Now we make the key observation that (\ref{nnR}) is {\it not\/} the expected answer for a generic non-integrable interacting quantum system 
at $T > 0$ for small $\omega$ and $k$. 
Upon applying an external perturbation which creates a local non-uniformity in density, we expect any such system to relax
towards the equilibrium maximum entropy state. This relaxation occurs via hydrodynamic diffusion of the conserved density.   Namely, we expect that on long time and length scales compared to $\hbar/k_{\mathrm{B}}T$ (or analogous thermalization length scale):  \begin{equation}
J_i \approx -D\partial_i \rho  + \mathrm{O}(\partial^3).
\end{equation}
We will give a more complete introduction to hydrodynamics in \S\ref{sec5}.  As described in some detail
by Kadanoff and Martin \cite{Kadanoff1963419}, the (very general) assumption that hydrodynamics is the correct description of the late dynamics forces the retarded density correlator at small $k$ and $\omega$ to be
\be
G_{\rho\rho}^{\mathrm{R}} = \chi \, \frac{ D k^2}{D k^2 - \mathrm{i} \omega}, \label{nnchi}
\ee
where $D$ is the diffusion constant, and $\chi$ is the susceptibility referred to as the compressibility. In terms of thermodynamic quantities, $\chi = \partial \rho/\partial \mu$, with $\mu$ the chemical potential. The Green's function (\ref{nnchi}) is found by directly solving the classical diffusion equation with proper initial conditions.  The disagreement between (\ref{nnR}) and 
(\ref{nnchi}) implies that all CFT2s have integrable density correlations, and do not relax to thermal equilibrium.    

The Green's function (\ref{nnchi}) is not quite right in CFT3s either; it ignores long-time tails at the lowest frequencies caused by hydrodynamic fluctuations.   As long-time tails are additionally subleading effects in $1/N$ in holography, we will mostly neglect long-time tails until \S\ref{sec:oneoverN}.  We will show in (\ref{eq:res}) that in $d=2$ these lead to $\log\omega$ corrections to correlators such as (\ref{nnchi}).  Long-time tails are weaker in higher dimensions, giving an $\omega^{1/2}$ correction in $d=3$. Finally,  CFT3s deformed by operators which break translation invariance will have (\ref{nnchi}) hold exactly, at the lowest frequencies, as long-time tails will then be suppressed by an additional power of $\omega$ as $\omega \rightarrow 0$ \cite{mhernst}. 

Let us consider an interacting CFT3 such as the Wilson-Fisher fixed point describing the
superfluid-insulator transition in \S \ref{sec:sit}. In this case (using a relativistic notation with $c=1$), we have a current $J_\mu$ ($\mu = 0, \ldots d$),
and its conservation and conformal invariance imply that in $d$ spatial dimensions and at zero temperature:
\be
\left\langle J_\mu (p) J_\nu ( - p) \right\rangle = - \mathcal{K} |p|^{d-1} \left( \delta_{\mu\nu} - \frac{p_\mu p_\nu}{p^2} \right) \,,\label{Kmn0}
\ee
where $p$ is a Euclidean spacetime momentum and $\mathcal{K}$ is a dimensionless number characteristic of the CFT. 
The non-analytic power of $p$ above follows from the fact that $\rho$ and $J_i$ have scaling dimension $d$.
Taking the $\mu=\nu=0$ component of the above, and analytically continuing to the retarded correlator, we obtain
\be
G_{\rho\rho}^{\mathrm{R}} = \mathcal{K} \frac{k^2}{\sqrt{k^2 - (\omega + \mathrm{i} \epsilon)^2}} \, , \qquad d=2, \, T=0. \label{nnR2}
\ee
While (\ref{nnR2}) is the exact result for all CFT3s, now we do {\it not\/} expect it to hold at $T>0$.
Instead, we expect that for $\omega, k \ll T$ CFT3s will behave like generic non-integrable quantum systems and relax diffusively
to thermal equilibrium. In other words, we expect a crossover from (\ref{nnR2}) at $\omega, k \gg T$ to 
(\ref{nnchi}) at $\omega, k \ll T$.   (At ultralow frequencies in CFT3s, hydrodynamics will break down due to long-time tails.) Furthermore, using the fact that $T$ is the only dimensionful parameter present, 
dimensional analysis from a comparison of these expressions gives us the $T$ dependence of the compressibility
and the diffusivity (diffusion constant)
\be 
\chi = \mathcal{C}_\chi T \, , \qquad D = \frac{\mathcal{C}_D}{T}, \label{CC}
\ee
where $\mathcal{C}_{\chi, D}$ are dimensionless numbers also characteristic of the CFT3. Note that there is no direct relationship
between $\mathcal{C}_{\chi, D}$ and $\mathcal{K}$ other than the fact that they are all numbers obtained from the same CFT3.

The computation of $\mathcal{C}_{\chi, T}$ for CFT3s is a difficult task we shall address by several methods in the following
sections. For now, we note that these expressions also yield the conductivity $\sigma$ upon using the Kubo formula \cite{mahan13}
\be
\sigma ( \omega ) = \lim_{\omega \rightarrow 0} \lim_{k \rightarrow 0} \frac{- \mathrm{i} \omega}{k^2} G^{\mathrm{R}}_{\rho\rho} (k , \omega). \label{Kubo}
\ee
From (\ref{nnR2}) we conclude that 
\be
\sigma (\omega \gg T ) = \mathcal{K}, \label{sigmaK}
\ee
while from (\ref{nnchi}) and (\ref{CC}) 
\be
\sigma (\omega \ll T ) = \mathcal{C}_\chi \mathcal{C}_D.
\ee
The crossover between these limiting values is determined by a function of $\omega/T$, and understanding the structure of this
function is an important aim of the following discussion.  In the application to the boson Hubbard model, we note that the above conductivity is measured in units of $(e^\ast)^2/\hbar$
where $e^\ast$ is the charge of a boson. 


Before embarking upon various explicit methods for the computation of $\sigma (\omega)$ and $G_{\rho\rho}^{\mathrm{R}} (k, \omega)$, it is useful state
two exact sum rules that are expected to be obeyed by $\sigma (\omega)$. These sum rules were first noticed in a holographic context, but with
the benefit of hindsight, strong arguments can now be made that they are very generally applicable at quantum critical points in $d=2$,
and even at those which are not conformally invariant, and at those which include quenched disorder. 
The first sum rule is the analog of the optical sum rule (also referred to as the $f$-sum rule). This states that the integral of $\mathrm{Re}[ \sigma (\omega) ]$
equals a fixed number proportional to the average kinetic energy of the system. For scale-invariant critical points, the large $\omega$ behavior is
given by (\ref{sigmaK}), and so the integral is divergent. This is not surprising, because the average kinetic energy of the critical field theory is also
an ultraviolet divergent quantity. However, it has been argued that after a simple subtraction of this divergence the integral is finite, and it
integrates to a vanishing value \cite{Gulotta:2010cu,WitczakKrempa:2012gn,Katz14,Lucas:2016fju}. So we have
\be
\int\limits_0^{\infty} \mathrm{d} \omega \left( \mbox{Re}\left[\sigma (\omega) \right] - \mathcal{K} \right) = 0 \label{sr1}
\ee
at all $T$.

The second sum rule is similar in spirit, but requires a much more subtle argument. As we will see in our discussion of holography below, but as can also be argued more generally \cite{Witten:2003ya},
CFT3s with a conserved U(1) are expected to have an `S-dual' formulation in which the U(1) current maps to the flux of a U(1) gauge field $A$ with
\be 
J_\mu = \frac{1}{2 \pi} \epsilon_{\mu\nu\lambda} \partial_\nu A_\lambda .
\ee
In the S-dual theory, conductivity of the dual particle current $\tilde J$ equals the resistivity of the original theory. Combining this fact with (\ref{sr1}),
the existence of the S-dual theory implies that \cite{WitczakKrempa:2012gn,Katz14}
\be
\int\limits_0^{\infty} \mathrm{d} \omega \left( \mbox{Re}\left[\frac{1}{\sigma (\omega)} \right] - \frac{1}{\mathcal{K}} \right) = 0. \label{sr2}
\ee
Explicit holographic computations verify this sum rule. As before, this sum rule is expected to be widely applicable, although it is 
difficult to come up with explicit non-holographic computations which fully preserve the non-perturbative S-duality.

\subsection{Standard approaches and their limitations}

We now describe a number of approaches applied to the computation of $\sigma (\omega)$ for CFT3s, and especially to the Wilson-Fisher fixed
point of the superfluid-insulator transition described by (\ref{Swf}). Similar methods can also be applied to other transport coefficients, and to other CFTs.
Each of the methods below has limitations, although their combination does yield significant insight.

\subsubsection{Quasiparticle-based methods}

We emphasized above that interacting CFT3s do not have any quasiparticle excitations. However, there are the exceptions of free CFT3s which do 
have infinitely long-lived quasiparticles. So we can hope that we could expand away from these free CFTs and obtain a controlled theory of interacting
CFT3s. After all, this is essentially the approach of the $\epsilon$ expansion in dimensionality 
by Wilson and Fisher for the critical exponents. However, we will see below
that the $\epsilon$ expansion, and the related vector $1/M$ expansion, have difficulty in describing transport because they do not hold uniformly in $\omega$.
Formally, the $\epsilon \rightarrow 0$ and the $\omega \rightarrow 0$ limits do not commute.

Let us first perform an explicit computation on a free CFT3: a CFT with $M$ two-component, massless Dirac fermions, $C$, with Lagrangian
\be
\mathcal{L} = \overline{C} \gamma^\mu \partial_\mu C.
\ee
We consider a conserved flavor U(1) current $J_\mu = - i \overline{C} \gamma_\mu \rho C$, where $\rho$ is a traceless flavor matrix normalized as $\mbox{Tr} \rho^2 = 1$. The $T=0$ Euclidean correlator of the currents is
\bea
\left\langle J_\mu (p) J_\nu ( - p) \right\rangle &=& - \int \frac{\mathrm{d}^3 k}{8 \pi^3} \frac{ \mbox{Tr} \left[ \gamma_\mu \gamma_\lambda k_\lambda 
\gamma_\nu \gamma_\delta ( k_\delta + p_\delta) \right] }{k^2 (p+k)^2} \nn\\
&=&- \frac{1}{2\pi} \left( \delta_{\mu\nu} - \frac{p_\mu p_\nu}{p^2} \right) 
\int\limits_0^1 \mathrm{d}x \sqrt{p^2 x (1-x)} \nn\\
&=& - \frac{1}{16} |p| \left( \delta_{\mu\nu} - \frac{p_\mu p_\nu}{p^2} \right).
\label{Kmn}
\eea
This is clearly of the form in (\ref{Kmn0}), and determines the value $\mathcal{K} = 1/16$.

The form of the current correlator of the CFT3 is considerably more subtle at $T>0$.  
In principle, this evaluation only requires replacing the frequency integral in (\ref{Kmn}) by a summation
over the Matsubara frequencies, which are quantized by odd multiples of $\pi T$. However, after performing this integration
by standard methods, the resulting expression should be continued carefully to real frequencies. 
We quote the final result for $\sigma(\omega)$, obtained from the current correlator via 
(\ref{Kubo})
\bea
\mbox{Re}[\sigma (\omega)]  &=& \frac{ T \log 2}{2} \, \delta (\omega) + \frac{1}{16} \tanh \left( \frac{|\omega|}{4T} \right) \,, \nonumber \\
\mbox{Im}[\sigma(\omega)] &=& \int\limits_{-\infty}^{\infty} \frac{\mathrm{d} \Omega}{\pi} \, \mathcal{P} \, 
\left( \frac{\mbox{Re}[\sigma (\Omega)] - 1/16}{\omega-\Omega} \right) ,
\label{drude}
\eea
where $\mathcal{P}$ is the principal part.
Note that in the limit $\omega \gg T$, (\ref{drude}) yields $\sigma (\omega) = \mathcal{K} = 1/16$, as expected.
However, the most important new feature of (\ref{drude}) is
the delta function at zero frequency in the real part of the conductivity, with weight proportional to $T$.
This is a consequence of the presence of quasiparticles, which have been thermally excited and can transport charge ballistically
in the absence of any collisions between them.

We can now ask if the zero frequency delta function is preserved once we move to an interacting CFT. A simple way to realize an interacting CFT3 
is to couple the fermions
to a dynamical U(1) gauge field, and examine the theory for large $M$. It is known that the IR physics is then described by a interacting CFT3 \cite{Appelquist:1988sr,Wen:1993zza,Chen:1993cd,Sachdev97,SVBSF,SBSVF,Strack13,Strack15} and its $T=0$ properties can be computed in a systematic $1/M$ expansion (we emphasize that this is a `vector' $1/M$ expansion, unlike the matrix large $N$ models considered elsewhere in this review). The same remains true at $T>0$ for the conductivity, provided we focus on the $\omega \gg T$ region of (\ref{drude}) 
(we will have more to say about this region in the following subsection). However, now collisions are allowed between the quasiparticles that were not present at $M=\infty$, and these collisions cannot be treated in a bare $1/M$ expansion. Rather, we have to examine the effects of repeated collisions, and ask if they lead to charge diffusion and a finite conductivity. This is the precise analog of the question Boltzmann asked for the classical ideal gas, and he introduced the Boltzmann equation to relate the long-time Brownian motion of the molecules to their two-particle collision cross-section. We can apply a quantum generalization of the Boltzmann equation to the
CFT3 in the $1/M$ expansion, where the quasiparticles of the $M=\infty$ CFT3 undergo repeated collisions at order $1/M$ by exchanging quanta of the U(1) guage field. The collision cross-section can be computed by Fermi's golden rule and this then enters the collision term in the quantum Boltzmann equation \cite{Damle97,Sachdev97}.

\begin{figure}
\centering
\includegraphics[width=3.4in]{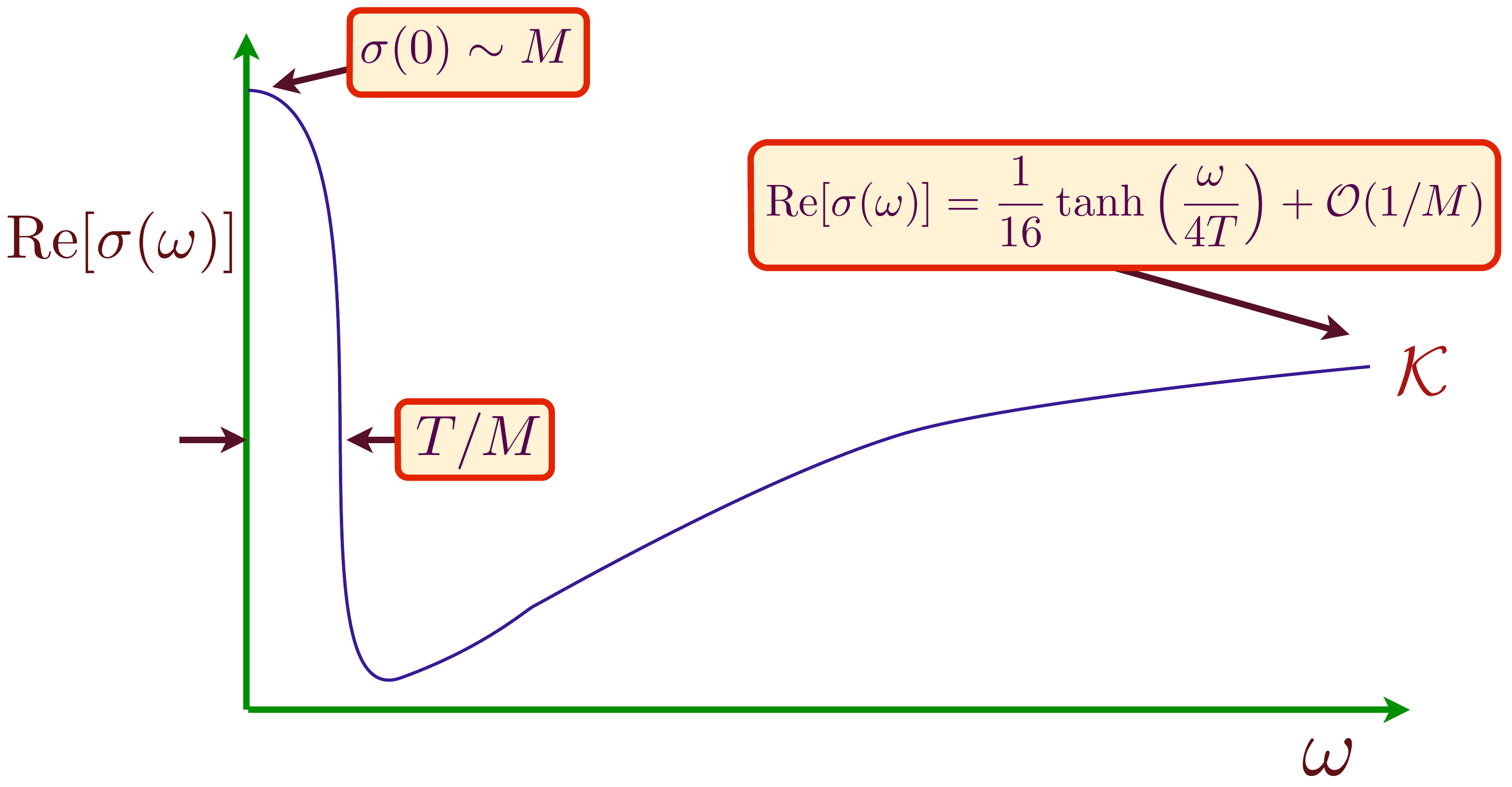}
\caption{\label{fig:damle} \textbf{Schematic of the real part of the conductivity of a CFT3 of $M$ Dirac fermions coupled to a U(1) gauge field
in the large $M$ limit.} Similar features apply to other CFT3s in the vector large $M$ limit. The peak at zero frequency is a 
remnant of the quasiparticles present at $M=\infty$, and the total area under this peak equals $(T \ln 2)/2$ as $M \rightarrow \infty$.}
\end{figure}

One subtlety should be kept in mind while considering the analogy with the classical ideal gas. The free CFT3 has both particle and anti-particle quasiparticles, and these move in opposite directions in the presence of an applied electric field. So the collision term must also consider collisions between particles and anti-particles. Such collisions have the feature that they can degrade the electrical current while conserving total momentum. Consequently, a solution of the Boltzmann equation
shows that the zero-frequency conductivity is {\em finite\/} even at $T>0$, after particle-anti-particle collisions have been accounted for \cite{Damle97,ssbook}. 
As in the usual Drude expression for the conductivity, the zero frequency conductivity is inversely proportional to the collision rate. As the latter is proportional
to $1/M$, we have $\sigma (0) \sim M$. Similarly, we can consider the frequency dependence of the conductivity at $\omega \ll T$, and find that the 
width of the peak in the conductivity extends up to frequencies of order $T/M$.
So the zero frequency delta function in (\ref{drude}) has broadened into peak of height $M$ and width $T/M$, as sketched in Figure \ref{fig:damle}. 
It required a (formally uncontrolled) resummation of the $1/M$ expansion using the Boltzmann equation to obtain this result.
And it should now also be clear from a glance at Fig.~\ref{fig:damle} that the $\omega \rightarrow 0$ and $M \rightarrow \infty$ limits do not commute
at $T>0$.

A similar analysis can also be applied to the Wilson-Fisher CFT3 described by (\ref{Swf}), using a model with a global O($M$) symmetry. 
The Boltzmann equation result for the zero frequency conductivity is $\sigma (0) = 0.523 M$, where the boson superfluid-insulator transition corresponds to the case $M=2$ \cite{Sachdev97,Krempa12}.

Our main conclusion here is that the vector $1/M$ expansion of CFT3s, which expands away from the quasiparticles present at $M=\infty$, 
yields fairly convincing evidence that $\sigma (\omega)$ is a non-trivial universal function of $\omega/T$. However, it does not appear to be a reliable
way of computing $\sigma (\omega)$ for $\omega < T$ and smaller values of $M$.

\subsubsection{Short time expansion}
\label{sec:OPE}

Transport involves the long time limit of the correlators of conserved quantities, but we can nevertheless ask if any useful
information can be obtained from short time correlators. For lattice models with a finite Hilbert space at each site, the short time expansion
is straightforward: it can be obtained by expanding the time evolution operator $\mathrm{e}^{-\mathrm{i} H t /\hbar}$ in powers of $t$. Consequently, the short time
expansion of any correlation function only involves integer powers of $t$. 

The situation is more subtle in theories without an ultraviolet cutoff because the naive expansion of the time evolution operator leads to ultraviolet divergencies. 
For CFTs these ultraviolet divergencies can be controlled by a renormalization procedure, and this is expressed in terms of a technology
known as the operator product expansion (OPE). Detailed reviews of the OPE are available elsewhere \cite{Rychkov,Rychkov2}, 
and here we only use a few basic results to examine
its consequences for the two-point correlator of the current operator at short times. After a Fourier transform from time to energy, the short time expansion
translates into a statement for the large frequency behavior of the conductivity of CFT3s, at frequencies much larger than $T$ \cite{Katz14}
\be
\sigma (\omega\gg T ) = \mathcal{K} - b_1 \left( \frac{T}{\mathrm{i} \omega} \right)^{3-1/\nu} - b_2 \left( \frac{T}{\mathrm{i} \omega} \right)^3 + \ldots .
\label{ope}
\ee
The leading term in (\ref{ope}) is clearly in agreement with (\ref{sigmaK}). The first sub-leading term, proportional to the numerical constant $b_1$, is determined by the OPE between the currents and a scalar operator with a scaling dimension given in (\ref{thermalop}). For the Wilson-Fisher CFT3, 
the exponent $\nu$ is the same as that determining the correlation length in the superfluid or insulator phases on either side of the critical point.
The value of $b_1$ is a characteristic property of the CFT3, and can be computed in a vector $1/M$ expansion, or by the numerical study to be described
in the following subsection. The final term displayed above is a consequence of the OPE of the currents with the stress-energy tensor (an operator of dimension 3). The term with coefficient $b_2$ will be purely imaginary.
Similarly, operators with higher dimensions can be used to obtain additional subleading corrections.   If we deform the CFT3 by adding a homogenenous source $g$ to the operator $\mathcal{O}$, then (\ref{ope}) generalizes to \cite{Lucas:2016fju} \begin{equation}
\sigma (\omega \gg T ) = \mathcal{K} - c_1  \frac{g}{(\mathrm{i} \omega)^{1/\nu}}   - c_2 \frac{\langle \mathcal{O}\rangle(g,T)}{(\mathrm{i} \omega)^{3-1/\nu}}  + \ldots.
\end{equation}
This formula remains true outside of the quantum critical fan in many circumstances.   Furthermore, the ratio $c_1/c_2$ is universal and depends only on $d$ and $\nu$.

\begin{figure}
\centering
\includegraphics[width=3.4in]{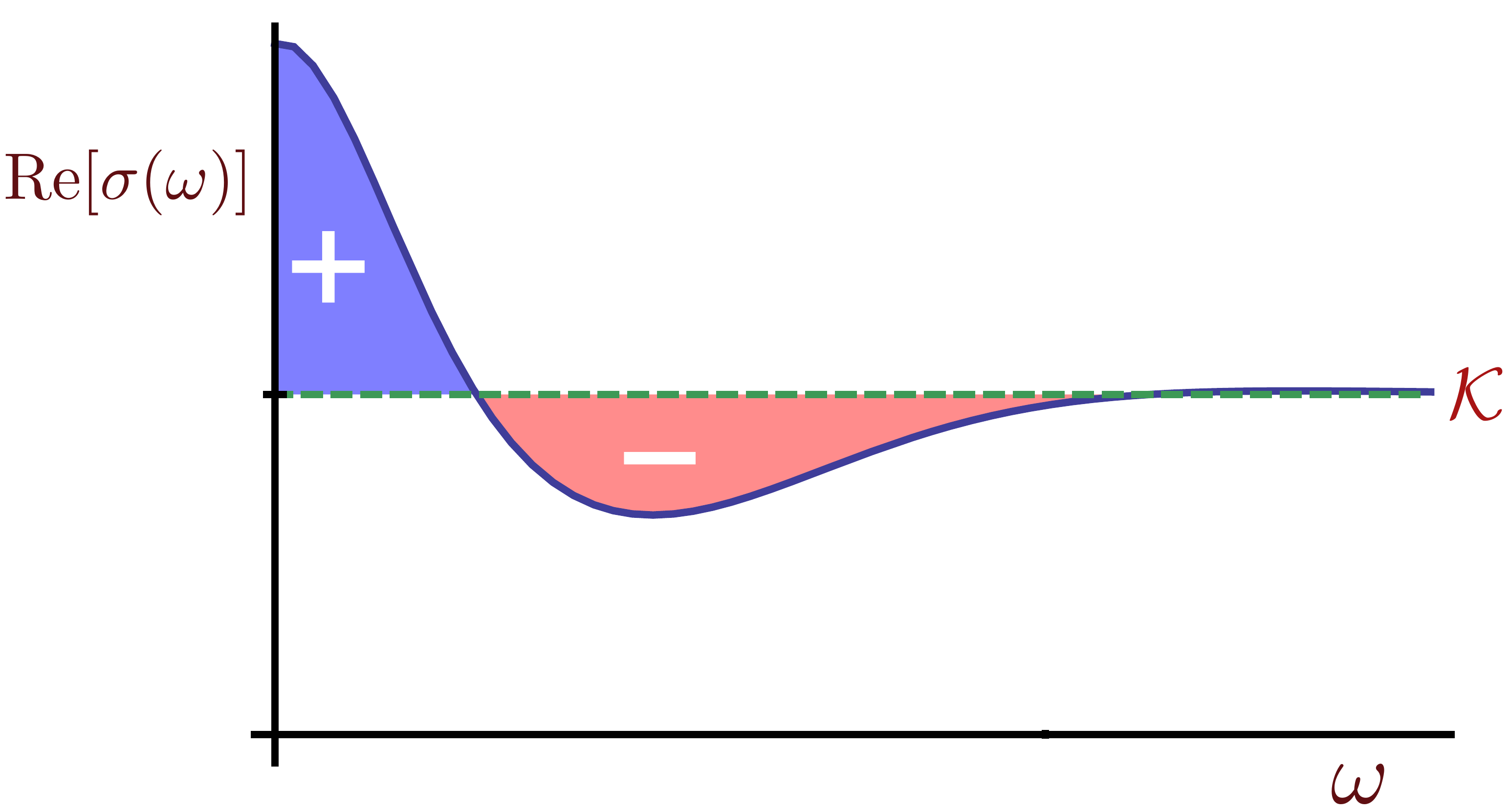}
\caption{\label{fig:sumrule} \textbf{Schematic of the real part of the conductivity of a CFT3} as constrained by the OPE in (\ref{ope}) 
and the sum rule in (\ref{sr1}). The areas of the shaded regions are equal to each other.}
\end{figure}
The high frequency expansion (\ref{ope}) does not directly place any constraints on the conductivity at low frequencies, $\omega \ll T$. However, in combination with the sum rule in (\ref{sr1}), we can make an interesting observation. 
The $1/M$ expansion for the Wilson-Fisher CFT, and the numerical results presented below, show that the constant $b_1 > 0$.
Consequently we have $\text{Re}[\sigma (\omega)] < \mathcal{K}$ at large enough $\omega$. The sum rule in (\ref{sr1}) now implies
that there must be a region at smaller $\omega$ where we have $\text{Re}[\sigma (\omega)] > \mathcal{K}$. If we make the
assumption that $\text{Re}[\sigma (\omega)] - \mathcal{K}$ crosses zero only once (this assumption is seen to fail in some of the holographic examples considered below), we can conclude that $\text{Re}[\sigma (\omega)]$ should have the form in 
Figure \ref{fig:sumrule}, with the two shaded regions having equal area. The qualitative form of $\text{Re}[\sigma (\omega)]$ has similarities
to the quasiparticle result in Figure \ref{fig:damle}, although we have not made any use of quasiparticles in the present argument. 

\subsubsection{Quantum Monte Carlo}

Quantum Monte Carlo studies perform a statistical sampling of the spacetime configurations of the imaginary time path integral
of quantum systems. So they are only able to return information on correlation functions defined at imaginary Matsubara frequencies.
However, even in such Euclidean studies there is the potential for a `sign problem', with non-positive weights, because of the presence of 
complex Berry phase terms \cite{ssbook}. 
For the CFT3s being studied here such sign problems are usually absent; however, the problems with a non-zero
chemical potential to be considered later in our review invariably do have a sign problem.

For the Wilson-Fisher CFT3 defined by (\ref{Swf}), there is an especially elegant model for efficient Monte Carlo studies. This is the lattice 
regularization provided by the Villain model. This model is defined in terms of integer valued variables, $j_{i,\mu}$, on the links of a cubic lattice.
Here $i$ labels the sites of a cubic lattice, and $\mu = \pm x, \pm y, \pm z$ represents the six links emerging from each site.
Each link has an orientation, so that $j_{i+\hat{\mu}, -\mu} = j_{i,\mu}$, just as conventional in lattice gauge theory. The $j_{i, \mu}$ represent
the currents of the complex scalar particle in (\ref{Swf}). These currents must be conserved, and so we have the zero divergence condition
\be
\Delta_\mu j_{i, \mu} = 0 \,,
\ee
where $\Delta_\mu$ is the discrete lattice derivative; this is the analog of Kirchoff's law in circuit theory.
Finally, the action of the Villain model is simply \cite{villain}
\be
S_{\mathrm{Villain}}= \frac{K}{2} \sum_{i , \mu} j_{i,\mu}^2 .
\ee
This model has a phase transition at a critical $K=K_c$, which is in the universality class of the O(2) Wilson-Fisher CFT3.

There have been extensive numerical studies of the Villain model, and sophisticated finite-size scaling methods have yielded detailed
information on $\sigma (i\omega_n)$ at the Matsubara frequencies, $\omega_n$, which are integer multiples of $2 \pi T$ \cite{smakov,krempa_nature,pollet,gazit13,gazit14,Katz14}. Note especially
that Monte Carlo does {\em not\/} yield any information at $\omega_n = 0$ (this is consequence of the structure of the Kubo formula for the 
conductivity, which is only defined at non-zero $\omega$, and reaches $\omega=0$ by a limiting process). So the smallest possible frequency at which we can determine
the conductivity is $2 \pi T \mathrm{i}$. The results at these, and higher Matsubara frequencies are all found \cite{Katz14} 
to be in excellent agreement with OPE
expansion in (\ref{ope}). So the conclusion is that the quantum Monte Carlo results are essentially entirely in the large $\omega$
regime, where the results of the OPE also apply. The simulations can therefore help determine the values of the numerical constants
$b_{1,2}$ in (\ref{ope}). While this agreement is encouraging, the somewhat disappointing conclusion is that quantum Monte Carlo
does not yield any more direct information on the low frequency behavior than is available from the OPE.

\subsection{Holographic spectral functions}
\label{sec:spectral}

In the next few sections we will cover holographic computations of the retarded Green's functions
for operators in quantum critical systems at nonzero temperature. In general we will be interested in multiple
coupled operators and so we write, in frequency space,
\be\label{eq:ho}
\delta \langle \ocal_A \rangle(\omega,k) = G^{\mathrm{R}}_{\ocal_A \ocal_B}(\omega,k) \delta h_B(\omega,k) \,.
\ee
A sum over $B$ is implied.
Here $\delta \langle \ocal_A \rangle$ is the change in the expectation value of the operator $\ocal_A$ due
to the (time and space dependent) change in the sources $\delta h_B$. From the holographic dictionary discussed in
previous sections, we know that expectation values and sources are given by the near boundary behavior
of the bulk fields $\{\phi_A\}$ dual to the operators $\{\ocal_A\}$. Therefore
\be\label{eq:GRAB}
G^{\mathrm{R}}_{\ocal_A \ocal_B} = \frac{\delta \langle \ocal_A \rangle}{\delta h_B}
= (2 \Delta - D_\text{eff.})\frac{\delta \phi_{A(1)}}{\delta \phi_{B (0)}} \,.
\ee
Here $D_{\mathrm{eff}} = d+z$, and $\phi_{(0)}$ and $\phi_{(1)}$ are as defined in (\ref{eq:boundaryL}) above. The near boundary behavior
of the fields has the same form at zero and nonzero temperature because, as we saw in \S\ref{sec:temp} above,
putting a black hole in the spacetime changes the geometry in the interior (IR), but not near the boundary (UV).

To compute the Green's function (\ref{eq:GRAB}) in a given background, one first perturbs the background fields
\be
\phi_A(r) \mapsto \phi_A(r) + \delta \phi_A(r) \mathrm{e}^{- \mathrm{i} \omega t + \mathrm{i} k \cdot x} \,.
\ee
Linearizing the bulk equations of motion leads to coupled, linear, ordinary differential equations for the perturbations $\delta\phi_A(r)$. Once these linearized equations are solved, the asymptotic behavior of fields can be read off from the solutions and the Green's function obtained from (\ref{eq:GRAB}). A crucial part of solving the equations for the perturbations is to impose the correct boundary conditions at the black hole horizon. One way to understand these boundary conditions is to recall that the retarded Green's function is obtained from the Euclidean (imaginary time) Green's function by analytic continuation from the upper half frequency plane (i.e. $\omega_n >0$). An important virtue of the holographic approach is that this analytic continuation can be performed in a very efficient way, as we now describe.

\subsubsection{Infalling boundary conditions at the horizon}
\label{sec:infal}

Starting with the Euclidean black hole spacetime (\ref{eq:BH2}), we described how to zoom in on the spacetime near the horizon in the manipulations following equation (\ref{eq:BHnearh}). To see how fields behave near the horizon, we can consider the massive wave equation in the black hole background (\ref{eq:thermalscreening}) at $r \sim r_+$. The equation becomes
\be
\phi'' + \frac{1}{r-r_+} \phi' - \frac{\omega_n^2}{(4 \pi T)^2 (r-r_+)^2} \phi = 0 \,.
\ee
The temperature $T$ is as given in (\ref{eq:T}). The two solutions to this equation are (recall that the coordinate range $r \leq 0 \leq r_+$)
\be
\phi_\pm = \left(r_+ - r \right)^{\pm \omega_n/(4 \pi T)} \,.
\ee
For $\omega_n >0$, it is clear that the regular solution is the one that decays as $r \to r_+$ (that is, the positive sign in the above equation). Once this condition is imposed on all fields, the solution is determined up to an overall constant, which will drop out upon taking the ratios in (\ref{eq:GRAB}) to compute the Green's function.

The real time equations of motion are the same as the Euclidean equations, with the substitution $\mathrm{i} \omega_n \to \omega$. In particular, the regular boundary condition at the horizon analytically continues to the ``infalling" boundary condition
\be\label{eq:infalling}
\phi_\text{infalling} = \left(r_+ - r \right)^{- \mathrm{i} \omega/(4 \pi T)} \,.
\ee
This behavior is called infalling because if we restore the time dependence, then
\be
\phi_\text{infalling} = \mathrm{e}^{- \mathrm{i} \omega(t + \frac{1}{4 \pi T} \log(r_+- r))} \,,
\ee
corresponding to modes that carry energy towards $r = r_+$ as $t \to \infty$. That is, they are modes that fall towards the black hole (rather than emerge out of the black hole). This is indeed the required behavior of physical excitations near an event horizon, and can also be derived directly from the real time, Lorentzian, equations of motion. Infalling modes are required for regularity in the Kruskal coordinates that extend the black hole spacetime across the future horizon (see e.g. \cite{Hartnoll:2009sz}). The derivation via the Euclidean modes makes clear, however, that infalling boundary conditions correspond to computing the retarded Green's function.

The infalling boundary condition (\ref{eq:infalling}) allows the retarded Green's function to be computed directly at real frequencies. This is a major advantage relative to e.g. Monte Carlo methods that work in Euclidean time. The infalling boundary condition is also responsible for the appearance of dissipation at leading order in the classical large $N$ limit, at any frequencies. This is also a key virtue relative to e.g. vector large $N$ expansions, in which $\omega \to 0$ and $N \to \infty$ do not commute at nonzero temperatures, as we explained above. Infalling modes have an energy flux into the horizon that is lost to an exterior observer. The black hole horizon has geometrized the irreversibility of entropy production. Physically, energy crossing the horizon has dissipated into the `order $N^2$' degrees of freedom of the deconfined gauge theory.

Infalling boundary conditions were first connected to retarded Green's function in \cite{Son:2002sd}. We derived the infalling boundary condition from the Euclidean Green's function. An alternative derivation of this boundary condition starts from the fact that the fully extended Penrose diagram of (Lorentzian) black holes has two boundaries, and that these geometrically realize the thermofield double or Schwinger-Keldysh approach to real time thermal physics \cite{Herzog:2002pc,Skenderis:2008dh}.

\subsubsection{Example: spectral weight $\text{Im}\, G^{\mathrm{R}}_{\ocal \ocal}(\omega)$ of a large dimension operator}
\label{sec:largeDelta}

We will now illustrate the above with a concrete example. We will spell out here in some detail steps that we will go over more quickly in later cases. To start with we will put $k=0$ and obtain the dependence of the Green's function on frequency. The scalar wave equation (\ref{eq:thermalscreening}) in the Lorentzian signature black hole background becomes
\be\label{eq:phiT}
\phi'' + \left(\frac{f'}{f} - \frac{z+d-1}{r} \right) \phi' - \frac{1}{f}\left( \frac{(mL)^2}{r^2} - \frac{r^{2z} \omega^2}{r^2 f}\right)\phi = 0 \,.
\ee
To develop an intuition for an equation like the above, it is often useful to put the equation into a Schr\"odinger form. To do this we set $\phi = r^{d/2} \psi$, so that (\ref{eq:phiT}) becomes
\be\label{eq:sch}
- \frac{f}{r^{z-1}} \frac{\mathrm{d}}{\mathrm{d}r} \left(\frac{f}{r^{z-1}} \frac{\mathrm{d}}{\mathrm{d}r} \psi \right) + V \psi = \omega^2 \psi \,.
\ee
If we define $r_*$ such that $\partial_{r_*} = r^{1-z}f \partial_r$, we immediately recognize the above equation as the time-independent Schr\"odinger equation in one spatial dimension.  All the physics is now contained in the Schr\"odinger potential
\be\label{eq:VV}
V(r) = \frac{f}{r^{2z}} \left((Lm)^2  + \frac{d \, (d+2z) f}{4} - \frac{d \, r f'}{2} \right) \,.
\ee

The Schr\"odinger equation (\ref{eq:sch}) is to be solved subject to the following boundary conditions. We will discuss first the near-horizon region, $r \to\ r_+$. The Schr\"odinger coordinate $r_\star = \int \mathrm{d}r \, \left( r^{z-1}/f \right) \to \infty$ as $r \to r_+$, so this is a genuine asymptotic region of the Schr\"odinger equation. The potential (\ref{eq:VV}) vanishes on the horizon because $f(r_+)=1$.  Therefore for any $\omega^2>0$, recall that $\omega^2$ plays the role of the energy in the Schr\"odinger equation (\ref{eq:sch}), the solution oscillates near the horizon. The infalling boundary condition is thus seen to translate into the boundary condition that modes are purely outgoing as $r_\star \to \infty$. This is a scattering boundary condition of exactly the sort one finds in elementary one dimensional scattering problems.

The boundary at $r=0$ is not an asymptotic region of the Schr\"odinger equation. There will be normalizable and non-normalizable behaviors of $\phi$ as $r \to 0$. To compute the Green's function we will need to fix the coefficient of the non-normalizable mode and solve for the coefficient of the normalizable mode.

A key physical quantity is the imaginary part of the retarded Green's function (also called the spectral weight) of the operator $\ocal$ dual to $\phi$. The imaginary part of the retarded Green's function is a direct measure of the entropy generated when the system is subjected to the sources $\delta h_B$ of equation (\ref{eq:ho}). Specifically, with an assumption of time reversal invariance (see e.g. \cite{Hartnoll:2009sz}), the time-averaged rate of work $w$ done on the system per unit volume by a spatially homogeneous source driven at frequency $\omega$ is given by
\be\label{eq:dissip}
\overline{\frac{\mathrm{d}w}{\mathrm{d}t}} = \omega \delta h^*_A \text{Im} G^{\mathrm{R}}_{AB}(\omega) \delta h_B \geq 0 \,.
\ee
We now illustrate how this spectral weight is given by the amplitude with which the field can tunnel from the boundary $r=0$ through to the horizon. Thus, we will exhibit the direct connection between dissipation in the dual QFT and rate of absorption by the horizon.

The analysis is simplest in the limit of large $mL \gg 1$. Recall that this corresponds to an operator $\ocal$ with large scaling dimension. In this limit the equation (\ref{eq:phiT}) can be solved in a WKB approximation. In the large mass limit, the potential (\ref{eq:VV}) typically decreases monotonically from the boundary towards the horizon. There is therefore a unique turning point $r_{\mathrm{o}}$ at which $V(r_{\mathrm{o}}) = \omega^2$. The infalling boundary condition together with standard WKB matching formulae give the solution \begin{widetext}
\be
\psi =
\begin{cases}
\displaystyle \exp\left\{ \mathrm{i} \int_{r_{\mathrm{o}}}^r \frac{r^{z-1}\mathrm{d}r}{f} \sqrt{\omega^2 - V(r)} - \frac{\mathrm{i} \pi}{4} \right\}& r_+ > r > r_{\mathrm{o}} \,, \\
\displaystyle  \exp\left\{ \int_{r}^{r_{\mathrm{o}}} \frac{r^{z-1}\mathrm{d}r}{f} \sqrt{V(r) - \omega^2} \right\} + \frac{\mathrm{i}}{2} \exp\left\{ - \int_{r}^{r_{\mathrm{o}}} \frac{r^{z-1}\mathrm{d}r}{f} \sqrt{V(r) - \omega^2} \right\} & r_0 > r \,.
\end{cases}
\ee
In this limit $V = (m L)^2 f/r^{2z}$. Expanding the above solution near the $r \to 0$ boundary, the retarded Green's function is obtained from the general formula (\ref{eq:GRAB}). In particular, the imaginary part is given by
\begin{align}
\chi''(\omega)  & \; \equiv \; \text{Im} G^{\mathrm{R}}_{\ocal \ocal}(\omega)  \propto \;  r_{\mathrm{o}}^{- 2 mL} \exp\left\{ - 2 \int_{0}^{r_{\mathrm{o}}} \frac{\mathrm{d}r}{r} \left(\sqrt{\frac{(mL)^2}{f} - \frac{r^{2z} \omega^2}{f^2}} - mL\right) \right\} \label{eq:chiWKB} \\[0.2cm]
& \; = \; \text{Probability for quanta of $\phi$ to tunnel from boundary into} \nonumber \\
& \qquad \text{the near horizon region (`transmission probability').} \nonumber
\end{align}
\end{widetext}
From the above expression the expected limits of high and low frequency are immediately obtained. Note that for $\omega r^z_+ \ll 1$ then $r_{\mathrm{o}} \sim r_+$, whereas for $\omega r^z_+ \gg 1$, $r_{\mathrm{o}} \sim \omega^{-1/z}$. Thus
\be
\chi''(\omega) \sim
\begin{cases}
\omega^{2 \Delta/z} & \omega \gg T \,, \\
T^{2 \Delta/z} & \omega \ll T \,.
\end{cases} \qquad \qquad \Delta \gg 1 \,. \label{eq:ld}
\ee
On general grounds $\chi''$ must be an odd function of $\omega$. This can be seen in a more careful WKB computation that includes the order one prefactor. Relatedly, to obtain (\ref{eq:ld}) we used the fact that the scaling dimension (\ref{eq:deltaL}) is given by $\Delta = m L$ in the limit $mL \gg 1$. We also used the expression (\ref{eq:Trp}) for the temperature. The full result (\ref{eq:chiWKB}) gives an explicit and relatively simple
crossover between the two limits in (\ref{eq:ld}).

More generally, away from the WKB limit, the equation (\ref{eq:phiT}) can easily be solved numerically. Later we will give an example of the type of \texttt{Mathematica} code one uses to solve such equations. We should note, however, that for general values of $\Delta$ (e.g. not integer, etc.) the boundary conditions often need to be treated to a very high accuracy to get stable results, as discussed in e.g \cite{Denef:2009tp}. Also, one should be aware that when the fast and slow falloffs differ by an integer, one can find logarithmic terms in the near boundary expansion. These must also be treated with care and, of course, correspond to the short distance logarithmic running of couplings in the dual QFT that one expects in these cases.

\subsubsection{Infalling boundary conditions at zero temperature}

In equation (\ref{eq:infalling}) above we obtained the infalling boundary condition for finite temperature horizons. Zero temperature geometries that are not gapped will also have horizons in the far interior (or, more generally, as we noted in our discussion of Lifshitz geometries in \S\ref{sec:scaling} above, mild null singularities). To compute the retarded Green's function from such spacetimes we will need to know how to impose infalling boundary conditions in these cases. This is done as before, by analytic continuation of the Euclidean mode that is regular in the upper half complex frequency plane. It is not always completely straightforward to find the leading behavior of solutions to the wave equation near a zero temperature horizon, especially when there are many coupled equations. The following three examples illustrate common possibilities. In each case we give the leading solution to $\nabla^2 \phi = m^2 \phi$ near the horizon. 
\begin{enumerate}
\item Poincar\'e horizon. Near horizon metric (as $r \to \infty$):
\be
\mathrm{d}s^2 = L^2 \left( \frac{-\mathrm{d}t^2 + \mathrm{d}r^2 + \mathrm{d}\vec x^2_d}{r^2} \right) \,.
\ee
Infalling mode as $r \to \infty$ and with $\omega^2 > k^2$
\be
\phi_\text{infalling} = r^{d/2} \mathrm{e}^{\mathrm{i} r \sqrt{\omega^2 - k^2}} \,.
\ee
\item Extremal $\mathrm{AdS}_2$ charged horizon (see \S \ref{sec:compressible}). Near horizon metric (as $r \to \infty$):
\be
\mathrm{d}s^2 = L^2 \left( \frac{-\mathrm{d}t^2 + \mathrm{d}r^2}{r^2}  + \mathrm{d}\vec x^2_d \right) \,.
\ee
Infalling mode as $r \to \infty$
\be\label{eq:ads2infall}
\phi_\text{infalling} = \mathrm{e}^{\mathrm{i} \omega r} \,.
\ee
\item Lifshitz `horizon'. Near horizon metric (as $r \to \infty$):
\be
\mathrm{d}s^2 = L^2 \left( \frac{-\mathrm{d}t^2}{r^{2z}} + \frac{\mathrm{d}r^2  + d\vec x^2_d}{r^2} \right) \,.
\ee
Infalling mode as $r \to \infty$, with $z > 1$,
\be
\phi_\text{infalling} = r^{d/2} \mathrm{e}^{\mathrm{i} \omega r^z/z} \,.
\ee
\end{enumerate}

Finally, occasionally one wants to find the Green's function directly at $\omega = 0$ (for instance, in our discussion of thermal screening around (\ref{eq:kWKB}) above). The equations at $\omega = 0$ typically exhibit two possible behaviors near the horizon, one divergent and the other regular. At finite temperature the divergence will be logarithmic. Clearly one should use the regular solution to compute the correlator.

\subsection{Quantum critical charge dynamics}
\label{sec:QCcharge}

So far we have discussed scalar operators $\ocal$ as illustrative examples. However, a very important set of operators are instead currents $J^\mu$ associated to conserved charges. The retarded Green's functions of current operators capture the physics of charge transport in the quantum critical theory. We explained around equation (\ref{eq:JA}) above that a conserved $\mathrm{U}(1)$ current $J^\mu$ in the boundary QFT is dually described by a Maxwell field $A_a$ in the bulk. Therefore, to compute the correlators of $J^\mu$, we need an action that determines the bulk dynamics of $A_a$.

\subsubsection{Conductivity from the dynamics of a bulk Maxwell field}
\label{sec:maxw}

As we are considering zero density quantum critical matter in this section, we restrict to situations where the Maxwell field is not turned on in the background (we will relax this assumption in \S \ref{sec:compressible} below). Therefore, in order to obtain linear response functions in a zero density theory, it is sufficient to consider a quadratic action for the Maxwell field about a fixed background. The simplest such action is the Maxwell action (with $F=dA$, as usual)
\be\label{eq:Maxwell}
S[A] = - \frac{1}{4 e^2} \int \mathrm{d}^{d+2}x \sqrt{-g} F^2 \,,
\ee
so we will start with this. We noted in \S\ref{sec:consistent} that Einstein-Maxwell theory can be obtained as a consistent truncation of explicit bulk theories with known QFT duals.

Before obtaining the Maxwell equations we should pick a gauge. A very useful choice for many purposes in holography is the `radial gauge', in which we put $A_r = 0$. Among other things, the bulk Maxwell field in this gauge has the same nonzero components as the boundary current operator $J^\mu$. The Maxwell equations about a black hole geometry of the form (\ref{eq:BH2}) can now be written out explicitly. Firstly, write the Maxwell field as
\be
A_\mu = a_\mu(r) \mathrm{e}^{- \mathrm{i} \omega t + \mathrm{i} k x}\,, \label{eq:aform}
\ee
where without loss of generality we have taken the momentum to be in the $x$ direction. The equations of motion,
$\nabla_a F^{ab} = 0$, become coupled ordinary differential equations for the $a_\mu(r)$. By considering the discrete symmetry $y \to -y$ of the background, where $y$ is a boundary spatial dimension orthogonal to $x$, we immediately see that the perturbation will decouple into longitudinal ($a_t$ and $a_x$) and transverse ($a_y$) modes. It is then useful to introduce the following `gauge invariant' variables \cite{Kovtun:2005ev}, that are invariant under a residual gauge symmetry $A_\mu \to A_\mu + \pa_\mu [\lambda \, \mathrm{e}^{- \mathrm{i} \omega t + \mathrm{i} k x}]$,
\be\label{eq:gaugeinvar}
a_\perp(r) = a_y(r) \,, \qquad a_\parallel(r) = \omega a_x(r) + k a_t(r) \,.
\ee
In terms of these variables, the Maxwell action becomes: \begin{widetext}\begin{equation}
S = -\frac{L^{d-2}}{2e^2}\int \mathrm{d}r\; \left[fr^{3-d-z}a_\perp^{\prime2} + a_\perp^2\left(k^2 r^{3-d-z}-\frac{\omega^2 r^{z+1-d}}{f}\right) + \frac{fr^{3-d-z}}{\omega^2-k^2fr^{2-2z}}a_\parallel^{\prime2} - \frac{r^{z+1-d}}{f}a_\parallel^2\right] \,,
\end{equation}
\end{widetext}
leading to the following two decoupled second order differential equations
\begin{subequations}\begin{align}
\left(\frac{fr^{3-d-z}}{\omega^2 - k^2 fr^{2-2z}} a_\parallel^\prime\right)^\prime &=  - \frac{r^{z+1-d}}{f}a_\parallel , \label{eq:ax} \\
\left(fr^{3-d-z}a_\perp^\prime\right)^\prime  &=  k^2 r^{3-d-z}a_\perp - \frac{\omega^2 r^{z+1-d}}{f}a_\perp . \label{eq:ay}
\end{align}\end{subequations}
When $k=0$ the transverse and longitudinal equations are the same, as we should expect. These equations with $z=1$ have been studied in several important papers \cite{Policastro:2002se,Herzog:2002fn, Kovtun:2005ev, Herzog:2007ij}, sometimes using different notation. The gauge-invariant variables avoid the need to solve an additional first order equation that constrains the original variables $a_x$ and $a_t$.

Near the boundary at $r\rightarrow 0$, the asymptotic behavior of $a_\mu$ is given by: \begin{subequations}\label{eq:34nearb}\begin{align}
a_t &= a_t^{(0)} + a_t^{(1)} r^{d-z} + \cdots, \label{eq:atlif} \\
a_i &= a_i^{(0)} + a_i^{(1)} r^{d+z-2} + \cdots.\label{eq:ailif}
\end{align}\end{subequations}
Note that when $z\ne 1$, as expected, the asymptotic falloffs are different for the timelike and spacelike components of $a_\mu$.   So long as $d>z$, it is straightforward to add boundary counterterms to make the bulk action finite. We can then, using the general formalism developed previously, obtain the expectation values for the charge and current densities:
 \begin{subequations}\begin{align}
\langle J^t \rangle &= -\frac{L^{d-2}}{e^2} (d-z)a_t^{(1)},  \label{eq:jt1} \\
\langle J^i \rangle &= \frac{L^{d-2}}{e^2} (d+z-2)a_i^{(1)}.  \label{eq:jx1}
\end{align}\end{subequations}
If instead $d<z$, the `subleading' term in (\ref{eq:atlif}) is in fact the largest term near the boundary. This leads to many observables having a strong dependences on short distance physics --- we will see an example shortly when we compute the charge diffusivity. A closely related fact is that when $d<z$ there is a relevant (double trace) deformation to the QFT given by $\int \mathrm{d}^{d+1}x \; \rho^2$. As discussed in \S \ref{sec:multitrace} above, and first emphasized in \cite{Hartnoll:2009ns}, this deformation will generically drive a flow to a new fixed point in which the role of $a_t^{(0)}$ and $a_t^{(1)}$ are exchanged. The physics of such a theory is worth further study, as there are known examples (nematic or ferromagnetic critical points in metals, at least when treated at the `Hertz-Millis' mean field level \cite{PhysRevB.14.1165,metlitski1}) of $d=2$ and $z=3$.

A quantity of particular interest is the  frequency dependent conductivity
\be\label{eq:optical}
\sigma(\omega)  = \frac{\langle J_x(\omega)\rangle}{E_x(\omega)} = \frac{\langle J_x(\omega)\rangle }{\mathrm{i}\omega a^{(0)}_x(\omega)} = \frac{G^{\mathrm{R}}_{J_xJ_x}(\omega)}{\mathrm{i} \, \omega} \,.
\ee
As we have noted above, the dissipative (real) part of the conductivity will control entropy generation due to Joule heating when a current is driven through the system. We shall compute the optical conductivity (\ref{eq:optical}) shortly, but first consider the physics of certain low energy limits.

\subsubsection{The dc conductivity}
\label{sec:dccond}

The dc conductivity
\be
\sigma = \lim_{\omega \to 0} \, \sigma(\omega) \,,
\ee
determines the dissipation (\ref{eq:dissip}) due to an arbitrarily low frequency current being driven through the system. It should therefore be computable in an effective low energy description of the physics. Following our discussion of Wilsonian holographic renormalization in \S\ref{sec:wilson}, this means that we might hope to obtain the dc conductivity from a computation in purely the near-horizon, far interior, part of the spacetime. Indeed this is the case. We will follow a slightly modernized (in the spirit of \cite{Donos:2014yya}) version of the logic in \cite{Iqbal:2008by}, which in turn built on \cite{Kovtun:2003wp}. The upshot is the formula (\ref{eq:sigma}) below for the dc conductivity which is expressed purely in terms of data at the horizon itself.

The dc conductivity can be obtained by directly applying a uniform electric field, rather than taking the $k,\omega \to 0$ limit of (\ref{eq:ay}). Consider the bulk Maxwell potential
\be\label{eq:unif}
A = (-Et + a_x(r)) \mathrm{d}x \,.
\ee
Given the near boundary expansion (\ref{eq:ailif}), the QFT source term is given by the non-normalizable constant term $a_x^{(0)}$ of the bulk field near the boundary $r \to 0$.   The boundary electric field is then obtained from (\ref{eq:unif}) as $E = - \pa_t a_x^{(0)} = - F_{tx}(r\rightarrow 0)$, and is uniform.
To obtain the dc conductivity we must now determine the uniform current response.

The bulk Maxwell equations can be written as $\partial_a(\sqrt{-g} F^{ab})=0$.\footnote{For vectors and antisymmetric tensors, covariant derivatives $\nabla_a X^{a\cdots} =  (-g)^{-1/2} \partial_a ( (-g)^{1/2} X^{a\cdots})$.}   For the field (\ref{eq:unif}) one immediately has that
\be\label{eq:Jcons}
\pa_r \left( \sqrt{-g} F^{rx} \right) = 0 \,.
\ee
We have thereby identified a radially conserved quantity. Furthermore, near the boundary
\be\label{eq:Jbdy}
\lim_{r \to 0} \sqrt{-g} F^{rx} = L^{d-2} (d+z-2)a_x^{(1)} = e^2 \langle J_x\rangle \,.
\ee
Here we used the near boundary expansion (\ref{eq:34nearb}), as well as (\ref{eq:jx1}) to relate $\langle J_x\rangle$ to $a_x^{(1)}$.

From (\ref{eq:Jcons}) and (\ref{eq:Jbdy}) it follows that we can obtain the field theory current response $\langle J_x \rangle$ if we are able to evaluate $\sqrt{-g} F^{rx}$ at any radius. It turns out that we can compute it at the horizon. It is a standard trick that physics near black hole horizons is often elucidated by going to infalling coordinates;  so we replace $t$ in favor of the `infalling Eddington-Finkelstein coordinate' $v$ by
\be
v = t + \int \frac{\sqrt{g_{rr}} \mathrm{d}r}{\sqrt{-g_{tt}}}\,.   \label{eq:eddfink}
\ee
It is easily seen that the (Lorentzian version of the) black hole spacetime (\ref{eq:BH2}) is regular at $r=r_+$ in the coordinates $\{v,r,\vec x\}$. Therefore, a regular mode should only depend on $t$ and $r$ through the combination $v$. It follows that, at the horizon,
\be\label{eq:EF}
\pa_r A_x\Big|_{r=r_+} =-  \sqrt{\frac{g_{rr}}{-g_{tt}}} \pa_t A_x\Big|_{r=r_+} \,.
\ee
This is exactly the maneuver outlined around equation (\ref{eq:firstin}) in the introductory discussion to obtain the `membrane paradigm' conductivity of an event horizon.   We find that 
\be
a_x^\prime(r_+) =  \frac{r_+^{z-1}}{f} E \,. \label{eq:JHor}
\ee
We are now able to compute the conductivity by evaluating 
\begin{align}
\sigma &= \frac{\langle J_x\rangle}{E} = \frac{1}{e^2} \lim_{r\rightarrow r_+} \sqrt{-g}g^{rr}g^{xx} a_x^\prime(r) \notag \\
&= \frac{L^{d-2}}{e^2}\left(f r_+^{3-d-z}\right) \frac{r_+^{z-1}}{f} = \frac{L^{d-2}}{e^2} r_+^{2-d} .\label{eq:sigma}
\end{align}
Using the relation (\ref{eq:Trp}) between $r_+$ and $T$, we obtain \begin{equation}
\sigma \sim T^{(d-2)/z}.
\end{equation}
Various comments are in order:
\begin{enumerate}
\item  (\ref{eq:sigma}) is an exact, closed form expression for the dc conductivity that comes from evaluating a certain radially conserved quantity at the horizon.
\item The temperature scaling of the conductivity is precisely that expected for a quantum critical system (without hyperscaling violation or an anomalous dimension for the charge density operator) \cite{Damle97}.
\item The derivation above generalizes easily to more complicated quadratic actions for the Maxwell field, such as with a nonminimal coupling to a dilaton \cite{Iqbal:2008by}.
\item Infalling boundary conditions played a crucial role in the derivation, connecting with older ideas of the `black hole membrane paradigm' discussed in \S\ref{sec:horizons} above \cite{Kovtun:2003wp, Iqbal:2008by}. Once again: this infalling boundary condition is the origin of nontrivial dissipation in holography.
\end{enumerate}

The same argument given above, applied to certain perturbations of the bulk metric rather than a bulk Maxwell field, gives a direct proof of a famous result for the shear viscosity $\eta$ over entropy density $s$ in a large class of theories with classical gravity duals \cite{Iqbal:2008by}:
\be
\frac{\eta}{s} = \frac{1}{4 \pi} \,.   \label{eq:KSSbound}
\ee
Here $\eta$ plays an analogous role to the dc conductivity $\sigma$ in the argument above. The shear viscosity was first computed for holographic theories in \cite{Policastro:2001yc} and the ratio above emphasized in \cite{Kovtun:2003wp,Kovtun:2004de}.

\subsubsection{Diffusive limit}\label{sec:diffusive}

The longitudinal channel includes fluctuations of the charge density. Because the total charge is conserved, this channel is expected to include a collective diffusive mode \cite{Kadanoff1963419}. This fact is why the longitudinal equation (\ref{eq:ax}) is more complicated than the transverse equation (\ref{eq:ay}). It is instructive to see how the diffusive mode can be explicitly isolated from (\ref{eq:ax}). We will adapt the argument in \cite{Starinets:2008fb}.

Diffusion is a process that will occur at late times and long wavelengths if we apply a source to the system to set up a nontrivial profile for the charge density, turn off the source, and then let the system evolve. Therefore diffusion should appear as a mode in the system that (i) satisfies infalling boundary conditions at the horizon and (ii) has no source at the asymptotic boundary. We will see in \S\ref{sec:quasinormal} below that these are the conditions that define a so-called quasinormal mode, which correspond precisely to the poles of retarded Green's functions in the complex frequency plane. More immediately, we must solve the longitudinal equation (\ref{eq:ax}) with these boundary conditions, and in the limit of small frequency and wavevector.

We cannot simply take the limit $\omega \to 0$ of the longitudinal equation (\ref{eq:ax}). This is because taking $\omega \to 0$ in this equation does not commute with the near horizon limit $r \to r_+$, at which $f \to 0$. As we take the low frequency limit, we need to ensure that the infalling boundary condition as $r \to r_+$ is correctly imposed. This can be achieved by writing
\be
a_\parallel(r) = f(r)^{- \mathrm{i} \omega/(4 \pi T)} S(r) \,. 
\ee
By extracting the infalling behavior (\ref{eq:infalling}) in this way, the equation satisfied by $S$ will no longer have a singular point at the horizon. $S$ must tend to a constant at the horizon. We can therefore safely expand $S$ in $\omega,k \to 0$.  
 We do this by setting $\omega = \epsilon\, \hat \omega$ and $k = \epsilon\, \hat k$ and then expanding in small $\epsilon$. With a little benefit of hindsight \cite{Starinets:2008fb}, we look for a solution of the form
\be
S(r) = \epsilon \, \hat \omega + \epsilon^2 s(r) + \cdots \,.
\ee
The resulting differential equation for $s$, to leading order as $\epsilon \to 0$, can be integrated explicitly. The solution that is regular at the horizon is \begin{widetext}
\be\label{eq:littles}
s(r) = - \int\limits_r^{r_+} \frac{\mathrm{d}r'}{f(r')} \left[  \frac{\mathrm{i} \hat \omega^2 f'(r')}{4 \pi T} -\frac{\mathrm{i} r'^{d-3-z} f'(r_+)}{4 \pi T r_+^{d-3+z}}  \left(\hat \omega^2 r'^{2z} - \hat k^2 f(r') r'^2 \right) \right]\,. 
\ee
\end{widetext}
To isolate the diffusive regime, consider $\omega \sim k^2$. This corresponds to taking $\hat \omega$ small. In this regime we can ignore the $\hat \omega^2$ terms in (\ref{eq:littles}). We had to keep these terms in the first instance in order to impose the regularity condition at the horizon and fix a constant of integration in the solution (\ref{eq:littles}). The full solution to the order we are working can now be written
\be\label{eq:bigS}
S(r) = \omega + \frac{\mathrm{i} k^2}{r_+^{d-2}} \int\limits^{r_+}_r \mathrm{d}r' r'^{d-z-1}  \,.
\ee
Here we used (\ref{eq:T}) to write $f'(r_+) = - 4 \pi T r_+^{z-1}$.
The remaining boundary condition to impose is the absence of a source $a_\parallel^{(0)}$ in the near boundary expansion (\ref{eq:34nearb}) of the Maxwell field. Thus we require $S(0) = 0$. Imposing this condition on (\ref{eq:bigS}) we obtain a diffusive relationship between frequency and wavevector:
\be\label{eq:D1}
\omega = - \mathrm{i} D k^2 \,, \qquad D = \frac{r_+^{2-z}}{d-z} \,.   
\ee
This is the anticipated diffusive mode. Some comments on this result:
\begin{enumerate}
\item Beyond finding the diffusive mode (\ref{eq:D1}), one can also find the full diffusive part of the longitudinal channel retarded Green's functions, giving a density Green's function of the form (\ref{nnchi}), as was originally done in \cite{Policastro:2002se,Herzog:2002fn}.
\item The diffusion constant in (\ref{eq:D1}), unlike the dc conductivity (\ref{eq:sigma}), is not given purely in terms of horizon data. To find the diffusive mode we had to explicitly solve the Maxwell equations everywhere. This is an example of the phenomenon of semi-holography mentioned in our discussion of Wilsonian holographic renormalization in \S\ref{sec:nongeom}. Because there is only one diffusive mode compared to the many (order `$N^2$') gapless modes of the IR theory, the diffusive mode is `not powerful' enough to backreact on the dynamics of the IR fixed point theory. To some extent this is an artifact of the large $N$ limit of holography. The consequence is that the diffusive mode is not fully described by the dynamics of the event horizon, but is rather a quasinormal mode in its own right, with support throughout the spacetime. A holographic Wilsonian discussion of the diffusive mode can be found in \cite{Iqbal:2008by,Faulkner:2010jy,Nickel:2010pr}.
\item When $d < z$,  the diffusivity is UV divergent and (\ref{eq:D1}) no longer holds \cite{Pang:2009wa}. This gives an extreme illustration of the previous comment: in this case the diffusivity is dominated by non-universal short distance physics, even while the dc conductivity (\ref{eq:sigma}) is captured by the universal low energy dynamics of the horizon.
\item A basic property of diffusive processes is the Einstein relation $\sigma = \chi D$, where $\chi$ is the charge susceptibility (compressibility). The susceptibility is obtained from the static, homogeneous two point function of $a_t$.  The easiest way to do this is to show that the following $a_t$ perturbation solves the linearized Maxwell equations (with $\omega=0$): \begin{equation}
a_t = \mu \left(1-\frac{r^{d-z}}{r_+^{d-z}}\right),  \label{eq:chi0}
\end{equation}
where $\mu$ is a small chemical potential.   The susceptibility is given by $\partial_\mu \rho(\mu,T)$ as $\mu\rightarrow 0$.   Combining (\ref{eq:chi0}) and (\ref{eq:jt1}), and using $\langle J^t\rangle=\rho$ we find  \begin{equation}
\chi = \frac{L^{d-2}}{e^2} \frac{d-z}{r_+^{d-z}}.  \label{eq:chi1}
\end{equation}
It is immediately verified from (\ref{eq:sigma}), (\ref{eq:D1}) and (\ref{eq:chi1}) that the Einstein relation holds.

\end{enumerate}

\subsubsection{$\sigma(\omega)$ part I: Critical phases}

We have emphasized in \S\ref{sec:cmtransport} and in (\ref{eq:dissip}) above that the real part of the frequency-dependent conductivity at zero momentum ($k=0$)
\be\label{eq:s1}
\text{Re} \, \sigma(\omega) = \frac{\text{Im} \, G_{J_xJ_x}^R(\omega)}{\omega} \,,
\ee
is a direct probe of charged excitations in the system as a function of energy scale. We also emphasized that this quantity is difficult to compute in strongly interacting theories using conventional methods, whereas it is readily accessible holographically. Two key aspects of the holographic computation are firstly the possibility of working directly with real-time frequencies, via the infalling boundary conditions discussed in  \S\ref{sec:infal}, and secondly the fact that dissipation occurs at leading, classical, order in the 't-Hooft large $N$ expansion, and that in particular the $N \to \infty$ and $\omega \to 0$ limits commute for the observable (\ref{eq:s1}).

Consider first a bulk Maxwell field in a scaling geometry with exponent $z$ as discussed in \S \ref{sec:maxw} above.
Putting $k=0$ in the equations of motion (\ref{eq:ax}) or (\ref{eq:ay}) for the Maxwell potential leads leads to
\be\label{eq:axx}
\left(f r^{3-d-z} a_x' \right)' = - \frac{\omega^2 r^{z+1-d}}{f} a_x \,.
\ee
We must solve this equation subject to infalling boundary conditions at the horizon. Given the solution, equations (\ref{eq:ailif}) and (\ref{eq:jx1}) for the near-boundary behavior of the field imply that the conductivity (\ref{eq:optical}) will be given by
\be\label{eq:ssw}
\sigma(\omega) = \frac{L^{d-2}}{e^2} \frac{1}{\mathrm{i} \omega} \lim_{r \to 0} \frac{1}{r^{d+z-3}} \frac{a_x'}{a_x} \,.
\ee

The equation (\ref{eq:axx}) can be solved explicitly in two boundary space dimensions, $d=2$. This is the most interesting case, in which the conductivity is dimensionless. The solution that satisfies infalling boundary conditions is
\be
a(r) = \exp\left( \mathrm{i} \omega \int_0^{r} \frac{s^{z-1} \mathrm{d}s}{f(s)} \right)\,.
\ee
The overall normalization is unimportant. It follows immediately from the formula for the conductivity (\ref{eq:ssw}), using only the fact that at the boundary $f(0) = 1$, that
\be\label{eq:Le}
\sigma(\omega) = \frac{1}{e^2} \,.
\ee
In particular, there is no dependence on $\omega/T$! In the language of \S \ref{sec:cmtransport}, this means that the, a priori distinct, constants characterizing the diffusive ($\omega \to 0$) and zero temperature ($\omega \to \infty$) limits are equal in this case: ${\mathcal K} = {\mathcal C}_\chi {\mathcal C}_D$, first noted in \cite{Herzog:2007ij}. We have obtained this result for all $z$ in $d=2$. We will see in \S \ref{sec:selfdual} that the lack of $\omega$ dependence is due to the electromagnetic duality enjoyed by the equations of motion of the bulk 3+1 dimensional Maxwell field, which translates into a self-duality of the boundary QFT under particle-vortex duality. Meanwhile, however, this means that in order to obtain more generic results for the frequency-dependent conductivity, we will need to depart from pure Maxwell theory in the bulk.

In the remainder of this subsection, we will restrict ourselves to the important case of CFT3s. That is, we put $d=2$ and $z=1$. The discussion in \S \ref{sec:OPE} showed that the existence of a relevant deformation of the quantum critical theory plays an important role in the structure of $\sigma(\omega)$, because it determines the leading correction to the constant $\omega/T \to \infty$ limit. Such an operator will always be present if the critical theory is obtained by tuning to a quantum critical point. However, a quantum critical phase, by definition, does not admit relevant perturbations. In such cases one may continue to focus on the universal sector described in the bulk by the metric and Maxwell field. It is an intriguing fact that the simplest holographic theories lead naturally to critical phases rather than critical points.

An especially simple deformation of Maxwell theory that does not introduce any additional bulk fields is
\be\label{eq:CFF}
S[A] = \frac{1}{e^2} \int \mathrm{d}^4x \sqrt{-g} \left( - \frac{1}{4} F^2 + \gamma L^2 C_{abcd} F^{ab} F^{cd} \right) \,.
\ee
Here $C_{abcd} = R_{abcd} - (g_{a[c}R_{d]b} - g_{b[c} R_{d]a}) + \frac{1}{3} g_{a[c}g_{d]b}$ is the Weyl curvature tensor. There is a new bulk coupling constant $\gamma$.
Aspects of transport in this theory have been studied in \cite{Ritz:2008kh, Myers:2010pk, WitczakKrempa:2012gn, Chowdhury:2012km, krempa_nature}. The theory (\ref{eq:CFF}) has several appealing features. In particular, it is still second order in derivatives of the Maxwell field. Therefore the formalism we have described in the last few sections for computing conductivities does not need to be modified. As we are interested in linear response around a zero density background (that is, with no Maxwell field turned on), for the present purposes it is sufficient to consider actions quadratic in the field strength and evaluated in geometric backgrounds that solve the vacuum Einstein equations (\ref{eq:vacE}). The action (\ref{eq:CFF}) is the unique such deformation of Maxwell theory that is of fourth order in derivatives of the metric and field strength, see e.g. \cite{Myers:2010pk,Chowdhury:2012km}. The reason to limit the number of total derivatives is that we can then imagine that the new term in (\ref{eq:CFF}) is the leading correction to Einstein-Maxwell theory in a bulk derivative expansion. Indeed, such terms will be generated by stringy or quantum effects in the bulk. See the references just cited for entry points into the relevant literature on higher derivative corrections in string theory. The effects on the optical conductivity of terms that are higher order yet in derivatives than those in (\ref{eq:CFF}) were studied e.g. \cite{Witczak-Krempa:2013aea, Bai:2013tfa}.

A word of caution is necessary before proceeding to compute in the theory (\ref{eq:CFF}). Generically, the bulk derivative expansion will only be controlled if the coupling constant $\gamma$ is parametrically small. Considering a finite nonzero $\gamma$ while neglecting other higher derivative terms requires fine tuning that may not be possible in a fully consistent bulk theory. Remarkably, it can be shown, both from the bulk and also from general QFT arguments, that consistency requires \cite{Hofman:2008ar, Hofman:2009ug,Myers:2010pk,Chowdhury:2012km,Hofman:2016awc}
\be\label{eq:gbound}
|\gamma| \leq \frac{1}{12} \,.
\ee
In the concrete model (\ref{eq:CFF}) we will see shortly that this bound has the effect of bounding the dc conductivity. However, physically speaking, and thinking of (\ref{eq:CFF}) as representative of a broader class of models, the bound (\ref{eq:gbound}) is a statement about short rather than long time physics \cite{Hofman:2008ar, Hofman:2009ug,Hofman:2016awc}. In particular, $\gamma$ is directly related to the $b_2$ coefficient that appears in the large frequency expansion (\ref{ope}) of the conductivity. As we have stressed, the $b_1$ term in (\ref{ope}) is absent in quantum critical phases. Thus the $b_2$ term is the leading correction to the asymptotic constant result. Because the $b_2$ term is pure imaginary, it does not appear in the sum rule (\ref{sr1}), although high frequency OPE data can appear on the right hand side of other conductivity sum rules, see e.g. \cite{Chowdhury:2016hjy}. Specifically, for the class of theories (\ref{eq:CFF}) one can show that as $\omega \to \infty$ \cite{Witczak-Krempa:2013aea}
\be
e^2 \, \sigma(\omega) = 1 - \mathrm{i} \frac{\gamma}{9} \left(\frac{4 \pi T}{\omega}\right)^3 +  \frac{\g (15 + 38 \g)}{324} \left(\frac{4 \pi T}{\omega}\right)^6 + \cdots \,.
\ee
This expansion is obtained by solving (\ref{eq:axweyl}) below in a WKB expansion. A slight correction to the $1/\omega^6$ term in \cite{Witczak-Krempa:2013aea} has been made.

Satisfying the bound (\ref{eq:gbound}) by no means guarantees that the bulk theory is a classical limit of a well defined theory of quantum gravity \cite{Camanho:2014apa}. However, it does mean that no pathology will arise at the level of computing the retarded Green's function for the current operator in linear response theory about the backgrounds we are considering. Therefore, we can go ahead and use the theory (\ref{eq:CFF}) as tool to generate Green's functions that are consistent with all necessary CFT axioms, that satisfy both of the sum rules (\ref{sr1}) and (\ref{sr2}), and for which we know the parameter $b_2$ appearing in the large frequency expansion (\ref{ope}).

The equations of motion for the Maxwell potential $A$ following from (\ref{eq:CFF}) are easily derived. The background geometry will be the AdS-Schwarzschild spacetime (i.e. the metric (\ref{eq:BH2}) with emblackening factor (\ref{eq:emblackfactor}), with $z=1$, $\theta= 0$ and $d=2$). As above, the perturbation takes the form $A = a_x(r) \mathrm{e}^{- \mathrm{i} \omega t} \mathrm{d}x$. The Weyl tensor term in the action (\ref{eq:CFF}) changes the previous equation of motion (\ref{eq:axx}) to \cite{WitczakKrempa:2012gn}
\be\label{eq:axweyl}
a_x'' + \left(\frac{f'}{f} + \g \frac{12 r^2}{r_+^3 + 4 \g r^3} \right) a_x' + \frac{\omega^2}{f^2} a_x = 0 \,.
\ee
The only effect of the coupling $\gamma$ is to introduce the second term in brackets. To obtain this equation of motion, one should use a \texttt{Mathematica} package that enables simple computation of curvature tensors (such as the Weyl tensor). Examples are the \texttt{RGTC} package or the \texttt{diffgeo} package, both easily found online.

Near the boundary, the spacetime approaches pure AdS$_4$, which has vanishing Weyl curvature tensor, and hence the new term in the action (\ref{eq:CFF}) does not alter the near boundary expansions of the fields. Therefore, from a solution of the equation of motion (\ref{eq:axweyl}), with infalling boundary conditions at the horizon, the conductivity is again given by (\ref{eq:ssw}). The following \texttt{Mathematica} code solves equation (\ref{eq:axweyl}) numerically. 
\begin{widetext}
\begin{quote} 
\% emblackening factor of background metric \\
\texttt{fs[r\_\,] = 1 - r\^\,3;}

\% equation of motion \\
\texttt{eq[$\omega$\_\,,$\gamma$\_\,] = 
 Ax''[r] + $\omega$\^\,2/f[r]\^\,2 Ax[r] + f'[r]/f[r] Ax'[r] \\ + (12r\^\,2 $\gamma$ Ax'[r])/(1+4r\^\,3 $\gamma$) /.\,f $\to$ fs;}
 
\%  series expansion of $A_x$ near the horizon. Infalling boundary conditions \\
\texttt{Axnh[r\_\,,$\omega$\_\,,$\gamma$\_\,] = (1-r)\^\,(-i$\omega$/3) (1+((3+8$\gamma$(6-i$\omega$)-2i$\omega$)$\omega$)/(3(1+4$\gamma$)(3i+2$\omega$))(1-r));}

\%  series expansion of $A_x'$ \\
\texttt{Axnhp[r\_\,,$\omega$\_\,,$\gamma$\_\,] = D[Axnh[r,$\omega$,$\gamma$],r];}

\%  small number for setting boundary conditions just off the horizon \\
\texttt{$\epsilon$\,=\,0.00001;}

\%  numerical solution \\
\texttt{Asol[$\omega$\_\,,$\gamma$\_\,] := 
 NDSolve[\{eq[$\omega$,$\gamma$]\,==\,0, Ax[1-$\epsilon$]\,==\,Axnh[1-$\epsilon$,$\omega$,$\gamma$], Ax'[1-$\epsilon$]\,\\==\,Axnhp[1-$\epsilon$,$\omega$,$\gamma$]\},Ax,\{r,0,1-$\epsilon$\}][[1]];}

\%  conductivity \\ 
\texttt{$\sigma$sol[$\omega$\_\,,$\gamma$\_\,] := 
 1/i/$\omega$ Ax'[0]/Ax[0] /.\,Asol[$\omega$,$\gamma$];}

\%  data points for $\sigma(\omega)$ \\
\texttt{tabb[$\gamma$\_\,] := 
 Table[\{$\omega$, Re[$\sigma$sol[$\omega$, $\gamma$]]\},\{$\omega$,0.001,4,0.05\}];}

\%  make table with different values of $\gamma$ \\
\texttt{tabs = Table[tabb[$\gamma$],\{$\gamma$,-1/12,1/12,1/12/3\}];}

\%  make plot \\
\texttt{ListPlot[tabs, Axes$\to$False, Frame$\to$True, PlotRange$\to$\{0, 1.5\}, Joined$\to$True, FrameLabel$\to$\{3$\omega$/(4$\pi$T),e\^\,2 Re $\sigma$\}, RotateLabel$\to$False, BaseStyle$\to$\{FontSize$\to$12\}]} 

\end{quote}
\end{widetext}

\begin{figure}[h]
\centering
\includegraphics[height = 0.21\textheight]{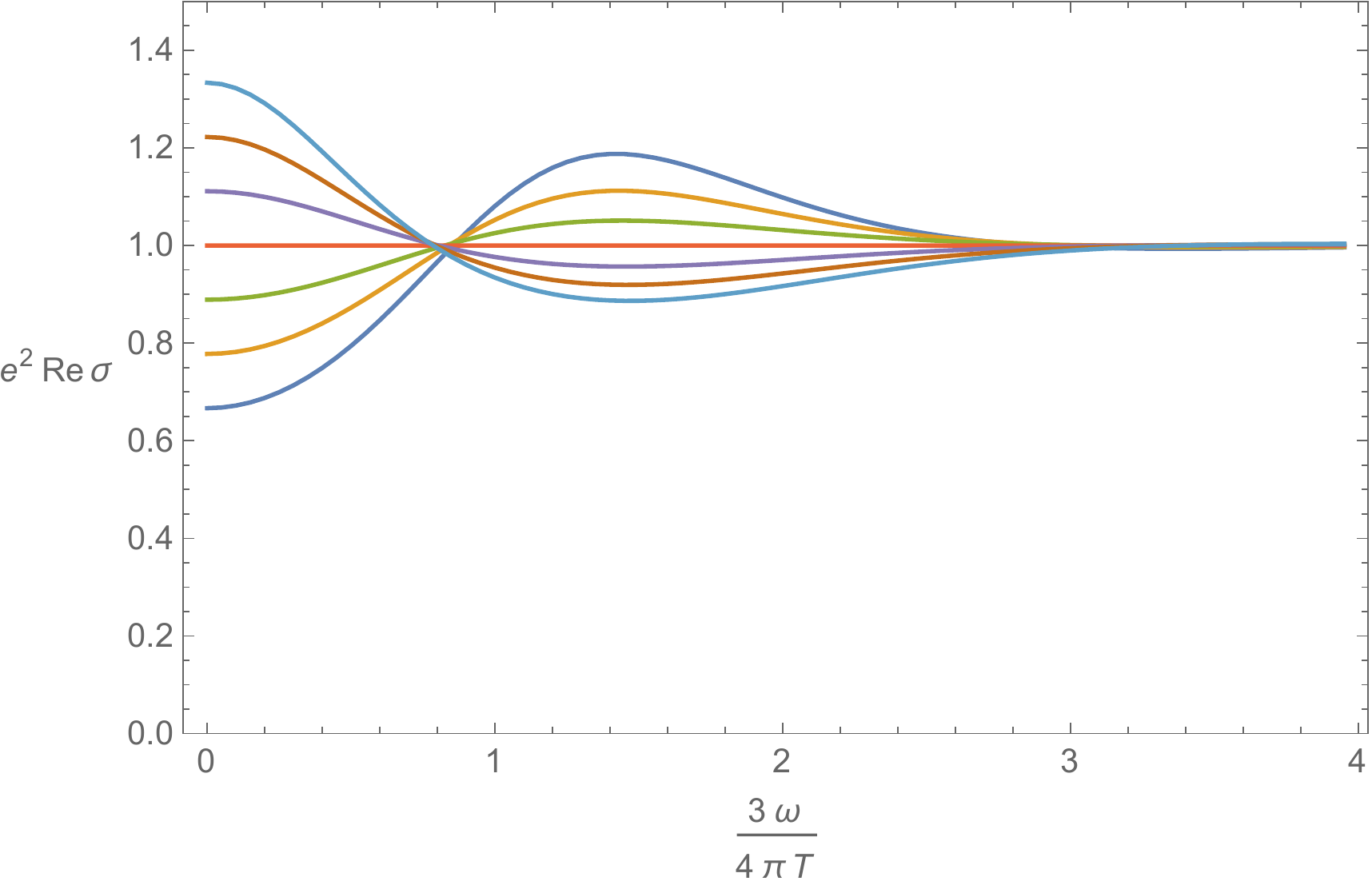}
\caption{\label{fig:CFF} {\bf Frequency-dependent conductivity} computed from the bulk theory (\ref{eq:CFF}).
From bottom to top, $\gamma$ is increased from $-1/12$ to $+1/12$.}
\end{figure}

The output is plots of the conductivity $\sigma(\omega)$ for different values of $\gamma$. These plots are shown in figure \ref{fig:CFF}. 
In performing the numerics, it is best to rescale the coordinates and frequencies by setting $r = r_+ \hat r$ and $\omega = \hat \omega/r_+$. In this way one can put the horizon radius at $r_+ = 1$ in doing calculations. Recall that $r_+ = 3/(4 \pi T)$ according to (\ref{eq:Tgo}). We must remember to restore the factor of $r_+$ in presenting the final result.
Figure \ref{fig:CFF} shows that for $\gamma$ nonzero, the conductivity acquires a nontrivial frequency dependence. This was the main reason to consider the coupling to the Weyl tensor in (\ref{eq:CFF}), which breaks the electromagnetic self-duality of the equations of motion. To leading order in small $\g$, electromagnetic duality now maps $\gamma \to - \gamma$ \cite{Myers:2010pk, WitczakKrempa:2012gn}. Furthermore we see that $\gamma > 0$ is associated to a `particle-like' peak at low frequencies whereas $\gamma < 0$ is associated with a `vortex-like' dip. In \S\ref{sec:quasinormal} below we will phrase these features in terms of the location of a `quasinormal pole' or zero in the complex frequency plane. We see in the plot that larger $|\gamma|$ results in larger deviations from the constant $\gamma=0$ result at low frequencies. The magnitude of this deviation is limited by the bound (\ref{eq:gbound}) on $\g$. Therefore, within this simple class of strongly interacting theories, the magnitude of the `Damle-Sachdev' \cite{Damle97} low frequency peak is bounded.

\subsubsection{$\sigma(\omega)$ part II: Critical points and holographic analytic continuation}
\label{sec:acsigma2}

It was emphasized in \cite{Katz14} that in order to describe quantum critical points rather than critical phases, it is essential to include the scalar relevant operator $\ocal$ that drives the quantum phase transition. A minimal holographic framework that captures the necessary physics was developed in \cite{Myers:2016wsu}. The action is
\begin{widetext}
\begin{equation}
S= - \int \mathrm{d}^4x \sqrt{-g} \left[\frac{1}{2 \k^2}\left( R + \frac{6}{L^2} + \left(\nabla \phi\right)^2 + m^2 \phi^2 - 2 \a_1 L^2 \phi \, C_{abcd} C^{abcd} \right) + \frac{1}{4e^2} \left(1 + \alpha_2 \phi\right) F^2 \right] \,. \label{eq:withrelevant}
\end{equation}
\end{widetext}
Here the new bulk field $\phi$ is dual to the relevant operator $\ocal$. Relevant means that the operator scaling dimension $\Delta < 3$. The scaling dimension is determined by the mass squared $m^2$ according to (\ref{eq:delta}) above.

The two couplings $\alpha_1$ and $\a_2$ in the theory (\ref{eq:withrelevant}) implement two important physical effects. The first, $\alpha_1$, ensures that $\ocal$ acquires a nonzero expectation value at $T>0$. This is of course generically expected for a scalar operator, but will not happen unless there is a bulk coupling that sources $\phi$. Because $C_{abcd} = 0$ in the $T=0$ pure AdS$_4$ background, this term does not source the scalar field until temperature is turned on. While the Weyl tensor thus is a very natural way to implement a finite temperature expectation value, the $C_{abcd} C^{abcd}$ term is higher order in derivatives and therefore should be treated with caution in a full theory. The second, $\alpha_2$, term ensures that there is a nonvanishing $C_{JJ\ocal}$ OPE coefficient. Both $\alpha_1$ and $\alpha_2$ must be nonzero for the anticiapted term to appear in the large frequency expansion of the conductivity -- i.e. for the coefficient $b_1$ in (\ref{ope}) to be nonzero. Plots of the resulting frequency-dependent conductivities can be found in \cite{Myers:2016wsu}.

An exciting use of the theory (\ref{eq:withrelevant}) is as a `machine' to analytically continue Euclidean Monte Carlo data \cite{krempa_nature, Katz14,Myers:2016wsu}. The conductivity of realistic quantum critical theories such as the $\mathrm{O}(2)$ Wilson-Fisher fixed point can be computed reliably along the imaginary frequency axis, using Monte Carlo techniques. However, analytic continuation of this data to real frequencies poses a serious challenge, as errors are greatly amplified. One way to understand this fact is that, as we will see in \S\ref{sec:quasinormal} shortly, the analytic structure of the conductivity in the complex frequency plane in strongly interacting CFTs at finite temperature is very rich \cite{WitczakKrempa:2012gn}. In particular there are an infinite number of `quasinormal poles' extending down into the lower half frequency plane.

The conductivity of holographic models, in contrast, can be directly computed with both Euclidean signature (imposing regularity at the tip of the Euclidean `cigar') and Lorentzian signature (with infalling boundary conditions at the horizon). Solving the holographic model thereby provides a method to perform the analytic continuation, which is furthermore guaranteed to satisfy sum rules and all other required formal properties of the conductivity. A general discussion of sum rules in holography can be found in \cite{Gulotta:2010cu}.
The model (\ref{eq:withrelevant}) allows an excellent fit to the imaginary frequency conductivity of the $\mathrm{O}(2)$ Wilson-Fisher fixed point. The fit fixes the two parameters $\a_1$ and $\a_2$ in the action. The dimension $\Delta$ of the relevant operator at the Wilson-Fisher fixed point is already known and is therefore not a free parameter. Once the parameters in the action are determined, the real frequency conductivity $\sigma(\omega)$ is easily calculated numerically using the same type of \texttt{Mathematica} code as described in the previous subsection.   

Further discussion of the use of sum rules and asymptotic expansions to constrain the frequency-dependent conductivity of realistic quantum critical points can be found in \cite{Witczak-Krempa:2015pia}.

\subsubsection{Particle-vortex duality and Maxwell duality}
\label{sec:selfdual}

All CFT3s with a global $\mathrm{U}(1)$ symmetry have a `particle-vortex dual' or `S-dual' CFT3, see e.g. \cite{Witten:2003ya}.
The name `particle-vortex' duality can be illustrated with the following simple example. Consider a free compact boson $\theta \sim \theta + 2\pi$,  with Lagrangian
\begin{equation}
\mathcal{L} = -\frac{1}{2}\partial_\mu \theta \partial^\mu \theta.  \label{eq:Lcompactboson}
\end{equation}
Vortices are (spatially) point-like topological defects where $\theta$ winds by $2\pi\times$ an integer, and so we write \begin{equation}
\theta = \tilde\theta + \theta_{\text{vortex}}  = \tilde\theta + \sum w_n \, \mathrm{arctan} \frac{y-y_n}{x-x_n} \,.
\end{equation}
Here $w_n\in \mathbb{Z}$ denotes the winding number of a vortex located at $(x_n,y_n)$, and $\tilde\theta$ is smooth.  We now perform a series of exact manipulations to the Lagrangian (\ref{eq:Lcompactboson}), understood to be inside a path integral.  We begin by adding a Lagrange multiplier $J^\mu$:\begin{equation}
\mathcal{L} =  -\frac{1}{2} J_\mu  J^\mu - J^\mu \partial_\mu (\tilde\theta + \theta_{\text{vortex}}).
\end{equation}
Now, we integrate out the smooth field $\tilde\theta$, which enforces a constraint $\partial_\mu J^\mu=0$.  The constraint is solved by writing $2J^\mu = \epsilon^{\mu\nu\rho}\partial_\nu A_\rho$, and after basic manipulations we obtain \begin{align}
\mathcal{L} &= -\frac{1}{4}F_{\mu\nu} F^{\mu\nu} + A^\rho \epsilon^{\rho\mu\nu}\partial_\mu \partial_\nu \theta_{\text{vortex}} \notag \\
&= -\frac{1}{4}F_{\mu\nu} F^{\mu\nu} + A_\rho(x) \sum_n w_n   \delta^{(2)}(x-x_n) n_n^\rho, \label{eq:Lchargedparticle}
\end{align}
with $F=\mathrm{d}A$ and $n_n^\rho$ a normal vector parallel to the worldine of the $n$th `vortex'.  Evidently, the compact boson is the same as a free gauge theory interacting with charged particles.   The vortices have become the charged particles, which leads to the name of the duality.

Above, we see that (\ref{eq:Lcompactboson}) and (\ref{eq:Lchargedparticle}) are not the same.  In some cases, such as non-compact $\mathbb{C}\mathrm{P}^1$ models \cite{vishwanath2003} or supersymmetric QED3 \cite{Kapustin:1999ha}, the theory is mapped back to iteslf by the particle-vortex transformation.  These theories can be called particle-vortex self-dual. We will see that the bulk Einstein-Maxwell theory (\ref{eq:Maxwell}) is also in this class, at least for the purposes of computing current-current correlators. We firstly describe some special features of transport in particle-vortex self-dual theories. 

In any system with current conservation and rotational invariance, the correlation functions of the current operators may be expressed as \cite{Herzog:2007ij}\begin{widetext}
\begin{equation}
G^{\mathrm{R}}_{\mu\nu}(\omega,k) = \sqrt{k^2-\omega^2} \left[\left(\eta_{\mu\nu} - \frac{p^\mu p^\nu}{p^2}\right) K^{\mathrm{L}}(\omega,k) - \left(\delta_{ij} - \frac{k_ik_j}{k^2}\right)(K^{\mathrm{L}}(\omega,k) - K^{\mathrm{T}}(\omega,k)\right]  \label{eq:KLKT0}
\end{equation}
\end{widetext}
with $p^\mu = (\omega,k)$. No Lorentz invariance is assumed. $K^L$ and $K^T$ determine the longitudinal and transverse parts of the correlator. The second term in the above equation is defined to vanish if $\mu=t$ or $\nu=t$.  In \cite{Herzog:2007ij} it was argued that particle-vortex self-duality implies \begin{equation}
K^{\mathrm{L}}K^{\mathrm{T}} = {\mathcal K}^2 \,,  \label{eq:KLKT}
\end{equation}
at all values of $\omega$, $k$ and temperature $T$.  ${\mathcal K}$ is the constant $\omega \to \infty$ value of the conductivity, as defined in \S \ref{sec:cmtransport} above, and must be determined for the specific CFT. The key step in the argument of \cite{Herzog:2007ij} is to note that particle-vortex duality is a type of Legendre transformation in which source and response are exchanged. This occurred in the simple example above when we introduced $2J^\mu = \epsilon^{\mu\nu\rho}\partial_\nu A_\rho$. When sources and responses are exchanged, the Green's functions relating them are then inverted. Self-duality then essentially requires a Green's function to be equal to its inverse, and this is what is expressed in equation (\ref{eq:KLKT}).

Setting $k\rightarrow 0$ in the above discussion, spatial isotropy demands that $K_{\mathrm{L}}(\omega)=K_{\mathrm{T}}(\omega) = \sigma(\omega)$.  Hence, (\ref{eq:KLKT}) in fact implies that for self-dual theories:
\begin{equation}\label{eq:cons}
\sigma\left(\frac{\omega}{T}\right) = {\mathcal K} \,,
\end{equation}
for any value of $\omega/T$!

The absence of a frequency dependence in (\ref{eq:cons}) mirrors what we found for Einstein-Maxwell theory in (\ref{eq:Le}). Let us rederive that result from the perspective of self-duality \cite{Herzog:2007ij}. Recall that in any background four dimensional spacetime, the Maxwell equations of motion are invariant under the exchange
\be\label{eq:emdual}
F_{ab} \leftrightarrow G_{ab} = \frac{1}{2} \epsilon_{abcd} F^{cd} \,.
\ee
This is not a symmetry of the full quantum mechanical Maxwell theory partition function, as electromagnetic duality inverts the Maxwell coupling. However, all we are interested in here are the equations of motion. To see how (\ref{eq:emdual}) acts on the function $\sigma(\omega)$ we can write
\begin{align}\label{eq:sinvert}
e^2 \sigma(\omega) &= \frac{1}{i \omega} \lim_{r \to 0} \frac{a_x'}{a_x} =  \lim_{r \to 0} \frac{F_{rx}}{F_{xt}} =  \lim_{r \to 0} \frac{G_{yt}}{G_{ry}} \notag \\
&=  \lim_{r \to 0} \frac{G_{xt}}{G_{rx}} = \frac{1}{e^2 \, \widetilde \sigma(\omega)} \,.
\end{align}
We have used isotropy in the second-to-last equality. We have written $\widetilde \sigma(\omega)$ to denote the conductivity of the dual theory, defined in terms of $G$ rather than $F$. However, electromagnetic duality (\ref{eq:emdual}) means that $G$ satisfies the same equations of motion as $F$. Therefore, we must have
\be
\sigma(\omega) = \widetilde \sigma(\omega) = \frac{1}{e^2} \,.
\ee
To paraphrase the above argument: from the bulk point of view, the conductivity is a magnetic field divided by an electric field. Electromagnetic duality means we have to get the same answer when we exchange electric and magnetic fields. This fixes the conductivity to be constant.

The argument of the previous paragraph can be extended to include the spatial momentum $k$ dependence and recover (\ref{eq:KLKT}) from bulk electromagnetic duality (\ref{eq:emdual}), allowing for the duality map to exchange the longitudinal and transverse modes satisfying (\ref{eq:ax}) and (\ref{eq:ay}) above \cite{Herzog:2007ij}. Furthermore, an explicit mapping between electromagnetic duality in the bulk and particle-vortex duality in the boundary theory can be shown at the level of the path integrals for each theory. See for instance \cite{Witten:2003ya, Marolf:2006nd, Herzog:2007ij}.

The inversion of conductivities (\ref{eq:sinvert}) under the duality map (\ref{eq:emdual}) is useful even when the theory is not self-dual. We noted above that for small deformations $\gamma$, the theory (\ref{eq:CFF}) is mapped back to itself with $\gamma \to -\g$ under (\ref{eq:emdual}). We can see the corresponding inversion of the conductivity clearly in Figure \ref{fig:CFF}.

\subsection{Quasinormal modes replace quasiparticles}
\label{sec:quasinormal}

\subsubsection{Physics and computation of quasinormal modes}

We have emphasized around  (\ref{Geta}) above that zero temperature quantum critical correlation functions are typically characterized by branch cut singularities. These branch cuts smear out the spectral weight that in weakly interacting theories is carried by dispersing quasiparticle poles, at $\omega = \epsilon(k)$, on or very close to the real frequency axis. Holographic methods of course reproduce this fact, as in equation (\ref{eq:easyG}) above. In contrast, at $T>0$, all strongly interacting holographic computations have found that the retarded Green's function is characterized purely by poles in the lower half complex frequency plane (causality requires the retarded Green's function to be analytic in the upper half plane):
\be\label{eq:sumofpoles}
G^{\mathrm{R}}_{\ocal\ocal}(\omega) = \sum_{\omega_\star} \frac{c_\star}{\omega - \omega_\star} \,.
\ee
The poles $\omega_\star$ are known as quasinormal modes. They are the basic `on shell' excitations of the system. Taking the Fourier transform of (\ref{eq:sumofpoles}), we see that the (imaginary part of the) quasinormal modes determine the rates at which the system equilibrates \cite{Horowitz:1999jd}.

Quasinormal modes do not indicate the re-emergence of quasiparticles at $T>0$. Instead, the imaginary and real parts of $\omega_\star$ are typically comparable, so that each modes does not represent a stable excitation. Long-lived quasinormal modes, close to the real axis, can generically only arise in special circumstances (as we shall see): as Goldstone bosons, Fermi surfaces and hydrodynamic modes. Occasionally, poles that are somewhat close to the real axis can produce features in spectral functions. More typically, the spectral density will be seen to receive contributions from a large number of quasinormal modes. Nonetheless, quasinormal modes have a rich and constrained mathematical structure and, in the bulk, capture essential features of gravitational physics.\footnote{A large part of the first gravitational wave signal detected by the LIGO collaboration is the `ringdown' of a quasinormal mode \cite{PhysRevLett.116.061102}.} As $T \to 0$, the poles coalesce to form the branch cuts of (\ref{Geta}). Some quasinormal modes can survive as poles in the $T = 0$ limit, as we shall see in our later descriptions of nonzero density states of holographic matter.

Perturbative computations at weak coupling lead to branch cuts in $T>0$ correlation functions. It is an open question whether these are artifacts of perturbation theory -- that the cuts break up nonpertubatively into finely spaced poles at arbitrarily weak nonzero coupling -- or whether the cuts become poles at some finite intermediate coupling or perhaps only in the infinite coupling limit \cite{Hartnoll:2005ju,Romatschke:2015gic, Grozdanov:2016vgg}. In any case, in this section we elaborate on the description of strongly interacting $T>0$ retarded Green's functions given holographically in the form (\ref{eq:sumofpoles}).

The quasinormal frequencies can be computed, in principle, from the formula for the holographic retarded Green's function in (\ref{eq:GRAB}) above. Specifically, poles of the retarded Green's function occur at frequencies $\omega_\star$ such that the corresponding perturbation $\delta \phi(r)$ of the bulk solution \cite{Birmingham:2001pj,Son:2002sd}
\begin{enumerate}
\item Satisfies infalling boundary conditions (\ref{eq:infalling}) at the horizon.
\item Is normalizable at the asymptotic boundary, so that $\delta \phi^{(0)} = 0$ in (\ref{eq:boundaryL}).
\end{enumerate}
Satisfying the two conditions above typically leads to an infinite discrete set of complex frequencies $\omega_\star$.
In particular, the infalling boundary conditions may be imposed for real or complex frequencies.

A technical challenge that arises in numerical studies is the following: in the lower half complex frequency plane, the infalling mode grows towards the horizon while the unphysical `outfalling' mode goes to zero (this fact is seen immediately from (\ref{eq:infalling})). This means that for frequencies with large imaginary parts, imposing infalling boundary conditions requires setting to zero a small correction to a large term, which can require high precision. This problem is especially severe for infalling boundary conditions at extremal horizons (as in e.g. (\ref{eq:ads2infall})), where one must set to zero an exponentially small correction to an exponentially large term. To circumvent this problem, various numerical methods have been devised to directly obtain the quasinormal modes in asymptotically AdS spacetimes. These include the use of truncated high order power series expansions \cite{Horowitz:1999jd} and a method using continued fractions \cite{Starinets:2002br}. For extremal horizons, a more general method due to Leaver is necessary \cite{PhysRevD.41.2986}, as described in \cite{Denef:2009yy}.

Analytic computations of quasinormal modes are possible using WKB methods, suitably adapted to deal with complex frequencies. These methods were pioneered in \cite{Motl:2003cd}. While formally capturing modes at large frequencies, the WKB formulae often give excellent approximations for $\omega_\star$ down to the lowest or second-lowest frequency. Comprehensive results for planar AdS black holes using this techniques can be found in e.g. \cite{Natario:2004jd,Festuccia:2008zx}. For example, for a scalar operator in a CFT3 with large scaling dimension $\Delta \gg 1$, cf. \S \ref{sec:largeDelta} above, one has \cite{Festuccia:2008zx}
\be
\omega_\star^{(n)} = \pi T \, \left(\pm \sqrt{3} - 3 \mathrm{i} \right) \left(n + \frac{1 - 2^{2/3}}{2} + \frac{2^{2/3}}{3} \Delta \right) + \ocal \left( \frac{1}{\Delta} \right) \,. \label{eq:qnwkb}
\ee
These are results for zero momentum, $k=0$. One corollary of this result is that there are infinitely many quasinormal frequencies, extending down at an angle in the complex frequency plane. The poles exist in pairs related by reflection about the imaginary frequency axis. This follows from time reversal invariance, according to which $G^{\mathrm{R}}(-\omega^*) = G^{\mathrm{R}}(\omega)^*$.

\begin{figure}
\centering
\includegraphics[height = 0.3\textheight]{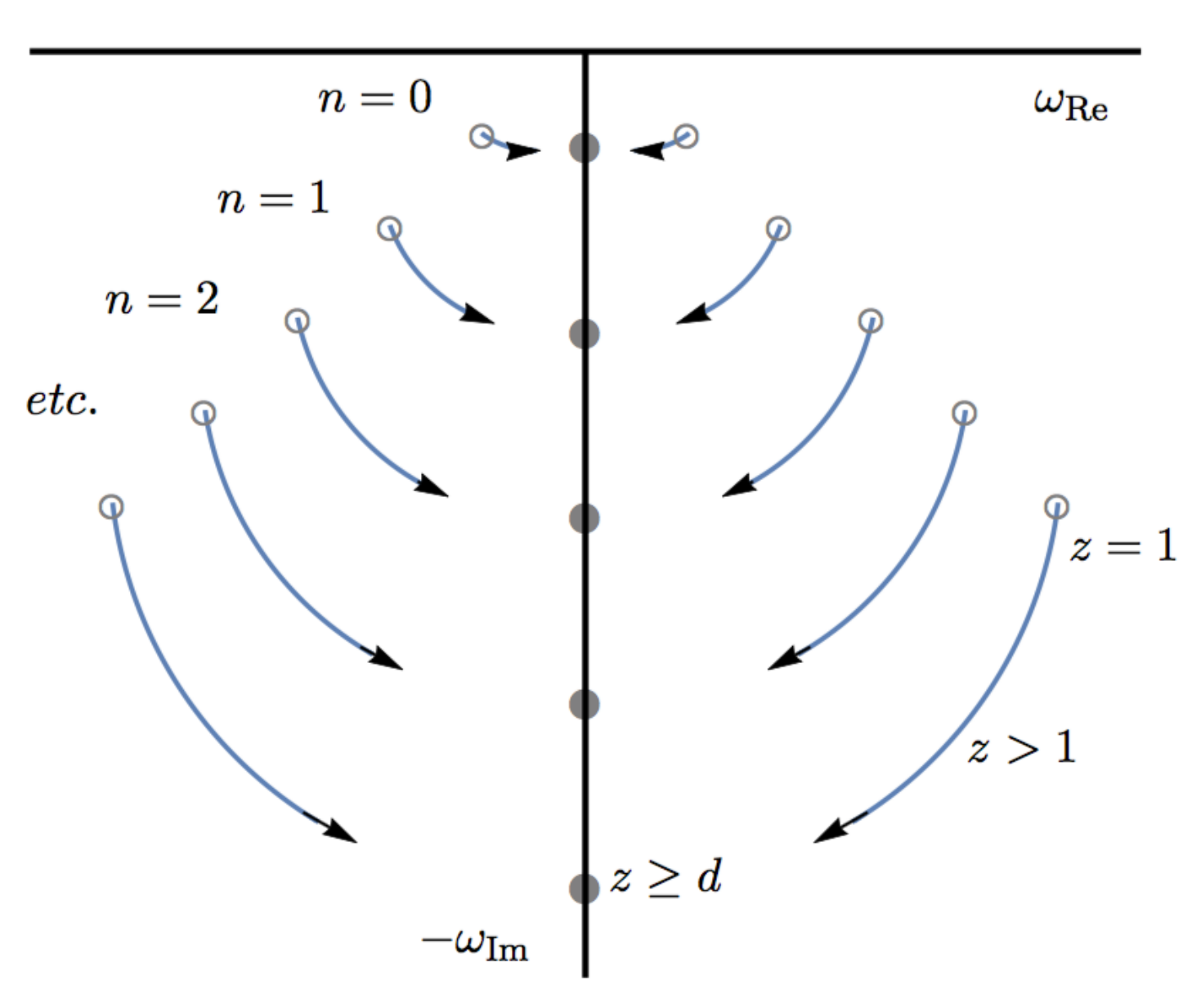}
\caption{\label{fig:LifQNM} {\bf Quasinormal poles as a function of z}. Hollow dots denote quasinormal frequencies at $z=1$. The arrows show the motion of the poles as $z$ is increased. For $z\geq d$ the modes are all overdamped, lying along the negative imaginary axis. All modes have momentum $k=0$. [Figure adapted with permission from \cite{Sybesma:2015oha}]}
\end{figure}

The quasinormal modes of scalar operators in quantum critical theories with $z>1$ have been studied in \cite{Sybesma:2015oha} as function of scaling dimension $\Delta$ of the operator and dimension $d$ of space. The modes have momentum $k=0$. When $z < d$, the quasinormal modes are qualitatively similar to those in CFTs (i.e. with $z=1$). That is, they extend down in the complex frequency plane similarly to (\ref{eq:qnwkb}). However, for $z \geq d$ all quasinormal frequencies are overdamped, with purely imaginary frequencies. This is illustrated in figure \ref{fig:LifQNM}.
The density of states in these quantum critical theories, on dimensional grounds, satisfies $\rho(\omega) \sim \omega^{(d-z)/z}$ \cite{Sybesma:2015oha}. Therefore, $d<z$ implies a large number of low energy excitations. Decay into these excitations may possibly be the cause of the overdamped nature of the modes.
Precisely at $z=d$, the quasinormal frequencies may be found analytically \cite{Sybesma:2015oha}. They have a similar integrable structure to those of CFT2s \cite{Birmingham:2001pj,Son:2002sd}. 

\begin{figure}
\centering
\includegraphics[width=3in]{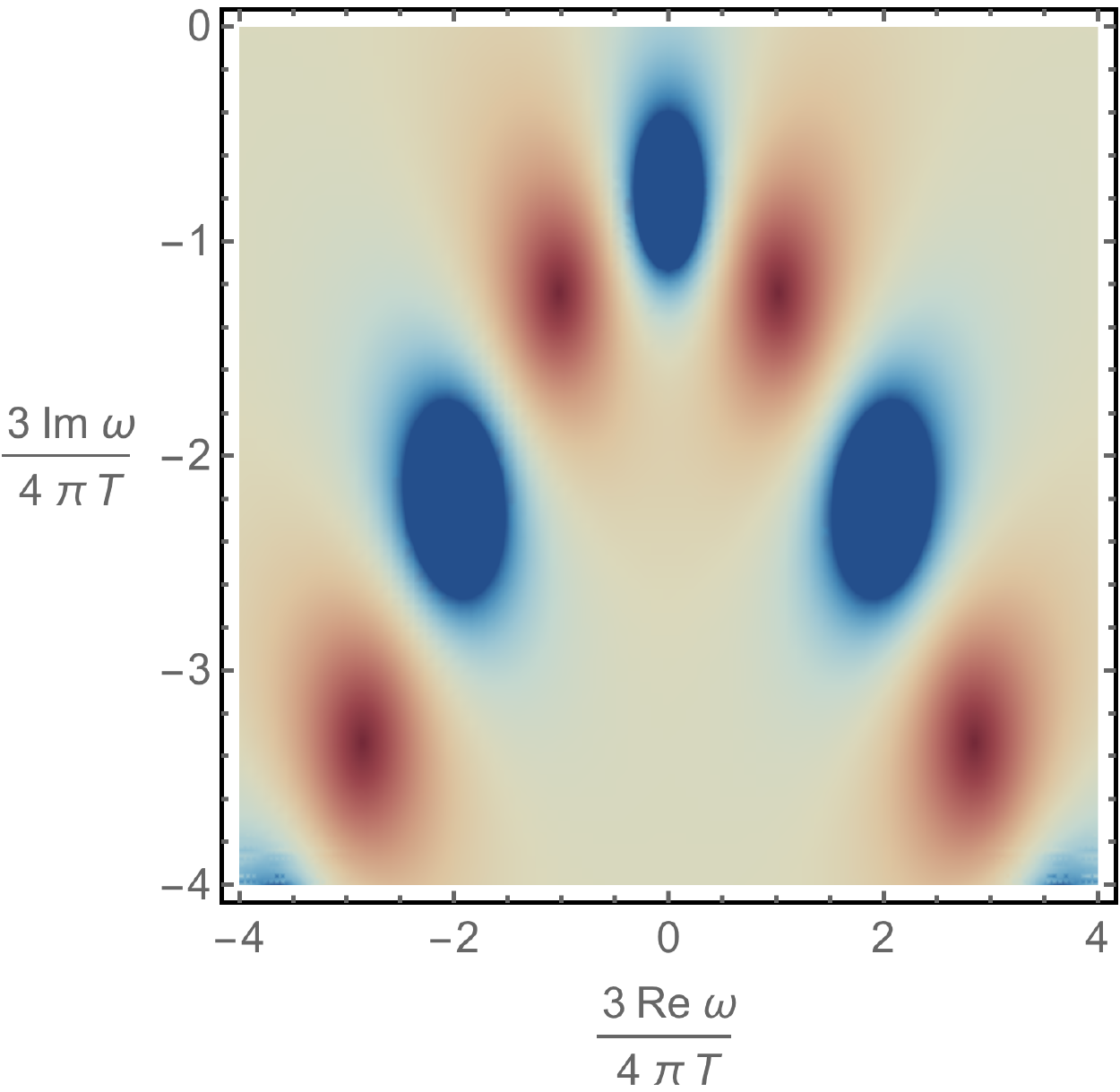} \includegraphics[width=3in]{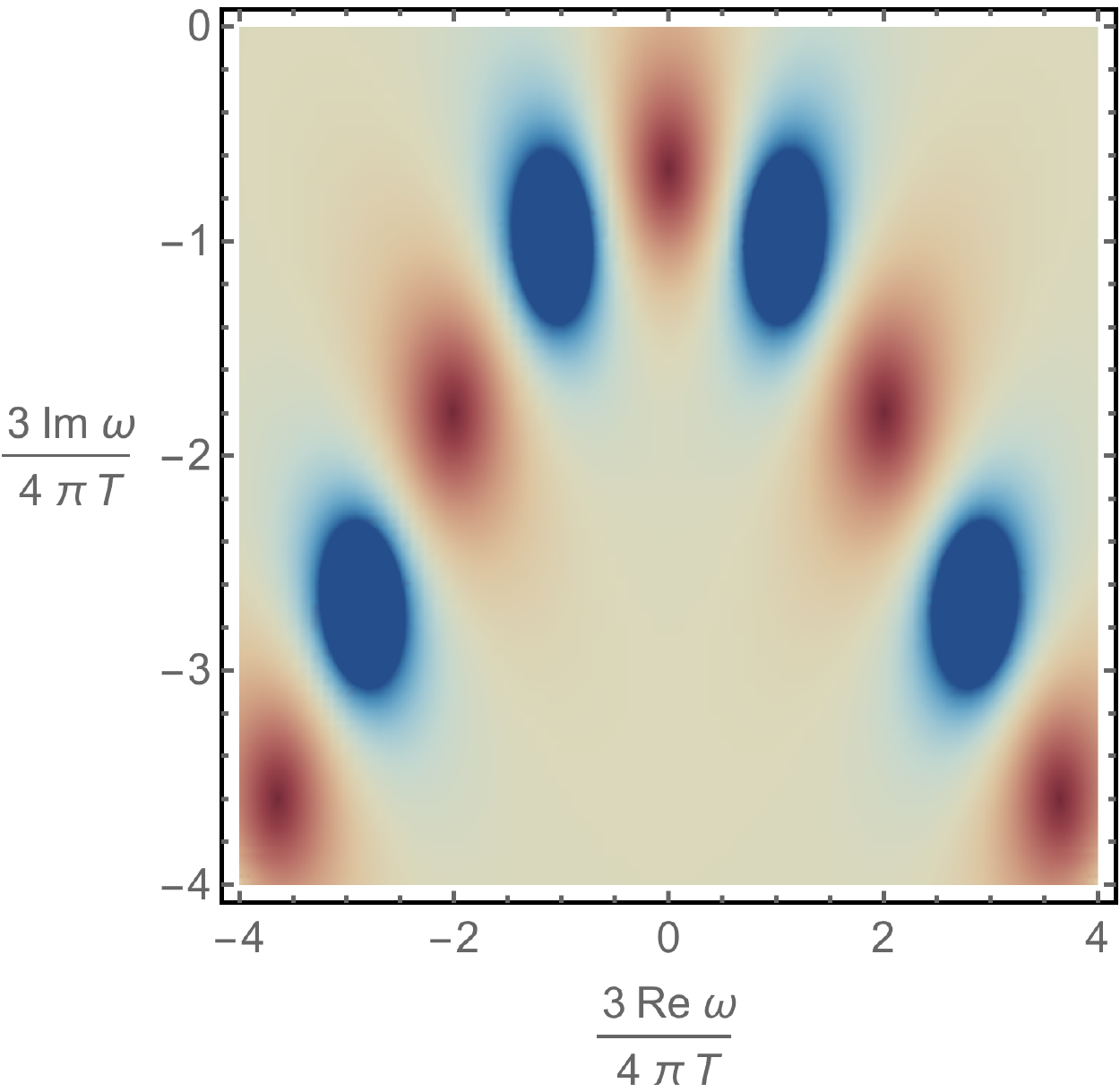}
\caption{\label{fig:WeylQNM} {\bf $|\sigma(\omega)|$ in the lower half plane}, with $\gamma = 1/12$ (top) and $\gamma = -1/12$ (bottom). The dominant features are poles (blue) and zeros (red). The plot has been clipped at $e^2 |\sigma| = 2$, the dark blue regions.}
\end{figure}

An example where a low-lying quasinormal mode imprints a feature on the conductivity along the real frequency axis is the theory (\ref{eq:CFF}), Maxwell theory modified by a coupling to the Weyl tensor, that we studied above. Figure \ref{fig:WeylQNM} plots $|\sigma(\omega)|$ in the lower half complex frequency plane. This plot is obtained using the same \texttt{Mathematica} code that produced Figure \ref{fig:CFF} above. A higher order series expansion has been used to set the initial conditions close to the horizon and a higher \texttt{WorkingPrecision} has been used in numerically solving the differential equation.
In the figure we see that $\gamma = 1/12$ leads to a pole on the negative imaginary axis whereas $\gamma = - 1/12$ leads to a zero on the negative imaginary axis. These are responsible for the peak and dip seen in Figure \ref{fig:CFF}, respectively. The poles and zeros are exchanged in the two plots of Figure \ref{fig:WeylQNM}. This is 
\begin{figure*}
\centering
\includegraphics[height = 0.4\textheight]{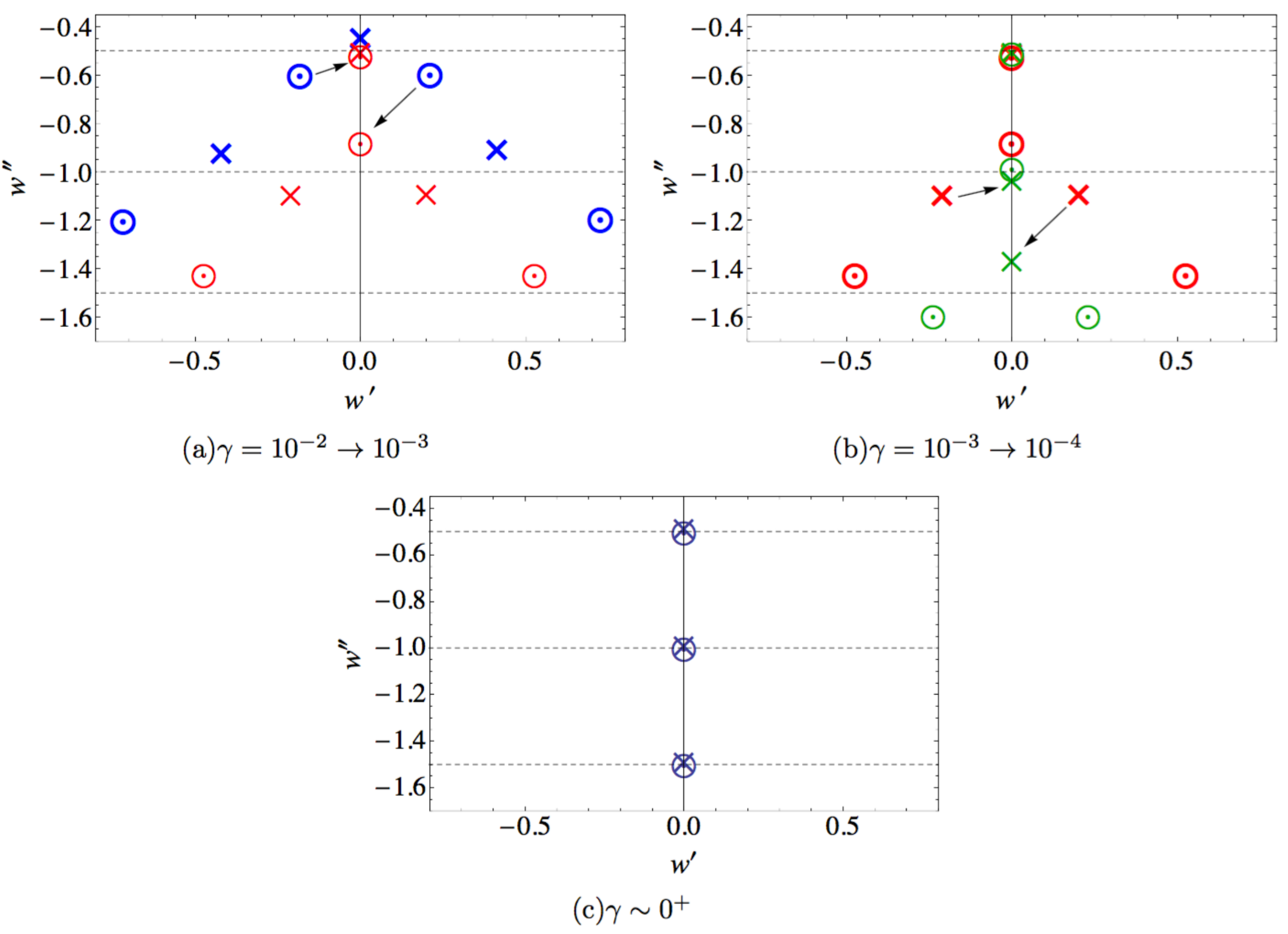}
\caption{\label{fig:dance1} {\bf Motion of poles and zeros as $\gamma$ is decreased}. Crosses are poles and circles are zeros. Pais of poles and zeros move to the imaginary frequency axis, and then move up and down the axis to annihilate with a zero or pole. In these plots $w = \omega/(4 \pi T)$. [Figure taken with permission from \cite{WitczakKrempa:2012gn}]}
\end{figure*}
consistent with the fact that electromagnetic duality inverts the conductivity (\ref{eq:sinvert}), and also, for $\gamma \ll 1$, sends $\gamma \to - \gamma$. We thus learn that the zeros of the conductivity in the complex frequency plane are also important, as they become the poles of the particle-vortex dual theory. A similar zero-pole structure to that of Figure \ref{fig:WeylQNM} is seen in other cases where particle-vortex duality acts in nontrivial way, such as in a background magnetic field \cite{Hartnoll:2007ip}. We will discuss magnetic fields and particle-vortex duality further in \S \ref{sec:magtrans} below.

Beyond peaks or dips caused by specific poles or zeros, one can use the lowest few poles and zeros to construct a `truncated conductivity' that accurately captures the low frequency behavior of the conductivity along the real frequency axis \cite{WitczakKrempa:2012gn}.

In the case of $\gamma = 0$, pure Maxwell theory in the bulk, there are no quasinormal modes. Recall that the conductivity (\ref{eq:Le}) is featureless in this case. It is instructive to see how the quasinormal modes plotted in the previous Figure \ref{fig:WeylQNM} annihilate in a `zipper-like' fashion as $\gamma \to 0$. Figure \ref{fig:dance1} shows 
that, as $\gamma$ is lowered, pairs of zero and poles move towards the imaginary frequency axis.
Once the pair reaches the axis, one moves up and the other moves down the axis. The pole or zero that moves up then annihilates, as $\gamma \to 0$, with a zero or pole that is higher up the axis.
The poles and zero annihilate precisely at the Matsubara frequencies $\omega_n^\text{zip} = -  2 \pi n T\mathrm{i}$, with $n=1,2,3, \ldots$ \cite{WitczakKrempa:2012gn}.

A very important class of quasinormal poles are those associated with hydrodynamic modes. Hydrodynamic modes have the property that $\omega_\star(k) \to 0$ as $k \to 0$. These modes are forced to exist on general grounds due to conservation laws (we will discuss hydrodynamics in \S\ref{sec5}) and can usually be found analytically. In fact, relativistic hydrodynamics can be derived in some generality from gravity using the `fluid/gravity' correspondence \cite{Bhattacharyya:2008jc}. We have already computed a hydrodynamic quasinormal mode in  \S\ref{sec:diffusive} above. In  (\ref{eq:D1}) we found the diffusion mode
\be
\omega_\star(k) =  - \mathrm{i} D k^2 + \cdots \,,
\ee
together with a formula for $D$. Important early holographic studies of hydrodynamic modes in zero density systems include \cite{Policastro:2002se, Policastro:2002tn,Kovtun:2005ev}, for CFT4s, and \cite{Herzog:2002fn,Herzog:2003ke} for CFT3s. In addition to diffusive modes there are sound modes, as we will discuss in \S\ref{sec:sound}.  Finally, an interesting perspective on hydrodynamic quasinormal modes of black holes can be obtained by taking the number of spatial dimensions $d \to \infty$. In this limit aspects of the gravitational problem simply and, in particular, the influence of black hole horizons on the geometry becomes restricted to a thin shell around the horizon \cite{Emparan:2013moa}. The hydrodynamic quasinormal modes become a decoupled set of modes existing in the thin shell around the horizon, and can be studied to high order in a derivative expansion \cite{Emparan:2015rva}.

In later parts of this review we shall come across other circumstances where $\omega_\star \to 0$: Goldstone modes, zero temperature sound modes and Fermi surface modes.

\subsubsection{$1/N$ corrections from quasinormal modes}
\label{sec:oneoverN}

We have said that quasinormal poles define the `on-shell' excitations of the strongly interacting, dissipative finite temperature system. Virtual production of these excitations in the bulk should be expected to contribute to observables at subleading order in the large $N$ expansion (which, we recall, is the semi-classical expansion in the bulk). Occasionally such a `one-loop' effect in the bulk can give the leading contribution to interesting observables. In the boundary theory, these effects either correspond to non-equilibrium thermodynamic fluctuations or else to physics that is suppressed because it involves a number of modes that remains finite in the large $N$ limit (e.g. there might only be one Goldstone boson).

The usual Euclidean determinant formulae for the one-loop correction to black hole backgrounds obscures the physical role of quasinormal modes. The determinant correction to the bulk partition function (\ref{eq:classicalsource}) takes the schematic form
\be
Z = \det \left( - \nabla^2_{g_\star} \right)^{\pm 1} \mathrm{e}^{-S_{\mathrm{E}}[g_\star]} \,.
\ee
Here $g_\star$ denotes the (Euclidean) saddle point and $\det \left( - \nabla^2_{g_\star} \right)$ schematically denotes the determinants corresponding to fluctuations about the background. Bosonic determinants are in the denominator and fermionic determinants in the numerator. In \cite{Denef:2009yy,Denef:2009kn} formulae for these determinants were derived in terms of the quasinormal modes of the fluctuations. For bosonic fields, for instance, the formula is
\be\label{eq:modGamma}
\frac{1}{\det \left( - \nabla^2_{g_\star} \right)} = \mathrm{e}^{\text{Pol}} \prod_{\omega_\star} \frac{|\omega_\star|}{4 \pi^2 T}
\left| \Gamma \left( \frac{\mathrm{i} \omega_\star}{2 \pi T} \right) \right|^2 \,.
\ee
Here $\omega_\star$ are the quasinormal frequencies. $\text{Pol}$ denotes a polynomial in the dimension $\Delta$ of the scalar field; this term is determined by short distance physics in the bulk and cannot contribute any interesting non-analyticities. One-loop effects are mainly interesting, of course, insofar as they can lead to non-analytic low energy (`universal') physics. This also simplifies the calculation -- we do not wish to compute the whole determinant, but just to isolate the important part.

The formula (\ref{eq:modGamma}) is useful for calculating the correction to thermodynamics quantities. One may also be interested in non-analytic corrections to Green's functions. Here quasinormal modes also play an essential role. We will sketch how this works, more details can be found in \cite{caronhuot, Hartman:2010fk, Anninos:2010sq, Faulkner:2013bna}. A typical one-loop correction $\Delta G$ to a Green's function will be given by the convolution of two bulk propagators. The finite temperature calculation is set up by starting with periodic Euclidean time. Fourier transforming in the boundary directions, but working in position space for the radial direction, the bulk Euclidean propagators take the form $G(\mathrm{i}\omega_n, k, r_1,r_2)$. Thus, schematically, \begin{widetext}
\be
\Delta G(\mathrm{i}\omega_n,k) \sim T \sum_m \int \frac{\mathrm{d}^dq}{(2\pi)^d} \int \mathrm{d}r \mathrm{d}r'  G_1(\mathrm{i} \omega_m,q,r,r') G_2(\mathrm{i} \omega_m + \mathrm{i} \omega_n,q+k,r',r) \; + \; (1 \leftrightarrow 2)  \,.
\ee
The second step is to re-express the correction in terms of retarded Green's functions. This is achieved using standard representations of the sum over Matsubara frequencies as a contour integral. See the references above. For bosonic fields, this leads to, again schematically,
\be\label{eq:lll}
\Delta G(\omega) \sim \int \frac{\mathrm{d}\Omega}{2 \pi} \int \frac{\mathrm{d}^dq}{(2\pi)^d} \int \mathrm{d}r \mathrm{d}r' \coth \frac{\Omega}{2T} \, \text{Im}\left[ G_1^{\mathrm{R}}(\Omega,q,r,r') \, G_2^{\mathrm{R}}(\Omega + \omega,q,r',r) 
 \right]  \; + \; (1 \leftrightarrow 2) \,.
\ee
\end{widetext}
We analytically continued the external frequency $i \omega_n \to \omega$ and for simplicity set the external $k=0$. We are interested typically in small $\omega \ll T$. The objective now is to identify the most IR singular part of these integrals. This looks challenging, in particular because of the integrals over the radial direction. However, if there is a quasinormal frequency $\omega_\star(k)$ that goes to zero, say as the momentum goes to zero, then one can isolate the singular contribution of this mode to the Green's function \cite{Anninos:2010sq}. Schematically,
\be\label{eq:zoomin}
G^{\mathrm{R}}(\omega,k,r,r') \sim \frac{\overline{\phi_\star(r)} \phi_\star(r')}{\omega - \omega_\star(k)} \,.
\ee
Here $\phi_\star(r)$ is the radial profile of the quasinormal mode with frequency $\omega_\star$. In the numerator $k$ can be set to zero (in some cases the correct normalization of $G^{\mathrm{R}}$ may require overall factors of $k$). Hence, using the above formula in (\ref{eq:lll}), the radial integrals simply lead to a numerical prefactor. The remaining integrals are then only in the field theory directions.

To illustrate how zooming in on the quasinormal mode (\ref{eq:zoomin}) picks out the singular physics of interest, consider the case of long-time tails that appear in the electrical conductivity. Here one is looking for a one-loop correction to the propagator of the bulk Maxwell field (photon). The one-loop diagram is one in which a photon and a graviton run in the loop (there are no photon self-interactions in Einstein-Maxwell theory). The retarded Green's functions of these bulk field each have a diffusive quasinormal mode, corresponding to diffusion of charge and momentum. Zooming in on these modes, and assuming that the relevant frequencies will have $\Omega \ll T$, the integrals in (\ref{eq:lll}) become, without keeping track of the overall numerical prefactor,
\bea
\lefteqn{\Delta G_{J^x J^x}(\omega)} \notag \\
&& \;\; \sim \int \frac{\mathrm{d}\Omega}{2 \pi} \int \frac{\mathrm{d}^dq}{(2\pi)^d} \frac{T}{\Omega} \text{Im}\left[ \frac{q^2}{\Omega + \mathrm{i} D_1 q^2} \frac{q^2}{\Omega  + \omega + \mathrm{i} D_2 q^2}  \right] \notag \\
&& \qquad \qquad \qquad \qquad \qquad \qquad \qquad \qquad + \; (1 \leftrightarrow 2) \nonumber \\
&&  \;\; \sim  \int \frac{\mathrm{d}^dq}{(2\pi)^d} \frac{q^2 \omega}{\omega^2 + D_2^2 q^4} \; + \; (1 \leftrightarrow 2) \nonumber \\
&& \;\; \sim \left\{
\begin{array}{cc}
\omega^{1/2} & d=1 \\
\omega \log \omega & d=2
\end{array}
\right. \,. \label{eq:res}
\eea

The results (\ref{eq:res}) are precisely the well-known singular `late time tail' contributions to $G^{\mathrm{R}}_{J^x J^x}(\omega)$ in one and two spatial dimensions. They lead to an infinite dc conductivity $\sigma_\text{dc} = \lim_{\omega \to 0} [G^{\mathrm{R}}_{J^x J^x}(\omega)/\mathrm{i} \omega]$. Because this effect has arisen from a quantum correction to the bulk, the singular late time tails are suppressed in holographic theories by factors of $1/N$. This fact was first noted in \cite{Kovtun:2003vj}. The logarithmic divergence in two spatial dimensions should be negligible for most practical purposes. For instance, weak translational symmetry breaking will lift the momentum diffusion mode that was important in the above argument. Finally, we noted in \S\ref{sec:cmtransport} above that in CFT2s there is no diffusion. Therefore the $d=1$ result in (\ref{eq:res}) does not apply to those cases.

The above outline is essentially enough to isolate the singular effect in many cases. To obtain the correct numerical prefactor, or to convince oneself that there are no other singular contributions, substantially more work is typically necessary \cite{caronhuot, Hartman:2010fk, Anninos:2010sq, Faulkner:2013bna}.

\section{Compressible quantum matter}
\label{sec:compressible}

The condensed matter systems described in \S \ref{sec:zerodensity} had the common feature of being at some
fixed density of electrons (or other quantum particles) commensurate with an underlying lattice: this was important in 
realizing a low energy description in terms of a conformal field theory. However, the majority of the experimental realizations of quantum
matter without quasiparticle excitations are not at such special densities. Rather, they appear at generic densities in a phase diagram
in which the density can be continuously varied. 

More precisely stated, the present (and following) section will consider systems with a global U(1) symmetry, with
an associated conserved charge density $\rho$ which can be varied smoothly as a function of system
parameters even at zero temperature. The simplest realization of such a system is the free electron gas,
and (as is well known) many of its properties extend smoothly in the presence of interactions to a state of
compressible quantum matter called the `Fermi liquid'. The Fermi liquid has long-lived quasiparticle excitations,
and almost all of its low temperature properties can be efficiently described using a model of a dilute {\em gas\/}
of fermionic quasiparticles. Indeed, the term Fermi {\em liquid\/} is a misnomer: the low energy excitations do not interact strongly as in a canonical classical liquid like water. 

Our interest here will be in realizations of compressible quantum matter which do not have quasiparticle
excitations. One can imagine reaching such a state by turning up the strength of the 
interactions between the quasiparticles until they reach a true liquid-like state in which the 
quasiparticles themselves lose their integrity. 

Any compressible state has to be gapless, otherwise the density would not vary as a function of the chemical potential.
The non-quasiparticle states of interest to us typically have a scale-invariant structure at low energies,
and so it is useful work through some simple scaling arguments at the outset. As in (\ref{eq:s}),
we write for the $T$ dependence of the entropy density
\beq
s \sim T^{(d-\theta)/z} \,, \label{eq:entT}
\eeq
where $d$ is the spatial dimensionality, $z$ determines the relative scaling of space and time and $\theta$ is the violation of hyperscaling exponent (discussed in \S \ref{sec:hsv1} above). In the arguments that follow only the combination $(d-\theta)/z$ appears.
In the presence of a global conserved charge density $\rho$, it is useful to introduce a conjugate chemical potential $\mu$, and consider the properties of the system as a function of both $\mu$ and $T$.

Thermodynamics is controlled by a `master function', given by the pressure $P(\mu,T)$ in the grand canonical ensemble. Scale invariance implies that, so long as the charge density operator does not acquire an anomalous dimension (which it can, see \S\ref{sec:anomalouscharge} below), $P$ is a function of the dimensionless ratio $\mu/T$: 
 \begin{equation}
P = T^{1+(d-\theta)/z} \mathcal{F}\left(\frac{\mu}{T}\right).  \label{eq:PTmu}
\end{equation}
The charge density $\rho$ and entropy density $s$ are defined as \begin{subequations}\begin{align}
\rho &= \frac{\partial P}{\partial \mu} = T^{(d-\theta)/z} \mathcal{F}^\prime \left(\frac{\mu}{T}\right), \\
s &= \frac{\partial P}{\partial T} = T^{(d-\theta)/z}\left[ \left(1+\frac{d-\theta}{z}\right)\mathcal{F} \left(\frac{\mu}{T}\right)- \frac{\mu}{T}\mathcal{F}^\prime \left(\frac{\mu}{T}\right)\right].
\end{align}\end{subequations}
Using the thermodynamic identity 
\begin{equation}
\epsilon+P = \mu \rho + T s,  \label{eq:gibbsduhem}
\end{equation}
where $\epsilon$ is the energy density, 
we obtain 
\begin{equation}
\epsilon = \frac{d-\theta}{z}P.   \label{eq:epsandP}
\end{equation}

Theories accessible with conventional methods will shortly be seen to have $\theta=0$ or $\theta=d-1$. The latter case arises when a Fermi surface contributes to the critical physics. A simple model for theories with more general $\theta$ consists of $N \gg 1$ species of free Dirac fermions with power-law mass distributions \cite{Karch:2015pha}. Other examples are non-conformal large $N$ supersymmetric gauge theories \cite{Dong:2012se}. We do not know of any non-large $N$ quantum critical model which has other $\theta$. Indeed, $\theta<0$ suggests that the low energy theory `grows' in dimensionality -- as the effective total number of spacetime dimensions with hyperscaling violation is $D_\text{eff.} = z + d - \theta$ -- which may require $N\rightarrow \infty$ degrees of freedom. The emergence of extra scaling spatial dimensions at low energies underlies the appearance of $\theta < 0$ in some holographic models with known large $N$ field theory duals, e.g. \cite{Gubser:2009qt, Gouteraux:2011ce}.


The first subsection below describes various condensed matter realizations of compressible quantum matter. 
The subsequent subsections take a holographic approach. The holographic realizations can be obtained by 
`doping' a strongly-coupled conformal field theory, although they are also well-defined low energy fixed point theories in their own right.
We will compare the properties of holographic compressible matter with those of the condensed matter systems in \S\ref{sec:nfl}.

\subsection{Condensed matter systems}
\label{sec:nfl}

We begin with a review of some basic ideas from 
the Fermi liquid theory \cite{pines1994theory} of interacting fermions in $d$ dimensions, 
expanding upon the discussion in \S \ref{sec:qmwqp}.
We consider spin-1/2 fermions $c_{k \alpha}$ with momentum $k$ and spin
$\alpha = \uparrow, \downarrow$ and dispersion $\varepsilon_k$. Thus the non-interacting
fermions are described by the action
\begin{equation}
S_{\mathrm{c}} = \int \mathrm{d}\tau \int \frac{\mathrm{d}^d k}{( 2\pi)^d} c_{k\alpha}^\dagger \left( \frac{\partial}{\partial \tau} + \varepsilon_k
\right) c_{k \alpha}\,. \label{sc}
\end{equation}
The fermion Green's function under the free fermion action $S_{\mathrm{c}}$ has the simple form
\begin{equation}
G_0 (k, \omega ) = \frac{1}{\omega - \varepsilon_k} \,.
\label{flt1}
\end{equation}
This Green's function has a pole
at energy $\varepsilon_k$ with residue 1. Thus there are quasiparticle excitations with residue $Z=1$,
with both positive
and negative energies, as $\varepsilon_k$ can have either sign; the Fermi surface is the locus of points
where $\varepsilon_k$ changes sign.
The positive energy qausiparticles
are electron-like, while those with negative energy are hole-like {\em i.e.\/} they correspond to the absence
of an electron. A perturbative analysis \cite{pines1994theory} describes the effect of interactions on this simple free fermion description.
A key result is that the pole survives on the Fermi surface: for small $|\varepsilon_k|$, the interacting 
fermion retarded Green's function takes the form
\beq
G(k, \omega) = \frac{Z}{\omega - \tilde{\varepsilon}_k + \mathrm{i} \, \alpha \, \tilde{\varepsilon}_k^2} + G_{\mathrm{inc}} \,,
\label{Gfl}
\eeq
where $\tilde{\varepsilon}_k$ is the renormalized quasiparticle dispersion, $\alpha$ is a numerical constant, $0 < Z < 1$
is the quasiparticle residue, and $G_{\rm inc}$ is the incoherent contribution which is regular as $\tilde{\varepsilon}_k \rightarrow 0$. The Fermi surface is now defined by the equation $\tilde{\varepsilon}_k = 0$, and we note that the lifetime of a
quasiparticle with energy $\tilde{\varepsilon}_k$ diverges as $1/\tilde{\varepsilon}_k^2$, indicating that the quasiparticles
are well-defined. Although the interactions can renormalize the shape of the Fermi surface, Luttinger's theorem states that
the volume enclosed by the Fermi surface remains the same as in the free particle case:
\beq
\int_{\tilde{\varepsilon}_k < 0} \frac{\mathrm{d}^d k}{(2 \pi)^d} = \rho \,.
\label{lutt1}
\eeq

In principle zeros of the Green's function 
can contribute to the Luttinger count \cite{2006PhRvL..96h6407K,2013PhRvL.110i0403D}. However, 
generally in gapless metallic systems they do not. This is because 
the Green's function is complex, and so it requires two conditions to find a zero of the Green's function in the complex plane. 
In $d$ spatial dimensions, the locus of zeros is a $(d-2)$-dimensional surface, and so makes a negligible contribution to the Luttinger computation. In contrast, the locus of poles is $(d-1)$-dimensional: the physical structure of the Feynman-Dyson
expansion for fermions ensures that the locus of points where the real part of $G^{-1}(k,\omega=0)$ vanishes coincides with the points where the imaginary part vanishes, as is clear from the structure of (\ref{Gfl}). Alternatively stated, the location
of the poles of $G$ in (\ref{Gfl}) is independent of $G_{\mathrm{inc}}$, while the location of the zeros is not.

A universal description of the low energy properties of a Fermi liquid is obtained by focusing in on each point
on the Fermi surface.
For each direction $\hat{n}$, we define the position
of the Fermi surface by the wavevector $\vec{k}_{\mathrm{F}} (\hat{n})$, so that $\hat{n} = \vec{k}_{\mathrm{F}} (\hat{n}) / | \vec{k}_F (\hat{n}) |$.
Then we identify wavevectors near the Fermi surface by
\begin{equation}
\vec{k} = \vec{k}_{\mathrm{F}}( \hat{n}) + k_\perp \, \hat{n} \,. \label{flt2}
\end{equation}
Now we should expand in small momenta $k_\perp$.  For this, we define the infinite set of fields $\psi_{\hat{n} a} (k_\perp)$,
which are labeled by the spin $a$ and the direction $\hat{n}$, related to the fermions $c$ by
\begin{equation}
c_{\vec{k}\alpha} = \frac{1}{\sqrt{S_{\mathrm{F}}}}\psi_{\hat{n} \alpha} (k_\perp), \label{flt3}
\end{equation}
where $\vec{k}$ and $k_\perp$ are related by (\ref{flt2}), and $S_{\mathrm{F}}$ is the area of the Fermi surface. 
Inserting
(\ref{flt3}) into (\ref{sc}), expanding in $k_\perp$, and Fourier transforming to real space $x_\perp$, we obtain
the low energy theory
\begin{equation}
S_{\rm FL} = \int \mathrm{d} \Omega_{\hat{n}} \int \mathrm{d} x_\perp \psi_{\hat{n} \alpha}^\dagger ( x_\perp ) \left( 
\frac{\partial}{\partial \tau} - \mathrm{i} v_{\mathrm{F}} (\hat{n}) \frac{\partial}{\partial x_\perp} \right) \psi_{\hat{n} \alpha} ( x_\perp ),
\label{flt4}
\end{equation}
where the Fermi velocity is the energy gradient on the Fermi surface 
$v_{\mathrm{F}} (\hat{n}) = | \nabla_k \varepsilon_{\vec{k}_{\mathrm{F}} (\hat{n})} |$.
For each $\hat{n}$, (\ref{flt4}) describes a fermion moving along the single dimension $x_\perp$ with the
Fermi velocity: this is a one-dimensional chiral fermion; the
`chiral' refers to the fact that the fermion only moves in the positive $x_\perp$ direction, and not 
the negative $x_\perp$ direction.
In other words, the low energy theory of the Fermi liquid is an infinite set of one-dimensional
chiral fermions, one chiral fermion for each point on the Fermi surface.
Apart from the free Fermi term in (\ref{flt4}), Landau's Fermi liquid theory also allows for
contact interactions between chiral fermions along different directions \cite{shankarrg, Polchinski:1992ed}.
These are labeled by the Landau parameters, and lead only to shifts in the quasiparticle
energies which depend upon the densities of the other quasiparticles. Such shifts are important
when computing the response of the Fermi liquid to external density or spin perturbations.

The chiral fermion theory in Eq.~(\ref{flt4}) allows us to deduce the values of the exponents characterizing
the low temperature thermodynamics of a Fermi liquid. We have
\beq
z = 1 \quad, \quad \theta = d-1 \,,
\eeq
from the linear dispersion and the effective dimensionality $d-\theta = 1$ of the chiral fermions.

For $d>1$, interactions are irrelevant at a Fermi liquid fixed point. The only exception to this last statement is, with time-reversal invariance (that guarantees that if a momentum $\vec k$ is on the Fermi surface then so is $- \vec k$), the marginally relevant BCS instability leading to superconductivity at exponentially low temperatures \cite{shankarrg, Polchinski:1992ed}. Hence, most ordinary compressible matter (such as the electron gas in iron, or liquid helium) is a Fermi liquid.  Nonetheless, it is possible to find routes to destabilize the Fermi liquid to obtain compressible states without quasiparticles, as we detail 
in the following \S\ref{sec:isingnematic}, \S\ref{sec:sdw} and \S\ref{sec:emerge2}.

\subsubsection{Ising-nematic transition}
\label{sec:isingnematic}

One route to non-Fermi liquid behavior is to study a quantum critical point associated with the breaking of a global
symmetry in the presence of a Fermi surface. This is analogous to our study in \S\ref{sec:honeycomb}, where we
considered spin rotation symmetry breaking in the presence of massless Dirac fermions whose energy vanished at isolated
points in the Brillouin zone. But here we have a $(d-1)$-dimensional
surface of massless chiral fermions, and so they can have a more singular influence.

We describe here the field theory for the simplest (and experimentally relevant) example of an Ising order parameter, 
represented by the real scalar field $\phi$, associated with the breaking of a point-group symmetry of the lattice. Other order
parameters which carry zero momentum have similar critical theories; the case of non-zero momentum order parameters
which break translational symmetry will be considered in \S \ref{sec:sdw}.

\begin{figure}
\centerline{\includegraphics[width=0.3 \textwidth]{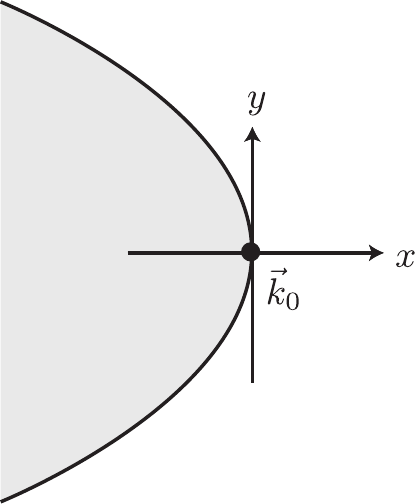}} 
\caption{\textbf{Fermionic excitations at the Ising-nematic critical point.} We focus on an extended patch of the
Fermi surface, and expand in momenta about the point $\vec{k}_0$ on the Fermi surface. 
This yields a theory of $d$-dimensional fermions $\psi$ in (\ref{flt6}).
The coordinate $y$ represents the $d-1$ dimensions parallel to the Fermi surface.} \label{fig:sungsik}
\end{figure}

Unfortunately, in this case the field theory cannot be obtained by simply coupling the field theory of $\phi$
to the chiral fermion theory in (\ref{flt4}), and keeping all terms consistent with symmetry.
A shortcoming of the effective action (\ref{flt4}) is that it only includes the dispersion of the fermions
transverse to the Fermi surface. Thus, if we discretize the directions $\hat{n}$, and pick a given point on 
the Fermi surface, the Fermi surface is effectively {\em flat\/} at that point. 
The curvature of the Fermi surface turns out to be important at the Ising-nematic critical point \cite{sungsik1}.
So we have to take the continuum scaling limit in a manner which keeps the curvature
of the Fermi surface fixed, and does not scale it to zero. For this, as shown in Fig.~\ref{fig:sungsik},
we focus attention on a single arc of the Fermi surface in the vicinity of any chosen point
$\vec{k}_0$. 
With $\vec{k}_0$ chosen as Fig.~\ref{fig:sungsik}, let us now define our low energy theory and scaling limit.
Unlike the one-dimensional chiral fermions which appeared in (\ref{flt4}), we will now use a $d$ dimensional
fermion $\psi_{\alpha} (x, y)$. Here $x$ is the one dimensional co-ordinate orthogonal to the Fermi surface,
and $\vec{y}$ represents the $(d-1)$-dimensional transverse co-ordinates; expanding the dispersion in the vicinity of $\vec{k}_0$ (contrast to the expansion away
from all points on the Fermi surface in (\ref{flt4})), we obtain the low energy theory \cite{sungsik1}
\begin{equation}
S_\psi = \int \! \mathrm{d} \tau \int \! \mathrm{d} x \int \! \mathrm{d}^{d-1} y \, \,  \psi_{\alpha}^\dagger \left(  \zeta_1 \frac{\partial}{\partial \tau} - \mathrm{i} v_{\mathrm{F}} \frac{\partial}{\partial x}
- \frac{\kappa}{2} \nabla_y^2 \right) \psi_{\alpha} . \label{flt6}
\end{equation}
We have added a co-efficient $\zeta_1$ to the temporal gradient term for convenience: we are interested in $\zeta_1=1$,
but the disappearance of quasiparticles in the ultimate theory is captured by the renormalization $\zeta_1 \rightarrow 0$.
Notice the additional second-order gradients in $y$ which were missing from (\ref{flt4}): the co-efficient $\kappa$
is proportional to the curvature of the Fermi surface at $\vec{k}_0$. There are zero energy fermion excitations when 
\begin{equation}
v_{\mathrm{F}} k_x + \kappa \frac{k_y^2}{2} = 0, \label{flt7}
\end{equation}
and so (\ref{flt7}) defines the position
of the Fermi surface, which is then part of the low energy theory including its curvature. Note that (\ref{flt6})
now includes an extended portion of the Fermi surface; contrast that with (\ref{flt4}), where  
the one-dimensional chiral fermion theory for each $\hat{n}$ describes only a single point on the Fermi surface.

The field theory for the Ising-nematic critical point now follows the traditional symmetry-restricted route described also in 
\S\ref{sec:honeycomb}. We restrict our subsequent attention to the important case of $d=2$, because this is the dimensionality in which quasiparticles are destroyed by the critical fluctuations. We combine a field theory like (\ref{Swf}) for $\phi$ with a `Yukawa' coupling $\lambda$ between the fermions and bosons  \begin{widetext}
\beq
S_{\rm IN} = S_\psi + \int \mathrm{d}x \mathrm{d}y \mathrm{d}\tau \left[\zeta_2 \left( \partial_\tau \phi \right)^2
+ \zeta_3 \left( \partial_x \phi \right)^2 + \left(\partial_y \phi \right)^2 + s \phi^2 + u \phi^4 + \lambda \phi \psi_{\alpha}^\dagger \psi_{\alpha} \right].
\label{inaction}
\eeq
\end{widetext}
As usual, the coupling $s$ is a tuning parameter across the quantum critical point, and we are interested here in the 
quantum state at the critical value $s=s_{\mathrm{c}}$.
The theory $S_{\rm IN}$ has been much studied in the literature by a variety of sophisticated field-theoretic 
methods \cite{sungsik1,metlitski1,mross,sungsik2}. 
For the single fermion patch theory as formulated above, exact results on the non-quasiparticle compressible state are available.
In the scaling limit, all the $\zeta_i \rightarrow 0$, and the fermionic and bosonic excitations are both characterized by
a common emergent dynamic critical exponent $z$. Their Green's functions do not have any quasiparticle poles (unlike (\ref{Gfl})), and instead we have the following scaling forms describing the non-quasiparticle excitations:
\begin{subequations}\begin{align}
G_\psi (k_x, k_y, \omega) &\sim \frac{\mathcal{F}_\psi \left( \frac{\omega}{(v_{\mathrm{F}} q_x + (\kappa/2) q_y^2)^z} \right)}{(v_{\mathrm{F}} q_x + (\kappa/2) q_y^2)^{1-\eta}} 
 \,,  \\
G_\phi (k_x, k_y, \omega) &\sim \frac{1}{q_y^{(2z-1)}} \mathcal{F}_\phi \left( \frac{\omega}{ q_y^{2z}} \right) \,,
\end{align}\end{subequations}
where $\eta$ is the anomalous dimension of the fermion, the boson does not have an independent anomalous dimension,
and $\mathcal{F}_{\psi,\phi}$ are scaling functions. Note that the curvature of the Fermi surface, $\kappa/v_{\mathrm{F}}$, does
not renormalize, and this follows from a Ward identity obeyed by $S_{\rm IN}$ \cite{metlitski1}. 
For the single-patch theory
described so far, the exact values of the exponents in $d=2$ are \cite{sungsik3}
\beq
z = 3/2  , \quad \theta = 1  , \quad \eta = 0 \,. \label{eq:isingexp}
\eeq

In the presence of additional symmetries in the problem, the single patch description is often not adequate. 
Many important cases have inversion symmetry, and then we have to include a pair of antipodal patches of the 
Fermi surface in the same critical theory \cite{metlitski1,mross}. 
The resulting two-patch theory is more complicated, and exact results are not available.
Up to three loop order, we find that $\eta \neq 0$, while the value $z=3/2$ is preserved. 
There is also a potentially strong instability to 
superconductivity in the two patch theory \cite{metlitski5}.
A strong superconducting instability is seen in Monte Carlo studies of
a closely related quantum critical point \cite{2016arXiv161201542L}.

\subsubsection{Spin density wave transition}
\label{sec:sdw}

Now we consider the case of an order parameter at non-zero momentum leading to symmetry breaking in the 
presence of a Fermi surface. The experimentally most common case is a spin density wave transition, represented by
a scalar field $\varphi^a$, with $a=1,2,3$, very similar to that found in \S \ref{sec:honeycomb} for the Gross-Neveu model.
The order parameter carries momentum $K$, and so the condensation of $\varphi^a$ breaks both spin rotation and translational
symmetries. The scalar will couple to fermion bilinears which also carry momentum $K$; as we are interested only in 
low energy fermionic excitations near the Fermi surface, we need to look for ``hot spots'' on the Fermi surface which 
are separated by momentum $K$. If no such hot spots are found, then the Fermi surface can be largely ignored in the theory
of the transition, as it is described by a conventional Wilson-Fisher theory for the scalar $\varphi^a$.
Here we consider the case where the hot spots are present, and then there is a relevant 
Yukawa coupling between $\varphi^a$ and fermion bilinears spanning the hot spots.

We consider the simplest case here of a single pair of hot spots, described by the fermion fields $\psi_{1\alpha}$ and $\psi_{2\alpha}$ representing
points on the Fermi surface at wavevectors $k_1$ and $k_2$. 
The low energy fermion action in (\ref{flt6}) is now replaced by 
\begin{figure}
\centering
 \includegraphics[width=0.47 \textwidth]{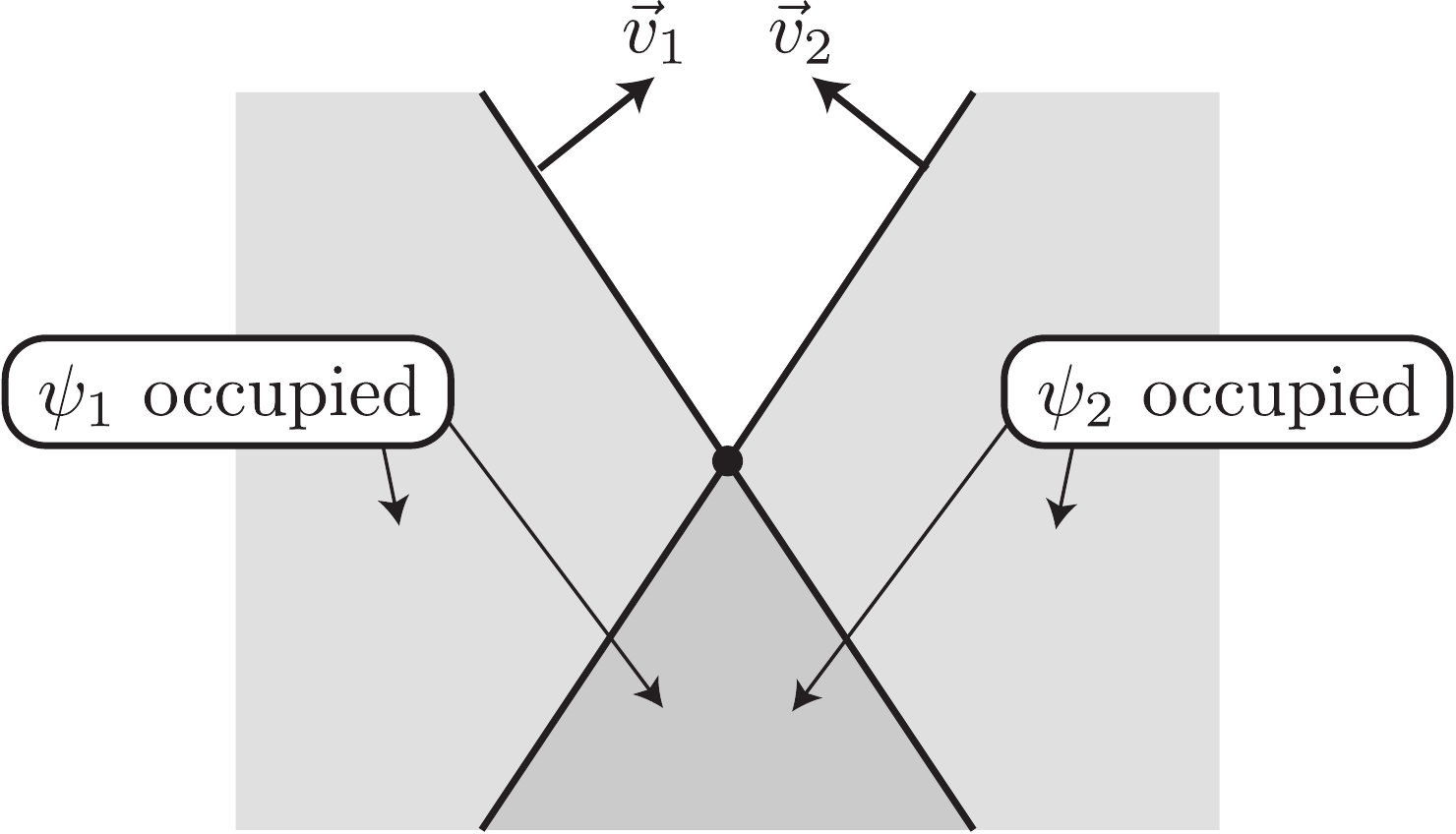}
 \caption{\textbf{Fermi surfaces of $\psi_1$ and $\psi_2$ fermions} in the plane defined by the Fermi velocities $\vec{v}_1$ 
 and $\vec{v}_2$. The gapless Fermi surfaces are one-dimensional, and are indicated by the full lines. The lines intersect at the hot spot, which is the filled circle at the origin.}
 \label{fig:sdwfs}
\end{figure}
\begin{widetext}
\begin{equation}
S_\psi = \int \mathrm{d} \tau \int \mathrm{d}^2 x \Biggl[ \psi_{1\alpha}^\dagger \left( \frac{\partial}{\partial \tau} - \mathrm{i} \vec{v}_1 \cdot \nabla_x \right) \psi_{1\alpha} + \psi_{2\alpha}^\dagger  \left( \frac{\partial}{\partial \tau} - \mathrm{i} \vec{v}_2 \cdot \nabla_x \right) \psi_{2\alpha} \Biggr] .
\label{sdw2}
\end{equation}
Here we are again restricting attention to the important case of spatial dimension $d=2$,
$\vec{v}_1 = \left. \nabla_k \varepsilon_{k} \right |_{k_1}$ is the Fermi velocity at ${k}_1$, and similarly for $\vec{v}_2$. 
In the $\psi_{1,2}$  formulation, the configurations of the Fermi surfaces and hot spots
are shown in Fig.~\ref{fig:sdwfs}.
The Fermi surface of the $\psi_1$ fermions is defined by $\vec{v}_1 \cdot \vec{k} = 0$,
and the Fermi surface of the $\psi_2$ fermions is defined by $\vec{v}_2 \cdot \vec{k} = 0$,
and the hot spot is at $k=0$.
We assume here and below that $\vec{v}_1$ and $\vec{v}_2$ are not collinear: the collinear case corresponds
to the ``nesting'' of the Fermi surfaces, and has to be treated separately.
Finally, as in \S\ref{sec:isingnematic}, we can couple these fermions to the scalar field $\varphi^a$ and obtain
the field theory for the onset of spin density wave order in a metal \cite{chubukov1}
\begin{equation}
S_{{\rm sdw}} = S_\psi + \int \mathrm{d}^2 x \int \mathrm{d} \tau \Biggl[  (\nabla_x \varphi^a)^2 + s \left(\varphi^{a}\right)^2
+ u \left( \left(\varphi^a \right)^2 \right)^2 + \lambda \varphi^a \sigma^{a}_{\alpha\beta} \left(
\psi_{1\alpha}^\dagger \psi_{2\beta} + \psi_{2\alpha}^\dagger \psi_{1\beta} \right) \Biggr]. \label{sdw3}
\end{equation}
\end{widetext}
Here $\sigma^a$ are the Pauli matrices. Notice the similarity of $S_{{\rm sdw}}$ to the theory for the onset of antiferromagnetism
in the honeycomb lattice in \S\ref{sec:honeycomb}: the main difference is that the massless Dirac fermions (which are gapless only at the isolated
point k=0) have been replaced by fermions with dispersion illustrated in Fig.~\ref{fig:sdwfs} (which are gapless on two lines in the Brillouin zone).

The non-Fermi liquid physics associated with the theory $S_{{\rm sdw}}$ has been studied in great detail in the literature. The expansion in powers 
of $\lambda$ is strongly  coupled, and a variety of large $N$ and dimensionality expansions have been 
employed \cite{metlitski2,sungsik4,strack2}. We will not enter into the details of this
analysis here, and only mention a few important points. The theory has a strong tendency to the appearance of $d$-wave superconductivity, and this has
been confirmed in quantum Monte Carlo simulations \cite{metlitski4,berg1,fawang1,fawang2,scalettar1}. The non-Fermi liquid behavior is characterized by a common dynamic critical exponent $z>1$ 
for both the fermions and bosons, but with no violation of hyperscaling $\theta = 0$ \cite{metlitski3,strack1}. The preservation of hyperscaling indicates that the critical non-Fermi
liquid behavior is dominated by the physics in the immediate vicinity of the hot spot; the gapless Fermi lines (responsible for violation of hyperscaling 
in a Fermi liquid) are subdominant in their non-Fermi liquid contributions.

\subsubsection{Emergent gauge fields}
\label{sec:emerge2}

\begin{figure}
\centering
\includegraphics[height = 0.3\textheight]{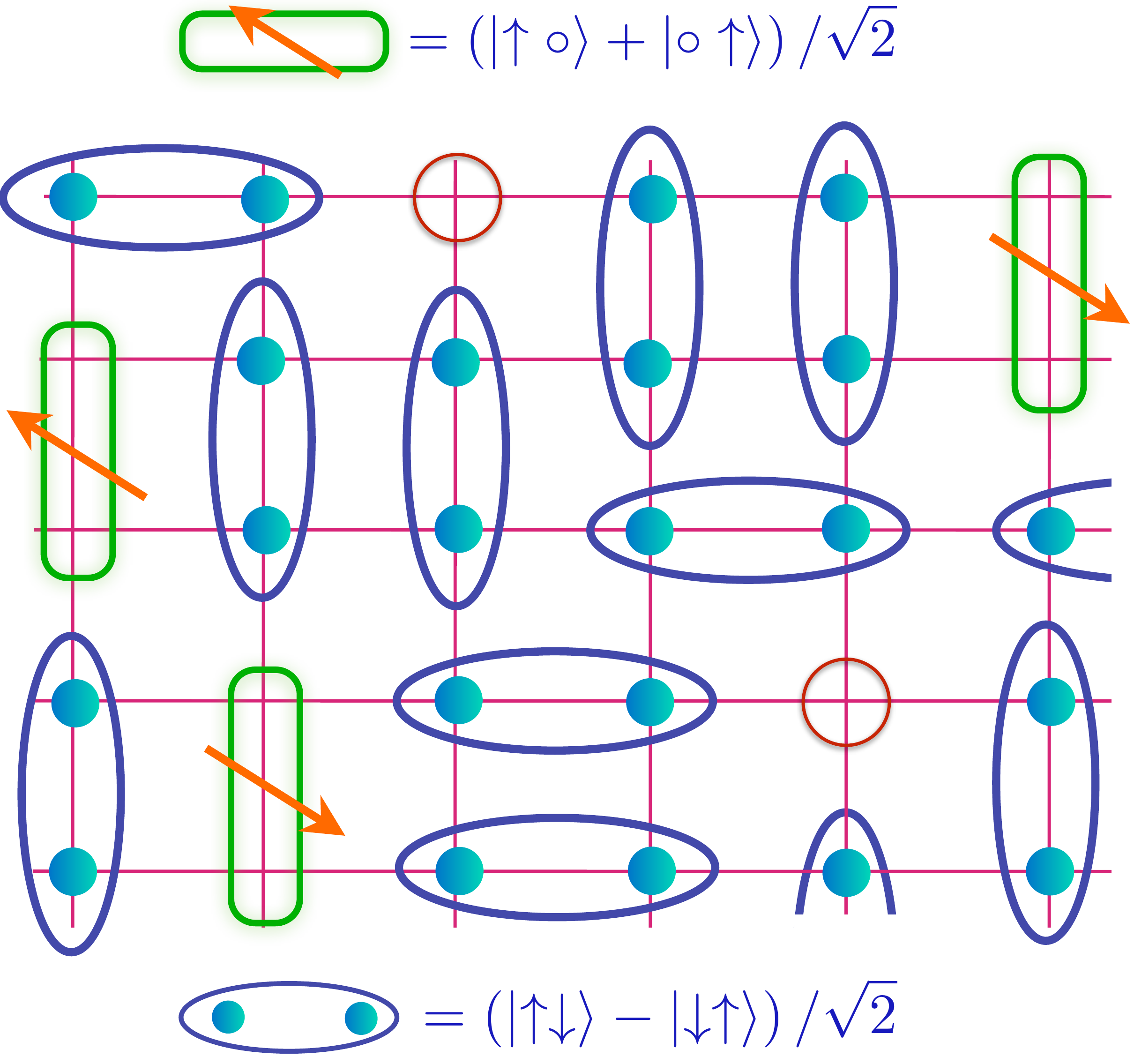}
\caption{\label{fig:flstar} \textbf{Schematic of a component of a state obtained by doping the resonating valence bond state} in Figure~\ref{fig:rvb}a by a density $p$
of holes. The orange circles represent spinless `holons' $h_q$ of electromagnetic charge $+e$, and emergent gauge charge $q=\pm 1$.
The green dimers, $d_\alpha$ are bound states of holons and the $z_\alpha$ quanta: the $d_\alpha$ are neutral under the emergent gauge field,
but carry electromagnetic charge $+e$ and spin $S=1/2$. Finally, the resonance of the blue and green dimers is captured by the emergent gauge field $A_\mu$.}
\end{figure}

Compressible quantum phases with topological order and emergent gauge fields are most directly realized by doping the insulating phases discussed in \S\ref{sec:emerge1}
by mobile charge carriers. We can dope the resonating valence bond state in Fig.~\ref{fig:rvb}a by removing a density $p$ of electrons. We are then left with a density
$1-p$ of electrons on the square lattice; this is equivalent to a density of $1+p$ holes relative
to the fully filled band. However, as long as the emergent gauge field of the parent resonating valence bond state remains deconfined, 
it is more appropriate to think of this system as having a density $p$ of positively charged carriers \cite{TSMVSS04,Punk:2015fha}. 
As illustrated in Fig.~\ref{fig:flstar}, these charge carriers can either
remain as spinless fermionic holons $h_q$, or bind with the quanta of the neutral spin-1/2 scalars $z_\alpha$ in Fig.~\ref{fig:rvb}b 
to form fermions, $d_\alpha$,  with electromagnetic charge $+e$
and spin $S=1/2$ \cite{RKK08}.

The final low energy description of this state is left with two kinds of fermions, both of which carry the Maxwell electromagnetic charge
(which is a global U(1) symmetry, as is conventional in the non-relativistic condensed matter contexts).
There are the `fractionalized' fermions, $h_q$, which carry charges $\pm 1$ under the emergent gauge field $A_\mu$ of (\ref{Sz}).
And we have the `cohesive' fermions, $d_\alpha$, which are neutral under $A_\mu$. The term `cohesive' was introduced in \cite{Hartnoll:2012ux} in opposition to `fractionalized'; we will see below that holographic descriptions
of compressible phases also lead naturally to such catergories of fermionic charge carriers.
So we can write down an effective low energy theory for the fermions in the spirit of (\ref{sc}): \begin{widetext}
\begin{equation}
S_{\mathrm{hd}} = \int \mathrm{d}\tau \int \frac{\mathrm{d}^2 k}{4 \pi^2} \Biggl[ d_{k \alpha}^\dagger \left( \frac{\partial}{\partial \tau}
+ E_{\mathrm{d}} (k) \right) d_{k \alpha} + \sum_{q = \pm 1} h_{k q}^\dagger \left( \frac{\partial}{\partial \tau} - \mathrm{i} A_\tau
+ E_{\mathrm{h}} (\vec{k} - q \vec{A}) \right) h_{k q} \Biggr] \,,
\end{equation}
\end{widetext}
where $E_{\mathrm{d}} (k)$ and $E_{\mathrm{h}} (k)$ are the dispersions of the cohesive and fractionalized fermions respectively. Note that the fractionalized fermions are minimally coupled to the emergent gauge field.

A modified version of the Luttinger relation (\ref{lutt1}) also applies in the presence of 
topological order \cite{TSMVSS04,APAV04}. As the fractionalized fermions can convert into cohesive fermions by binding with the $z_\alpha$, the relation only applies to the sum of the volumes enclosed by all the Fermi surfaces \cite{powell1,coleman1}
\beq
\sum_\alpha \int_{E_{\mathrm{d}} (k) < 0} \frac{\mathrm{d}^2 k}{4 \pi^2} + \sum_q \int_{E_{\mathrm{h}} (k) < 0} \frac{\mathrm{d}^2 k}{4 \pi^2}= p.
\label{lutt2}
\eeq
This is expected to be an exact relationship, satisfied by the fully renormalized excitations in the presence of the gauge
excitations, as long as the gauge field does not undergo a confinement transition and preserves topological order. In the confining case, the right hand side of (\ref{lutt2}) would be $1+p$, as we would have to count
all the holes, not just those relavtive to the resonating valence bond state.

Actually, there is an important feature we have glossed over in the discussion so far of (\ref{lutt2}).
The $h_q$ fermions are minimally coupled to a U(1) gauge field, and this coupling leads to a breakdown of the quasiparticle
physics near the Fermi surface. Indeed, the theory of the Fermi surface coupled to a gauge field \cite{metlitski1,mross} is nearly identical to that Ising-nematic
transition discussed in \S \ref{sec:isingnematic}. The field theory (\ref{inaction}) applies also to the gauge field problem,
with the scalar field $\phi$ representing the component of the gauge field normal to the Fermi surface (in the Coulomb gauge).
Consequently, 
much of the analysis for the structure of the excitations near the Fermi surface can be adapted from the Ising-nematic analysis
summarized in \S \ref{sec:isingnematic}. Despite the absence of quasiparticles, however, the position of the Fermi surface
in Brillouin zone can be defined precisely. Now, rather then the quasiparticle condition $E_{\mathrm{h}} (k) = 0$ used naively in (\ref{lutt2}),
we have the more general condition on the $h_q$ fermion Green's function \cite{sssolvay}
\beq
G_{\mathrm{h}}^{-1} (k, \omega = 0 ) = 0. \label{Gzero}
\eeq
With this modification, the Luttinger relation in (\ref{lutt2}) continues to apply to this realization of compressible quantum matter
without quasiparticles (similar relations also apply to the examples discussed in \S \ref{sec:isingnematic} and \S \ref{sec:sdw}). As discussed below (\ref{lutt1}), we need not consider the zeros of $G_{\mathrm{h}}$.

The reader is referred to separate reviews \cite{2016RSPTA.37450248S,SSDCNambu} for discussions of the applications of models of Fermi surfaces coupled to emergent gauge fields to the physics of the optimally doped cuprates. 

\subsection{Charged horizons}
\label{sec:chargedhor}

A simple way to obtain compressible quantum phases in holography is by adding a chemical potential $\mu$ to the strongly interacting holographic scale invariant theories described in  \S\ref{sec:scaling}. We are ultimately interested in the universal low energy dynamics of these compressible phases. For such questions it does not matter precisely which high energy starting point we use to obtain the compressible matter. It is convenient, then, to take CFTs ($z=1$ and $\theta=0$) as the UV completion of the IR compressible phase. This is analogous, for instance, to obtaining a Fermi liquid by `doping' the particle-hole symmetric point of graphene.

The chemical potential $\mu$ is the source for the charge density operator $\rho = J^t$ in the field theory. As explained around equation (\ref{eq:JA}) above, and elaborated in \S\ref{sec:QCcharge}, the bulk field dual to the charge current $J^\mu$ is a bulk Maxwell field $A_a$. In particular, $\rho$ is dual to $A_t$ in the bulk. 
The holographic dictionary implies that $\mu$ and $\rho$ are contained in the near boundary ($r \to 0$) expansion of $A_t$ according to
\be\label{eq:atnearb}
A_t = \mu - \frac{e^2 \rho}{L^{d-2} (d-1)} r^{d-1} + \cdots \,.
\ee
This expression -- which combines (\ref{eq:atlif}) and (\ref{eq:jt1}) above -- has assumed that near the boundary the Maxwell field is described by the Maxwell action (\ref{eq:Maxwell}) in an $\mathrm{AdS}_{d+2}$ background. This is typically the case for the class of doped holographic CFTs we will be considering. Away from the boundary, the Maxwell field will backreact on the geometry. This is the difference with \S\ref{sec:QCcharge}, where the Maxwell field was treated as a perturbation in order to obtain correlation functions in the zero density theory.

The asymptotic behavior (\ref{eq:atnearb}) implies the presence of a bulk electric field in the radial direction. Near the boundary
\be\label{eq:Eandrho}
E^r = \sqrt{-g} F^{tr} = - \sqrt{-g} g^{tt} g^{rr} \pa_r A_t = e^2 \rho \,.
\ee
Thus the charge density of the field theory is precisely given by the asymptotic electric flux in the bulk.
This bulk electric field is subject to Gauss's law (in the bulk). Therefore it must either originate from charged matter in the interior of the bulk geometry or the electric field lines must emanate from a charged horizon. We will see that these two possibilities respectively correspond to the `cohesive' (a term coined by \cite{Hartnoll:2012ux}) and `fractionalized' charge densities discussed briefly in \S \ref{sec:emerge2}.

\begin{figure}
\centering
\includegraphics[height = 0.22\textheight]{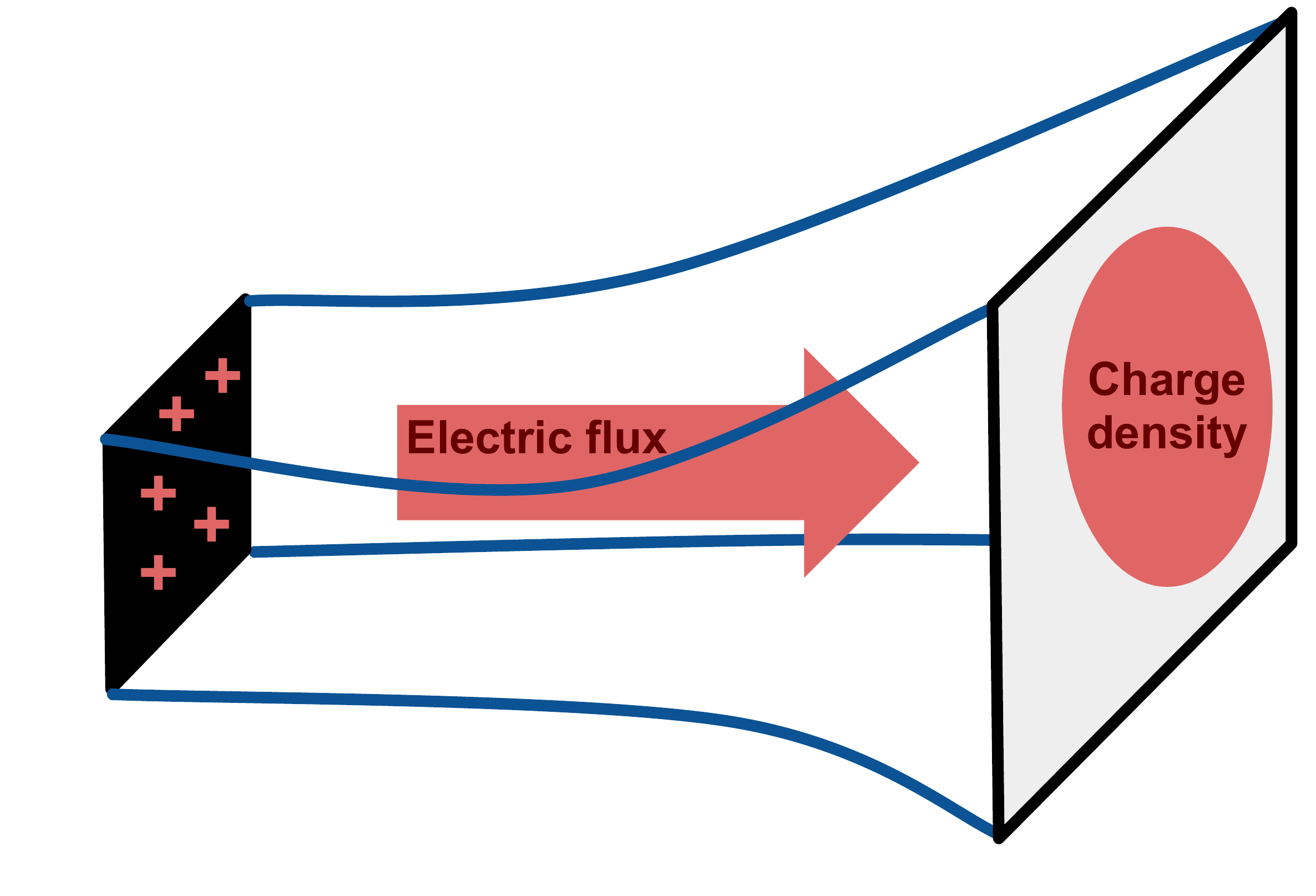}\\
\includegraphics[height = 0.22\textheight]{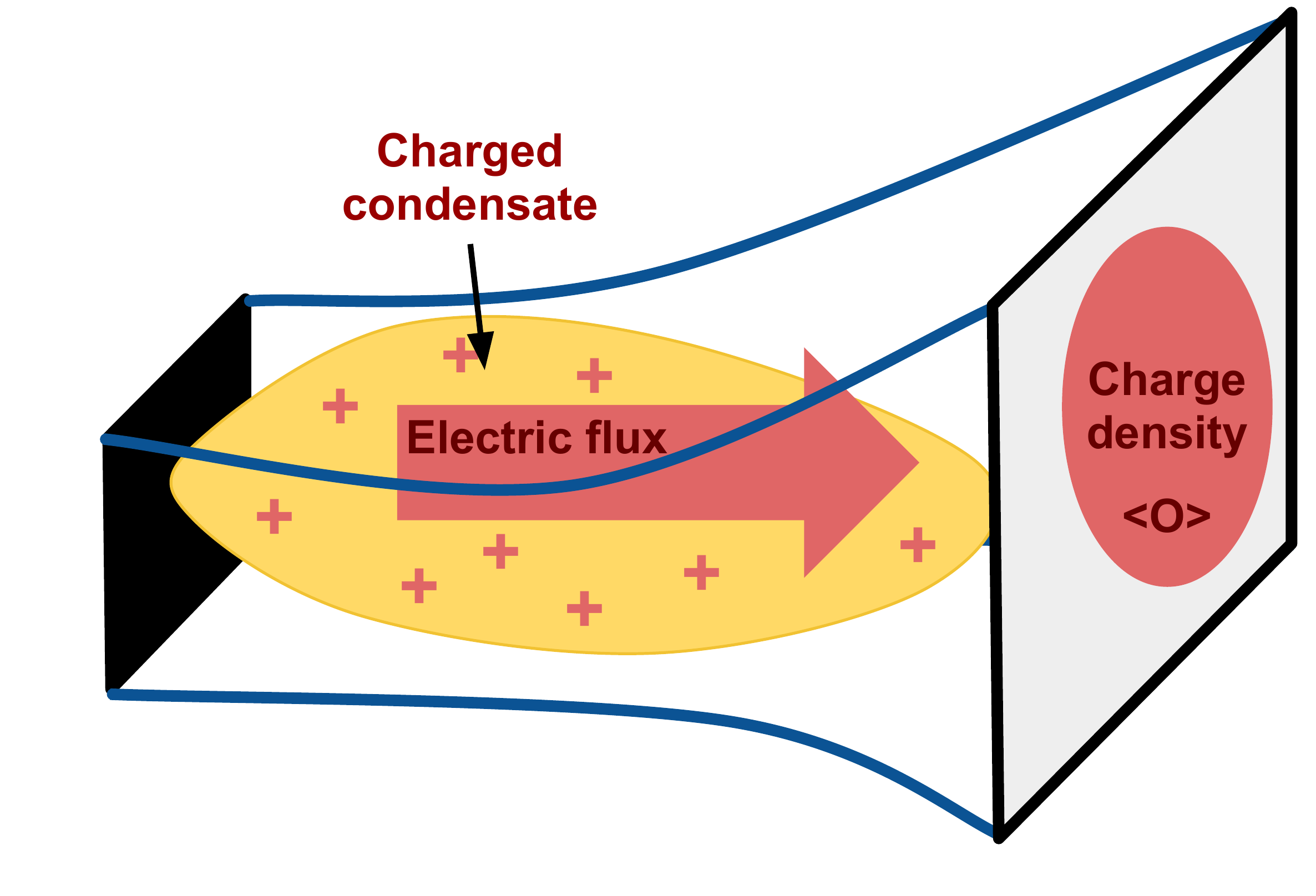}
\caption{\label{fig:charge} \textbf{Fractionalized vs. cohesive charge.} Top: The electric flux dual to the field theory charge density emanates from a charged event horizon. The charge is said to be `fractionalized'. Bottom: The electric flux is instead sourced by a condensate of charged matter in the bulk. The charge is said to be `cohesive'. Figures taken from \cite{Hartnoll:2011fn} with permission.}
\end{figure}

We saw in previous sections that much of the technical novelty of holographic quantum matter has to do with the geometrization of dissipation by classical event horizons.
Similarly, the most novel -- and indeed the simplest -- compressible phases arising holographically will be those in which the interior IR geometry is described by a charged horizon. We will start with these cases. These horizons can suffer instabilities leading to the spontaneous emission of charged matter from the black hole. Fermionic charged matter in the bulk will be discussed in \S \ref{sec:bulkF} below whereas bosonic charged matter will be the subject of \S \ref{sec:holoS}. The different scenarios are illustrated in figure \ref{fig:charge}.

\subsubsection{Einstein-Maxwell theory and $\mathrm{AdS}_2 \times \R^d$ (or, $z = \infty$)}
\label{sec:ads2rd}

The simplest holographic theory with the required ingredients to describe compressible states of matter is Einstein-Maxwell-AdS theory. This theory describes the interaction of the metric $g$ with a Maxwell field $A$ via the action
\be\label{eq:fullEM}
S = \int \mathrm{d}^{d+2}x \sqrt{-g}  \left[ \frac{1}{2 \k^2} \left(R + \frac{d(d+1)}{L^2} \right) - \frac{1}{4 e^2} F^2 \right] \,.
\ee
Einstein-Maxwell theory will be seen to describe a compressible phase with quite remarkable low temperature properties. Some of these remarkable properties have proven unpalatable to many, for reasons we shall see, leading to the introduction of the more complicated theories considered in later subsections. Nonetheless, the nontrivial and exotic physics captured by this simplest of holographic theories may yet contain clues about the dynamics of strongly interacting compressible phases of electronic matter.

The charged black hole solutions of (\ref{eq:fullEM}) were studied in some detail shortly after the discovery of holography \cite{Chamblin:1999tk}, but gained a new lease of life in the application of holographic methods to quantum matter. Both in early and more recent work, Einstein-Maxwell theory is sometimes studied as a special case of a more general classes of gravitational theories that are dual to specific supersymmetric field theories such as ${\mathcal N}=4$ SYM theory at a nonzero chemical potential, e.g. \cite{Cvetic:1999ne, Cvetic:1999xp, Son:2006em}. This is because ${\mathcal N}=4$ SYM admits three commuting global $\mathrm{U}(1)$ symmetries, and Einstein-Maxwell theory describes the case in which the corresponding three chemical potentials are all equal.

The charged black hole solutions to (\ref{eq:fullEM}) are relatively simple and well known. We shall simply quote the black hole metric and Maxwell field. In the following subsection we will describe in detail how to find more general charged black hole solutions, and the present planar Reissner-Nordstr\"om-AdS (AdS-RN) geometry is a special case of that discussion.  We will restrict to $d>1$ for simplicity. The metric is
\be\label{eq:gRN}
\mathrm{d}s^2 = \frac{L^2}{r^2} \left(- f(r) \mathrm{d}t^2 + \frac{\mathrm{d}r^2}{f(r)} + \mathrm{d} \vec x^2_d \right) \,,
\ee
where the radial function is
\be
f(r) = 1 - \left( 1 + \frac{r_+^2 \mu^2}{\g^2} \right) \left(\frac{r}{r_+}\right)^{d+1} + \frac{r_+^2 \mu^2}{\g^2} \left(\frac{r}{r_+} \right)^{2d} \,.
\ee
The spacetime therefore has a horizon at $r=r_+$, and tends to $\mathrm{AdS}_{d+2}$ near the boundary as $r \to 0$.
We have grouped various coefficients into
\be
\g^2 = \frac{d}{d-1} \frac{e^2 L^2}{\k^2} \,.
\ee
This constant determines the relative strengths of bulk gravitational and electromagnetic interactions. The background Maxwell potential is
\be\label{eq:athor}
A_t = \mu \left[1 - \left(\frac{r}{r_+} \right)^{d-1} \right] \,.
\ee
Note that $A_t$ vanishes on the horizon. This boundary condition is necessary for the Wilson loop of the Maxwell field around the Euclidean thermal circle to be regular as $r \to r_+$, where the thermal circle shrinks to zero (see figure \ref{fig:cigar} above).
 Comparing with the near-boundary expansion (\ref{eq:atnearb}), we see that the black hole carries a charge density
\be\label{eq:rrr}
\rho = \frac{ (d-1) L^{d-2}}{e^2 r_+^{d-1}} \mu \,.
\ee
Finally, the temperature is found following the procedure described in \S \ref{sec:thermoBH} to be
\be
T = \frac{1}{4 \pi r_+} \left(d+1 - \frac{(d-1) r_+^2 \mu^2}{\g^2} \right) \,.
\ee
This expression is readily inverted to obtain an explicit but slightly messy formula for $r_+$ in terms of $T$ and $\mu$. The limiting behaviors are
\be
r_+ \approx \left\lbrace
\begin{array}{ll} \displaystyle \dfrac{d+1}{4\pi T} &\ \mu \ll T \\[1em]
\displaystyle  \sqrt{\dfrac{d+1}{d-1}} \dfrac{\g}{\mu} \quad &\ \mu \gg T 
\end{array} \right. \label{eq:hsize}
\ee
From the high temperature behavior in the first line above, it is straightforward to see (from the fact that $r_+$ is becoming small, i.e. close to the boundary at $r=0$) that the solution becomes the AdS-Schwarzchild black brane geometry together with an electric field. In contrast, in the second line we see that the horizon remains of nonvanishing size as the temperature is taken to zero. That behavior is very different from the neutral AdS-Schwarzchild solution.

The entropy density is obtained from the horizon area according to (\ref{eq:s}). Using the horizon radius (\ref{eq:hsize}), in the low temperature limit one obtains
\be\label{eq:sss}
s = \frac{2 \pi L^d}{\k^2 r_+^d} \sim \mu^d \qquad \text{as} \quad  T \to 0 \,. 
\ee
In particular, the entropy density is finite at $T = 0$. This is a violation of the third law of thermodynamics; mollifying this feature was an important goal driving the study of the more general Einstein-Maxwell-dilaton models discussed below. See however the final paragraph of this section: the solvable SYK model has lead to some intuition for how the absence of quasiparticles can lead to a proliferation of low energy states. In the remainder of this subsection we will trace the origin of the zero temperature entropy density to the fact that this theory has an emergent IR scaling symmetry with $z = \infty$. Taking $z \to \infty$ (with $\theta$ finite) indeed leads to a zero temperature entropy density. Recall that according to (\ref{eq:s}) or (\ref{eq:entT}), $s \sim T^{(d-\theta)/z} \to \text{const.}$ as $z \to \infty$.

In later sections we will see how $z=\infty$ describes a phase of quantum critical matter with fascinating properties. Because $z = \infty$ implies that time scales but space does not, momentum is dimensionless under the scaling symmetry. Thus low energy critical excitations are present at all momenta, not just small momenta (as would be the case for any finite $z$, wherein $\omega \sim k^{z}$). This fact allows fermions living at a Fermi momentum $k_{\mathrm{F}}$ to interact strongly with the critical sector (see \S\ref{sec:bulkfermions}). It also allows momentum non-conserving scattering due to a lattice at wavevector $k_{\mathrm{L}}$ to participate strongly in the critical dynamics (see \S\ref{sec:holexample}). More generally, $z=\infty$ criticality is compatible, in principle, with arbitrary spatial inhomogeneity. Both of these features -- quantum criticality compatible with a Fermi surface and with momentum relaxation -- are desirable and difficult to achieve in other kinematic frameworks.

For completeness, note that with the entropy density (\ref{eq:sss}) and charge density (\ref{eq:rrr}) at hand, the energy density and pressure can be obtained from the general thermodynamic relations (\ref{eq:gibbsduhem}) and (\ref{eq:epsandP}). In the latter equation, relating energy and pressure, one should put $z=1$ and $\theta = 0$, as this is a UV relationship that holds for all states in the doped CFT. We will expand more on this point in the last paragraph of  \S\ref{sec:EMDgeom} below. While the low temperature scaling of the entropy density is determined purely by IR data (the horizon), this is not the case for the energy density and pressure, that receive contributions from all energy scales.

\begin{figure}
\centering
\includegraphics[width=3.3in]{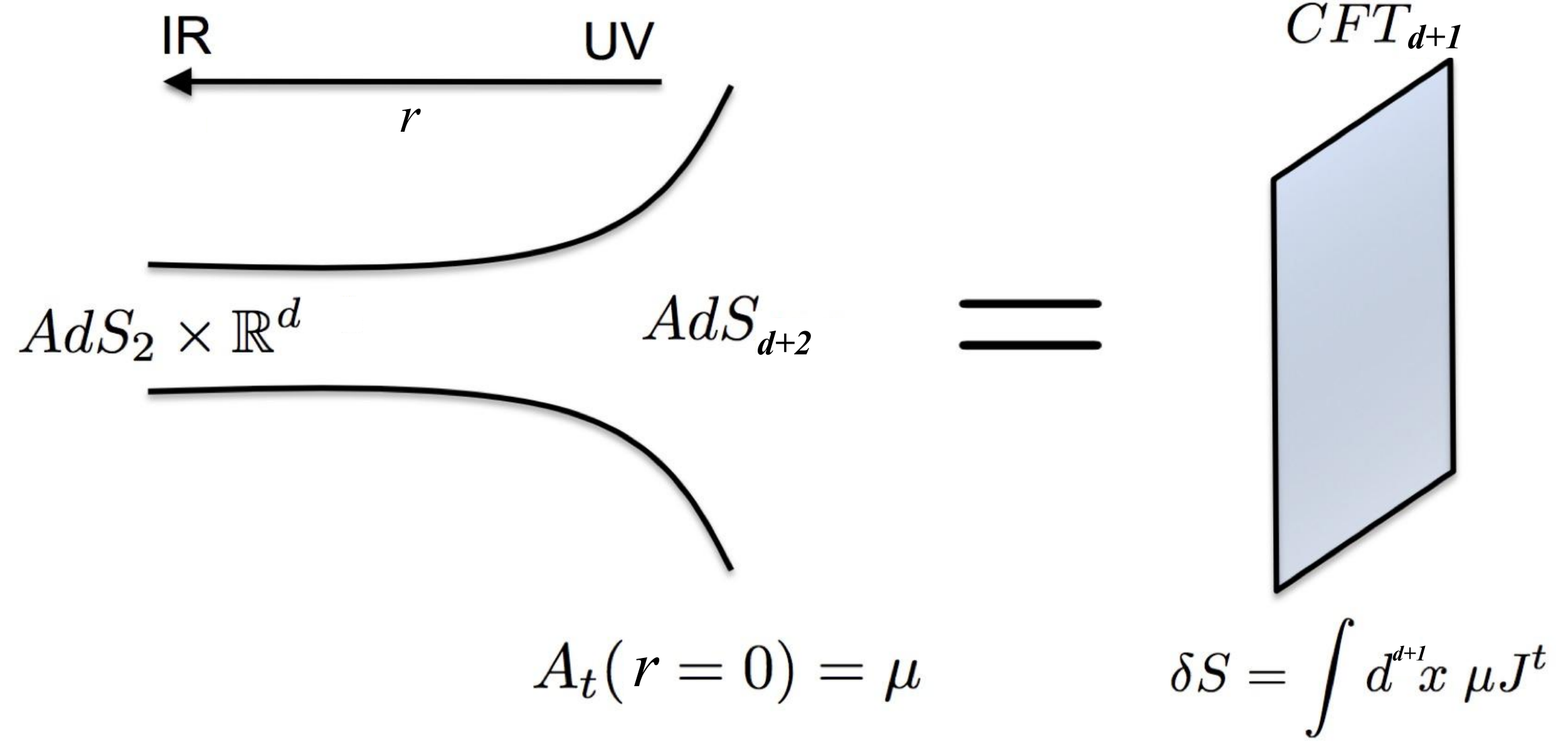}
\caption{\label{fig:ads2} \textbf{Emergence of $\mathrm{AdS}_2$.}  A CFT, dually described by Einstein-Maxwell theory in the bulk, is placed at a nonzero chemical potential. This induces a renormalization group flow in the bulk, leading to the emergence of semi-locally critical $\mathrm{AdS}_2 \times \mathbb{R}^{d}$ in the far IR, near horizon region. Figure adapted with permission from \cite{Iqbal:2011ae}.}
\end{figure}

The near-horizon regime of the black hole spacetime will be seen to control all low energy dissipative processes. It is therefore very important to determine exactly what this geometry is.   To do so, let us switch to the new coordinate \begin{equation}
\zeta = \frac{1}{r_*-r} \frac{r_*^2}{d(d+1)} ,
\end{equation} 
where we have defined \begin{equation}
r_* = \sqrt{\frac{d+1}{d-1}} \frac{\gamma}{\mu} \,, \label{eq:rrstar}
\end{equation}
as the $T=0$ limit of the horizon location.   Let us temporarily fix $T=0$ exactly.   Then $\zeta\rightarrow\infty$ as we approach the horizon.  Switching from the $r$ coordinate to $\zeta$, as $\zeta \rightarrow \infty$ the geometry (\ref{eq:gRN}) becomes: \begin{equation}
\mathrm{d}s^2 = L_2^2 \frac{- \mathrm{d}t^2 + \mathrm{d}\zeta^2 }{\zeta^2} + \frac{L^2}{r_*^{2}}\mathrm{d}\vec x_d^2 \,. \label{eq:nhmetric}
\end{equation}
This near-horizon geometry is an $\mathrm{AdS}_2 \times \mathbb{R}^{d}$ spacetime \cite{Gibbons,Faulkner:2009wj}. The radius of the emergent AdS$_2$ is \begin{equation}
L_2 \equiv \frac{L}{\sqrt{d(d+1)}} < L,
\end{equation}
and we will see in \S\ref{sec:IRinstab} that this fact has implications for possible instabilities of the near-horizon geometry.   Furthermore, we see that the emergent scaling regime in the IR does not involve the spatial coordinates! The scaling symmetry is simply $\zeta \to \l \zeta, t \to \l t$. Following the discussion around (\ref{eq:Lif}) above, we obtain a dynamic critical exponent $z=\infty$;  such a theory is called locally critical, or more properly, as we will see, semi-locally critical \cite{Iqbal:2011in}. As we have mentioned, semi-local criticality leads to interesting physics, that we will return to later in this chapter. Figure \ref{fig:ads2} illustrates the emergence of the semi-local criticality from a doped CFT.

Finally, let us add a small nonzero temperature $T$.   The location of the horizon shifts to \begin{equation}
r_+ = r_*\left[1-\frac{2\pi \g^2 T}{(d-1)\mu^2 r_*}\right]  + \mathrm{O}\left(T^2\right) \equiv r_*(1-\delta).
\end{equation}   In this case, we change the definition of $\zeta$ slightly:  \begin{equation}
\zeta = \frac{1}{r_*(1-\delta(d-1)/d)-r} \frac{r_*^2}{d(d+1)}.  \label{eq:zetafiniteT}
\end{equation}
The near-horizon metric now becomes \begin{equation}
\mathrm{d}s^2 \approx \frac{L_2^2}{\zeta^2} \left(-f_2(\zeta) \mathrm{d}t^2 + \frac{\mathrm{d}\zeta^2}{f_2(\zeta)} \right) + \frac{L^2}{r_*^{2}}\mathrm{d}\vec x_d^2 \,,  \label{eq:smallTgeom}
\end{equation}
where \begin{equation}
f_2(\zeta) = 1 - \frac{\zeta^2}{\zeta_+^2} \quad \text{and} \quad \zeta_+ \equiv \frac{1}{2\pi T}.
\end{equation}
Indeed, as one might have guessed,  the emergent geometry is $\mathbb{R}^d$ times an $\mathrm{AdS}_2$-Schwarzchild black brane. That is to say, the event horizon is contained within the emergent low energy $\mathrm{AdS}_2$ geometry. The nonzero temperature near horizon geometry (\ref{eq:smallTgeom}) will control all of the dissipative physics of the theory at low temperatures and low energies $\{\omega,T\} \ll \mu$. Once $T\sim \mu$,  this emergent scaling geometry disappears, swallowed by the horizon, and we transition back to the $\mathrm{AdS}_{d+2}$ black brane.

The spacetime $\mathrm{AdS}_2 \times \mathbb{R}^{d}$ is a solution of Einstein-Maxwell theory in its own right. While we have obtained the geometry as the low energy description of a CFT deformed by a chemical potential, it should more properly be considered as a self-contained fixed point theory. The $\mathrm{AdS}_2 \times \mathbb{R}^{d}$ solution by itself will capture all of the universal low energy physics. Indeed, for instance, a small temperature leads to $\text{Schwarzschild-}\mathrm{AdS}_2 \times \mathbb{R}^{d}$, which is also a solution on its own. We will see many examples of how the IR fixed point geometry controls the low energy physics below.

The emergent semi-locally critical regime described above is a critical phase. No parameter has been tuned to obtain the quantum criticality. Indeed, within the bulk Einstein-Maxwell theory there is no extra parameter to tune. We simply started with a CFT in the UV and deformed by a chemical potential. In more general bulk theories, with extra fields, there can be relevant deformations of the locally critical IR geometry.

It must be kept in mind that a non-zero entropy in the zero temperature limit (which appears in locally critical theories
with $z=\infty$) does {\em not\/} imply an exponentially large ground
state degeneracy. In a system with $N$ degrees of freedom, there are $e^{\alpha N}$ states (for some constant $\alpha$),
and so the typical energy level spacing is of order $\mathrm{e}^{- \alpha N}$. In systems with quasiparticles, the spacing of the lowest energies near the ground state is no longer exponentially small in $N$, and the zero temperature limit of the entropy vanishes.
In the present holographic system, the energy level spacing remains of order $\mathrm{e}^{-\alpha N}$ even near the ground state:
this understanding has emerged from studies of the SYK models, which we will describe in \S \ref{sec:SYK}.
These models, and their holographic duals, allow a systematic study of $1/N$ corrections \cite{Almheiri:2014cka,WFSS16,Polchinski:2016xgd, Maldacena:2016hyu, Maldacena:2016upp, Engelsoy:2016xyb}, and the locally critical scaling holds down
to the smallest values of (total energy)/$N$.


\subsubsection{Einstein-Maxwell-dilaton models}
\label{sec:EMD}

The next simplest set of holographic models are Einstein-Maxwell-dilaton (EMD) theories. The bulk action takes the general form
\begin{align}
S &= \int \mathrm{d}^{d+2}x\sqrt{-g} \left[\frac{1}{2\kappa^2}\left(R - 2(\partial \Phi)^2 - \frac{V(\Phi)}{L^2}\right) \right. \notag \\
&\left.\;\;\;\;\;- \frac{Z(\Phi)}{4e^2}F^2\right] \,. \label{eq:EMDaction}
\end{align}
In particular, the Einstein-Maxwell action (\ref{eq:fullEM}) has been enhanced to include an additional scalar field $\Phi$, the `dilaton' (the term is used loosely here). In consistent truncations of string theory backgrounds, such scalars are ubiquitous. We gave some examples in \S\ref{sec:consistent} above. The general class of bulk theories (\ref{eq:EMDaction}) is divorced from any particular string theoretic realization, and is instead used to explore the space of possible behaviors of the bulk spacetime. Early papers that studied this class of theories as holographic models of compressible matter include \cite{Taylor:2008tg, Goldstein:2009cv, Gubser:2009qt, Charmousis:2010zz, Iizuka:2011hg}. In the following subsection we will see that 
these theories lead to a wealth of new emergent critical compressible phases, now with a range of possible values of the exponents $z$ and $\theta$. Before that, we illustrate how one goes about finding the bulk equations of motion from an action such as (\ref{eq:EMDaction}).

When looking for rotationally and translationally invariant states, the bulk fields will only have a nontrivial dependence on the radial coordinate, $r$. Therefore, the bulk equations of motion reduce to ordinary rather than partial differential equations. To find these differential equations we make the following ansatz for the dilaton, electrostatic potential and metric
\begin{align}
\Phi &= \Phi(r) \,, \quad A=p(r)\mathrm{d}t \,, \notag \\
\mathrm{d}s^2 &= \frac{L^2}{r^2}\left[\frac{a(r)}{b(r)}\mathrm{d}r^2 - a(r)b(r)\mathrm{d}t^2 + \mathrm{d}\vec x_d^2\right] \,.  \label{eq:abmetric}
\end{align}
There is some choice in how the metric components are parametrized. The choice made above (see e.g. \cite{Lucas:2014sba}) proves convenient as near a black hole horizon at $r=r_+$, \begin{equation}
b(r) \approx 4\pi T(r_+-r),   \label{eq:b4pT}
\end{equation}
regardless of other fields. In the coordinates in which the metric (\ref{eq:abmetric}) has been written, spatial infinity is at $r=0$. If we are working in an asymptotically AdS spacetime, dual to a `doped CFT' as described at the start of this \S\ref{sec:chargedhor}, then $a(0)=b(0)=1$.

On the ansatz (\ref{eq:abmetric}), the Einstein-Maxwell-dilaton equations of motion become
\begin{subequations}\label{eq:EMDEOM}\begin{align}
-\frac{da^\prime}{ra} &= 2\Phi^{\prime 2}, \label{eq:EMDEOM1} \\
r^da\left(\frac{(ab)^\prime}{ar^d}\right)^\prime &=   \frac{\kappa^2}{e^2L^2} \, 2 r^2 p^{\prime2}, \label{eq:EMDEOM2} \\
\left(\frac{Zp^\prime}{ar^{d-2}}\right)^\prime &= 0, \\
4\left(\frac{b}{r^d}\Phi^\prime\right)^\prime &= \frac{a}{r^{d+2}}\frac{\partial V}{\partial \Phi} - \frac{\kappa^2}{e^2L^2} \, \frac{p^{\prime 2}}{a r^{d-2}} \frac{\partial Z}{\partial \Phi}.
\end{align}\end{subequations}
In order, these equations come from the $rr$ and $tt$ components of Einstein's equation,  the $tt$ and $ii$ components of Einstein's equation,   the $t$ component of Maxwell's equation, and the dilaton equation.   In fact, the Maxwell equation above is simply Gauss' Law. After integrating this equation, from (\ref{eq:Eandrho}) we can interpret the integration constant as the charge density $\rho$ of the black hole horizon (and the dual field theory): \begin{equation}
\rho= - \frac{L^{d-2}}{e^2} \frac{Zp^\prime}{ar^{d-2}}.  \label{eq:gausslaw}
\end{equation}
Here $a$ is directly determined by the dilaton profile using (\ref{eq:EMDEOM1});  and then the gauge field $p$ is determined from $a$ and $\Phi$ through Gauss's law. The coefficient $b$ is related to the gauge field through (\ref{eq:EMDEOM2}) -- in the case where the black hole is uncharged, then we find that the computation of $b(r)$ for a black hole geometry reduces to a first order differential equation.     

Although (\ref{eq:EMDEOM}) can be derived by hand, in practice it is helpful to use differential geometry packages to evaluate the Ricci tensor in Einstein's equations.  For convenience, we provide below the \texttt{Mathematica} code necessary to obtain (\ref{eq:EMDEOM}),  employing the \texttt{RGTC} package.   The code below will only solve the equations in a specific spacetime dimension $d=2$, but is easily changed for any $d$ of interest: \begin{widetext}

\begin{quote} 

\% define the coordinate labels 

\texttt{coords = \{r,t,x,y\};}   

 \% define the metric ansatz 

\texttt{g = DiagonalMatrix[\{a[r]*L\^ \,2/(r\^ \,2*b[r]), -a[r]*b[r]*L\^ \,2/r\^ \,2, L\^ \,2/r\^ \,2, L\^ \,2/r\^ \,2\}];} 

\%  generate $R_{MN}$,  $R_{MNRS}$ etc.

\texttt{RGtensors[g, coords, \{0, 0\}];}  

\%  define the ansatz for $\Phi$

\texttt{Phi := phi[r];}  

\%  define the 1-form $\nabla_M \Phi$.

\texttt{dPhid := covD[Phi];}  

\%  define the ansatz for $A_M$;  the d label reminds us that the index is down

\texttt{Ad := \{0, p[r], 0, 0\};}  

\%  define $F_{MN}$

\texttt{Fdd := Transpose[covD[Ad]] - covD[Ad];}  

\%  define $F^{MN}$; raise indices

\texttt{FUU := Raise[Fdd, 1, 2];}  

\%   $\Phi$ contributions to the right hand side of Einstein's equations;   multiplying tensors together

\texttt{PhiTdd := 2*Outer[Times,dPhid,dPhid] - (Contract[Outer[Times,dPhid,gUU,dPhid], \{1,2\},\{3,4\}] + V[Phi])*gdd/2;}  

\%   $F^2$ contributions to the right hand side of Einstein's equations

\texttt{F2scalar := Contract[Outer[Times, Fdd, FUU], \{1, 3\}, \{2, 4\}];}

\texttt{F2Tdd := Z[Phi]*(Contract[Outer[Times, Fdd, gUU, Fdd], \{2, 3\}, \{4, 6\}] - F2scalar*gdd/4);}

\%   Einstein's equations; the output is a $(0,2)$ tensor, and vanishes on-shell

\texttt{EOM1 := Rdd - R*gdd/2 - k\^ \,2/e\^ \,2 * F2Tdd - PhiTdd;} 

\%   Maxwell's equations; the output is a $(1,0)$ tensor 

\texttt{EOM2 := covDiv[Z[r]*FUU, \{1, \{2\}\}];}  

\%   dilaton equation

\texttt{EOM3 := 4*covDiv[Raise[dPhid,1],1] - V'[Phi]/L\^ \,2 - Z'[Phi] * F2scalar * k\^ \,2/(2*e\^ \,2);}  
\end{quote}
\end{widetext}

Given this code, one finds a messy set of equations. With a little practice one can gain intuition for 
re-organizing them into more transparent forms such as (\ref{eq:EMDEOM}).

\subsubsection{Critical compressible phases with diverse $z$ and $\theta$} \label{sec:EMDgeom}

In Einstein-Maxwell theory we explained that the universal low energy and low temperature physics is determined
by an emergent near horizon $\mathrm{AdS}_2 \times \mathbb{R}^{d}$ spacetime. We also noted that this scaling geometry
was a solution of the Einstein-Maxwell equations of motion on its own. The same phenomenon occurs for
an important class of potentials $V$ and $Z$ in Einstein-Maxwell-dilaton theory (\ref{eq:EMDaction}). In this section
we describe the IR scaling solutions that arise, leading to critical phases of compressible holographic matter.
Following the presentation of \cite{Huijse:2011ef}, we parametrize these solutions in terms of ``arbitrary" $z$ and $\theta$, up to some constraints which we note below. 

Start at zero temperature. We look for a scaling solution by plugging the ansatz
\begin{subequations}\label{eq:EMDmetric42}\begin{align}
a &=a_\infty \, r^{-(d(z-1)-\theta)/(d-\theta)} \,, \\
b &= b_\infty \, r^{-(d(z-1)+\theta)/(d-\theta)} \,,
\end{align}\end{subequations}
into the equations of motion (\ref{eq:EMDEOM}). This ansatz corresponds to the same form of metric as considered in the hyperscaling-violating solutions of \S\ref{sec:hsv1} above, albeit in a different coordinate system. Namely,
\begin{align}
\mathrm{d}s^2 &= \frac{L^2}{r^2}\left( - a_\infty b_\infty r^{-2d(z-1)/(d-\theta)}\mathrm{d}t^2 \right. \notag \\
&\left.\;\;\;\;\; +
\frac{a_\infty}{b_\infty} r^{2 \theta/(d-\theta)} \mathrm{d}r^2 + \mathrm{d}\vec x_d^2\right) \,. 
\end{align}
This metric can be put into the form (\ref{eq:hsv}) through the coordinate change $r \to r^{1-\theta/d}$. Thus we see that $z$ is the dynamic exponent and $\theta$ the hyperscaling violation exponent. The difference with the discussion in \S\ref{sec:hsv1} is the presence of a Maxwell field. From (\ref{eq:EMDEOM1}) we find that the dilaton
\begin{equation}
\Phi = \Phi_0 \log\frac{r}{r_0} \,, \label{eq:ppsol}
\end{equation}
with $\Phi_0^2 = d[d(z-1)-\theta]/[2(d-\theta)]$ and $r_0$ a constant related to the crossover out of the IR scaling region. The presence of the dimensionful quantity $r_0$ in the solution is nontrivial, as we discuss below.  As $\Phi(r)$ must be real, we see that \begin{equation}
\frac{d(z-1)-\theta}{d-\theta}\ge 0.
\end{equation}
This constraint is also implied by the null energy condition, as we noted in \S\ref{sec:hsv1}.

One can now use the remainder of the equations of motion to show that
\begin{subequations}\begin{align}
V(\Phi) & = -V_0 \, \mathrm{e}^{-\beta\Phi}, \\
Z(\Phi) & = Z_0 \, \mathrm{e}^{\alpha \Phi},
\end{align}\label{eq:VZ}\end{subequations}
with \begin{subequations}\begin{align}
\alpha^2 &= \frac{8 (d(d-\theta)+\theta)^2}{d(d-\theta)(dz-d-\theta)} \,, \\
\beta^2 &=  \frac{8 \theta^2}{d(d-\theta)(dz-d-\theta)} \,.
\end{align}\label{eq:albe}\end{subequations}
The exact prefactors $V_0$ and $Z_0$ can be found in \cite{Huijse:2011ef, Lucas:2014sba}.   Finally, one finds that the scalar potential
\begin{equation}
p = p_\infty \, r^{-d-dz/(d-\theta)} \,, \label{eq:pEMD}
\end{equation}
with the prefactor fixed by (\ref{eq:EMDEOM2}).

Logically speaking, of course, the theory with potentials $V$ and $Z$ is prior to any specific solution. With this in mind, the results in the previous paragraph can be rephrased as follows: If the potentials $V$ and $Z$ in the Einstein-Maxwell-dilaton action behave as (\ref{eq:VZ}) to leading order as $\Phi \to \infty$, then the theory admits the above emergent low energy scaling solution in the far interior as $r \to \infty$ (according to (\ref{eq:ppsol}), $\Phi \to \infty$ as $r \to \infty$ ). The critical exponents $z$ and $\theta$ are determined by $\a$ and $\b$ according to (\ref{eq:albe}).  

If we start with a doped CFT in the UV ($r \to 0$), the above $(z,\theta)$ scaling solutions will emerge in the far interior of the spacetime ($r \to \infty$), in just the way that $\mathrm{AdS}_2 \times \mathbb{R}^{d}$ emerged in Einsten-Maxwell theory. Generically the full interpolating geometry -- solving the equations of motion (\ref{eq:EMDEOM}) at all $r$ -- can only be found numerically. Let us stress again, however, that all of the universal low energy physics will be determined by the scaling solution. We will see many examples of this shortly. At $T=0$, the only scale in the theory is the charge density $\rho$ induced by doping the CFT. This determines a radial scale
\begin{equation}
r^* = \hat{\rho}^{-1/d} \,, \label{eq:rstar}
\end{equation}
with \begin{equation}
\hat{\rho} \equiv \frac{e \, \kappa}{L^{d-1}} \rho \,. \label{eq:rhorescale}
\end{equation}
One expects to find that near the boundary, for $0 < r \lesssim r^*$, the geometry will be approximately AdS, whereas in the interior, for $r^* \lesssim r$, the geometry will be of the $(z,\theta)$ scaling form. Thus the charge density sets the UV cutoff for the emergent scaling region. The factors of $e,\k,L$ in (\ref{eq:rhorescale}) are obtained as follows. If we perform the rescaling $p=(eL/\kappa) \hat p$, and switch to the coordinate $\hat r = r/r^*$,  then all dimensionful quantities drop out of the equations of motion (\ref{eq:EMDEOM}).

The $T=0$ scaling geometries are generally singular in the far IR, as $r \to \infty$. The metrics typically have naked or null curvature singularities. Furthermore, the blowup of the dilaton field indicates the onset of quantum gravity or string theory effects at large $r$. For discussions of the singular nature of these solutions from various angles see \cite{Charmousis:2010zz, Iizuka:2011hg, Horowitz:2011gh, Shaghoulian:2011aa, Dong:2012se, Bao:2012yt,Harrison:2012vy}. As we shall see, the singularities are sufficiently mild that infalling boundary conditions can consistently be imposed at the $T=0$ `horizons', and the computations of thermodynamic and linear response quantities in these backgrounds make sense. Nonetheless, the singularities mean that the low energy limit ultimately does not commute with the large $N$ classical limit for these backgrounds. While at large $N$, then, classical gravity describes scale-invariant compressible phases of matter over a parametrically large range of energy scales, the 't Hooft limit is not strong enough to uncover new strictly controlled finite density $T=0$ fixed points in these cases. An exception seems to be when $z=1+\theta/d$, in which case the IR geometry is not singular \cite{Shaghoulian:2011aa, Lei:2013apa}. These regular geometries saturate the first of the null energy constraints in (\ref{eq:null}) and deserve further attention.

A class of black hole solutions to (\ref{eq:EMDEOM}) can be found analytically. These will describe the universal low energy physics of the critical phase at temperatures $0 < T \ll \k/(eL) \, \mu$. The second inequality corresponds to the temperature below which the bulk event horizon will be well within the IR scaling region -- see the discussion below (\ref{eq:rhorescale}) above. The solution is found by noting that (\ref{eq:EMDEOM2}) is satisfied (in the IR scaling region) by \begin{equation}
b(r) = b_{T=0}(r) + \text{constant}\times r^{d+1}.
\end{equation}
Hence, we can add a black hole horizon by setting \begin{equation}
b(r) = b_\infty r^{-(d(z-1)+\theta)/(d-\theta)} \left[1 - \left(\frac{r}{r_+}\right)^{d(1+z/(d-\theta))}\right] \,.
\end{equation}
Using (\ref{eq:b4pT}), the horizon radius is related to the temperature by  \begin{equation}
r_+ = \left(\frac{4\pi T}{db_\infty} \frac{d-\theta}{d-\theta+z} \right)^{-(1-\theta/d)/z},  \label{eq:rPEMD}
\end{equation}
and hence the entropy density \begin{equation}
s \sim T^{(d-\theta)/z}.  \label{eq:Sthetaz}
\end{equation}
This is the expected scaling (\ref{eq:entT}). In particular, for general values of $d-\theta>0$ and $z$, the entropy $s \to 0$ as $T\to 0$ and the third law of thermodynamics holds true. Furthermore, at any finite $T>0$ the geometric singularities are hidden behind a black hole horizon, and so in this sense the singular nature of the IR geometry is `good' \cite{Gubser:2000nd}.

An interesting limit of this class of solutions is to take \cite{Hartnoll:2012wm, Anantua:2012nj}
\begin{equation}\label{eq:etalimit}
z\rightarrow \infty \quad \text{with} \quad  -\frac{\theta}{z} \equiv \eta > 0 \quad \text{fixed}.
\end{equation}
This limit retains the interesting phenomenology of $z = \infty$ fixed points without a zero temperature entropy density ($s\sim T^\eta \to 0$). Furthermore, the simplest string theory realizations of Einstein-Maxwell-dilaton theory turn out to realize this limit \cite{Cvetic:1999xp, Gubser:2009qt}.

The following sections will characterize the spectrum of fluctuations about these backgrounds in some detail. We can make a quick prior comment. In hyperscaling-violating geometries ($\theta \neq 0$) a dimensionful scale does not decouple at low energy (e.g. $r_0$ in (\ref{eq:ppsol}) or $R$ in (\ref{eq:hsv})). This means that observables computed in these backgrounds are not guaranteed to take a scale-invariant form. For instance, consider a scalar field $\varphi$ in the bulk (not the dilaton, an additional scalar dual to some scalar operator in the boundary QFT). The quadratic action for the scalar field might take the general form
 \begin{equation}
S_\phi = -\frac{1}{2}\int \mathrm{d}^{d+2}x\; \sqrt{-g}\left((\partial \varphi)^2 + B(\Phi)\varphi^2\right) \,. \label{eq:4phi}
\end{equation}
For a simple mass term, $B(\Phi) = m^2/L^2$, it is straightforward to show that for $\theta \ne 0$ these dual operators are gapped, with a large separation Green's function $G(x)\sim \mathrm{e}^{-mx}$ \cite{Dong:2012se}. 
However, if $B(\Phi) \sim \mathrm{e}^{-\beta \Phi}$ as $\Phi \rightarrow \infty$, so that on the background scaling solution
\begin{equation}
B(\Phi(r)) = \frac{B_0 b(r)}{L^2a(r)} \,,
\end{equation}
then in the IR, this scalar has Green's functions of the form $G(x)\sim x^{-2\Delta^\prime}$, with the dimension $\Delta^\prime$ given by  \cite{lucas1401}
\begin{equation}\label{eq:Btheta}
B_0 = \frac{d^2}{(d-\theta)^2} \left[\left(\Delta^\prime - \frac{d+z}{2}\right)^2 - \left(\frac{d-\theta+z}{2}\right)^2 \right]\,.
\end{equation}
In the following sections we will see that interesting operators such as the electric current (whose correlators determine the electrical conductivity) are like this second case: the fluctuations inherit scaling properties from the background. This will allow us to discuss power laws in transport in terms of the exponents $\{z,\theta\}$.

Finally, a comment about the thermodynamical relation (\ref{eq:PTmu}) in a theory with exponents $z$ and $\theta$. We are considering the case in which the $(z,\theta)$-scaling is obtained as the IR limit of an asymptotically AdS spacetime in the UV (a doped CFT). If we apply the holographic dictionary to obtain the thermodynamic variables, we will find that (\ref{eq:PTmu}) is satisfied, but with $z=1$ and $\theta=0$, independently of the emergent IR scaling. However, (\ref{eq:PTmu}) is not optional for the IR theory --- in particular, (\ref{eq:epsandP}) follows from a Ward identity in a Lifshitz theory \cite{Hoyos:2013qna}. The resolution to this apparent tension is that there will be an `emergent stress tensor' $\mathcal{T}_{\mu\nu}$ (generally with $\mathcal{T}_{ti} \ne \mathcal{T}_{it}$), distinct from from the UV $T_{\mu\nu}$, and this will obey the emergent Lifshitz Ward identities: see e.g. \cite{Korovin:2013bua}.

\subsubsection{Anomalous scaling of charge density}
\label{sec:anomalouscharge}

To fully characterize the scaling backgrounds described in the previous paragraph, and other closely related solutions, a third critical exponent -- beyond $z$ and $\theta$ -- turns out to be necessary \cite{Gouteraux:2012yr, Gath:2012pg, Gouteraux:2013oca, Gouteraux:2014hca, Karch:2014mba}. In terms of the bulk solution, while $\theta$ determines the anomalous power with which the overall metric scales, the new exponent $\Phi$ will determine the anomalous scaling of the bulk Maxwell field, relative to that expected given $z$ and $\theta$. For instance, if we put $\theta = 0$ in the near-horizon scaling of the EMD metric components (\ref{eq:EMDmetric42}) and Maxwell field (\ref{eq:pEMD}) then we find $g_{tt} \sim r^{-2z}$ as expected, but $A_t \sim r^{-d-z}$. This expression for the bulk Maxwell field has an extra factor of $r^{-d}$ relative to that required by a strict Lifshitz scale invariance. Clearly, then, an additional exponent is necessary to characterize the solution.

A natural expectation is that this extra exponent leads to an anomalous scaling dimension for the charge density operator, so that (see e.g. the discussion in  \cite{Hartnoll:2015sea})
\be
[\rho] = d-\theta+\Phi \,. \label{eq:rhoanomdim}
\ee
Here $[x]$ is the momentum dimension of $x$. Conventional wisdom is that $J^\mu$ can pick up no anomalous dimensions \cite{xgwen1992, subir1994}; the arguments essentially amount to the fact that the gauge invariant derivative in ordinary QFT is $\nabla_\mu - \mathrm{i}A_\mu$, implying that $A$ can pick up no anomalous dimensions. However, the argument is a little too fast in suitably general scaling theories (while it is provably true in a CFT using the conformal algebra). A first indication of this fact is that a hyperscaling violation exponent $\theta$ already leads to an anomalous scaling dimension for the energy density which is also associated with a conserved current.  While $\theta \ne 0$ also shifts  $[\rho]$ from its canonical value in (\ref{eq:rhoanomdim}), it is $\Phi$ that is associated with an anomalous dimension for $A$.

Field theoretic models for $\Phi \neq 0$ are in short supply, beyond a model involving a large number $N$ of free fields with variable masses and charges \cite{Karch:2015pha}, related to ``unparticle" physics \cite{Phillips:2013tya}. Models with temperature-dependent charge fractionalization \cite{Hartnoll:2015sea} or strong non-locality due to charge screening \cite{Limtragool:2016gnl} can avoid the non-renormalization theorems on anomalous dimensions for $J^\mu$.

The most common way that $\Phi$ is detected in holography is through the conductivity $\sigma$. We will explore the conductivity in some detail in later sections; it is a subtle observable in nonzero density systems. The upshot is that a certain `incoherent' conductivity is expected to scale as \cite{Davison:2015taa}
\begin{equation}
\sigma \sim T^{(d-2-\theta+2\Phi)/z}. \label{eq:sQT}
\end{equation}
For the EMD models we have just discussed, one finds $\Phi=z$. Interestingly, this corresponds to the charge density operator becoming marginal in the emergent scaling theory \cite{Davison:2015taa, Hartnoll:2012rj}. This is the way that the IR fixed point manages to retain a nonzero charge density consistently with scaling properties. Another possibility is that the charge density is irrelevant about the IR fixed point, in this case one can find more general values of $\Phi$ \cite{Gouteraux:2012yr}. Finally, another possibility for realizing the exponent $\Phi$ in holographic models involves the use of probe brane physics \cite{Karch:2014mba}.  We will discuss probe brane models in \S \ref{sec:probebrane}.

Additional models which exhibit this anomalous scaling include Proca bulk actions, in which the `Maxwell' field in the bulk has a mass \cite{Gouteraux:2012yr}. However, in this case the bulk Proca field is not dual to a conserved current (or, it is dual to a conserved current in a symmetry broken phase) -- recall from (\ref{eq:JA}) that broken gauge invariance in the bulk implies that in the boundary theory,  $\partial_\mu J^\mu \ne 0$. While the exponent in this case can determine, for instance, the scaling of the conductivity with temperature in a symmetry broken phase, as we will see in \S\ref{sec:condphase}, it does not correspond to an anomalous dimension for a conserved current.

At the time of writing, it is still not known the extent to which $\Phi$ imprints itself on thermodynamic quantities, and if it does so in a manner consistent with the 3 emergent scaling exponents $z$, $\theta$ and $\Phi$.

%
%

\subsection{Low energy spectrum of excitations}
\label{sec:lowEspect}

The compressible states of matter accessible to conventional field theoretic techniques, described in \S \ref{sec:nfl}, all contained a Fermi surface. Such a Fermi surface can be defined even in cases without quasiparticles, as we discussed around
(\ref{Gzero}) in
\S \ref{sec:emerge2}. Moreover, there was a Luttinger sum rule on the volumes of all the Fermi surfaces.
In systems without quasiparticles,  Fermi surfaces are often associated with gauge-charged particles, 
and we expect this to be the case in holography. The 
two-point correlators of such particles are not accessible in holographic models, which only allow computation of gauge-invariant obervables.
The robustness of Fermi surfaces, embodied by the Luttinger theorem, behoves us to search for the imprint of the Fermi surface in gauge-invariant observables -- thermodynamics, transport, entanglement -- of the new strongly interacting compressible phases that we have constructed holographically.


In this section we characterize the low energy excitations of fractionalized holographic phases with a charged horizon. We take this opportunity to introduce important holographic techniques for computing spectral densities. In the following \S\ref{sec:bulkfermions} we will study cohesive phases in which the charge is explicitly carried by fermions. One upshot of these two sections will be that various properties of Fermi surfaces that necessarily come together at weak coupling --- such as spectral weight at nonzero momentum and the exponent $\theta = d-1$ --- can be dissociated with strong interactions. Some holographic phases have certain features that are intriguingly suggestive of Fermi surfaces-like dynamics but not others, while other holographic phases seem to have no indications of a Fermi surface at all. It has been argued \cite{Polchinski:2012nh,Faulkner:2012gt,Sachdev:2012tj} that 
bulk quantum corrections will be needed to resurrect the Fermi  surface in these cases.

\subsubsection{Spectral weight: zero temperature}
\label{sec:zeroTspectral}

The low energy spectral weight of an operator ${\mathcal{O}}$ as a function of momentum $k$ is characterized by
\be\label{eq:rhok}
\rho(k) \equiv \lim_{\omega \to 0} \frac{\text{Im} \, G^{\mathrm{R}}_{{\mathcal{O}}{\mathcal{O}}}(\omega,k)}{\omega} \,.
\ee
The division by $\omega$ is generally required: recall that the imaginary part of the Green's function is typically odd under $\omega \to - \omega$ (see e.g. \cite{Hartnoll:2009sz}).   Hence at nonzero temperature, the limit (\ref{eq:rhok}) is generally finite and nonzero. Furthermore, the imaginary part of the Green's function appears divided by $\omega$ in the Kramers-Kronig relations and in several of the formulae that will appear below. As noted in (\ref{eq:dissip}) above, the spectral weight is a direct measurement of the spectrum of excitations in the theory. For free fermions, for example, $\rho(k)$ for the charge density ${\mathcal{O}} = J^t$ shows a sharp non-analyticity at $2 k_{\mathrm{F}}$. Two electrons can move charge around the Fermi surface at zero energy cost, but the largest momentum transfer possible is if both electrons move from one side of the Fermi surface to the other. Hence $\rho(k)$ vanishes for $k > 2 k_{\mathrm{F}}$.

Let us first show that the extremal (zero temperature) AdS-RN black hole of \S \ref{sec:ads2rd} is dual to a compressible phase with spectral weight at nonzero momentum.  We will focus on a neutral scalar operator of UV dimension $\Delta$, and will comment on other operators later.  As in \S \ref{sec:spectral}, to find the retarded Green's function we need to solve the equations of motion for a massive scalar field in the AdS-RN background.     Our previous experience suggests that the low energy spectral weight will be completely fixed, at least the singular properties, by the IR scaling geometry ($\mathrm{AdS}_2 \times \mathbb{R}^d$).  The nonzero momentum spectral weight can then be understood as a consequence of the emergent $z = \infty$ scaling symmetry, in which momentum is dimensionless, as anticipated in \S \ref{sec:ads2rd}. We will go ahead and assume that we can restrict attention to the IR geometry, justifying this approach at the end of the calculation.

In the near-horizon $\mathrm{AdS}_2 \times \mathbb{R}^d$ geometry (\ref{eq:nhmetric}), the equation of motion for a massive scalar field $\varphi$ at spatial momentum $k$ and frequency $\omega$ is \begin{align}
&\frac{1}{\sqrt{-g}}\partial_M \left(\sqrt{-g}g^{MN} \partial_N\varphi\right) \notag \\ &= \frac{\zeta^2}{L_2^2} \partial_\zeta^2 \varphi + \frac{\zeta^2}{L_2^2} \omega^2 \varphi - k^2 \frac{r_*^2}{L^2}\varphi = m^2 \varphi.  \label{eq:ads2rdeom}
\end{align}
Consider first $\omega=0$.   The two linearly independent solutions to (\ref{eq:ads2rdeom}) are power laws: \begin{equation}
\varphi_0(k,\zeta) = A_1 \zeta^{\frac{1}{2}-\nu_k} + A_2\zeta^{\frac{1}{2}+\nu_k},\label{eq:varphi0zeta}
\end{equation}
where the IR scaling exponent $\nu_k$ is given by
\begin{equation}
\nu_k \equiv \sqrt{\frac{1}{4} + m^2 L_2^2 + \frac{1}{d(d-1)}\frac{\g^2 k^2}{\mu^2} }.   \label{eq:nuk}
\end{equation}
This expression reveals a remarkable feature of the emergent $\mathrm{AdS}_2\times\mathbb{R}^d$ geometry  \cite{Faulkner:2009wj}: 
a continuum of operator dimensions labelled by the momentum $k$. These decoupled operators at each momentum indicate a large number of degrees of freedom at low energy, as we will see shortly once we have computed the actual Green's functions. Note that in comparing the behavior (\ref{eq:varphi0zeta}) of the field to the more general expression (\ref{eq:boundaryL}) we see that for $\mathrm{AdS}_2\times\mathbb{R}^d$ geometries, we should take $D_\text{eff.}=1$ in (\ref{eq:boundaryL}), i.e. only counting the dimensions involved in scaling by putting $d=0$ and $z=1$ in (\ref{eq:deff}).

At nonzero $\omega$, the solution of (\ref{eq:ads2rdeom}) that satisfies infalling boundary conditions (\ref{eq:ads2infall}) is
\be\label{eq:hankel}
\varphi(k,\omega,\zeta) = \mathrm{H}^{(1)}_{\nu_k}(\omega \zeta) \,.
\ee
Here $\mathrm{H}^{(1)}_{\nu_k}$ is a Hankel function (a linear combination of Bessel functions). We can now define an `IR Green's function' by expanding (\ref{eq:hankel}) near the boundary of the IR region, $\zeta \to 0$. In this limit, the solution returns to the form (\ref{eq:varphi0zeta}). Adapting the usual holographic dictionary (\ref{eq:GRAB}) we can then write (the prefactor will not be important)
\be\label{eq:IRG}
\left. G^{\mathrm{R}}_{\mathcal{OO}}(\omega,k) \right|_\text{IR} \equiv \frac{A_2}{A_1} \propto \omega^{2 \nu_k} \,.
\ee
This is really just the definition of the IR Green's function. We now show that this quantity can be related to the actual Green's function that is defined via (\ref{eq:GRAB}) at the UV boundary of the full AdS-RN spacetime, $r \to 0$. The following matching argument is general and important.

\begin{figure}[h]
\centering
\includegraphics[height = 0.15\textheight]{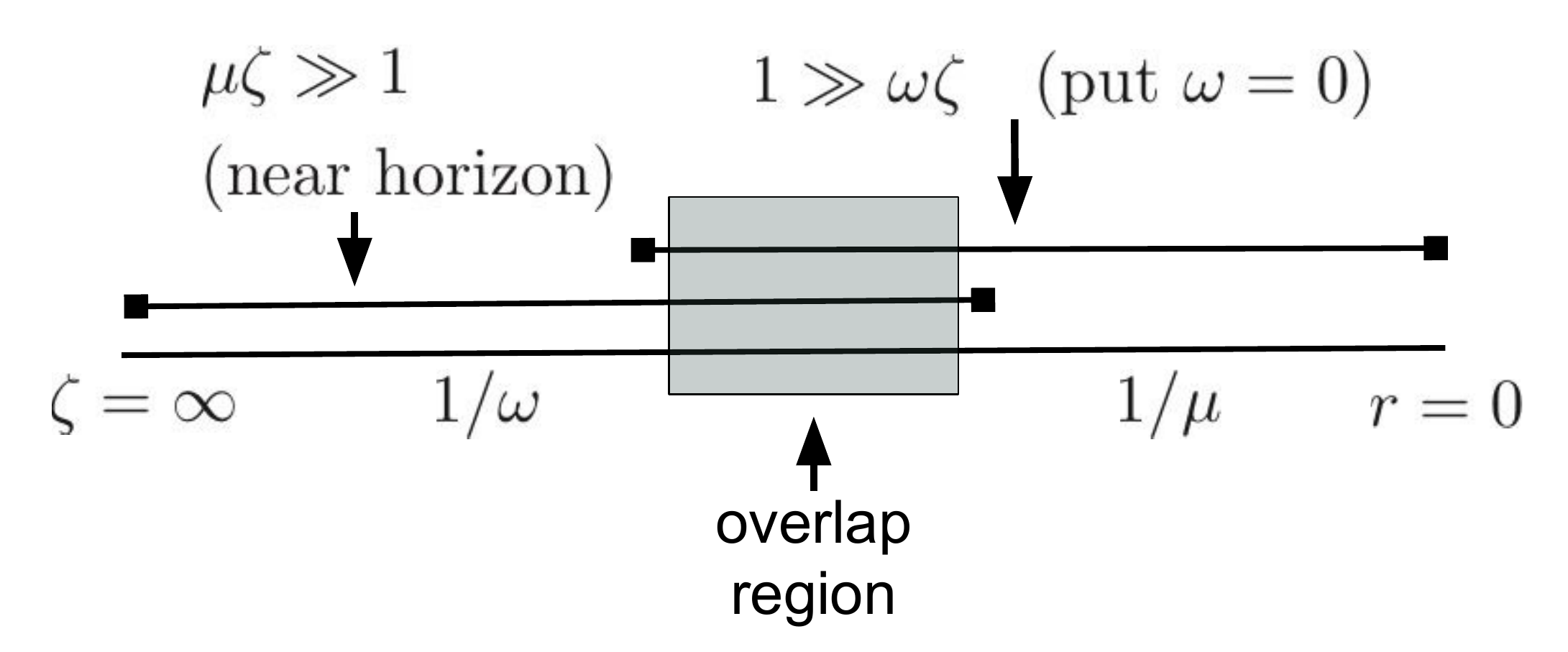}
\caption{\label{fig:overlap} \textbf{The matching argument.} When $\omega \ll \mu$, then there is a parametrically large region of overlap between the range where the near-horizon solution (\ref{eq:hankel}) is valid, and the asymptotic range where the frequency may be set to zero. This allows the solution satisfying infalling boundary conditions to be extended to the boundary of the full spacetime.}
\end{figure}

The complication with taking the low energy ($\omega \to 0$) limit  of the full Green's function is that, in the wave equation for $\varphi$ in the full geometry, taking $\omega \to 0$ does not commute with the very near horizon limit $\zeta \to \infty$. This can be seen already in the near horizon wave equation (\ref{eq:ads2rdeom}), due to the $\zeta^2 \omega^2$ term. The low frequency Green's function can be found using a matching argument, as we now explain. The matching argument works provided that
\be\label{eq:smallw}
\omega \ll \mu \,.
\ee
At these low frequencies, the near horizon condition $\mu \zeta \gg 1$ (cf. the discussion around (\ref{eq:rrstar}) above) is compatible
with the condition to drop the frequency dependence in the wave equation, $1 \gg \omega \zeta$. The overlap of the two regions is illustrated in figure \ref{fig:overlap}.
In the near horizon regime we know the full solution to the wave equation, it is given by (\ref{eq:hankel}). Now we solve the equation for $\varphi$ away from the horizon, i.e. from $1 \gg \omega \zeta$ in the near horizon region, all the way out to the boundary of the full AdS-RN geometry at $r \to 0$. In this range we can safely set $\omega = 0$ to leading order in the limit (\ref{eq:smallw}). Let $\varphi(k, r)$ be the solution to the scalar equation of motion in this range which asymptotes to the form (\ref{eq:varphi0zeta}) at the boundary of the near-horizon region, parametrized by the two constants $A_{1,2}$.  This solution further extends to the UV of the full spacetime as $r\rightarrow 0$, where it can be expanded as (\ref{eq:boundaryexpand}):
\begin{equation}
\varphi = \frac{B_{(0)}}{L^{d/2}} \, r^{d+1-\Delta} + \frac{B_{(1)}}{(2\Delta-d-1)L^{d/2}} \, r^\Delta + \cdots.
\end{equation}
The extra factor of $(2\Delta-d-1)$ cleans up the expression for the Green's function below, using (\ref{eq:nicevev}).
Because the wave equation for the scalar is linear and real, the coefficients $B_{(0)}$ and $B_{(1)}$ are linear combinations of $A_{1,2}$ with real coefficients. Therefore the full Green's function can be written
\be\label{eq:GRrelate}
G^{\mathrm{R}}_{\mathcal{OO}} = \frac{B_{(1)}}{B_{(0)}} = \frac{b^1_{(1)} A_1 + b^2_{(1)} A_2}{b^1_{(0)} A_1 + b^2_{(0)} A_2}
= \frac{b^1_{(1)} + b^2_{(1)} \left. G^{\mathrm{R}}_{\mathcal{OO}} \right|_\text{IR}}{b^1_{(0)} + b^2_{(0)} \left. G^{\mathrm{R}}_{\mathcal{OO}} \right|_\text{IR}}  \,.
\ee
Here the $b^i_{(j)}$ are all real and independent of $\omega$ (because the equation has $\omega  = 0$ over the coordinate range relating $A_1$ and $A_2$ to $B_{(0)}$ and $B_{(1)}$). In the final step we used the definition (\ref{eq:IRG}) of the IR Green's function.

Taking the imaginary part of (\ref{eq:GRrelate}), and noting that $\nu_k > 0$ in (\ref{eq:IRG}) so that $\left. G^{\mathrm{R}}_{\mathcal{OO}}(\omega,k) \right|_\text{IR} \to 0$ as $\omega \to 0$, we obtain to leading order at low energies
\be
\text{Im} \, G^{\mathrm{R}}_{\mathcal{OO}}(\omega,k) \; \propto \; \text{Im} \left. G^{\mathrm{R}}_{\mathcal{OO}}(\omega,k) \right|_\text{IR}  \;  \propto \; \omega^{2 \nu_k} \,. \label{eq:lowIm}
\ee
Thus, up to a frequency-independent constant, the low frequency spectral weight of the system is completely determined by the near horizon Green's function and is independent of the form of the spacetime away from the horizon. This simplification is physically intuitive and very helpful in practice -- for many purposes it will not be necessary to explicitly solve the wave equation in the full geometry. Versions of the matching argument above have appeared in many papers, including \cite{Gubser:2008wz, Horowitz:2009ij, Goldstein:2009cv, Edalati:2009bi, Faulkner:2009wj, Hartnoll:2010gu, Donos:2012ra}. In fact, such matching in black hole Green's functions has an older history, predating AdS/CFT and stretching back into the astrophysical literature.

A similar computation to the above goes through for the retarded Green's function of the charge and energy currents, $J^\mu$ and $T^{\mu\nu}$. The calculation is a little more complicated because there are multiple coupled fields in the bulk that are excited (different components of the metric and Maxwell field). The equations can be turned into decoupled second order equations analogous to (\ref{eq:ads2rdeom}) using `gauge-invariant variables'. This is similar to what we did in equation (\ref{eq:gaugeinvar}) above, but with more variables. For $d=2$ (i.e. for a 2+1 dimensional dual field theory) extremal AdS-RN these equations were found and solved in \cite{Edalati:2009bi, Edalati:2010hk, Edalati:2010pn}, building on \cite{Kodama:2003kk}. The low energy Green's functions all take the form (\ref{eq:lowIm}), but now the exponents have a more complicated $k$ dependence
\be\label{eq:RNnu}
\nu_k^{\pm}  =  \frac{1}{2} \sqrt{5 + 2 \frac{\g^2 k^2}{\mu^2} \pm 4 \sqrt{1 + 2 \frac{\g^2 k^2}{\mu^2}}},\;\;\; (d=2).
\ee
The exponents turn out to be the same in the transverse and longitudinal channels in this case, but
note that there are still two different exponents. The most singular will dominate generic correlators \cite{Donos:2012ra}. In fact, if all that is required are the exponents $\nu_k$ --- that, we have seen, control low energy dissipation --- then these can be found without using gauge-invariant variables. It suffices to make a power law ansatz for the near-horizon fields with $\omega = 0$, to find the solutions analogous to (\ref{eq:varphi0zeta}) above.
 
A generalization of the result (\ref{eq:RNnu}) for the exponents controlling the retarded Green's functions of charge and energy currents can be found for the EMD theories with $z \to \infty$ in the limit described in (\ref{eq:etalimit}) above. One finds \cite{Hartnoll:2012wm, Anantua:2012nj} \begin{widetext}
\begin{subequations}\begin{align}
\nu_k^{\perp,\pm} & =  \frac{1}{2} \sqrt{5 + \eta(2+\eta) + 4 (1 + \eta)^2 \frac{k^2}{k_0^2} \pm 4 (1+ \eta) \sqrt{1 + 2 (1+\eta) \frac{k^2}{k_0^2}}} \,, \\
\nu_k^{\parallel,\pm} & =  \frac{1+\eta}{2 \sqrt{2 + \eta}} \sqrt{10 + \eta + 4 (2 + \eta) \frac{k^2}{k_0^2} \pm 8 \sqrt{1 + (2+\eta) \frac{k^2}{k_0^2}}} \,, \\
\nu_k^{\parallel,0} & =  \frac{1+\eta}{2} \sqrt{1 + 4 \frac{k^2}{k_0^2}} \,.
\end{align}\label{eq:etanu}\end{subequations}\end{widetext}
Here there are two transverse exponents and three longitudinal exponents. Recall that $\eta$ was defined in (\ref{eq:etalimit}). The momentum scale $k_0$ depends on details of how the IR geometry crosses over to the full spacetime. In particular, the fact that the Green's function takes the form (\ref{eq:lowIm}) shows that the charge and energy currents are `nice' operators whose fluctuations inherit scaling properties in these $z = \infty$ hyperscaling violating backgrounds, in the sense discussed around equation (\ref{eq:4phi}) above.

The power-law form of the spectral weight (\ref{eq:lowIm}) in all the cases above shows that $z=\infty$ scaling geometries (including those with additional hyperscaling violation) admit nonzero momentum excitations at all energy scales. This can be contrasted with the behavior we will find shortly for $z < \infty$, in which the nonzero momentum spectral weight is exponentially suppressed as $\omega \to 0$. A power law spectral weight is the sort of behavior one might expect for a `critical' Fermi surface -- there is no longer a sharp Fermi momentum, or `$2 k_{\mathrm{F}}$' singularity, but instead the low energy spectral weight spreads over all of momentum space, with a continuously weakening power law at large momentum. This behavior is perhaps illustrated most clearly by the $T>0$ result. We will derive the corresponding nonzero temperature result rigorously below, but meanwhile, the reader should not be surprised to know that the result for $\rho(k)$ in (\ref{eq:rhok}) is
\be\label{eq:Trhonu}
\rho(k) \; \propto \; T^{2 \nu_k - 1} \,.
\ee
It seems quite remarkable that these completely non-quasiparticle $z = \infty$ compressible phases exhibit behavior reminiscent of Pauli exclusion/ Fermi surface physics, in which low energy excitations are pushed out to nonzero momenta. The expression (\ref{eq:Trhonu}) diverges as $T \to 0$ if $2 \nu_k < 1$, as can occur for some of the cases discussed above \cite{Anantua:2012nj}.

Given formulae such as (\ref{eq:nuk}), (\ref{eq:RNnu}) and (\ref{eq:etanu}) for an exponent $\nu_k$ appearing in the spectral weight (\ref{eq:lowIm}), one can ask if there is any preferred momentum $k_\star$ that could play a role analogous to a Fermi momentum. A characteristic feature of the above formulae is the presence of branch cuts. The (complex!) momenta $k_\star$ of the branch points are special, in that they are the singular points of the Green's function in the $k$-plane. In particular, the imaginary part of $k_\star$ determines the exponential decay of correlators with distance \cite{Iqbal:2011in, Blake:2014lva}. This is the essence of semi-local criticality as defined by \cite{Iqbal:2011in}: there is an infinite correlation length in time, but finite correlation in space. In this regard one should note that strictly local criticality, in which there is no $k$ dependence at all in the low energy spectral weight, is fine-tuned. Once $k$ is dimensionless with $z = \infty$ scaling, the dimensions of operators can acquire $k$ dependence and generically they will.

In contrast to the preceding discussion, we have already noted around equation (\ref{eq:imglif}) above
that in scaling geometries with $z < \infty$ the nonzero momentum spectral weight is exponentially
suppressed as $\omega \to 0$, like $\mathrm{e}^{- A (k^z/\omega)^{1/(z-1)}}$. No analogue of Fermi surface-like physics is visible. We will give a derivation of the exponential suppression -- allowing for hyperscaling violation -- in the following finite temperature \S\ref{sec:SWT}. The exponential suppression was first noted for fermions in \cite{Faulkner:2010tq}, electric and energy currents in \cite{Hartnoll:2012wm} and scalars in \cite{Keeler:2015afa}. A special case is that of an emergent pure CFT ($z=1, \theta = 0$), wherein $G^{\mathrm{R}}_{\mathcal{OO}} \sim (k^2-\omega^2)^{2\Delta-d-1}$, which is completely real for $\omega < k$.  In this case, the low spectral weight trivially vanishes at nonzero momenta.

There are several indications that the inclusion of effects beyond classical gravity may reveal more conventional Fermi-surface like behavior. Classical but string-theoretic $\a^\prime$ corrections were shown to lead to non-analyticities in momentum space correlators in \cite{Polchinski:2012nh}. Faulkner and Iqbal \cite{Faulkner:2012gt} computed quantum corrections on AdS$_3$, 
and found that inclusion of monopoles in the bulk gauge field led to $2 k_{\mathrm{F}}$ Friedel oscillations in the boundary
quantum liquid in one spatial dimension. Bulk monopole contributions were argued in \cite{Sachdev:2012tj} to be necessary for signatures of the
Fermi surface to appear for two-dimensional quantum liquids. 
And, we will see in \S\ref{sec:probebrane} below that probe branes, that include the resummation of a class of $\a^\prime$ effects, lead to slightly more conventional -- albeit still strongly interacting -- analogues of Fermi-surface like dynamics. The inclusion of $\a^\prime$ physics retains the benefits of the large $N$ 't Hooft limit, and is relatively underexplored.

\subsubsection{Spectral weight: nonzero temperature}\label{sec:SWT}

In this section we obtain a rather general and powerful expression (\ref{eq:SWT}) for the nonzero temperature spectral weight (\ref{eq:rhok}). It is given purely in terms of the behavior of the fluctuating field at $\omega = 0$ and on the horizon itself. An early use of these kind of formulae is \cite{Chakrabarti:2010xy}. We will follow the elegant Wronskian derivation of \cite{Lucas:2015vna}.

Consider the equation of motion for a neutral scalar (with action (\ref{eq:4phi})) in the background (\ref{eq:abmetric}):
\begin{align}
& L^2 B(\Phi)\varphi(k,\omega,r) + k^2 r^2  \varphi(k,\omega, r) - \omega^2\frac{r^2}{ab}  \varphi(k,\omega, r) \notag \\
& = \frac{r^{d+2}}{a} \left( \frac{b}{r^d} \varphi(k,\omega,r)^\prime \right)^\prime.  \label{eq:442eom}
\end{align}
Recall that $a$ and $b$ are functions of $r$.
The crucial observation is that this equation only depends on $\omega^2$, not $\omega$.   Hence, to linear order in $\omega$, we may write \begin{equation}
\varphi(k,\omega,r) = \varphi_0(k,r) + \omega \varphi_{\mathrm{I}}(k,r) + \mathcal{O}\left(\omega^2\right) \,, \label{eq:0I}
\end{equation}
where $\varphi_0(k,r)$ is the regular solution with $\omega = 0$ and $\varphi_{\mathrm{I}}(k,r) $ will also be determined from the $\omega = 0$ equation, as we describe shortly. The zero of $b(r)$ at the horizon means that we cannot ignore $\omega$ when solving (\ref{eq:442eom}). This is the noncommutativity of limits that we have encountered previously.  However, upon taking $r\rightarrow r_+$,  infalling boundary conditions (\ref{eq:infalling}) amount to \begin{equation}
\varphi(r\rightarrow r_+, \omega) = \text{constant} \times \left(1-\frac{r}{r_+}\right)^{-\mathrm{i}\omega/4\pi T}.
\end{equation}
So if we work at fixed small $r_+-r$, and take $\omega \rightarrow 0$, we may safely expand \begin{equation}
\left(1-\frac{r}{r_+}\right)^{-\mathrm{i}\omega/4\pi T} = 1 + \frac{\mathrm{i}\omega}{4\pi T} \log \frac{r_+}{r_+-r} + \mathcal{O}\left(\omega^2\right).   \label{eq:omegaexpand}
\end{equation}
Comparing (\ref{eq:omegaexpand}) with (\ref{eq:0I}), we only need to find a solution to the $\omega =0$ equation which has this precise logarithmic singularity near the horizon to capture the $\mathcal{O}(\omega)$  response.

Suppose that we know the solution $\varphi_0$ to the wave equation (\ref{eq:442eom}) at $\omega = 0$ that is regular on the horizon, and that we have normalized it so that near the asymptotic boundary
\be\label{eq:zerbdy}
\varphi_0(k,r\rightarrow 0) = \frac{r^{d+1-\Delta}}{L^{d/2}} + \cdots \,,
\ee
i.e. it is sourced by unity.  We can construct the singular solution as an integral (we have conveniently normalized it): \begin{equation}
\varphi_1(k,r) = \frac{\varphi_0(k,r)}{L^d} \int \limits_0^r \mathrm{d}s \frac{s^d}{b(s)\varphi_0(k,s)^2}.  \label{eq:vphi1}
\end{equation}
This is a standard result for ordinary differential equations that is obtained using the Wronskian.
Expanding near the boundary $r=0$ we see that the second solution is normalizable:
\begin{equation}
\varphi_1(k,r\rightarrow 0) = \frac{1}{2\Delta-d-1}\frac{r^\Delta}{L^{d/2}} + \cdots \,.   \label{eq:vphi2}
\end{equation}
Near $r=r_+$, however, this solution is singular.   The singular behavior comes from the factor of $b(s)$ in the integral expression (\ref{eq:vphi1}):
\begin{equation}
\varphi_1(k, r\rightarrow r_+) = \frac{r_+^d}{4\pi TL^d \varphi_0(k,r_+)}\log \frac{r_+}{r_+-r} + \text{regular}. \label{eq:loglog}
\end{equation}

Putting together the small $\omega$ behavior (\ref{eq:0I}), the near horizon behavior (\ref{eq:omegaexpand}) and the logarithmic singular behavior (\ref{eq:loglog}) we can conclude that the solution to linear order in $\omega$ must be
\begin{equation}
\varphi(k,r,\omega) = \varphi_0(k,r) + \frac{\mathrm{i}\omega L^d}{r_+^d} \varphi_0(k,r_+)^2 \varphi_1(k,r) + \mathcal{O}\left(\omega^2\right).  \label{eq:vphi3}
\end{equation}
This solution can now be expanded near the boundary, using the near boundary behavior of $\varphi_0$ (in (\ref{eq:zerbdy})) and $\varphi_1$ (in (\ref{eq:vphi2})). Using (\ref{eq:vphi3}) together with the basic holographic dictionary, we obtain the spectral weight
\begin{equation}
\rho(k) = \lim_{\omega \rightarrow 0} \frac{\mathrm{Im}\, G^{\mathrm{R}}_{\mathcal{OO}}(\omega,k)}{\omega} = \frac{L^d}{r_+^d} \varphi_0(k,r_+)^2.   \label{eq:SWT}
\end{equation}
Two very useful aspects of this expression bear repeating. Firstly, it gives the low energy spectral weight in terms of data directly at $\omega = 0$; it is not necessary to perform nonzero $\omega$ computations and then take the limit. Secondly, it gives the answer in terms of data on the horizon $r=r_+$. Sometimes, it is possible to determine this data without having to solve the equations of motion at all radii explicitly. For these reasons, (\ref{eq:SWT}) will play a key role in the theory of holographic metallic transport developed in \S\ref{sec5}.

To illustrate the use of (\ref{eq:SWT}), we can return to the example of the spectral weight of an operator in the AdS-RN background. Now
we turn on a small temperature $T\ll \mu$. The matching argument has essentially the same structure as the one we used previously around (\ref{eq:smallw}). We set $\omega = 0$ and use the small temperature as the parameter that guarantees the existence of two parametrically overlapping regions. The near horizon region is again $\mu \zeta \ll 1$. The near horizon geometry is a black hole in $\mathrm{AdS}_2 \times {\mathbb R}^2$, given in (\ref{eq:smallTgeom}) above. To simplify the argument even further, we will use the scaling symmetry to avoid solving the wave equation explicitly in this black hole geometry (it can be done in terms of hypergeometric functions). If we solve the wave equation with $\omega = 0$ in the ($T>0$) near horizon background, the solution that is regular at the horizon must take the form
\be\label{eq:phinear}
\varphi^\text{near}_0(k,\zeta) = \varphi_0(k,r_+) f\left(k,\frac{\zeta}{\zeta_+} \right) \,.
\ee
We have used the fact the wave equation is only a function of $\zeta/\zeta_+$ when there is no time dependence. Recall that $\zeta_+ \propto 1/T$ in (\ref{eq:smallTgeom}). Regularity at the horizon requires that $f(k,1)$ is a constant, that without loss of generality we have set to $1$. The solution (\ref{eq:phinear}) can be expanded for small $\zeta$ (the boundary of the near horizon region) to give the zero temperature answer (\ref{eq:varphi0zeta})
\begin{align}\label{eq:phinearbdy}
&\varphi^\text{near}_0(k,\zeta) = \\
& \;\; \varphi_0(k,r_+) \left(\a_1(k) \left(\zeta/\zeta_+\right)^{\half - \nu_k} + \a_2(k) \left(\zeta/\zeta_+\right)^{\half + \nu_k} \right) \,. \notag
\end{align}
Importantly, the $\alpha_i(k)$ do not depend on temperature (i.e. $\zeta_+$).

At small temperatures $T \ll \mu$, the near horizon solution has an overlapping regime of validity with a `far' solution, valid for $1 \ll \zeta/\zeta_+ \sim \zeta T$, all the way out to the asymptotic boundary of the full spacetime. This is entirely analogous to the situation depicted in figure \ref{fig:overlap}, replacing $\omega \to T$. In the far regime, we can safely set $T = 0$ in the wave equation, similarly to how we set $\omega = 0$ away from the horizon in our discussion around (\ref{eq:smallw}). But this means that upon integrating out (\ref{eq:phinearbdy}) to the boundary, no additional factors of temperature will appear. Furthermore, at the boundary we have imposed in (\ref{eq:zerbdy}) that $\varphi_0$ tend to a temperature independent constant. Given that the equation is linear, this will only happen (generically) if the temperature dependence drops out to leading order at low temperatures in (\ref{eq:phinearbdy}). This requires $\varphi_0(k,r_+) \propto \zeta_+^{\half - \nu_k} \propto T^{\nu_k - \half}$. Using (\ref{eq:SWT}), the low temperature spectral weight is immediately found to be
\begin{equation}
\rho(k) \sim T^{2\nu_k-1} \,. \label{eq:zinfSW}
\end{equation}
In applying (\ref{eq:SWT}), we used the fact (from (\ref{eq:hsize})) that at low temperatures $r_+ \sim 1/\mu$,   independent of temperature. The result above is that anticipated in (\ref{eq:Trhonu}), confirming
that there is a power law spectral weight at nonzero momentum in these locally critical theories. It should be clear in the derivation above that almost no details about the full spacetime geometry are needed. The only step where the equation of motion was necessary was in determining the exponents $\nu_k$ in the IR scaling region.

The matching arguments are more complicated in theories with finite $z$ and $\theta$. The starting point is the wave equation (\ref{eq:442eom}). There are now two dimensionless combinations of parameters that can be made, $k T^{1/z}$ and $k r^{d/(d-\theta)}$ (or alternatively $r/r_+$). The spectral weight can be determined by simple arguments in the limits $k \ll T^{1/z}$ and $k \gg T^{1/z}$.

When $k \ll T^{1/z}$  we can put $k=0$ in the equation of motion. The dimensionless combination of $k$ and $r$, $k \, r^{d/(d-\theta)} < k \, r^{d/(d-\theta)}_+ \sim k/T^{1/z} \ll 1$, is small everywhere. Recall that $r_+$ is related to $T$ via (\ref{eq:rPEMD}) in an EMD background. With $k=0$, the only dimensionless ratio left is $r/r_+$, and the matching proceeds exactly as in the $z=\infty$ cases discussed above. In particular, the solution to (\ref{eq:442eom}) in the matching region --- i.e. putting $\omega = T = 0$ ---  is a power law,
\begin{align}
& \varphi_0(r) = \\
& \;\; r^{- d \theta/(2 (d-\theta))} \left( A_1 \, r^{d(d+z - \Delta')/(d-\theta)} + A_2 \, r^{d \Delta'/(d-\theta)}  \right) \,, \notag
\end{align}
analogously to (\ref{eq:varphi0zeta}) above.
The scaling dimension $\Delta'$ is defined in (\ref{eq:Btheta}). Using the same argument as previously, the spectral weight is then found from (\ref{eq:SWT}) to be \cite{lucas1401}
\begin{equation}
\rho(k) \; \propto \; T^{(2\Delta'-2z-d)/z} \qquad \; (k \ll T^{1/z}) \,.\label{eq:GOOT}
\end{equation}
This is a power law spectral weight, but only for `thermally populated' small momenta.

The zero temperature limit is instead obtained by considering $k \gg T^{1/z}$. In this limit, to find the matching solution we need to solve the equation of motion (\ref{eq:442eom}) with $k \, r^{d/(d-\theta)} \gg 1$. This approximation now holds from the horizon at $r=r_+$, all the way out to the matching region at $r_+ \ll r$. As in \cite{Faulkner:2010tq, Hartnoll:2012wm, Keeler:2015afa}, the large $k$ solution can be found using WKB. The WKB solution is
\be\label{eq:varWKB}
\varphi_0(r) = \exp \left( - k \int\limits_0^r \mathrm{d}r  \sqrt{\frac{a}{b}}  \right)\,.
\ee
A short WKB connection argument is necessary to convince oneself that it is this mode -- decaying towards the horizon -- that corresponds to the regular solution at the horizon. A growing mode would lead to a nonsensical exponentially large spectral weight below. 
The solution has been normalized so that it carries no temperature dependence into the matching region ($r \ll r_+$). As previously, this is necessary to ensure that the asymptotic UV behavior of the field has no temperature dependence, as required by the boundary condition (\ref{eq:zerbdy}). Evaluating (\ref{eq:varWKB}) on the horizon -- one does not need to perform the integral, it is sufficient to rescale $r \to r_+ \hat r$ in order to move all of the factors of $r_+$, and hence temperature, out of the integrand -- and plugging the expression into 
(\ref{eq:SWT}), we find the spectral weight
\begin{equation}
\rho(k) \propto \exp\left( - C \, \frac{k}{T^{1/z}}\right) \,. \label{eq:IRlif2}
\end{equation}
Here $C$ is a temperature-independent constant, depending on the UV scale (typically charge density $\rho$) which supports the emergent scaling geometry. The scaling with $k$ is a little different to the zero temperature suppression quoted in (\ref{eq:imglif}) above.
We see that for any finite $z$ and $\theta$,  as $T\rightarrow 0$ all nonzero momentum spectral weight is exponentially suppressed. In terms of spectral weight, then, there is no indication of Fermi surface-like physics in Einstein-Maxwell-dilaton models with $z < \infty$.

\subsubsection{Logarithmic violation of the area law of entanglement}

In compressible phases with a Fermi surface, the nonlocality (in space) of the many low-lying degrees of freedom at the Fermi surface leads to a logarithmic violation of the `area law' of the entanglement entropy of a region. For free fermions the entanglement entropy of a spatial region of size $R$ has, in addition to the usual area law, a contribution \cite{Wolf:2006zzb, klich2006} \begin{equation}
S \sim \left(R \, k_{\mathrm{F}} \right)^{d-1} \log (R \, k_{\mathrm{F}}).
\end{equation}
This result can intuitively be understood by thinking of the low energy excitations near the Fermi surface as a collection of chiral fermions forming CFT2s at each point on the Fermi surface \cite{Swingle:2009bf}.  Hence,  the logarithm is related to (\ref{eq:CFT2EE}), and the area prefactor counts the number of these CFT2s.

We saw in \S\ref{sec:nfl} above that Fermi surfaces lead to the thermodynamic hyperscaling violation exponent $\theta = d - 1$. \cite{Huijse:2011ef} suggested that  logarithmic area law violation may be a generic property of compressible phases with  $\theta=d-1$. In particular, building on \cite{Ogawa:2011bz}, they showed that this is the case in holographic models. We will outline this result.

It is convenient to write the hyperscaling violating metric as (\ref{eq:hsv}) in the calculation below. 
Using a holographic calculation similar to that presented in \S\ref{sec:EE}, one finds that the entanglement entropy of a sphere of radius $R$ is given by \begin{equation}
S = S_{\mathrm{UV}} + \frac{L^d \Omega_{d-1}}{4G_{\mathrm{N}}} \int \frac{\mathrm{d}r \; }{r^{d-\theta}} \frac{\rho^{d-1}}{R_*^\theta } \sqrt{1+\left(\frac{\mathrm{d}\rho}{\mathrm{d}r}\right)^2} . \label{eq:433integral}
\end{equation}
$S_{\mathrm{UV}}$ is the contribution from the asymptotically AdS region, and will have a characteristic area law:  $S_{\mathrm{UV}}\sim(R/\epsilon)^{d-1}$. To evaluate the integral (\ref{eq:433integral}) one first needs to find the surface $\rho(r)$ that minimizes the above functional. This can be done in an expansion about $r=0$ (the boundary of the IR region). This is where one expects any singular contribution to originate from. It turns out that $\rho$ is constant to leading order in this limit \cite{Huijse:2011ef}, and so the scaling of the integral is determined the measure $\mathrm{d}r/r^{d-\theta}$. If $d-\theta>1$,  the integral in the hyperscaling violating region is dominated by small $r$, leading to a further (finite) contribution to the area law. However, if $\theta=d-1$,  then the integral scales as $\mathrm{d}r/r$.   This is just like the CFT2 case that we previously studied, which had a logarithmic area law violation.  So in this case (schematically):  \begin{equation}
S \sim  \left(\frac{R}{\epsilon}\right)^{d-1} + \left(\frac{R}{R_*}\right)^{d-1} \log \frac{R}{R_*} \,.                 \label{eq:logviolate}
\end{equation}
It was furthermore shown in \cite{Huijse:2011ef} that the prefactor of the logarithmic term in (\ref{eq:logviolate}) has $R_* \propto \rho^{-1/d}$, with no other dimensionful scales appearing, and with a prefactor that is independent of the shape of the entangling surface. These results make nontrivial use of Gauss's law in the bulk \cite{Huijse:2011ef}. If $\theta > d-1$, then  the integral (\ref{eq:433integral}) is dominated in the IR, and we find power law violations of the area law. This suggests imposing the condition
\begin{equation}
\theta \le d-1 \,.
\end{equation}
This is a stronger condition than that required for the entropy density in (\ref{eq:entT}) not to diverge as $T \to 0$. It is also independent of the null energy conditions in (\ref{eq:null}) above.

The fact that backgrounds with $\theta = d-1$ share both the thermodynamics and entanglement entropy of Fermi surfaces is suggestive of a common physics at work.
String theoretic constructions of such backgrounds may shed further light on the microscopic origin of the logarithmic entanglement in holographic models \cite{Narayan:2012hk}. As we have seen above, however, if $z$ is finite these backgrounds will not have any spectral weight at nonzero momentum (at least in the simple operators considered above). Conversely, the $z= \infty$ geometries discussed above --- that do have finite momentum spectral weight --- do not have logarithmic violations of the entanglement entropy area law \cite{Swingle:2009wc}.

The upshot is that EMD backgrounds can have important Fermi-surface-like physics if $z = \infty$ or if $\theta = d-1$. However, each case only shares some features with weakly interacting Fermi liquids. When neither of these two relations hold, as can certainly be the case in microscopic models, then there is little evidence for Fermi-surface like dynamics of the charged horizon.

Finally, we can note a tantalizing connection to the condensed matter systems discussed in \S\ref{sec:nfl} above. Putting $\theta = d-1$, and furthermore restricting to the unique regular geometries with $z = 1 + \theta/d$, discussed at the end of \S\ref{sec:EMDgeom}, leads to $z = 1+ (d-1)/d$. For the two dimensional case $d=2$ this leads to
\be
z = 3/2 \,, \qquad \theta = 1 \,,
\ee
in agreement with the exact exponents (\ref{eq:isingexp}) for the single patch Ising-nematic theory.

\subsection{Fermions in the bulk I:  `classical' physics}
\label{sec:bulkfermions}

The most direct signature of a Fermi surface is a zero in the inverse fermionic Green's function;
this is expected to be present even in cases where fermionic quasiparticles are absent.
We have stated previously that the microscopic electron operators are not readily accessible
from the standpoint of emergent universal low energy physics, and certainly do not
help to organize that physics. The low energy effective description, however,
will typically also contain gauge-invariant fermionic operators. In this section we describe
the correlation functions of such fermionic operators in holographic compressible phases.
These computations are `classical' in the sense that they involve solving the classical Dirac equation
in various backgrounds.
The pioneering works in this endeavor were \cite{Lee:2008xf, Liu:2009dm, Cubrovic:2009ye, Faulkner:2009wj}.
The Fermi surfaces found below in such computations are analogs of the $d_\alpha$ fermions in \S\ref{sec:emerge2} \cite{Huijse:2011hp,SS10}.

\subsubsection{The holographic dictionary}
\label{sec:holodicF}

A few words are necessary about the holographic dictionary for fermions. A simple quadratic action for bulk fermion fields is the Dirac action with minimal coupling to a background metric and Maxwell potential: \begin{widetext}
\begin{equation}
S_{\text{bulk fermions}} = -\mathrm{i}\int \mathrm{d}^{d+2}x\sqrt{-g} \; \left\lbrace\bar\Psi \Gamma^M \left(\partial_M + \frac{1}{4} \omega_{M AB}\Gamma^{AB} - \mathrm{i}qA_M\right) \Psi - m \bar\Psi \Psi \right\rbrace.  \label{eq:fermionaction}
\end{equation}\end{widetext}
Here $\lbrace \Gamma^M, \Gamma^N\rbrace = 2\delta^{MN}$ are $d+2$ dimensional (bulk) Gamma matrices, and $\omega$ is the spin connection, necessary to describe fermions on curved spaces. A further important type of bulk fermions are spin-$3/2$ Rarita-Schwinger fields.

Unlike the bosonic fields we have studied so far, the Dirac action is manifestly first order, and so we must revisit the basic holographic dictionary \cite{Mueck:1998iz, Iqbal:2009fd}.   As the computations are similar to those for e.g. scalar fields, we will skip most details and outline the necessary changes.   Crudely speaking, we think of writing down a second order equation for half of the components of the bulk spinor.  It is this half of the bulk spinor which is dual to a fermionic operator in the boundary theory.  If the bulk spacetime dimension $d+2$ is even, this is consistent with the dual operator also being a Dirac spinor. This is because a Dirac spinor in $d+1$ odd dimensions has half the number of components of a Dirac spinor in $d+2$ even dimensions. However, if the bulk spacetime dimension is odd, then we are left with ``half" of a Dirac spinor in the boundary theory -- namely, a Weyl spinor. We can now be more precise -- see \cite{Mueck:1998iz, Iqbal:2009fd} for more details on the following.

We will assume that the bulk metric depends only on $r$ and is diagonal. This is the case for all the spacetimes we have considered so far. In this case, it is convenient to make the change of variable \cite{Faulkner:2009wj} \begin{equation}
\Psi = (-g g^{rr})^{-1/4}\mathcal{X},
\end{equation}in which case the equation of motion following from (\ref{eq:fermionaction}) becomes \begin{equation}
e^a_A \Gamma^A (\partial_a - \mathrm{i}qA_a) \mathcal{X} = m\mathcal{X} .
\end{equation}
For a diagonal metric, the vielbein can be chosen as $e^a_A = \sqrt{|g^{aa}|}$.  Near the asymptotically AdS boundary, this equation reads \begin{equation}
\Gamma^r \frac{r}{L}\partial_r \mathcal{X}  = m\mathcal{X} + \cdots \,,
\end{equation}
which is solved by \begin{equation}
\Psi = \psi_+(x)r^{(d+1)/2+mL}  +  \psi_-(x) r^{(d+1)/2-mL}.
\end{equation}
Here $\Gamma^r \psi_\pm = \pm \psi_\pm$ -- this is the ``splitting" of the bulk spinor in half, as advertised.   One can further show that 
the dimension $\Delta$ of the fermionic boundary operator is
\begin{equation}
\Delta = \frac{d+1}{2} + m L \,.
\end{equation}
To apply holographic renormalization to bulk fermions, a boundary counterterm proportional to $\bar\Psi \Psi$ is required.  At the end of the day, one finds that $\psi_+$ is the source term for the boundary fermion, and $\psi_-$ is proportional to the expectation value of the boundary fermionic operator. For the range of masses $0 \leq m < \half$, an alternate quantization is also possible, analogous to that discussed in \S\ref{sec:multitrace} above for scalars. In this quantization the dimension $\Delta = (d+1)/2 - m L$.

\subsubsection{Fermions in semi-locally critical ($z=\infty$) backgrounds}
\label{sec:fermionz}

An important understanding that emerged from holographic studies of fermionic correlators is that semi-locally critical degrees of freedom (as discussed in \S \ref{sec:ads2rd} and \S \ref{sec:zeroTspectral} above) with $z=\infty$ can efficiently scatter low energy fermions at nonzero momentum $k \approx k_{\mathrm{F}}$. This leads to broad non-Fermi liquid fermionic self-energies at low energies and temperatures, $\omega, T \ll \mu$. This was first understood in detail from the AdS-RN background, as we now describe \cite{Faulkner:2009wj,Faulkner:2011tm,Iqbal:2011ae}. The same analysis applies to fermions in the more general $z= \infty$ geometries described around equation (\ref{eq:etalimit}) above \cite{Gubser:2012yb, DeWolfe:2013uba}.

The matching computation for the zero temperature Green's functions proceeds exactly as for the bulk scalar fields studied in \S\ref{sec:zeroTspectral}. Thus the result for the low energy Green's function takes the form
\begin{equation}
G^{\mathrm{R}}(\omega, k) = \frac{\mathcal{B}_1(\omega,k) + \omega^{2\nu_k} \mathcal{B}_2(\omega,k)}{\mathcal{B}_3(\omega,k) + \omega^{2\nu_k}\mathcal{B}_4(\omega,k)} \,, \label{eq:BB}
\end{equation}
where now the critical exponents from the $\mathrm{AdS}_2$ IR geometry are
\begin{equation}
\nu_k =  \sqrt{m^2 L_2^2 + \frac{1}{d(d-1)}\frac{\g^2 \, k^2}{\mu^2} - \frac{q^2e^2}{\k^2} L_2^2} \,. \label{eq:nukferm}
\end{equation}
As previously for the scalar fields, these exponents control the momentum-dependent scaling dimensions of operators in the semi-locally critical IR theory. The dimension is again given by $\nu_k + 1/2$. Note that if $qe$ is large enough,  $\nu_k$ can become imaginary. This indicates an instability of the black hole towards pair production of fermions, which we will return to and resolve in \S \ref{sec:bulkF}.

In a slight generalization of (\ref{eq:GRrelate}) we have allowed the $\mathcal{B}$ functions in (\ref{eq:BB}) to have regular frequency dependence (i.e. positive integer powers of $\omega$ starting at $\omega^0$). This comes from perturbatively re-inserting the frequency dependence in the wave equation in the far region. Recall that away from the horizon, the $\omega \to 0$ limit is analytic. Unlike the exponents in (\ref{eq:nukferm}), the $\mathcal{B}$ functions are not universal IR data, but depend upon the entire bulk geometry. The functions $\mathcal{B}_1$ and $\mathcal{B}_3$ are real, following the arguments given in \S\ref{sec:zeroTspectral}.

Fermi surfaces occur if the denominator of (\ref{eq:BB}) vanishes at $\omega=0$ for some $k=k_{\mathrm{F}}$. That is, if $\mathcal{B}_3(0,k_{\mathrm{F}}) = 0$. This may or not occur, depending on the full background. Typically the Fermi momenta must be found by solving the equations of motion numerically (see \cite{Faulkner:2009wj} for systematic numerical results in AdS-RN). Occasionally the full $\omega=0$ wave equation can be solved analytically, as in e.g. \cite{Hartman:2010fk, Gubser:2012yb}. Once $k_{\mathrm{F}}$ is found, the Green's function can be expanded around the Fermi sufrace.
Denoting $k_\perp=k-k_{\mathrm{F}}$ to be the momentum transverse to the Fermi surface (as the theory is isotropic), one finds
\begin{equation}
G^{\mathrm{R}}(\omega,k_\perp) \approx \frac{Z}{\omega - v_{\mathrm{F}} k_\perp + \omega^{2\nu_k}\widetilde\Sigma}+\cdots \,.  \label{eq:AdSRNprop}
\end{equation}
The Fermi velocity $v_{\mathrm{F}}$ and weight $Z$ are real whereas $\widetilde\Sigma$ is complex. They are all constant in the limit $k_\perp,\omega\rightarrow 0$. All of $Z, k_{\mathrm{F}},v_{\mathrm{F}},\widetilde\Sigma$ depend on the full spacetime, not just the IR scaling geometry.

Equation (\ref{eq:AdSRNprop}) is indeed the two point function of a fermionic particle, with a Fermi surface at momentum $k_{\mathrm{F}}$, and with self-energy $\Sigma(k,\omega)=\omega^{2\nu_k} \widetilde \Sigma$. The imaginary part of $\Sigma$ corresponds to the decay rate of this excitation. Close to the Fermi surface, the decay rate is determined by the scaling dimension $\nu_k$ with $k \approx k_{\mathrm{F}}$.  We can, however, contrast (\ref{eq:AdSRNprop}) with the Fermi liquid result (\ref{Gfl}):
\begin{enumerate}
\item  If $2\nu_{k_{\mathrm{F}}} >1$,  then as $\omega \rightarrow 0$, (\ref{eq:AdSRNprop}) may be approximated as $G^{\mathrm{R}} = Z(\omega-v_{\mathrm{F}}k_\perp)^{-1}$.     The fermion propagator genuinely resembles that of the Fermi liquid.   In this case,  there is a long-lived quasiparticle, but its decay occurs via interactions with a locally critical bath and is not governed by the standard Fermi liquid phenomenology.

\item When $2\nu_{k_{\mathrm{F}}}=1$, then (\ref{eq:AdSRNprop}) becomes \begin{equation}
G^{\mathrm{R}}(\omega,k_\perp)  \approx \frac{Z}{\omega - v_{\mathrm{F}}k_\perp - \widetilde\Sigma \, \omega \log (\omega/\mu)} + \cdots.
\end{equation}
It was found in 1989 that such a Green's function explains multiple experimentally anomalous properties of the cuprate strange metal \cite{varma}, and is called the `marginal Fermi liquid'.  This possibility is fine tuned in the AdS-RN black hole, because the charge and mass need to be just right for $2 \nu_{k_{\mathrm{F}}} = 1$, but it is interesting to see these connections with a much older model.

\item If $2\nu_{k_{\mathrm{F}}} <1$, then as $\omega \rightarrow 0$,  (\ref{eq:AdSRNprop}) is $G^{\mathrm{R}}\approx Z(\widetilde\Sigma \omega^{2\nu_k} - v_{\mathrm{F}}k_\perp)^{-1}$,  and does not look like a long lived quasiparticle.
\end{enumerate}
 
Plots of $\nu_{k_{\mathrm{F}}}$ in the AdS-RN background as a function of the mass and charge of the fermion operator can be found in  \cite{Faulkner:2009wj}. It was also found in \cite{Faulkner:2009wj} that whenever Fermi surfaces appear in AdS-RN, then as $k \rightarrow 0$ (i.e. at small momenta, not close to $k_{\mathrm{F}}$), $\nu_k$ becomes imaginary.   This indicates an instability of the low energy $\mathrm{AdS}_2\times \mathbb{R}^d$ geometry in all of these cases, as we will discuss below.

The following sections will consider the physical interpretation of these Green's functions, as well as their implications for the charged horizon background. Further work on the Fermi surfaces themselves includes the addition of bulk dipole couplings between the spinor and the background Maxwell field \cite{Edalati:2010ww, Edalati:2010ge, Guarrera:2011my} --- such couplings are common in consistent truncations, e.g. \cite{Bah:2010yt, Bah:2010cu, Liu:2010pq, DeWolfe:2011aa, DeWolfe:2012uv} --- as well as an understanding of the physics of spin-orbit coupling \cite{Alexandrov:2012xe, Herzog:2012kx}. Sum rules for holographic fermionic Green's functions have been discussed in \cite{Gursoy:2011gz}.  Holographic fermions in periodic potentials were discussed in \cite{Liu:2012tr}. For massless fermions, exchanging the choice of quantization -- as described at the end of \S \ref{sec:holodicF} -- inverts the fermion Green's function \cite{Faulkner:2009wj}. With a bulk dipole coupling, this can lead to zeros in the Green's functions that are `dual' to Fermi surface poles \cite{Alsup:2014uca,Vanacore:2014hka}.

\subsubsection{Semi-holography: one fermion decaying into a large $N$ bath}
\label{sec:semiholog}

The Green's function (\ref{eq:AdSRNprop}) can be understood from what was termed a `semi-holographic' perspective in \cite{Faulkner:2010tq}. This perspective leads to an intuitive picture that will furthermore clarify the nature of the large $N$ physics going into (\ref{eq:AdSRNprop}).

The semi-holographic picture is motivated as follows. The Fermi surface pole in (\ref{eq:AdSRNprop}) is nothing other than a quasinormal mode of the Dirac equation in the charged black hole background; recall the discussion in \S \ref{sec:quasinormal} above. As $k \to k_{\mathrm{F}}$, the frequency of the mode goes to zero. At $\w = 0$ exactly, it is a zero energy normal mode (with no imaginary part) in the background. The radial profile of this mode is peaked at some radius set by the crossover from the IR to the UV region \cite{Faulkner:2010tq}, consistent with the fact that $k_{\mathrm{F}}$ comes from solving the Dirac equation in the `far' region, away from the horizon. This mode is therefore an example of the phenomenon described in \S \ref{sec:nongeom} above: it is a low energy excitation that is not geometrized. That is,
it is not captured (at $\omega = 0$) by the near-horizon region of the geometry. The reason this can occur is that the Green's function (\ref{eq:AdSRNprop}) is describing a single fermion at a Fermi surface whereas geometrized gravitational dynamics necessarily describes -- as we have seen above -- a large $N$ number of degrees of freedom.

At nonzero $\omega$, the decay of this mode is determined by the rate at which it falls through the black hole horizon.
The near horizon region describes the low energy dynamics of a strongly interacting large $N$ quantum critical phase. To describe the decay, one can write down an action in which a `free fermion' $\psi$ is coupled to a fermion $\chi$ that is part of a strongly interacting large $N$ sector:
\begin{align}
S &= S_{\mathrm{strong}}[\chi,\ldots] + \int \mathrm{d}^{d+1}x \Big[\psi^\dagger\left(\mathrm{i}\partial_t - \varepsilon(\partial_i)\right)\psi  \notag \\
&+ \lambda \psi^\dagger \chi  + \lambda \chi^\dagger \psi\Big].
\end{align}
The strongly interacting sector $S_{\mathrm{strong}}[\chi,\ldots]$ above is the near horizon region of the AdS-RN geometry.  In particular, $\chi$ has a dual field in this near horizon spacetime. At large $N$, the only fermion correlation function of the strongly interacting sector which is not
suppressed by a power of $N$ is the two point function $\langle \chi^\dagger \chi\rangle $ -- in the bulk this manifests itself in the fact that the action for the field dual to $\chi$ is quadratic up to $1/N$ corrections.  
Hence, employing the bare Green's functions \begin{equation}
G^0_{\psi\psi} = \frac{1}{\omega - \varepsilon(k)}, \;\;\;\;\; G^0_{\chi\chi} = c \, \omega^{2\nu_k},
\end{equation}
with the latter Green's function coming from the holographic IR expression (\ref{eq:IRG}), the full Green's function for $\psi$ may be computed exactly as a geometric sum of diagrams:
 \begin{align}
G_{\psi\psi} &= \sum_{n=0}^\infty G^0_{\psi\psi} \left(\lambda^2 G^0_{\chi\chi}G^0_{\psi\psi}\right)^n+ \mathcal{O}\left(N^{-1}\right)\notag \\
&= \frac{1}{\omega - \varepsilon_k - \lambda \, c \, \omega^{2\nu_k}} \,.
\end{align}
This matches the qualitative form of (\ref{eq:AdSRNprop}) up to the normalization coefficient, and upon `zooming in' near a Fermi surface where $\varepsilon_k \approx v_{\mathrm{F}}(k-k_{\mathrm{F}})$.

The argument above can be turned into a precise derivation using the renormalization group flow of double trace operators at large $N$ \cite{Faulkner:2010jy, Heemskerk:2010hk}. This semi-holographic perspective gives a field-theoretic understanding of why we were able to compute the Green's function (\ref{eq:AdSRNprop}) with only a small amout of information coming from the full black hole geometry (such as the value of $k_{\mathrm{F}}$). The non-trivial input comes from (\emph{i}) the large $N$ suppression of higher order correlation functions, (\emph{ii}) the existence of a single fermion normal mode at $\omega = 0$ and (\emph{iii}) the semi-locally critical form of the Green's functions in the large $N$ critical sector. Any model with these features, holographic or not, will lead to (\ref{eq:AdSRNprop}).

The semi-holographic perspective also applies when the IR critical sector has $z < \infty$. In these cases, the imaginary part of the IR Green's function at low temperatures and energies, but nonzero $k \approx k_{\mathrm{F}}$, takes the same exponentially suppressed form as we have discussed previously for bosons in (\ref{eq:imglif}) and (\ref{eq:IRlif2}). See e.g. \cite{Faulkner:2010tq, Iizuka:2011hg, Hartnoll:2011dm, Iqbal:2011in, Gursoy:2012ie, Kim:2016nqo}. The decay into the horizon is very ineffective for these bulk fermion normal modes, a geometrization of the fact that there are no low energy excitations at nonzero momenta when $z < \infty$.

The semi-holographic derivation emphasizes the separation between the single bulk fermion zero mode and the remaining large $N$ low energy degrees of freedom that are described by the near horizon dynamics. Consistent with this picture, we will see in \S \ref{sec:luttinger} below that the holographic Fermi surface comes with an order one amount of charge carried outside of the black hole horizon, by the bulk fermion field. This can be contrasted with the large $N$ amount of charge behind the horizon, suggesting that the Fermi surface is only capturing the dynamics of a very small part of the system \cite{Hartnoll:2011fn, Huijse:2011ef, Iqbal:2011bf}. An alternative interpretation starts by noting that the bulk fermion is dual to a gauge-invariant composite fermionic operator in the boundary \cite{Gubser:2009qt, DeWolfe:2011aa, Cosnier-Horeau:2014qya}. One can roughly think of the composite operator as the product of a colored (gauge-charged) boson with a colored fermion. In a generic state the colored bosons can be condensed, this then allows the composite operator to have an overlap with a large $N$ number of colored `gaugino' fermions. The Fermi surfaces detected holographically would then count the large $N$ amount of charge carried by these gaugino Fermi surfaces.

Independently of the interpretation of the bulk Fermi surface, the presence of a small amount of charge in a Fermi surface outside of the horizon is responsible for important bulk quantum mechanical effects.  These effects are subleading in large $N$. This means firstly that the properties of the Fermi surface do not imprint themselves onto leading order in $N$ observables. For instance, the electrical conductivity at leading order in large $N$ will be determined by fluctuations of the bulk Maxwell field (as we have discussed above) and will not notice the existence or not of a Fermi surface. This last perspective emphasizes how in strongly coupled dynamics one must be prepared to disassociate physics (such as Fermi surfaces and conductivities) that is closely connected at weak coupling. Of course, in a microscopically grounded holographic dual of a given boundary theory, the single theory fixes the background and the equation for the Maxwell field and for the Dirac field, and therefore the conductivity and the Fermi surfaces arise from a common underlying structure. 

Secondly, it turns out that many of the quantum mechanical effects associated with Fermi surfaces, as is well known in condensed matter, are singular at low energies. These include quantum oscillations, Cooper pairing and singular contributions to observables such as conductivities. The presence of quantum mechanical effects in the bulk that can dominate observables at low energies amounts to a breakdown of the large $N$ expansion. In the following \S\ref{sec:bulkF} we will describe some of these effects, including the quantum mechanical instability of the near horizon region in the presence of fermionic operators with imaginary scaling dimension.

\subsection{Fermions in the bulk II:  quantum effects}
\label{sec:bulkF}

\subsubsection{Luttinger's theorem in holography}
\label{sec:luttinger}

Towards the end of the previous section we noted that the presence of Fermi surface poles in Green's functions is suggestive of the presence of charge in the bulk. Because the bulk fermions obey Pauli exclusion, they cannot carry the charge by condensing into a macroscopically occupied ground state. Instead they will fill up a Fermi surface in the bulk, and outside of the black hole event horizon. A derivation of this fact necessarily requires quantum mechanics.
In this section we will sketch the result that
\be\label{eq:bulklut}
\rho = \rho_\text{hor.} + \sum_i q_i \, \frac{V^\text{FS}_i}{(2 \pi)^d} \,.
\ee
Here $\rho$ is, as always, the charge density of the boundary QFT, $\rho_\text{hor.}$ is the electric flux coming through the charged horizon and the sum is over the ($d$-dimensional) momentum-space volumes $V^\text{FS}_i$ of Fermi surfaces in the bulk, due to fermionic fields with charges $q_i$. The result (\ref{eq:bulklut}) was first proven in a WKB limit in the bulk in \cite{Hartnoll:2011dm, Iqbal:2011in}, away from this limit in \cite{Sachdev:2011ze} and in complete generality (allowing for horizons in the bulk) in \cite{Iqbal:2011bf}.

The expression (\ref{eq:bulklut}) is strongly reminiscent of the modified Luttinger relation (\ref{lutt2}), in which the charge density is made up of a sum of gauge-neutral `cohesive' Fermi surfaces and gauge-charged `fractionalized' Fermi surfaces \cite{Huijse:2011hp, Hartnoll:2011fn, Hartnoll:2011pp, Huijse:2011ef}. At large $N$, `gauge-charged' operators are more rigorously thought of as those built up as very long traces of the fundamental fields (and hence with a large scaling dimension). For instance, the deconfinement transition can be profitably discussed in this language \cite{Aharony:2003sx}. Fractionalization in this context is an analogue of such large $N$ deconfinement in which the charge is carried by long rather than short operators. The large degeneracy of long operators at large $N$ leads to this charge being hidden behind an event horizon. Equation (\ref{eq:bulklut}) does not say whether the charge behind the horizon is carried by Fermi surfaces or not. The reader is referred to our discussion in \S\ref{sec:lowEspect} that bulk quantum corrections could lead to signatures of a Fermi surface \cite{Polchinski:2012nh,Faulkner:2012gt,Sachdev:2012tj} from the charge behind the horizon. 
The work of \cite{Aharony:2003sx} has not yet been generalized to a discussion of possible large $N$ fractionalization in weakly interacting theories in the 't Hooft limit. An attempt to find a holographic order parameter for charge fractionalization can be found in \cite{Hartnoll:2012ux}.

To prove (\ref{eq:bulklut}), consider the free bulk Dirac fermion (\ref{eq:fermionaction}), of charge $q$,  coupled to  e.g. EMD theory. We can sketch the derivation in \cite{Sachdev:2011ze, Iqbal:2011bf}, and for simplicity work at zero temperature. The fermion can be integrated out, leading to the effective bulk action
\begin{equation}
S = S_{\mathrm{EMD}} - \mathrm{tr} \log \left(\mathrm{i}\Gamma^\mu \mathrm{D}_\mu + m\right).
\end{equation} 
Employing standard many-body manipulations (and assuming the only nontrivial dependence of the background fields is in the radial direction), one obtains
\begin{equation}
S = S_{\mathrm{EMD}} - V_d\sum_l \int \frac{\mathrm{d}^dk}{(2\pi)^d} \epsilon_l(k) \Theta(-\epsilon_l(k)),  \label{eq:fermiongs}
\end{equation}
where $V_d$ is the spatial volume of the boundary theory, $\epsilon_l(k)$ are the eigenvalues of the bulk Dirac equation, with eigenspinors $\chi_l(k,r)$, subject to appropriate boundary conditions \cite{Sachdev:2011ze, Iqbal:2011bf}. The effective action (\ref{eq:fermiongs}) shows that the free energy due to fermions is the energy of all the occupied sites in a bulk Fermi sea.   $\epsilon_l(k)$ is implicitly a function of the background EMD fields.
As advertised, equation (\ref{eq:fermiongs}) shows the Pauli exclusion principle at work. The bulk geometry contains fermionic matter placed into its lowest energy states.

The equations of motion for the metric and Maxwell fields originating from the effective action (\ref{eq:fermiongs}) are nonlocal. This makes the effects of the backreaction of the Fermi surfaces on the spacetime difficult to study. A backreacting and gravitating Fermi surface is nothing other than a star. For this reason, the backreacted solutions were called `electron stars' in \cite{Hartnoll:2010gu}, by analogy to astrophysical neutron stars. The asymptotically Anti-de Sitter boundary conditions provide an additional gravitational potential well that allows a charged star to exist, despite the repulsion between the individual charged fermions. Indeed, taking a cue from astrophysics, the backreacted equations are most easily solved in a WKB approximation for the fermions. This is a generalization of the Oppenheimer-Volkoff-Tolman approximation \cite{Oppenheimer:1939ne, Tolman:1939jz} for neutron stars to include charge and a negative cosmological constant. Neutron stars themselves have also been considered in Anti-de Sitter spacetime \cite{deBoer:2009wk}. In a condensed matter context, the WKB approximation for many electrons interacting with an electromagnetic field is called the Thomas-Fermi approximation \cite{PSP:1732980, Fer00}. For discussion on the connection between WKB fermion wavefunctions and the effective `fluid' descriptions of these solutions, see e.g. \cite{Ruffini:1969qy, Arsiwalla:2010bt}.
WKB electron star solutions are studied in \S \ref{sec:thomasfermi} below.

Away from the WKB limit, the backreaction problem is very challenging. Important progress has been made in \cite{Allais:2012ye, Allais:2013lha}. One result of these papers in that the WKB approximation works well somewhat beyond its naive regime of applicability. Without solving the full backreaction equations, however, we proceed to show how the result (\ref{eq:bulklut}) can be obtained exactly from one of the bulk Maxwell equations.

Assuming the usual ansatz $A = p(r)\mathrm{d}t$ for the Maxwell field, one obtains \cite{Sachdev:2011ze, Iqbal:2011bf}
\begin{align}
& \frac{1}{e^2}\partial_r \left(\sqrt{-g} g^{rr}g^{tt} \partial_r p \right) \notag \\
& = q \sum_l  \int \frac{\mathrm{d}^dk}{(2\pi)^d} \Theta(-\epsilon_l(k)) \chi^\dagger_l(k,r) \chi_l(k,r) \,.
\end{align}
The right hand side follows from the Feynman-Hellmann theorem and the action (\ref{eq:fermiongs}).  Integrating over $r$ from the AdS boundary to the horizon gives
\begin{align}
& \frac{1}{e^2} \left.\left(\sqrt{-g} g^{rr}g^{tt} \partial_r p \right)\right|_0^{r_+} \equiv \rho - \rho_{\mathrm{hor}} \notag \\
& = q\sum_l  \int \frac{\mathrm{d}^dk}{(2\pi)^d} \Theta(-\epsilon_l(k))  \equiv q \frac{V^{\mathrm{FS}}}{(2 \pi)^d}  \,, \label{eq:luttads}
\end{align}
where we have used the AdS/CFT dictionary at the boundary, orthonormality of $\chi_l$,  defined $\rho_{\mathrm{hor}}$ as a (normalized) radial electric flux across the horizon, and defined $V^{\mathrm{FS}}$ as the total volume of the bulk Fermi surfaces. In this way we establish (\ref{eq:bulklut}), the holographic analogue of the Luttinger relation (\ref{lutt2}) in the presence of both cohesive and fractionalized charge carriers.
$qV^{\mathrm{FS}}$ counts the cohesive charge which can detected by a gauge-neutral fermionic operator (for instance, via Fermi surface singularities).  $\rho_{\mathrm{hor}}$ counts the fractionalized charge which is not directly detected by such gauge-neutral operators, but remains hidden behind the horizon. We note in passing that  (\ref{eq:luttads}) has a nonzero temperature generalization \cite{Sachdev:2011ze, Iqbal:2011bf}. Prior to the fermions considered in this subsection, the solutions we have considered have $\rho=\rho_{\mathrm{hor}}$ and all the charge was fractionalized. Cohesive charge can also be carried by condensed bosons, as we will see in \S \ref{sec:holoS} below.

The semi-holographic description of \S \ref{sec:semiholog} makes clear that while cohesive fermions obey a conventional Luttinger relation, their decay is not that of a Fermi liquid. Rather, the fermions decay into a large $N$ critical sector. An ordinary Landau Fermi liquid is obtained holographically if the critical sector is removed, by considering a gapped geometry of the kind discussed in \S\ref{sec:gapped}. In these spacetimes there is no horizon for the bulk fermions to fall into. An explicit toy model of this situation, using a `hard wall' in AdS, was considered in \cite{Sachdev:2011ze}. As anticipated from \S\ref{sec:gapped}, at a classical level the imaginary part of the fermion self-energy vanishes. One expects quantum bulk computations to follow standard many-body treatments, leading to $\mathrm{Im}(G^{\mathrm{R}}(\omega\rightarrow0)) \sim \omega^2$.  That is because, in the absence of a horizon, one is simply studying interacting fermions in a fancy (i.e. curved spacetime) box. There are no geometrized low energy degree of freedom and much of the interesting ingredients of holography are gone. The situation is analogous to confinement in QCD. The strongly interacting dynamics leads to a mass gap, and the remaining `semi-holographic' degrees of freedom (hadrons and mesons in the case of QCD) interact weakly at low energies.

\subsubsection{$1/N$ corrections}\label{sec:qo}

We now will explore three consequences of the bulk fermion matter whose existence has been revealed by the holographic Luttinger theorem (\ref{eq:bulklut}). These will be IR singular quantum mechanical effects in the bulk. The study of these effects (quantum oscillations, Cooper pairing and contributions to the conductivity) will follow very closely the conventional treatment in standard weakly interacting fermion systems. In this sense, bulk fermion matter is fairly conventional. The two important differences are that (\emph{i}) we must learn to do the computations in a curved spacetime and (\emph{ii}) the presence of an event horizon means that the tree level fermion propagators already have significant imaginary self-energy terms, as shown in \S\ref{sec:fermionz} above. As explained in \S\ref{sec:semiholog}, this latter effect is a consequence/artifact of having a large $N$ critical bath into which the fermions can decay.

The computations outlined below follow the logic described in \S\ref{sec:oneoverN}. The reader should read that section before continuing. The Fermi surface pole in the fermion Green's function corresponds to a quasinormal mode with complex frequency approaching zero. As we explained in \S\ref{sec:oneoverN}, the non-analytic one-loop effects of such poles can be extracted zooming in on the portion of the propagator controlled by this pole, as in (\ref{eq:modGamma}) or (\ref{eq:zoomin}) above.

\paragraph{Quantum oscillations} 
\label{sec:qoqo}

We begin with quantum oscillations.  These are a characteristic phenomenon of a Fermi surface in a magnetic field:  the magnetic susceptibility (the second derivative of the free energy with respect to the magnetic field) oscillates like $\cos (\pi k_{\mathrm{F}}^2/B)$. We are using units with $\hbar = c  = e = k_{\mathrm{B}} = 1$. These oscillations are resonances that occur when the cyclotron orbit of the fermion (in momentum space) coincides with a cross section of the Fermi surface, so that there are many low energy excitations that can participate in the motion. The existence of the oscillations is very robust and independent of the form of the self-energy of the fermions, so long as the fermion Green's function has a pole along a Fermi surface.  More details of the following computation can be found in \cite{Denef:2009yy, Denef:2009kn, Hartnoll:2009kk}.

The amplitude of the quantum (or de Haas-van Alphen) oscillations is a function of temperature. This temperature dependence is strongly sensitive to the self-energy of the fermions. In a Fermi liquid, the temperature dependence of the amplitude takes a celebrated form called the Lifshitz-Kosevich formula \cite{kosevich}. In particular, at large temperatures compared to the magnetic field, the amplitude decays exponentially as, schematically, $A(T) \sim \mathrm{e}^{- T m_\star/B}$. The bulk fermions in a holographic semi-locally critical background will lead to quantum oscillations. We now outline the computation of $A(T)$ in this case, restricting attention to the simplest case of $d=2$ space dimensions. 

The quantity to be computed is the determinant of the Dirac operator in the bulk fermion action (\ref{eq:fermionaction}). The logarithm of this determinant is the one loop contribution of the fermions to the free energy. Using methods similar to those leading to (\ref{eq:modGamma}) above, a general formula can be derived for the `oscillatory' part of this fermion contribution to the free energy \cite{Hartnoll:2009kk} (see also \cite{wasserman1996influence} for a more diagrammatic derivation -- this is not a holographic formula per se)
\be\label{eq:fosc}
f_\text{osc} =  \frac{B T}{\pi} \text{Re} \sum_{n=0}^\infty \sum_{k=1}^\infty \frac{1}{k} \mathrm{e}^{2 \pi \mathrm{i} \, k \, \ell_\star(n)} \,.
\ee
Here $B$ is the magnetic field and $\ell_\star(n)$ is defined by the singular locus of the fermion retarded Green's function
\be\label{eq:Ginf}
G^{\mathrm{R}}(\omega_n,\ell_\star(n)) = \infty \,.
\ee
The fermionic Matsubara frequencies are $\omega_n = 2 \pi \mathrm{i} T (n + \half)$. The second argument of the Green's function in (\ref{eq:Ginf}) is the Landau level. To leading order in a small magnetic field, the effects of the magnetic field are captured by starting with the zero magnetic field holographic Green's function and replacing the momentum dependence by $k^2 \to 2 \ell B$.

To evaluate (\ref{eq:fosc}) for semi-locally critical fermions, it is important to use the full nonzero temperature IR Green's function for the fermions. For the case of AdS-RN, this can be found by solving the Dirac equation in the AdS$_2$-Schwarzschild near-horizon geometry (\ref{eq:smallTgeom}). The upshot is that in the expression (\ref{eq:AdSRNprop}) for the fermion retarded Green's function, the frequency dependence of the self-energy is generalized according to \cite{Faulkner:2011tm}
\be
\omega^{2 \nu_k} \to T^{2 \nu_k} \frac{\G(\half + \nu_k - \frac{\mathrm{i} \omega}{2 \pi T} + \frac{\mathrm{i} q e}{\sqrt{2} \k})}{\G(\half - \nu_k - \frac{\mathrm{i} \omega}{2 \pi T} + \frac{\mathrm{i} q e}{\sqrt{2} \k})} \,. \label{eq:IRTfinite}
\ee
The form of this Green's function is determined by an emergent $\mathrm{SL}(2,\R)$ symmetry of the AdS$_2$ IR theory \cite{Faulkner:2011tm}. This is an additional symmetry that the theory enjoys and is logically distinct from the fact that $z=\infty$.

Inserting the IR Green's function (\ref{eq:IRTfinite}) into the full Green's function (\ref{eq:AdSRNprop}), the oscillatory part of the free energy (\ref{eq:fosc}) can be evaluated. The full expression is a little complicated, see \cite{Hartnoll:2009kk}, but a distinctive behavior emerges in the limit of large temperatures compared to the magnetic field, where the magnetic susceptibility
\be\label{eq:Bsuscept}
\chi \sim \mathrm{e}^{- c \, T^{2 \nu}/B}\cos \frac{\pi k_{\mathrm{F}}^2}{B} \,, \qquad (\text{for} \; \nu < \half)
\ee
Here $c$ is a constant set by the Fermi momentum and chemical potential, and $\nu$ is the exponent $\nu_k$ evaluated at the Fermi momentum $k = k_{\mathrm{F}}$. In this non-Fermi liquid case where $\nu < \half$ (case 3 in \S\ref{sec:fermionz} above) the large temperature falloff of the amplitude of the oscillations is distinct from the Kosevich-Lifshitz form and depends upon the dimension $\nu$ of the fermionic operator in the low energy critical theory.

The oscillatory susceptibility in (\ref{eq:Bsuscept}) is a quantum effect in the bulk and is therefore suppressed by factors of $N$ compared to the leading classical contribution, that does not oscillate.
It is notable that none of the holographic models of compressible matter discussed so far exhibit quantum oscillations in the leading order in $N$ magnetic susceptibility, despite some backgrounds manifesting Fermi-surface like physics as discussed in \S\ref{sec:lowEspect}. The only known, somewhat artificial, way to make the oscillations contribute at leading order is in the semiclassical electron star discussed in \S\ref{sec:thomasfermi} below. 

\paragraph{Cooper pairing}
\label{sec:cooper}

Fermi surfaces are well known to be unstable at low temperatures towards condensation of Cooper pairs. This condensation spontaneously breaks the $\mathrm{U}(1)$ symmetry. Holographic Fermi surfaces can undergo Cooper pair condensation following the usual BCS mechanism adapted to curved spacetime.  More details of the following computation can be found in \cite{Hartman:2010fk}.

Cooper pairing requires an attractive force between the fermions. In \cite{Hartman:2010fk} the following bulk contact interaction was considered
\be\label{eq:finteract}
S_\text{int} = \frac{1}{M_{\mathrm{F}}^2} \int \mathrm{d}^4x \sqrt{-g} \left(\bar \Psi_{\mathrm{c}} \Gamma^5 \Psi\right)\left(\bar \Psi \Gamma^5 \Psi_{\mathrm{c}}\right) \,.
\ee
Here $\Gamma^5 = \mathrm{i} \G^0 \G^1 \G^2 \G^3$, $M_{\mathrm{F}}$ determines the mass scale of the interaction and $\Psi_{\mathrm{c}}$ is the charge conjugate fermion (see \cite{Hartman:2010fk}). The action (\ref{eq:finteract}) can be considered a toy model for exchange interactions between the fermions mediated by other bulk fields. The interaction has been chosen to decouple nicely in the `Cooper channel'. This decoupling is achieved by introducing a charged Hubbard-Strotonovich field $\Delta$ so that the interaction is written
\be\label{eq:bogoliubov}
S_\text{int} = \int \mathrm{d}^4x \sqrt{-g} \left(\bar \Psi_{\mathrm{c}} \Gamma^5 \Psi \Delta + \bar \Psi \Gamma^5 \Psi_{\mathrm{c}} \Delta^* - M_{\mathrm{F}}^2 |\Delta|^2 \right) \,.
\ee

The next step is to integrate out the fermions to obtain the one-loop effective action for $\Delta$. An instability arises when the effective mass squared for $\Delta$ becomes negative. Restricting to configurations of $\Delta = \Delta(r)$ that only depend on the radial direction, one finds the quadratic effective Euclidean action to be
\begin{widetext}
\be
S^{(2)}_\text{eff}[\Delta] = \frac{V_2}{T} \left[ M_{\mathrm{F}}^2  \int \mathrm{d}r \sqrt{g(r)} |\Delta(r)|^2 + \int \mathrm{d}r \mathrm{d}r' \sqrt{g(r) g(r')} \Delta(r) \Delta^*(r') F(r,r') \right] \,.\label{eq:HS}
\ee
Here $V_2$ is the volume of the boundary spatial directions and $T$ is the temperature (inverse period of the Euclidean time circle). After some standard manipulations with fermionic Green's functions and with Gamma matrices and, crucially, using (\ref{eq:zoomin}) to zoom into the singular part of the Green's function, one obtains
 \cite{Hartman:2010fk},
\be
F(r,r') = \mathrm{i} \int_{-\infty}^\infty \frac{k_{\mathrm{F}} \mathrm{d}k_\perp}{2 \pi} \int \frac{\mathrm{d} \Omega}{\pi} \tanh \frac{\Omega}{2T} \,  G^{\mathrm{R}}(\Omega,k_\perp)^* G^{\mathrm{R}}(-\Omega,k_\perp) |\psi_{0}(r)|^2 |\psi_{0}(r')|^2 \,. \label{eq:Frr}
\ee
\end{widetext}
Here $G^{\mathrm{R}}(\Omega,k_\perp)$ is the fermion propagator (\ref{eq:AdSRNprop}), at low energies and with momenta close to the Fermi surface, with the self energy generalized to the nonzero temperature expression (\ref{eq:IRTfinite}). The Green's functions describe the fermion loop sourced by two insertions of $\Delta$ (as specified in (\ref{eq:HS}) above). The momentum integral has also been restricted to the contribution close to the Fermi surface. The hyperbolic tangent arises from standard manipulations in which a sum over fermionic Matsubara frequencies is expressed as a contour integral. Finally, $\psi_0(r)$ is the radial profile of the normalizable fermion zero mode at $\omega = T = 0$ and $k = k_{\mathrm{F}}$. These last terms arise from the numerator of (\ref{eq:zoomin}).

The IR singular contribution to the integrals in (\ref{eq:Frr}) comes from the frequency range $T \ll \Omega \ll \mu$ and is readily evaluated. If $\nu > \half$ (as in the discussion of quantum oscillations, $\nu$ denotes the exponent $\nu_k$ at $k = k_{\mathrm{F}}$) one finds \cite{Hartman:2010fk}
\be
F(r,r') \sim |\psi_{0}(r)|^2 |\psi_{0}(r')|^2 \, \log \frac{T}{\mu}  \,.
\ee
This is essentially the standard BCS logarithmic IR divergence. As $T \to 0$, the logarithm leads to a negative mass squared term in the effective potential (\ref{eq:HS}) and condensation of $\Delta$. The instability onsets below the critical temperature at which the two terms in (\ref{eq:HS}) are equal. The critical temperature is given, very schematically, by $T_{\mathrm{c}} \sim \mathrm{e}^{- M_{\mathrm{F}}^2} \mu$. Recall that $M_{\mathrm{F}} \gg 1$ was the coefficient of the bulk interaction (\ref{eq:finteract}). From the perspective of the dual compressible phase of matter, this implies that the critical temperature is determined by the magnitude of a certain four point correlation function of the composite fermion operator $\Psi$ dual to $\psi$ in the bulk.

Recall that $\nu > \half$ corresponds to long-lived `quasiparticle' fermion excitations, according to the discussion in \S\ref{sec:fermionz} above. In contrast, in the non-quasiparticle cases where $\nu \leq \half$, no IR divergence is found in $F(r,r')$ and there is no pairing instability \cite{Hartman:2010fk}. There is not enough fermionic spectral weight at the Fermi surface in these strongly damped cases. In this instance, the non-Fermi liquid behavior impacts superconductivity negatively. This is ultimately due to the semi-holographic nature of the fermions. 
At leading order in large $N$, the only physics going into the non-Fermi liquid behavior is the strong decay 
into a large $N$ bath. There are no interesting vertex corrections or long range attractive forces at this leading order in $N$ level.

The condensation of $\Delta$ discussed above only impacts an order one part (that in bulk Fermi surfaces) of a large $N$ amount of charge (mostly behind the horizon). Most large $N$ observables, therefore, will not notice that the $\mathrm{U}(1)$ has been spontaneously broken. In contrast, in \S\ref{sec:holoS} below we will describe a distinct, fully holographic and leading order in large $N$ mechanism for holographic superconductivity. 

\paragraph{Corrections to the conductivity}

The charge carried by the bulk Fermi surface will, of course, move when an electric field is applied and will hence contribute to the conductivity.  The conductivity is obtained holographically as described around  (\ref{eq:optical}) above, by solving Maxwell's equations in the bulk.   More details for the following computation can be found in \cite{Faulkner:2010zz,Faulkner:2011tm, Faulkner:2013bna}.  The fermion contribution is a quantum correction to this process, whereby the fermions `screen' the bulk Maxwell field. That is, there is a fermion one loop correction to the Maxwell field propagator in the bulk. This one loop process is evaluated using standard methods, together with (\ref{eq:zoomin}) to zoom into the singular part of the Green's function, and the result takes the form \cite{Faulkner:2010zz, Faulkner:2013bna} \begin{widetext}
\be\label{eq:condd}
\sigma(\omega) = \frac{1}{\mathrm{i}\omega} \int \frac{d^2k}{(2\pi)^2} \frac{\mathrm{d}\omega_1}{2 \pi} \frac{\mathrm{d}\omega_2}{2 \pi} \frac{f(\w_1) - f(\w_2)}{\w_1 - \w_2 - \w - \mathrm{i} \epsilon} \text{Im} G^{\mathrm{R}}(\w_1,k) \, \text{Im} G^{\mathrm{R}}(\w_2,k) \,\Lambda(k)^2 \,.
\ee
\end{widetext}
Here $f(\w) = 1/(\mathrm{e}^{\w/T} + 1)$ is the Fermi-Dirac distribution function, $G^{\mathrm{R}}(\omega,k)$ is once again the fermion propagator (\ref{eq:AdSRNprop}), at low energies and with momenta close to the Fermi surface, with the self energy generalized to the nonzero temperature expression (\ref{eq:IRTfinite}), and $\Lambda(k)$ is a smooth function of $k$ that describes (\emph{i}) the vertex coupling the fermions to the Maxwell field, (\emph{ii}) the `bulk to boundary' propagator of the Maxwell field (and metric) and (\emph{iii}) integrals over the radial profile $\psi_0(r)$ of the normalizable fermion zero mode at $\omega = T = 0$ and $k = k_{\mathrm{F}}$. In general the function $\Lambda$ is frequency and temperature dependent, but these dependences drop out at low energies and temperatures and close to the Fermi surface.

The IR singular contribution of the integrals in (\ref{eq:condd}) is readily extracted. The low temperature dc conductivity is found to be \cite{Faulkner:2010zz, Faulkner:2013bna}
\be\label{eq:sigT}
\sigma \sim T^{- 2 \nu} \,,
\ee
where, as previously, $\nu = \nu_{k_{\mathrm{F}}}$ is the exponent at the Fermi momentum. The resistivity is linear in temperature when $\nu = \half$. The frequency-dependent conductivity can also be computed and is particularly interesting for the `non-quasiparticle' case (in the sense discussed in \S\ref{sec:fermionz} above) of $\nu < \half$. In these cases there is a broad non-Drude scaling feature in the conductivity for $T \ll \omega \ll \mu$ with \cite{Faulkner:2010zz, Faulkner:2013bna}
\be\label{eq:sigw}
\sigma(\w) \sim (\mathrm{i}\w)^{-2 \nu} \,.
\ee

As with previous bulk quantum effects, the contributions (\ref{eq:sigT}) and (\ref{eq:sigw}) will be strongly suppressed compared the leading order in large $N$ conductivities that will be the main topic of \S\ref{sec5} below. The computation of the fermion conductivity outlined above avoids many of the subtleties typically associated with computing conductivities in quantum critical theories. This is because of the semi-holographic coupling to a large $N$ critical bath. The fermions acquire nontrivial self-energies at leading order in $N$ due to dissipation into the black hole, but there are no vertex corrections at this leading order. Furthermore, because the momentum of the order one `probe' fermions is not conserved at large $N$ -- it is lost to the quantum critical bath -- one does not need to worry about the momentum-conservation delta functions that will be discussed at depth in \S\ref{sec5}.

\subsubsection{Endpoint of the near-horizon instability in the fluid approximation} \label{sec:thomasfermi}

We explained in \S\ref{sec:luttinger} above that the backreaction of bulk fermions on the geometry is most easily captured in the WKB approximation. The individual wavefunctions of the Dirac field become sufficiently localized in space that they are insensitive to variations in the metric, Maxwell or other fields. We noted that this amounts to a conflation of the Thomas-Fermi approximation in condensed matter physics with the Oppenheimer-Volkoff-Tolman approximation of astrophysics. Namely, the fermions are treated as a charged, gravitating ideal fluid. This limit was first studied in 
\cite{Hartnoll:2009ns} and further developed in \cite{Hartnoll:2010gu}. In this subsection we overview the physics of 
the resulting semiclassical `electron stars'.

In this subsection we will restrict attention to $d=2$ space dimensions. This follows most of the literature on this topic.
The wavefunctions of fermions in a background coming from the Einstein-Maxwell-dilaton action (\ref{eq:EMDaction}) enter a WKB regime when (e.g. \cite{Hartnoll:2011dm})
\be\label{eq:WKBlimit}
\frac{q e L}{\k} \gg 1 \,.
\ee
In this limit the mass of the fermion appears in Dirac equation in the combination $(\k m)/(q e)$. This ratio quantifies the relative strength of gravitational and electric interactions between fermions. For the ratio to be order one, the dimension of the dual QFT operator $\Delta \sim (mL)^2 \sim (q e L)^2/\k^2 \gg 1$. This large operator dimension is also the condition for the quantum wavelength of the bulk fermions to be small compared to the background AdS curvature scale. This latter condition however is not strictly implied by the limit (\ref{eq:WKBlimit}). Further discussion of the conditions for the WKB limit to hold can be found in \cite{deBoer:2009wk, Hartnoll:2010gu, Arsiwalla:2010bt, Allais:2013lha}. The WKB limit is not necessarily natural from the point of the view of the dual compressible phase of matter. Its function instead is to obtain a tractable bulk description of the fermion backreaction and some intuition for the resulting physics.

In \S\ref{sec:luttinger} we saw that every bulk Fermi surface comes with an associated bulk charge carried by fermions. For typical parameter values (order one charges and scaling dimensions) we have noted that the amount of charge carried by such bulk Fermi surfaces is subleading compared to charge behind the black hole horizon. In the large $N$ limit, then, the gravitational and electric effects of the backreaction of this charge is negligible. In addition, because these Fermi surfaces are semi-holographic and localized away from the event horizon, their backreaction does not alter the low energy quantum critical dynamics.

\begin{figure}
\centering
\includegraphics[height = 0.25\textheight]{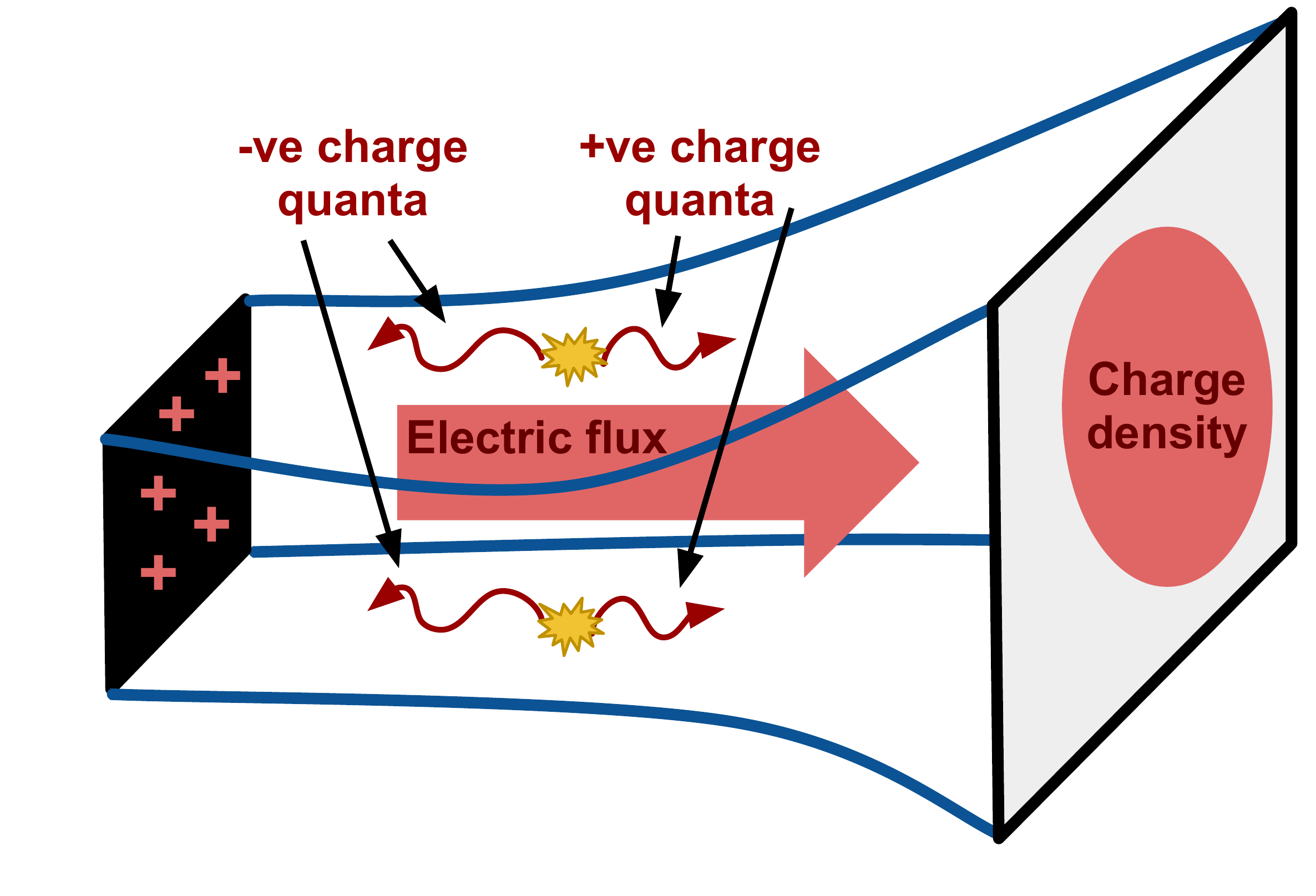}
\caption{\label{fig:gversuse} \textbf{Pair production in the near horizon region} leads to a discharging of the black hole if electromagnetic screening overcomes gravitational clumping. Figure taken with permission from \cite{Hartnoll:2011fn}.}
\end{figure}

The more dramatic effect of backreaction arises instead from the near-horizon instability of modes with imaginary scaling dimension under the IR critical scaling. From (\ref{eq:nukferm}) we see that such an instability occurs in the AdS-RN background (over some range of low momenta) whenever
\be\label{eq:instabF}
\k m  < q  e \,.
\ee
This inequality coincides with the condition for Schwinger fermion pair production to occur in $\mathrm{AdS}_2 \times \R^d$ \cite{Pioline:2005pf}. A fruitful way to think of (\ref{eq:instabF}) is as a competition between electromagnetic screening and gravitational anti-screening. Suppose a fermion anti-fermion pair is spontaneously produced in the near-horizon region. The electromagnetic interaction will act to screen the charge of the black hole: the fermion with an opposite sign charge to the horizon will be attracted towards the horizon, while the fermion with the same sign charge will be pushed away. This effect tends to discharge the black hole and populates the bulk outside of the black hole with charged fermions. The gravitational interaction, in contrast, famously anti-screens or clumps. Both of the charges will be gravitationally attracted towards the horizon and the pair production will have no net effect.  (\ref{eq:instabF}) therefore expresses the fact that the horizon will discharge if the charge of the fermions is large relative to their mass, so that the electromagnetic screening effect wins out. This process is illustrated in Figure \ref{fig:gversuse}.

The fate of the fermionic charge in the bulk can be determined, in the WKB approximation, by solving the Einstein-Maxwell-(dilaton) equations of motion coupled to a charged, gravitating fluid. At zero temperature and with no rotation, such a fluid can be described by a generalization \cite{Hartnoll:2010gu} of the Schutz action \cite{Schutz:1970my}
\begin{equation}
S = S_{\mathrm{EMD}} + \int \mathrm{d}^{d+2}x \sqrt{-g} \; P_{\mathrm{f}}(\mu_{\mathrm{f}}),
\end{equation}
where $\mu_{\mathrm{f}}$ is the local chemical potential:  \begin{equation}
\mu_{\mathrm{f}} = \frac{A_t}{\sqrt{-g_{tt}}},  \label{eq:muf}
\end{equation}
and $P_{\mathrm{f}}$ is the pressure of the fermions:  \begin{equation}\label{eq:pressure}
P_{\mathrm{f}}(\mu_\text{f}) = \int\limits_m^{q \mu_{\mathrm{f}}} \mathrm{d}E \; \nu(E) \, (q \mu_{\mathrm{f}} - E) \,,
\end{equation}
where $\nu(E) = E \sqrt{E^2 - m^2}/\pi^2$ is the density of states of the bulk fermions in flat space.
The equations of motion following from this action are those of an ideal fluid minimally coupled to the EMD theory.
The origin of the gravitational redshift in (\ref{eq:muf}) can be thought of as follows. To make the metric look locally flat, one should use the locally rescaled time $\mathrm{d}\hat t = \sqrt{-g_{tt}}\mathrm{d}t$; in this time coordinate, local thermal equilibrium will require $A_{\hat t} = \text{constant}$.  To include the effects of nonzero temperature and rotation in the fluid, see \cite{Schutz:1970my, Hartnoll:2010gu, Hartnoll:2011fn}.

From the expression (\ref{eq:pressure}) for the pressure, it is clear that a fluid of fermions will be present in the geometry wherever $q \mu_\text{f} > m$. In the semiclassical limit, this is where the local chemical potential is large enough to overcome the rest mass energy and populate a local Fermi surface. Near the asymptotically AdS boundary, since $g_{tt} \rightarrow \infty$ and $A_t \rightarrow \mu$, then $\mu_{\mathrm{f}}\rightarrow 0$.  Hence, at any nonzero mass $m$, the bulk fermion fluid can only appear towards the IR, if at all. In the case in which the fluid forms, it is therefore present beyond some radius $r_s$ at which $q \mu_\text{f}(r_s) = m$. This is the boundary of the electron star.

\begin{figure}
\centering
\includegraphics[height = 0.25\textheight]{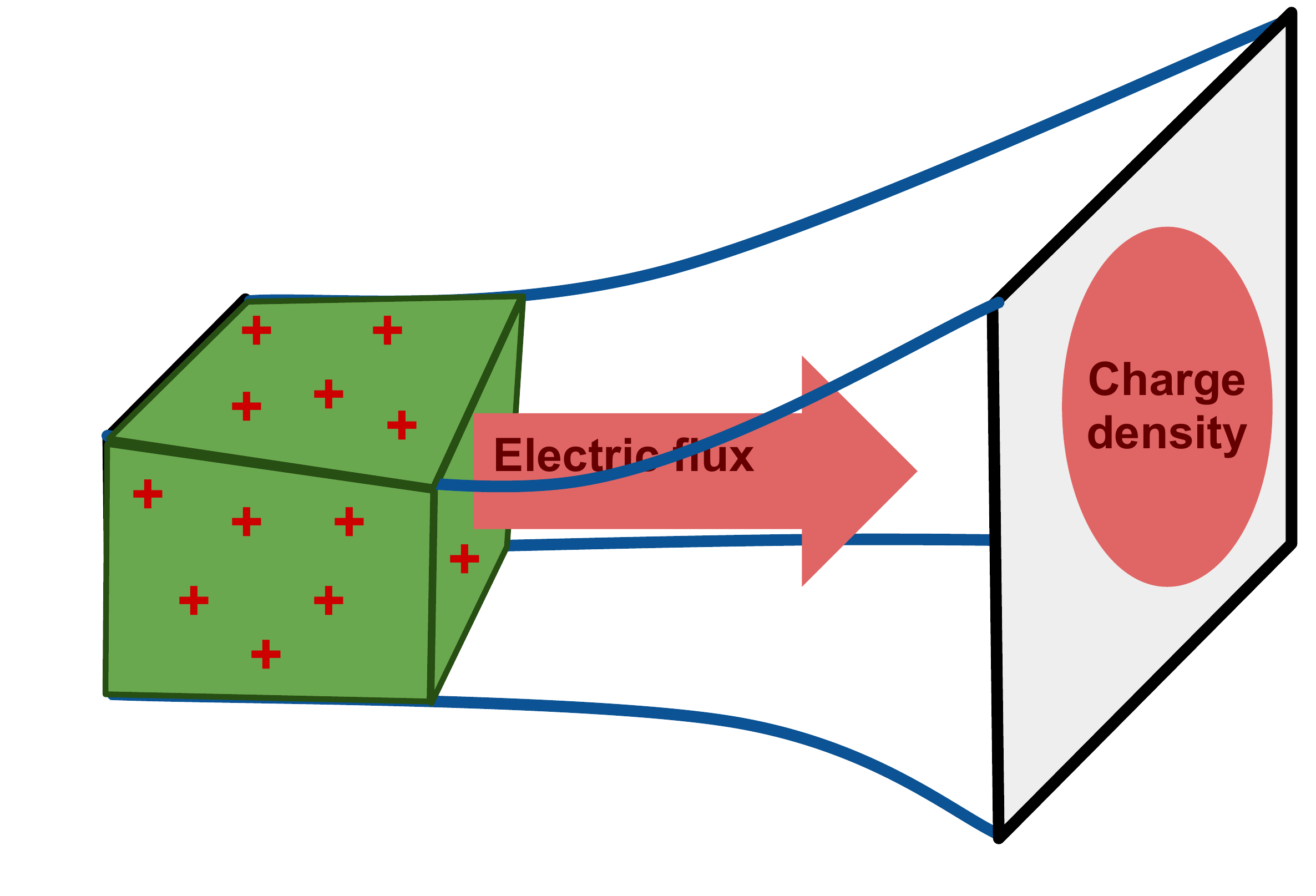}
\caption{\label{fig:star} \textbf{The electron star.} A charge, gravitating fluid is present in the spacetime for $\infty < r < r_s$. All the charge is carried by the cohesive fermion fluid. Figure taken with permission from \cite{Hartnoll:2011fn}.}
\end{figure}

The most important question for universal low energy physics is the fate of the geometry in the far interior, as $r \to \infty$. One can note that if there is an emergent Lifshitz scaling invariance (cf. (\ref{eq:Lif})) so that $g_{tt} \sim r^{-2z}$ and $A_t \sim r^{-z}$, then the local chemical potential (\ref{eq:muf}) remains constant as $r \to \infty$. If this constant is large enough compared to the mass, then this emergent scaling will be consistent with the presence of a fluid all the way to $r \to \infty$. Precisely this scenario is realized in Einstein-Maxwell theory, whenever the criterion (\ref{eq:instabF}) for instability is satisfied \cite{Hartnoll:2009ns, Hartnoll:2010gu}. A fluid extends from $\infty < r < r_s$, all the charge is carried by fluid, and the far interior is given by a Lifshitz geometry:
\begin{align}
\mathrm{d}s^2 &= L^2 \left( - \frac{\mathrm{d}t^2}{r^{2z}} + g_\infty \frac{\mathrm{d}r^2}{r^2} + \frac{\mathrm{d}\vec x^2_2}{r^2} \right) , \notag \\
A &= h_\infty \frac{\mathrm{d}t}{r^z} \,, \qquad P_{\mathrm{f}} = p_\infty \,. \label{eq:lifstar}
\end{align}
While the full solution is found numerically, the emergent scaling solution in the interior can be found analytically.
The value of $z$ depends on the mass and charge of the fermion, it can take any value from $1 < z < \infty$.
The solution is illustrated in Figure \ref{fig:star}.

The IR fixed point solution (\ref{eq:lifstar}) is somewhat remarkable as a gravitational background.
Typically a spatially uniform and static matter density is impossible in a theory of gravity, as the energy density causes the universe itself to contract. The solution (\ref{eq:lifstar}) gets around this in an intrinsically relativistic way: a scale-invariant gravitational redshift allows the gravitational and electric interactions to consistently and stably balance themselves.

The discussion so far has been at $T=0$. At sufficiently high temperatures, one expects the star to collapse to a black hole. That is because as the temperature is increased at fixed chemical potential, the mass of the black hole increases relative to its charge and hence gravitational clumping is increasingly favored over electric screening. Eventually, pair production is no longer possible and the high temperature black hole is stable. This scenario has been studied quantitatively in \cite{Hartnoll:2010ik, Puletti:2010de}. The main points are as follows. Firstly, at any $T>0$, then the local chemical potential $\mu_{\mathrm{f}} \rightarrow 0$ near the nonzero temperature horizon. This follows from the definition of $\mu_{\mathrm{f}}$ and the fact that near the horizon $A_t \sim g_{tt} \sim r-r_{\mathrm{h}}$. This behavior of $A_t$ at a horizon was noted below (\ref{eq:athor}) above. Thus, even at very low temperatures, the fermion fluid is pushed away from the horizon, and exists only in an intermediate region in the bulk $r_{\mathrm{s},2} < r < r_{\mathrm{s},1}$. The range of radii over which this `electron cloud' exists narrows as the temperature is increased, until the fluid disappears altogether in a third order phase transition at a critical temperature $T_{\mathrm{c}}$. This is, then, a temperature-driven fractionalization phase transition: At $T=0$ all charge is cohesive, at nonzero temperature some amount of the charge is fractionalized, and at $T> T_{\mathrm{c}}$, all of the charge is fractionalized.

Fractionalization transitions can also occur at zero temperature as parameters in the theory are varied. Quantum fractionalization transitions were studied in \cite{Hartnoll:2011pp} in a class of Einstein-Maxwell-dilaton theories coupled to a charged fluid. The novel feature compared to the Einstein-Maxwell case is that in cases where $Z(\Phi) \rightarrow \infty$ in the IR, then some of the charge remains on the horizon (and hence fractionalized, in the dual theory).   The EMD fields look similar to the theories described in \S\ref{sec:EMDgeom} in this case.  In particular,  note from (\ref{eq:EMDmetric42}) and (\ref{eq:pEMD}) that
\be\label{eq:muzero}
\mu_{\mathrm{f}} \sim r^{-d + \frac{\theta}{\theta-d}} \rightarrow 0 \,,
\ee
as $r\rightarrow \infty$ in the IR. This behavior -- a non-constant $\mu_{\mathrm{f}}$ in the scaling regime -- is a direct consequence of the anomalous scaling dimension of the charge density, as discussed in the first paragraph of \S \ref{sec:anomalouscharge}. According to (\ref{eq:muzero}), if a fermion fluid forms at all in the bulk,  it can only be over an intermediate range of radii. Depending on the details of the model, $\mu_{\mathrm{f}}$ may be so small that no Fermi surfaces are populated, and all the charge is fractionalized. The quantum phase transitions between fractionalized, partially fractionalized and cohesive compressible phases can be first order or continuous in these models.

As the bulk fermionic fluid describes cohesive boundary charge, then, according to the holographic Luttinger theorem (\ref{eq:bulklut}), there must be poles in fermionic Green's functions at nonzero momenta.  The Dirac equation was solved in the electron star backgrounds in \cite{Hartnoll:2011dm, Iqbal:2011in, Cubrovic:2011xm}. In the WKB limit (\ref{eq:WKBlimit}) a large number of closely spaced Fermi surfaces are present. These essentially correspond to the distinct bulk Fermi surfaces at each radius of the star. The electron star is therefore a distinctly holographic phase of matter, in which the distribution of Fermi surfaces captures the radial distribution of charge in the bulk. While the many Fermi surfaces should presumably be treated as a large $N$ `artifact', the corresponding smearing of the fermionic spectral weight over a range of momenta does have some interesting phenomenological consequences. For instance, the closely spaced Fermi surfaces lead to cancellations in the quantum oscillations of \S\ref{sec:qo} so that only a single `extremal' oscillation survives \cite{Hartnoll:2010xj}.


Observables such as the charge density spectral weight and electrical conductivities are computed in the same way as in other holographic backgrounds. Namely, by solving the bulk Maxwell equations in the presence of the charged fluid. 
These quantities turn out to be controlled by the properties of the IR scaling solutions such as (\ref{eq:lifstar}), much as in the fractionalized cases \cite{Hartnoll:2010gu, Gouteraux:2013oca}. The full hydrodynamic behavior of these geometries remains to be investigated.

\subsection{Magnetic fields}
\label{sec:magfields}

Magnetic fields are an important probe of compressible phases. The boundary QFT has a global $\mathrm{U}(1)$ symmetry, and we can put this
theory in a background (non-dynamical) magnetic field. Per the essential holographic dictionary applied to conserved currents in \S\ref{sec:global} above,
the source $A_\mu^{(0)}(x)$ for a current in the QFT is the non-normalizable mode of the bulk Maxwell field as $r \to 0$ at the boundary
\be
A_\mu(x,r) = A_\mu^{(0)}(x) + \cdots \,.
\ee
The gauge symmetry $A \to A + \mathrm{d}\l$ ensures that the leading behavior near the boundary is a constant in $r$. The boundary magnetic field is then
\be
B = F^{(0)}_{xy} = \pa_x A_y^{(0)} -  \pa_y A_x^{(0)} \,.
\ee
To determine the effects of such a magnetic field, we need to solve the bulk equations of motion subject to this boundary condition.
We will start by considering the case of $d=2$ spatial boundary dimensions, so that $F^{(0)}_{xy}$ is the only component of the boundary magnetic field. Later we will also consider the case of $d=3$, in which a magnetic field necessarily breaks isotropy by pointing in a specific direction.

\subsubsection{$d=2$: Hall transport and duality}
\label{sec:hall}

We will limit ourselves to constant magnetic fields so that we can write $A^{(0)} = B x \mathrm{d}y$. In $d=2$ dimensions we can then search for a bulk solution of the same form (\ref{eq:abmetric}) as previously, except that now we allow the bulk Maxwell field to take the more general form
\be
A = A_t(r) \mathrm{d}t + B(r) x \mathrm{d}y \,.
\ee 
The near boundary behavior of $A_t(r)$ controls the boundary chemical potential and charge density following (\ref{eq:atnearb}) and (\ref{eq:Eandrho}) above, whereas the near boundary behavior of $B(r)$ controls the boundary magnetic field. Note a certain asymmetry: the asymptotic electric flux in the bulk is an expectation value in the dual QFT (charge density) whereas the asymptotic magnetic flux in the bulk is a source in the QFT (background magnetic field).

An especially simple solution can be found in ($d=2$) Einstein-Maxwell theory (\ref{eq:fullEM}). This is the dyonic AdS-RN solution. The Maxwell potential takes the form
\be
A = \mu \left[ 1 - \frac{r}{r_+} \right] \mathrm{d}t+ B x \mathrm{d}y \,,
\ee
generalizing (\ref{eq:athor}) above. The charge density $\rho$ is still given by (\ref{eq:rrr}). The metric still takes the form (\ref{eq:gRN}), but now with the redshift given by
\begin{align}
f(r) &= 1 - \left(1 + \frac{r_+^2 \mu^2 + r_+^4 B^2}{\g^2} \right) \left(\frac{r}{r_+} \right)^3 \notag \\
&+\frac{r_+^2 \mu^2 + r_+^4 B^2}{\g^2}  \left(\frac{r}{r_+} \right)^4 \,.
\end{align}
Thus we see that the radial dependence of the metric is virtually unchanged from the purely electric case. Similarly to the purely electric case, the $T = 0$ limit of this solution has an $\mathrm{AdS}_2 \times \R^2$ near horizon geometry, very similar to (\ref{eq:nhmetric}) above, and associated zero temperature entropy. Details of the thermodynamics of this solution can be found in \cite{Hartnoll:2007ai, hkms}. In particular, the magnetic susceptibility $\chi = \pa^2 P/\pa B^2$  remains a smooth negative function at $T=0$, showing no quantum oscillations as a function of the magnetic field and qualitatively rather similar to the magnetic susceptibility for free bosons \cite{Denef:2009yy}. Recall that we discussed possible subleading in $1/N$ quantum oscillations in \S\ref{sec:qo} above.

A magnetic field introduces qualitatively new features into transport. In particular, because the magnetic field breaks time reversal invariance, Hall conductivities are now allowed. Hall conductivities are zero in a time reversal invariant and isotropic state because then the retarded Green's function
\begin{align}
\langle [J_i(t), J_j(0)] \rangle &= \langle \mathcal{T}  [J_i(t), J_j(0)] \mathcal{T} \rangle^* = \langle [J_j(0), J_i(-t)] \rangle \notag \\
&= \langle [J_j(t), J_i(0)] \rangle \,,
\end{align}
which contradicts the fact that in an isotropic system one must have $\langle [J_i(t), J_j(0)] \rangle \sim \epsilon_{ij}$, unless $\langle [J_i(t),J_j(0)]\rangle = 0$. We will proceed to discuss aspects of the electrical conductivities following from the dyonic AdS-RN solution of Einstein-Maxwell theory above. We follow the presentation of \cite{Hartnoll:2007ip}.

With time reversal invariance broken, there is a matrix of electrical conductivities
\be
\left(
\begin{array}{c}
J_x \\
J_y
\end{array}
\right) = \left(
\begin{array}{cc}
\s_{xx} & \s_{xy} \\
- \s_{xy} & \s_{xx}
\end{array}
\right)
\left(
\begin{array}{c}
E_x \\
E_y
\end{array}
\right) \,.
\ee
We have assumed isotropy. This information is usefully packaged into a complex conductivity. If we define
\be
E_\pm \equiv E_x \pm \mathrm{i} E_y \,, \qquad J_\pm \equiv J_x \pm \mathrm{i} J_y \,,
\ee
we can then write Ohm's law as
\be\label{eq:ohmpm}
J_\pm = \mp \mathrm{i} \sigma_\pm E_\pm \,, \quad \text{where} \qquad \sigma_\pm = \sigma_{xy} \pm \mathrm{i} \sigma_{xx} \,.
\ee
Note that in frequency space, $\sigma_{xx}(\omega)$ and $\sigma_{xy}(\omega)$ are already complex functions. The rate of work done by an electric field in the presence of Hall conductivities is
\be
\dot w = \int \mathrm{d}\omega E_i^*(\omega) \chi''_{ij} E_j(\omega) \,,
\ee
where
\be
\chi''(\omega) =  \left(
\begin{array}{cc}
\text{Re}\, \s_{xx}(\omega) & \mathrm{i} \, \text{Im} \, \s_{xy}(\omega) \\
- \mathrm{i} \, \text{Im} \, \s_{xy}(\omega) & \text{Re} \, \s_{xx}(\omega)
\end{array}
\right) \,.
\ee
This result is obtained from manipulations similar to those leading to (\ref{eq:dissip}) above.

Transport in compressible phases will be the topic of \S\ref{sec5}. We will be able to discuss aspects of transport in magnetic fields prior to our more general exposition below. This is because it turns out that a magnetic field avoids the complications related to conservation of momentum that will be a central issue later. Technically this is due to fact while the magnetic field itself is uniform, the vector potential (such as $A^{(0)} = B x \mathrm{d}y$) necessarily breaks translation invariance.  As a consequence, there is a Lorentz force on the resulting system which violates momentum conservation.

As in \S\ref{sec:QCcharge} above, the conductivities $\sigma_{ij}(\omega)$ are obtained by perturbing the bulk Maxwell field about the dyonic black hole background above.
Because of the background charge density and magnetic field, the perturbations of the Maxwell field couple to metric perturbations. At $k=0$, a self-consistent set of modes to perturb are $\delta A_x, \delta A_y, \delta g_{tx}, \delta g_{ty}$.
Let us define the bulk electric and magnetic fields to be \begin{subequations}
\bea
e_i & \equiv & \pa_t \delta A_i - \frac{B L^2}{r^2} \epsilon_{ij} \delta g_{tj} \,, \\
b_i & \equiv & - f(r) \epsilon_{ij} \pa_r \delta A_j \,.
\eea
\end{subequations}
If we define
\be
e_\pm = e_x \pm \mathrm{i} e_y \,, \qquad b_\pm = b_x \pm \mathrm{i} b_y \,,
\ee
then the linearized Einstein-Maxwell equations about the dyonic background can be written as \cite{Hartnoll:2007ip} \begin{subequations}
\bea
\omega \pa_r e_\pm + \frac{\omega^2}{f(r)} b_\pm & =  & \frac{4 r^2}{\g^2} \left(B^2 b_\pm \pm e^2 \r B e_\pm\right) \,, \label{eq:eb1} \\
\omega \pa_r b_\pm - \frac{\omega^2}{f(r)} e_\pm & = & -  \frac{4 r^2}{\g^2} \left(e^4 \r^2 e_\pm \pm e^2 \r B b_\pm\right) \,.  \label{eq:eb2}
\eea
\end{subequations}
These two equations are exchanged under electromagnetic duality:
\be\label{eq:emB}
e_\pm \to b_\pm\,, \quad b_\pm \to - e_\pm\,,\quad e^2 \rho \to B\,,\quad B \to - e^2 \rho \,,
\ee
This transformation will be important shortly.

As usual, we must solve the above equations of motion subject to infalling boundary conditions at the horizon. The conductivities (\ref{eq:ohmpm}) are then extracted from the boundary behavior according to
\be\label{eq:spm}
\sigma_\pm = \lim_{r \to 0} \frac{1}{e^2} \frac{b_\pm}{e_\pm} \,.
\ee
This expression comes from the fact that the boundary current $e^2 \langle J_\pm \rangle = \mp i \lim_{r \to 0} b_\pm$, from the definitions above and the holographic dictionary (\ref{eq:jx1}). The boundary value of $e_\pm$ is precisely the boundary electric field $E_\pm$. An immediate consequence of the expression (\ref{eq:spm}) is that electromagnetic duality (\ref{eq:emB}) implies that
\be\label{eq:dualitysigma}
2 \pi \sigma_+^{(e^2 \rho,B)}(\omega) = \frac{-1}{2 \pi \sigma_+^{(B, - e^2 \rho)}(\omega)} \,.
\ee
That is: if we exchange the charge and magnetic field, the complex conductivity (\ref{eq:ohmpm}) is inverted. This is of course a generalization of the
particle-vortex duality (\ref{eq:sinvert}) to the case with magnetic field, charge density and Hall conductivities. The factors of $2\pi$ appear because under the duality the electromagnetic coupling is inverted as $2\pi/e^2 \to e^2/2\pi$. We will see this duality from a more hydrodynamic perspective in \S\ref{sec:magtrans} below.

In fact, the duality map (\ref{eq:dualitysigma}) is part of a larger $\mathrm{SL}(2,\Z)$ group of dualities. See \cite{Witten:2003ya} and references therein. Electromagnetic duality is the $S$ generator of the group. The $T$ generator is understood by adding a topological theta term to the bulk Einstein-Maxwell theory (\ref{eq:fullEM}) in $d=2$:
\be\label{eq:Stheta}
S_\theta = \frac{\theta}{8 \pi^2} \int F \wedge F \,.
\ee
Here $\theta$ is a constant. This term has no effect on the bulk dynamics. Its only effect is that under a shift $\theta \to \theta + 2\pi$, the Hall conductivity shifts so that
\be
2 \pi \sigma_+ \to 2 \pi \sigma_+ + 1\,.
\ee
This follows from the boundary contribution that (\ref{eq:Stheta}) makes to the on-shell action bulk action, which determines the current expectation value according to (\ref{eq:vevfull}).

To evaluate the conductivities, one must solve the equations of motion (\ref{eq:eb1}) and (\ref{eq:eb2}). The $\omega = 0$ limit is especially simple; $b_+ = - e^2 \rho/B e_+$ and hence from (\ref{eq:spm}), as first found in \cite{Hartnoll:2007ai},
\be
\sigma_{xy} = \sigma_+ = \frac{\rho}{B} \,.
\ee
This is a general result that will be rederived in \S\ref{sec:magtrans} from relativistic hydrodynamics.

A second limit in which the equations of motion can be solved analytically is at small frequencies, with $\rho^2 \sim B^2 \sim \omega$ also taken to be small. The small frequency expansion is implemented similarly to in \S\ref{sec:diffusive} above, by first factoring out the non-analytic frequency dependence of the infalling boundary condition at the horizon,
\be
e_+(r) = \mathrm{e}^{\mathrm{i} \omega \int^r \frac{\mathrm{d}s}{f(s)}} \left(s_0(r) + \omega s_1(r) + \cdots \right) \,.
\ee
Solving for $s_0(r)$ and $s_1(r)$ one finds \cite{Hartnoll:2007ip} from (\ref{eq:spm}) that
\begin{align}\label{eq:aabb}
e^2 \sigma_+ &= \frac{b_+(0)}{e_+(0)} = \frac{1}{\omega} \frac{e_+'(0)}{e_+(0)} = \frac{s_1'(0)}{s_0(0)} + \mathrm{i} \notag \\
&= \mathrm{i} \frac{\omega + \mathrm{i} \omega_c^2/\Gamma + \omega_c}{\omega + \mathrm{i} \Gamma - \omega_c} \,.
\end{align}
That is, there is a cyclotron pole with frequency and decay rate given by
\be\label{eq:ccdd}
\omega_c - \mathrm{i} \Gamma = \frac{\rho B}{\epsilon + P} -\mathrm{i} \frac{1}{e^2} \frac{B^2}{\epsilon + P} \,.
\ee
It is simple to check that the expression (\ref{eq:aabb}) is consistent with the duality (\ref{eq:dualitysigma}).
This collective cyclotron mode can also be obtained from hydrodynamics, as discussed in \S\ref{sec:magtrans} below. We will elaborate more on its physics in that later section and will give the separate expressions for $\sigma_{xx}$ and $\sigma_{xy}$. The cyclotron mode is just the longest lived of many quasinormal modes that extend into the lower half complex frequency plane. In fact, the analytic structure of the conductivity obtained in this case is similar to that shown in Figure \ref{fig:WeylQNM} above. Now the magnetic field and charge density are the parameters that deform away from the self-dual point. Poles and zeros alternate and are exchanged under electromagnetic duality.

For a discussion of how the physics described above generalizes to Einstein-Maxwell-dilaton theories, see \cite{Goldstein:2010aw, Lindgren:2015lia}.

\subsubsection{$d=3$: Chern-Simons term and quantum phase transition}
\label{sec:CSterm}

The above results have been for $d=2$. The case of $d=3$ has been shown to contain rich physics. The role of charge density $\rho$ and magnetic field $B$ are quite different in this case, and a quantum phase transition can occur as a function of the dimensionless ratio $B/\rho^{2/3}$. This system has been studied in a series of papers including \cite{D'Hoker:2009mm, D'Hoker:2009bc, D'Hoker:2010rz, D'Hoker:2010ij}. These papers and others are summarized in \cite{D'Hoker:2012ih}. We will note a few salient results.

Firstly, in $4+1$ bulk dimensions (corresponding to $d=3$) there is a natural Chern-Simons term that often appears in Einstein-Maxwell theory. This is
\be\label{eq:AFF}
S_k = \frac{2 k}{3 \k^2} \int A \wedge F \wedge F \,.
\ee
This interaction term is not a total derivative, and alters the Einstein-Maxwell equations. It leads to an $F \wedge F$ term on the right hand side of Maxwell's equations. Thus the electromagnetic field itself now carries charge. This fact leads to instabilities in this theory as will be discussed in \S\ref{sec:holoS} below. It allows leads to the possibility of cohesive phases within Einstein-Maxwell-Chern-Simons theory, with all charge being carried by the bulk electromagnetic fields. The coupling (\ref{eq:AFF}) is also closely connected to an anomaly in the global symmetry dual to the bulk Maxwell field \cite{Witten:1998qj}.  Such relativistic anomalies are not unheard of in condensed matter systems, as we will see in \S \ref{sec:weylSM}.

We will not write down the form of the bulk fields. These are still homogeneous -- only depending on a radial coordinate $r$ -- but no longer isotropic, because the magnetic field singles out a direction. This leads to a greater number functions of $r$ (metric components etc.) that must be solved for. Let us describe the $T=0$ solutions for different values of $B/\rho^{2/3}$. The asymptotic near boundary geometry is fixed to be $\mathrm{AdS}_5$. The most important question for low energy observables, as we have seen, is the near horizon geometry. At $B = 0$ the near horizon geometry is just the $\mathrm{AdS}_2 \times \R^3$ of the five dimensional planar AdS-RN solution. However, at $\rho = 0$ the near horizon geometry is $\mathrm{AdS}_3 \times \R^2$ \cite{D'Hoker:2009mm}. This asymmetry arises because in the electrically charged case the bulk Maxwell field is in the $t,r$ directions (that define the $\mathrm{AdS}_2$) while in the magnetic case the bulk Maxwell field is in the $x,y$ directions, that define the $\R^2$. Given these different IR geometries, it is clear that a phase transition will need to occur at intermediate values of $B/\rho^{2/3}$. Indeed this is the case \cite{D'Hoker:2010rz, D'Hoker:2010ij}, and the phase diagram is shown in Figure \ref{fig:kd} below.

This phase diagram is best understood when the Chern-Simon coupling in (\ref{eq:AFF}) is sufficiently large, $k>1/2$. In this case the $T=0$ near-horizon geometry is known at all magnetic fields above the critical field \cite{D'Hoker:2010ij}. It is given by
\begin{align}\label{eq:nhs}
& \mathrm{d}s^2 = \frac{L^2}{3}\frac{\mathrm{d}r^2 - 2 \mathrm{d}x^+ \mathrm{d}x^- }{r^2} \notag \\ & - \left(\frac{\a_{\mathrm{o}}}{r^2} + \frac{2 e_{\mathrm{o}}^2}{k(2k-1)} \frac{1}{r^{4k}} \right) (\mathrm{d}x^+)^2 + \mathrm{d}x^2 +\mathrm{d}y^2 \,.
\end{align}
This geometry is supported by an electric field with
\be
A_+ = \frac{e_{\mathrm{o}}}{k} \frac{1}{r^{2k}} \,.
\ee
Now, for $\a_{\mathrm{o}} > 0$, the second term in brackets in (\ref{eq:nhs}) is subleading in the far interior as $r \to \infty$. In this case the near horizon geometry reduces to $\mathrm{AdS}_3 \times \R^2$ written in null coordinates, as in the pure magnetic case. There is no electric flux through the horizon. The quantum critical point, however, is characterized by a value of the magnetic field at which $\a_{\mathrm{o}} = 0$ in the near-horizon geometry (\ref{eq:nhs}). In this case the second term in the brackets must be kept and the scaling symmetry of the IR geometry is changed to
\begin{align}
r \to \l r \,, \quad x^+ &\to \l^{2k} x^+ \,, \quad x^- \to \l^{2-2k} x^- \,, \notag \\
 \{x,y\} &\to \{x,y\} \,.
\end{align}
The critical IR geometry is of the Schr\"odinger form mentioned briefly in \S\ref{sec:gal} above, and may be the most microscopically well-grounded and simple of the known realizations of this geometry in string theory backgrounds. The IR geometry determines the temperature scaling in the quantum critical regime  \cite{D'Hoker:2010ij, D'Hoker:2012ih}. Note that the quantum critical point separates two phases that are in themselves critical (i.e. gapless).

The phase transition in Figure \ref{fig:kd} is `metamagnetic', that is to say, no symmetries are broken across the transition. Instead, this transition is a fractionalization quantum phase transition, analogous to those discussed for charged fermions towards the end of \S\ref{sec:thomasfermi} above. Below the critical magnetic field, some of the asymptotic electric field emanates from behind the horizon, corresponding to fractionalized charge, and the rest is sourced by the bulk electromagnetic field via the Chern-Simons interaction (\ref{eq:AFF}), corresponding to cohesive charge. Above the critical magnetic field, all of the charge is cohesive and there is no electric flux through the horizon.

The low temperature thermodynamics of these solutions admits a physical interpretation \cite{D'Hoker:2012ih}. At large magnetic fields, the behavior $s \sim T$ describes the excitations of a chiral CFT that propagate in the direction of the magnetic field. At small magnetic fields there remains a zero temperature ground state entropy density. While a compelling connection has yet to be made, it is worth noting that a ground state entropy density does in fact arise for free relativistic fermions in a magnetic field: the `zero point' energy of the lowest Landau level is precisely cancelled by the Zeeman splitting energy gain of the spin up electrons. Therefore, the entire lowest Landau level is at zero energy.

\begin{figure}
\centering
\includegraphics[height=6cm]{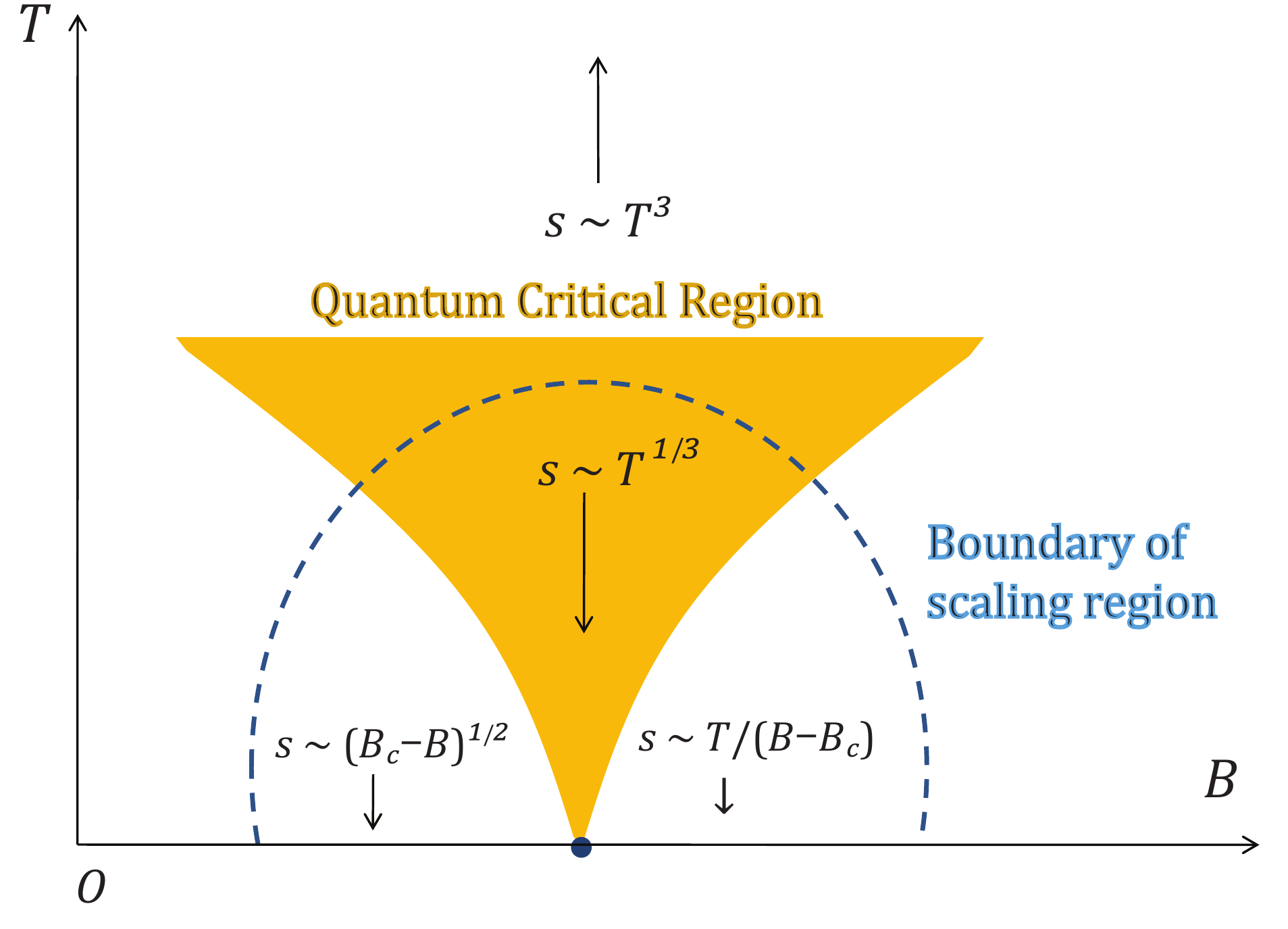}
\caption{\label{fig:kd} \textbf{Phase diagram of Einstein-Maxwell-Chern-Simons theory} as a function of temperature and magnetic field, at fixed nonzero charge density. The Chern-Simons coupling $k>3/4$ (for $3/4>k>1/2$ the temperature scaling of $s$ in the critical region is different). Figure taken with permission from \cite{D'Hoker:2012ih}. This phase diagram ignores certain spatial modulation instabilities that will be discussed in \S\ref{sec:holoS}.}
\end{figure}

\section{Metallic transport without quasiparticles}
\label{sec5}

\subsection{Metallic transport with quasiparticles}
\label{sec5a}

We begin this section with a historical perspective on metallic transport in metals {\it with\/} quasiparticle excitations. Essentially all
of the condensed matter literature on quantum transport in metals is built on a theory of the scattering of quasiparticles formulated using the
quantum Boltzmann equation, and its Baym-Kadanoff-Keldysh extensions \cite{baymkad,kamenev}. 

In the simplest and most common cases, the scattering of quasiparticles is dominated by elastic scattering off impurities in the crystal. 
We can then define an impurity mean-free path, $\ell_{\mathrm{qp:imp}}$, and a corresponding impurity scattering time
$\tau_{\mathrm{qp:imp}}$, related by $\ell_{\mathrm{qp:imp}} = v_{\mathrm{F}} \tau_{\mathrm{qp:imp}}$, where $v_{\mathrm{F}}$ is the Fermi velocity
(the cumbersome notation is needed to distinguish other impurity-related times and lengths we define in subsequent subsections).
It is assumed that the elastic impurity scattering time is much shorter than the quasiparticle lifetime from inelastic processes, 
including those due to electron-phonon ($\tau_{\mathrm{qp:phonon}}$) and electron-electron
($\tau_{\mathrm{qp:ee}}$) interactions
\beq
\tau_{\mathrm{qp:imp}}\, \ll \, \tau_{\mathrm{qp:phonon}}, \tau_{\mathrm{qp:ee}}\,.
\label{eq:tau}
\eeq
Such a theory of quasiparticle transport has a number of well-known characteristics, amply verified by experimental observations:
\begin{enumerate}
\item
The low temperature resistivity $\rho = \rho_0 + A T^2$, where $\rho_0$ is the impurity-induced `residual' resistivity at zero temperature,
and $A$ arises from electron-electron interactions which relax momentum, often including umklapp processes.
\item The ratio of the electrical, $\sigma$, and thermal, $\kappa$, conductivities obeys the Wiedemann-Franz law
\beq
\frac{\kappa}{T \sigma} = \frac{\pi^2}{3} \frac{k_{\mathrm{B}}^2}{e^2} \,. \label{eq:WF}
\eeq
\item Identifying a quasiparticle requires a length scale larger than the typical wavelength of a quasiparticle excitation $\sim k_{\mathrm{F}}^{-1}$.
Consequently, the value of $\ell_{\mathrm{qp:imp}}$ is only meaningful when it is larger that $k_{\mathrm{F}}^{-1}$. The
inequality, $k_{\mathrm{F}} \ell_{\mathrm{qp:imp}} > 1$, then leads to the proposal
of the Mott-Ioffe-Regel lower bound on the conductivity of a metal \cite{gunnarsson}
\beq\label{eq:MIR}
\sigma \sim \frac{e^2}{h} k_{\mathrm{F}}^{d-1} \ell_{\mathrm{qp:imp}}\,  > \, \frac{e^2}{h} k_{\mathrm{F}}^{d-2} \,.
\eeq
We will return to this bound in \S \ref{sec:in1}.
\end{enumerate}

\subsection{The momentum bottleneck}
\label{sec:mombottle}

Our interest here is in metals without quasiparticles where, as we will describe below, all three of the quasiparticle transport characteristics highlighted
above do not apply. But it is instructive to first discuss their breakdown in a situation which arises already within the quasiparticle framework.

Consider a clean metal where $\tau_{\mathrm{qp:imp}}$ is very long, and the dominant scattering of electronic quasiparticles is off
the phonons \cite{ziman}. Indeed, just such a situation was considered in the classic work by Bloch \cite{Bloch1,Bloch2} in 1929, 
in which he derived `Bloch's law',
stating that at low $T$ the resistivity of metals from electron-phonon scattering varies with temperature as $T^{d+2}$.
However, in 1930, Peierls wrote insightful papers \cite{Peierls1,Peierls2} pointing out a crucial error in Bloch's argument. In the electron-phonon scattering event, 
the momentum carried by the quasiparticles is deposited into the phonons, and the total
momentum of the electron+phonon system is conserved (umklapp events freeze out at low $T$). So after multiple scattering events, the combined electron+phonon system
will reach a state of maximum entropy consistent with the conservation of total momentum. In a generic state of quantum matter 
at non-zero density, the two-point correlator between the momentum and current is non-zero (although, such a correlator vanishes in CFTs),
and consequently such an equilibrated state has a non-zero electrical current, proportional to the initial total momentum. 
So the electrical current does not decay to zero,
and the resistivity vanishes. This is the `phonon drag' effect, in which the mobile quasiparticles drag the phonons along with them, and the combined
system then flows without decay of the electrical current.

In practice, however, it is found that Bloch's law works well in most metals, and Peierls' phonon drag has turned out to be difficult to observe.
This is due to a combination of two factors: ({\em i\/}) the electron-phonon coupling is weak, and so it takes a long time
for the electron-phonon system to reach local thermal equilibrium; ({\em ii\/}) impurities are invariably present in experiments,
and they can absorb the momentum deposited into the phonons before the electron-phonon system has equilibrated.
We need metals with exceptional purity, and with appreciable electron-phonon (or electron-electron) interactions, to observe
the long-lived flow of electrical current in a thermally equilibrated electronic system with momentum conservation.
Only recently have such experiments become possible \cite{Molenkamp95,Girvin96,Bandurin1055,Moll16}.

We turn, finally, to metals without quasiparticle excitations. We can imagine reaching such a state by turning up the strength
of the electron-electron interactions in a quasiparticle system, and so reducing the value of $\tau_{\mathrm{qp:ee}}$.
In a moderately clean sample, (\ref{eq:tau}) will be violated before the quasiparticles become ill-defined. Consequently,
the `drag' mechanism applies forcefully to metals without quasiparticles, and we should describe transport in terms of a quantum fluid
which has locally thermally equilibrated. A number of well-known computations of non-Fermi liquid transport \cite{Lee89,IoffeWiegmann90,IoffeKotliar90,Metzner07}
use a Boltzmann-Baym-Kadanoff-Keldysh framework
to describe scattering of charged excitations around a Fermi surface by their strong coupling to a neutral bosonic excitation,
and then relate this scattering rate to the transport co-efficients. However, these computations ignore the fact that the bosonic excitations
will rapidly equilibrate with the charged fermionic excitations in a short time of order $\tau_{\mathrm{eq}} \sim 1/T$, as we discussed in \S\ref{sec:qmwqp}, and so yield incorrect
results for the transport coeffecients for the models studied \cite{MaslovChubukov11}. 


The holographic perspective on strange metals has the advantage of naturally building in the momentum bottleneck. Such
a formulation of transport makes no direct reference to the Fermi surface, or quasiparticles of any kind, and instead describes a thermally equilibrated fluid
of fermions and bosons which obeys all the important global conservation laws. There are important connections between the holographic
approach to such liquids, and more traditional hydrodynamic and memory function approaches. We will review these connections
below, and show that such methods lead to a unified theory of metals without quasiparticles with mutual consistency in overlapping
regimes of validity. We also note that hydrodynamic perspectives on electronic transport have been suggested for ultra-clean metals
with well-defined quasiparticles \cite{Gurzhi63,Gurzhi68,KS06,andreev2011}.

A further possibility afforded by non-quasiparticle transport is that momentum may also be strongly degraded. This occurs if the effects of disorder or lattice scattering are large. With quasiparticles, this readily results in localization. Without quasiparticles, one seems more likely to enter an incoherent metallic regime instead (see \S \ref{sec:in1}). Incoherent metals are again locally equilibrated and hence hydrodynamic, but there is no (approximate) sound mode and hence the only collective motion of charge is diffusive. Incoherent metals can have large resistivities, in contrast to `momentum drag' transport in which the conductivity is large. They are thus a natural framework for describing `bad metals' that violate the Mott-Ioffe-Regel bound (\ref{eq:MIR}), as discussed in \cite{Hartnoll:2014lpa}.
Incoherent metals appear readily in holographic theories, 
and will be discussed in \S \ref{sec:in2} and \S \ref{sec:in3}; explicit quantum matter realizations of incoherent metals are harder to find, and are present in the
SYK models discussed in \S \ref{sec:SYK}.

\begin{figure}[h]
\centering
\includegraphics[height = 0.23\textheight]{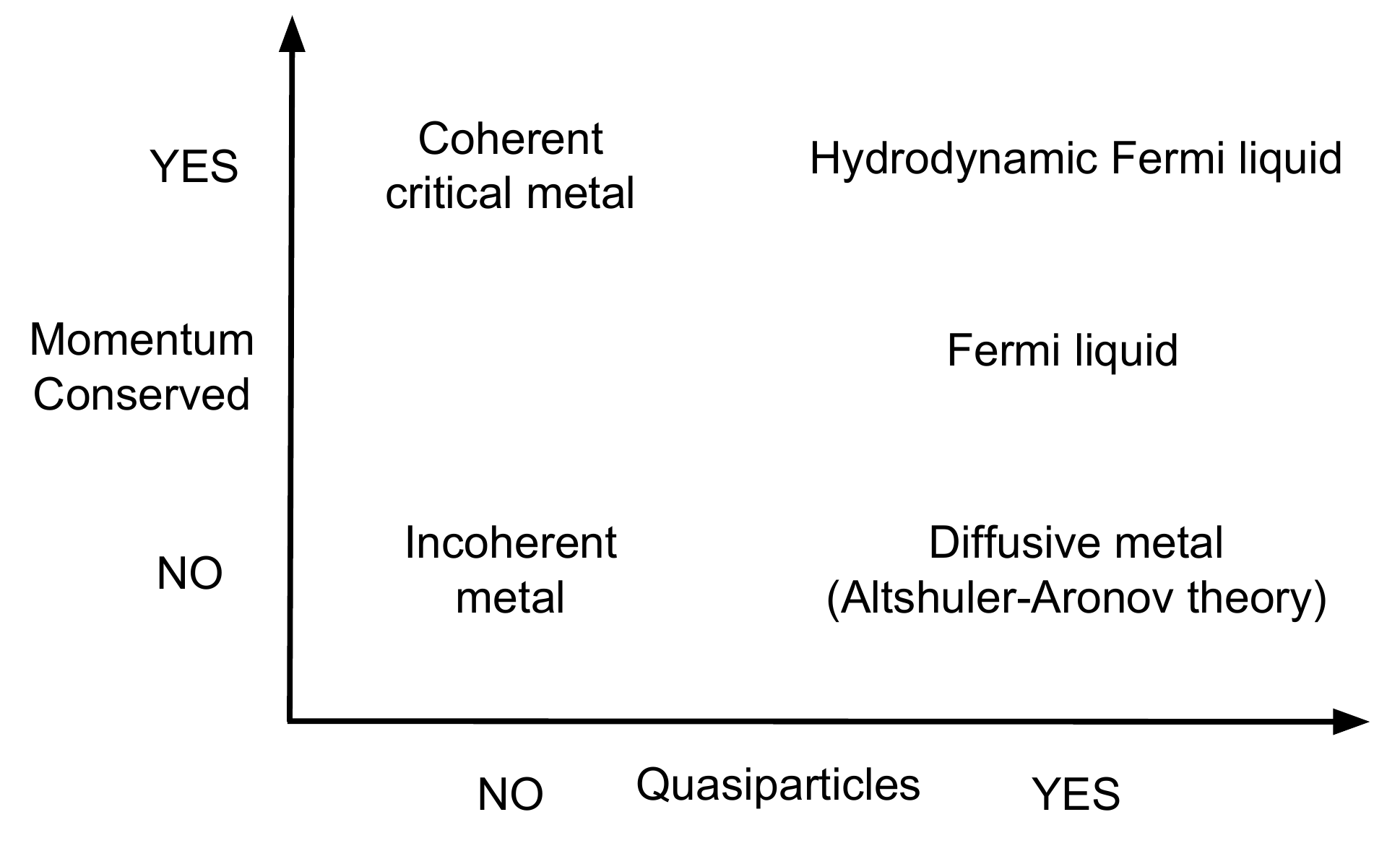}
\caption{\label{fig:allmetals} \textbf{A `phase space' of metals.} In this section we will discuss the coherent and incoherent non-quasiparticle metals that appear on the left column. Figure developed in discussion with Aharon Kapitulnik.}
\end{figure}

Figure \ref{fig:allmetals} summarizes how the various regimes discussed in the previous paragraphs are related to each other.

In the first half of the remaining section, we will detail the major developments in the theory of transport in strongly interacting metals with weak momentum relaxation.  The latter part of this section consists of holographic approaches to transport that are non-perturbative in the strength of momentum relaxation. We postpone our review of non-holographic condensed matter models to \S\ref{sec:transcondmat} and \S \ref{sec:SYK}, until after we have had a chance to lay out the formalisms which are relevant for the discussion.

\subsection{Thermoelectric conductivity matrix}

Electric and thermal transport generally couple together in charged quantum matter.   Hence, we will want to compute not just the electrical conductivity $\sigma$, as in \S\ref{sec:qctransport}, but a more general matrix of thermoelectric conductivities.    We will need to consider the transport coefficients
\begin{equation}
\left(\begin{array}{c} J^i \\ Q^i  \end{array}\right) = \left(\begin{array}{cc} \sigma^{ij} &\ T\alpha^{ij} \\ T\bar\alpha^{ij} &\ T \bar\kappa^{ij} \end{array}\right) \left(\begin{array}{c} E_j \\ \zeta_j  \end{array}\right) \,, \label{eq:thermoelectric}
\end{equation} 
where we have defined the heat current \begin{equation}
Q^i \equiv T^{ti} - \mu J^i.
\end{equation}
The heat current naturally couples to a temperature gradient,
\begin{equation}
\delta \zeta_i \equiv -\frac{\partial_i T}{T} \,,
\end{equation}
as we will demonstrate below, following \cite{Hartnoll:2009sz}. The classic discussion can be found in \cite{PhysRev.135.A1505}. The matrix of thermoelectric conductivities defined above is guaranteed to be  symmetric, assuming time-reversal symmetry -- this is called Onsager reciprocity.

We want to impose a uniform electric field $\delta E_i$, and a uniform temperature gradient $\delta \zeta_i$. For simplicity, we suppose that the field theory lives on flat space.  An electric field is imposed through an external gauge field
\begin{equation}
\delta A = \frac{\mathrm{e}^{-\mathrm{i}\omega t}}{\mathrm{i}\omega} \delta E_i  \mathrm{d}x^i.   \label{eq:Aperturb}
\end{equation}
Imposing a temperature gradient is more subtle.  A clean way to do this is to think about Euclidean time in the rescaled coordinate $\tilde t$: \begin{equation}
t = \frac{\tilde t}{T}.
\end{equation}
This Euclidean coordinate has periodicity $\tilde t \sim \tilde t + 1$, while the metric
\begin{equation}
g_{\tilde{t}\tilde{t}} = \frac{1}{T^2}.
\end{equation}
Imposing a constant temperature gradient $\delta \zeta_i$ therefore leads to
\begin{equation}
\delta g_{\tilde{t}\tilde{t}} = \frac{2 \mathrm{e}^{-\mathrm{i}\tilde \omega \tilde t}}{T^2} \delta \zeta_i x^i \, ,
\end{equation}
with $\tilde\omega = \omega/T$.  We now make a coordinate change \begin{equation}\label{eq:cchange}
\tilde{t} \rightarrow \tilde{t}+ \xi^{\tilde{t}}, \qquad   \xi^{\tilde t} = \mathrm{i}x^i \delta \zeta_i \frac{ \mathrm{e}^{-\mathrm{i}\tilde{\omega}\tilde{t}}}{\tilde\omega T^2} \,.
\end{equation}
In general, the effect of an infinitesimal coordinate change is to generate the perturbation
\begin{align}
\delta g_{\mu\nu} &\rightarrow \delta g_{\mu\nu} + \partial_\mu \xi_\nu + \partial_\nu \xi_\mu , \notag \\
\delta A_\mu &\rightarrow \delta A_\mu + \delta A_\nu \partial_\mu \xi^\nu + \xi^\nu \partial_\nu \delta A_\mu .
\end{align}
Thus the coordinate change (\ref{eq:cchange}) leads to the following set of perturbations describing the temperature gradient: 
\begin{equation}
\delta g_{\tilde{t}\tilde{t}}=0,\;\; \delta g_{\tilde{t}i} = -\mathrm{i} \delta \zeta_i \frac{ \mathrm{e}^{-\mathrm{i}\tilde{\omega}\tilde{t}}}{\tilde\omega T^2},\;\; \delta A_i = \mathrm{i} \delta\zeta_i \frac{\mathrm{e}^{-\mathrm{i}\tilde{\omega}\tilde{t}} }{\tilde\omega T^2}A_{\tilde{t}} \, .  \label{eq:zetaperturb}
\end{equation}

Let us now collect our results and rescale back to the original time coordinate $t$.   If we perturb the metric and gauge field, using (\ref{eq:Aperturb}) and (\ref{eq:zetaperturb}) the action of the field theory is deformed to \begin{align}
\delta S &= \int \mathrm{d}^{d+1}x \left[T^{\mu\nu}\delta g_{\mu\nu} + J^\mu \delta A_\mu \right] \notag \\
&= \int \mathrm{d}^{d+1}x \left[\frac{J^i}{\mathrm{i}\omega} \delta E_i + \frac{Q^i}{\mathrm{i}\omega} \delta \zeta_i   \right] \,. \label{eq:TEcoupling}
\end{align}
Hence, temperature gradients indeed couple to the heat current, and the natural thermoelectric transport coefficients to discuss are given by (\ref{eq:thermoelectric}).  In holography, we have also learned how to encode a thermal gradient by perturbing $g_{ti}$ and $A_i$ at the boundary.   For simplicity assuming that the boundary is asymptotically AdS: \begin{subequations}\begin{align}
\delta g_{ti}(r=0) &= \frac{L^2}{r^2} \frac{\mathrm{e}^{-\mathrm{i}\omega t}}{\mathrm{i}\omega} \delta \zeta_i, \\
\delta A_{i}(r=0) &= - \frac{\mathrm{e}^{-\mathrm{i}\omega t}}{\mathrm{i}\omega} \delta \zeta_i A_t(r=0).
\end{align}\end{subequations}

The coupling (\ref{eq:TEcoupling}) implies that the thermoelectric conductivities (\ref{eq:thermoelectric}) are given by retarded Green's functions of the currents $J^i$ and $Q^i$. Recall that these Green's functions relate sources and expectation values, as in e.g. (\ref{eq:easyG}) above. The appearance of time derivatives (factors of $\omega$) in (\ref{eq:TEcoupling}) means that some care is required in the manipulation of the Green's functions. The answer is:
\begin{subequations}\begin{align}
\s^{ij}(\omega) &= \frac{G^{\mathrm{R}}_{J^i J^j}(\omega) - \chi_{J^i J^j}}{ \mathrm{i} \omega}, \\
\bar \k^{ij}(\omega) &= \frac{G^{\mathrm{R}}_{Q^i Q^j}(\omega) - \chi_{Q^i Q^j}}{ \mathrm{i} \omega T}, \label{eq:skG} \\
\a^{ij}(\omega) &= \frac{G^{\mathrm{R}}_{Q^i J^j}(\omega) - \chi_{Q^i J^j}}{ \mathrm{i} \omega T}, \\
 \bar \a^{ij}(\omega) &= \frac{G^{\mathrm{R}}_{J^i Q^j}(\omega) - \chi_{J^i Q^j}}{ \mathrm{i} \omega T} \,. \label{eq:aaG}
\end{align}\end{subequations}
The susceptibilities $\chi_{AB}$ are equal to the nonzero temperature Euclidean Green's function at Matsubara frequency $\omega_n = 0$. The susceptibilities in the above expressions are often zero by gauge invariance at $T>0$ (because they would generate mass terms if the theory is coupled to dynamical photons or gravitons). We will return to these quantities later. The derivation of (\ref{eq:skG}) and (\ref{eq:aaG}) from (\ref{eq:TEcoupling}) can be found in \cite{PhysRev.135.A1505}. In \cite{PhysRev.135.A1505} the answer is given in terms of Kubo functions. We will relate these to the quantities appearing in  (\ref{eq:skG}) and (\ref{eq:aaG}) later.

There is a zoo of thermoelectric coefficients used in the condensed matter literature.   All such coefficients are related to the matrices defined in (\ref{eq:thermoelectric}), and are defined by changing the ``boundary conditions" (instead of measuring $J_i$ and $Q_i$ given $E_i$ and $\zeta_i$, we choose to fix $J_i$ and $\zeta_i$, for example).   Let us note two famous ones for convenience.  The thermal conductivity at vanishing electric current is:
\begin{equation}
\left.Q_i\right|_{J_i=0} \equiv T\kappa_{ij} \zeta_j \;\; \Rightarrow \;\; \kappa_{ij} = \bar\kappa_{ij} - T\bar\alpha_{ik}(\sigma^{-1})_{kl}\alpha_{lj}.   \label{eq:kappadef}
\end{equation}
The Seebeck coefficient $\mathfrak{s}_{ij}$ is the ratio of the voltage drop to the temperature drop, again when $J_i=0$: \begin{equation}
\left.E_i \right|_{J_i=0} \equiv -\mathfrak{s}_{ij}T\zeta_j \;\; \Rightarrow \;\; \mathfrak{s}_{ij} = (\sigma^{-1})_{ik}\alpha_{kj} \, .
\end{equation}
Observables with $J_i = 0$ arise naturally in experimental situations that typically work with open circuit boundary conditions. Charge accumulates at one end of the sample and the resulting electric field precisely cancels out any net electric current.

\subsection{Hydrodynamic transport (with momentum)}
\label{sec:hydromomentum}

Before discussing holographic transport, it is important to understand what features of holography are genuinely novel, and which features are already expected on general field theoretic grounds.   In fact, we will see that the transport problem is tightly constrained by hydrodynamics.

Hydrodynamics is the effective theory describing the relaxation of an interacting classical or quantum system towards thermal equilibrium.   
The key assumption of hydrodynamics is that the field theory has locally reached thermal equilibrium.  In thermal equilibrium, we assign to a QFT with conserved charge and energy-momentum a chemical potential $\mu$, and a temperature $T$ and four-velocity $u^\mu$,  defined such that the local density matrix $\rho$ is ``approximately" \begin{equation}
\rho \sim \exp\left[\frac{u^\mu P_\mu + \mu Q}{T}\right] \,,   \label{rho52}
\end{equation}
with $P_\mu$ the total energy-momentum in a volume of linear size $l_{\mathrm{th}}$, where $l_{\mathrm{th}}$ is the ``thermalization length scale", and $Q$ the charge.   More precisely, (\ref{rho52}) is true so long as we only ask for the expectation value of products of local operators.   The equations of hydrodynamics describe the dissipative dynamics under which a theory with long wavelength inhomogeneity in $\mu$, $T$ and $u^\mu$ slowly relaxes to global equilibrium (or as close to it as boundary conditions allow).

Hydrodynamics characterizes the dynamics perturbatively in powers of $l_{\mathrm{th}}/\xi$, where $\xi$ is the scale over which there is spatial variation.   We will focus, to start with, on the case of relativistic dynamics, and so after trivial multiplication by the (effective) speed of light $c$ the same considerations apply to time scales:   the local thermalization time is fast compared to the scale of hydrodynamic phenomena.  The hydrodynamic expansion is achieved by expanding $T^{\mu\nu}$ and $J^\mu$ as functions of $T$, $\mu$ and $u^\mu$, and their derivatives.   We will perform this task explicitly in the next subsection.   But beforehand let us emphasize that the hydrodynamic limit is parametrically opposite to the standard limit of quantum field theory computations -- $\xi/l_{\mathrm{th}}$ crudely counts the ``number of collisions" (in a quasiparticle framework), which apparently must be large for the hydrodynamic limit to be sensible.   Hence, recovery of hydrodynamics from quasiparticle approaches such as kinetic theory requires some conceptual `care', although it is well-known how to do so \cite{Arnold:1997gh, Arnold:2000dr}.  In kinetic theory, dissipative coefficients such as diffusion and viscosity are non-perturbative in the coupling constant $\lambda$, typically scaling as $1/\lambda^2$ as $\lambda\rightarrow 0$.

\subsubsection{Relativistic hydrodynamics near quantum criticality}
\label{sec:relativhydro}

We start by considering relativistic hydrodynamics. This is constrained by additional symmetries, and  is directly applicable to a CFT deformed by temperature and chemical potential. As we have discussed, this is a natural framework for holography and for compressible matter more generally.

The equations of motion of hydrodynamics read \begin{subequations}\begin{align}
\partial_\mu T^{\mu\nu} &= F^{\mu\nu}J_\mu, \label{eq:dmuTmu}\\
\partial_\mu J^\mu &=  0,  \label{eq:dmujmu}
\end{align}\label{eq:heom}\end{subequations}
where $T^{\mu\nu}$ is the expectation value of the local stress tensor,  $J^\mu$ is the charge current, and $F^{\mu\nu}$ is an external electromagnetic field tensor.   Our goal is to construct $T^{\mu\nu}$ and $J^\mu$ for a relativistic theory, following \cite{hkms}.    

The expansion in $l_{\mathrm{th}}/\xi$ will be an expansion in derivatives $\partial_\mu$.    So we begin by constructing $T^{\mu\nu}$ and $J^\mu$ at zeroth order in derivatives.   The most general possible answer is 
\begin{subequations}\begin{align}
T^{\mu\nu} &=  (\epsilon+P)u^\mu u^\nu + P \eta^{\mu\nu}, \label{eq:tmunu0} \\
J^\mu &= \rho \, u^\mu.  \label{eq:jmu0}
\end{align}\label{eq:orderzero}\end{subequations}
We were only able to use $\eta^{\mu\nu}$ and $u^\mu$ to construct these tensors, since $F_{\mu\nu} = \partial_\mu A_\nu - \partial_\nu A_\mu$ is first order in derivatives.  In the rest frame of the fluid, we readily interpret $\epsilon$ as the energy density, $P$ as the pressure, and $\rho$ as the charge density.    

There is a further conservation law at this order in derivatives: \begin{equation}
\partial_\mu s^\mu = 0,   \label{eq:dmusmu}
\end{equation}
where \begin{equation}
s^\mu = s u^\mu \,,
\end{equation}
and $s$ is the entropy density.   We can derive (\ref{eq:dmusmu}) by using that at leading order in derivatives, taking the divergence of (\ref{eq:tmunu0}), \begin{equation}
-\partial^\mu P = u^\mu u^\nu \partial_\nu (\epsilon+P) + (\epsilon+P) u^\mu \partial_\nu u^\nu + (\epsilon+P) u^\nu \partial_\nu u^\mu.
\end{equation}
Contracting with $u_\mu$ and employing $u_\mu u^\mu = -1$ and the fact that $\mathrm{d}P = \rho \mathrm{d}\mu + s \mathrm{d}T$, we obtain: 
\begin{equation}
\partial_\mu \left((\epsilon+P)u^\mu\right) =  u^\nu\partial_\nu P =  \rho u^\nu \partial_\nu \mu + s u^\nu \partial_\nu T.
\end{equation}
Further employing (\ref{eq:dmujmu}), (\ref{eq:jmu0}), along with the thermodynamic identity (\ref{eq:gibbsduhem}), we obtain (\ref{eq:dmusmu}).   (\ref{eq:dmusmu}) asserts that entropy is conserved at leading order, and is the second law of thermodynamics for a non-dissipative system.  At higher orders in hydrodynamics, we will require that a certain entropy current satisfy \begin{equation}
\partial_\mu s^\mu \ge 0,  \label{eq:hydro2law}
\end{equation}
so that the fluid is dissipative, and entropy increases over time locally.

The positivity constraints coming from (\ref{eq:hydro2law}) are important. However, the need to construct an entropy current is a weakness of the conventional formulation of hydrodynamics, as identifying the correct entropy current can be subtle. While we will follow the conventional description \cite{LLfluid}, the reader may be interested in more modern treatments that dispense of (\ref{eq:hydro2law}) as an additional postulate \cite{Jensen:2012jh, Banerjee:2012iz, Haehl:2015foa, Crossley:2015evo}.

The next step is to construct the first order derivative corrections to $T^{\mu\nu}$ and $J^\mu$.   A priori, we have to add  terms containing all possible combinations of $\partial_\mu \mu$, $\partial_\mu T$ and $\partial_\mu u_\nu$.   However, we immediately run into an ambiguity with how to define the fluid variables in an inhomogeneous (out of equilibrium) background \cite{Kovtun:2012rj}.   This is the issue of picking a ``fluid frame":   a simultaneous re-definition of $u^\mu$, $T$ and $\mu$ at first order in derivatives is allowed.   We will work in the Landau frame, which chooses \begin{equation}\label{eq:landauframe}
0 = u_\nu T_{(1)}^{\mu\nu} = u_\nu J_{(1)}^\nu \,,
\end{equation}
where $T_{(1)}^{\mu\nu}$ and $J_{(1)}^\nu$ are the first order in derivative terms in the conserved currents.     

Now, we use the equations of motion to construct the most general $T^{\mu\nu}_{(1)}$ and $J^\mu_{(1)}$ consistent with positivity of entropy production (\ref{eq:hydro2law}) and the Landau frame conditions.   We follow the steps we used above to derive the zeroth order conservation of entropy (\ref{eq:dmusmu}), but now include first order corrections.   We find \begin{equation}
\partial_\mu \left(s u^\mu\right) = \frac{\mu}{T}\partial_\nu J^\nu_{(1)} - \frac{1}{T}F^{\mu\nu}u_\mu J_{(1)\nu} - \frac{1}{T}T^{\mu\nu}_{(1)} \partial_\mu u_\nu,
\end{equation}
and hence
\begin{align}
& \partial_\mu \left(s u^\mu - \frac{\mu}{T} J^\mu_{(1)}\right) \notag \\
& = -J^\nu_{(1)} \left[\partial_\nu \left(\frac{\mu}{T}\right) + \frac{1}{T}F^{\mu\nu}u_\mu\right] - \frac{1}{T}T^{\mu\nu}_{(1)} \partial_\mu u_\nu. \label{eq:ttt}
\end{align}
The entropy current at first order in derivatives is the quantity appearing on the left hand side of the above equation,
\begin{equation}
T s^\mu =  s T u^\mu - \mu J_{(1)}^\mu \,.
\end{equation}
This is just the heat current, up to a factor of $T$:  $Q^\mu  =Ts^\mu$.
With this identification of $s^\mu$, the most general form of $J_{(1)}^\nu$ and $T_{(1)}^{\mu\nu}$ consistent with the second law (\ref{eq:hydro2law}) and the Landau frame is, from (\ref{eq:ttt}),
 \begin{subequations}\begin{align}
J_{(1)}^\mu &= -\sigma_{\textsc{q}} \mathcal{P}^{\mu\nu}\left(\partial_\nu \mu - \frac{\mu}{T} \partial_\nu T + F_{\rho \nu}u^\rho\right), \\
T_{(1)}^{\mu\nu} &= -\eta \mathcal{P}^{\mu\rho}\mathcal{P}^{\nu\sigma}\left( \partial_\rho u_\sigma + \partial_\sigma u_\rho- \frac{2}{d}\eta_{\rho\sigma} \partial_\lambda u^\lambda \right) \notag \\
& \;\;\;\; - \zeta\mathcal{P}^{\mu\nu} \partial_\rho u^\rho \,,
\end{align}\label{eq:consti}\end{subequations}
where $\sigma_{\textsc{q}}$, $\eta$ and $\zeta$ are all positive, and we have defined the projector \begin{equation}
\mathcal{P}^{\mu\nu} = \eta^{\mu\nu} + u^\mu u^\nu \,.
\end{equation}
$\eta$ and $\zeta$ are the shear and bulk viscosity of the fluid.   $\sigma_{\textsc{q}}$ is a ``quantum critical conductivity" which plays a very important role in transport, as we will see. In the zero density limit, $\sigma_{\textsc{q}}$ reduces to the ``relativistic'' conductivity described in \S\ref{sec:qctransport}. The equations (\ref{eq:consti}) are called the first order constitutive relations. The zeroth order constitutive relations were (\ref{eq:orderzero}). Combining the constitutive relations with the conservation laws (\ref{eq:heom}) leads to a set of dynamical equations for the conserved densities $\epsilon,\rho,u^\mu$.

As we have emphasized previously, the black hole geometries employed in holography are dual to field theories with consistent thermodynamics. Therefore, it must be that when these black holes are perturbed on very long wavelengths, hydrodynamic behavior arises. In fact, there is a mapping between solutions of hydrodynamics and those of charged black branes, perturbed on very long length scales. We will not discuss this ``fluid-gravity correspondence" further. For original papers see \cite{Bhattacharyya:2008jc, Banerjee:2008th, Erdmenger:2008rm}, and for a review see \cite{Hubeny:2011hd}.

\subsubsection{Sound waves}\label{sec:sound}

With the hydrodynamic equations set up, the next step is to solve the equations and find the hydrodynamic modes. We got a taste of this in \S \ref{sec:diffusive}, where we found a diffusive mode carrying charge fluctuations in an overall neutral system. What, however, are the hydrodynamic modes of a charged fluid? The most important feature of charged hydrodynamics in a medium with a conserved momentum is the presence of sound waves that carry charge. We will see that these modes have a huge effect on charge transport.

To find the collective modes of linear transport, we linearize the hydrodynamic equations about a fluid at rest:  \begin{equation}
u^\mu = (1, \delta v^i), \;\;\;\; \mu = \mu_0+\delta\mu,\;\;\;\; T = T_0+\delta T.  \label{eq:deltav}
\end{equation}
To linear order in the perturbations, the stress tensor and charge current are
\begin{align}
& \delta J^t = \delta \r, \;\;\;\; \delta J^i = \r \, \delta v^i, \;\;\;\; \delta T^{tt} = \delta \epsilon,\\
 & \delta T^{ij}  = \delta^{ij}\left( \partial_\epsilon P \delta \epsilon + \partial_\r P \delta \r\right), \;\;\;\;  \delta T^{ti} = (\epsilon+P)\delta v^i. \notag
\end{align}
Following \cite{Kovtun:2012rj}, we have used $\delta \epsilon$ and $\delta \r$ as hydrodynamic variables instead of $\delta \mu$ and $\delta T$. Assuming that $\delta v^i$, $\delta \epsilon$ and $\delta \r$ have $x$ and $t$ dependence given by $\mathrm{e}^{\mathrm{i} (k x - \omega t)}$, it is straightforward to use the conservation laws (\ref{eq:heom}) and the constitutive relations (\ref{eq:orderzero}) and (\ref{eq:consti}) to obtain a set of algebraic equations.  The condition that these equations admit non-vanishing solutions defines the hydrodynamic normal modes. 

The hydrodynamic modes are easily found. The modes decompose into transverse and longitudinal sectors.
The perpendicular components of the velocity field obey a diffusion equation with dispersion relation \begin{equation}
\omega = -\mathrm{i} \frac{\eta}{\epsilon+P} k^2.  \label{eq:momdiff}
\end{equation}
That is, the shear viscosity controls transverse momentum diffusion. The longitudinal modes are as follows:  there is a diffusive mode with with dispersion relation $\omega = -\mathrm{i}Dk^2$, with diffusivity 
\begin{widetext}\begin{equation}
D = \sigma_{\textsc{q}} \frac{(\partial_\epsilon P)_\rho[ (\partial_\rho \mu)_\epsilon - (\mu/T)(\partial_\rho T)_\epsilon] - (\partial_\rho P)_\epsilon [ (\partial_\epsilon \mu)_\rho - (\mu/T)(\partial_\epsilon T)_\rho ]}{(\partial_\epsilon P)_{\rho/s}}. \label{eq:Dinc}
\end{equation}
\end{widetext}
This is the generalization of the diffusive mode derived in \S \ref{sec:diffusive}. This `incoherent' mode does not carry pressure or momentum and is discussed in some detail in \cite{Davison:2015taa}. Finally, there is a sound mode \begin{equation}
\omega = \pm v_{\mathrm{s}} k - \mathrm{i}\Gamma_{\mathrm{s}} k^2\label{eq:soundwave} \,,
\end{equation}
with speed of sound
\begin{equation}
v_{\mathrm{s}} = \sqrt{(\partial_\epsilon P)_{\rho/s}} \,,
\end{equation}and attenuation constant
\begin{align}
\Gamma_{\mathrm{s}} & = \frac{1}{\epsilon+P}\left(\frac{2d-2}{d}\eta + \zeta\right) \notag \\
& + \frac{\sigma_{\textsc{q}}(\partial_\rho P)_\epsilon}{v_{\mathrm{s}}^2}\left[(\partial_\epsilon \mu)_{\rho/s} - \frac{\mu}{T}(\partial_\epsilon T)_{\rho/s}\right] \,.
\end{align}
The sound mode carries momentum, heat and charge.

In holographic models, the sound modes can be found explicitly by perturbing the background black hole. The general computational strategy is identical to that in \S \ref{sec:diffusive} above, in which we found the diffusive mode of a neutral black hole. The sound modes are also low lying quasinormal modes of the black hole. The technical difficulty is that there are now more bulk fields that are coupled and it is more difficult to find gauge-invariant variables analogous to (\ref{eq:gaugeinvar}) that can lead to (ideally decoupled) second order differential equations. It is often useful to fix the gauge $\delta g_{rM}=0$ and $\delta A_r = 0$. To find the diffusive momentum mode (\ref{eq:momdiff}), one must then solve for the bulk metric perturbations $\delta g_{ty}$ and $\delta g_{xy}$. For the remaining longitudinal modes one must couple $\delta A_t$, $\delta A_x$, $\delta g_{ii}$, $\delta g_{xx}$, $\delta g_{tx}$, $\delta g_{tt}$, and any further scalar modes such as a dilaton.

Once the fluctuation equations have been obtained, the procedure is the same as in \S \ref{sec:diffusive}. That is, one has to integrate the equations from the horizon to the boundary and demand the absence of a source at the boundary. This integration can be done perturbatively in $\omega ,k \rightarrow 0$, with $\omega/k$ fixed. The analysis was done for neutral AdS-Schwarzschild black holes in \cite{Policastro:2002tn, Herzog:2003ke, Kovtun:2005ev} and for charged AdS-RN black holes in \cite{Edalati:2010pn, Davison:2011uk}. Modes are found at precisely the frequencies (\ref{eq:soundwave}) predicted by hydroynamics. Decoupled gauge-invariant equations of motion for both longitudinal and transverse channels have also been found for some Einstein-Maxwell-dilaton theories in \cite{Anantua:2012nj}, although the sound modes have not been studied.

Hydrodynamics requires $\omega,k \ll T$ in order for the derivative expansion to hold. Perhaps surprisingly, holographic sound modes were found to exist even at $T=0$ in the AdS-RN black hole \cite{Davison:2013bxa}, and obey the expected dispersion relations. While reminiscent of `zero sound' modes in Fermi liquids, the raison d'\^etre of these gapless collective modes is not clear at the time of writing. A linearly dispersing collective mode at zero temperature has also been found in an EMD model without a ground state entropy, although still with $z=\infty$ \cite{Davison:2013uha}. It would be interesting to see if the existence of these modes could be directly tied to the presence of low energy, nonzero momentum spectral weight when $z=\infty$. The two properties (spectral weight and zero sound modes) are tied together in weakly interacting Fermi surfaces. We shall come across similar modes in \S\ref{sec:zeros} below on probe brane models.

\subsubsection{Transport coefficients}
\label{sec:transcoeff}

We are now almost ready to use hydrodynamics to compute the transport coefficients $\{\s,\a,\bar\a,\bar \k\}$ defined in (\ref{eq:thermoelectric}).   However, there is an immediate problem which arises:  at nonzero $\rho$ and $s$, the transport coefficients are all divergent as $\omega\rightarrow 0$ in any fluid with translation invariance! We shall prove this result in due course. To get a feel for where it comes from, consider the following easy way to generate nonzero currents $J_i$ and $Q_i$:  start with a fluid at rest, $u^\mu = (1,0,\ldots)$, and perform a small `Galilean' boost of velocity $v_x$.  Going to a moving frame leads to electric and heat currents $J_x = \rho v_x$ and $Q_x =  s T v_x$, respectively.   But there is no temperature gradient and no electric field.    Evidently, all of the coefficients in (\ref{eq:thermoelectric}) are divergent in any direction with translation invariance.  This is not inconsistent with our results for $\sigma$ from \S\ref{sec:qctransport}, because for those systems $\rho=0$.  While it might appear from this argument that boost rather than translation invariance is the symmetry at work here, we will see shortly that this is not the case. Boost invariance gives extra structure: it fixes the coefficients of the delta functions in equations (\ref{eq:Cleansigma2}) -- (\ref{eq:hkmskappa}) below.

Nonetheless, let us now compute $\sigma(\omega)$ at $\omega>0$ from hydrodynamics;  for simplicity, we assume the theory is isotropic.  We shall do this first by directly solving the hydrodynamic equations of motion in the presence of a uniform electric field and temperature gradient. We will freely use the conservation equations (\ref{eq:heom}) and constitutive relations (\ref{eq:orderzero}) and (\ref{eq:consti}). Consider first a uniform electric field $E$, so $F^{xt} = -E$, which is infinitesimally small.   This will induce a perturbatively small velocity field $v$, and hence a perturbatively small momentum current $T^{tx} = (\epsilon+P)v$. Integrating the Lorentz force (\ref{eq:dmuTmu}) over space gives\begin{equation}
\partial_t T^{ti} = -\mathrm{i}\omega (\epsilon+P)v = E\rho,   \label{eq:HKMStau2}
\end{equation}
since $J_t = -\rho$ up to spatial derivatives.  Solving (\ref{eq:HKMStau2}) for $v$ and using the leading order constitutive relation
\begin{equation}
J^x = \rho v + \sigma_{\textsc{q}}E \,,
\end{equation}
we obtain \begin{equation}
\sigma(\omega) = \sigma_{\textsc{q}} + \frac{\rho^2}{\epsilon +P } \frac{\mathrm{i}}{\omega}.  \label{eq:Cleansigma}
\end{equation}
In fact, exactly at $\omega=0$, analyticity properties of $\sigma(\omega)$ -- in particular the fact $\sigma(\omega)$ is a retarded Green's function (\ref{eq:skG}) and hence analytic in the upper half plane and subject to the Kramers-Kronig relations \cite{Hartnoll:2009sz} -- demand that there be a $\delta$ function: \begin{equation}
\sigma(\omega) = \sigma_{\textsc{q}} + \frac{\rho^2}{\epsilon +P } \left[\frac{\mathrm{i}}{\omega} + \pi \delta(\omega)\right].   \label{eq:Cleansigma2}
\end{equation}
We will see in later sections more ways to understand the emergence of this $\delta$ function.
Using the constitutive relation for $Q_x$, we also obtain \begin{equation}
\alpha(\omega) = -\frac{\mu}{T} \sigma_{\textsc{q}}+ \frac{s \r}{\epsilon +P }  \left[\frac{\mathrm{i}}{\omega} + \pi \delta(\omega)\right].  \label{eq:hkmsalpha}
\end{equation}

To compute $\bar\kappa$ we need an expression for the force due to a thermal gradient.  A thermal 
`drive' is equivalent to a background metric and Maxwell field according to (\ref{eq:zetaperturb}).
The hydrodynamic equation of motion (\ref{eq:dmuTmu}) is placed in a nontrivial background metric by making the derivatives covariant and contracting indices with the background metric. In a background given by flat spacetime perturbed by (\ref{eq:zetaperturb}) one obtains
\begin{equation}
\partial_\mu T^{\mu i} =  s\zeta^i \,.
\end{equation}
This is now the starting point analogous to (\ref{eq:HKMStau2}) above. Following the same steps as previously leads to the thermal conductivity
\begin{equation}
\bar\kappa(\omega) = \frac{\mu^2}{T}\sigma_{\textsc{q}} + \frac{Ts^2}{\epsilon+P}  \left[\frac{\mathrm{i}}{\omega} + \pi \delta(\omega)\right].   \label{eq:hkmskappa}
\end{equation}

A very systematic way to obtain transport coefficients from hydrodynamic equations of motion was developed by
Kadanoff and Martin \cite{Kadanoff1963419}. The result of that analysis is as follows \cite{Kadanoff1963419, Kovtun:2012rj}. Write the hydrodynamic equations of motion in the matrix form
\be\label{eq:hydroeqs}
\dot \phi_A + M_{AB}(k) \l^B = 0\,,
\ee
where, for example, $\phi_A = \{\delta \r, \delta s, \delta T_{0i}\}$ are the fluctuating hydrodynamic variables and $\lambda^A = \{\d \mu, \d T, \delta v^i\}$ are fluctuations of the corresponding sources. The fluctuations of sources and hydrodynamic variables are related by the static susceptibilities
\be
\phi_A = \chi_{AB} \l^B \,.
\ee
The matrix of hydrodynamic Green's function for the $\phi_A$ variables can then be shown to be given by
\be
G^{\mathrm{R}}(\omega,k) = M \frac{1}{i \omega \chi - M} \chi \,.
\ee
These are Green's functions for the conserved densities. They clearly have poles on the hydrodynamic modes where (\ref{eq:hydroeqs}) is satisfied. The Green's functions for the currents $J_A = \{J^i, Q^i, T^{ij}\}$ -- needed to compute the matrix of conductivities using formulae such as (\ref{eq:skG}) and (\ref{eq:aaG}) -- is obtained using the conservation equations, which hold as operator relations. Thus, for instance, $\omega^2 G^{\mathrm{R}}_{\r\r} = k^2 G^{\mathrm{R}}_{J^x J^x}$. The Kadanoff-Martin method was used in \cite{hkms} to first obtain (\ref{eq:Cleansigma2}), (\ref{eq:hkmsalpha}) and (\ref{eq:hkmskappa}).

The results (\ref{eq:Cleansigma2}), (\ref{eq:hkmsalpha}) and (\ref{eq:hkmskappa}) obey the following relations that can be obtained as relativistic Ward identities \cite{Hartnoll:2007ip, Herzog:2009xv}
\be
(T\a + \mu \s) i \omega = - \rho \,, \qquad (\bar \k + \mu \a) i \omega = - s \,.
\ee
These relations are valid beyond hydrodynamics. In a relativistic theory, it is therefore sufficient to compute the electrical conductivity $\sigma$ to obtain the entire matrix of thermoelectric conductivities. This is why there is only one independent dissipative coefficient $\sigma_{\textsc{q}}$.

Given that (\ref{eq:Cleansigma2}), (\ref{eq:hkmsalpha}) and (\ref{eq:hkmskappa}) follow from general considerations, they must of course be true in holographic models. This can be verified explicitly from bulk computations \cite{Hartnoll:2007ip}. Furthermore, in holographic models the coefficient $\sigma_{\textsc{q}}$ will acquire a particular form. In Einstein-Maxwell-dilaton theories one obtains \cite{Hartnoll:2007ip,Jain:2010ip,Chakrabarti:2010xy,Davison:2015taa}
\be\label{eq:incclean}
\sigma_{\textsc{q}} = \frac{Z_+}{e^2} \left(\frac{s}{4 \pi} \right)^{(d-2)/d} \left(\frac{sT}{\epsilon + P} \right)^2 \,.
\ee
Here $Z_+$ is the value of $Z(\Phi)$ in the Einstein-Maxwell-dilaton action (\ref{eq:EMDaction}) evaluated on the horizon. We will discuss the temperature scaling of $\sigma_{\textsc{q}}$ shortly.

In experiments, we have explained that it is often convenient to measure $\kappa$, defined in (\ref{eq:kappadef}).    Unlike $\sigma$, $\alpha$ and $\bar\kappa$, there is no divergence in $\kappa$ as $\omega \rightarrow 0$. One finds \begin{equation}
\kappa = \sigma_{\textsc{q}}\frac{(\epsilon+P)^2}{T\rho^2}.
\end{equation}
The cancellation of the divergence between the various terms in (\ref{eq:kappadef}) has a simple physical explanation \cite{Mahajan:2013cja}, as we now explain. $\k$ is the thermal conductivity with open circuit boundary conditions. With such boundary conditions, as we noted, electric current does not flow. However, in a charged system, momentum necessarily transports charge (the sound mode carries charge). Therefore with no electric current, there can be no momentum and the sound wave is not excited. Thus there is no divergence. The association between the sound mode and the divergent conductivities will be seen very explicitly below. The fact that $\kappa$ is finite but $\sigma$ diverges in a fully hydrodynamic and nonzero density theory leads to a strong violation of the Wiedemann-Franz law (\ref{eq:WF}) because the Lorenz ratio $L \equiv \k/(T\sigma) \to 0$.

A further quantity that remains finite as $\w \to 0$ is the conductivity of the `incoherent current' \cite{Davison:2015taa}
\be\label{eq:Jinc}
\vec J^\text{inc} \equiv \frac{sT \, \vec J - \rho \, \vec Q}{\epsilon + P} \,.
\ee
This is the current corresponding to the incoherent combination of charges. Recall that this combination does not couple to the velocity and hence diffuses, with diffusivity (\ref{eq:Dinc}). This mode that is decoupled from momentum behaves in much the same way as charge in a neutral relativistic system, described in \S\ref{sec:qctransport}.
In fact, using (\ref{eq:Jinc}) together with (\ref{eq:Cleansigma2}), (\ref{eq:hkmsalpha}) and (\ref{eq:hkmskappa}), it is clear that the incoherent conductivity is nothing other than
\be\label{eq:sigmainc}
\sigma_{J^\text{inc}J^\text{inc}}(\omega) = \sigma_{\textsc{q}} \,.
\ee

In Einstein-Maxwell-dilaton theories, the incoherent conductivity has the low temperature scaling \cite{Davison:2015taa}
\be
\sigma_{\textsc{q}} \sim T^{2 + (d - 2 - \theta)/z} \,.
\ee
As noted in the discussion around equation (\ref{eq:sQT}) above, this corresponds to an anomalous dimension for the charge density operator with $\Phi = z$, making the operator marginal in the IR. The incoherent conductivity is sensitive to the IR physics precisely because it has decoupled from the `dragging' effect of the sound mode.

\subsubsection{Drude weights and conserved quantities}\label{sec:drudeweights}

Delta functions in the conductivity, such as those appearing in (\ref{eq:Cleansigma2}), (\ref{eq:hkmsalpha}) and (\ref{eq:hkmskappa}), always arise when a conserved operator overlaps with the current operator. In this section we will prove this fact. Let us focus on the electrical conductivity for concreteness. The conductivity can be decomposed into a `Drude weight' part and a regular part
\be
\s(\w) = {\mathcal D} \left(\delta(\w) + \frac{1}{\pi} \frac{\mathrm{i}}{\omega} \right) + \s_\text{reg}(\w) \,. \label{eq:sreg}
\ee
The Drude weight is ${\mathcal D}$. From the expression for the conductivity in terms of Green's functions (\ref{eq:skG}) it is clear that
\be
\frac{{\mathcal D}}{\pi} = \chi_{J^x J^x} - G^{\mathrm{R}}_{J^x J^x}(0) \,. \label{eq:Dweight}
\ee

We will now relate the Drude weight (\ref{eq:Dweight}) to conserved quantities. This argument is originally due to Mazur \cite{mazur1969non} and Suzuki \cite{suzuki1971ergodicity}. We follow the presentation in \cite{Gur-Ari:2016xff}. Both of the two terms in (\ref{eq:Dweight}) admit a spectral representation. While they are rather similar, 
$\chi_{J^x J^x}$ is a zero frequency Euclidean Green's function and $G^{\mathrm{R}}_{J^x J^x}(0)$ a zero frequency retarded Green's function. Zero energy excitations contribute to the former and not the latter, so that
\be
\frac{{\mathcal D}}{\pi} = \frac{1}{V T} \frac{1}{Z} \sum_{
  \begin{smallmatrix}
  m,n \\
  E_m=E_n
  \end{smallmatrix}
  }
 \mathrm{e}^{-\beta E_n} 
\vert \langle n \vert J^{x} \vert m\rangle \vert^2 \,. \label{eq:Ddiff}
\ee
Here $Z$ is the partition function and $V$ the volume. The expression (\ref{eq:Ddiff}) is furthermore the time-averaged correlation function, so that
\be\label{eq:D2}
\frac{{\mathcal D}}{\pi} = \frac{1}{T} \lim_{t_0 \to \infty} \frac{1}{t_0} \int\limits_0^{t_0} \mathrm{d}t \langle J^x(0) J^x(t) \rangle \,.
\ee
This can again be seen by using the spectral representation of the correlator.
Formula (\ref{eq:D2}) is intuitive. The delta function in the conductivity is due to current that does not relax.

Suppose that there are conserved Hermitian operators $Q_A$ in the system. Without loss of generality we can take these to be orthogonal with respect to an inner product on the space of operators:
\be\label{eq:ip}
(A|B) \equiv \langle A^\dagger B \rangle = T \chi_{AB} \,.
\ee
The latter inequality -- relating a correlation function of operators at equal time to the susceptibility defined
below (\ref{eq:aaG}) -- is not obvious but can be shown, see \cite{forster1995} and also \S\ref{sec:memorymatrix} below. Using this inner product we can write
\be
J^x = \sum_A \frac{\langle Q_A J^x \rangle Q_A}{\langle Q_A Q_A \rangle} + \widetilde J^x \,,
\ee
where $\widetilde J^x$ is orthogonal to all of the conserved operators with respect to (\ref{eq:ip}). Inserting this decomposition into (\ref{eq:D2}) one obtains
\begin{align}
\frac{{\mathcal D} T}{\pi}  & =   \sum_A \frac{| \langle Q_A J^x \rangle|^2}{\langle Q_A Q_A \rangle} + \lim_{t_0 \to \infty} \int\limits_0^{t_0} \mathrm{d}t \langle \widetilde J^x(0) \widetilde J^x(t) \rangle \notag \\
 & \geq \sum_A \frac{| \langle Q_A J^x \rangle|^2}{\langle Q_A Q_A \rangle} \; = \; \sum_A \frac{T \chi_{Q^A J^x}^2}{\chi_{Q^A Q^A}} \,. \label{eq:ineq}
\end{align}
To obtain the last line we used the representation (\ref{eq:Ddiff}) of the time averaged correlator which is manifestly positive. This is the advertized result: conserved charges that overlap with the current operator lead to delta functions in the conductivity.

In the hydrodynamic results (\ref{eq:Cleansigma2}), (\ref{eq:hkmsalpha}) and (\ref{eq:hkmskappa}) the inequality (\ref{eq:ineq}) is saturated by the overlap of the conserved momentum with the electric and thermal current operators. That is
\be
\chi_{PP} = \epsilon + P\,, \quad \chi_{JP} = \rho \,, \quad \chi_{QP} = s T \,.
\ee
These relations encode the intuitive facts that a net density causes overlap between the current and momentum, while a net entropy causes an overlap between heat and momentum.

In this chapter the long-lived operator will always be the momentum. Another important case is when the long-lived operator is the supercurrent \cite{Davison:2016hno}. Some aspects of superfluid transport will be considered in \S\ref{sec:holoS}.
 
\subsubsection{General linearized hydrodynamics}
\label{sec:genhydro}

Let us briefly mention the extension of the (linearized) hydrodynamics derived above to models which are not Lorentz invariant.  A powerful approach to this more general problem is the memory matrix formalism described in \S\ref{sec:memorymatrix} below. However, a hydrodynamic perspective is also instructive. We must again choose a frame to fix the ambiguity in the definition of $\mu,T$ and $v_i$ out of equilibrium. A convenient choice, which can be thought of as a generalization of the Landau frame conditions (\ref{eq:landauframe}), is as follows. Require that the thermodynamic expressions for the momentum density $\mathfrak{g}^i$ (while $\mathfrak{g}^i = T^{ti}$ in a relativistic theory, we prefer to use manifestly non-relativistic notation here) and energy and charge densities $\epsilon$ and $\rho$ are obeyed to all orders in the derivative expansion. Thus in particular for the momentum density at linear order
\begin{equation}
\mathfrak{g}^i = \mathcal{M}v^i \,,
\end{equation}
and the parameter $\mathcal{M}$ is the equilibrium susceptibility $\mathcal{M} = \chi_{PP}$, as will be discussed in more detail just above (\ref{eq:memform1}) below. It is analogous to the ``mass density". Finally recall that in a non-relativistic theory $T^{ti} \neq T^{it}$.

In the non-relativistic notation, the hydrodynamic equations become \begin{subequations}\label{eq:NReom}\begin{align}
\partial_t \rho + \partial_i J^i &= 0, \\
\partial_t \epsilon + \partial_i J^i_{\mathrm{E}} &=  E_i J^i, \\
\partial_t \mathfrak{g}^i + \partial_j T^{ij} &= \rho(E^i + v_k F^{ki}).
\end{align}\end{subequations}
We now need the constitutive relations for $T^{ij}$, $J^{\mathrm{E}}_i$ and $J_i$. This is done following the same logic as previously for the relativistic case. The spatial components of the stress tensor are found to be identical to before: \begin{widetext}
\begin{equation}\label{eq:Tconsgen}
T^{ij} = P\delta^{ij} - \eta \left(\partial^i v^j + \partial^j v^i\right) + \left(\zeta-\frac{2\eta}{d}\right)\partial_k v^k \delta^{ij} + \cdots \,.
\end{equation}
The  energy current, however, is now given by
\begin{equation}\label{eq:JEconsgen}
J^{\mathrm{E}}_i = (\epsilon+P) v_i - (T \alpha_{\textsc{q}} - \mu \sigma_{\textsc{q}})(\partial_i \mu - E_i) - (\bar\kappa_{\textsc{q}} - \mu \bar\alpha_{\textsc{q}}) \partial_i T + \cdots,
\end{equation}\end{widetext}
with $E$ the external electric field,  and $\alpha_{\textsc{q}}$, $\bar\alpha_{\textsc{q}}$ and $\bar\kappa_{\textsc{q}}$ new dissipative coefficients.     We will see that they play the same roles as $\sigma_{\textsc{q}}$, as microscopic ``quantum critical" thermoelectric conductivities.   While the energy current looks similar to the (linearized when $v\ll 1$) relativistic energy current,  for a non-relativistic fluid $J^{\mathrm{E}}_i \ne \mathfrak{g}_i$, even when neglecting dissipative effects.  The charge current is now \begin{equation}
J_i = \rho v_i - \sigma_{\textsc{q}}(\partial_i \mu - E_i) - \bar\alpha_{\textsc{q}} \partial_i T + \cdots. \label{eq:Jconsgen}
\end{equation}
Note that the non-dissipative terms in these constitutive relations (e.g. $J_i = \rho v_j$) are no longer fixed by Lorentz invariance, but by the absence of entropy production. Let us see how this works by computing the entropy production to leading order in derivatives. Because entropy production occurs at quadratic order away from equilibrium, it is necessary here to keep a $v_i \mathrm{d} \mathfrak{g}^i$ term in the first law of thermodynamics for $\mathrm{d}\epsilon$. Everywhere else in our discussion we have dropped velocity contributions to thermodynamic relations, as they are quadratic in fluctuations about the equilibrium state (which we are taking to have $v_i = \mathfrak{g}_i = 0$) and are hence irrelevant for linear response. Thus we have
\bea
\pa_t s & = & \frac{\pa_t \epsilon - \mu \pa_t \rho - v_i \pa_t \mathfrak{g}^i}{T} = - \frac{\pa_i J_E^i - \mu \pa_i J^i - v_i \pa_j T^{ij}}{T}\notag  \\
 & = & - \pa_i (Q^i/T) - \frac{Q^i}{T^2} \pa_i T - \frac{J^i}{T} \pa_i \mu + \frac{v_i}{T} \pa_j T^{ij} \notag  \\
 & = & - \pa_i (Q^i/T) - \frac{s}{T} v^i \pa_i T - \frac{\rho}{T} v^i \pa_i \mu + \frac{1}{T} v^i \pa_i P + \cdots \notag \\
 & = & - \pa_i (Q^i/T) + \cdots \,.
\eea
The first line uses the first law of thermodynamics and the conservation laws (\ref{eq:NReom}). The
second line collects an overall total derivative term involving the entropy current. The third line uses the 
constitutive relations (\ref{eq:Tconsgen}), (\ref{eq:JEconsgen}) and (\ref{eq:Jconsgen}) to leading order (recall $Q^i = J_{\mathrm{E}}^i - \mu J^i)$. The final line shows that the non-total derivative terms cancel, using $\mathrm{d}P = \rho \mathrm{d}\mu + s \mathrm{d}T$. This cancellation required the leading terms in the constitutive relations to involve $P, \rho$ and $\epsilon + P = \mu \rho + sT + \mathcal{O}(v^2)$ in precisely the way they appear in (\ref{eq:Tconsgen}), (\ref{eq:JEconsgen}) and (\ref{eq:Jconsgen}). If we were to keep the next order in derivatives term in the constitutive relations, positivity of entropy production would be seen to require the viscosities and the matrix of critical conductivities $\sigma_{\textsc{q}},\alpha_{\textsc{q}},\bar \kappa_{\textsc{q}}$ to be positive definite, analogous to our discussion around (\ref{eq:consti}).

The transport coefficients are found by solving the hydrodynamic equations exactly as we did previously, the results are now
\begin{subequations}\label{eq:524results}\begin{align}
\sigma &= \sigma_{\textsc{q}} + \frac{\r^2}{ \mathcal{M}}\left[\frac{\mathrm{i}}{\omega} + \pi \delta(\omega)\right], \\
\alpha &= \alpha_{\textsc{q}} + \frac{\r s}{ \mathcal{M}}\left[\frac{\mathrm{i}}{\omega}  + \pi \delta(\omega)\right], \\
\bar\kappa &= \bar\kappa_{\textsc{q}} + \frac{T s^2}{\mathcal{M}}\left[\frac{\mathrm{i}}{\omega}  + \pi \delta(\omega)\right].
\end{align}\end{subequations}
Also as previously, $\kappa$ is finite: \begin{equation}
\kappa  = \bar\kappa_{\textsc{q}} - \frac{2T s}{\r}\alpha_{\textsc{q}} + \frac{T\s^2}{\r^2}\sigma_{\textsc{q}}.  \label{eq:kappagen}
\end{equation}
One can compute $\alpha$ by either applying a temperature gradient and computing $J_i$, or applying an electric field and computing the heat current.   Demanding Onsager reciprocity (that these two computations of $\alpha$ yield the same result) gives \begin{equation}
\alpha_{\textsc{q}} = \bar\alpha_{\textsc{q}}.
\end{equation}

In the special case of Galilean-invariant field theories, it follows from the Galilean Ward identities that the momentum density $\mathfrak{g}$ is (see e.g. \cite{Jensen:2014aia, cha95})
\begin{equation}
\mathfrak{g}^i = m J^i .
\end{equation}
In this case, we find that \begin{equation}
\sigma_{\textsc{q}} = \alpha_{\textsc{q}} = \bar\alpha_{\textsc{q}}=0.
\end{equation}
In addition to the two viscosities, there is a single dissipative coefficient $\bar\kappa_{\textsc{q}}$, and from (\ref{eq:kappagen}) we see that $\kappa = \bar\kappa_{\textsc{q}}$.

In the case where Lorentz invariance is restored, we have \begin{equation}
\mathcal{M} = \epsilon+P, \;\;\; \alpha_{\textsc{q}} = -\mu \sigma_{\textsc{q}}, \;\;\; \bar\kappa_{\textsc{q}} = \frac{\mu^2}{T}\sigma_{\textsc{q}}.  \label{eq:385}
\end{equation}
While at the linearized level the velocity $v_i$ defined in this subsection is the same as $\delta v_i$ defined in (\ref{eq:deltav}), further corrections are required at the nonlinear level to restore full Lorentz invariance.   We do not have a systematic understanding of how this arises.   Indeed, to the best of our knowledge, there has not been a systematic development of hydrodynamics for theories which are neither Galilean nor Lorentz invariant.

\subsection{Weak momentum relaxation I: inhomogeneous hydrodynamics}
\label{sec:weakI}

To obtain finite transport coefficients we must break translation invariance.  In this section we describe how this
can be done, in a certain limit, within hydrodynamics. The simplest hydrodynamic approach to translation symmetry breaking is a ``mean field" approximation, where momentum is relaxed at some constant (small) rate $\tau$. The spatial components of the hydrodynamic `Newton's law' (\ref{eq:dmuTmu}) are modified to \cite{hkms}:
\begin{equation}
\partial_t \mathfrak{g}^i + \partial_j T^{ij} = -\frac{\mathfrak{g}^i}{\tau_{\mathrm{imp}}} + F^{i\nu}J_\nu.  \label{eq:HKMStau}
\end{equation}
Here $\mathfrak{g}^i$ is the momentum density, as in the previous subsection.
Following the previous section, we may derive the electrical conductivity  \begin{equation}
\sigma(\omega) = \sigma_{\textsc{q}} + \frac{\rho^2}{\mathcal{M}} \frac{1}{\tau_{\mathrm{imp}}^{-1}-\mathrm{i}\omega}.  \label{eq:hkmssigma}
\end{equation}
Indeed, as is manifest from (\ref{eq:HKMStau}), we may simply set $-\mathrm{i}\omega \rightarrow \tau_{\mathrm{imp}}^{-1} - \mathrm{i}\omega$ in (\ref{eq:524results}). The $\omega \to 0$ dc conductivity obtained from (\ref{eq:hkmssigma}) is now finite. We can directly see that this is due to the fact that momentum is no longer conserved.

In the quasiparticle limit, $\rho \rightarrow ne$ and $\mathcal{M} \rightarrow nm$, with $n$ the number density of quasiparticles of charge $e$ and mass $m$. Then (\ref{eq:hkmssigma}) is simply the classical Drude formula. Thus a generalized Drude formula (\ref{eq:hkmssigma}) is valid for any model with a clean (momentum-relaxation dominated) hydrodynamic limit of transport. Note that in a quasiparticle metal away from this clean hydrodynamic limit (i.e. most conventional metals), one should strictly not use the Drude formula but rather a Boltzmann equation that allows for the fact that momentum is not a privileged observable, but rather decays at the same long timescale as generic quasiparticle excitations $\delta n_k$ \cite{ziman}.

To leading order in perturbation theory in $\tau_{\mathrm{imp}}^{-1}$ (i.e. slow momentum relaxation), the thermodynamic quantities $\rho$ and $\mathcal{M}$ appearing in (\ref{eq:hkmssigma}) can be evaluated in the clean theory, with no momentum relaxation. Thus all the physics of momentum relaxation is captured by the parameter $\tau$. However, corrections to the these thermodynamic quantities of order $\tau_{\mathrm{imp}}^{-1}$ give contributions to the dc conductivity of the same size as  $\sigma_{\textsc{q}}$ (as is immediately seen from (\ref{eq:hkmssigma})). Thus, in this framework, $\sigma_{\textsc{q}}$ cannot reliably be computed in the clean theory without taking into account such `spectral weight transfer' from the coherent to the incoherent part of the conductivity.  Holography gives a particularly convenient way to account for these corrections, as we will see.    

To go further,  we need a more `microscopic' approach in order to derive an expression for $\tau$.   A physically transparent way to account for the presence of $\tau$ is to directly solve the hydrodynamic equations described in previous sections, but to explicitly break translational symmetry in the background fluid with long wavelength, hydrodynamic modes. These modes could be disordered or could form a long wavelength lattice. We will refer to the position dependence of the sources as `disorder' for convenience.   A particularly simple way this can be done is to source the fluid with charged disorder.  Although the resulting inhomogeneous hydrodynamic equations must in general be solved numerically, this approach is non-perturbative in the amplitude of (long-wavelength) disorder  \cite{Lucas:2015lna, Lucas:2015sya}. In the remainder of this section we consider a simpler limit in which the disorder is weak and can be treated perturbatively. In this limit of weak, long wavelength disorder we will obtain an explicit formula for $\tau$ from hydrodynamics alone. Similar ideas, of self-consistently describing momentum relaxation within hydrodynamics, have been developed in \cite{andreev2011, Balasubramanian:2013yqa, Davison:2013txa}. The discussion below follows \cite{Lucas:2015lna, Lucas:2015sya}.

Suppose that we have a translationally invariant fluid at nonzero temperature and density.  The translation-invariant Hamiltonian $H_0$ is perturbed by a time-independent, spatially-varying source which linearly couples to a scalar operator $O$: \begin{equation}
H = H_0 - \int \mathrm{d}^dx \; h_0(x) O(x).   \label{eq:Hbroken}
\end{equation}
Suppose further for simplicity that the equation of state is symmetric under $h \to -h$, so that the local expressions for $\rho$ and $s$ are only corrected at $\mathcal{O}(h^2)$. This assumption can be relaxed \cite{Lucas:2015sya}, leading to more complicated expressions for $\tau$ that hide the simple physics at work. There is the following Ward identity (see \cite{Kim:2016hzi} for a field theoretic derivation)\begin{equation}
\partial_\mu T^{\mu \nu} =  F^{\nu\mu} J_\mu+ \langle O\rangle \partial^\nu h_0 \,,  \label{eq:scalarward}
\end{equation}
with $\langle O\rangle$ the thermal expectation value of $O$; for simplicity henceforth we drop the angled brackets. To solve the hydrodynamic equations of motion, we must also account for the dependence of pressure on $h$.  The pressure is microscopically $P=W/V_d$, with $W$ the generator of correlation functions in the QFT. By definition $\partial W/\partial h(k) = O(k)$. Therefore
\begin{equation}\label{eq:dPdh}
\frac{\partial P}{\partial h} = O \,.
\end{equation}
This allows us to find an exact, non-dissipative solution of the hydrodynamic equations with $\mu$ and $T$ uniform, $u^\mu=(1,0,\ldots)$, and $h=h_0$. The Ward identity (\ref{eq:scalarward}), in the absence of an electric field, together with (\ref{eq:dPdh}) implies that the stress tensor takes the required ideal fluid form $T^{ij} = P \delta^{ij}$, since
\begin{equation}
\partial_i T^{ij} = O\partial_j h_0 = O \partial_j h = \partial_j P \,. \label{eq:exactsol}
\end{equation}

Now, we linearly perturb  this background with an external force. For simplicity, we apply an electric field $E_i$. A thermal gradient works similarly.   The argument below is not long, but is somewhat subtle.   We are performing a strict linear response calculation in $E$.   Only after taking this limit, we then impose the perturbative limit that the strength of the background disorder $h_0\rightarrow 0$.   We also turn on a small frequency $\omega$. The interesting frequency scales will be $\omega \sim h_0^2$.

We make an ansatz that the only two fields which pick up corrections due to the external electric field are a constant shift to the velocity, $v_i \sim h_0^{-2}$,  and a perturbation to $h$, $\delta h\sim h^{-1}_0$.  The velocity induced is large when inhomogeneities are small. This makes sense because in the translationally invariant case there is no equilibrium velocity in the presence of an electric field. Given the solution we construct below, it is straightforward to check that this ansatz is consistent, with all other perturbations $\sim h^0_0$.  The momentum conservation equation from (\ref{eq:scalarward}) reads
\begin{align}
-\mathrm{i}&\omega \mathcal{M} v_j + \partial_j P = -\mathrm{i}\omega\mathcal{M} v_j  +  (O + \delta O) \partial_j (h_0+\delta h) \notag \\
&= \rho E_j + (O+\delta O)\partial_j h_0 + \text{subleading in }h_0.   \label{eq:233}
\end{align}
In the above formula $\rho, v_j$ are uniform, that is to say, evaluated at zeroth order in a derivative expansion. Inhomogeneities are induced at higher order in $h_0$, and these are the terms that are dropped in (\ref{eq:233}).
At linear order in perturbations, averaging (\ref{eq:233}) over space leads to
\begin{equation}\label{eq:ddd}
\frac{1}{L^d}\int \mathrm{d}^dx\; O\partial_j \delta h = \rho E_j + \mathrm{i}\omega\mathcal{M} v_j . 
\end{equation}
The objective now is to solve this equation to obtain a relationship between the electric field $E_i$ and the induced velocity $v_i$ and hence obtain the conductivity.

In order to evaluate (\ref{eq:ddd}), it turns out to be easier to evaluate $\delta O$ than $\delta h$.  Happily, these can be related by a static susceptibility, or retarded Green's function, in Fourier space: \begin{equation}
\delta O(k) = \chi_{OO}(k) \delta h(k) = G^{\mathrm{R}}_{OO}(\omega=0,k) \delta h(k).   \label{eq:chiOO}
\end{equation}
So long as perturbations of the operator $O$ decay at late times, there is no associated Drude weight and hence from (\ref{eq:Dweight}) the susceptibility and zero frequency Green's functions are equal. To evaluate the integral in (\ref{eq:ddd}) we need to obtain the $\delta O$ that is induced due to a flow $v_i$ (that is itself induced by the electric field). Because $\delta O$ is a scalar but $v_i$ is a vector, $\delta O$ can only be sourced in the presence of an inhomogeneities that result in nonzero gradients. Therefore we expect $\delta O \sim v_i \partial_i h_0$. To obtain the precise relation it is easiest to work in the fluid rest frame. In this case the inhomogeneous source (\ref{eq:Hbroken}) is
\be
\delta H = - \int \mathrm{d}^dx h_0(x-vt) O(x) \,.
\ee
This coupling leads to the response (taking a Fourier transform)
\begin{align}
& O(k) + \delta O(k)  = G^{\mathrm{R}}_{OO}(- k \cdot v,k) h_0(k) \notag \\*
& \approx \chi_{OO}(k) h_0(k) - k \cdot v \left. \frac{\partial G^{\mathrm{R}}_{OO}(\omega,k)}{\partial \omega}\right|_{\omega = 0} h_0(k) \,, 
\end{align}
where we have only kept terms to linear order in our perturbative expansion in $E$, as $v\sim E$.   Hence from (\ref{eq:ddd}), we obtain
\begin{align}
& \mathrm{i}\omega \mathcal{M} v_j +\rho E_j = \int \frac{\mathrm{d}^dk}{(2\pi)^d} O(- k) \mathrm{i}k_j \frac{\delta O(k)}{\chi_{OO}(k)}  \notag \\
& = \int \frac{\mathrm{d}^dk}{(2\pi)^d} \frac{O(-k)}{\chi_{OO}(-k)} \mathrm{i}k_j \left(-k_i v_i \left. \frac{\partial G^{\mathrm{R}}_{OO}(\omega,k)}{\partial \omega}\right|_{\omega = 0}\right) h_0(k) \notag \\
&= \int \frac{\mathrm{d}^dk}{(2\pi)^d}  |h_0(k)|^2 k_i k_j \lim_{\omega\rightarrow 0} \frac{\mathrm{Im}\, G^{\mathrm{R}}_{OO}(\omega,k)}{\omega} v_i \notag \\
& \equiv \frac{\mathcal{M}}{\tau_{\mathrm{imp}}} v_j \,,
\end{align}
where we have used isotropy and (\ref{eq:chiOO}) on the background, along with analyticity properties of Green's functions.   Assuming that the source $h_0$ is isotropic, then we may replace $k_i k_j \rightarrow (k^2/d) \delta_{ij}$ above.   This leads to the explicit formula for $\tau$: \begin{equation}
\frac{1}{\tau_{\mathrm{imp}}} = \frac{1}{d\mathcal{M}}  \int \frac{\mathrm{d}^dk}{(2\pi)^d} |h(k)|^2 k^2 \lim_{\omega\rightarrow 0} \frac{\mathrm{Im} \, G^{\mathrm{R}}_{OO}(\omega,k)}{\omega}.  \label{eq:memmatrixtime}
\end{equation}
Now, using that \begin{equation}
v_i = \frac{\rho E_i}{\mathcal{M}(\tau_{\mathrm{imp}}^{-1}-\mathrm{i}\omega)},
\end{equation} 
along with the constitutive relation $J_i \approx \rho v_i$ at this order in perturbation theory, we obtain the Drude formula
\begin{equation}
\sigma(\omega) = \frac{\rho ^2}{\mathcal{M}(\tau_{\mathrm{imp}}^{-1}-\mathrm{i}\omega)},  \label{eq:memdrude}
\end{equation}
which is what we obtained from hydrodynamics, but without the $\sigma_{\textsc{q}}$ contribution, which is generically subleading at this order in perturbation theory.

\subsection{Weak momentum relaxation II:  the memory matrix formalism}\label{sec:memorymatrix}

Let us now re-derive the results of the previous section, and more, using the memory matrix formalism. This will require a couple of pages of rather formal manipulations and definitions, but the payback will be worth the effort (the pragmatic reader can skip to the results in the subsections below). Specifically, the memory matrix extends the results of the previous section beyond the hydrodynamic limit. That is, there will be no assumption that the disorder is slowly varying. For instance, the memory matrix can describe generalized umklapp scattering that occurs at microscopic distances.  The memory matrix is the toolkit of choice for describing non-quasiparticle physics in which a small number of slowly decaying variables dominate the dynamics. Our formal development will follow \cite{forster1995}.

Define the following correlation function between two Hermitian operators $A$ and $B$: 
\begin{align}
\mathcal{C}_{AB}(t) &\equiv T\int\limits_0^{1/T} \mathrm{d}\lambda \left\langle A(t)^\dagger B(\mathrm{i}\lambda)\right\rangle_T \notag \\
&= T\int\limits_0^{1/T} \mathrm{d}\lambda \left\langle A \mathrm{e}^{-\mathrm{i}Lt} B(\mathrm{i}\lambda)\right\rangle_T \,,
\end{align}
where $\langle \cdots \rangle_T$ denotes an average over quantum and thermal fluctuations, $\langle \cdots\rangle$ will denote the average just over quantum fluctuations, and $L=[H,\circ]$ is the Liouvillian operator.   What is the usefulness of this object?   We take a time derivative and perform some basic manipulations: 
\begin{align}
\partial_t \mathcal{C}_{AB} &= -\mathrm{i}T\int\limits_0^{1/T}\mathrm{d}\lambda \left\langle AL\mathrm{e}^{-\mathrm{i}Lt} B(\mathrm{i}\lambda)\right\rangle_T \notag \\
& = -\mathrm{i}T\int\limits_0^{1/T}\mathrm{d}\lambda \left\langle A\mathrm{e}^{-\mathrm{i}Lt} L \mathrm{e}^{-\lambda L} B \mathrm{e}^{- H/T}\right\rangle \notag \\
&= \mathrm{i}T \left\langle A \mathrm{e}^{-\mathrm{i}Lt} \left(\mathrm{e}^{-L/T}-1\right) B \mathrm{e}^{-H/T}\right\rangle \notag \\ 
& = \mathrm{i}T \left\langle A \left(\mathrm{e}^{-H/T}B(-t) \mathrm{e}^{H/T} - B(-t)\right)\mathrm{e}^{-H/T}\right\rangle \notag \\
&= -\mathrm{i}T\left\langle [A(t),B] \right\rangle_T,
\end{align}
where in the last step we have used time translation invariance. The essential point of $\mathcal{C}_{AB}(t)$ is this fact that it represents the integral of a commutator. The commutator of course plays a central role in the definition of the retarded Green's function, so that now we have that
\begin{equation}
\Theta(t)\partial_t \mathcal{C}_{AB}(t) = -T G^{\mathrm{R}}_{AB}(t) \,.
\end{equation}
Integrating $t$ from $-\infty$ to $+\infty$:
\begin{equation}\label{eq:Cchi}
\frac{1}{T}\mathcal{C}_{AB}(t=0) = G^{\mathrm{R}}_{AB}(\omega=0) = \chi_{AB}.
\end{equation}
In the last step we assume that the operators $A$ and $B$ decay at late times so that the corresponding Drude weight vanishes in (\re{eq:Dweight}). The result (\ref{eq:Cchi}) is somewhat counterintuitive in that it relates a fixed time ($t=0$) quantity to an integral over all time ($\omega=0$).

We may also do the `Laplace transform' integral:
\begin{align}
\int\limits_0^\infty \mathrm{d}t &\mathrm{e}^{\mathrm{i}zt} \Theta(t)\partial_t \mathcal{C}_{AB}(t) = -\mathcal{C}_{AB}(z=0) - \mathrm{i}z\mathcal{C}_{AB}(z) \notag \\
&= -T\int\limits_{-\infty}^\infty \mathrm{d}t G^{\mathrm{R}}_{AB}(t) \mathrm{e}^{\mathrm{i}zt} = -TG^{\mathrm{R}}_{AB}(z),
\end{align}
and so we conclude \begin{equation}
\mathcal{C}_{AB}(z) = \frac{T}{\mathrm{i}z}\left(G^{\mathrm{R}}_{AB}(z) - G^{\mathrm{R}}_{AB}(0)\right) \,. \label{eq:lapp}
\end{equation}
In the above expressions $z$ is a complex frequency in the upper half complex plane, so that the integral above converges. Results at real frequencies are obtained by setting $z = \omega + \mathrm{i} 0^+$. Note that we are using the same symbol for $\mathcal{C}$ and its Laplace transform. We will keep the argument explicit in order to distinguish them.

Suppose that we pick $A=B=J_x$, and that we have a metal with finite electrical conductivity (so that $G^{\mathrm{R}}_{JJ}(0) = \chi_{JJ}$). Up to a factor of the temperature, the Laplace transform $\mathcal{C}_{JJ}(z)$ in (\ref{eq:lapp}) is nothing other than the electrical conductivity (\ref{eq:skG}), so that
\be\label{eq:sC}
\sigma(\omega) = \frac{1}{T} \mathcal{C}_{JJ}(\omega+ \mathrm{i} 0^+) \,.
\ee
We will leave the $+\mathrm{i}0^+$ implicit in expressions below.
Analogous identities hold for the other thermoelectric conductivities in (\ref{eq:skG}) and (\ref{eq:aaG}).
The identification of the conductivity with the correlation function $\mathcal{C}$ is the starting point for the memory matrix method. We now proceed to perform manipulations on $\mathcal{C}$.

As we have seen in previous sections, the conductivity and hence $\mathcal{C}_{AB}$ is parametrically large when translation symmetry breaking is weak.   The goal of the memory matrix formalism is to re-organize the computation of $\mathcal{C}_{AB}$ so that we can instead compute parametrically small quantities, which can then be analyzed easily perturbatively.   In order to do this, we will introduce a further abstraction:  an inner product on an ``operator Hilbert space" \begin{equation}
(A(t)|B) \equiv \mathcal{C}_{AB}(t).
\end{equation}
This inner product obeys all axioms of complex linear algebra, and further obeys \begin{equation}
\frac{1}{T}(A(0)|B) = \frac{1}{T}(A|B) = \chi_{AB},
\end{equation}
which will soon come in handy.   We can now again perform the Laplace transform, this time writing the answer as \begin{equation}
\mathcal{C}_{AB}(z) = \int\limits_0^\infty \mathrm{d}t\; \mathrm{e}^{\mathrm{i}zt} \left( A\left| \mathrm{e}^{-\mathrm{i}Lt}\right|B\right) = \left(A\left| \mathrm{i}(z-L)^{-1} \right|B\right).   \label{eq:MMformal1}
\end{equation}

We now perform a series of manipulations on (\ref{eq:MMformal1}) in order to cleanly extract out the perturbatively small ``denominator" to the conductivity $\mathcal{C}_{AB}$.    We pick a selective set of ``long-lived" modes $A,B\cdots$.
These will be the hydrodynamic degrees of freedom in a nearly clean fluid. More generally they can be any set of operators we wish, although they must include any operators whose correlation functions we wish to compute. The strategy will be to treat separately the time dependence of these operators and the remaining ``fast decaying'' operators. To this end we introduce the projection operator (that acts on the space of operators)
\begin{equation}
\mathfrak{q} \equiv 1-\mathfrak{p} \equiv  1  -\frac{1}{T} \sum_{AB} |A)\chi^{-1}_{AB}(B|,
\end{equation} 
with $\chi$ the reduced susceptibility matrix, only including the long-lived modes.  It is simple to check that $\mathfrak{q}^2=\mathfrak{q}$, and hence $\mathfrak{q}$ projects out the slow degrees of freedom.   We strongly emphasize that we are free to choose any set of operators to call the ``long-lived" modes, at this point, so long as we only compute correlation functions between this set of privileged operators.  Only at the end of the calculation will it be clear why to choose the modes which are, in fact, long-lived to be privileged.

Our goal is now to separate $z-L$ in (\ref{eq:MMformal1}) into the contributions from fast and slow modes.   Note the exact operator identities \begin{align}
(z-L)^{-1} & = (z-L\mathfrak{p}-L\mathfrak{q})^{-1} \notag \\
& = (z-L\mathfrak{q})^{-1}\left(1 + L\mathfrak{p}(z-L)^{-1}\right) \,.
\end{align}
Now, since by definition $\mathfrak{q}|B)=0$: \begin{equation}
(A|(z-L\mathfrak{q})^{-1}|B) = \left(A\left|\frac{1}{z} + \frac{L\mathfrak{q}}{z^2} + \cdots \right|B\right) = \frac{\chi_{AB}}{Tz},
\end{equation}
and \begin{align}
\mathrm{i}\mathfrak{p}(z-L)^{-1}|B) & = \frac{\mathrm{i}}{T}\sum_{CD}|C)\chi^{-1}_{CD}(D|(z-L)^{-1}|B) \\
& = \frac{1}{T}\sum_{CD} |C) \chi^{-1}_{CD}\mathcal{C}_{DB}(z),
\end{align}we find that
\begin{align}
& \mathcal{C}_{AB}(z) - \frac{\mathrm{i}\chi_{AB}}{z} = \frac{1}{T}\sum_{CD} (A|(z-L\mathfrak{q})^{-1} L |C)\chi^{-1}_{CD}\mathcal{C}_{DB}(z) \notag \\*
& =  \frac{1}{T} \sum_{CD}  \left(A \left| \left(\frac{1}{z} + \frac{L\mathfrak{q}}{z}(z-L\mathfrak{q})^{-1}\right) L \right|C\right) \chi^{-1}_{CD}\mathcal{C}_{DB}(z). \label{eq:MMgross}
\end{align}

To make the contents of (\ref{eq:MMgross}) more transparent, introduce the following two matrices on the space of long-lived operators. The eponymous memory matrix is
\be\label{eq:MAB}
M_{AC}(z) \equiv \frac{\mathrm{i}}{T} (\dot{A}|\mathfrak{q}(z-\mathfrak{q}L\mathfrak{q})^{-1}\mathfrak{q}|\dot C) \,,
\ee
and then the matrix
\begin{equation}\label{eq:NAB}
N_{AB} \equiv \chi_{A\dot{B}} = \frac{1}{T}(A|\dot{B}) = \frac{\mathrm{i}}{T}(A|L|B) \,.
\end{equation}
Note that $M$ is a symmetric matrix, while $N$ is antisymmetric. Further, $N=0$ in a time-reversal symmetric theory if the operators $A$ and $B$ transform identically under time reversal, as will often be the case. More generally note that $N$ is a constant whereas $M$ depends on $z$. The $N$ matrix is clearly directly related to the first term in brackets in (\ref{eq:MMgross}). To see the appearance of the memory matrix itself from the second term in (\ref{eq:MMgross}), note that
\begin{align}
& (A| L\mathfrak{q}(z-L\mathfrak{q})^{-1}L|C) = \left(\dot{A} \left| \frac{\mathfrak{q}}{z} + \frac{\mathfrak{q}L\mathfrak{q}}{z^2} + \frac{\mathfrak{q}(L\mathfrak{q})^2}{z^3}+\cdots\right|\dot C\right) \notag \\&  = (\dot{A}|\mathfrak{q}(z-\mathfrak{q}L\mathfrak{q})^{-1}\mathfrak{q}|\dot C) = -\mathrm{i}TM_{AC},  \label{eq:memform3}
\end{align}
where we have used the property $\mathfrak{q}^2=\mathfrak{q}$ in this chain of equations.
Then, (\ref{eq:MMgross}) becomes \begin{equation}
 \mathcal{C}=\frac{\mathrm{i}T\chi - \mathrm{i}(N+M)\chi^{-1}\mathcal{C}}{z},
\end{equation}
which can be further rearranged to obtain the key result that: \begin{align}
\frac{1}{T}\mathcal{C}_{AB}(\w) & \equiv \sigma_{AB}(\w) \label{eq:memmatrixgeneral} \\
& = \sum_{C,D} \chi_{AC}\left(\frac{1}{M(\w)+N-\mathrm{i}\omega \chi}\right)_{CD}\chi_{DB},  \notag
\end{align}
with $\sigma_{AB}$ the generalized conductivity between operators $A$ and $B$, as we noted in (\ref{eq:sC}) above.  We have analytically continued to real frequencies. 
Of particular relevance to us will be the thermoelectric conductivities (cf. (\ref{eq:skG}) and (\ref{eq:aaG}) above):  \begin{equation}
\sigma_{J_i J_j} = \sigma_{ij}, \;\;\;\;\; \sigma_{J_iQ_j} = T\alpha_{ij}, \;\;\;\;\; \sigma_{Q_iQ_j} = T\bar\kappa_{ij}.
\end{equation}

The expression (\ref{eq:memmatrixgeneral}) for the conductivity has a lot of physical information which will be elaborated upon in the following subsections. The basic idea is that the susceptibilities $\chi_{AC}, \chi_{DB}$ determine the overlap of the current operators with the long-lived modes, as in equation (\ref{eq:ineq}) above. This overlap can be thought of as a fast process. Then the extended Drude-peak structure of the rest of the formula will describe on how the long-lived operators themselves decay. Both $M$ and $N$ are proportional to time derivatives of the long-lived operators. Therefore, if these time derivatives are small, $M$ and $N$ will also be small, and the Drude-like peak will be sharp. In this way the formula 
(\ref{eq:memmatrixgeneral}), with no assumptions whatsoever at this point, has split the computation of the conductivity into a thermodynamic piece (the susceptibilities) and a piece that only depends on the dynamics of long-lived operators. The formula will be useful when a small number of parametrically long-lived operators exist.

\subsubsection{The Drude conductivites}
\label{sec:memdrude}

The results of the previous section can now be used to re-derive the results (\ref{eq:memmatrixtime}) and (\ref{eq:memdrude}) for the conductivities. As advertised previously, the memory matrix derivation does not require momentum relaxation to be a long wavelength process. We again consider a clean and isotropic fluid,  weakly perturbed by an inhomogeneous field coupling to a scalar operator $O$, as in (\ref{eq:Hbroken}).   If this breaking of translation invariance is small enough,  then the longest-lived mode that will have substantial overlap with the (vector) current $J_x$ is the momentum $P_x$.   Hence, we will allow the long-lived modes in the memory matrix formalism to be $J_x$ and $P_x$.     These are both time-reversal odd and so, assuming a time reversal invariant state, we may set the matrix $N=0$.    The next key observation is that as $\omega \rightarrow 0$,  $M$ has a zero eigenvalue associated with $M_{PP}$ in the clean limit.   This follows from the definition of $M_{PP}$ because in a clean metal $\dot{P}=0$ as an \emph{operator} equation.   Indeed, with the coupling (\ref{eq:Hbroken}) breaking translation invariance
\begin{equation}\label{eq:Pdot}
\dot{P}  = \mathrm{i}[H,P] = -\int\mathrm{d}^dx\; h(x) (\nabla O)(x).
\end{equation}

With (\ref{eq:Pdot}) at hand we can evaluate the memory matrix. The basic idea is that the inhomogeneous coupling will lift the zero eigenvalue of the clean theory and replace it by a small nonzero eigenvalue, which we can calculate. This eigenvalue will control momentum relaxation and hence the conductivities. At first nontrivial order we will see that (\emph{i}) the computation can be reduced to evaluating correlators in the clean theory and (\emph{ii}) the projection operators that appear in the definition of the memory matrix can be set to unity. Both of these facts offer substantial simplifications that will allow the momentum relaxation rate to be evaluated explicitly in later sections. The derivation below follows \cite{Hartnoll:2012rj}, that built on \cite{hkms,Hartnoll:2008hs}.

The operator $|\dot{P})$ coming from (\ref{eq:Pdot}) is a sum of nonzero momentum operators (the homogeneous part drops out upon taking the gradient), and hence in the clean theory has no overlap with any zero momentum operator, such as $J$. Recall that $J$ is the total, $k=0$, current.  Hence,  $\mathfrak{q}|\dot{P}) = |\dot{P})$ in the clean theory.  The following manipulations aim to exploit this fact. Firstly write 
 \begin{align}
& T M_{PP} = (\dot{P}| \mathfrak{q} (\omega-L+L\mathfrak{p})^{-1}|\dot{P}) \notag \\
& \;\; = (\dot{P}|  \mathfrak{q} (\omega-L)^{-1}[1 - L\mathfrak{p}(\omega-L)^{-1}+\cdots] |\dot{P})  \notag \\
& \;\; = (\dot{P}| \mathfrak{q}  (\omega-L)^{-1}|\dot{P}) \notag \\*
& \;\; + \frac{1}{T}\sum_{A,B} (\dot{P}|  \mathfrak{q} (\omega-L)^{-1}|\dot{A})\chi^{-1}_{AB}(B| (\omega-L)^{-1}|\dot{P}) + \cdots \notag \\
& \; \; = (\dot{P}| (\omega-L)^{-1}|\dot{P}) + \mathcal{O}(h^3) \,. \label{eq:setq1}
\end{align}
The first term in (\ref{eq:setq1}) is $\mathcal{O}(h^2)$. We must first explain why the second term after the third equality is subleading. The sum in that term is over $A,B\in \lbrace J,P\rbrace$. That term includes the expression $(B| (\omega-L)^{-1}|\dot{P})$ where $(B|$ is a zero momentum operator.  This will be zero unless we include $\mathcal{O}(h)$ corrections to $L$ and the inner product, associated with the breaking of translation invariance.   Accounting for $\dot{P}\sim h$, we conclude that (at least) $(B| (\omega-L)^{-1}|\dot{P})\sim h^2$.  
Noting the additional factor of $(\dot{P}|$, which contributes an additional factor of $h$, it follows that the second term (and higher order terms) in (\ref{eq:setq1}) may be neglected compared to the first term, at small $h$. Finally, in the last line of (\ref{eq:setq1}) we have used $\mathfrak{q}|\dot{P}) = |\dot{P})$. This is true in the clean theory. However, this term is already $\mathcal{O}(h^2)$ because of the two $\dot{P}$s, so any corrections to the relation $\mathfrak{q}|\dot{P}) = |\dot{P})$ due to the inhomogeneous coupling necessarily lead to terms of $\mathcal{O}(h^3)$ or higher.

Using the expression (\ref{eq:setq1}) for $M_{PP}$ in the definition of $ \mathcal{C}(z)$ in (\ref{eq:MMformal1}), as well as the connection to the retarded Green's function in (\ref{eq:lapp}), we can obtain the following important result at leading order in $h$ \cite{Hartnoll:2008hs, Hartnoll:2012rj}:
\begin{align}
\frac{\mathcal{M}}{\tau_{\mathrm{imp}}} & \equiv M_{PP}(\omega = 0) = \lim_{\omega\rightarrow 0} \frac{\mathrm{Im}\left(G^{\mathrm{R}}_{\dot{P}\dot{P}}(\omega)\right)}{\omega} \notag \\
& =  \int \frac{d^dk}{(2\pi)^d} |h(k)|^2 k_x^2 \lim_{\omega\rightarrow 0} \frac{\mathrm{Im}\left(G^{\mathrm{R}}_{OO}(\omega,k)\right)}{\omega}.  \label{eq:memform2}
\end{align}
The Green's functions in the above formula may be evaluated \emph{in the clean theory}. The final step in the above uses the explicit expression (\ref{eq:Pdot}) for the operator $\dot P$. We have used the fact that the real part of $G^{\mathrm{R}}_{\dot{P}\dot{P}}(\omega)$ is even in $\omega$ to focus on the imaginary part as $\omega \to 0$.

Two aspects of (\ref{eq:memform2}) are worth highlighting. This quantity will shortly become the width of the Drude peak in the frequency dependent conductivity, controlling the current and momentum decay rate. Firstly, the expression (\ref{eq:memform2}) is fairly intuitive. It can be thought of as a non-quasiparticle version of Fermi's Golden rule. The rate of decay of the momentum is determined by the number of low energy excitations (measured by the spectral density, the imaginary part of the Green's function) that overlap with the operator $\dot P$. Secondly, because the decay rate is given in terms of a low energy spectral density, it is purely a property of the low energy critical theory. It can be calculated within the low energy effective theory. In particular, in order for perturbation theory in $h$ to be valid, it is sufficient that the inhomogeneous coupling be small in the effective theory. If the coupling is irrelevant at low energies, perturbation theory can be used at low temperatures even if the inhomogeneities are large at short distance and high energy scales.

Similar arguments to the above show that $M_{JP}\sim h^2$. One factor of $h$ comes from $\dot P$, while the other comes from the need to break translation invariance in order to achieve an overlap between operators at zero and nonzero momentum \cite{Hartnoll:2012rj}. Finally, one expects that generically (i.e. in the absence of additional almost conserved operators) $M_{J_xJ_x}\sim h^0$. These facts together imply that that $(M^{-1})_{PP} = 1/M_{PP} \sim 1/h^2$ will dominate the inverse of the memory matrix that appears in the expression (\ref{eq:memmatrixgeneral}) for the conductivity.

To evaluate the conductivity we also need the susceptibilities. At leading order in weak momentum relaxation these may also be evaluated in the clean theory. Note, however, that the susceptibility is then the susceptibility computed within the low energy theory. This can have a nontrivial relationship in general with the microscopic susceptibility if disorder is important at high energies. In the clean theory, the susceptibility $\chi_{J_xP_x}$ is the charge density $\rho $.   To see this, recall that the velocity is thermodynamically conjugate to momentum, the same way that $\mu$ is conjugate to $\rho $. Thus $\chi_{JP} \equiv J_x/v_x = \rho$. We noted in our discussion of zeroth order hydrodynamics around (\ref{eq:Jconsgen}) above that the relation $J_x = \rho v_x$ is determined by thermodynamics and does not require any kind of boost invariance.
 We define further $\chi_{P_xP_x}\equiv \mathcal{M}$. In  a relativistic fluid,  $\mathcal{M}  = \epsilon+P$, the enthalpy.   In a Galilean-invariant fluid, $\mathcal{M}$ is the mass density.   At leading order in $h$, as noted in the previous paragraph, we may neglect $M_{JP}$ and $M_{JJ}$ when inverting $M-\mathrm{i}\omega \chi$.   (\ref{eq:memmatrixgeneral}) becomes \begin{equation}
\sigma(\omega) = \sigma_{JJ}(\omega) = \frac{\chi_{JP}^2}{M_{PP}(\omega) - \mathrm{i}\omega \chi_{PP}} \,.  \label{eq:memform1}
\end{equation}
Upon plugging in the parameters above and restricting to low frequencies $\omega \sim h^2$, where in particular we can approximate the memory matrix by its zero frequency limit (\ref{eq:memform2}), we obtain
\begin{equation}\label{eq:swcoh}
\sigma(\omega) = \frac{\rho ^2}{\mathcal{M}(\tau_{\mathrm{imp}}^{-1}-\mathrm{i}\omega)}.
\end{equation}  
In this way we have recovered the hydrodynamic expression (\ref{eq:memdrude}) for the Drude peak. The formula for $\tau_{\mathrm{imp}}$ is the same as we previously obtained in (\ref{eq:memmatrixtime}), but in now in significantly more general circumstances.

\subsubsection{The incoherent conductivities}
\label{sec:incformal}

It is also possible to account for the `incoherent' contributions $\sigma_{\textsc{q}}$ to the conductivities (\ref{eq:hkmssigma}) from the memory function formalism. This requires some care about which quantities are small or large in perturbation theory, and teaches us a limit in which the issues of spectral weight transfer discussed in the paragraphs below (\ref{eq:hkmssigma}) can be avoided, so that both terms in (\ref{eq:hkmssigma}) can be trusted.
This was first discussed in  \cite{Lucas:2015pxa}.  Here we present a cleaner derivation, following \cite{Davison:2016hno}. 

As a warm up, consider first the case with $\rho =0$. Here we expect that $\sigma=\sigma_{\textsc{q}}$ at low frequencies.   In this limit $\chi_{JP}=0$ and hence  the memory matrix formalism immediately gives us
\begin{equation}
\sigma_{JJ} = \frac{\chi_{JJ}^2}{M_{JJ}(\w) - \mathrm{i}\omega \chi_{JJ}}.
\end{equation}
In fact, this expression is always exactly true, as we are free to consider the set of slow operators to contain only $J$. Setting $\w = 0$ we obtain
\begin{equation}
\sigma_{\textsc{q}} = \frac{\chi_{JJ}^2}{M_{JJ}(0)} \,.  \label{eq:MJJsigmaQ}
\end{equation}

Next, let us turn on a nonzero charge density $\rho $.   The conductivity is
\begin{equation}
\sigma(\omega) = \sum_{A,B \in \lbrace J,P \rbrace} \chi_{JA} \left(M(\omega) - \text{i} \omega \chi\right)^{-1}_{AB} \chi_{BJ} \,.
\end{equation}
Recall that in the perturbative limit $M_{PP} \sim M_{JP}\sim h^2$. Let us zoom in on frequencies that saturate the Drude peak, so that $\omega \sim h^2$. In order for an `incoherent' frequency-independent term to contribute at leading order as $h \to 0$, on equal footing with the Drude term, it is necessary for the `incoherent susceptibility' to be large.
We define the incoherent susceptibility as \cite{Davison:2016hno}
\be
\chi^\text{inc}_{JJ} \equiv \chi_{JJ} - \frac{\rho^2}{\mathcal{M}} \,,
\ee
being the susceptibility of the incoherent current
\be
J^\text{inc.} \equiv J - \frac{\rho}{\mathcal{M}} P \,,
\ee
that we encountered previously in equation (\ref{eq:Jinc}) above in the relativistic case. Now suppose that $\chi^\text{inc}_{JJ} \sim 1/h$ as $h \to 0$. In this scaling limit, the conductivity becomes
 \begin{align}
\sigma(\omega) &= \frac{\left(\chi^\text{inc}_{JJ}\right)^2}{M_{JJ}(0)} + \frac{\chi_{JP}^2}{M_{PP}(0)-\mathrm{i}\omega \chi_{PP}} \notag \\
& = \sigma_{\textsc{q}} + \frac{\rho ^2}{\mathcal{M}} \frac{1}{(\tau_{\mathrm{imp}}^{-1}-\mathrm{i}\omega)} \,,
\end{align}
in agreement with (\ref{eq:hkmssigma}). In the final line we obtained an expression for the incoherent conductivity
\be\label{eq:sQinc}
\sigma_{\textsc{q}} = \frac{\left(\chi^\text{inc}_{JJ}\right)^2}{M_{J^\text{inc}J^\text{inc}}(0)} \,.
\ee
There is in fact some freedom with how the scaling limit is taken and how this formula is written. We could equally well have taken $\chi_{JJ} \sim 1/h$ and obtained precisely the result (\ref{eq:MJJsigmaQ}) above for $\sigma_{\textsc{q}}$, with no reference to incoherent currents.
The expression (\ref{eq:sQinc}) is natural given that that $\sigma_{\textsc{q}}$ in hydrodynamics is the dc conductivity of the incoherent current, according to (\ref{eq:sigmainc}), and therefore (\ref{eq:sQinc}) is the natural nonzero density generalization of (\ref{eq:MJJsigmaQ}).

The crucial point, however, about the above derivation is that we see that to perturbatively recover $\sigma_{\textsc{q}}$ in an exact formalism, we have to treat $\sigma_{\textsc{q}}$ as anomalously large in perturbation theory.  In general this will not be the case. We will observe corrections to (\ref{eq:hkmssigma}) at the order of $\sigma_{\textsc{q}}$ in \S \ref{sec:562} below.

To account for the incoherent thermal conductivities $\alpha_{\textsc{q}}$ and $\bar\kappa_{\textsc{q}}$, we must add the heat current $Q$ as an index to the memory matrix. We will take the simpler limit for the current susceptibilities
$\chi_{JQ}\sim \chi_{QQ} \sim \chi_{JJ} \sim 1/h$, note the discussion below (\ref{eq:sQinc}) above, and not worry about working with incoherent currents (note, however, that an incoherent thermal current can also be defined as $Q^\text{inc.} \equiv Q - \frac{s T}{\mathcal{M}} P$). Analogous to (\ref{eq:MJJsigmaQ}),  we define $\sigma_{\textsc{q}}$,  $\alpha_{\textsc{q}}$ and $\bar\kappa_{\textsc{q}}$ within the memory matrix formalism via  \begin{align}
& \left(\begin{array}{cc} \sigma_{\textsc{q}} &\  T\alpha_{\textsc{q}} \\ T\alpha_{\textsc{q}} &\ T\bar\kappa_{\textsc{q}} \end{array}\right)  \label{eq:MJJ2} \\
& =  \left(\begin{array}{cc} \chi_{JJ} &\  \chi_{JQ} \\ \chi_{JQ} &\ \chi_{QQ} \end{array}\right) \left(\begin{array}{cc} M_{JJ} &\  M_{JQ} \\ M_{JQ} &\ M_{QQ} \end{array}\right)^{-1} \left(\begin{array}{cc} \chi_{JJ} &\  \chi_{JQ} \\ \chi_{JQ} &\ \chi_{QQ} \end{array}\right)  \,. \notag
\end{align}
The net $\sigma$, $\alpha$ and $\bar\kappa$ are then found using the matrices (third row/column in matrix corresponds to the momentum $P$ index): \begin{subequations}\begin{align}
\chi &= \left(\begin{array}{ccc} \chi_{JJ} &\  \chi_{JQ} &\ \rho  \\ \chi_{JQ} &\ \chi_{QQ} &\ T s \\ \rho  &\ T s &\ \mathcal{M} \end{array}\right), \\
M &\approx \left(\begin{array}{ccc} M_{JJ} &\  M_{JQ} &\ 0 \\ M_{JQ} &\ M_{QQ} &\ 0 \\ 0 &\ 0 &\ \mathcal{M} \tau^{-1}_{\mathrm{imp}} \end{array}\right).
\end{align}\end{subequations}
Here $\chi_{QP} = T s$ gives the entropy density in terms of microscopic Green's functions. As we noted previously with $\chi_{JP} = \rho$, this is fixed by the absence of entropy production at zeroth order the hydrodynamic derivative expansion. We have only included leading order contributions to the memory matrix $M$.
 
Accounting for slow momentum relaxation as before, we recover the thermoelectric conductivities from $\sigma = \chi (M-\mathrm{i}\omega \chi)^{-1}\chi$, and keeping only the leading order contributions in the scaling limit as $h \rightarrow 0$: \begin{subequations}\label{eq:allsigma}\begin{align}
\sigma = \sigma_{JJ} &= \sigma_{\textsc{q}} + \frac{\rho ^2}{\mathcal{M}(\tau^{-1}_{\mathrm{imp}}-\mathrm{i}\omega)}, \\
\alpha = \frac{\sigma_{JQ}}{T} &= \alpha_{\textsc{q}} + \frac{\rho s}{\mathcal{M}(\tau^{-1}_{\mathrm{imp}}-\mathrm{i}\omega)}, \\
\bar\kappa = \frac{\sigma_{QQ}}{T} &= \bar\kappa_{\textsc{q}} + \frac{T s^2}{\mathcal{M}(\tau^{-1}_{\mathrm{imp}}-\mathrm{i}\omega)}.
\end{align} \label{eq:alltrans}
\end{subequations}
This is exactly (\ref{eq:524results}),  upon setting $\mathrm{i}\omega \rightarrow \mathrm{i}\omega - \tau^{-1}_{\mathrm{imp}}$.

Note that $M_{JJ}$ changes (but $\sigma_{\textsc{q}}$ does not) depending on whether or not $Q$ is included as an index in the memory matrix, due to the change in the projection operator $\mathfrak{q}$.     In general, computing $\chi_{JJ}$, $M_{JJ}$ etc. will require a microscopic computation more complicated than the perturbative computation of $\tau_{\mathrm{imp}}$, which is readily expressed in terms of microscopic Green's functions as in (\ref{eq:memform2}).

Finally, it is worth noting that in a relativistic theory,  $Q=P-\mu J$ 
holds as an operator equation.   In this case, $Q$ should not be a separate index in the memory matrix, and instead one computes $\sigma_{JQ} = \sigma_{JP} - \mu \sigma_{JJ}$, e.g.    Similarly, in a Galilean invariant fluid, the momentum $P = m J$,  and hence $J$ should be removed as an independent index in the memory matrix. Of course, an inhomogeneous coupling breaks both Lorentz and Galilean invariance, so these simplifications are likely only possible at leading order in a small $h$ expansion.

\subsubsection{Transport in field-theoretic condensed matter models}
\label{sec:transcondmat}

With the general form of the thermoelectric transport response functions at hand in (\ref{eq:524results}) via hydrodynamics, and also in 
(\ref{eq:alltrans}) via memory matrices, we now return to the condensed matter models of compressible quantum matter
in \S\ref{sec:isingnematic}-\ref{sec:emerge2}, and discuss the frequency and temperature dependence of their transport properties.

First, we have the `diffusive' or `incoherent' components, $\sigma_{\textsc{q}}$, $\alpha_{\textsc{q}}$, and $\bar\kappa_{\textsc{q}}$.
These are not expected to experience special constraints from total momentum conservation, and so should be finite scaling functions of $\omega/T$. Applying a naive scaling dimensional
analysis to these terms, we can expect that they obey scaling forms similar to those discussed in \S\ref{sec:qctransport}:
\begin{equation}
\sigma_{\textsc{q}} \sim \alpha_{\textsc{q}} \sim \frac{\bar\kappa_{\textsc{q}}}{T} \sim T^{(d-2)/z} \Upsilon (\omega/T),
\label{eq:sigmakappascale}
\end{equation}
with different scaling functions $\Upsilon$ for each of the transport coefficients. As discussed at the beginning of \S\ref{sec:compressible}, 
thermodynamic quantities related to the free energy density typically violate hyperscaling. This violation amounts to a replacement  $d \rightarrow d - \theta$ in scaling forms. So we might similarly expect that (\ref{eq:sigmakappascale}) is replaced more generally by
\begin{equation}
\sigma_{\textsc{q}} \sim \alpha_{\textsc{q}} \sim \frac{\bar\kappa_{\textsc{q}}}{T} \sim T^{(d-2-\theta)/z} \, \Upsilon (\omega/T).
\label{eq:sigmakappascale2}
\end{equation}
This issue has been investigated in recent works in the models of non-Fermi liquids described in \S\ref{sec:nfl} for the case of 
$\sigma_{\textsc{q}}$, and it has been found that (\ref{eq:sigmakappascale2}) is indeed obeyed. The Ising-nematic transition of \S\ref{sec:isingnematic} was investigated in \cite{Eberlein16}, and $\sigma_{\textsc{q}}$ obeyed (\ref{eq:sigmakappascale2}) in $d=2$
with $\theta = 1$. Similarly, for the spin density wave transition of \S \ref{sec:sdw}, the expected value, $\theta = 0$, appeared
also in $\sigma_{\textsc{q}}$ in $d=2$ \cite{metlitski3,strack1}. These computations do not show evidence for the separate anomalous dimension for the charge density discussed in \S\ref{sec:anomalouscharge}. 

Next, let us consider the ``momentum drag'' terms in (\ref{eq:524results}) and (\ref{eq:alltrans}). A key characteristic of the theories
of non-Fermi liquids in \S\ref{sec:isingnematic} and \S\ref{sec:sdw} is that they can all be formulated in a manner in which
the singular low-energy processes (responsible for the breakdown of quasiparticles) obey a Lagrangian which has continuous
translational symmetry. Consequently, the resistance of such theories is strictly zero, contrary to computations in the literature \cite{Lee89,IoffeWiegmann90,IoffeKotliar90,Metzner07}. Let us emphasize, again, that here it is the emergent translation invariance of the low energy critical theory that matters. Following the discussion in \S\ref{sec:mombottle}, we need  additional perturbations which relax momentum, and these can be inserted in  (\ref{eq:memform2}) and then (\ref{eq:alltrans}) to compute the transport coefficients.

The thermodynamic parameters $\rho$ and $\mathcal{M}$ are invariably dominated by non-critical `background' contributions of the compressible state. So we can take them to 
be non-singular $T$- and $\omega$-independent constants. The singular behavior of the entropy density, $s$, was discussed already
at the beginning of \S\ref{sec:compressible}. So it only remains to discuss the singular behavior of $\tau_{\mathrm{imp}}$.

The memory matrix result for $\tau_{\mathrm{imp}}$ in (\ref{eq:memform2}) was expressed
in terms of the spectral density of the operator $O$ coupling to a space-dependent source, $h(x)$, in (\ref{eq:Hbroken}). 
If we assume that average of $|h(k)|^2$ is $k$-independent (this is the case for a Gaussian random source which is
uncorrelated at different spatial points), and the scaling dimension $\mbox{dim}[O] = \Delta$ in the theory without the random source, 
then a scaling analysis applied to (\ref{eq:memform2}) yields \cite{hkms,Hartnoll:2008hs,Hartnoll:2014gba, lucas1401}
\begin{equation}
\frac{1}{\tau_{\mathrm{imp}}} \sim h^2 T^{2(1+\Delta-z)/z} \,. \label{eq:lss14}
\end{equation}
at small disorder. 

The expression (\ref{eq:memform2}) has been evaluated for specific models of disorder in the
non-Fermi liquids discussed above. For the Ising-nematic critical point, \cite{Hartnoll:2014gba} argued that the most relevant disorder
was a `random field' term that coupled linearly to the Ising-nematic order parameter $\phi$ in \S\ref{sec:isingnematic}.
However, it was found that the order parameter spectral weight, {\em i.e.\/} $\text{Im}\,G_{\phi\phi}$, violated the expected scaling form
because of an emergent gauge-invariance in the underlying critical theory. Consequently, (\ref{eq:lss14}) is not obeyed in this 
specific model. It was found that, in $d=2$, ${1}/{\tau_{\mathrm{imp}}} \sim T^{-1/2}$ (up to logarithms) at low $T$. The fact that $\tau_{\mathrm{imp}}$ becomes large at low temperatures in this model is a reflection of the fact that disorder is strongly relevant at the critical point. The computation is only self-consistent, then, if the disorder is tuned to be very small, so that the clean fixed point can still control the physics down to low temperatures. Later parts of this section and \S \ref{sec:disfpt} will consider the physics of strongly disordered fixed points, leading to incoherent transport. One can also consider a higher temperature regime in the pure Ising-nematic fixed point, where 
the Landau damping of the order parameter fluctuations
is negligible and so they obey $z=1$ scaling \cite{1995PhRvB..52.9520S,Fitzpatrick:2013rfa}: here, a linear-in-$T$ resistivity was 
found \cite{Hartnoll:2014gba,Lucas:2014sba}
via ${1}/{\tau_{\mathrm{imp}}} \sim T$.

For the spin-density-wave transition of \S\ref{sec:sdw}, we have to consider an additional subtlety before evaluating
(\ref{eq:memform2}). The critical theory (\ref{sdw2}) is expressed in terms of fermions at hot spots on the Fermi surface
which couple strongly to the spin-density-wave order parameter. However, fermions away from the hot spots can
also carry charge and energy and hence contribute to the transport co-efficients. In \cite{Patel:2014jfa}, it was assumed that
higher-order corrections to the critical theory \cite{metlitski3} 
are strong enough to rapidly equilibrate momentum around the entire Fermi surface. With this assumption, the slowest
process is the impurity-induced momentum relaxation, and this is dominated by the contributions of the hot spot fermions,
and can be computed \cite{Patel:2014jfa} from (\ref{eq:memform2}).
Here, the disorder was assumed to be associated with local variations in the position of the critical point, and so coupled linearly to
$(\varphi^a)^2$. In this case, the scaling law (\ref{eq:lss14}) was found to apply, and yields ${1}/{\tau_{\mathrm{imp}}} \sim T$ (up to logarithms),
implying a resistivity linear in $T$.

Finally, we consider the scaling of the shear viscosity $\eta$. The viscosity is well-defined in these critical theories with an emergent translation invariance. Using the same scaling arguments leading to 
(\ref{eq:sigmakappascale}) and (\ref{eq:entT}), or from the holographic result in (\ref{eq:KSSbound}), we expect that
\begin{equation}
\eta \sim s \sim T^{(d-\theta)/z}.
\label{eq:scaleviscosity}
\end{equation}
However, explicit computations \cite{Patel:2016ymd} on non-Fermi liquid models of a Fermi surface coupled to Ising-nematic or gauge fluctuations lead to a rather different conclusion. Although strong interactions can destroy quasiparticles
near the Fermi surface, they do leave the Fermi surface intact, as we discussed in \S \ref{sec:isingnematic}. 
So far, we have found that this residual Fermi surface leads to the same scaling of transport properties as 
might be expected in a critical and isotropic quantum fluid in $d-\theta$ spatial dimensions. 
But for the shear viscosity, the anisotropic structure of momentum space near each point on the Fermi surface
turns out to have important consequences, among which is the breakdown of (\ref{eq:scaleviscosity}):
computations on the non-Fermi liquid model show that the ratio $\eta/s$ diverges \cite{Patel:2016ymd}
\begin{equation}
\eta/s \sim T^{-2/z}. \label{eq:etasbreakdown}
\end{equation}
The result (\ref{eq:etasbreakdown}) does not emerge from holography using classical gravity. It depends strongly on the Fermi surface structure of the field theory. See the discussion in \S \ref{sec:lowEspect} concerning Fermi surface physics in holographic strange metals.
  
 \subsubsection{Transport in holographic compressible phases}
\label{sec:holexample}

Let us now go through a few examples of the holographic application of the memory matrix formula (\ref{eq:memform2}) for the momentum relaxation rate.  We have really done the hard work in \S\ref{sec:SWT} by computing the spectral weight of scalar operators in various critical holographic phases. The ease with which the results below are obtained --- in particular, without solving any partial differential equations in the bulk spacetime, corresponding to the inhomogeneous sources ---  reflects two powerful features of the approach. Firstly, the memory matrix has isolated the universal low energy data, the low energy spectral weight in (\ref{eq:memform2}), that determines the momentum relaxation rate. Secondly, in \S \ref{sec:SWT} we showed how the low temperature scaling of the low energy spectral weight could be computed from near-horizon data. Combining these two facts, the momentum relaxation rate and hence the transport can be obtained purely from the geometrical properties of the emergent critical phase in the far IR of the spacetime.

\begin{figure}
\centering
\includegraphics[width = 7.5cm]{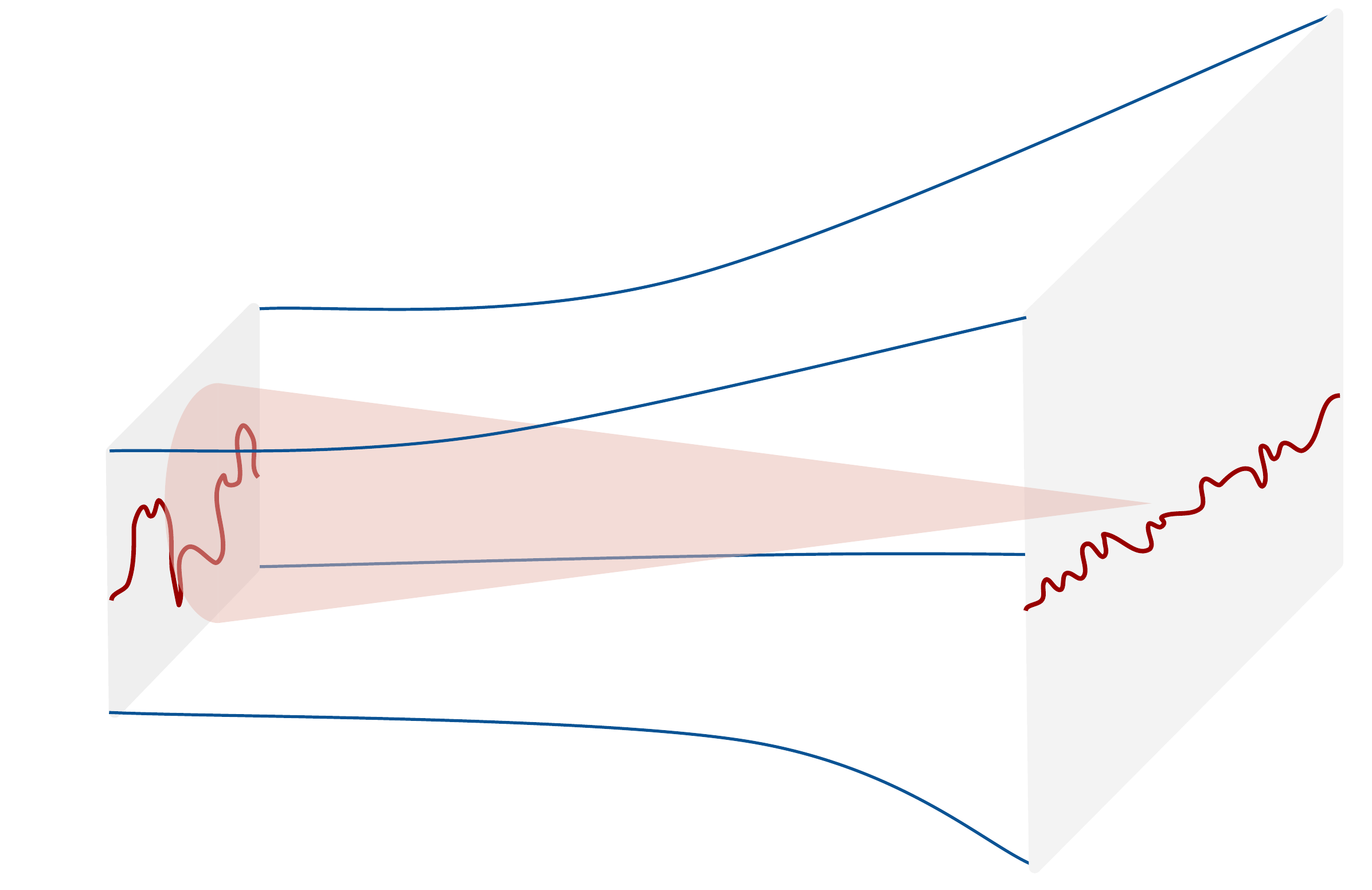}\\
\includegraphics[width = 7.5cm]{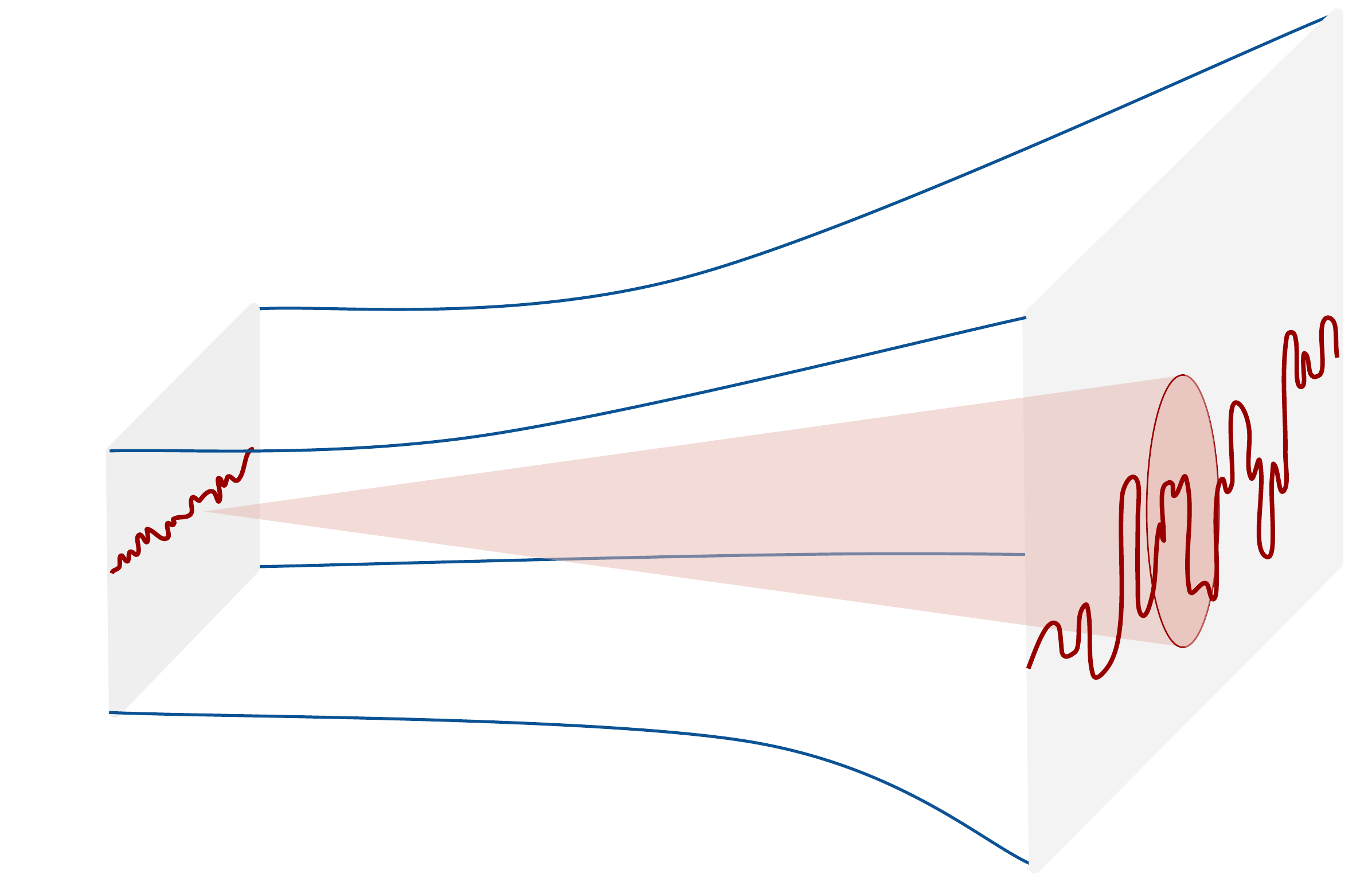}
\caption{{\bf Relevant and irrelevant inhomogeneities}. In the top plot the inhomogeneities grow towards the interior of the spacetime whereas in the bottom plot they decay towards the interior. In the irrelevant case one case use perturbation theory in the interior, almost homogeneous, geometry to describe momentum relaxation. In the relevant case one must find the new inhomogeneous interior spacetime, typically by solving PDEs.}
\label{fig:inhomog}
\end{figure}

Throughout this section we are assuming that the effects of translation symmetry breaking are weak in the IR, so that perturbation theory is possible.  Holographically speaking, this means that inhomogeneities induced by the boundary source (\ref{eq:Hbroken}) decay towards the interior of the spacetime. If, instead, the inhomogeneities are relevant and grow towards the interior, then a new, intrinsically inhomogeneous IR emerges. This will lead to incoherent hydrodynamic transport, without momentum, that is the subject of  \S\ref{sec:in1}--\ref{sec:in3} below. For generic situations without extra symmetries (to be discussed later), there is no way around solving bulk PDEs in those cases. The cases of relevant and irrelevant disorder are illustrated in Figure \ref{fig:inhomog}.
 
There are two important distinctions in the holographic models: firstly, whether momentum relaxation occurs due to a periodic lattice or a random potential, and secondly, whether the dynamic critical exponent $z$ is finite or infinite. Let us discuss these in turn. Note first that the memory matrix expression (\ref{eq:memform2}) for the momentum relaxation rate depends only on the $k$-dependent spectral weight $\rho(k)$ in (\ref{eq:rhok}), so that we can write
\be\label{eq:taurho}
\frac{{\mathcal M}}{\t_\text{imp}} = \int \frac{\mathrm{d}^dk}{(2\pi)^d} |h(k)|^2 k_x^2 \, \rho(k) \,.
\ee
The susceptibility ${\mathcal M}$ will generically be constant in a compressible phase at $T \ll \mu$. This formula will thus determine the temperature dependence of $\t_\text{imp}$ from the temperature dependence of $\rho(k)$.

Consider first the case of a regular, periodic lattice. For simplicity, focus on the periodicity in one direction and take the lattice potential to be a single cosine mode with lattice wavevector $k_{\mathrm{L}}$. These are not crucial assumptions for the physics we are about to describe. Then in (\ref{eq:taurho}) we have
\be
|h(k)|^2 \to \bar h^2 \delta(k - k_{\mathrm{L}})  \qquad \Rightarrow \qquad \frac{{\mathcal M}}{\t_\text{imp}} \sim \bar h^2 k_{\mathrm{L}}^2 \, \rho(k_{\mathrm{L}}) \,.
\ee
The essential point here is that the momentum relaxation rate is given in terms of the low energy spectral density at a nonzero wavevector. The lattice can only relax momentum efficiently if there exist low energy degrees of freedom that the lattice can scatter into. This again brings out the analogy to Fermi's Golden rule.

If $z < \infty$, then we saw in (\ref{eq:IRlif2}) above that low energy, nonzero wavevector spectral weight is exponentially suppressed. This reflects the intuitive fact that if $z < \infty$ then low energy excitations scale towards the origin of momentum space. A spatially periodic lattice is highly irrelevant with such scaling. It follows that momentum relaxation is very inefficient and momentum is exponentially long-lived
\be
\t_\text{imp} \sim \mathrm{e}^{+ C k_{\mathrm{L}}/T^{1/z}} \,.
\ee
An example of this long timescale, which leads to an exponentially large dc conductivity in (\ref{eq:swcoh}), was found numerically in the holographic computations of \cite{Horowitz:2013jaa}, for the normal component of the conductivity in the superconducting phase. It reflects the emergence (up to logarithmic corrections) of a $z=1$, $\mathrm{AdS}_4$ geometry in the far interior of the spacetime, and the corresponding inefficient scattering by a periodic lattice. Exponentially large conductivities can occur in conventional metals in a `phonon drag' regime, where the bottleneck for momentum relaxation is phonon umklapp scattering.  Evidence for this effect was presented in \cite{2012PhRvL.109k6401H}. The kinematic cause is the same: phonons do not have low energy spectral weight at nonzero wavevector and hence cannot scatter efficiently off a lattice.

As was emphasized in \S\ref{sec:zeroTspectral} and \S\ref{sec:SWT}, a key aspect of $z= \infty$ scaling is the existence of low energy, nonzero momentum spectral weight. In particular, at low temperatures we can use (\ref{eq:zinfSW}) in the formula (\ref{eq:taurho}) for the relaxation rate to obtain \cite{Hartnoll:2012rj}
\be\label{eq:tauinfty}
\frac{{\mathcal M}}{\t_\text{imp}} \sim \bar h^2 k_{\mathrm{L}}^2 \, T^{2 \nu_{k_{\mathrm{L}}} - 1} \,.
\ee
Recall that this results holds for $\mathrm{AdS}_2\times\mathbb{R}^d$ and also other geometries, conformal to $\mathrm{AdS}_2\times\mathbb{R}^d$, with $z = \infty$. The above expression (\ref{eq:tauinfty}) leads to a power law in temperature dc resistivity via (\ref{eq:swcoh}), with the power determined by the scaling dimension $\nu_{k_{\mathrm{L}}}$. This power law can be seen explicitly in full blown numerical results \cite{Horowitz:2012ky}. Perturbation theory is valid so long as the lattice is irrelevant and the resistivity goes to zero at $T=0$, which requires $\nu_{k_{\mathrm{L}}} > \half$. In later sections we will discuss the relevant case, leading to insulators, as well as the case of $\nu_k$ imaginary, leading to instabilities.

The discussion in \S\ref{sec:zeroTspectral} and \S\ref{sec:SWT} emphasized that the existence of low energy spectral weight was a property that $z=\infty$ theories shared with Fermi liquids. Similarly, the result (\ref{eq:tauinfty}) parallels the $T^2$ power law resistivity found from umklapp scattering by a lattice in Fermi liquids. Umklapp scattering is an interaction in which charge density is moved around the Fermi surface (at zero energy cost) with a net change in momentum by $k_{\mathrm{L}}$. A unified description of umklapp, holographic and conventional, is developed in \cite{Hartnoll:2012rj}.

The simplest model of weak disorder is Gaussian short range disorder. In this model the inhomogeneous source $h(x)$ of (\ref{eq:Hbroken}) is drawn randomly from the space of functions with average and variance
\be\label{eq:randomh}
\overline{h(x)} = 0 \,, \qquad \overline{h(x) h(y)} = \bar h^2 \delta^{(d)}(x-y) \,.
\ee
We will discuss the physics of random disorder further in \S\ref{sec:disfpt} below. For the moment we can simply use (\ref{eq:randomh}) in the formula (\ref{eq:taurho}) for the momentum relaxation rate, where it amounts to
\be\label{eq:disrho}
|h(k)|^2 \to \bar h^2  \;\; \Rightarrow \;\; \frac{{\mathcal M}}{\t_\text{imp}} \sim \bar h^2 \int \frac{\mathrm{d}^dk}{(2\pi)^d} k_x^2 \, \rho(k) \,.
\ee
Unlike the case of a periodic lattice, the random disorder contains modes of all wavelengths, and this is why an integral over $k$ survives. This means that disorder can scatter long wavelength modes and can efficiently relax momentum in theories with $z < \infty$.

For the Einstein-Maxwell-dilaton models with $z < \infty$, we saw in (\ref{eq:GOOT}) and (\ref{eq:IRlif2}) above that the spectral weight is exponentially suppressed for $T^{1/z} \lesssim k$, and a constant power of the temperature for $k \lesssim T^{1/z}$. Thus from (\ref{eq:disrho}) we obtain the resistivity \cite{hkms, Hartnoll:2008hs, lucas1401}
\begin{align}
\rho &= \frac{1}{\sigma} = \frac{{\mathcal M}}{\t_\text{imp}} \sim \bar h^2 \int^{T^{1/z}} \mathrm{d}k k^{d+1} T^{(2 \Delta' - 2 z - d)/z} \notag \\ &\sim \bar h^2 T^{2(1 + \Delta' - z)/z} \,. 
\end{align}
This is in agreement with the general scaling expectation (\ref{eq:lss14}) quoted above, as it should be. The disorder leads to a resistivity determined by the scaling dimension of the disordered operator. The perturbative computation is controlled so long as the resistivity goes to zero as $T \to 0$. We will discuss relevance and irrelevance of disorder in \S\ref{sec:disfpt}.

Finally, we can consider disorder in holographic models with $z = \infty$. The spectral weight is again (\ref{eq:zinfSW}) and hence the momentum relaxation rate and resistivity \cite{Hartnoll:2012rj, Anantua:2012nj, Lucas:2014sba}
\be
\rho = \frac{1}{\sigma} = \frac{{\mathcal M}}{\t_\text{imp}} \sim \bar h^2 \int \mathrm{d}k k^{d+1} T^{2 \nu_k - 1} \,.
\ee
Assuming that $\nu_k$ grows at large $k$, as it typically does, at low temperatures this integral is dominated by small $k$. At low $T$ we can write the integrand as a exponential, which then has saddle point
\be
k_\star^{2} \sim \frac{1}{\log \frac{1}{T}} \,.
\ee
He we used the fact that generically $d\nu_k/dk \sim k$ as $k \to 0$. See for instance (\ref{eq:RNnu}) above for the AdS-RN case (in some interesting cases there are cancellations such that $\mathrm{d}\nu_k/\mathrm{d}k \sim k^3$, as in for instance \cite{Davison:2013txa}, in such cases $k_\star^{4} \sim 1/\log \frac{1}{T}$).
Thus the resistivity
\be\label{eq:logrho}
\rho  = \frac{{\mathcal M}}{\t_\text{imp}} \sim \bar h^2 \frac{T^{2 \nu_0 - 1}}{\left(\log \frac{1}{T}\right)^{1+d/2}}  \,.
\ee
In the AdS-RN case, $2 \nu_0 - 1 = 0$ from (\ref{eq:RNnu}) and hence the resistivity is purely given by the `universal' logarithmic correction.
Note that the result (\ref{eq:logrho}) only depends on the dimension of the disordered operator as $k \to 0$ in the semi-locally critical IR theory.

\subsubsection{From holography to memory matrices}
\label{sec:memmatrixderivation}

The memory matrix results for the Drude conductivities, (\ref{eq:memform2}) and (\ref{eq:swcoh}), follow from general principles and apply to all systems. In the previous section we have used these results to obtain the conductivity in holographic models. It is also instructive to see how the memory matrix formulae themselves can be directly obtained holographically, and how the holographic computation of $\sigma(\omega)$ proceeds in a momentum-relaxing geometry. We will present the derivation in this section, following \cite{Lucas:2015vna}.  The derivation will be perturbative in the disorder strength, which is required to be small at all energy scales.   

Consider a homogeneous background EMD geometry, of the kind discussed in \S \ref{sec:EMD}. This background is now perturbed by an additional scalar field $\varphi$, that we can take to have action (\ref{eq:4phi}). We choose $\varphi$ to be dual to an operator of dimension $\Delta>(d+1)/2$. The asymptotic boundary conditions as $r \to 0$ corresponding to the inhomogeneous source (\ref{eq:Hbroken}) are
\begin{equation}
\varphi(x,r) = \frac{r^{d+1-\Delta}}{L^{d/2}} \int \frac{\mathrm{d}^dk}{(2\pi)^d} h(k) \mathrm{e}^{\mathrm{i} k \cdot x} + \cdots .
\end{equation}
Then at leading (linear) order in $h$, the bulk $\varphi$ will take the form
\begin{equation}\label{eq:scalars}
\varphi(x,r) = \int \frac{\mathrm{d}^dk}{(2\pi)^d} h(k) \varphi_0(k,r) \mathrm{e}^{\mathrm{i} k \cdot x},
\end{equation}
where $\varphi_0$ solves the wave equation of motion for $\varphi$ in the homogeneous bulk background, is regular at the black hole horizon and satisfies $\varphi_0(k,r\rightarrow 0) = r^{d+1-\Delta}L^{-d/2} + \cdots$.

Our goal now is to compute the conductivity in the homogeneous background perturbed by the scalar field (\ref{eq:scalars}). We will find that the
conductivity $\sigma(\omega)$ contains a Drude peak so long as $h$ is perturbatively small, and we will recover (\ref{eq:memform2}).  The Drude peak is a perturbative limit of a double expansion in small $\omega$ and small $h$.   From (\ref{eq:swcoh}), $\sigma_{\mathrm{dc}} \sim \tau \sim h^{-2}$;  evidently we must treat $h^{2}$ and $\omega$ as small parameters on the same footing.  Now, the backreaction of the inhomogeneous field (\ref{eq:scalars}) onto the background means that the original EMD fields are altered at order $h^2$. However, these perturbative corrections to the background cannot lead to a singular conductivity as $h, \omega \rightarrow 0$. We will see that the leading order effect of inhomogeneity --  smearing out the $\delta$ function in $\sigma(\omega)$ -- is achieved by the inhomogeneous scalar (\ref{eq:scalars}) directly.  We may thus treat the EMD fields as uniform, and only consider the nonzero momentum response in the $\varphi$ field \cite{Blake:2013owa}.   We further need only compute the holographic linear response problem to linear order in $\omega$ \cite{Lucas:2015vna}.

We first identify the perturbations which will be excited by an infinitesimal uniform electric field.   Since the electric field is uniform, the lowest order processes in $h$ occur when a linear perturbation $\delta\varphi(k)$ scatters off of the background $\varphi(-k)$.     Hence the perturbations of interest will be $\delta \varphi(k)$, which furthermore couples to spatially uniform spin 1 perturbations under the spatial $\mathrm{SO}(d)$ group: $\delta A_x$ and $\delta \tilde{g}_{tx} \equiv r^2 \delta g_{tx}/L^2$  (these perturbations both tend to constants at $r=0$).   We use the `axial' gauge $\delta g_{rx}=0$, as previously.    The linearized equations of motion for these fields read (as throughout, it is recommendable to use symbolic manipulation programs such as \texttt{Mathematica}, along with a differential geometry/general relativity package, to perform many tedious steps in such computations): \begin{widetext}
\begin{subequations}\label{eq:mmh1}\begin{align}
\frac{L^d}{2\kappa^2} \frac{\delta \tilde{g}_{tx}^\prime}{ar^d} &= \rho  \, \delta A_x - \frac{L^d b}{r^d\omega} \int \frac{\mathrm{d}^dk}{(2\pi)^d} k_x |h(k)|^2 \varphi_0(k,r)^2 \left(\frac{\delta\varphi(k,r)}{\varphi_0(k,r)}\right)^\prime, \\
\frac{e^2}{L^{d-2}} \rho \, \delta \tilde{g}_{tx}^\prime &= \left(br^{2-d}Z\delta A_x^\prime\right)^\prime + \frac{r^{2-d}Z}{b}\omega^2 \delta A_x, \\
-k_x\omega  \frac{\varphi_0(k,r)^2\delta \tilde{g}_{tx}^\prime}{br^d} &= \left(\frac{b\,\varphi_0(k,r)^2}{r^d} \left(\frac{\delta \varphi(k,r)}{\varphi_0(k,r)}\right)^\prime\right)^\prime + \frac{\omega^2}{br^d} \varphi_0(k,r)\delta\varphi(k,r). 
\end{align}\end{subequations}\end{widetext}
Recall that $a$ and $b$ are functions that appear in the background metric (\ref{eq:abmetric}), $Z$ appears in the EMD action (\ref{eq:EMDaction}) and $\rho$ is defined in (\ref{eq:gausslaw}).

Terms $\sim \omega^2$ in the above equations are beyond the order in perturbation theory to which we work, so we may drop them.   Now (\ref{eq:mmh1}) collapses to a set of 3 differential equations: \begin{subequations}\label{eq:mmh2}\begin{align}
\frac{L^d}{2\kappa^2} \frac{\delta \tilde{g}_{tx}^\prime}{ ar^d} &= \rho \, \delta A_x - L^d \delta\mathcal{P}_x, \label{eq:mmh2a} \\
\frac{e^2}{L^{d-2}} \rho \,  \delta \tilde{g}_{tx}^\prime&= \left(br^{2-d}Z\delta A_x^\prime\right)^\prime, \label{eq:mmh2b} \\
\delta \mathcal{P}_x^\prime &= -\frac{\delta \tilde{g}_{tx}}{b} \int \frac{\mathrm{d}^dk}{(2\pi)^d} \frac{k^2 |h(k)|^2 \varphi_0(k,r)^2}{dr^d}  \,, \label{eq:mmh2c}
\end{align}\end{subequations}
where in the last line above we have assumed isotropy to replace $k_x^2 \rightarrow k^2/d$ (specifically, we assume for
simplicity that the modes $h(k)$ are chosen so that $\sigma_{ij} = \sigma \delta_{ij}$; the theory is microscopically disordered but macroscopically isotropic), and defined the new field\begin{equation}
\delta \mathcal{P}_x \equiv \frac{b}{r^d\omega} \int \frac{\mathrm{d}^dk}{(2\pi)^d} k_x |h(k)|^2 \varphi_0(k,r)^2 \left(\frac{\delta \varphi(k,r)}{\varphi_0(k,r)}\right)^\prime.
\end{equation}

We now proceed to directly construct (at leading order) the bulk response to the applied electric field.  Counting derivatives in (\ref{eq:mmh2}) shows there are 4 linearly independent modes.   With the benefit of hindsight, none are singular in $h$, and so we may look for these modes assuming $h=0$.  The first such mode -- and the only one containing $\delta\mathcal{P}_x$, is: \begin{equation}
\rho \, \delta A_x = L^d \delta\mathcal{P}_x = \text{constant}, \;\;\; \delta\tilde{g}_{tx} = 0.   \label{eq:mmh3}
\end{equation}
The remaining three are the modes of the translation invariant black hole, and so have $\delta \mathcal{P}_x =0$.   There is a ``diffeomorphism mode" $\delta\tilde{g}_{tx} = 1$,  a ``Galilean boost" mode: \begin{equation}
\delta\tilde{g}_{tx}=1-ab,\;\;\; \delta A_x = p + \frac{Ts}{\rho },   \label{eq:mmh4}
\end{equation}
and a third mode (explicitly given in \cite{Lucas:2015vna}) which is the ``Wronskian" partner to this Galilean boost mode; it is logarithmically divergent at the horizon.   (\ref{eq:mmh2a}) near $r=r_+$ allows us to fix the constant term in $\delta A_x$ in (\ref{eq:mmh4}). Recall that $p$ is the background gauge field in (\ref{eq:abmetric}).

What is the bulk response if we impose the boundary condition that $\delta A_x(\omega,r=0) =1$ and infalling boundary conditions at the horizon?  We do not source a heat current, and so $\delta \tilde{g}_{tx}(r=0)=0$ and we do not source the diffeomorphism mode. We are looking for the response at small frequency and small inhomogeneity: $\w \sim h^2 \to 0$. We will see that the response is not singular in this limit, and so we need only compute the response at  $\mathcal{O}(\omega^0)$. However, the response does contain a term proportional to $\omega/h^2$ that we must keep track off.  Strictly at
$\omega=0$, $\sigma(0) = \sigma_\text{dc}$ remains finite, but there is no electric field in this limit (the electric field being the time derivative of $\delta A_x$), and hence there can be no current sourced. The upshot is that at $\omega = 0$ we source only the mode (\ref{eq:mmh3}). 

Next consider a nonzero but small $\omega$. We are after the $\mathcal{O}(\omega)$ correction to the response of the previous paragraph. At nonzero frequency we must impose infalling boundary conditions (\ref{eq:infalling}) at the horizon. Recalling the behavior (\ref{eq:b4pT}) of the function $b$ near the horizon, we find that equation (\ref{eq:mmh2c}) requires the near horizon behavior
\begin{equation}\label{eq:tgx}
\delta \tilde{g}_{tx} = -\frac{\mathrm{i}\omega \rho \tau}{\epsilon+P} (r_+ - r)^{- \mathrm{i} \omega/(4 \pi T)} \left( 1 + \cdots \right) \,,
\end{equation}
where
\begin{equation}
\frac{\epsilon +P}{\tau} \equiv \frac{L^d}{r_+^d} \int \frac{\mathrm{d}^dk}{(2\pi)^d} \frac{k^2}{d}|h(k)|^2 \varphi_0(k,r_+)^2.  \label{eq:tauhol}
\end{equation}
To get the factor of $\rho$ in (\ref{eq:tgx}), we use the $\omega = 0$ result from (\ref{eq:mmh3}) that $\delta {\mathcal{P}}_x = \rho/L^d \, \delta A_x = \rho/L^d$. The last step here uses the boundary condition $\delta A_x = 1$. Note that in the scaling limit $\omega \sim h^2$, then $\delta \tilde{g}_{tx}$ in (\ref{eq:tgx}) is 
actually $\mathcal{O}(\omega^0)$  in perturbation theory. At this order in perturbation theory we can set the $\omega$ in the exponent to zero, effectively removing the oscillating part for most purposes.

The lifetime defined in (\ref{eq:tauhol}) is in fact precisely that obtained from the memory matrix in (\ref{eq:memform2}). To see this, recall the formula derived in (\ref{eq:SWT}) above, in which the low energy spectral weight is given in terms of the value of the field on the horizon, which is precisely what appears in (\ref{eq:tauhol}).

To obtain the Drude peak itself, with the lifetime (\ref{eq:tauhol}), we turn to the perturbation $\delta A_x$. Regularity at the horizon requires that the only mode sourced -- upon including the metric perturbation (\ref{eq:tgx}) at order $\omega/h^2$ -- is the Galilean boost mode (\ref{eq:mmh4}).  Adding this term to the $\omega = 0$ solution of $\delta A_x = 1$, we obtain
\begin{equation}\label{eq:axxx}
\delta A_x(\omega, r) \approx 1 -\frac{\mathrm{i}\omega \rho \tau}{\epsilon+P} \left(p+\frac{Ts}{\rho }\right).
\end{equation}
We have neglected the oscillatory part $(r_+ - r)^{- \mathrm{i} \omega/(4 \pi T)}$ that is unimportant at small $\omega$, except very close to the horizon. The Drude form (\ref{eq:swcoh}) for the conductivity now follows from (\ref{eq:axxx}) using basic entries of the holographic dictionary: The conductivity is given by (\ref{eq:optical}), with the current given by (\ref{eq:jx1}), with $z=1$, and using (\ref{eq:jt1}) to relate the subleading behavior in $p$ to $\rho $ (recall $p(0)=\mu$).

The following comment on the above result is useful. We saw that the bulk response is dominated by a `Galilean boost'.  This is the holographic analogue of the derivation of the Drude conductivities from hydrodynamics in (\ref{eq:memmatrixtime}). In the holographic argumentation, however, we did not need to assume that the inhomogeneities were slowly varying. We will see other instances below of how holographic models often admit a `hydrodynamic' description beyond its naive regime of validity.

\subsection{Magnetotransport}
\label{sec:magtrans}
   
This section will discuss transport in a magnetic field as well as disorder. We will restrict to $d=2$;  in higher dimensions, adding a magnetic field breaks isotropy.   As previously, we start with a discussion of the predictions of hydrodynamics. We will generalize both the hydrodynamic discussion of \S\ref{sec:weakI} as well as the memory matrix approach of \S\ref{sec:memorymatrix}.

The setup is similar to before: a fluid with translational symmetry broken weakly by an inhomogeneous field $h(x)$ coupled to a scalar operator $\mathcal{O}$, and now additionally in a uniform magnetic field $B$. In the case in which the inhomogeneities are slowly varying, the entire discussion can be made within hydrodynamics, as in \S\ref{sec:weakI} above. The `momentum conservation' equation now reads
\begin{equation}\label{eq:lore}
\partial_t  \mathfrak{g}^i + \partial_j T^{ij} = O \partial^i h_0 + F^{i\mu}J_\mu,
\end{equation}
with $F_{\mu\nu}$ the (externally imposed) gauge flux;  in our case, we will have $F_{xy}=B$, as well as $F^{it} = -E$.   We assume that $B\sim \omega \sim \mathcal{O}(h_0^2)$.  The magnetic field must be small else it will introduce a scale that takes us outside of the hydrodynamic regime. The `force balance' equation (\ref{eq:233}) for the fluctuations now becomes \begin{align}
-\mathrm{i}\omega \mathcal{M} v_j &+ (O+\delta O) \partial_j (h_0 + \delta h) \notag \\
&= \rho E_j + (O+\delta O) \partial_j h_0 + B \varepsilon_{jk} J_k.
\end{align} 
The treatment of $\delta h$ is identical to before -- $B$-dependent corrections to $\tau$ are subleading, just as $\omega$-dependent corrections were.  Thus upon spatial averaging we obtain
\begin{equation} \label{eq:HKMSB0}
-\mathrm{i}\omega \mathcal{M} v_j = \rho E_j + \rho B\varepsilon_{jk}v_k - \frac{\mathcal{M}}{\tau_{\mathrm{imp}}}v_j,
\end{equation} 
with an identical $\tau$ to previously, given in (\ref{eq:memmatrixtime}).  Solving for the velocities in terms of the electric field, and using $J_i = \rho v_i$, the longitudinal and Hall conductivities are obtained as
 \begin{subequations}\label{eq:HKMSB1}\begin{align}
\sigma_{xx} &= \sigma_{yy} = \frac{\rho ^2\mathcal{M}(\tau_{\mathrm{imp}}^{-1}-\mathrm{i}\omega)}{(\rho B)^2 + \mathcal{M}^2(\tau_{\mathrm{imp}}^{-1}-\mathrm{i}\omega)^2} \,, \\
\sigma_{xy} &=  -\sigma_{yx}  = \frac{\rho ^3B}{(\rho B)^2 + \mathcal{M}^2(\tau_{\mathrm{imp}}^{-1}-\mathrm{i}\omega)^2}.
\end{align}\end{subequations}

The incoherent conductivity can be added into the above derivation by using $J_i = \sigma_{\textsc{q}} (E_i + B \epsilon_{ij} v_j) + \rho v_i$ in (\ref{eq:lore}) and in the definition of the conductivities, instead of only including the latter term.   One finds \cite{hkms}:
\begin{subequations}\label{eq:HKMSB}\begin{align}
\sigma_{xx} &= \sigma_{\textsc{q}} \frac{(\tau_{\mathrm{imp}}^{-1}-\mathrm{i}\omega)(\tau_{\mathrm{imp}}^{-1} + \omega_c^2/\Gamma + \Gamma-\mathrm{i}\omega)}{\omega_c^2 + (\tau_{\mathrm{imp}}^{-1}+\Gamma-\mathrm{i}\omega)^2} \,, \\
\sigma_{xy} &=  \frac{\r}{B}\frac{2(\tau_{\mathrm{imp}}^{-1}-\mathrm{i}\omega)\Gamma + \omega_c^2 + \Gamma^2}{\omega_c^2 + (\tau_{\mathrm{imp}}^{-1}+\Gamma-\mathrm{i}\omega)^2}.
\end{align}\end{subequations}
Thermoelectric conductivities have also been obtained in \cite{hkms}.
In the above formulas we introduced the hydrodynamic cyclotron frequency and decay rate
\be\label{eq:spole}
\omega_c \equiv \frac{\rho B}{\mathcal{M}} \,, \qquad \Gamma \equiv \frac{\sigma_Q B^2}{\mathcal{M}} \,.
\ee
Upon setting $\tau_{\mathrm{imp}}^{-1} = 0$, the formulae in (\ref{eq:HKMSB}) and (\ref{eq:spole}) agree exactly with the holographic results in equations (\ref{eq:aabb}) and (\ref{eq:ccdd}) above. In this identification $\sigma_{\textsc{q}} = 1/e^2$, which is the limit of the incoherent conductivity (\ref{eq:incclean}), in $d=2$ and with $Z_+ = 1$, when $\rho$ is small (so that $sT = \epsilon + P$), which was assumed in the holographic derivation of (\ref{eq:aabb}) and (\ref{eq:ccdd}). We will discuss the physics of this cycotron mode below. 

The agreement with the holographic results implies that the hydrodynamic expression (\ref{eq:HKMSB}) will exhibit a particle-vortex duality, dual to the Maxwell duality in the bulk. The transformation (\ref{eq:dualitysigma}) of the conductivities is manifested as follows. Under the transformations \begin{equation}
\rho  \leftrightarrow B \quad  \text{ and } \quad \sigma_{\textsc{q}} \leftrightarrow 1/\sigma_{\textsc{q}} \label{eq:selfdual2}
\end{equation}then \begin{equation}
\sigma_{xx} \leftrightarrow \rho_{xx} \quad \text{ and } \quad \sigma_{xy} \leftrightarrow -\rho_{xy}
\end{equation} 
where $\rho  = \sigma^{-1}$ is the resistivity matrix. 

It is instructive to derive (\ref{eq:HKMSB1}) from the memory matrix formalism.   To do so, we must employ the full expression (\ref{eq:memmatrixgeneral}) for the conductivity, including the matrix $N$ which breaks time reversal symmetry.  We will outline the computations contained in \cite{Lucas:2015pxa, Davison:2016hno}. For simplicity we will focus on the case $\t_\text{imp} = \infty$, that is, with no inhomogeneities. Momentum is relaxed by the magnetic field. The operator equation for momentum relaxation is now
\be\label{eq:PB}
\dot P^i = B \epsilon^{ij} J^j  \,.
\ee
The space of `slow operators' will now be $\{P_x,P_y,J_x,J_y\}$.

To evaluate the memory matrix expression for the conductivity, firstly we need the susceptibilities. These are symmetric in the operators and hence, using isotropy, the nonzero components are
\be
\chi_{J^i J^j} = \chi_{J^x J^x} \delta^{ij} , \;\; \chi_{J^i P^j} = \rho \delta^{ij} , \;\; \chi_{P^i P^j} =  {\mathcal M} \delta^{ij} \,.
\ee
The reason that $\chi_{J_x P_y}$ vanishes is that, using (\ref{eq:PB}) and (\ref{eq:NAB}), $\chi_{J_x P_y} = - N_{P_y P_y}/B = 0$, because the $N$ matrix is antisymmetric. The nonzero components of $N$ are evaluated using (\ref{eq:NAB}) to be
\be
N_{J^i J^j} = N_{J^x J^y} \epsilon^{ij} \,, \; N_{J^i P^j} = - B \chi_{J^xJ^x} \epsilon^{ij} \,, \; N_{P^i P^j} = - B \rho \epsilon^{ij} \,.
\ee
Finally we also need the nonzero components of the memory matrix $M$ in (\ref{eq:MAB}). These are simplified by the fact that the projection operator $\mathfrak{q}$ in the memory matrix projects out $J^i$, and these same $J^i$ appear in the expression (\ref{eq:PB}) for $\dot P^i$. Thus one finds
\be
M_{J^i J^j} \neq 0 \,, \qquad M_{J^i P^j} = 0 \,, \qquad M_{P^i P^j} = \frac{\mathcal{M}}{\tau_{\mathrm{imp}}} \,.
\ee

If the above expressions for $\chi,N,M$ are used in the memory matrix conductivity (\ref{eq:memmatrixgeneral}), one recovers precisely the conductivities (\ref{eq:HKMSB}) in the scaling limit $\omega \sim B$. In order for the incoherent conductivity $\sigma_{\textsc{q}}$ to contribute at leading order in this limit one must furthermore take $\chi_{J^x J^x} \sim 1/\sqrt{\omega}$. See the discussion in \S \ref{sec:incformal} above. The incoherent conductivity is now given by
\be
\sigma_{\textsc{q}} = \frac{\chi_{J^x J^x}^2 M_{J^x J^x}}{M_{J^x J^x}^2+(M_{J^x J^y} + N_{J^x J^y})^2} \,.
\ee
This is the generalization of (\ref{eq:sQinc}) above. As discussed in that section, one can express $\sigma_{\textsc{q}}$ in terms of incoherent currents, but the results are not distinguishable at leading order in this scaling limit. In the present setup there is also an incoherent Hall conductivity $\sigma_{\textsc{q}}^{\mathrm{H}}$, but this is expected to be subleading in powers of $B$ in the scaling limit and hence we have not included it.

One of the interesting features of (\ref{eq:HKMSB}) is a violation of Kohn's theorem in the clean limit $\tau_\text{imp} = \infty$.  Suppose we take a translationally invariant fluid with $\sigma_{\textsc{q}}=0$.  Then we see that the conductivities given by (\ref{eq:HKMSB1}) diverge at $\pm \omega = \omega_{\mathrm{c}}$. This is called the cyclotron resonance, and is associated with exciting a collective mode which uniformly rotates the momentum of the fluid. Kohn's theorem states that this cyclotron resonance frequency is robust against electron-electron interactions in a Galilean-invariant fluid \cite{Kohn1961}.  In a quantum critical fluid with $\sigma_{\textsc{q}}\ne0$, but still $\tau_\text{imp}=\infty$, we see that there is no longer any divergence, as the cyclotron resonance is damped by the particle-hole ``friction". The loophole in Kohn's theorem is that we have both electrons and holes, and that a generic quantum critical system does not conserve the number of electrons (due to electron-hole creation/annihilation processes). It is not Galilean invariant.

Another interesting feature of (\ref{eq:HKMSB}) is the possibility for anomalous scaling in the Hall angle, defined as \begin{equation}
\tan \theta_{\mathrm{H}} = \frac{\sigma_{xy}}{\sigma_{xx}}.
\end{equation}
The present discussion is at $\omega=0$.  If $\sigma_{\textsc{q}} = 0$,  then 
\begin{equation}
\tan \theta_{\mathrm{H}} \sim \frac{B\rho}{\mathcal{M}} \tau_{\mathrm{imp}} \sim \sigma_{xx}.  \label{eq:tanth}
\end{equation}
The last relation is concerned with the temperature scaling, which comes from $\tau_{\mathrm{imp}}$.
Mechanisms that evade the connection (\ref{eq:tanth}) have been of interest since it has been known for some time that (\ref{eq:tanth}) is violated in the cuprates \cite{Chien1991}. A nonzero
$\sigma_{\textsc{q}}$ in the formulae (\ref{eq:HKMSB}) provides such a mechanism \cite{Blake:2014yla}:  suppose that $\sigma_{xx}(B=0)\approx\sigma_{\textsc{q}}$ (recall from (\ref{eq:hkmssigma}) that $\sigma_{xx}$ is a sum of two terms).   Using (\ref{eq:HKMSB}), the scaling $\tan \theta_{\mathrm{H}} \sim \tau_\text{imp}$ is unchanged. It follows that if $\tau_{\mathrm{imp}}$ and $\sigma_{\textsc{q}}$ have unrelated $T$-scaling,  then the scaling $\tan\theta_{\mathrm{H}} \sim \sigma_{xx}$ need not hold. This conclusion requires  $\sigma_{\textsc{q}}$ to dominate the dc conductivity in (\ref{eq:hkmssigma}). If, instead, $\tau_\text{imp}$ is very large, perhaps in an ultra-clean sample, then $\tan \theta_{\mathrm{H}} \sim \sigma_{xx}$ is recovered. 

As we will later discuss below (\ref{eq:sigma15}),  at higher orders in perturbation theory in $1/\tau_{\mathrm{imp}}$, corrections appear to the hydrodynamic theory of \cite{hkms}, and to the equations (\ref{eq:HKMSB}).  In particular, if $\sigma_{\textsc{q}} \sim \tau^0_{\mathrm{imp}}$, corrections arise to (\ref{eq:HKMSB}) at the same order as terms involving $\sigma_{\textsc{q}}$  \cite{Blake:2015hxa}.   In the dc limit $\omega=0$, holographic mean-field models will be seen to lead to an exact formula that looks similar to the dc limit of (\ref{eq:HKMSB}). Similarly to in (\ref{eq:axionsigma}) below, however, the quantities that appear are not in fact 
the same $\sigma_{\textsc{q}}$ and $\tau$ that appear in the $\omega$ dependent conductivities. In the thermoelectric conductivities $\alpha_{ij}$ and $\bar\kappa_{ij}$ the exact holographic mean-field results are qualitatively different to the theory
of \cite{hkms} even at $\omega=0$ \cite{Amoretti:2015gna, Blake:2015ina, Kim:2015wba}.    

\subsubsection{Weyl semimetals: anomalies and magnetotransport}
\label{sec:weylSM}

Let us also briefly discuss transport in models of 3+1 dimensional metals with approximate axial anomalies: Weyl semimetals.  At the level of band theory, there are many predictions and experimental realizations \cite{2015PhRvX...5c1013L,2015Sci...349..613X,2015Sci...349..622L} of metallic systems with `Weyl points' in the Brillouin zone where the low energy degrees of freedom exhibit a chiral (axial) anomaly.  This axial anomaly is an emergent IR effect: (\emph{i})  high energy processes allow for scattering between fermions located near distinct Weyl points, and (\emph{ii}) there is a theorem which states that the total axial anomaly of lattice fermions must vanish \cite{ninomiya}.  Temporarily putting aside this theorem, at the hydrodynamic level, the axial anomaly causes the breakdown of charge conservation \cite{Son:2009tf, Neiman:2010zi}: \begin{equation}
\partial_\mu J^\mu = -\frac{C}{8} \epsilon^{\mu\nu\rho\sigma}F_{\mu\nu}F_{\rho\sigma} \,. \label{eq:anomhydro}
\end{equation}
The right hand side of (\ref{eq:anomhydro}) is proportional to $\vec E \cdot \vec B$.
In a background magnetic field, the transport problem for charge is ill-posed with such a violated conservation law, so clearly the microscopic processes which restore the conservation of global charge will play an important role. 

A hydrodynamic approach to thermoelectric transport in such anomalous theories was developed in \cite{Lucas:2016omy}, by coupling together chiral relativistic fluids, allowing for the exchange of energy and charge between the fluids.  The key result is that, if the magnetic field is oriented in the $z$-direction, there is an additional contribution to the conductivity $\sigma_{zz}$ as a consequence of the axial anomaly: \begin{equation}
\Delta \sigma_{zz}(\omega) = \frac{C^2B^2}{\Gamma_{\mathrm{anom}}-\mathrm{i}\omega},  \label{eq:sigmazz}
\end{equation}
where $\Gamma_{\mathrm{anom}}$ is a coefficient related to the rate at which the chiral fluids exchange charge and energy.  It can be defined using the memory matrix formalism, and is small in the limit where the `Weyl' behavior of the semimetal is pronounced, and hence $\Delta \sigma_{zz}$ will be large.  The $B$-dependence of this signal, only present parallel to $\vec B$, is called negative magnetoresistance (NMR).  NMR in $\sigma_{zz}$ was predicted earlier using kinetic approaches in \cite{2013PhRvB..88j4412S}, and hydrodynamic/holographic methods in  \cite{Landsteiner:2014vua}. Experimental signatures for electrical NMR were first definitively observed in \cite{2015ong,2015PhRvX...5c1023H}, along with many additional experiments since. 

It was further noted in \cite{Lucas:2016omy} that there is NMR in $\alpha$ and $\bar\kappa$ if there is a non-vanishing axial-gravitational anomaly.  (This distinct anomaly implies an $R_{\mu\nu\rho\sigma}^2$ term on the right hand side of (\ref{eq:anomhydro}), and anomalous violation of energy-momentum conservation as well.)  The link between NMR and the axial-gravitational anomaly is independent of the strong coupling limit, and can be observed in the weakly coupled limit accessible in experiment.  Hence, thermal transport measurements could provide a remarkable demonstration of this novel anomaly.

There has been recent development of holographic approaches to studying the strongly coupled analogues of Weyl semimetals \cite{Landsteiner:2014vua, Jimenez-Alba:2015awa, Landsteiner:2015pdh, Landsteiner:2016stv}.  The analogue of $\Gamma_{\mathrm{anom}}$ is added by coupling the axial gauge field to a certain background scalar \cite{Jimenez-Alba:2015awa}.   When $\Gamma_{\mathrm{anom}}=0$, one might expect (\ref{eq:sigmazz}) to hold even beyond the hydrodynamic limit as the consequence of a new `Ward identity', analogous to the case of a conserved momentum (\ref{eq:Cleansigma}). A holographic calculation seems to confirm this conjecture \cite{Sun:2016gpy}.  In the future, it would be interesting to understand the interplay of NMR with strong disorder.   It would also be interesting to study thermal transport holographically.  However, following the lines of \cite{Lucas:2016omy} would require a model with two separate `energy-momentum tensors', only one of which is conserved.  This may require the addition of a massive spin-2 field in the bulk, which is a delicate matter.

\subsection{Hydrodynamic transport (without momentum)}
\label{sec:in1}

So far, we have discussed hydrodynamic transport assuming that (up to weak disorder) energy, momentum and charge are all good conserved quantities.   If translational symmetry is very strongly broken -- for example, the amplitude of short wavelength disorder is large -- then there is no reason to expect that the dynamics associated with momentum relaxation are slower than other microscopic processes.   In this limit, the only conserved quantities are charge and energy \cite{Hartnoll:2014lpa}. (If inelastic phonon scattering is important, one should consider the combined electron-phonon system.) Therefore,
the hydrodynamic description should only include fluctuations of the chemical potential $\mu$ and temperature $T$.  
We will call such a system an ``incoherent metal".  

In the absence of external sources the two hydrodynamic equations of motion are conservation of charge and energy
\be\label{eq:inccons}
\frac{\pa \rho}{\pa t} + \nabla \cdot J = 0 \,, \qquad \frac{\pa \epsilon}{\pa t} + \nabla \cdot J_{\mathrm{E}} = 0 \,.
\ee
There is no longer a conserved momentum. This furthermore means that velocity is no longer a hydrodynamic variable, and hence the constitutive relations for the charge and heat currents, to first order in the derivative expansion, are simply
\be\label{eq:inccons2}
\left(\begin{array}{c} J  \\ \frac{1}{T} Q \end{array}\right) = \left(\begin{array}{cc} \sigma_0 &\ \alpha_0 \\ \alpha_0 &\ \frac{1}{T} \bar\kappa_0 \end{array}\right) \left(\begin{array}{c} - \nabla \mu  \\ - \nabla T \end{array}\right) \,.
\ee
As previously the heat current $Q = J_{\mathrm{E}} - \mu J$. It is immediate from the above expression that $\s_0,\a_0$ and $\k_0$ are nothing other than the thermoelectric conductivities. Recall that the matrix of thermoelectric susceptibilities is
\be\label{eq:incchi}
\left(\begin{array}{c} \nabla \rho  \\ \nabla s \end{array}\right) = \left(\begin{array}{cc} \chi &\ \zeta \\ \zeta &\ c/T \end{array}\right) \left(\begin{array}{c} \nabla \mu  \\ \nabla T \end{array}\right) \,.
\ee
Combining the conservation laws (\ref{eq:inccons}) with the constitutive relations (\ref{eq:inccons2}) and susceptibilities (\ref{eq:incchi}) one obtains coupled diffusion equations for entropy and charge
\be\label{eq:incdiff}
 \left(\begin{array}{c} {\pa \rho}/{\pa t}  \\ {\pa s}/{\pa t} \end{array}\right) = - \left(\begin{array}{cc} \sigma_0 &\ \alpha_0 \\ \alpha_0 &\ \frac{1}{T} \bar\kappa_0 \end{array}\right) \left(\begin{array}{cc} \chi &\ \zeta \\ \zeta &\ c \end{array}\right)^{-1} \left(\begin{array}{c} \nabla^2 \rho  \\  \nabla^2 s \end{array}\right) \,.
\ee
We used the fact that the conservation laws (\ref{eq:inccons}) imply that within linear response: $\dot s + \nabla \cdot (Q/T) = 0$. The above equations can also be expressed as the coupled diffusion of energy and charge \cite{Hartnoll:2014lpa}. The diffusion equations reveal the generalized Einstein relation according to which the matrix of conductivities is equal to the matrix of diffusivities times the matrix of susceptibilities.

Thus incoherent metals are characterized by two independent diffusive modes, the eigenvectors of the coupled diffusion equations above. If thermoelectric effects are weak (as is the case in e.g. conventional metals), then heat and charge diffusion are directly decoupled. Charge diffusion alone controls the electrical conductivity. The loss of Lorentz or Galliliean invariance means that we can proceed no further on symmetries alone -- unlike the hydrodynamics described previously.  There are few non-trivial constraints beyond the second law of thermodynamics, which enforces that the matrix of conducitivities is positive definite. 

Until recently, most understanding of hydrodynamic transport in this incoherent limit has been phenomenological and non-rigorous.  There is a large literature on `phases' of transport in disordered media.   Various methods for tackling such problems have been developed, including effective medium theory (EMT) \cite{Landauer1952} and analogies with resistor networks \cite{Kirkpatrick1971}.    For simple inhomogeneous metals,  EMT can work quite well.  A simple model of the metal-insulator transition  beyond EMT consists of a resistor network with a random fraction $p$ of resistors deleted (so the resistance $R=\infty$ for that link);  the classical percolation transition of these deleted links then becomes the metal-insulator transition \cite{Kirkpatrick1971}.  Some of these techniques generalize to diffusive transport controlled by (\ref{eq:incdiff}), and indeed to hydrodynamic transport with momentum \cite{Lucas:2015lna}.

Ultimately to use the approaches mentioned in the previous paragraph, we need knowledge of the coefficients in (\ref{eq:inccons2}) or (\ref{eq:incdiff}), and in this regard progress has been more limited. In the absence of useful symmetries, a compelling approach is to ask whether any of these coefficients are subject to universal bounds following from general principles of quantum mechanics and statistical physics. We will briefly discuss some (still speculative) ideas in this direction. The remaining sections below will discuss specific, calculable models of incoherent metals that are obtained in holography.

 A simple argument leads to the so-called Mott-Ioffe-Regel bound in quasiparticle theories of transport \cite{gunnarsson}, in the presence of a sharp Fermi surface (in momentum space, so that excitations have a well defined momentum). There are several ways to formulate this bound but the one that is most relevant to our discussion is the following.  A well-defined quasiparticle must have a mean free path $l_{\mathrm{mfp}} = v_{\mathrm{F}} \tau$ larger than its wavelength $\sim k_{\mathrm{F}}^{-1}$.   Along with the definition of quasiparticle mass $m=\hbar k_{\mathrm{F}}/v_{\mathrm{F}}$, we obtain from the Drude formula \begin{equation}
\sigma = \frac{ne^2\tau}{m} \sim \frac{e^2}{\hbar} \frac{k_{\mathrm{F}}^{d}v_{\mathrm{F}}\tau}{\hbar k_{\mathrm{F}}} = k_{\mathrm{F}}l_{\mathrm{mfp}} \times \frac{e^2}{\hbar} k_{\mathrm{F}}^{d-2} \gtrsim \frac{e^2}{\hbar} k_{\mathrm{F}}^{d-2}.  \label{eq:MIRbound}
\end{equation}
This bound is violated in many strange metals \cite{gunnarsson}, and if these systems lack quasiparticles this is not surprising.   Nonetheless, these compounds do seem to saturate the bound \cite{Bruin2013} \begin{equation}
\sigma = \frac{ne^2\tau}{m} \sim \frac{\tau E_{\mathrm{F}}}{\hbar} \frac{e^2}{\hbar} k_{\mathrm{F}}^{d-2}  \gtrsim \frac{e^2}{\hbar} k_{\mathrm{F}}^{d-2} \frac{E_{\mathrm{F}}}{k_{\mathrm{B}}T},  \label{eq:MIR2}
\end{equation}
where we have used the Heisenberg uncertainty principle to bound the scattering lifetime:  $\tau \gtrsim \hbar/(k_{\mathrm{B}}T)$. Such a lifetime bound is of course in the same spirit \cite{ssbook,ZaanenNature04} as the bound (\ref{SSbound}). It is desirable, however, to formulate a potential bound directly in the language of incoherent transport, namely, diffusion.

A precedent for the reformulation of uncertainty principle bounds as diffusivity bounds already exists. An argument identical to that of (\ref{eq:MIR2}) can be applied to the ratio of shear viscosity to entropy density \cite{Kovtun:2004de}. In a CFT at nonzero temperature, this ratio is precisely the transverse diffusivity of momentum \cite{Son:2007vk}, so that the famous conjectured viscosity bound of \cite{Kovtun:2004de} can be written 
\be
D_{\pi_\perp} \gtrsim \frac{\hbar c^2}{k_{\mathrm{B}} T} \,.
\ee
We will be agnostic about the numerical prefactor in this and (most) other bounds \cite{Cremonini:2011iq}.
Similarly, the bound in (\ref{eq:MIR2}) can be reformulated, using the Einstein relation $\sigma = D\chi$ and assuming decoupled heat and charge diffusion for simplicity, as a bound on charge diffusion  \cite{Hartnoll:2014lpa}
\begin{equation}
D_\rho \gtrsim \frac{\hbar v_{\mathrm{F}}^2}{k_{\mathrm{B}}T}\, ,  \label{eq:MIR3}
\end{equation}
under plausible assumptions about the charge susceptibility $\chi$.  Such a bound could provide a simple explanation for the ubiquitous scaling $\sigma \sim 1/T$ observed in incoherent metals \cite{Bruin2013}.

Various aspects of the proposed bound (\ref{eq:MIR3}) are unsatisfactory. In particular, $v_{\mathrm{F}}$ is a weakly coupled notion, whereas the objective of formulating diffusivity bounds is to move away from weak coupling. Relatedly, clearly the diffusivity goes to zero at low temperatures in an insulator. Indeed, even the constant $T=0$ diffusivity of free electrons in a disordered potential in three dimensions violates (\ref{eq:MIR3}) if $v_{\mathrm{F}}$ is interpreted too literally. An interesting recent proposal is that the velocity that appears in diffusivity bounds such as (\ref{eq:MIR3}) is the  `butterfly velocity' $v_{\textsc{b}}$ \cite{Blake:2016wvh, Blake:2016sud}. This is appealing because the butterfly velocity can be defined in any system, with or without quasiparticles.

In weakly coupled models the butterfly velocity is close to the Fermi velocity \cite{Swingle:2016jdj, Aleiner:2016eni}. In certain strongly coupled models, in contrast, it is found to be temperature dependent \cite{Gu:2016oyy,Patel:2016wdy,GuRichard16}.
Recent results have furthermore suggested that the diffusive process most directly related to the butterfly velocity is in fact thermal diffusion \cite{Patel:2016wdy,Blake:2016jnn,GuRichard16,Gouteraux16}, with the potential bound
\be\label{eq:MIR4}
D_\epsilon \gtrsim \frac{\hbar v_{\textsc{b}}^2}{k_{\mathrm{B}}T} \,.
\ee
The butterfly velocity characterizes the spread of entanglement and chaos in quantum systems \cite{Shenker:2013pqa, Roberts:2014isa}, and such phase randomization is connected to energy
fluctuations (and hence thermal diffusivity) because the Schr\"odinger equation 
maps the time derivative of the
phase to the energy.
In contrast, many different physical effects, unrelated to scrambling, 
can contribute to the charge diffusivity. The rate of growth of quantum chaos has recently been bounded \cite{Maldacena:2015waa},
as discussed in \S \ref{sec:qmwqp}.
A bound in the spirit of (\ref{eq:MIR4}) has recently been used to understand anomalous aspects of thermal transport in a cuprate \cite{Zhang:2016ofh}.
In \S \ref{sec:SYKh}, we will present solvable models for which the equality (\ref{eq:Dbutterfly}) is consistent with (\ref{eq:MIR4}).   

Much of the evidence for a relation of the form (\ref{eq:MIR3}) or (\ref{eq:MIR4}) between $D$ and $v_{\textsc{b}}$ comes from studying homogeneous momentum-relaxing models, discussed in \S \ref{sec:in2} below.  However, upon generalizing to inhomogeneous models, strong violations of (\ref{eq:MIR4}) have been observed \cite{Gu:2017ohj}.  In a holographic setting, inhomogeneity affects $D_{\rho}$ and $v_{\textsc{b}}^2$ in different ways, violating (\ref{eq:MIR3}) with $v_{\mathrm{F}} \to v_{\textsc{b}}$ \cite{Lucas:2016yfl}.  

Although we have framed the discussion here around diffusion bounds, we will discuss specific holographic models in \S \ref{sec:in3} where electrical and thermal conductivity (but not diffusion) can be rigorously bounded from below.
The correct formulation of possible universal, model-independent bounds on transport (of energy, charge and momentum) remains a fascinating question at the time of writing. Proofs of appropriately formulated transport bounds may be within reach.

\subsection{Strong momentum relaxation I:  `mean field' methods}
\label{sec:in2}

The simplest way to model an incoherent metal holographically is through a `mean field' approach,  where translational symmetry has been broken,  but the spacetime geometry is homogeneous.    Although this might seem contradictory, it is not.
The essential `trick' is to use a source that breaks translations but preserves some combination of translations with a different (often internal) symmetry. We will see that this leads to a homogeneous energy-momentum tensor and hence a homogeneous spacetime. These are not realistic models of translation symmetry breaking, but are often the simplest to work with:  the bulk geometry, and response functions, are computed using ODEs rather than PDEs. Some results obtained have been found to be qualitatively similar to those following from more generic breaking of translation invariance.

Several versions of `mean field' translation symmetry breaking have been developed. In
holographic `Q-lattices' \cite{Donos:2013eha}, the background is invariant under a combination of translation combined with rotation under an internal $\mathrm{U}(1)$ symmetry. With `helical lattices' \cite{Donos:2012js}, the preserved symmetry is geometric; translation in a specific spatial direction (in 3 boundary spatial dimensions) is combined with a rotation in the plane perpendicular to the translation. This symmetry leads to a homogeneous but anisotropic spacetime and is 
an instance of the Bianchi classification of spacetimes with homogeneous boundary sources \cite{Iizuka:2012iv}. In `linear axion' models \cite{Andrade:2013gsa}, translations are combined with an internal shift symmetry. 

We will give some explicit computations in linear axion models \cite{Andrade:2013gsa}, which are perhaps the simplest of all. Furthermore, as elaborated in \cite{Taylor:2014tka, Baggioli:2014roa, Alberte:2014bua, Alberte:2015isw}, this broad class of models -- where translations are broken by a linear source that preserves a combination of translations and shift symmetry -- is precisely the St\"uckelburg formulation of a bulk `massive gravity' theory dual to broken translation invariance. Momentum relaxation through massive gravity was initially formulated on its own terms without gauge invariance \cite{Vegh:2013sk, Davison:2013jba}. Results obtained with Q-lattices and helical lattices are qualitatively similar to those we shall derive for linear axion models. We will gives references below. Let us emphasize again that there is no mystery concerning how these models are relaxing momentum. They have explicit non-translationally invariant sources (periodic or linear in a spatial coordinate).

We will illustrate the physics of mean field models with the following action \cite{Andrade:2013gsa}: \begin{equation} \label{eq:EMDA}
S = S_{\mathrm{EMD}} - \frac{1}{2}  \sum_{I=1}^d \int \mathrm{d}^{d+2}x \sqrt{-g} \; Y(\Phi) (\partial \chi^I)^2 \,.
\end{equation}
Note that the number of scalar fields $\chi$ is equal to the number of spatial dimensions.
We will be considering solutions in which the EMD fields depend only on the radial direction (as in all the backgrounds we have considered thus far), while the scalar fields
\begin{equation}
\chi^I =  \frac{mx^I}{\sqrt{2}\kappa},\;\;\;\; (I=1,\ldots,d).   \label{eq:axionMF}
\end{equation}
The parameter $m$ quantifies the strength of translation symmetry breaking.
The $\chi$ fields are often called axions, as they have a shift symmetry, just like an axion in particle physics.   In a homogeneous background for the other EMD fields, it is easy to see that (\ref{eq:axionMF}) solves the $\chi$ equation of motion.    Now, note that the dilaton and graviton equations of motion depend only on derivatives of $\chi^I$, and that the derivatives of $\chi^I$ are constants, and thus homogeneous in the spatial directions.   Hence, the assumption that the EMD fields are homogeneous is justified.

In mean field models, an explicit formula for $\sigma_\text{dc}$ in terms of horizon data can be obtained. There are a variety of methods to do this.   Let us short-cut to the `membrane paradigm' technique developed in \cite{Blake:2013bqa}. This is closely related to the method we used in \S \ref{sec:dccond} for the simpler case of a translationally invariant and zero density theory. For the present EMD-axion theory, we have essentially already computed the equations of motion in  \S\ref{sec:memmatrixderivation}:  in particular, (\ref{eq:mmh2}) is valid for any homogeneous geometry, and if we rewrite (\ref{eq:mmh2c}) in position space, and use the axion profile (\ref{eq:axionMF}), we obtain
\begin{equation}
\delta \mathcal{P}_x^\prime = \frac{-\delta \tilde{g}_{tx}}{b} \sum_I \frac{|\nabla \chi^I|^2}{dr^d} = -\frac{Ym^2}{2\kappa^2 br^d}\delta \tilde{g}_{tx} \,,  \label{eq:axiondp}
\end{equation}
with \begin{equation}
\delta \mathcal{P}_x = \frac{\mathrm{i}bm}{\sqrt{2}\kappa \omega r^d} Y \,  \delta \chi^x{}^\prime \,.
\end{equation}
Now,  we can carry through the same calculation as previously,  but taking $\omega\rightarrow 0$ while not taking a perturbative limit $m\rightarrow 0$.  (\ref{eq:mmh2b}) implies that
\be\label{eq:iwsdcA}
br^{2-d}Z\delta A_x^\prime - \frac{e^2\rho }{L^{d-2}}\delta \tilde{g}_{tx} = C_0,
\ee
with $C_0$ a constant.   We can evaluate $C_0$ both at the boundary and at the horizon. Using the near-boundary asymptotics of $\delta A_x$, $Z$, $b$ and $\delta \tilde{g}_{tx}$ we can conclude that as $r\rightarrow 0$ only the $\delta A_x^\prime$ term contributes. In this way the constant is obtained in terms of the dc conductivity as
\be\label{eq:iwsdcB}
C_0 = -\mathrm{i}\omega \sigma_{\mathrm{dc}} \frac{e^2}{L^{d-2}} \,.
\ee

 Near the horizon, infalling boundary conditions require that 
 \begin{equation}
\delta A_x^\prime \approx -\frac{\mathrm{i}\omega}{b} \delta A_x,  \label{eq:infallb}
\end{equation}and similarly for other fields.   (\ref{eq:mmh2a}) implies that near the horizon $\rho \delta A_x = L^d \delta \mathcal{P}_x$, and combining this fact with (\ref{eq:axiondp}), evaluating the constant in (\ref{eq:iwsdcA}) on the horizon and setting it equal to (\ref{eq:iwsdcB}) we obtain \begin{equation}
\sigma_{\mathrm{dc}} = \frac{Z_{+} L^{d-2}}{e^2r_{+}^{d-2}} + \frac{2\kappa^2 \rho ^2 r^d_{+}}{m^2 L^dY_{+}} =  \frac{Z_{+} L^{d-2}}{e^2r_{+}^{d-2}} + \frac{4\pi \rho ^2}{s \, m^2Y_{+}}  \,, \label{eq:axionsigma}
\end{equation}
where $Z_+$ and $Y_+$ are the values of $Z(\Phi)$ and $Y(\Phi)$ on the horizon, respectively.  An expression analagous to (\ref{eq:axionsigma}) was found in very similar models by  \cite{Donos:2014uba, Gouteraux:2014hca}.

Many models lead to results similar to (\ref{eq:axionsigma}), where the conductivity is a sum of two terms. The split is rather suggestive. While the formula (\ref{eq:axionsigma}) is nonperturbative in the disorder strength,  there is an obvious comparison to make with the hydrodynamic prediction (\ref{eq:hkmssigma}): the zeroth order term in $\rho $ analogous to the incoherent `$\sigma_{\textsc{q}}$' -- compare also the result (\ref{eq:sigma}) --  and the quadratic term in $\rho $ the analogue of the Drude term.    We will see shortly that this comparison is not quite right.  
Let us point out three further important points:  \begin{enumerate}
\item As $m\rightarrow 0$, wherein translational symmetry is weakly broken, we recover the perturbative memory matrix regime \cite{Blake:2013owa, lucas1401}.
\item As the translation symmetry breaking parameter $m\rightarrow \infty$, with other quantities held fixed,  the conductivity is not driven to zero but rather saturates at the value set by the first term in (\ref{eq:axionsigma}).   This is rather interesting
from a condensed matter perspective, as it is suggestive of the phenomenological `parallel resistor' formula
used to model resistivity saturation, see e.g. \cite{gunnarsson}.
 We will see that this is a generic prediction of many holographic models in $d=2$.
\item $\sigma_{\mathrm{dc}}$ is a sum of two contributions which can have different low temperature $T$-scalings. While the temperature dependence of $Z_+, Y_+$ and $s$ can be expected to follow the quantum critical dimensional analysis developed in \S\ref{sec:compressible}, $\rho$ and $m$ are temperature independent, and hence the dimensions to be soaked up by powers of temperature are different.
\end{enumerate} 

The last of the above points indicates a limitation of the mean field approach. Because the background geometry is homogeneous despite breaking translation invariance, thermodynamic variables can have conventional critical behavior even though there is a scale $m$ in the problem. In more generic models of strongly inhomogeneous fixed points, the nature of any emergent scaling regime will have to be strongly tied up with the nature of translation symmetry breaking. Disordered fixed points will be discussed in \S\ref{sec:disfpt}. Another interesting possibility is that $z=\infty$ scaling is a natural home for inhomogeneous fixed points with strong momentum relaxation, because space is not involved in the critical scaling \cite{Hartnoll:2014gaa}.

One should also take into account the possibility that as $m \to \infty$ a quantum phase transition to a different (e.g. gapped) phase might occur before the momentum relaxation rate becomes too large, cf. \cite{Adams:2014vza}. In our opinion this possibility is relatively underexplored.  Recently it was noted that some of these linear axion models are unstable at large $m$ \cite{Caldarelli:2016nni}:  this instability is signalled by the energy density becoming negative.  As $\epsilon<0$ is a common feature of many of the simplest models \cite{Davison:2014lua}, this may have important consequences for the existence of some of the simplest holographic incoherent metals.

\subsubsection{Metal-insulator transitions}
\label{sec:MIT}

The ability to perform controlled calculations with very strong momentum relaxation makes the mean field models a natural framework for metal-insulator transitions. We should first describe insulators. In an insulator the resistivity diverges rather than going to zero as $T \to 0$. Holographic models allow for gapless as well as gapped insulators, as we will discuss in turn. Gapless holographic insulators were constructed in \cite{Donos:2012js, Donos:2014uba, Gouteraux:2014hca, Donos:2014oha, Baggioli:2016oqk, Gouteraux:2016wxj}. The essential part of the computation is to find near-horizon scaling solutions to the equations of motion in the presence of fields (such as the linear axion) that break translation invariance. The geometry itself is homogeneous and so still has the form (\ref{eq:abmetric}). Nonetheless, solving the equations of motion can be tricky, even if they are only ODEs. We shall not discuss the explicit backgrounds here -- the reader is referred to the papers for details. With the solutions at hand, the temperature dependence of the resistivity is obtained from formulae such as (\ref{eq:axionsigma}). The background is an insulator if both of the terms in (\ref{eq:axionsigma}) go to zero as $T \to 0$.  Typically in these cases, the first term -- that does not depend explicitly on the charge density -- is the larger one at low temperatures.

The holographic insulators obtained in this way are quite different from the more familiar classes of insulators in condensed matter physics: band insulators, Anderson insulators and Mott insulators \cite{dobro}. Band and Anderson insulators make essential reference to single-particle properties whereas a Mott insulator requires commensurability leading to an emergent particle-hole symmetry. Instead the holographic insulators described in this section are best understood as the consequence of relevant (i.e. strong in the IR) generalized umklapp scattering from a periodic potential. That is to say, the scattering from the potential is simply too strong to allow the current to flow as $T \to 0$. This physics allows for anisotropic systems that are insulating in one direction but conductive in another \cite{Donos:2012js}.
\begin{figure*}
\centering
\includegraphics[width = 5in]{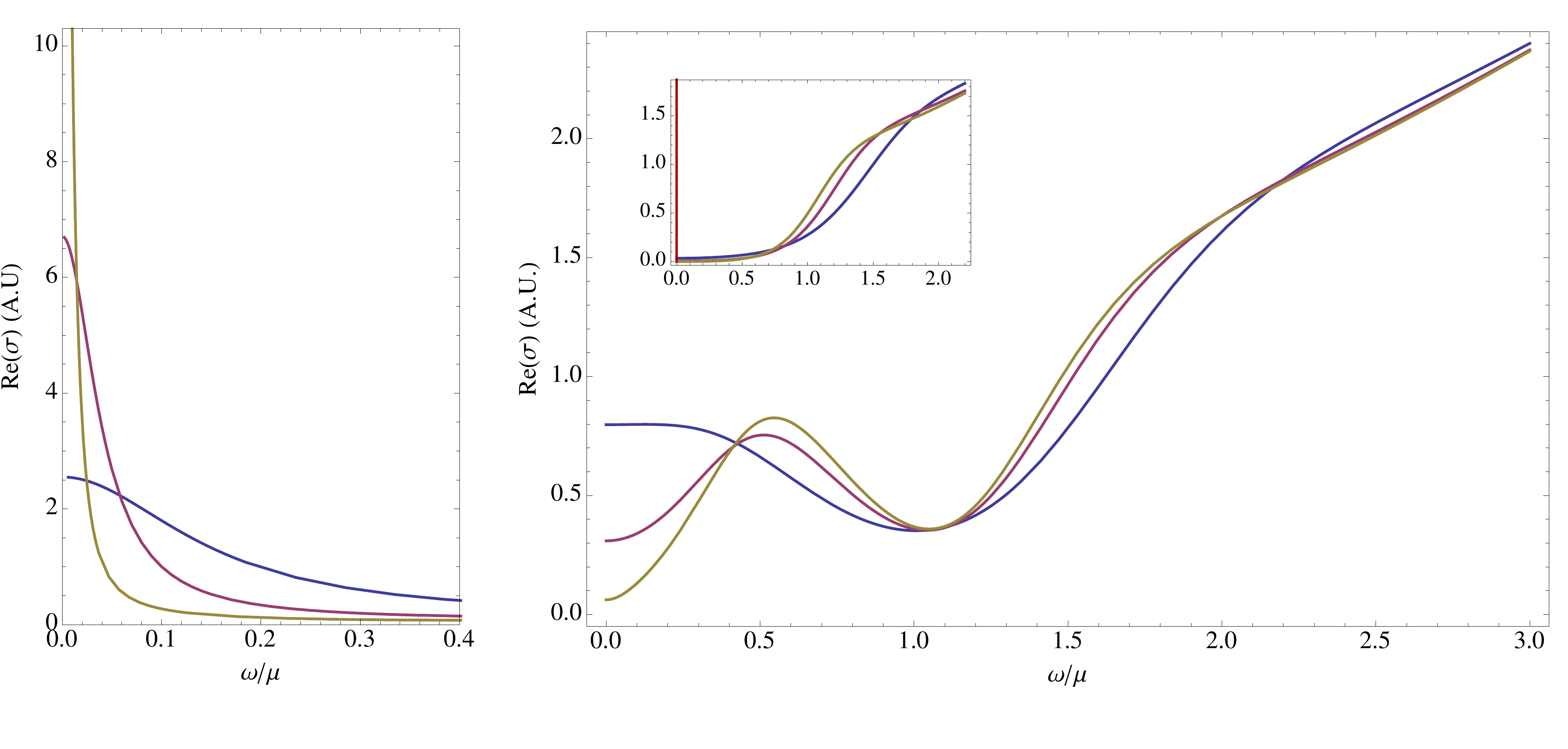}
\caption{{\bf Spectral weight transfer:} Optical conductivity in the metallic (left) and insulating (right) phases of the model from \cite{Donos:2012js}. Highest temperature is blue and lowest temperature is yellow. Inset described in \cite{Donos:2012js}. Figure used with permission.}
\label{fig:MIT}
\end{figure*}

In a conventional Fermi liquid metal, umklapp processes are irrelevant \cite{Hartnoll:2012rj} and lead to the well known $T^2$ resistivity mentioned in \S\ref{sec5a} and \S\ref{sec:holexample} above. In one spatial dimension, it is known that umklapp can become relevant in a Luttinger liquid \cite{Emery:1976zz}. In holographic models, the most natural starting point for relevant umklapp scattering (i.e. for an operator in the theory that has a periodic space dependence) is a $z=\infty$ compressible phase. We have already noted around equation (\ref{eq:tauinfty}) that $z = \infty$ fixed points are especially susceptible to spatially periodic deformations. In equation (\ref{eq:tauinfty}) it is explicitly seen that if the exponent of the latticized operator $\nu_{k_{\mathrm{L}}} < \half$ then the resistivity diverges at low temperatures. This divergence indicates that at sufficiently low temperatures the lattice will backreact strongly on the metric. In general, the field $\phi$ describing the amplitude of the lattice will grow like
\be\label{eq:ads2nu}
\phi(r) \sim r^{-\half + \nu_{k_{\mathrm{L}}}} \,.
\ee
as $r \to 0$ in the IR of the $z=\infty$ geometry. This is the behavior we have already seen in e.g. (\ref{eq:varphi0zeta}). The growth of the field shows that eventually backreaction will be important. In the cases studied in \cite{Donos:2012js, Donos:2014uba, Gouteraux:2014hca, Donos:2014oha}, this backreaction induces an RG flow to a new, insulating interior geometry.

If the dimension $\nu_{k_{\mathrm{L}}}$ of the lattice deformation in the low energy compressible phase can be continuously tuned at $T=0$ from being irrelevant to relevant -- for instance by varying the charge density at fixed lattice wavevector --  then a metal-insulator transition results \cite{Donos:2012js}. This appears to be an infinite order quantum phase transition, similar to the quantum BKT transitions that we discuss in \S \ref{sec:bkt}.

Once the background is understood, the temperature and frequency dependent conductivities are computed using the same holographic methods we have described previously. In Figure \ref{fig:MIT} we show the low temperature optical conductivity in the metallic and insulating phases of the model described in \cite{Donos:2012js}. These plots show the canonical spectral weight transfer expected through a metal-insulator transition. The spectral weight in the Drude peak is transferred to `interband' energy scales of order the chemical potential.

As can be seen in Figure \ref{fig:MIT}, the insulating phase only has a `soft' power law gap in the spectral weight. For the particular model shown, in fact, $\sigma \sim \omega^{4/3}$. Relatedly, the resistivity $\rho_\text{dc} \sim T^{-4/3}$. The soft gap is due to the persistence of a black hole horizon in the solution with a scaling near-horizon geometry. There exist quantum critical excitations, but they are not able to conduct efficiently due to strong interactions with the lattice.

A gapped insulator is instead obtained by starting with a geometry that is gapped to charged excitations (cf. \S\ref{sec:gapped} above), and then breaking translation invariance in a manner that does not destroy the charge gap \cite{Charmousis:2010zz, Balasubramanian:2010uw,Kiritsis:2015oxa}. The role of strong interactions here is not to have strong momentum relaxation but instead to gap out the charge sector in a situation where there is no translational zero mode.

At zero charge density, the second term in (\ref{eq:axionsigma}) is zero. In this case a vanishing conductivity at $T=0$ can also be obtained if $Z_+ = 0$ on the zero temperature horizon \cite{Mefford:2014gia}. In such zero density systems, it is not necessary to break translation invariance.

\subsubsection{ac transport}
\label{sec:562}

The low frequency conductivity $\sigma(\omega)$ can also be computed in mean field models. In particular, by including next-to-leading order effects in the  translational symmetry breaking parameter $m^2$, taken to be small, one can see the relationship between the two terms in the exact result (\ref{eq:axionsigma}) and the two terms in the dc limit of the hydrodynamic formula (\ref{eq:hkmssigma}). For a simple linear axion model of the form (\ref{eq:EMDA}) in $d=2$, the conductivity at small $m$ and $\omega$ was obtained in \cite{Davison:2015bea, Blake:2015epa} as:
\begin{equation}
\sigma(\omega) = \sigma_{\textsc{q}} + \frac{1}{1-\mathrm{i}\omega \tau} \left(\frac{1}{e^2}-\sigma_{\textsc{q}} + \frac{4 \pi \rho^2}{s m^2}\right) + \mathcal{O}\left(\omega, \frac{1}{\tau}\right),   \label{eq:sigma15}
\end{equation}
with $m^2 \sim 1/\tau$ a perturbatively small parameter, 
\begin{equation}\label{eq:tauofm}
\frac{1}{\tau} = \frac{s m^2}{4\pi (\epsilon+P)} \left(1+\lambda m^2 + \cdots\right) \,,
\end{equation}
and $\sigma_{\textsc{q}}$ is the clean incoherent conductivity given in (\ref{eq:incclean}) above. The temperature dependent quantity $\l$ is given in  \cite{Davison:2015bea, Blake:2015epa}.

To leading order at small $m$, and using (\ref{eq:tauofm}), the weight of the Drude peak in (\ref{eq:sigma15}) is seen to agree with the hydrodynamic result (\ref{eq:hkmssigma}). However, in the dc limit $\w \to 0$, the subleading terms in the spectral weight of the Drude peak contribute at the same $m^0$ order as the clean incoherent contribution. In fact, we see that there is a cancellation so that $\sigma_{\textsc{q}} \to \frac{1}{e^2}$. Thus the first term in the exact dc conductivity (\ref{eq:axionsigma}) is unrelated, and generically has different temperature scaling to, the clean incoherent conductivity $\sigma_{\textsc{q}}$. This spectral weight shift in the Drude peak is not visible in the hydrodynamic arguments leading to (\ref{eq:hkmssigma}).   This should not be particularly surprising.   As we made clear in our more formal derivations of (\ref{eq:hkmssigma}) at weak disorder in \S\ref{sec:incformal}, the $\sigma_{\textsc{q}}$ contribution was generically a subleading term in perturbation theory.   Hence, disorder-induced corrections to both `$\tau$' (though this may lose a sharp definition) and the equations of state, lead to corrections as relevant as $\sigma_{\textsc{q}}$.  By considering $\alpha(\omega)$ and $\bar\kappa(\omega)$ in addition to $\sigma(\omega)$, one sees that the hydrodynamic results \S\ref{sec:genhydro} and \S\ref{sec:weakI} do not have enough parameters to reproduce the full answers at order $m^0$ \cite{Davison:2015bea, Blake:2015epa}.

At higher orders in perturbation theory, \cite{Davison:2014lua} has demonstrated the loss of a Drude peak in $\bar\kappa(\omega)$ once $m\sim T$ in this same axion model, in agreement with the qualitative predictions of incoherent transport in \cite{Hartnoll:2014lpa}.    

 \subsubsection{Thermoelectric conductivities}
 \label{sec:thermoelectric}
 
 So far, we have focused on computing the electrical conductivity $\sigma$.   Computations of $\alpha$ and $\bar\kappa$ can be  a bit more involved, because fluctuations of the bulk metric participate, but let us sketch out the procedure \cite{Amoretti:2014zha, Donos:2014cya, Amoretti:2014mma}.   We follow a similar approach to that used in \S \ref{sec:dccond}; again, we assume $d=2$ for simplicity, and use the axion model from before.   This time, the bulk perturbations take the form \begin{subequations}\begin{align}
 \delta A_x &= t (A_t \zeta - E) + \delta \tilde A_x(r), \\
 \delta g_{tx} &= tg_{tt}\zeta + \delta \tilde g_{tx}(r), 
 \end{align}\end{subequations}
 along with time-independent perturbations $\delta \chi^x$ and $\delta g_{rx}$.  The time-dependent parts of the perturbations above are fixed by employing the $tx$ component of the linearized Einstein's equations and the $x$-component of the Maxwell equations, and showing the $t$-linear piece only vanishes when the above choices are made.   Asymptotic analysis near $r=0$ confirms that these perturbations encode the electric field $E$ and thermal ``drive" $\zeta$.  
 
 The linearized Maxwell equations can be shown to imply that $\partial_r J = 0$,  where \begin{align}
 e^2 J \equiv &\sqrt{-g}  Z(\Phi) F^{rx} = \sqrt{-\frac{g_{tt}}{g_{rr}}} Z(\Phi) \times \notag \\ &\left[\partial_r \delta \tilde{A}_x +| g^{tt}| (\partial_r A_t) \delta \tilde{g}_{tx} \right] \,.
 \end{align}
The asymptotic behavior as $r\rightarrow 0$ implies that $J$ is nothing more than the expectation value of the electric current in the dual field theory.   Next, the $t$-independent part of the $tx$-Einstein equation implies that $\partial_r Q=0$, where
\begin{equation}\label{eq:QQ}
Q \equiv \frac{1}{2\kappa^2} \sqrt{-\frac{g_{tt}^3}{g_{rr}}}\partial_r \left(\frac{\delta \tilde{g}_{tx}}{g_{tt}}\right) -  J A_t,
\end{equation}
and again boundary asymptotics imply $Q$ is the boundary heat current. The strategy is now to evaluate the constants $J$ and $Q$ at the horizon and in this way relate the currents to the electric field and thermal drive and hence obtain the conductivities.

  Demanding regularity near the horizon, one finds that
  \begin{equation}\label{eq:Qreg}
Q = \left.\frac{1}{2\kappa^2}\sqrt{-\frac{g_{tt}^3}{g_{rr}}}\partial_r \frac{\delta g_{rx}}{\sqrt{-g_{rr}g_{tt}}}\right|_{r=r_{+}}.
\end{equation}
Note that the $J$ term in (\ref{eq:QQ}) drops out because $A_t = 0$ on the horizon. Furthermore, solving the
linearized $rx$-Einstein equation, along with the background equations of motion, pertubatively near the horizon enforces
\begin{equation}
\delta g_{rx} =  \frac{1}{2 \pi T (r_{+}-r)}\frac{\kappa^2}{Y_{+}m^2} \left(s T \zeta + \rho E \right)+ \cdots \,.
\end{equation}
Plugging this expression into (\ref{eq:Qreg}) and using the definitions of $T\alpha$ and $T\bar\kappa$ as the coefficients of $\partial_E Q$ and $\partial_\zeta Q$, respectively, one finds: \begin{subequations}\label{eq:thermoelectricaxion}\begin{align}
 \alpha &= \frac{4\pi \rho }{Y_{+}m^2}, \\
 \bar\kappa &= \frac{4\pi Ts}{Y_{+}m^2}.
 \end{align}\end{subequations}
 Using the expression for $J$, we can independently compute $\sigma$, as given in (\ref{eq:axionsigma}), and confirm that Onsager reciprocity holds by an independent computation of $\alpha$.   

When $m \rightarrow 0$,  we recover the Drude predictions for thermoelectric conductivities as derived from hydrodynamics.   As $m$ gets larger, we see deviations from the hydrodynamic predictions -- there is no $\sigma_{\textsc{q}}$ dependence in $\alpha$ or $\bar\kappa$, in contrast to (\ref{eq:hkmsalpha}) and (\ref{eq:hkmskappa}).    This is the same issue that we discovered in \S \ref{sec:562} -- namely, the transport theory of \S\ref{sec:genhydro} and \S\ref{sec:weakI} is not correct beyond leading order in perturbation theory in $m$.    
  
\subsection{Strong momentum relaxation II: exact methods}
\label{sec:in3}

\subsubsection{Analytic methods} \label{sec:analyticstrongII}

Recent remarkable results have in a certain sense `solved' the problem of computing the dc thermoelectric conductivities in holographic models with
arbitrary inhomogeneous sources. These results go well beyond the mean field models considered thus far, and amount to a substantial generalization of the mean field expressions (\ref{eq:axionsigma}) and (\ref{eq:thermoelectricaxion}). Specifically, it was shown in \cite{Donos:2015gia, Banks:2015wha} that the dc transport problem can be reduced to the solution of some `hydrodynamic' PDEs on the black hole horizon. The two key facts here are (\emph{i}) even if the boundary sources themselves vary over microscopic distance scales and with large amplitudes, there exists an effective hydrodynamic description that computes the dc conductivities and (\emph{ii}) this hydrodynamic description depends only on the far IR geometry of the spacetime, the black hole event horizon. In these results the old black hole membrane paradigm discussed in \S\ref{sec:horizons} has finally found its holographic home.

The derivation of the dc conductivities in \cite{Donos:2015gia, Banks:2015wha} is essentially a significantly souped up version of the membrane paradigm calculation of thermoelectric conductivities that we just reviewed in the previous section for the axion model, and so we simply state the results.   We assume that the black hole horizon is at constant temperature $T>0$ and is connected, with the topology of a torus.  The answer is simplest  if we employ Gaussian normal coordinates near the horizon \cite{Medved:2004tp},  in which case the background EMD fields have near-horizon expansion (switching to radial coordinate $\tilde{r}$ where the horizon is at $\tilde{r}=0$): \begin{subequations}\begin{align}
\mathrm{d}s^2 &= \mathrm{d}\tilde{r}^2 - \left(2\pi T\tilde{r} + \frac{F(x)}{6}\tilde{r}^3 + \cdots\right)^2 \mathrm{d}t^2 \notag \\
& + \left(\gamma_{ij}(x) + \frac{h_{ij}(x)}{2}\tilde{r}^2+\cdots\right)\mathrm{d}x^i\mathrm{d}x^j, \\
A &=  \left(\pi T p(x) \tilde{r}^2 +\cdots\right) \mathrm{d}t , \\
\Phi &=  \phi(x) + \cdots.
\end{align}\end{subequations}
Here $\gamma_{ij}$ is the induced metric on the black hole horizon. The horizon data itself is three functions of $x$: $\{\gamma_{ij}(x), \phi(x), p(x)\}$. One then writes down the following linear partial differential equations for variables $\{v^i(x), \delta \mu(x), \delta \Theta(x)\}$ that depend on these background fields as well as two constant sources $E_i$ and $\zeta_i$ (note that with indices raised by $\g^{ij}$, $E^i$ and $\zeta^i$ will often not be constant):
 \begin{subequations}\label{eq:horfluid}\begin{align}
0 &= \nabla_i \left(s_{\mathrm{h}} T v^i\right), \\
0 &= \nabla_i \left(\rho _{\mathrm{h}}v^i + \Sigma_{\mathrm{h}} \left(E^i - \nabla^i \delta \mu\right)\right), \label{eq:fluid2} \\
- \nabla_i \phi \nabla_j \phi \, v^j & = \rho _{\mathrm{h}}(\nabla_i \delta \mu - E_i) + s_{\mathrm{h}}(\nabla_i \delta \Theta - T\zeta_i) \notag \\
& -2\eta_{\mathrm{h}} \nabla^j\nabla_{(j}v_{i)} \,,
\end{align}\end{subequations}
where covariant derivatives are taken with the induced horizon metric, and we defined
\be\label{eq:horfluideos}
\eta_{\mathrm{h}} = \frac{s_{\mathrm{h}}}{4\pi} = 1\,, \;\; \Sigma_{\mathrm{h}}(x)  = Z(\phi(x)) \,, \;\; \rho _{\mathrm{h}}(x) = Z(\phi(x))p(x) \,.
\ee
With solutions to the horizon fluid equations (\ref{eq:horfluid}) at hand, one can show that the spatially averaged
charge and heat currents in the boundary theory are given by
\begin{subequations}\label{eq:jqint}\begin{align}
J^i &= \frac{1}{L^d}\int \mathrm{d}^dx\; \sqrt{\gamma} \left(\rho _{\mathrm{h}}v^i + \Sigma_{\mathrm{h}} \left(E^i - \nabla^i \delta \mu\right)\right) \,, \\
Q^i &= \frac{1}{L^d}\int \mathrm{d}^dx\; \sqrt{\gamma} T s_{\mathrm{h}} v^i. \label{eq:Qb}
\end{align}\end{subequations}
Solving the fluid equations amounts to obtaining a linear expression for $\{v^i(x), \delta \mu(x), \delta \Theta(x)\}$ in terms of the sources $E_i$ and $\zeta_i$. Therefore, with such a solution, the formulae (\ref{eq:jqint}) give expressions for $J$ and $Q$ that are linear in $E$ and $\zeta$, and hence it is straightforward to read off the thermoelectric conductivities (\ref{eq:thermoelectric}).

The `horizon fluid' equations (\ref{eq:horfluid}) look like the equations describing transport in a relativistic fluid on a curved space, with inhomogeneous background fields such as the chemical potential \cite{Lucas:2015lna}, but with strange equations of state.   In particular, the thermodynamic Maxwell relations are not obeyed for any sensible equation of state.\footnote{Thermodynamic Maxwell relations require the entropy $s_\text{h}$ and charge $\rho_\text{h}$ to be obtained as derivatives of the pressure with respect to temperature and chemical potential. This is only possible in (\ref{eq:horfluideos}) if the pressure is a separable function of $\mu$ and $T$.}    In a simple axion model, \cite{Blake:2015epa, Blake:2015hxa} has perturbatively shown that these equations of state can be understood from a hydrodynamic perspective, by choosing a curious fluid frame where the fluid velocity is proportional to the heat current (cf.  (\ref{eq:Qb}) above). This seems to be a convenient frame in holography.   One can gain further intuition by sprinkling factors of $\sqrt{\gamma}$ in front of each term in (\ref{eq:horfluideos}) -- then $s_{\mathrm{h}}$ follows from the Bekenstein-Hawking formula, and $\rho _{\mathrm{h}}$ is the local charge density on the horizon \cite{Lucas:2015lna}.

In general, the horizon fluid equations are a dramatic simplification from a numerical point of view -- given a complicated black hole geometry,  one can solve a lower dimensional PDE to compute the dc transport coefficients.   Remarkably, there are some powerful statements that can also be made analytically.   For example, Onsager reciprocity immediately follows from the hydrodynamic form of these equations \cite{Lucas:2015lna}.   

One can analytically solve the transport equations if translational symmetry is only broken in a single direction.   In fact this was noted earlier \cite{Donos:2014yya, Rangamani:2015hka}.   To understand why this is so, consider the following simple limit.  If we have a charge-neutral system, then the hydrodynamic equations reduce to a simple Laplace-like equation,  which governs steady-state diffusion of charge.   Discrete diffusion equations are resistor networks (one thinks of $Z \sqrt{\g} \g^{ij}$ in (\ref{eq:fluid2}) as a local conductance matrix).   The resistance of a 1d resistor network is just the sum of all the individual resistances (in this case given by $\sum \g_{xx}/(Z \sqrt{\g})$, cf. equation (\ref{eq:invone}) below).    We can compute this resistance by just computing the net power dissipated in the network, which is a sum of a local quantity.   This latter perspective readily generalizes to the full equations (\ref{eq:horfluid}).   One finds using this approach that the inverse conductivity matrix is expressed as a single integral over \emph{local} quantities on the horizon \cite{Banks:2015wha}, just as the power dissipated in the resistor network is the sum of local Joule heating:    \begin{subequations}\label{eq:invS}\begin{align}
\frac{T\bar\kappa }{T\sigma \bar\kappa- (\alpha T)^2} &= \frac{1}{L} \int \mathrm{d}x \; \frac{\gamma_{xx}}{\sqrt{\gamma} Z(\phi)}, \label{eq:invone}\\
-\frac{T\alpha}{T\sigma \bar\kappa- (\alpha T)^2} &= -\frac{1}{4\pi TL} \int \mathrm{d}x \; \frac{\gamma_{xx} p}{\sqrt{\gamma}}, \\
\frac{\sigma}{T\sigma \bar\kappa- (\alpha T)^2} &= \frac{1}{(4\pi T)^2L}\int \mathrm{d}x \left[\frac{(\partial_x \phi)^2}{\sqrt{\gamma}} + \frac{g_{xx}Z(\phi)p^2}{\sqrt{\gamma}} + \right. \notag \\
&\left.\;\;\;\;\;\; \frac{(\partial_x \log \sqrt{\gamma \gamma ^{xx}})^2 + \gamma^{ik}\gamma^{jl}\partial_x \gamma_{ij}\partial_x \gamma_{kl}}{2\sqrt{\gamma}}\right] \,,
\end{align}\end{subequations}
where we have assumed that translation invariance is broken in the $x$-direction, and assumed that $\gamma_{xj}=0$ for $x\ne j$, for simplicity. The combinations of conductivities appearing in (\ref{eq:invS}) are the components of the inverse of the thermoelectric conductivity matrix $\s$ from (\ref{eq:thermoelectric}). This determines Joule heating through $(\sigma^{-1})_{AB} J^A J^B$, with $J^A$ being the electric and thermal current operators. It is straightforward to obtain an expression for $\sigma, \bar \kappa$ and $\alpha$ from (\ref{eq:invS}), but the inverse matrix is more physically transparent.

While the horizon fluid equations cannot be solved analytically when translation invariance is broken in more than one direction, they do allow for the proof of non-trivial lower bounds on conductivities for bulk theories dual to $d=2$ metals.  For simplicity, we assume that the metal is isotropic.   The intuition behind this approach again follows from resistor network tricks \cite{Lucas:2015lna}. That is, given any arbitrary conserved flows of heat and charge in the horizon fluid,  the one which dissipates the minimal power solves the equations of motion.  Hence, the power dissipated on other configurations gives us lower bounds on the conductivities.   The calculations are a bit technically involved, so let us state without proof the results, which hold for any isotropic theory with a connected black hole horizon.   The first is a bound on the electrical conductivity in the Einstein-Maxwell system \cite{Grozdanov:2015qia}: \begin{equation}
\sigma \ge \frac{1}{e^2}.   \label{eq:boundsigma}
\end{equation}
This bound generalizes in a nautral way to any EMD theory \cite{Fadafan:2016gmx}. In that more general setting the thermal conductivity $\kappa$ can also be bounded \cite{Grozdanov2}: \begin{equation}
\kappa = \bar\kappa - \frac{T\alpha^2}{\sigma} \ge \frac{L^2}{16\pi G_{\mathrm{N}}} \frac{8\pi^2 T}{|\min(V(\Phi))|}.   \label{eq:boundkappa}
\end{equation}
We have written the coupling in the Einstein-Hilbert action as $2 \k^2 = 16\pi G_\text{N}$ to avoid the symbol $\k$ meaning two different things in the same equation. For theories with bounded potentials (which can generally occur for models with $\theta \ge 0$ in the clean limit) we hence find a strictly finite thermal conductivity at finite $T$ (the factor of $T$ is trivially necessary by dimensional analysis).

Both of these bounds are saturated in some of the simplest mean field models, implying that these mean field approaches to breaking translational symmetry are -- at least in part -- quantitatively sensible.   In particular, in a simple Einstein-Maxwell model with massive gravity \cite{Blake:2013bqa} or axions \cite{Andrade:2013gsa}, one finds the dc conductivity (\ref{eq:axionsigma}) with $Z_{+}=Y_{+}=1$, which we readily see saturates (\ref{eq:boundsigma}) either when $m\rightarrow \infty$ or $\rho \rightarrow 0$.    In this same model, the thermal conductivity is given by \cite{Donos:2014cya} \begin{equation}
\kappa = \frac{L^2}{16\pi G_{\mathrm{N}}} \frac{4\pi T s}{m^2 + \mu^2} \ge  \frac{L^2}{16\pi G_{\mathrm{N}}} \frac{4\pi^2 T}{3},
\end{equation}
where the inequality follows from bounding the entropy density in this particular model by $s \geq s(T=0) = \pi (2 m^2 + \mu^2)/3$.   This again agrees with (\ref{eq:boundkappa}) upon using that $V=-6$ for the Einstein-Maxwell system.

The bounds on $\s$ and $\k$ above are powerful non-localization theorems in specific strongly interacting metals. However much the strength of disorder is increased, the conductivity is never driven to zero. These results amount to a no-go theorem on the possibility of realizing a many-body localized state in  certain phases of certain holographic models. A many-body localized state is non-ergodic, with the absence of thermalization and transport coefficients vanishing at finite energy density \cite{nandkishore}.   This is in contrast with (\ref{eq:boundsigma}) and (\ref{eq:boundkappa}) -- hence we conclude that at least some holographic models with connected black hole horizons do not realize such many-body localization.  We will discuss disconnected horizons shortly, in \S\ref{sec:num510}. Another possibility to keep in mind are possible first order phase transitions to geometries with no horizon. We have already noted in \S\ref{sec:MIT} that insulators are obtained from inhomogeneous geometries with a charge gap.

A bound away from zero on the electrical conductivity such as (\ref{eq:boundsigma}) is evaded even with horizons in the gapless (mean field) insulating geometries discussed in \S\ref{sec:MIT} above. In an insulator the conductivity goes to zero at zero temperature. Those models are either anisotropic or have additional fields that effectively drive $Z_+$ to zero in the horizon formula (\ref{eq:axionsigma}) for the dc conductivity (in addition to translation symmetry breaking being strong enough that the second term in (\ref{eq:axionsigma}) is absent at $T=0$).

The essence of the challenge of bad metals (that violate the MIR bound (\ref{eq:MIR})) is not to show that insulators cannot exist, which they clearly do, but rather to demonstrate that it is possible for strongly interacting non-quasiparticle systems to remain metallic even in the face of strong disorder. From this perspective, a bound such as (\ref{eq:boundsigma}) on the conductivity in a specific system (Einstein-Maxwell theory) is an exciting development.

It has recently been shown that the dc magnetotransport problem in EMD models reduces once again to solving fluid-like equations on a black hole horizon  \cite{Donos:2015bxe}.  Interestingly, particle-vortex duality in holography is robust to disorder \cite{Grozdanov:2015qia, Donos:2015bxe}, and the bound (\ref{eq:boundsigma}) can partially be understood as a consequence of this duality \cite{Grozdanov:2015qia}.

\subsubsection{Numerical methods}\label{sec:num510}

To employ the horizon fluid equations (\ref{eq:horfluid}) in a specific model one needs, of course, to first construct an inhomogeneous black hole background.  In essentially all cases, this must be done numerically -- a fact that has spurred several recent developments in numerical general relativity \cite{Dias:2015nua}.    Furthermore, to compute conductivities at finite frequency, we need to linearize around the inhomogeneous black hole background.  Both of these procedures are computationally intensive. We will briefly discuss some classes of inhomogeneous backgrounds that have been studied numerically.

Breaking translation invariance in a single direction, $x$, minimizes the number of inhomogeneous dimensions and hence offers a significant saving in computational cost while still capturing important momentum-relaxing physics. For this reason, this case has been widely studied. A minimal way to do this is allow an inhomogeneous chemical potential of the form \cite{Horowitz:2012gs, Ling:2013nxa} \begin{equation}
\mu = \mu_0 + V\cos(k_{\mathrm{L}} x).  \label{eq:hollat}
\end{equation}
This can also be done with scalar fields \cite{Horowitz:2012ky}.   These potentials often go by the name ``holographic lattices". From a microscopic perspective, this is a little different to an actual ionic lattice in a traditional condensed matter model: there is no relation between the lattice spacing and the charge density, a point we will return to later. However, as we have repeatedly stressed above, the real objective is to understand the universal low energy physics, and from this point of the view one should consider (\ref{eq:hollat}) as simply a way to source a spatially periodic structure. Furthermore, given that cases with an irrelevant lattice can be understood from perturbation theory as in \S\ref{sec:holexample} above, the most important objective of numerical studies is to identify new intrinsically inhomogeneous IR geometries that are beyond perturbation theory.

As described in \S\ref{sec:holexample} above, a periodic source such as (\ref{eq:hollat}) is not expected to have strong effects in $z < \infty$ geometries. Each lattice mode decays exponentially towards the interior of the spacetime. Indeed, when $\mu_0=0$, numerical study shows that even a strong periodic deformation of an AdS spacetime simply flows back to a rescaled AdS in the far IR \cite{Chesler:2013qla}.   Even when $\mu_0 \ne 0$, in $z=\infty$ spacetimes (as would arise in Einstein-Maxwell theory), the amplitude of the lattice scales as (\ref{eq:ads2nu}) -- that behavior is not specific to homogeneous lattices. In Einstein-Maxwell theory, this scaling corresponds to a decay towards the interior of the spacetime because for all $k$ one has $\nu_k \geq \half$ from (\ref{eq:RNnu}). Thus one anticipates that lattice deformations (\ref{eq:hollat}) will also not lead to strongly inhomogeneous low temperature horizons in nonzero density Einstein-Maxwell theory (which has an $\mathrm{AdS}_2 \times \R^2$ IR geometry prior to deformation by a lattice) \cite{Hartnoll:2012rj}. Numerical work supports this conclusion \cite{Donos:2014yya}, though other numerical results suggest that nonlinear effects might lead to the survival of inhomogeneities as $T \to 0$ \cite{Hartnoll:2014gaa} (this evidence is most compelling with a large amplitude lattice).  The $T\rightarrow 0$ limit is challenging to precisely address numerically with present day methods, so it is worth looking for new numerical or analytical techniques to conclusively settle this issue.

Beyond Einstein-Maxwell theory (in more general EMD models) the exponents $\nu_k$ can be such that the lattice grows towards the interior of a $z=\infty$ IR spacetime. This is what is going on in the paper \cite{Rangamani:2015hka}, who have shown a metal-insulator transition as a function of the wavelength of a periodic potential source for the dilaton (at fixed charge density). The insulating phase has (presumably) an inhomogeneous low temperature horizon. This transistion is entirely analogous to the original metal-insulator transition of \cite{Donos:2012js}, discussed in \S\ref{sec:MIT} above, and confirms the persistence of that physics in a more generic (inhomogeneous) setting.

\begin{figure*}
\centering
\includegraphics[width = \textwidth]{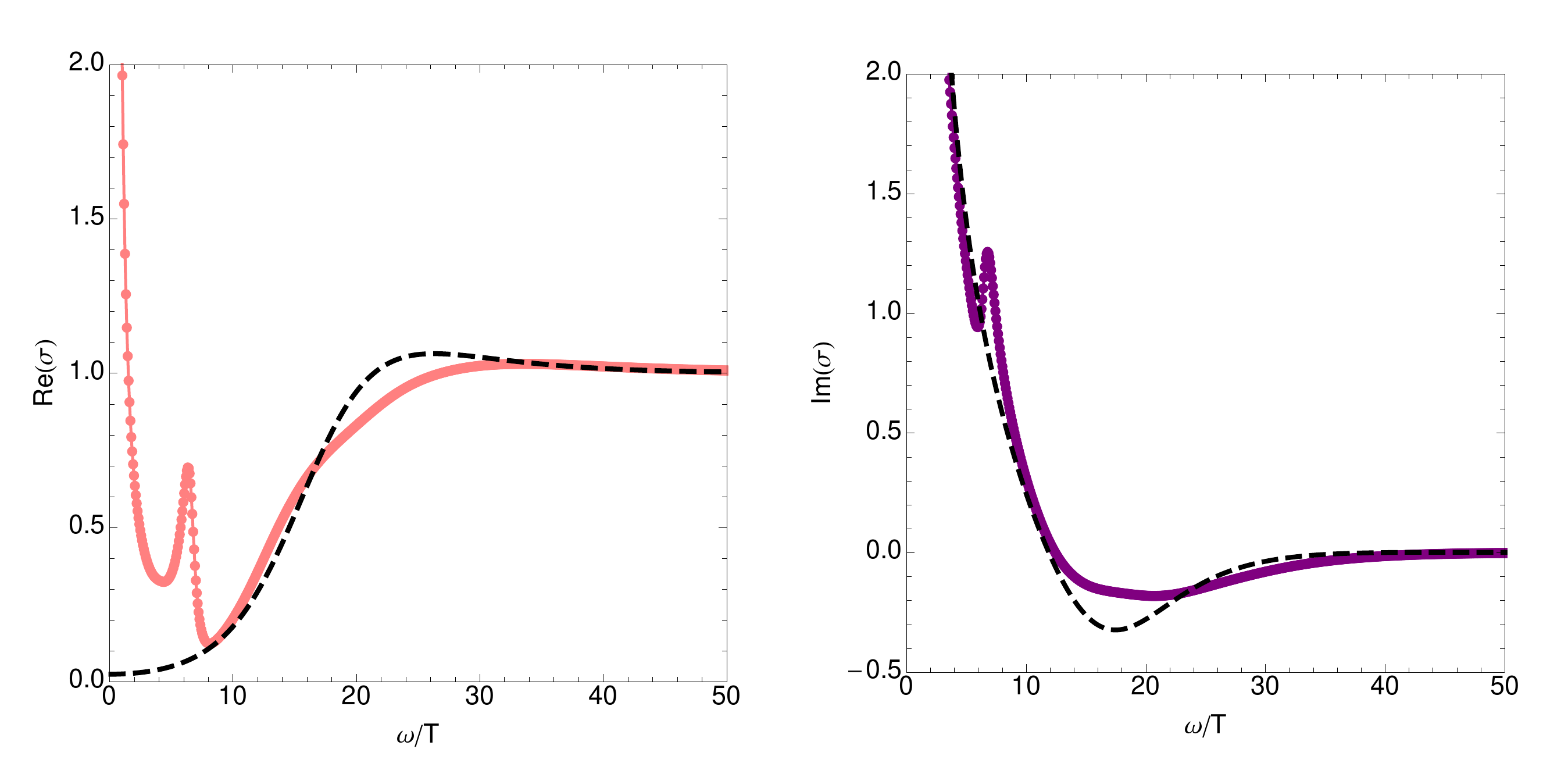}
\caption{\textbf{ac conductivity in an inhomogeneous background.} Illustrative plot of the real and imaginary parts of $\sigma(\omega)$ in a holographic model deformed by the periodic chemical potential (\ref{eq:hollat}). The dashed lines show the corresponding conductivities without the lattice deformation. The lattice plot shows a Drude peak at small frequencies, a resonance due to scattering off the lattice with $\omega \sim k_{\mathrm{L}}$ and the asymptotic behavior $\sigma \to 1$. Figure taken from \cite{Horowitz:2012gs} with permission.}
\label{fig:572}
\end{figure*}

Beyond universal IR physics, the numerically constructed solutions allow one to compute the ac conductivities at all frequencies.   
This is a probe of higher energy physics. An example of the kind of results one obtains in $d=2$ is shown in Figure \ref{fig:572}. This is for relatively weak momentum relaxation, so that a Drude peak (\ref{eq:memdrude}) controls the low frequency conductivity. An additional structure is seen at intermediate frequencies $\omega \sim k_{\mathrm{L}}$,  associated with resonances with the modulating potential \cite{Horowitz:2012gs, Chesler:2013qla}.  At high frequencies $\omega$, which are much larger than all other scales in the problem,  one finds that $\sigma(\omega)\rightarrow 1$ (in units with the coupling $e=1$), in agreement with (\ref{eq:Le}).

In addition to lattices, a natural class of inhomogeneous sources to consider are random potentials. These describe random disorder. As we have discussed in \S\ref{sec:holexample}, random disorder contains inhomogeneities at all wavelengths and is therefore capable of having a strong effect on the IR physics even for $z < \infty$. We will discuss the IR geometry of disordered black holes in \S \ref{sec:disfpt}.

A final class of inhomogeneities that has revealed some rather interesting physics are pointlike defects \cite{Horowitz:2014gva, Horowitz:2016ezu}. A sufficiently strong charged defect can lead to a deformation of the horizon even in the zero temperature limit. Furthermore, it was found that as the temperature is lowered the defect can cause part of the horizon to split off, leading to a localized `hovering' black hole above the deep IR black brane horizon. This suggests a novel kind of `topological order' associated with the defect. The hovering black holes are challenging to find numerically, but can be understood heuristically as the statement that some black hole spacetimes admit static, charged geodesics.   It is natural to imagine placing a point particle of charge $q$ and mass $m$ at such a geodesic, and `growing it' into a tiny black hole.  The existence of such hovering defect solutions suggests the possible existence of disordered or lattice black holes with many such `hovering' horizons.  In this `point mass' approximation, geometries with many hovering black holes have been proposed \cite{Anninos:2013mfa}.  Such geometries with a density of disconnected horizons may avoid the `no-go theorems' on many-body localization that we have mentioned above \cite{Grozdanov:2015qia, Grozdanov2}.

\subsection{SYK models}
\label{sec:SYK}

A class of random fermion Sachdev-Ye-Kitaev (SYK) models \cite{SY92,kitaev2015talk} have recently emerged as important
solvable models of strange metal states that are stable even at $T=0$. 
An explicit derivation of a holographic gravitational dual of the long time energy and number fluctuations can be established, with properties closely
related to the $\mathrm{AdS}_2$ horizons of charged black holes discussed in \S \ref{sec:ads2rd}. 
A class of higher-dimensional SYK models \cite{Gu:2016oyy} exhibit diffusion of energy and number density at $T=0$ in a metallic state
without quasipartices: thus, they provide an explicit quantum matter realization of the holographic models with strong momentum
relaxation that were examined in \S \ref{sec:in1}, \S \ref{sec:in2}, and \S \ref{sec:in3}, and of the holographic disordered physics discussed
in \S \ref{sec:disfpt}. Finally, as we noted in \S \ref{sec:qmwqp},
these models saturate the bound in (\ref{MSSbound}) for the Lyapunov time to quantum chaos \cite{kitaev2015talk,Maldacena:2016hyu}.

The simplest SYK model of direct relevance to strange metals
has the nonlocal Hamiltonian \cite{SS15}
\beq
H_{\mathrm{SYK}} = \frac{1}{(2 N)^{3/2}} \sum_{i,j,k,\ell=1}^N J_{ij;k\ell} \, c_i^\dagger c_j^\dagger c_k^{\vphantom \dagger} c_\ell^{\vphantom \dagger} 
- \mu \sum_{i} c_i^\dagger c_i^{\vphantom \dagger}. \label{HSYK}
\eeq
A local generalization will be considered in \S\ref{sec:SYKh}, enabling the realization of incoherent transport with strong momentum relaxation.
Here, the $c_i$ are fermion operators on sites $i=1 \ldots N$, which obey the standard fermion anti-commutation relations
\beq
c_i c_j + c_j c_i = 0 \quad, \quad c_i^{\vphantom \dagger} c_j^\dagger + c_j^\dagger c_i^{\vphantom \dagger} = \delta_{ij}.
\eeq
The physical properties evolve smoothly with changes in the chemical potential, $\mu$, which determines the average fermion
density 
\beq
\rho = \frac{1}{N} \sum_i c_i^\dagger c_i^{\vphantom \dagger}. 
\eeq
The properties described below hold for all $0 < \rho < 1$.
The couplings $J_{ij;k\ell}$ are independent random variables with $\overline{J_{ij;k\ell}} = 0$ and $\overline{|J_{ij;k\ell}|^2} = J^2$. A key property is that this disorder largely self-averages in the limit of large $N$: in particular, the on-site Green's function
on any site is the same, and equal to the disorder-averaged value.

We now describe the critical strange metal obtained in the $N \rightarrow \infty$ limit.
The simplest way to generate the large $N$ saddle point equations is to simply perform a 
Feynman graph expansion in the $J_{ijk\ell}$, followed by a graph-by-graph average.
Then we obtain the following equations for the on-site fermion Green's function \cite{SY92} \begin{subequations}
\begin{align}
G(i\omega) = \frac{1}{\mathrm{i} \omega + \mu - \Sigma (\mathrm{i}\omega)} \quad &, \quad \Sigma (\tau) = -  J^2 G^2 (\tau) G(-\tau) \,, \label{SYeqns} \\
G(\tau = 0^-) & = \rho;  \label{SYeqns2}
\end{align}\end{subequations}
here $\omega$ is a frequency and $\tau$ is imaginary time.
There is a remarkably rich structure in the solution of these equations, and we will present a few highlights here. 
It is not difficult to establish that any solution of these equations must be gapless \cite{SY92}: briefly, if there were a gap,
the first equation in (\ref{SYeqns}) would imply that $G$ and $\Sigma$ must have the same gap, while the second equation
implies that the gap in $\Sigma$ is 3 times the gap in $G$, which is a contradiction. The low frequency structure of the
Green's function can be determined analytically \cite{SY92}
\begin{align}
\Sigma (z) = \mu - \frac{1}{A} \sqrt{z} + \ldots \quad , \quad G(z) = \frac{A}{\sqrt{z}} \,, \label{SYG}
\end{align}
for some complex number $A$, where $z$ is a frequency in the upper-half of the complex plane. 
The result for the local Green's function in (\ref{SYG}) indicates a divergence in the local density of states at the Fermi level, 
and the absence of quasiparticles in a compressible quantum state with a continuously variable density $\rho$.

We now itemize some further properties of $H_{\mathrm{SYK}}$ that were obtained in early work.
\begin{enumerate}
\item
The imaginary time Green's function at $T=0$ can be written as 
\beq
G (\tau)  \sim \left\{ \begin{array}{c} -1/\sqrt{\tau}~~~~ \mbox{for $\tau > 0$} \\ \mathrm{e}^{-2 \pi \mathcal{E}}/\sqrt{\tau} ~~\mbox{for $\tau < 0$} \end{array} \right. \,. \label{SYGtau}
\eeq
Here $\mathcal{E}$ is a real number related to the phase of the complex number $A$ in (\ref{SYG}). It is clear that $\mathcal{E}$
is a measure of the particle-hole asymmetry in the fermion density of states: for $\mathcal{E} > 0$ ($\mathcal{E} < 0$), the
density of states for inserting a particle (hole) is larger. The value of $\mathcal{E}$ can be analytically related to the value
of $\rho$ via an argument similar to that required to prove the Luttinger theorem \cite{GPS01}.
\item 
It is also possible to solve (\ref{SYeqns}) for the Green's function at small non-zero temperatures $T \ll J$. It was found 
that $G = g_{\mathrm{s}}$ with \cite{PG98}
\beq
g_{\mathrm{s}} (\tau) \sim \frac{\mathrm{e}^{-2 \pi \mathcal{E} T \tau}}{(\sin (\pi T \tau))^{1/2}}. \label{GPGtau}
\eeq
This form resembles the local Green's function of a conformal field theory.
\item The entropy, $S$, of the large $N$ state was computed in \cite{GPS01}, and it was found that this entropy did not
vanish in the low temperature limit $S(T \rightarrow 0) = N s_0$. An expression was obtained for $s_0$ as a function of  the density $\rho$. Note that a non-zero $s_0$ does not imply an exponentially large ground state degeneracy; rather it is a consequence of 
the exponentially small level spacing of the many-body spectrum, generically found in states with finite energy density, 
extending all the way down to the ground state \cite{WFSS16}.
\end{enumerate}

After the appearance of black hole models of strange metals, it was realized in \cite{SS10} that all three properties of the SYK
model listed above match precisely with those of the charged black holes we examined in \S \ref{sec:ads2rd} and \S \ref{sec:fermionz}: specifically $g_{\mathrm{s}} (\tau)$ in (\ref{GPGtau}) is the Fourier transform 
of the IR Green's functions of $\mathrm{AdS}_2$ horizons \cite{Faulkner:2009am}
examined in \S \ref{sec:fermionz} (specifically, the nonzero temperature Green's function of equation (\ref{eq:IRTfinite})), while the entropy $N s_0$ of \cite{GPS01} matches the non-zero Bekenstein-Hawking
entropy of the $\mathrm{AdS}_2$ horizon in (\ref{eq:sss}). More recently \cite{SS15}, it was noted that the match between the Green's functions in (\ref{GPGtau})
and (\ref{eq:IRTfinite}) implied that the particle-hole asymmetry parameter, $\mathcal{E}$, was equal to the near-horizon electric field of the black hole, which can be determined from (\ref{eq:athor}). 
Furthermore, it was found that \cite{SS15} both the SYK model, and the black hole solution in \S \ref{sec:ads2rd}
obeyed the relationship \cite{Sen05,SS15}
\beq
\frac{\partial s_0}{\partial \rho} =  2 \pi \mathcal{E} \quad, \quad T \rightarrow 0, \label{dsdrho}
\eeq
where the density 
$\rho$ in the black hole case is given by (\ref{eq:rrr}). This result is further evidence for the identification of the entropy of the SYK model \cite{GPS01}
with the Bekenstein-Hawking entropy of the black hole, and of the close connection between SYK models and AdS$_2$ \cite{SS10,SS15}.

\subsubsection{Fluctuations}
\label{sec:SYKf}

Recent work has made the holographic connection between SYK models and $\mathrm{AdS}_2$ horizons quite explicit by deriving
a common effective action for $T>0$ energy and density fluctuations in both models. Here, we will outline the derivation
of this action starting from the SYK side. 

The basic structure of the effective action can be largely deduced by a careful considerations of symmetries. 
While solving the equations for the Green's function and the self energy, (\ref{SYeqns}), we found that, at $\omega, T \ll J$, 
the $\mathrm{i} \omega + \mu$ term in the inverse Green's function could be ignored in determining the low energy structure of the solution. 
The $\mu$ cancels with the leading term in $\Sigma(z)$ in (\ref{SYG}), while the $\mathrm{i} \omega$ is less important than the $\sqrt{z}$ term
in $\Sigma (z)$.
After dropping the $\mathrm{i} \omega + \mu$ term, it is not difficult to show that (\ref{SYeqns}) have remarkable, emergent,
time reparameterization and U(1) gauge invariances. This is clearest if we write the Green's function in a two-time notation, $G(\tau_1 - \tau_2) = G(\tau_1, \tau_2)$; then (\ref{SYeqns}) are invariant under \cite{kitaev2015talk,SS15} \begin{subequations}\label{repara}
\begin{align}
\tau &= f (\sigma) \,, \\
G(\tau_1 , \tau_2) &= \left[ f' (\sigma_1) f' (\sigma_2) \right]^{-1/4} \frac{ g (\sigma_1)}{g (\sigma_2)} \, G(\sigma_1, \sigma_2) \,, \\
{\Sigma} (\tau_1 , \tau_2) &= \left[ f' (\sigma_1) f' (\sigma_2) \right]^{-3/4} \frac{ g (\sigma_1)}{g (\sigma_2)} \, {\Sigma} (\sigma_1, \sigma_2) \,,
\end{align} \end{subequations}
where $f(\sigma)$ and $g(\sigma)$ are arbitrary functions representing the reparameterizations and U(1) gauge transformations respectively. 
Furthermore, it can be shown that not just the saddle-point equations, but also the entire action functional for the bilocal Green's function \cite{1995PhRvB..52..384R,GPS01,SS15,Jevicki16}
is invariant under (\ref{repara}) at low energies. 

The second key aspect is that the large set of (approximate) symmetries in (\ref{repara}) are spontaneously broken 
by the saddle point solution in (\ref{SYGtau}) and (\ref{GPGtau}); the saddle point is only invariant under a small subgroup
of these transformations. For the reparameterizations, we can only choose the SL(2, $\mathbb{R}$) subgroup \cite{kitaev2015talk}
\beq
f(\tau) = \frac{a \tau + b}{c \tau + d} \quad, \quad ad -bc =1 \,, \label{sl2r}
\eeq
so that $G(\tau = \tau_1 - \tau_2)$ in (\ref{SYGtau}) is invariant under (\ref{repara}). Similarly,  we can only 
choose the global U(1) transformation for $g(\tau)=\mathrm{e}^{-\mathrm{i} \phi}$. (At $T>0$, (\ref{GPGtau}) implies that there is subtle intertwining
of the SL(2, $\mathbb{R}$) and U(1) symmetries, for which we refer the reader to \cite{GuRichard16}.)

The reader should now note the remarkable match between the low energy symmetries of the SYK model,
and those of the Einstein-Maxwell theory of $\mathrm{AdS}_2$ horizons: this is ultimately the reason for the mapping between
these models. Like the SYK model, the Einstein-Maxwell theory has a reparameterization and U(1) gauge invariance,
and SL(2, $\mathbb{R}$) is the isometry group of $\mathrm{AdS}_2$. 

We now present the effective action for the analog of the Nambu-Goldstone
modes associated with the breaking of the approximate reparameterization and U(1) gauge symmetries to SL(2, $\mathbb{R}$)
and a global U(1) respectively. To this end, we focus only a small portion of the fluctuations of the bilocal $G$ by writing
\beq
G(\tau_1, \tau_2) = [f'(\tau_1) f'(\tau_2)]^{1/4} g_{\mathrm{s}} (f(\tau_1) - f(\tau_2)) \mathrm{e}^{\mathrm{i} \phi (\tau_1) - \mathrm{i} \phi (\tau_2)}, \label{GGs}
\eeq
and similarly for $\Sigma$; here $g_{\mathrm{s}}$ is the $T>0$ 
saddle-point solution in (\ref{GPGtau}).  Now we need an effective action for $f(\tau)$ and $\phi (\tau)$ which obeys
the crucial constraint that the action vanishes exactly for all $f(\tau)$ and $\phi (\tau)$ which leave (\ref{GGs}) invariant,
{\em i.e.\/} the $G(\tau_1, \tau_2)$ on the l.h.s. equals $g_{\mathrm{s}}$ in (\ref{GPGtau}); for such $f(\tau)$ and $\phi (\tau)$
there is no change in the Green's function, and hence the action must be zero. 
The action is obtained by a careful examination of this constraint, and also by explicit computation from the SYK 
model \cite{Maldacena:2016hyu,GuRichard16}; after writing
\beq
f(\tau) \equiv \tau + \epsilon (\tau),
\eeq
the action is\begin{widetext}
\beq
\frac{S_{\phi,\epsilon}}{N} = \frac{K}{2} \int\limits_0^{1/T} \mathrm{d} \tau \left[ \partial_\tau \phi + \mathrm{i} (2 \pi \mathcal{E} T) \partial_\tau \epsilon \right]^2 -\frac{\gamma}{4 \pi^2} \int\limits_0^{1/T} \mathrm{d} \tau \, \{\tan(\pi T (\tau + \epsilon(\tau)), \tau\}.
\label{pf1}
\eeq
\end{widetext}
The curly brackets in the last term represent a Schwarzian
\beq
\{f, \tau\} \equiv \frac{f'''}{f'} - \frac{3}{2} \left( \frac{f''}{f'} \right)^2 ,
\eeq
which has the important property of vanishing under the SL(2, $\mathbb{R}$) transformation in (\ref{sl2r}).
The couplings in (\ref{pf1}) are fully determined by thermodynamics: 
in terms of (\ref{eq:incchi}), the matrix of thermodynamic susceptibilities is 
\beq
 \left(\begin{array}{cc} \chi &\ \zeta \\ \zeta &\ c/T \end{array}\right)  =  \left( \begin{array}{cc}
K &  2 \pi K \mathcal{E} \\
2 \pi K \mathcal{E}  & (\gamma + 4 \pi^2 \mathcal{E}^2 K ) 
\end{array} \right),
\label{chi2}
\eeq
where $c$ is the specific heat at fixed chemical potential, $\chi = (\partial \rho/\partial \mu)_T$ is the compressibility,
and $\zeta = (\partial \rho/\partial T)_\mu$.

We can obtain correlators of the heat and density fluctuations in the SYK model by the identifications
\beq
\delta E(\tau) - \mu \delta \rho(\tau) = \frac{\delta S_{\phi,\epsilon}}{\delta\epsilon' (\tau)} \quad , \quad 
 \delta \rho(\tau) = \mathrm{i}\frac{\delta S_{\phi,\epsilon}}{\delta\phi' (\tau)}, \label{dsde}
\eeq
where $E$ is the energy operator. We define
\beq
\widetilde{\phi} (\tau) = \phi (\tau) +  2 \pi \mathrm{i} \mathcal{E} T \epsilon (\tau)
\eeq
and expand (\ref{pf1}) to quadratic order in $\phi$ and $\epsilon$ to obtain the Gaussian action \begin{widetext}
\beq
S_{\phi,\epsilon} = \frac{K T}{2} \sum_{\omega_n \neq 0} \omega_n^2 \left| \widetilde{\phi} (\omega_n) \right|^2 
+ \frac{T \gamma}{8 \pi^2} \sum_{|\omega_n| \neq 0, 2 \pi T} \omega_n^2 (\omega_n^2 - 4 \pi^2 T^2) |\epsilon (\omega_n)|^2 + \ldots
\label{pf2}
\eeq
\end{widetext}
where $\omega_n$ is a Matsubara frequency. Note the restrictions on $n=0,\pm 1$ frequencies in (\ref{pf2}), which are needed to eliminate
the zero modes associated with SL(2, $\mathbb{R}$) and U(1) gauge invariances. Computing correlators under (\ref{pf2}) for the observables
in (\ref{dsde}) we obtain results consistent with the fluctuation-dissipation theorem and the susceptibilities in (\ref{chi2}): indeed, 
this computation is a derivation of (\ref{chi2}).

Finally, a key observation is that the action (\ref{pf1}) can also be obtained from a gravitational theory of a black hole. 
This has been established so far for the case without density fluctuations represented by $\phi (\tau)$, although we do expect that the gravitational
correspondence is more general. 
Integrating out the bulk modes from a gravitational theory in $\mathrm{AdS}_2$ 
leads to an effective action for the boundary fluctuations of gravity, represented by $f(\tau)$, which coincides with the Schwarzian term in (\ref{pf1}):
we refer the reader to \cite{Maldacena:2016upp,Jensen:2016pah,Engelsoy:2016xyb,Cvetic:2016eiv} for details. This result confirms the intimate connection between
SYK and holographic theories of strange metals.

\subsubsection{Higher dimensional models}
\label{sec:SYKh}

\begin{figure*}
[t]
\center
\begin{tikzpicture}[scale=1.6,baseline={(current bounding box.center)}]
\draw[xshift=0pt,   fill=white] (-20pt,-33pt) rectangle (20pt,7pt);
\filldraw[xshift=0pt] (-12pt,-14pt) circle (0.6pt);
\filldraw[xshift=0pt] (-10pt,0pt) circle (.6pt);
\filldraw[xshift=0pt] (-15pt,-24pt) circle (0.6pt);
\filldraw[xshift=0pt] (15pt,-24pt) circle (0.6pt) ;
\filldraw[xshift=0pt] (-11pt,-28pt) circle (0.6pt);
\filldraw[xshift=0pt] (2pt,-22pt) circle (0.6pt);
\filldraw[xshift=0pt] (-4pt,-16pt) circle (0.6pt);
\filldraw[xshift=0pt] (-3pt,-6pt) circle (0.6pt);
\filldraw[xshift=0pt] (5pt,-8pt) circle (.6pt) ;

\filldraw[xshift=0pt] (12pt,-14pt) circle (0.6pt)
node[left] {\scriptsize $k$};
\filldraw[xshift=0pt] (12pt,-2pt) circle (.6pt) node[left]{\scriptsize $j$};

\node at (30pt,17pt) {$J'_{jklm}$};
\draw [dashed] (12pt,-14pt)--(30pt,12pt);
\draw [dashed] (12pt,-2pt)--(30pt,12pt);
\draw [dashed] (50pt,0pt)--(30pt,12pt);
\draw [dashed] (48pt,-14pt)--(30pt,12pt);

\draw[xshift=60pt, ] (-20pt,-33pt) rectangle (20pt,7pt);
\filldraw[xshift=60pt] (-3pt,-6pt) circle (0.6pt);
\filldraw[xshift=60pt] (5pt,-8pt) circle (.6pt);
\filldraw[xshift=60pt] (12pt,-2pt) circle (.6pt);
\filldraw[xshift=60pt] (12pt,-14pt) circle (0.6pt);
\filldraw[xshift=60pt] (-15pt,-24pt) circle (0.6pt);
\filldraw[xshift=60pt] (15pt,-24pt) circle (0.6pt);
\filldraw[xshift=60pt] (-11pt,-28pt) circle (0.6pt);
\filldraw[xshift=60pt] (2pt,-22pt) circle (0.6pt);
\filldraw[xshift=60pt] (-4pt,-16pt) circle (0.6pt);
\filldraw[xshift=60pt] (-12pt,-14pt) circle (0.6pt) node[right]{\scriptsize $m$};
\filldraw[xshift=60pt] (-10pt,0pt) circle (.6pt) node[right]{\scriptsize $l$};
\draw[xshift=-60pt,   fill=white] (-20pt,-33pt) rectangle (20pt,7pt);
\filldraw[xshift=-60pt] (-3pt,-6pt) circle (0.6pt) node[left]{\scriptsize $k$};
\filldraw[xshift=-60pt] (5pt,-8pt) circle (.6pt) node[right]{\scriptsize $l$};
\filldraw[xshift=-60pt] (-10pt,0pt) circle (.6pt) node[left]{\scriptsize $j$} ;
\filldraw[xshift=-60pt] (12pt,-2pt) circle (.6pt) node[right]{\scriptsize $m$};
\draw [xshift=-60pt,dashed] (12pt,-2pt)--(0pt,12pt);
\draw [xshift=-60pt,dashed] (-10pt,0pt)--(0pt,12pt);
\draw [xshift=-60pt,dashed] (5pt,-8pt)--(0pt,12pt);
\draw [xshift=-60pt,dashed] (-3pt,-6pt)--(0pt,12pt);
\filldraw[xshift=-60pt] (-12pt,-14pt) circle (0.6pt);
\filldraw[xshift=-60pt] (12pt,-14pt) circle (0.6pt);
\filldraw[xshift=-60pt] (-15pt,-24pt) circle (0.6pt);
\filldraw[xshift=-60pt] (15pt,-24pt) circle (0.6pt);
\filldraw[xshift=-60pt] (-11pt,-28pt) circle (0.6pt);
\filldraw[xshift=-60pt] (2pt,-22pt) circle (0.6pt);
\filldraw[xshift=-60pt] (-4pt,-16pt) circle (0.6pt);
\node[xshift=-90pt] at (0pt,17pt){ $J_{jklm}$};
\draw (-100pt,-13pt)--(-80pt,-13pt);
\draw (100pt,-13pt)--(80pt,-13pt);
\draw[dashed] (-100pt,-13pt)--(-120pt,-13pt);
\draw[dashed] (100pt,-13pt)--(120pt,-13pt);
\draw (-20pt,-13pt)--(-40pt,-13pt);
\draw (20pt,-13pt)--(40pt,-13pt);
\end{tikzpicture}
\caption{\textbf{A chain of coupled SYK sites} with complex fermions: each site contains $N\gg 1$ fermions with on-site
interactions as in (\ref{HSYK}). The coupling between nearest neighbor sites are four fermion interaction with two from each site.   Figure adapted from \cite{Gu:2016oyy} with permission.}
\label{fig:chainSYK}
\end{figure*}
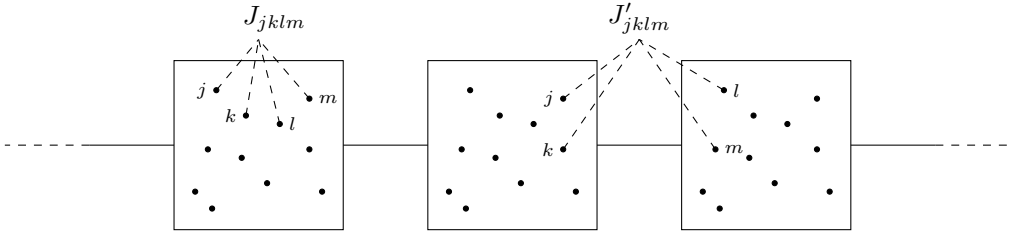

Gu {\em et al.\/} have defined a set of higher-dimensional SYK models \cite{Gu:2016oyy} which turn out to match the 
holographic transport results in \S \ref{sec:thermoelectric}, in the case of a `maximally incoherent' Einstein gravity model with axions.   For the complex SYK model in (\ref{HSYK}), the one-dimensional
Hamiltonian on sites, $x$, can be written as
\begin{equation}
H_{\mathrm{SYK}}^{\prime} =\sum_{x} (H_{x}+\delta H_x) \,.
\end{equation}
The on-site term $H_x$ is equivalent to a copy of (\ref{HSYK}) on each site
\begin{equation}
H_x= \sum_{\substack{1\leq j < k \leq N,\\ 1\leq l < m \leq N}} J_{jklm,x} c^\dagger_{j,x} c^\dagger_{k,x} c_{l,x} c_{m.x} - \sum_{j=1}^N \mu c_{j,x}^\dagger c_{j,x}.
\end{equation}
The nearest neighbor coupling term $\delta H_x$ denotes nearest-neighbor interaction as shown in Fig~\ref{fig:chainSYK}:
\begin{equation}
\delta H_x= \sum_{\substack{1\leq j<k\leq N \\ 1\leq l<m\leq N}} J'_{jklm,x}c^\dagger_{j,x+1}c^\dagger_{k,x+1}c_{l,x}c_{m,x} + \mbox{H.c.}
\end{equation}
The coupling constants $\{J_{jklm,x} \}$ and $\{  J'_{jklm,x}\}$ are all independent random coefficients with zero mean and the variances: 
\begin{equation}
\overline{J^2_{x,jklm}}=\frac{2J_0^2}{N^3},\quad \overline{J'^2_{x,jklm}}=\frac{J_1^2}{N^3}.
\end{equation}

This model can be analyzed along the lines of \S \ref{sec:SYKf}. The diagonal single 
fermion Green's functions retain the form in
(\ref{SYG}) of the original SYK model: in the context of the higher-dimensional models, the spatial independence of the 
Green's function implies that they obey a `locally critical' $z=\infty$ scaling. Along with this $z=\infty$ scaling,
the non-vanishing $T \rightarrow 0$ limit of the entropy also remains unchanged in the higher dimensional case.
However, there is an important difference for the effective action for the density and heat fluctuations in (\ref{pf2});
for the higher dimensional case, this becomes an action for {\em diffusive\/} density and heat modes \begin{widetext}
\begin{equation}
S_{\phi,\epsilon}^\prime = \frac{K T}{2} \sum_{q,\omega_n \neq 0} |\omega_n| (D_1 q^2 + |\omega_n|) \left| \widetilde{\phi} (q,\omega_n) \right|^2 + \frac{T \gamma}{8 \pi^2} \sum_{q,|\omega_n| \neq 0, 2 \pi T} |\omega_n| (D_2 q^2 + |\omega_n|) (\omega_n^2 - 4 \pi^2 T^2) |\epsilon (q,\omega_n)|^2.
\label{pf3}
\end{equation}
\end{widetext}
The fluctuation fields $\phi$ and $\epsilon$ are functions of Matsubara frequency $\omega_n$ and wavevector $q$. 
The action (\ref{pf3}) contains two temperature-independent diffusion constants $D_{1,2}$ which control the transport of heat and number density;
their values are determined by the off-site couplings $J'$ in the Hamiltonian. 

We can now proceed as for (\ref{pf2}), and compute correlators $\phi$ and $\epsilon$, 
and hence the full thermoelectric conductivity matrix in (\ref{eq:inccons2}).
We obtain \cite{GuRichard16}
\beq
 \left( \begin{array}{cc}
 \sigma   & \alpha  \\
 \alpha T &  \bar{\kappa} \end{array} \right) = 
 \left( \begin{array}{cc}
 D_1 K & 2 \pi K \mathcal{E} D_1 \\
 2 \pi K \mathcal{E} D_1 T & (\gamma D_2 + 4 \pi^2 \mathcal{E}^2 K D_1) T 
 \end{array} \right) \,. \label{pf4}
\eeq
We also have the thermal conductivity
\beq
\kappa = \bar{\kappa} - \frac{T \alpha^2}{\sigma} = \gamma D_2  T. \label{pf5}
\eeq

We now compare the transport coefficients in (\ref{pf4}), (\ref{pf5}), and the thermodynamic susceptibilities in (\ref{chi2}), with the 
holographic results in \S \ref{sec:thermoelectric}, and find an excellent match. Equation (\ref{pf4}) shows that the SYK model realizes the incoherent transport of \S\ref{sec:in1}; the three transport coefficients $\sigma$, $\alpha$ and $\kappa$ are determined by just two diffusivities, $D_{1,2}$, and the thermodynamic susceptibilities. In the SYK case, this further leads to the 
following relationship between the thermoelectric and electric conductivities \cite{GuRichard16}
\beq
\mathfrak{s} \equiv \frac{\alpha}{\sigma} = \frac{\partial s_0}{\partial \rho} \label{alphasigma} \,, \quad \text{as} \quad T \rightarrow 0,
\eeq
which can be obtained from (\ref{pf4}) and (\ref{dsdrho}). It can be verified that (\ref{alphasigma}) is also obeyed exactly by the black holes with AdS$_2$ horizons and mean-field disorder, where $\alpha$ is of the form discussed in \S \ref{sec:thermoelectric}. The match between the higher-dimensional SYK and holographic models also extends to the facts
that they obey $z=2$ scaling for the heat and number diffusivities (as implied by their $T$ independence),
while obeying $z=\infty$ scaling for the fermion Green's functions along with non-zero entropies in the $T\rightarrow 0$ limit.

Recent work has computed properties associated with 
many-body quantum chaos, which were discussed briefly in \S \ref{sec:qmwqp} and \S \ref{sec:in1}.
It was found that the diffusivity, $D_2$, and the butterfly velocity, $v_{\textsc{b}}$, are related
by
\begin{equation}
D_2 = \frac{v_{\textsc{b}}^2}{2 \pi T},  \label{eq:Dbutterfly} 
\end{equation}
in both the SYK models and momentum-dissipating holographic theories with AdS$_2$ horizons \cite{Gu:2016oyy,Blake:2016jnn,GuRichard16}. This equality is consistent with the bound (\ref{eq:MIR4}),
as we noted in \S \ref{sec:in1}.    However, generalizing these higher-dimensional SYK models by allowing $J_1^2$ to vary in space, one finds that (\ref{eq:Dbutterfly}) no longer holds, and $D_2 \le v_{\textsc{b}}^2/2\pi T$ \cite{Gu:2017ohj}.

\section{Symmetry broken phases}
\label{sec:holoS}

\subsection{Condensed matter systems}
\label{sec:sccondmat}


The metallic Fermi liquid states with quasiparticles, reviewed in \S \ref{sec:nfl}, are well-known to be unstable to BCS superconductivity
in the presence of an arbitrarily weak attractive interaction. This is a consequence of the finite density of quasi-particle states at the Fermi 
surface, and the consequent logarithmic divergence of the Cooper pair propagator. Moreover, this instability is present even in systems with a bare repulsive interaction: it was argued \cite{KohnLuttinger} that the renormalized interaction eventually becomes negative in a high angular momentum channel, leading to superconductivity by the condensation of Cooper pairs with non-zero internal angular momentum,
albeit at an exponentially small temperature.

Turning to non-Fermi liquid metallic states, discussed in \S \ref{sec:isingnematic}--\ref{sec:emerge2}, the instability to superconductivity
is usually present, and often at a reasonably high temperature. All of these theories have critical bosonic excitations which
are responsible for the disappearance of quasiparticle excitations at the Fermi surface. The same bosons can also induce a strong
attractive interaction between the Fermi surface excitations, leading to Cooper pair formation and the appearance of superconductivity.
However, the absence of quasiparticles also implies that the logarithmic divergence of the Cooper pair propagator is not present.
Therefore, there is a subtle interplay between the strong critical attractive interaction, which promotes superconductivity,
and the absence of quasiparticles, which is detrimental to superconductivity: the reader is referred to the literature
\cite{ACF01,She:2011cm,metlitski5,2015PhRvL.114i7001L,Raghu:2015sna,2016PhRvL.117o7001W} for studies examining
this interplay in a renormalization group framework. One these works \cite{Raghu:2015sna} 
has studied a $T=0$ BKT transition between a superconductor
and a critical metal, similar to that found in holography in \S \ref{sec:bkt}.

The case for high temperature superconductivity in a non-quasiparticle metal is best established for the case of the spin density
wave critical point examined in \S \ref{sec:sdw}. Here the attractive interaction induced by the boson fluctuations leads to $d$-wave
superconductivity for Fermi surfaces with the topology of those found in the cuprates. Sign-problem-free quantum Monte Carlo
simulations find $d$-wave superconductivity \cite{metlitski4,berg1,fawang1,2016arXiv160909568W} 
at temperatures in reasonable agreement with those predicted by the
quantum critical theory \cite{ACS03,2016PhRvL.117o7001W}. 

The quantum critical metal with nematic fluctuations, studied in \S \ref{sec:isingnematic}, is unstable to pairing in all spin-singlet
even parity channels \cite{metlitski5,2015PhRvL.114i7001L,Raghu:2015sna}. This insensitivity to angular momenta arises from the 
fact that the nematic boson is near zero lattice momentum, and so only couples electrons independently on small antipodal patches of the
Fermi surface. In practice, other UV features of the model will select a particular angular momentum as dominant. We have already mentioned above the strong superconducting instability seen in Monte Carlo studies of
a closely related quantum critical point \cite{2016arXiv161201542L}.

Finally, the critical metals with emergent gauge fields in \S \ref{sec:emerge2} are also usually unstable to superconductivity
via a route similar to that for the nematic fluctuations. With non-Abelian gauge fields, there is always a channel
with attractive interactions, and this has been discussed in the context of color superconductivity in the quark-gluon plasma \cite{Son:1998uk}. With a U(1) gauge field, superconductivity appears if there are charged excitations with opposite gauge charges
(as was the case with the model of Figure \ref{fig:flstar}).
The single exception is the case with a U(1) gauge field in which all the fermions carry the same gauge charge: now the singular
interaction between antipodal fermions is repulsive, and a non-superconducting critical state can be stable \cite{metlitski5}.

\subsection{The Breitenlohner-Freedman  bound and IR instabilities}
\label{sec:IRinstab}

The holographic IR scaling geometries that we have discussed have a built-in mechanism for instability towards ordered phases.
In the best studied case of AdS spacetimes, this is called the Breitenlohner-Freedman bound \cite{Breitenlohner:1982bm, Mezincescu:1984ev,Klebanov:1999tb}. A field becomes unstable in the IR scaling geometry if its scaling dimension $\Delta_\text{IR}$ becomes complex. For minimally coupled scalar fields with dimension given by (\ref{eq:deltaL}), this occurs if\footnote{As noted below (\ref{eq:nuk}), in $z= \infty$ geometries such as $\mathrm{AdS}_2\times\mathbb{R}^d$ one must take $D_\text{eff.} = 1$, counting the number of scaling boundary dimensions.}
\be\label{eq:BFbound}
m^2 L_\text{IR}^2 < - \frac{D_\text{eff.}^2}{4} \,.
\ee
We had noted in \S \ref{sec:multitrace} above that a negative mass squared does not in itself lead to an instability of these geometries. That is because the negative curvature of the background effectively acts like a box that cuts off long wavelength instabilities. However, the criterion (\ref{eq:BFbound}) shows that the mass squared cannot be too negative. Note that the Breitenlohner-Freedman bound is a bound on the bulk mass squared, and is different from the unitarity bound on the scaling dimension (also discussed in \S \ref{sec:multitrace}, the unitary bound is the lower limit of $\Delta_\text{IR}$ for which alternate quantization is allowed). In cases with hyperscaling violation, or with a logarithmically running scalar field that violates true scale invariance, the criterion for instability can be more complicated. That is because the correlation functions of fields need not have a simple scaling form in the IR regime. Some issues arising in these cases are discussed in \cite{Cremonini:2016bqw}. In general, the most robust indication of an instability will be from the behavior of correlations functions in the leading far IR limit $\omega \to 0$. A complex scaling exponent in that limit indicates an instability.

If the instability (\ref{eq:BFbound}) were to occur in the near-boundary region of the spacetime, it would indicate a sickness of the underlying theory. However, if the instability occurs in the interior of the spacetime, it can be resolved by condensation of the unstable mode. The backreaction of the condensate then alters the interior geometry, self-consistently removing the instability. We shall discuss some examples shortly. There are various holographic mechanisms that can lead to an IR mass that is unstable due to (\ref{eq:BFbound}) while the UV mass is stable. Indeed, understanding the different such mechanisms amounts to understanding the different types of ordering instability that can arise in holography. The canonical mechanism is that of a charged scalar field. The minimal Lagrangian for such a field is
\be\label{eq:Scs}
S = - \int \mathrm{d}^{d+2}x \sqrt{-g} \left(|\nabla \phi - \mathrm{i} q A \phi|^2 + m^2 |\phi|^2 + V(|\phi|) \right) \,.
\ee
Here $q$ is the charge of the field. The effective mass squared of the field gets a negative contribution from the coupling to a background Maxwell scalar potential $A_t$, so that \cite{Gubser:2008px}
\be\label{eq:meff}
m_\text{eff}^2 = m^2 - q^2 |g^{tt}| A_t A_t\,.
\ee
For a compressible holographic phase $|g^{tt}|A_t^2$ will decay towards the UV boundary (recall (\ref{eq:atnearb})), and hence in the UV $m_{\mathrm{eff}}^2 \approx m^2$. However, in a $T=0$ interior scaling geometry the Maxwell term often leads to a constant or stronger negative shift to the effective mass squared. In such cases, an instability occurs whenever the effective IR mass in (\ref{eq:meff}) obeys the instability criterion (\ref{eq:BFbound}), verified explicitly in \cite{Hartnoll:2008kx, Gubser:2008pf, Denef:2009tp}.

\begin{figure}
\centering
\includegraphics[height = 0.25\textheight]{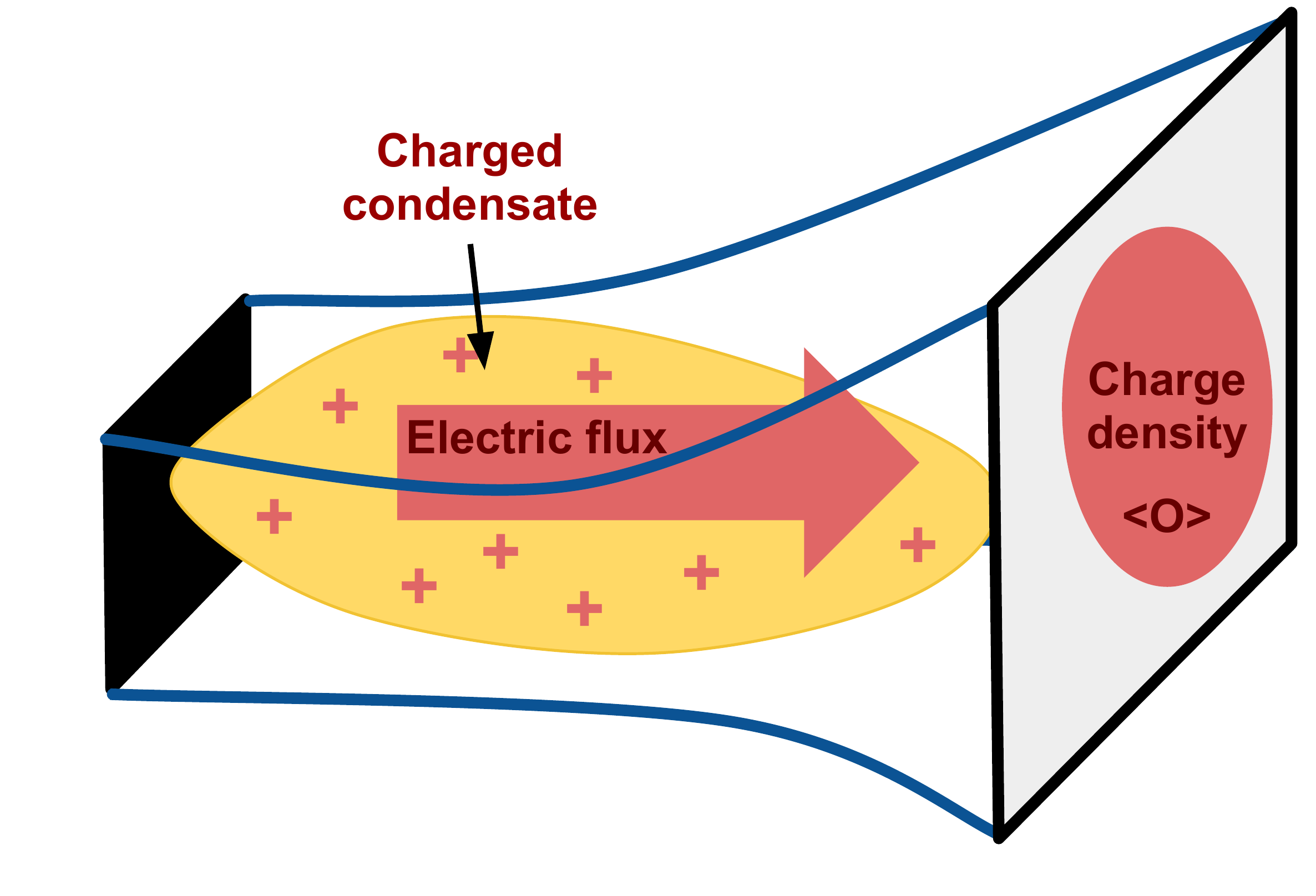}
\caption{\label{fig:endpt} \textbf{The zero temperature endpoint of the instability}: a charged bosonic condensate in the spacetime. Figure taken with permission from \cite{Hartnoll:2011fn}.}
\end{figure}

The physics at work behind the imaginary scaling dimension for a charged scalar field is the same as that discussed for charged fermions in \S \ref{sec:thomasfermi} above. That is to say, the competition between $m$ and $q$ in (\ref{eq:meff}) is between electromagnetic screening and gravitational anti-screening. If the electromagnetic term wins out, then pair production in the near-horizon geometry will discharge the event horizon. See Figure \ref{fig:gversuse} above. The endpoint of the instability will be a charged condensate in the near horizon geometry, illustrated in Figure \ref{fig:endpt}.
The charged condensate `hair' on the black hole is held in by the gravitational potential of the asymptotically anti-de Sitter spacetime and is pushed away from the black hole by any remaining charge behind the horizon (or, strictly at $T=0$, by the spatial geometry collapsing to zero size in the interior). This is the manner in which such backgrounds evade the `no hair' theorems of conventional general relativity. Usually, one expects any matter to either fall into the black hole or radiate out to infinity, but both avenues of escape are closed off here.
The resulting solutions -- black holes with a hovering charged condensate outside the horizon -- are of course closely analogous to the electron stars depicted in Figure \ref{fig:star} above. There is no need to take a WKB limit with bosons in order to obtain classical soltuions.

There are two important differences between bosonic and fermionic instabilities. These are both due to the fact that bosons 
can macroscopically occupy quantum states. Firstly, the bosonic pair production instability can be seen at the level of solutions to the classical equations of motion following from (\ref{eq:Scs}). See the discussion below and in \cite{Faulkner:2009wj}. Secondly, the condensate will have a definite phase and hence spontaneously break the electromagnetic symmetry \cite{Gubser:2008px, Hartnoll:2008vx, Hartnoll:2008kx}. Such spontaneous symmetry breaking is the main topic of this section.

The discussion above pertains to $T=0$. That is where there is a true IR fixed point that can be discussed without reference to the UV completion of the geometry. As temperature is turned on, the horizon appears as an IR cutoff on the spacetime. As temperature is increased, more and more of the interior scaling regime is swallowed up by the black hole horizon, and eventually most of the background geometry is given by the asymptotic regime where the negative Maxwell contribution to (\ref{eq:meff}) is small. The scalar field will therefore be stable at such high temperatures. It follows that whenever (\ref{eq:BFbound}) is satisfied, so that the $T=0$ interior scaling geometry is unstable, there exists a critical temperature $T_{\mathrm{c}}$ above which symmetry is restored. Increasing the temperature effectively makes the box in which the scalar propagates narrower in the radial direction, ultimately cutting off the long wavelength instability entirely. This restoration of symmetry with temperature is of course what one would expect to find on general grounds.

Symmetry can also be restored remaining at $T = 0$ by tuning sources in the boundary field theory, such as magnetic fields. This leads to a very interesting class of infinite order quantum phase transitions that will be discussed in \S\ref{sec:bkt} below.

Our focus in the following will be on instabilities of low energy compressible phases, of the sort outlined above. A few qualifications are in order. Firstly, there can also be instabilities for which the unstable mode is not localized in the near horizon region \cite{Faulkner:2010gj}. These are bosonic analogues of Fermi surfaces located away from the IR geometry, discussed in \S \ref{sec:semiholog}. While such instabilities certainly break the symmetry, they are not generated by the universal low energy dynamics and are probably best understood semi-holographically, as in \S \ref{sec:semiholog}. Secondly, continuous phase transitions triggered by the zero mode instability can sometimes be pre-empted by first order transitions to the symmetry broken phase \cite{Franco:2009yz}. We will discuss the computation of the relevant free energy below, but will mostly focus on the case of continuous phase transitions. Thirdly, neutral fields can also condense by violating the IR Breitenlohner-Freedman bound but not the UV bound \cite{Hartnoll:2008kx}. This leads to a phase transition that is not necessarily an ordering transition. A proper physical understanding of these transitions has not been achieved at the time of writing. It is possible that they are simply large $N$ `artifacts', in which a thermal vacuum expectation value changes in a non-analytic way at some temperature $T_c$. If decorated with additional indices or with a $\Z_2$ symmetry, such neutral fields can act as proxies for magnetization transitions \cite{Iqbal:2010eh, Cai:2014oca}.

\subsection{Holographic superconductivity}

\subsubsection{The phase transition}
\label{sec:transition}

The first step is to diagnose the instability of the normal, symmetric phase. Following the above discussion we start with a zero temperature perspective and explain why a complex imaginary scaling dimension $\Delta$ in the IR scaling geometry leads to an instability. An instability is found by showing that there is a pole in the upper half complex frequency plane of the retarded Green's function of the charged scalar operator $\ocal$ in the boundary field theory, dual to the charged field $\phi$ with action (\ref{eq:Scs}) in the bulk. Such poles are disallowed by causality and lead to exponentially growing perturbations upon taking the Fourier transform. The low frequency, zero temperature retarded Green's function of $\ocal$ can be computed using the same matching procedure described in \S \ref{sec:zeroTspectral} and \S\ref{sec:fermionz}.  Thus, as in (\ref{eq:GRrelate}) and (\ref{eq:BB}), we have
\be\label{eq:GRrelate2}
G^{\mathrm{R}}_{\mathcal{OO}}(\omega) = \frac{b^1_{(1)} + b^2_{(1)} \omega^{2 \nu}}{b^1_{(0)} + b^2_{(0)} \omega^{2 \nu}} \,, \qquad \Delta_\text{IR} = \frac{D_\text{eff.}}{2} + \nu  \,.
\ee
Recall again that $D_\text{eff.} = 1$ for the case of $\mathrm{AdS}_2\times\mathbb{R}^d$.
Here -- unlike in the earlier sections -- we are interested in zero momentum, $k=0$ modes, and hence the above expression holds also for IR scaling geometries with $z < \infty$. With a complex dimension $\Delta_\text{IR}$, the exponent $\nu$ is pure imaginary. For scalar fields, it can be shown that the Green's function (\ref{eq:GRrelate}) then has an infinite number of poles in the upper half complex frequency plane, accumulating at $\omega = 0$ \cite{Faulkner:2009wj, Iqbal:2011ae}. This is the superconducting instability.

For example, for the minimal charged scalar field (\ref{eq:Scs}) in the extremal AdS$_2 \times \R^d$ background (\ref{eq:nhmetric}), one has
\be\label{eq:nuchargedboson}
\nu = \sqrt{\frac{1}{4} + m^2 L_2^2 - \frac{q^2 e^2}{\k^2} L_2^2} \,.
\ee
This is the charged, zero momentum version of (\ref{eq:nuk}) above. It is analogous to the exponent (\ref{eq:nukferm}) found earlier for charged fermions. From (\ref{eq:nuchargedboson}) we see that the IR geometry becomes unstable whenever the charge of the boson is sufficiently large compared to the mass
\be\label{eq:qm}
q^2 e^2 > \k^2 \left( m^2 + \frac{1}{4 L_2^2} \right) \,.
\ee
This formula is the bosonic analogue of the condition (\ref{eq:instabF}) for fermionic pair production in the IR spacetime. Similar inequalities relating the charge and mass of the field will exist for more general IR scaling geometries. The key point is to know the effective mass (\ref{eq:meff}) of the charged field in the IR, as well the condition (\ref{eq:BFbound}) for instability.

Zero temperature is typically deep in the regime where the normal phase is unstable. We have discussed in the previous section how increasing the temperature works to stabilize the solution. At the critical temperature, the unstable mode becomes a zero mode. The critical temperature $T_{\mathrm{c}}$ itself is therefore found by looking for a normalizable solution to the scalar equation of motion with $\omega = k = 0$. In a general background of the form (\ref{eq:abmetric}), the linearized zero frequency and momentum equation for the scalar is, from the action (\ref{eq:Scs}),
\be\label{eq:zeromode}
- \frac{\mathrm{d}}{\mathrm{d}r}\left(\frac{b}{r^d} \phi^\prime\right) + \frac{1}{r^{2+d}} \left( (mL)^2 a - \frac{q^2 r^2 p^2}{b} \right) \phi = 0\,.
\ee
Normalizable means that the field is not sourced at the asymptotic boundary, i.e. $\phi_{(0)}=0$ in (\ref{eq:boundaryexpand}), and is regular at the event horizon, so that $\phi \to \text{const.}$ on the horizon as in \S \ref{sec:SWT} above. The equation is to be solved through the entire spacetime, not just in the IR region. Equation (\ref{eq:zeromode}) has the form of a Schr\"odinger equation. The zero mode occurs when there is a zero energy bound state of the Schr\"odinger equation. The critical temperature $T_{\mathrm{c}}$ is the highest temperature at which such a bound state appears. To find the bound state of the Schr\"odinger equation, one must typically use numerics. 
Shooting or spectral methods have both been widely used.

\begin{figure}
\centering
\includegraphics[width=3.2in]{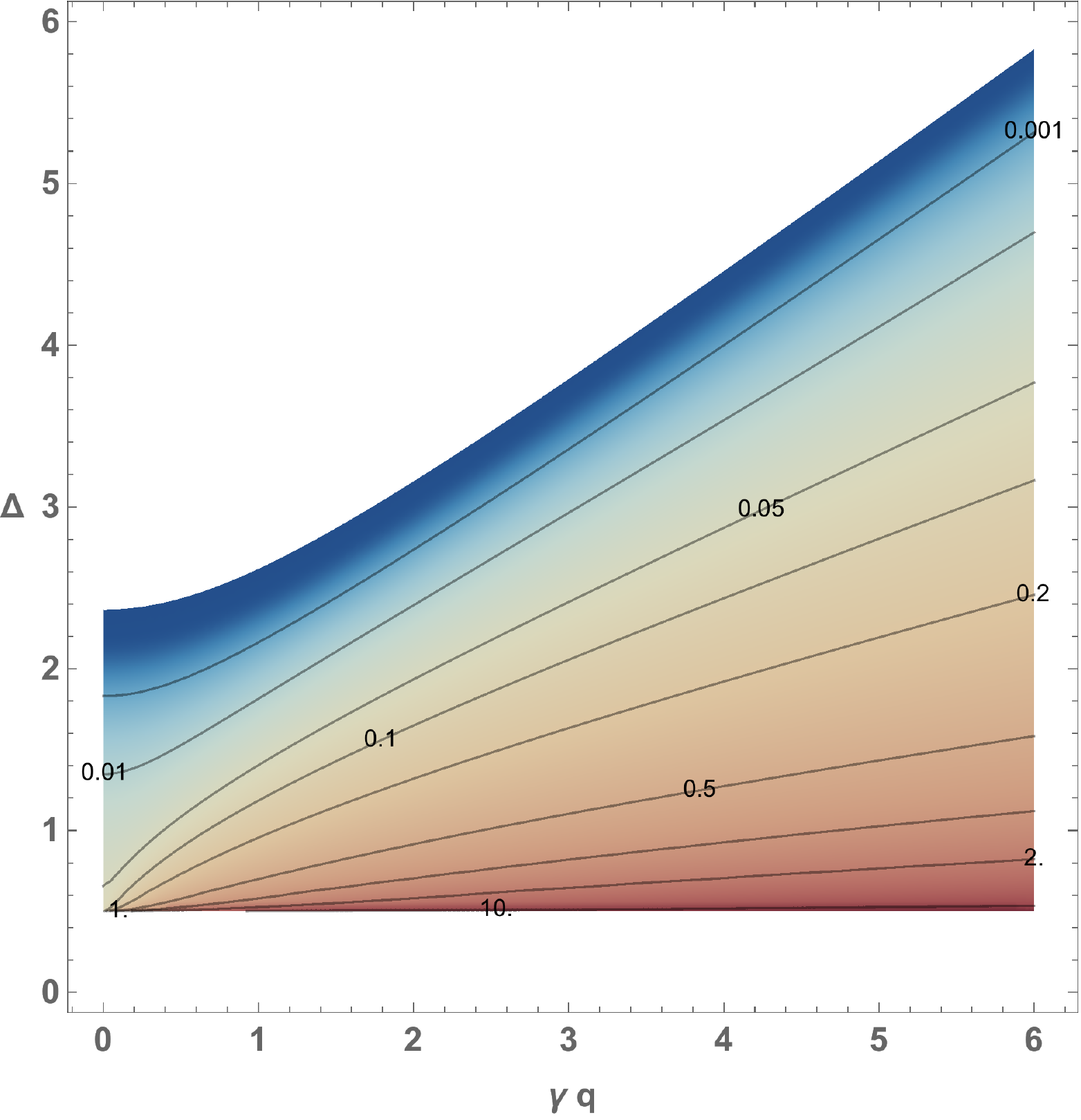}
\caption{\label{fig:tcplot} \textbf{Critical temperature} as a function of UV scaling dimension $\Delta$ and charge $\g q$ of the operator that condenses. Contours show values of $\gamma T_{\mathrm{c}}/\mu$. Figure adapted with permission from \cite{Denef:2009tp}.}
\end{figure}

Because the zero temperature instability is localized in the scaling regime, one expects that the normal phase will become stable once the nonzero temperature horizon has `eaten up' the scaling regime. In the simplest holographic phases, obtained by doping a strong interacting CFT as described in \S \ref{sec:compressible} above, this will imply that $T_{\mathrm{c}} \sim \mu$ is set by the chemical potential. For the example of a charged scalar field (\ref{eq:Scs}) in the Reissner-Nordstr\"om-AdS background of Einstein-Maxwell theory, discussed in \S\ref{sec:ads2rd} above, $T_{\mathrm{c}}/\mu$ is a function of the UV scaling dimension $\Delta$ and charge $\gamma q$ of the scalar field  \cite{Denef:2009tp}. The dependence is shown in Figure \ref{fig:tcplot}.
The boundary of the unstable region in this plot is precisely described by the curve (\ref{eq:qm}), as we should
expect. A second feature of the plot is that as the UV unitarity bound $\Delta = \half$ is approached, the critical temperature diverges.
See \cite{Denef:2009tp} for a possible interpretation of this fact.

When the critical temperature is low, $T_{\mathrm{c}} \ll \mu$, then the dynamics of the instability is fully captured by the emergent IR scaling theory.
In particular, this means that the low energy ($\omega \ll \mu$) spectral weight of the order parameter at temperatures $T_{\mathrm{c}} < T \ll \mu$ is sensitive the IR scaling dynamics and can be described using the matching procedures of \S\ref{sec:lowEspect} so that
\be\label{eq:relate3}
G^{\mathrm{R}}_{\mathcal{OO}}(\omega,T) = \frac{b^1_{(1)} + b^2_{(1)} \omega^{2 \nu} F(\frac{\omega}{T})}{b^1_{(0)} + b^2_{(0)} \omega^{2 \nu} F(\frac{\omega}{T})} \,, 
\ee
Here the $b^i_{(j)}$ can have analytic (i.e. series expansion) dependence on $\omega$ and $T$. More nontrivial is the scaling dependence of the function $F$ on $\omega/T$. Such an anomalous `pair susceptibility' above $T_{\mathrm{c}}$ is an interesting signature for unconventional mechanisms of superconductivity \cite{She:2011cm}. For the particular case of the AdS$_2 \times \R^d$ background (\ref{eq:nhmetric}) of Einstein-Maxwell theory, an extra emergent $\mathrm{SL}(2,\R)$ symmetry fixes the form of $F(\omega/T)$ to be a ratio of gamma functions \cite{Faulkner:2011tm}, very similar to the fermionic Green's functions given in (\ref{eq:IRTfinite}) above.

As $T \to T_{\mathrm{c}}$ from above it can be shown that a quasinormal mode of the system moves upwards in the complex frequency plane, towards $\w = 0$ \cite{Amado:2009ts, Plantz:2015pem}. Close to the critical temperature this leads to a pole in the retarded Green's function
\be
G^{\mathrm{R}}_{\mathcal{OO}}(\omega,k) = \frac{1}{a\omega + b k^2 + c (T-T_{\mathrm{c}})} \,,
\ee
where $a$ is complex, such that the pole is in the lower half complex frequency plane for $T>T_{\mathrm{c}}$, while $b$ and $c$ are real. The divergence as $T \to T_{\mathrm{c}}$ is the generic behavior expected from time-dependent Landau-Ginzburg theory, with dissipation of the order parameter.

\subsubsection{The condensed phase}

So far we have characterized the onset of the instability form the point of view of the normal phase. We can now turn to the ordered dynamics at temperatures $T < T_{\mathrm{c}}$.

Once the scalar field has condensed, a new background geometry must be found in which $\phi$ is nonzero. This requires solving the coupled equations of motion following from adding the charged scalar (\ref{eq:Scs}) to the Einstein-Maxwell-dilaton (or more general) theory (\ref{eq:EMDaction}) that described the normal state. This task is not substantially different
to the construction of the Einstein-Maxwell dilaton backgrounds of \S \ref{sec:EMD} or the electron star of \S \ref{sec:thomasfermi}.
A characteristic of the spontaneous symmetry breaking case is that a solution with the appropriate boundary conditions for the scalar field (an expectation value but no source) will only exist below the critical temperature.

As always, the universal low energy and low temperature dissipative dynamics of the symmetry broken phase is captured by the zero temperature IR geometry. The bulk physics is very similar to that of the electron star discussed in \S \ref{sec:thomasfermi}. The only real difference is that the charge-carrying field in the bulk is a boson rather than a fluid of fermions. In particular, as with the electron star in (\ref{eq:lifstar}), a common scenario is that at zero temperature all of the charge is outside of the horizon, enabling an emergent Lifshitz scaling
\begin{align}
\mathrm{d}s^2 &= L^2 \left(- \frac{\mathrm{d}t^2}{r^{2z}} + g_\infty \frac{\mathrm{d}r^2}{r^2} + \frac{\mathrm{d}\vec x^2_d}{r^2} \right) \,, \notag \\
 A &= h_\infty \frac{\mathrm{d}t}{r^z} \,, \qquad \phi = \phi_\infty \,.
\label{eq:IRlifSC}
\end{align}
The simplest instance of this behavior may be Einstein-Maxwell-charged scalar theory, where the scalar has $m^2 > 0$ and no potential \cite{Horowitz:2009ij, Gubser:2009cg}, but it has also been found more widely in e.g. \cite{Hartnoll:2011fn, Hartnoll:2012pp, Gouteraux:2012yr}.  The value of $z$ depends on the details of the theory. An interesting phenomenon that appears to be fairly common is that the charge density operator can become irrelevant in the new IR scaling theory. 
Irrelevance of the charge density operator leads to an emergent IR Lorentz invariance with $z=1$ \cite{Gubser:2008wz, Gauntlett:2009dn, Gubser:2009gp,Gouteraux:2013oca}. This means that the bulk scalar potential vanishes with a faster power of $r$ towards the interior than shown in (\ref{eq:IRlifSC}), i.e.
\be\label{eq:emergezone}
A =  h_\infty \frac{\mathrm{d}t}{r^{1 + \alpha}} \qquad \qquad (z=1, \alpha > 0)\,. 
\ee
This behavior is less exotic than the anomalous dimension of a conserved current discussed in \S\ref{sec:anomalouscharge}, because now the U(1) symmetry is spontaneously broken at the IR fixed point. The Maxwell field is Higgsed in the near horizon geometry by the charged scalar condensate. Equation (\ref{eq:emergezone}) is just the behavior of a massive vector field in Anti-de Sitter spacetime.

More generally if the charged scalar runs in the IR -- playing a role similar to the logarithmically running dilaton in (\ref{eq:ppsol}) above --  rather than tending to a constant as in (\ref{eq:IRlifSC}), hyperscaling violation is also possible \cite{Gouteraux:2012yr,Gouteraux:2013oca}. This more general class of behaviors allows for zero temperature transitions between solutions will all, some or none of the charge in the scalar field condensate outside the black hole \cite{Adam:2012mw}. Such transitions are analogous to the fermionic fractionalization transitions discussed in \S \ref{sec:thomasfermi}. A particular feature of bosonic fractionalization transitions in large $N$ holographic models is that the fractionalized phase cannot break the symmetry spontaneously. That is, if all the charge is behind the horizon, the symmetry is unbroken.

Emergent scaling geometries such as (\ref{eq:IRlifSC}) mean, as we have explained in \S\ref{sec:thermoBH}, that in general the specific heat at low temperatures in the symmetry broken phase goes like
\be\label{eq:csc}
c = T \frac{\pa s}{\pa T} \sim T^{(d -\theta_{\textsc{s}})/z_{\textsc{s}}} \,.
\ee
We have added the subscript \textsc{s} to distinguish the superconducting values of the exponents $\theta_{\textsc{s}}$ and $z_{\textsc{s}}$ from their values in the unstable normal state. We will refer to those as $\theta_{\textsc{n}}$ and $z_{\textsc{n}}$ in the next few sentences. The power law behavior (\ref{eq:csc}) shows that, contrary to a conventional s-wave superconductor, the ordered state is not gapped. The persistence of a neutral $T=0$ `horizon' in the IR shows that while the charge has become cohesive, neutral degrees of freedom remain deconfined. This may connect with some of the topologically ordered phases discussed in \S\ref{sec:sccondmat}. The neutral degrees of freedom can be gapped by triggering a confinement transition, as in \S \ref{sec:gapped}. However, gapping the deconfined critical excitations in the normal state will also remove the IR superconducting instability. The interplay of holographic superconductivity and confinement is studied in \cite{Nishioka:2009zj,Horowitz:2010jq}.

While general results do not exist at the time of writing, in various examples with Lifshitz scaling in the normal and superconducting states (i.e. $\theta_{\textsc{s}} = \theta_{\textsc{n}} = 0$), it was found that $z_{\textsc{s}} < z_{\textsc{n}}$ \cite{Hartnoll:2012pp}. This implies that while a gap has not formed, the ordered state does have fewer low energy degrees of freedom than the unstable normal state. This is broadly to be expected from the perspective of `entropy balance' as follows. At the critical temperature $T_{\mathrm{c}}$ of a second order phase transition the free energy and entropy of the two phases are equal $f_{\textsc{s}}(T_{\mathrm{c}}) = f_{\textsc{n}}(T_{\mathrm{c}})$, $s_{\textsc{s}}(T_{\mathrm{c}}) = s_{\textsc{n}}(T_{\mathrm{c}})$, but the specific heat jumps so that $c_{\textsc{n}}(T_{\mathrm{c}}) < c_{\textsc{s}}(T_{\mathrm{c}})$. The sign of the jump is fixed by the fact that the free energy of the superconducting state must be higher just below $T_{\mathrm{c}}$. Using these facts and integrating up the relation $c = T \pa s/\pa T$ in both phases
\be
\int_0^{T_{\mathrm{c}}} \frac{c_{\textsc{n}}(T)}{T} \mathrm{d}T + s_{\textsc{n}}(0) = \int_0^{T_{\mathrm{c}}} \frac{c_{\textsc{s}}(T)}{T} \mathrm{d}T \,.
\ee
We have allowed for a zero temperature entropy density in the normal phase. This is the entropy balance equation.
The fact that $c_{\textsc{s}} > c_{\textsc{n}}$ close to the upper limit of these integrals, i.e. the ordered state has more degrees of freedom at energies just below the transition temperature, must then be balanced by the normal state having more low energy degrees of freedom. This balance was explored in various holographic models in \cite{Hartnoll:2012pp}, which also discusses experimental realizations of the balance in unconventional superconductors.

So far we have discussed the universal low temperature dynamics of the ordered phase. A further universal regime emerges at temperatures just below $T_{\mathrm{c}}$. Here, on general grounds, the system should be described by Landau-Ginzburg theory. Indeed this is the case, and the only input of the full bulk solution is to fix the various phenomenological parameters in the Landau-Ginzburg action \cite{Maeda:2009wv, Herzog:2010vz, Plantz:2015pem}. These coefficients then determine various properties of the state, such as the response to an external magnetic field. At leading order in large $N$ there are no quantum corrections to these mean field results for the phase transition, although unconventional exponents are possible \cite{Franco:2009yz, Franco:2009if}, even in low space dimension.
In \S\ref{sec:coleman} below we show how bulk quantum fluctuations, suppressed by large $N$, lead to algebraic long range order in two boundary space dimensions.

There is a large literature characterizing the non-universal regimes at intermediate temperatures in the ordered phase.\footnote{In approaching the literature, one should be aware that many computations are done in a `probe limit' in which the backreaction of the Maxwell field and charged scalar on the metric can be neglected \cite{Hartnoll:2008vx, Gubser:2008zu}. While this limit has the advantage that some interesting behavior can be exhibited with minimal computational effort, and was historically important, it is unable to capture the universal low temperature dynamics in which the metric plays a key role, and also cannot describe the interplay of superconductivity with momentum conservation.} By numerically constructing the backgrounds one can obtain thermodynamic quantities and the expectation value of the charged operator as a function of temperature. One can study the Meissner effect and the destruction of superconductivity by a magnetic field. The main conclusion of these works is that, away from the gapless low temperature limit (\ref{eq:csc}) and away from the unconventional nature of the superconducting instability as revealed in e.g. (\ref{eq:relate3}), the basic phenomenology at intermediate temperatures is just that of a conventional s-wave superconductor \cite{tinkham}. See e.g. \cite{Hartnoll:2008kx, Horowitz:2010gk, Musso:2014efa}. By constructing backgrounds with inhomogeneous sources, which requires solving PDEs, one can furthermore recover the conventional physics of Josephson junctions \cite{Horowitz:2011dz} and vortices \cite{Albash:2009iq, Montull:2009fe, Domenech:2010nf, Keranen:2009re, Iqbal:2011bf,Dias:2013bwa}. Unconventional and universal low energy physics does emerge in the zero temperature limit of holographic vortices. The vortices interact with the emergent gapless degrees of freedom described by the interior scaling geometry (\ref{eq:IRlifSC}), appearing as a defect in the low energy scale invariant theory. This interaction leads to a `rigorous' computation of drag forces on a vortex \cite{Dias:2013bwa}: a feat which may prove useful in our discussion of superfluid turbulence in \S \ref{sec:turb}.

\subsection{Response functions in the ordered phase}

\subsubsection{Conductivity}
\label{sec:condphase}

The most characteristic response of a superconductor is, of course, the infinite dc conductivity.
Formally speaking, this arises in the same way as the infinite conductivities due to translational invariance that we encountered in formulae such as (\ref{eq:Cleansigma2}). As explained in \S \ref{sec:drudeweights}, whenever an exactly conserved operator overlaps with the total electrical current operator $\vec J$, the real part of the low frequency optical conductivity is given by (\ref{eq:sreg})
\be\label{eq:resig}
\text{Re} \, \sigma(\omega) = {\mathcal D} \delta(\omega) + \s_\text{reg}(\omega) \,,
\ee
with the weight ${\mathcal D}$ determined by the overlap of the conserved operator with $\vec J$. For the hydrodynamic cases considered above, the conserved operator was the total momentum. In a superconductor, the conserved operator is the gradient of the Goldstone boson $\vec J_\varphi = \int \mathrm{d}^dx \nabla \varphi$. We use $\varphi$ for the Goldstone phase to differentiate from the charged bulk field $\phi$. This operator can be called the total supercurrent or total superfluid velocity.\footnote{Superconductors and superfluids differ by the presence of a dynamical gauge field in superconductors. In this respect, what we are describing is a superfluid, as there is no dynamical Maxwell field in the boundary QFT. However, for many purposes including computing $\sigma(\omega)$, we can pass back and forth between the language of superfluidity and superconductivity without problem. This is because in any charged system coupled to a dynamical Maxwell field,
when we write Ohm's law $j = \sigma E$, $E$ is the total electric field, the external field plus that generated by polarization of the medium (screening), and hence $\sigma$ is given by the current-current Green's functions of the ungauged theory, without a dynamical Maxwell field.}

As with the weight of the translational invariance delta function in (\ref{eq:Cleansigma2}), the strength of a superconducting delta function is determined by thermodynamic susceptibilities that appear in hydrodynamics. In the case in which momentum is not exactly conserved, so that the supercurrent is the only infinitely long-lived operator, then from  \S \ref{sec:drudeweights}
\be\label{eq:rhosonly}
\frac{{\mathcal D}}{\pi} = \frac{\chi_{J^x_\varphi J^x}^2}{ \chi_{J^x_\varphi J_\varphi^x}} \equiv \frac{\rho_{\mathrm{s}}}{m} \,.
\ee
Here we defined the superfluid density $\rho_{\mathrm{s}}$ as well as mass scale $m$. See e.g. \cite{Davison:2016hno} for more details (and a slightly different normalization of $\rho_{\mathrm{s}}$). In the case of superconductivity emerging from a compressible phase obtained by doping a CFT, as in \S \ref{sec:compressible} above, the translational and superconducting delta functions combine to give \cite{Dinprogress}
\be\label{eq:rhoandrhos}
\frac{{\mathcal D}}{\pi} = \frac{\rho_{\mathrm{n}}^2}{\mu \rho_{\mathrm{n}} + s T} + \frac{\rho_{\mathrm{s}}}{\mu} \,.
\ee
Here $\rho_{\mathrm{n}}$ and $\rho_{\mathrm{s}}$ are the normal and superfluid components of the charge density. These can be defined from the overlap of momentum and supercurrent operators with the total current operator, respectively. The total charge density $\rho = \rho_{\mathrm{n}} + \rho_{\mathrm{s}}$.
In the zero temperature limit, therefore, (\ref{eq:rhoandrhos}) simplifies to ${\mathcal D}/\pi = \rho/\mu$. In general,
the denominators in (\ref{eq:rhoandrhos}) follow from relativistic superfluid hydrodynamics \cite{Herzog:2011ec, Bhattacharya:2011eea, Dinprogress}, in an analogous way to how the denominators in (\ref{eq:Cleansigma2}) followed from ordinary relativistic hydrodynamics. In particular, the relation $m = \mu$ going from (\ref{eq:rhosonly}) to (\ref{eq:rhoandrhos}) is fixed by Lorentz invariance.

To obtain the weight of the delta function in a holographic superconductor it is therefore sufficient to compute the thermodynamic susceptibilities above. These are purely static quantities, and so they can be obtained without needing to solve any nonzero frequency equations. Furthermore, despite controlling the response at zero frequency, the susceptibilities are not IR dissipative variables but will instead depend on the spectrum of the theory at all energies. Thus the zero frequency computation to be done will depend on the entire bulk geometry, not just the near horizon region. To obtain ${\mathcal D}$ using the definition (\ref{eq:Dweight}) above one must solve the perturbation equation for the vector potential at $\omega = k = 0$, that is $\delta A_x = a_x(r)$. This perturbation will mix with metric perturbations $\delta g_{tx}$. We saw this, for instance, in equation (\ref{eq:mmh1}) above. Consider the case of the Einstein-Maxwell-charged scalar model.
After solving for $\delta g_{tx}(r)$,\footnote{In order to obtain the first order constraint equation for $\delta g_{tx}(r)$ it is necessary to allow the fields to be time dependent and then set the time dependence to zero after integrating the constraint equation in time. This is important because we will ultimately want to calculate $G^R_{J^x J^x}(0) = \lim_{\omega \to 0} G^R_{J_xJ_x}(\omega)$ to get the `Drude' weight in (\ref{eq:Dweight}), while the susceptibility $\chi_{J^xJ^x}$ vanishes. To obtain the equation in the text it is also necessary to use the equations of motion for the background.} and eliminating it from the equations, the resulting equation for $a_x(r)$ is \cite{Hartnoll:2008kx}
\begin{align}
0 &= - a_x^{\prime\prime} + \left(\frac{d-2}{r} - \frac{b'}{b} \right) a_x^\prime \notag \\
&+ \frac{2}{b} \left(e^2 q^2 L^2 \frac{a \phi^2}{r^2}  + \frac{\k^2}{e^2 L^2} \frac{r^2 p'^2}{a} \right) a_x  \,.
\label{eq:Deqn}
\end{align}
Here we have expressed the background metric and Maxwell field as in (\ref{eq:abmetric}), and the charged background scalar field profile is given by $\phi(r)$. Note there are two contributions to the final `mass' term for $a_x$. One comes from the background charge density -- proportional to $p'^2$ -- and the other comes from the background condensate -- proportional to $\phi^2$. Roughly speaking, these lead to the two contributions in the weight of the delta function in (\ref{eq:rhoandrhos}). To calculate the weight ${\mathcal D}$ we most solve equation (\ref{eq:Deqn}), imposing regularity on the horizon. In practice, $a_x$ will have two behaviors towards the horizon as $r \to r_+$ and we must keep the less singular one. With the solution $a_x(r)$ at hand, the weight of the delta function is then given by
\be\label{eq:DD}
\frac{{\mathcal D}}{\pi} = - G^R_{J^x J^x}(0) =  - \frac{L^{d-2}}{e^2} \lim_{r \to 0} r^{d+z-3}  \frac{a_x'(r)}{a_x(r)} \,.
\ee
For the above Einstein-Maxwell model, $z=1$. Here we have used the holographic dictionary (\ref{eq:ailif}) and (\ref{eq:jx1}) and also the formula (\ref{eq:Dweight}) for the `Drude' weight with $\chi_{J^xJ^x} = 0$.

The weight of the delta function has been computed numerically in many models of holographic superconductivity. Of particular interest are cases in which translational symmetry has been broken. In these cases the delta function is purely due to superconductivity and the expression (\ref{eq:rhosonly}) holds. Early papers that computed the superfluid density in non-translationally invariant setups are \cite{Horowitz:2013jaa, Zeng:2014uoa, Ling:2014laa}. While one expects that the superfluid density is associated to the symmetry-breaking charge that is condensed outside of the event horizon, a quantitive relation between ${\mathcal D}$ and the total charge outside the horizon has not been established at the time of writing. With weak momentum relaxation, the superconducting delta function sits on top of a Drude peak of the type described in \S \ref{sec5}. More generally, with weak or strong momentum relaxation, one expects that in the low frequency limit the divergent zero frequency conductivity is accompanied by a constant `incoherent' conductivity, analogous to $\sigma_{\textsc{q}}$ that appears in (\ref{eq:Cleansigma2}) and (\ref{eq:sigmainc}). The temperature dependence of this `normal' component will then depend on the properties of the near horizon scaling geometry as well as
the relevance or irrelevance of translational symmetry breaking. 

A second feature of conventional BCS superconductors is a gap in the optical conductivity $\sigma(\omega)$. This gap is the energy required to excite a conducting particle-hole pair from the Cooper pair condensate. Thus, while a conventional superconductor has infinite conductivity at $\omega = 0$, at zero temperature it has vanishing conductivity for small but nonzero $\omega$. At nonzero temperature, $\sigma_\text{reg} \sim \mathrm{e}^{-2\Delta/T}$, where $\Delta$ is the energy gap for a single particle excitation. The low frequency behavior of the optical conductivity is dissipative and determined by low energy degrees of freedom. Therefore $\sigma_\text{reg}(\omega)$ should depend only on the near horizon scaling geometry (\ref{eq:IRlifSC}). The absence of a gap suggests that this contribution will be power law in temperature or frequency rather than exponentially suppressed. Indeed the matching arguments of \S \ref{sec:compressible}, here applied in the simpler case of $k=0$, imply that in general at zero temperature \cite{Gubser:2008wz, Horowitz:2009ij, Gouteraux:2013oca}
\be\label{eq:softgap}
\sigma_\text{reg}(\omega) \sim \omega^{(d-2-\theta+2\Phi)/z} \,.
\ee
Recall that the exponent $\Phi$ was discussed in \S\ref{sec:anomalouscharge}. One similarly expects the temperature dependence $\sigma_\text{reg} \sim T^{(d-2-\theta+2\Phi)/z}$ as in (\ref{eq:sQT}). The appendix of \cite{Davison:2015taa} is useful for relating various different scaling exponents that have been used to describe the conductivity in scaling geometries.

\begin{figure}[t]
\centering
\includegraphics[height = 0.2\textheight]{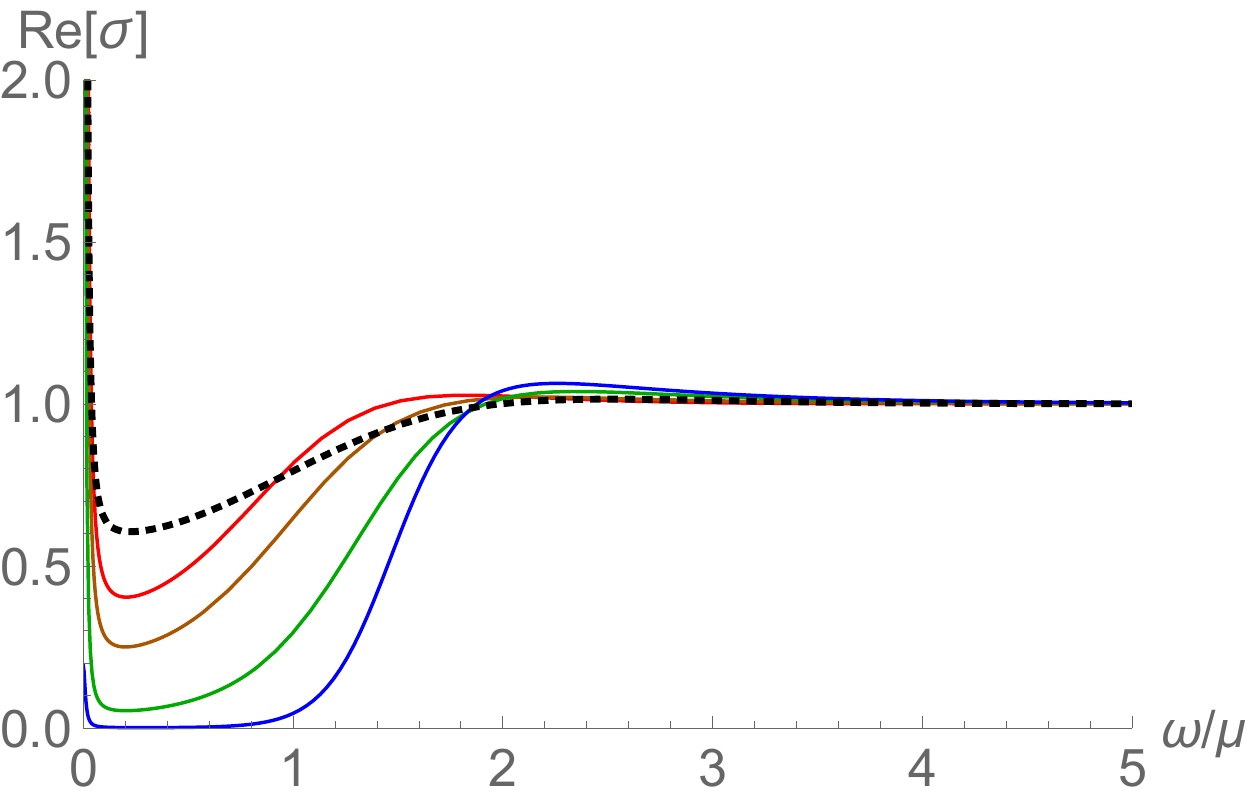}\hspace{0.25cm}\includegraphics[height = 0.2\textheight]{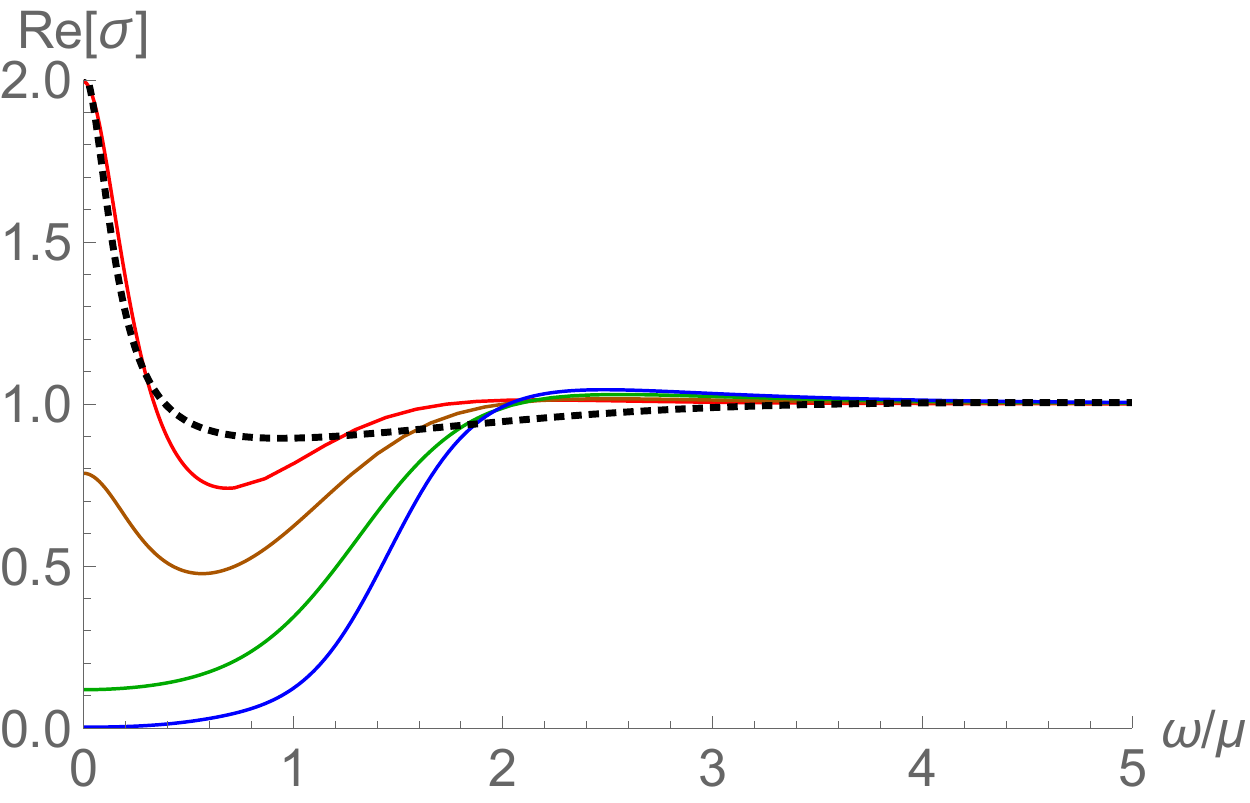}
\caption{\label{fig:scsw} \textbf{Optical conductivity in a holographic superconductor with broken translation invariance.} Dashed line is above $T_{\mathrm{c}}$, red line is at $T_{\mathrm{c}}$, subsequent lines are at progressively lower temperatures. In the top plot, translations are weakly broken, and the sharp Drude peak in the normal state persists into the superconducting state. In the bottom plot, translations are strongly broken. The weak Drude peak in the normal state eventually disappears at low temperatures in the superconducting state. All plots exhibit the loss of spectral weight into the superconducting delta function as temperature is lowered. Figures taken with permission from \cite{Kim:2015dna}.}
\end{figure}

Equation (\ref{eq:softgap}) shows that generally there is a soft gap in the optical conductivity. On general grounds one expects  a reduction of low energy spectral weight at low and zero temperature in the superconducting phase, relative to the normal phase. This is because of the Ferrel-Glover-Tinkham sum rule:
\be\label{eq:fgt}
\int_{0^+}^\infty \mathrm{d}\omega \, \text{Re} \, \Big(\s_{\textsc{n}}(\omega) - \s_{\textsc{s}}(\omega) \Big) = \frac{{\mathcal D}}{2} \,.
\ee
Here the normal and superconducting optical conductivities can be at any temperature (in the normal and superconducting phases, respectively). The sum rule can be explicitly verified in particular calculations, for a general holographic discussion see \cite{Gulotta:2010cu}. It is clear in (\ref{eq:fgt}) that the presence of the superfluid density in ${\mathcal D}$ means the nonzero frequency spectral weight is depleted in the superconducting state. Of course, at low but nonzero temperatures, in addition to the behavior (\ref{eq:softgap}) there may also be a Drude peak at low frequencies if translations are weakly broken. These behaviors are illustrated in Figure \ref{fig:scsw} for a holographic model of superconductivity with broken translation invariance.

\subsubsection{Superfluid hydrodynamics}

In the ordered phase, the new dynamical ingredient is the superfluid velocity or supercurrent, which is the gradient of the phase of the Goldstone boson $\xi_\mu = \pa_\mu \phi$. A homogeneous supercurrent flows without decay and is responsible
for the divergent conductivity discussed in the previous subsection. The conserved supercurrent is a new thermodynamic variable and can be taken to be large, leading to new thermodynamic states. In a Lorentz invariant theory the first law becomes, see e.g. \cite{Herzog:2008he},
\be\label{eq:depsilon}
\mathrm{d}\epsilon = T \mathrm{d}s + \mu \mathrm{d} \rho + \frac{\rho_{\mathrm{s}}}{2 \mu} \mathrm{d} (\xi^2 + \mu^2) \,.
\ee
Here $\rho_{\mathrm{s}}$ is the superfluid charge density, that appeared in the previous \S\ref{sec:condphase}.

Bulk gravitational solutions with a nonlinear supercurrent have been constructed in the probe limit in \cite{Basu:2008st, Herzog:2008he, Arean:2010xd, Arean:2010zw} and with full backreaction on the metric in \cite{Sonner:2010yx}. This last paper derives the equations of non-dissipative superfluid hydrodynamics from gravity. Because the solution involves a flow, it is stationary but not static. This requires a more general form of the background fields than we have considered so far \begin{widetext}
\bea
\mathrm{d}s^2 & = & \left(f(r) \eta_{ab} + g(r) u_a u_b + h(r) n_a n_b + k(r) u_{(a} n_{b)} \right) \mathrm{d}x^a \mathrm{d}x^b + (t(r) u_a + u(r) n_a ) \mathrm{d}x^a \mathrm{d}r \nonumber \\
A & = & - (p(r) u_a + q(r) n_a) \mathrm{d}x^a - s(r) \mathrm{d}r \,, \qquad \phi = \xi(r) \,.
\eea
\end{widetext}
Here $u^a$ is a timelike boost velocity and $n^a$ is spacelike vector. A gauge choice has been made to put the supercurrent into the Maxwell field $A$ rather than as an $x$-dependent condensate $\phi$. The various radial functions we have introduced above are not all independent. See \cite{Sonner:2010yx} for further details.

The papers in the previous paragraph all show that if the supercurrent becomes large, then eventually
the charged condensate is driven to zero and the normal state is recovered. Energetic computations in those papers furthermore suggest that sometimes the continuous transition is pre-empted by a first order transition to the normal state. However it is subtle to compare the free energies of the normal state and the superfluid state with a supercurrent because the supercurrent is a thermodynamic variable that does not exist in the normal state. It has been noted in \cite{Amado:2013aea} that before the condensate is driven to zero by the supercurrent, the Goldstone boson mode develops an instability at a nonzero wavevector $k>0$. The endpoint of this instability is not known at the time of writing. It may restore the normal state or it may lead to a spatially modulated superfluid state. The second option would add a further example to spatially modulated phases discussed in \S \ref{sec:inhomogS} below.

A finite non-dissipative supercurrent can furthermore flow at zero temperature. In this case, the destruction of superfluidity caused by increasing the supercurrent causes a quantum phase transition \cite{Arean:2011gz}.

In a strongly coupled, non-quasiparticle, superfluid state, the appropriate framework to understand transport
is superfluid hydrodynamics. This amounts to adding an additional hydrodynamic field, the superfluid velocity
$\pa \phi$, to the various hydrodynamic theories of \S \ref{sec5}. The most important consequence of the new
hydrodynamic variable is the appearance of a new dispersive mode, second sound. All of the modes can be found
and characterized from perturbations of the bulk Einstein equations about the ordered state, much as we have done previously for some of the normal state hydrodynamic modes. In particular, properties such as the second sound speed and parameters characterizing dissipative effects in inhomogeneous superfluid flow can be computed from the bulk.
Important papers in this endeavor include \cite{Amado:2009ts, Herzog:2009ci, Herzog:2009md, Herzog:2011ec, Bhattacharya:2011eea, Bhattacharya:2011tra, Bhaseen:2012gg}. As with other hydrodynamics modes, these
excitations appear as quasinormal modes with frequencies that become small at low momenta. Certain non-hydrodynamic
quasinormal modes close to the real frequency axis have also been calculated and shown to influence 
the late time relaxation to equilibrium in a holographic superfluid state \cite{Bhaseen:2012gg}: see \S \ref{sec:quench}.

\subsubsection{Destruction of long range order in low dimension}\label{sec:coleman}

In 1+1 dimensions at zero temperature, or in 2+1 dimensions at nonzero temperature, global
continuous symmetries cannot be spontaneously broken. This is known as the
Coleman-Mermin-Wagner-Hohenberg theorem. Quantum or thermal fluctuations of the 
phase of the order parameter lead to infrared divergences that wash out the classical
expectation value of the order parameter. There is, however, a remnant of the ordering in the form
of `algebraic long-range order'. This means that the correlation function of the order parameter
has the large distance behavior
\be\label{eq:powerlaw}
\lim_{|x| \to \infty} \langle \ocal(x)^\dagger \ocal(0) \rangle \sim \frac{1}{|x|^K} \,.  
\ee
In a truly long range ordered state, this correlator would tend to a constant, while in a conventional
disordered state it would decay exponentially rather than as a power law.

The ordered holographic phases we are discussing do not exhibit this expected low dimensional behavior at leading order in large $N$. This is because there are order $N^2$ (say) degrees of freedom, but only a single Goldstone boson.
Fluctuation effects of the Goldstone mode will then only show up at subleading order at large $N$. The upshot is that
the decay power $K$ in (\ref{eq:powerlaw}) goes like e.g. $1/N^2$ (cf. \cite{Witten:1978qu}). Thus the large $N$ and long distance limits do not commute. This is part of our comment in \S\ref{sec:nongeom} that such Goldstone modes are
`non-geometrized' low energy degrees of freedom in holographic models. To see the destruction of long range order,
one-loop effects need to be considered in the bulk. This can be done using the methods we have already
discussed in \S\ref{sec:oneoverN} and \S\ref{sec:qo}. We now outline the computation, following \cite{Anninos:2010sq}.

The first step is to identify a quasinormal mode in the bulk that corresponds to the Goldstone or `second sound' mode in the superfluid phase. This mode will then give the dominant low energy contribution to the Green's function for the scalar field, using the formula (\ref{eq:zoomin}). It will be this Green's function running in a loop that causes the quantum disordering of the phase. The mode is found starting from a rotation of the profile $\phi(r)$ of the background scalar field
\be\label{eq:deltaphi}
\delta \phi(r) = \mathrm{i} \phi(r) \,.
\ee
This just a gauge transformation of the background: $\phi \to \mathrm{e}^{\mathrm{i} \lambda} \phi$, with $\lambda$ constant. However, it is a `large' gauge transformation that doesn't vanish on the boundary (because $\lambda$ is constant in $r$). This means that the mode generated in (\ref{eq:deltaphi}) is a normalizable and physical $\omega = k = 0$ excitation. Starting from this solution, one can solve the bulk equations of motion perturbatively in small $\omega$ and $k$. Doing this turns out to be essentially independent of the form of the background solution. One finds that a normalizable solution continues to exist so long as \cite{Iqbal:2010eh, Anninos:2010sq}
\be\label{eq:ssound}
\omega^2 = c_{\textsc{s}}^2 k^2 \,,
\ee
with the speed of second sound
\be
c_{\textsc{s}}^2 = \lim_{r \to 0} \frac{\pa_r \d A_x^{\omega = k = 0}}{\pa_r \delta A_t^{\omega = k = 0}} = - \frac{\pa^2_\xi P}{\pa^2_\mu P} \,.
\ee
Here $\d A_x^{\omega = k = 0}$ and $\d A_t^{\omega = k = 0}$ are solutions to the equations of motion for perturbations about the background with $\omega= k = 0$ that are normalizable at the boundary. In the final expression $P$ is the pressure, which is the thermodynamic potential that is a function of $T, \mu$ and superfluid velocity $\xi$, cf. (\ref{eq:depsilon}) above. This final result can be derived directly from superfluid hydrodynamics \cite{Herzog:2008he}.

The above paragraph establishes the second sound quasinormal mode (actually a normal mode to the order we have worked, dissipation will appear at higher orders in the dispersion relation (\ref{eq:ssound})). The algebraic decay (\ref{eq:powerlaw}) is now obtained as follows. In the bulk we can calculate
\begin{align}
\langle \phi(r,x)^\dagger \phi(r',0) \rangle &= \phi(r) \phi(r') \langle \mathrm{e}^{- \mathrm{i} (\theta(r,x) - \theta(r',0))} \rangle \notag \\
&= \phi(r) \phi(r')  \mathrm{e}^{- \half \langle (\theta(r,x) - \theta(r',0))^2 \rangle} \,. \label{eq:ttto}
\end{align}
The last equality is a standard result following from Wick contracting. Here $\phi(r)$ is the classical background
profile and $\theta$ is the phase. The phase is related to the perturbation (\ref{eq:deltaphi}) as $\delta \phi = \phi(r) \theta$. Using (\ref{eq:zoomin}), then, the singular part of the phase two point function is
\be
G^{\mathrm{R}}_{\theta\theta}(r,r') = \frac{\Theta}{\omega^2 - c_{\textsc{s}}^2 k^2} + \cdots \,.
\ee
The numerical prefactor $\Theta$ is obtained in \cite{Anninos:2010sq}. The prefactor scales like an inverse power of large $N$. This singular contribution does not depend on $r$ or $r'$. The one loop correlation function in the exponent of (\ref{eq:ttto}) is given, using standard methods to turn the sum over Matsubara frequencies into an integral along the real frequency axis, by 
\begin{align}
& \langle \theta(x) \theta(0) \rangle - \langle \theta(0) \theta(0) \rangle = \\
& - \int\limits_{-\infty}^\infty \frac{\mathrm{d}\Omega}{2 \pi} \int \frac{\mathrm{d}^2k}{(2\pi)^2} \coth \frac{\Omega}{2T} \, \text{Im} \, G^R_{\theta\theta}(\Omega,k) \left(\mathrm{e}^{\mathrm{i} k \cdot x} - 1 \right) \,. \notag 
\end{align}
At large spatial separation the above integral gives
\be
\lim_{|x| \to \infty} \left(\langle \theta(x) \theta(0) \rangle - \langle \theta(0) \theta(0) \rangle \right) = - \frac{\Theta T}{2 \pi c_{\textsc{s}}^2} \log |x| + \cdots \,,
\ee
and hence, taking (\ref{eq:ttto}) to the boundary $r,r' \to 0$, we obtain (\ref{eq:powerlaw}) with \begin{equation}
    K = \frac{\Theta T}{2\pi c_{\textsc{s}}^2}
    \end{equation}

\subsubsection{Fermions}
\begin{figure*}
\centering
\includegraphics[height = 0.21\textheight]{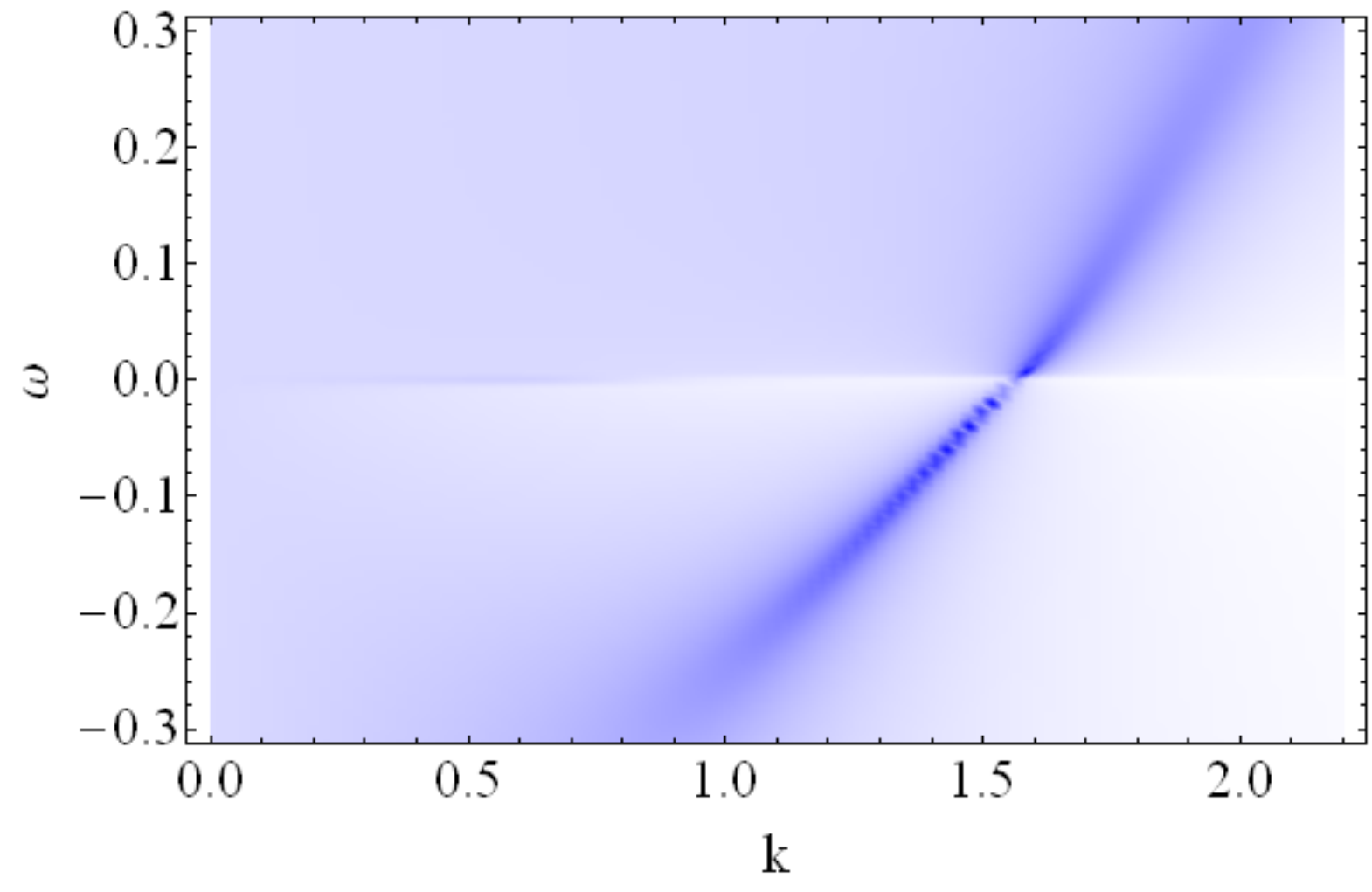}\includegraphics[height = 0.21\textheight]{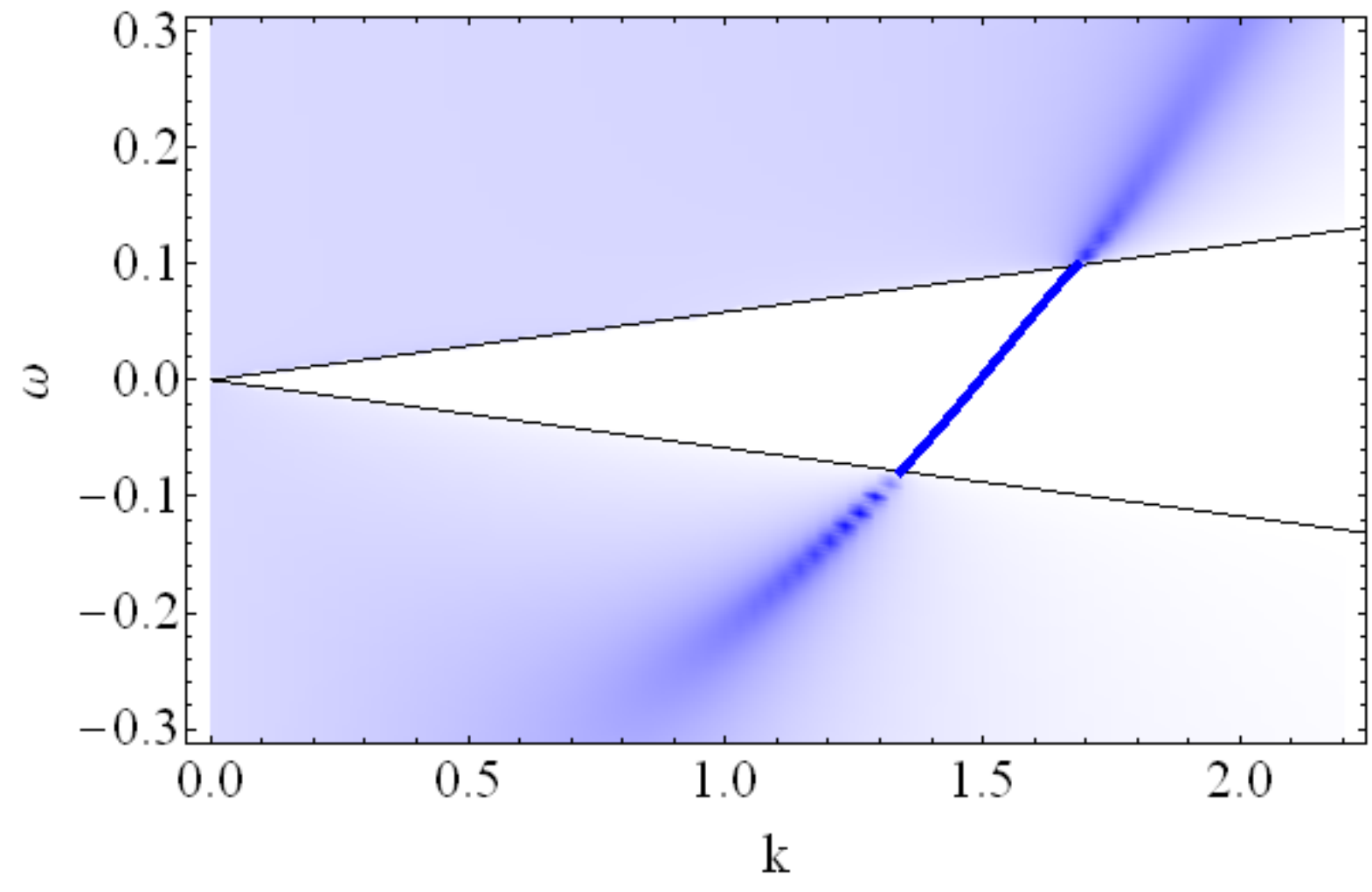}\\
\includegraphics[height = 0.21\textheight]{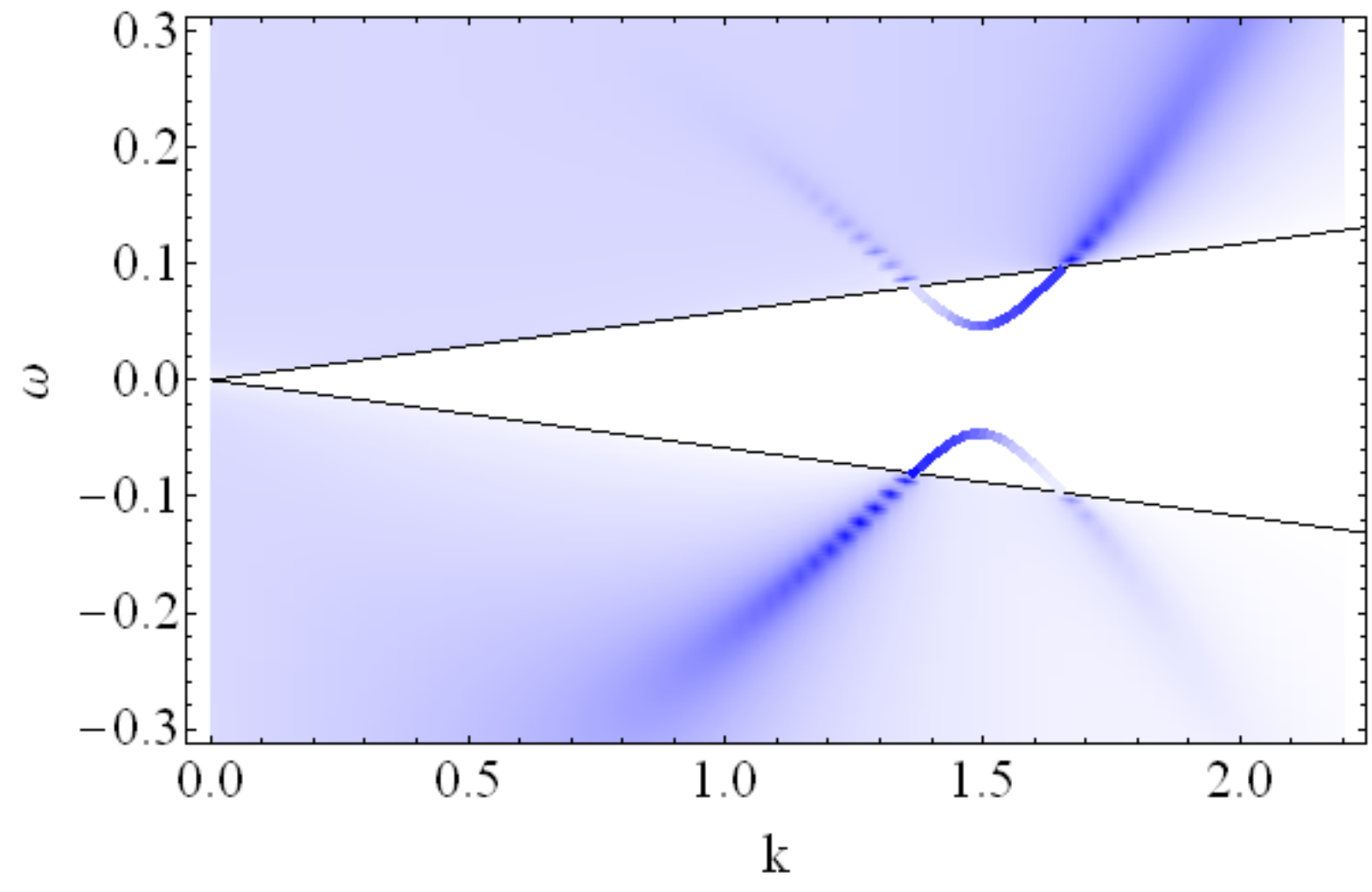}\includegraphics[height = 0.21\textheight]{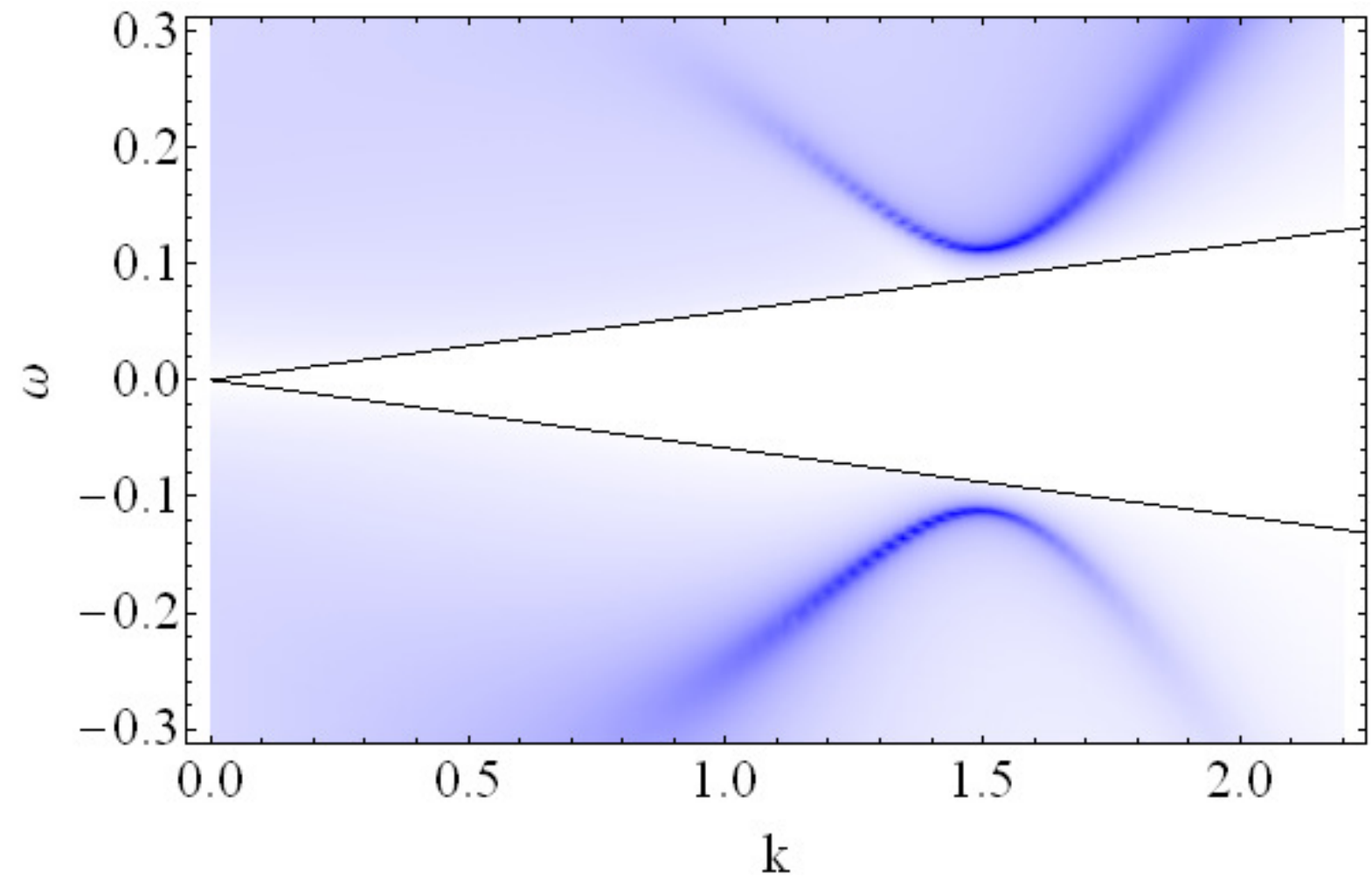}
\caption{\label{fig:fz} \textbf{Zero temperature fermionic spectral densities} as a function of $\omega$ and $k$. Top left: Normal state extremal RN-AdS background, exhibiting gapless but strongly scattered fermions. Top right: Zero temperature superconducting state (with $z=1$) with $\eta_5 = 0$. Gapless fermions still exist but are now very weakly scattered within the $z=1$ lightcone. Bottom left: Small nonzero $\eta_5$ in the superconducting state. The fermions are now gapped, but still long lived within the lightcone. Bottom right: Larger nonzero $\eta_5$. The fermions are both gapped and short lived.   Figures taken with permission from \cite{Faulkner:2009am}.}
\end{figure*}

The spectral functions of fermionic operators computed in \S\ref{sec:bulkfermions}
can be revisited in the ordered background. It was quickly recognized that there are two
important new ingredients in the superconducting phase \cite{Faulkner:2009am, Gubser:2009dt}.
Firstly, the zero temperature emergent IR scaling geometry (\ref{eq:IRlifSC}) in the superconducting phase typically
has $z < \infty$. As explained in \S\ref{sec:semiholog}, low energy fermionic excitations near some nonzero $k \approx k_{\mathrm{F}}$ are kinematically unable to decay into such a quantum critical bath and hence will have exponentially long
lifetimes. Secondly, there are couplings that can appear generically in the bulk action between the charged
fermion and charged scalar field. In the presence of a scalar condensate, certain of these couplings
have the effect of gapping out low energy fermionic excitations.

The combination of the two effects outlined in the previous paragraph
can then lead to well defined (long lived) fermionic excitations with a small gap in the superconducting phase,
emerging from a normal $z=\infty$ phase in which there were gapless but very short lived fermions, as
described in \S\ref{sec:fermionz} above. We will now explain this scenario in more detail. The scenario itself
is semi-holographic in the sense of \S\ref{sec:semiholog}. It can be described in terms of conventional field theoretic
fermions coupled to a large $N$ quantum critical bath, with the nature of the bath changing between the normal ($z=\infty$) and superconducting ($z<\infty$) states.

If the charge of the condensed scalar is twice that of the fermion operator, then the following `Majorana'
coupling \cite{Faulkner:2009am} can be added to the fermion action (\ref{eq:fermionaction}) 
\be\label{eq:eta5}
S_{\phi\psi} = \int \mathrm{d}^{d+2}x \sqrt{-g} \left(\eta_5 \,  \phi \, \Psi_{\mathrm{c}} \Gamma^5 \overline \Psi + \text{h.c.} \right) \,.
\ee
In fact this coupling only works for even $d$, for odd $d$ two different bulk Dirac fields are needed.
This coupling is the relativistic version of the conventional coupling between Bogoliubov quasiparticles in the
superconducting state with the condensate. In fact, we have encountered this coupling already in (\ref{eq:bogoliubov})
above, where we defined $\Gamma_5$ and the charge conjugate fermion $\Psi_{\mathrm{c}}$. The difference is that whereas $\Delta$ previously was a condensate of a bulk fermion bilinear, $\phi$ is dual to a single-trace scalar operator in the dual field theory.

A nonzero $\eta_5$ coupling gaps any low energy fermionic excitations in the superconducting state \cite{Faulkner:2009am}. This is illustrated in the Figure \ref{fig:fz}, which shows the zero temperature spectral density of the fermionic operators as a function of $\omega$ and $k$ at different values of $\eta_5$. This figure also shows the effects of the emergent $z=1$ scaling, as described in the figure caption.
`Effective lightcones' similar to those seen in Figure \ref{fig:fz} -- i.e. within which fermions are inefficiently scattered by the large $N$ critical bath --  also arise for $1 < z < \infty$ scaling geometries \cite{Hartnoll:2011dm}.

Fermionic spectral functions have been studied in geometries with a charged scalar background that originate from
a microscopic string theoretic construction \cite{DeWolfe:2015kma, DeWolfe:2016rxk}. The effective bulk theories that arise in this case are more complicated, involving several coupled fermionic fields. With the full microscopically determined interactions, all fermions are found to be gapped, some by couplings analogous to (\ref{eq:eta5}).

Fermion spectral functions can also be computed in backgrounds with a $p$-wave or $d$-wave superconducting condensate, to be discussed in \S\ref{sec:homogS} below. In these cases couplings analogous to (\ref{eq:eta5}) lead to anistropic gapping of the Fermi surface, due to the anisotropy of the underlying condensate. This leads to versions of `Fermi arcs'
\cite{Benini:2010qc, Vegh:2010fc}.

\subsection{Beyond charged scalars}

The minimally coupled charged scalar field (\ref{eq:Scs}) is perhaps the simplest model that
can lead to a sufficiently negative effective mass squared (\ref{eq:meff}) in the interior scaling geometry
such that an instability is triggered according to (\ref{eq:BFbound}). In this case, the global $\mathrm{U}(1)$ symmetry of the boundary theory is spontaneously broken by condensation of the charged operator dual to the bulk field.
This general mechanism can be greatly generalized to allow for other patterns of symmetry breaking.
We will first discuss more general instabilities that preserve spatial homogeneity in \S \ref{sec:homogS},
while \S \ref{sec:inhomogS} will discuss the spontaneous breaking of spatial homogeneity.
With many possible types of instability, it is natural to expect competition between the different
orders to lead to rich phase diagrams. We will not discuss these phase diagrams here but refer the reader to 
works such as \cite{Basu:2010fa, Donos:2012yu, Cai:2015cya, Kiritsis:2015hoa}. 

\subsubsection{Homogeneous phases}
\label{sec:homogS}

\paragraph{$p$-wave superconductors from Yang-Mills theory}

A rather constrained model is given by gravity coupled to a non-abelian gauge field in the bulk \cite{Gubser:2008zu, Gubser:2008wv, Roberts:2008ns,Basu:2009vv,Ammon:2009xh}:
\be\label{eq:Sp}
S = \int \mathrm{d}^{d+2}x \sqrt{-g} \left[ \frac{1}{2 \k^2} \left( R + \frac{d(d+1)}{L^2} \right) - \frac{1}{4e^2} F_{ab}^A F^{A\,ab}  \right] \,.
\ee
Here $F_{ab}^A$ is the field strength of an $\mathrm{SU}(2)$ gauge field. That is
\be\label{eq:Fp}
F^A_{ab} = \pa_a A^A_b - \pa_b A^A_a + \epsilon^{ABC} A^B_a A^C_b \,.
\ee
The three generators $\tau^A$ of the $\mathrm{SU}(2)$ algebra satisfy $[\t^B,\t^C] = \t^A \epsilon^{ABC}$. Take the $\mathrm{U}(1)$ subgroup of $\mathrm{SU}(2)$ generated by $\tau^3$ to be the electromagnetic $\mathrm{U}(1)$. The action (\ref{eq:Sp}) together with the expression for the field strength (\ref{eq:Fp}) implies that the $A^1$ and $A^2$ components of the gauge potential, the `W-bosons', are charged under this $\mathrm{U}(1)$. Therefore, this Einstein-Yang-Mills theory has a similar to structure to the Einstein-Maxwell-charged scalar theory discussed in the previous few sections. In particular, in a background carrying $\mathrm{U}(1)$ charge, the W-bosons will acquire a negative mass squared analogous to (\ref{eq:meff}). They can therefore be expected to condense below some $T_{\mathrm{c}}$ if the gauge coupling $e$ is large compared to the gravitational coupling $\kappa$.

There are several new ingredients with a Yang-Mills condensate relative to the charged scalar case. Firstly, the condensate is a spatial vector, and hence should be called a $p$-wave (rather than s-wave) superconductor \cite{Gubser:2008wv}. Secondly, the condensate is the spatial component of gauge potential, which is dual to a conserved current in the boundary field theory, which therefore breaks time reversal invariance. Thirdly, the zero temperature limit of the backgrounds turns out to have qualitative differences with the scalar cases discussed above \cite{Basu:2009vv, Basu:2011np}, realizing a hard rather than soft gap towards charged excitations. Let us discuss these in turn.

Generically, the vector condensate means that the condensed phase will be anisotropic.
The gauge potential \cite{Gubser:2008zu, Gubser:2008wv} and metric \cite{Basu:2009vv,Ammon:2009xh} take a form such as 
\begin{align}
A &= p(r) \tau^3 \mathrm{d}t + \phi(r) \tau^1 \mathrm{d}x \,, \notag \\
\mathrm{d}s^2 &= - f(r) \mathrm{d}t^2 + g(r) \mathrm{d}r^2 + h(r) \left(c(r) \mathrm{d}x^2 + \mathrm{d}\vec x^2_{d-1} \right) \,. 
\label{eq:anisA}
\end{align}
The notation above emphasizes the fact that $A^3$ plays a role analogous to the Maxwell field previously, and $A^1_x$ is analogous to the charged scalar. The anisotropy arises because, in this case, the $x$ direction is singled out by the condensed vector. In fact, in $d=2$ space dimensions it is also possible to achieve isotropic condensates. This occurs if the $\mathrm{U}(1)$ gauge and $\mathrm{U}(1)$ spatial rotational symmetries are locked, so that
\be
A = p(r) \t^3 \mathrm{d}t + \phi(r) \left(\tau^1 \mathrm{d}x + \t^2 \mathrm{d}y \right) \,.
\ee
This case is called a $p+\mathrm{i} p$ superconductor, and the corresponding background metric is isotropic, with $c(r)=1$ in (\ref{eq:anisA}). The form of the condensate is determined by finding the configuration that minimizes the free energy.
In the Einstein-Yang-Mills theory, the anisoptropic phase (\ref{eq:anisA}) is dominant \cite{Gubser:2008wv}.

The breaking of time reversal is most interesting in the isotropic $p + \mathrm{i} p$ case. We explained in \S\ref{sec:hall} above how in an isotropic system, breaking time reversal allowed a nonzero Hall conductivity $\sigma_{xy}$. Indeed, such a Hall conductivity arises in symmetry broken holographic $p+\mathrm{i} p$ states \cite{Roberts:2008ns}. No external magnetic field is needed. The Hall conductivity remains finite in the $\omega =0$ limit.
 
In the zero temperature limit, the holographic $p$-wave superconductor exhibits an emergent $z=1$
scaling symmetry in the far interior, similarly to the discussion between equations (\ref{eq:IRlifSC}) and (\ref{eq:emergezone}) above. That is, the charge density operator is irrelevant in the emergent IR scaling theory. However,
in contrast to equation (\ref{eq:emergezone}) for the s-wave case, at least for sufficiently large gauge field coupling $e$, the electrostatic potential now vanishes exponentially
fast towards the IR \cite{Basu:2009vv, Basu:2011np}
\be\label{eq:expsmall}
A^3 = h_\infty \mathrm{e}^{- \alpha \, r} \mathrm{d}t \,, \qquad \qquad (r \to \infty, \a > 0) \,.
\ee
The technical reason for this different behavior is that in the emergent IR Anti-de Sitter spacetime,
the Higgsing of the electric field by a charged vector is stronger than the Higgsing by
a charged scalar. This is because the effective mass for the photon involves the contraction $m^2_{A^3} \sim g^{xx} A^1_x A^1_x \sim r^2 (A^1_x)^2$, which is stronger as $r \to \infty$ with $A^1_x$ constant in the IR than the corresponding scalar term $m^2_A \sim \phi^2$ with $\phi$ constant in the IR. Upon heating up the zero temperature solution, it is clear that (\ref{eq:expsmall}) will lead to an exponentially small amount of electric flux through the horizon at low temperatures. Relatedly, while the emergent $z=1$ scaling geometry means that there are gapless neutral degrees of freedom in the system, the regular contribution to the conductivity is exponentially small at low frequencies and temperatures $\sigma_\text{reg} \sim \mathrm{e}^{- \Delta/T}$. That is, the charged sector is gapped. Indeed at zero temperature the regular part of the conductivity satisfies
\be
\text{Re} \, \sigma_\text{reg}(\omega) = 0 \quad \text{for} \quad \omega < \omega_0 \,.
\ee
The conductivity vanishes as $\omega \to \omega_0$ from above according to $\text{Re} \, \sigma_\text{reg}(\omega) \sim \sqrt{1 - \omega_0^2/\omega^2}$.
 
 \paragraph{Challenges for $d$-wave superconductors}
 
A $d$-wave condensate is described by a charged, spin two field in the bulk \cite{Chen:2010mk}. While a charged spin one field admits a simple description via Yang-Mills theory, the same is not true for a charged spin two field. The dynamics of spin two fields is highly constrained by the need to reduce the degrees of freedom of a complex, symmetric tensor field $\varphi_{ab}$, which has $(d+2)(d+3)/2$ components, down to the $d(d+3)/2$ dimensions of the irreducible representation of the relevant little group, $\mathrm{SO}(d-1)$. If the extra modes are not eliminated by suitable constraints, they tend to lead to pathological dynamics. In the context of applied holography, these issues are discussed in \cite{Benini:2010pr,Kim:2013oba}. Those papers include references to the earlier gravitational physics literature.

With a dynamical complex symmetric tensor $\varphi_{ab}$ at hand, a $d$-wave condensate is described by a nonvanishing profile for
$\varphi_{xx}(r) = - \varphi_{yy}(r)$ or $\varphi_{xy}(r) = \varphi_{yx}(r)$. Unlike a $p$-wave condensate, a single component $d$-wave condensate does not break time reversal invariance. However, time reversal is broken in the presence of a complex superposition of the two profiles above \cite{Chen:2011ny}.

A theory of a charged spin two field that is consistent on a general background is not known. The current state of the art is the following quadratic action for $\varphi_{ab}$ that can be added to Einstein-Maxwell theory \cite{Benini:2010pr} \begin{widetext}
\bea
\lefteqn{S = \int \mathrm{d}^{d+2}x \sqrt{-g} \left( - |D_a \varphi_{bc}|^2 + 2 |D_a \varphi^{ab} |^2 + |D_a \varphi|^2 - [D_a \varphi^{*ab} D_b \varphi + \text{c.c.}] \right.} \nonumber \\
&& \left. - m^2 \left(|\varphi_{ab}| - |\varphi|^2 \right) + 2 R_{abcd} \varphi^{* ac} \varphi^{bd} - \frac{1}{d+2} R |\varphi|^2 - 2 g \mathrm{i} F_{ab} \varphi^{* a c} \varphi_c{}^b \right) \,.
\eea \end{widetext}
 Here $\varphi = \varphi^a{}_a$ and $D_a = \nabla_a - \mathrm{i} A_a$. There is a parameter $g$, the gyromagnetic ratio \cite{Kim:2013oba}. This Lagrangian is ghost free and has the correct number of degrees of freedom, but only with a fixed background Einstein geometry, satisfying $R_{ab} = - (d+1)/L^2 \, g_{ab}$ and with $g=1/2$. Even in such cases, for large enough values of the electromagnetic field strength, or gradients thereof, the equations of motion are either non-hyperbolic or lead to acausal propagation. Therefore, while the model can capture some of the expected dynamics of $d$-wave superconductivity close to the transition temperature, it will not be able to access the universal low temperature regimes where backreaction on the metric and often large Maxwell fluxes are important.
 
It has been emphasized in \cite{Kim:2013oba} that a natural way to obtain fully consistent dynamics for a massive, charged spin two degree of freedom is to consider Kaluza-Klein modes in the bulk, discussed in \S \ref{sec:consistent}. In a Kaluza-Klein construction there is an internal manifold that has a $\mathrm{U}(1)$ symmetry. A perturbation of the bulk metric with `legs' in the noncompact holographic dimensions (the `AdS' directions) but that depends on the coordinates of the internal manifold will be charged under this $\mathrm{U}(1)$ symmetry. It will be a massive, charged spin two field. In the language of \S \ref{sec:consistent}, the difficulty of constructing a theory for just this single excitation is the difficulty of finding a consistent truncation that includes this mode. This single mode instead couples to the infinite tower of Kaluza-Klein modes with increasing mass and charge. That is to say, one must solve the full problem of the inhomogeneities in the internal dimension, rather than restricting to a single Fourier component. While technically challenging, this is possible in principle and will lead to a completely well-behaved holographic background with $d$-wave condensate(s).

At the time of writing it is not known whether the additional bulk structure that seems to be required for a $d$-wave condensate is a reflection of interesting field theoretic aspects of such symmetry breaking. The presence of internal manifolds in the bulk is something of an undesirable feature of holographic models, as it implies the emergence of an additional locality in the dual field theory that is presumably an artifact of large $N$. The extra locality is hidden from sight in consistent truncations. On the other hand, the ability at large $N$ to focus on a single operator $\ocal$ without having to worry about the multitrace operators $\ocal^2$ etc, is also an artifact of large $N$. So, the need to consider many operators (the Kaluza-Klein modes) in the discussion of $d$-wave condensates is also, in some sense, more realistic.

Finally, we can note that while several experimentally observed unconventional superconductors are $d$-wave,
the underlying effective low energy dynamics is often s-wave. Specifically, for superconductivity emerging from a spin density wave critical point, mentioned in \S \ref{sec:sccondmat}, the $d$-wave condensate arises from pairing between fermions in distinct hot spots. In the effective theory, the different hot spots appear as a `flavor' label for the fermions, as in \S \ref{sec:sdw}. Thus, the microscopic orientation dependence of the pairing is simply described by the flavor index structure of an s-wave interaction.

\subsubsection{Spontaneous breaking of translation symmetry}
\label{sec:inhomogS}

Spontaneous breaking of translation symmetry is common in condensed matter physics.  This subsection will review
holographic models where translation symmetry is spontaneously broken. The underlying mechanism is the same as 
that described in \S \ref{sec:IRinstab} and the dynamics of the transition are again those described in \S\ref{sec:transition}. The important difference with cases we have considered so far is that the modes that satisfy the instability criterion (\ref{eq:BFbound}) have a nonzero spatial momentum, $k \ne 0$. 

Recall that the instability criterion (\ref{eq:BFbound}) corresponds to an operator with a complex exponent in the IR scaling geometry. In \S \ref{sec:zeroTspectral} and elsewhere above we have emphasized a special feature of $z=\infty$ scaling: the resulting semi-local criticality means that operators typically have $k$-dependent scaling exponents. This allows, in principle, operators with a range of momenta with $k \neq 0$ to acquire complex exponents while the homogeneous $k=0$ operators remain stable. This will be the origin of all the inhomogeneous instabilities discussed below. In \S \ref{sec:zeroTspectral} we also emphasized that $z = \infty$ scaling shared an important property of Fermi surface physics:
the presence of low energy spectral weight at nonzero momenta. The instabilities we are about to discuss
depend on the presence of this spectral weight in $z= \infty$ IR scaling geometries. In this loose sense they
can be considered strongly coupled analogues of the nonzero wavevector instabilities of Fermi surfaces.

\paragraph{Helical instabilities} 

Helical order breaks translation symmetry in a homogeneous way, as we have noted in our discussion of transport in \S \ref{sec:in2}.  A  simple holographic model leading to helical order is Einstein-Maxwell theory in $d=3$ with a Chern-Simons term of coupling constant $\alpha$ \cite{Nakamura:2009tf}: \begin{align}
S &= \int \mathrm{d}^5x \sqrt{-g}\left(\frac{1}{2\kappa^2}\left(R+\frac{12}{L^2}\right) - \frac{F^2}{4e^2} \right.\notag \\
&\left. \;\;\;\; - \frac{\alpha}{6}\epsilon^{abcde}A_a F_{bc}F_{de}\right) \,.  \label{eq:64action1}
\end{align}
We have encountered this theory previously in \S\ref{sec:CSterm}. We will now see that the symmetric background of the theory has an instability. As we explained in \S\ref{sec:CSterm}, the Maxwell field itself becomes charged due to the nonlinear Chern-Simons term. Heuristically, this is why the theory can have instabilities of the `pair production' kind that we have discussed above.


To investigate possible instabilities of the Chern-Simons theory, it is instructive to start with five dimensional Maxwell-Chern-Simons theory in flat space. Since the Chern-Simons theory is non-linear, we will need to turn on a background electromagnetic field for the term to have an effect on linearized perturbations.  Denote the 5 dimensions with $(t,r,x,y,z)$. Here $r$ will become the bulk radial dimension shortly.   Turning on an electric field $F_{rt}=E$, the linearized equations of motion for the transverse modes $f_i = \frac{1}{2}\epsilon_{ijk}\partial_j A_k$ are \cite{Nakamura:2009tf}: \begin{equation}
\left(-\partial_t^2 + \partial_r^2 + \partial_j\partial_j\right) f_i = 4\alpha  E \epsilon_{ijk}\partial_j f_k \,,
\end{equation}
with $\partial_i f_i = 0$.   Looking for modes with only $\omega$ and $k_x$ non-zero, we find a pair of circularly polarized modes $f_y \pm \mathrm{i}f_z$ with dispersion relation \begin{equation}
0 = \omega^2 - (k_x\mp 2\alpha E)^2 + 4\alpha^2E^2 \,.   \label{eq:4a2E2}
\end{equation}
Clearly, for $0<\pm k_x<4 |\alpha E|$ (with appropriate choice of sign), there is an instability. In particular, the most unstable mode is at $|k_x| = 2 |\alpha E|$.

The story is similar in a holographic context, though a little more complicated.  The background is the $\mathrm{AdS}_5$-RN black hole, which has a bulk electric field with $r$-dependence, as in (\ref{eq:athor}).  At low enough temperatures, and for large enough values of $|\a| > \a_{\mathrm{c}}$ there is a continuous phase transition to a phase with spontaneous spatial modulation, analogous to that described in the previous paragraph \cite{Nakamura:2009tf}.\footnote{It is possible that this phase transition is actually first order once quantum effects in the bulk are accounted for -- see the arguments in \cite{Ooguri:2010kt}.}   As with superfluid instabilities, these low temperature instabilities correspond to an ``effective mass" of the perturbations that is below the bound (\ref{eq:BFbound}) for the near-horizon, zero temperature $\mathrm{AdS}_2 \times \R^3$ spacetime.  This analysis gives the critical Chern-Simons coupling to be 
$\alpha_{\mathrm{c}} =  (2\k^2/e^2)^{3/2} \times 0.2896\ldots$. The unstable modes have the same form as those found in the previous paragraph -- the gauge field will break translation symmetry in a boundary direction ($x$, without loss of generality), and the spatial components of the gauge field will be of the form
\begin{equation}\label{eq:AxAy}
 A_y + \mathrm{i}A_z = h(r)\mathrm{e}^{\mathrm{i}kx}.
\end{equation}
In the boundary, there will be a similar helical pattern of electric current. The condensates $\langle J_y \rangle$ and $\langle J_z \rangle$ appear at $T < T_{\mathrm{c}}$, similarly to in the holographic superconductors above. The critical temperature associated with such instabilities appears to be reduced by external magnetic fields \cite{Ammon:2016szz}. The condensed phase was constructed numerically in a probe limit in \cite{Ooguri:2010kt}; the backreaction of gravity was accounted for in \cite{Donos:2012wi}. The backreaction is tractable without solving PDEs because the helical condensate (\ref{eq:AxAy}) leads to a homogeneous geometry with Bianchi VII symmetry \cite{Iizuka:2012iv}. The zero temperature limit of the fully backreacted solution exhibits an emergent anisotropic IR scaling symmetry with entropy density vanishing like $s \sim T^{2/3}$ \cite{Donos:2012js}.

Supergravity fixes the value of $\alpha = (2\k^2/e^2)^{3/2} \times 1/(2\sqrt{3})$ \cite{Gunaydin:1983bi}. As was noted in \cite{Nakamura:2009tf}, this happens to be just below the critical value for this spatially modulated instability.

\paragraph{Striped order}

This section describes the development of ``striped" order which breaks translation symmetry in a single spatial direction, in an inhomogeneous fashion.   We focus on the endpoint of an instability identified in \cite{Donos:2011bh}.   This task requires solving coupled nonlinear PDEs and has only been possible numerically.   The first such models were of the form \cite{Rozali:2012es, Rozali:2013ama, Donos:2013wia, Withers:2013loa} \begin{align}
S & = \int\mathrm{d}^4x \sqrt{-g} \Bigg(\frac{R - 2\Lambda}{2\kappa^2} - Z(\phi) \frac{F^2}{4e^2} \notag  \\
& -\frac{1}{2}(\partial \phi)^2 - V(\phi) - \frac{Y(\phi)}{\sqrt{-g}} \epsilon^{abcd}F_{ab}F_{cd}\Bigg),  \label{eq:64section2}
\end{align}
with $Y(\phi\rightarrow 0) = c\phi + \cdots$.    Like (\ref{eq:64action1}), this action breaks parity and time-reversal. The important point about the final term in this action is that, in a background electric field, this term gives a derivative coupling between the pseudoscalar $\phi$ and magnetic fluctuations of the Maxwell potential. This term has a similar structure to the Chern-Simons coupling in (\ref{eq:64action1}).

The symmetric nonzero density background is the $\mathrm{AdS}_4$-RN charged black hole.  Below a critical temperature $T_{\mathrm{c}}$, there is a continuous phase transition to a striped phase.  As in the previous subsection, the existence of the instability -- supported at nonzero wavevector -- can be seen by analyzing the effective mass squared of fluctuations about the zero temperature $\mathrm{AdS}_2 \times \R^2$ near horizon geometry. In numerics, one looks for striped black holes with a fixed wavevector $k$; in practice, the black hole which forms will have whatever $k$ leads to the lowest free energy. Numerical results confirm that the striped phase is thermodynamically preferred below the critical temperature. 

\begin{figure*}
\centering
\includegraphics[width = \textwidth]{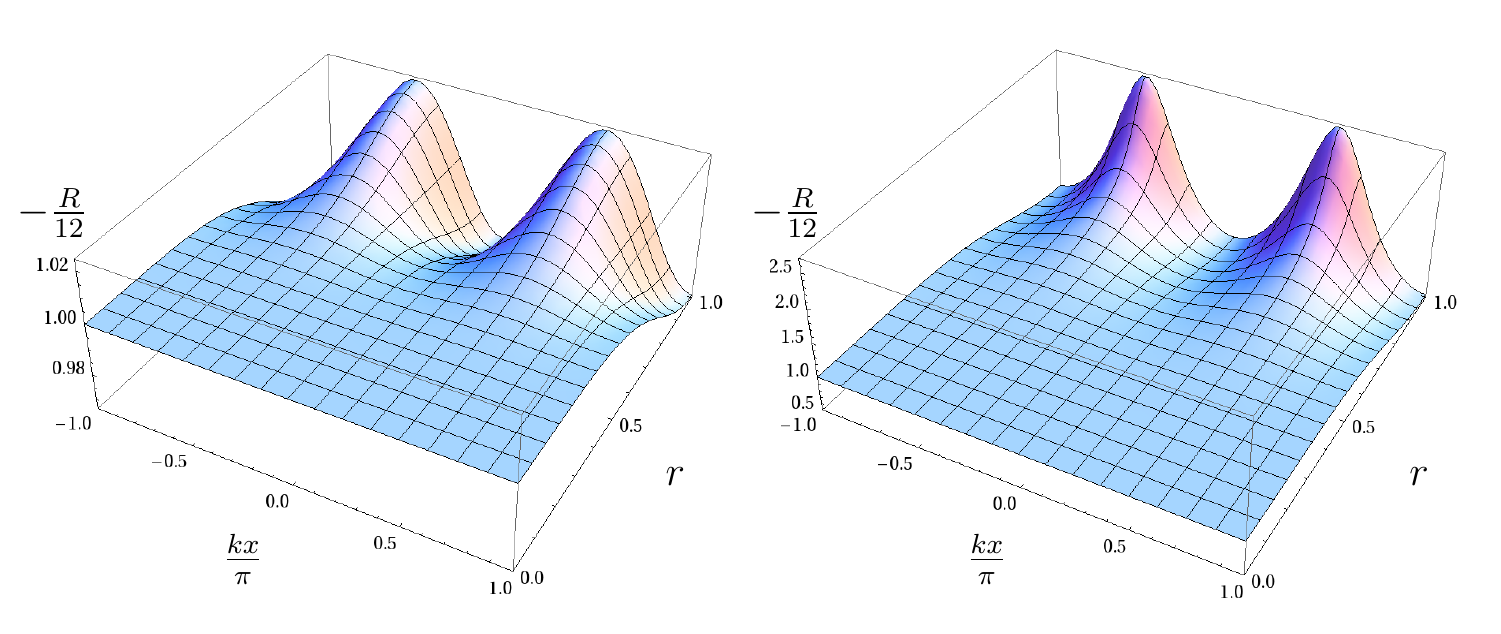}
\caption{\textbf{The profile of the Ricci scalar (normalized to the AdS value) in the striped phase} for $T=0.87T_{\mathrm{c}}$ (left) and $T=0.04T_{\mathrm{c}}$ (right).  Figure taken from \cite{Withers:2013loa} with permission.}
\label{sec64fig1}
\end{figure*}

Figure \ref{sec64fig1} shows surface plots of the curvature of the
resulting striped geometries. Here $x$ denotes the boundary direction along which stripes form and $r$ is the holographic direction.  As $T/T_{\mathrm{c}}$ becomes small, the curvature modulations become more pronounced.   In the boundary theory, in addition to the operator dual to $\phi$ picking up a spatially modulated expectation value, one finds an electric current $J_y$ which flows normal to the modulation direction.   The resulting striped phase appears to have vanishing zero-temperature entropy.  The identification of possible inhomogeneous zero temperature, near horizon geometries that would control the low energy and low temperature physics, potentially exhibiting emergent scaling of some kind, is a challenging open problem. It was emphasized in \cite{Hartnoll:2014gaa} that $z=\infty$ scaling is in principle compatible with strong spatial inhomogeneities.

In all the examples mentioned above, the bulk models which spontaneously break translation symmetry have broken either parity or time-reversal symmetry.   \cite{Donos:2013gda} presents a linear stability analysis of two systems preserving these symmetries, in which spontaneous translation symmetry breaking is possible.    More complicated models than (\ref{eq:64section2}) likely exhibit first order transitions to striped phases \cite{Withers:2013loa}.  String theoretic constructions of holographic models with closely related spatially modulated instabilities were studied in \cite{Donos:2011qt, Donos:2011bh}.

Stripe instabilities can also arise in holographic CFTs at zero charge density but in background magnetic fields \cite{Donos:2011qt, Donos:2011pn, Cremonini:2012ir}. These instabilities do not require the final interaction in (\ref{eq:64section2}) but rather occur already in the EMD type models (\ref{eq:EMDaction}) that we have studied in detail. The only requirement is that $Z'(0) \neq 0$ or $V'(0) \neq 0$ in (\ref{eq:EMDaction}) so that the dilaton is forced to be nonzero. Fluctuations about a magnetic background with a $Z(\Phi) F_{ab} F^{ab}$ interaction are the same as fluctuations about an electric background with the $Y(\phi) \epsilon^{abcd}F_{ab}F_{cd}$ interaction of (\ref{eq:64section2}). Hence the magnetic instabilities are closely related to the instabilities above. Backreacted solutions of the corresponding ordered phase showing magnetization density waves were constructed in \cite{Donos:2016hsd}.

\paragraph{Crystalline order} 

\begin{figure}
\centering
\includegraphics[width = 2.9in]{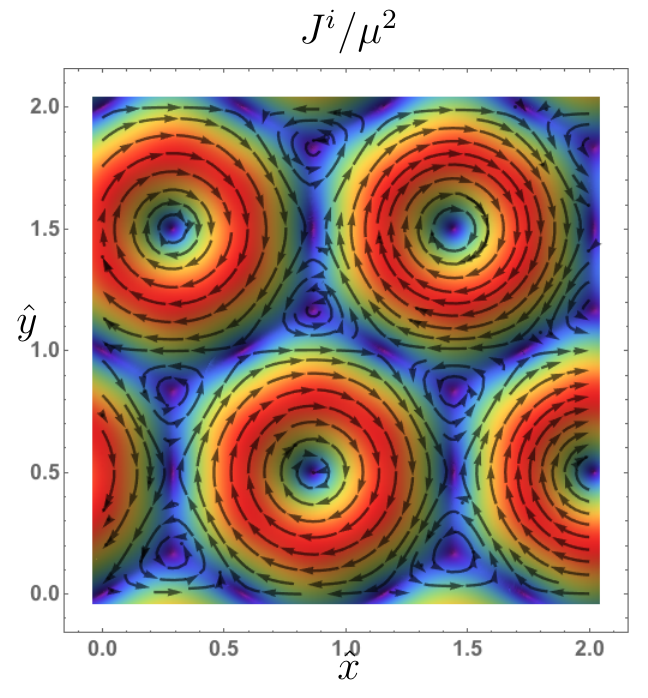}
\caption{\textbf{Triangular lattice.}  The expectation value of the current in the boundary theory, in a black hole with a triangular lattice `ground state'.   Red (purple) colors imply larger (smaller) magnitudes of $J_i$. Figure taken from \cite{Donos:2015eew} with permission.}
\label{figtriangular}
\end{figure}

In addition to striped phases, one can also look for ``crystalline" phases in holography, where translation symmetry is broken in all spatial directions.   As a first step, very close to the critical temperature, we can imagine superimposing stripes of specific wavevectors to form a Bravais lattice.  Of course, we wish to determine whether or not the true endpoint of these instabilities corresponds to stripes or lattices (and if so, what kind of lattice).   A first step towards such a calculation occurred in the probe limit of a bulk SU(2) Yang-Mills theory with a spatially modulated instability.   This model suggested that the triangular lattice was indeed the endpoint of this instability \cite{Bu:2012mq}.     More recently, the gravitational backreaction has also been taken into account using models similar to (\ref{eq:64section2}).   As in the discussion on stripes, we focus on boundary theories in two spatial dimensions.   In \cite{Withers:2014sja}, black holes with a rectangular ``lattice" of stripes were constructed;  later in \cite{Donos:2015eew}, it was shown that triangular lattices are often preferred in a background magnetic field.   Figure \ref{figtriangular} shows the formation of `magnetization' currents in the boundary theory in the triangular lattice phase.

Perhaps surprisingly, it seems to be the case that often the crystalline phases are thermodynamically disfavored compared to the simple striped phase \cite{Withers:2014sja}.   It is possible to choose boundary conditions which make the rectangular lattice preferred \cite{Withers:2014sja}, but we do not have a clear understanding of when or why different symmetry breaking patterns are preferred. As with the inhomogeneous striped geometries discussed in the previous section, the construction of inhomogeneous zero temperature IR scaling geometries -- if such geometries exist -- remains a challenging open problem.

In doped CFT3s, the spontaneous formation of a lattice admits an S-dual description. We discussed this S-duality (or, particle-vortex duality) of CFT3s in \S \ref{sec:selfdual} and \S \ref{sec:magfields}. The charge density dualizes to a background magnetic field and hence the lattice can be dually understood as a generalized Abrikosov flux lattice. In confining phases obtained by condensing vortices, these lattices obey a commensurability relation on the magnetic flux (and hence charge density in the original desciption) per unit cell of the lattice \cite{Sachdev:2012tj}. Such commensuration is not seen in the holographic models discussed above. This pictured motivated the study of an Abrikosov lattice in  \cite{Bao:2013fda}, following a perturbative technique used for vortex lattices in holographic superconductors \cite{Maeda:2009vf}. It was argued in \cite{Maeda:2009vf} that this lattice will be triangular, based on Ginzburg-Landau arguments, but as noted in \cite{Banks:2015aca}, these arguments do not correctly predict the order of the holographic phase transition, so it is unclear whether such arguments can reliably predict the final shape of the lattice.

\paragraph{Conductivity} 

In the presence of spontaneous symmetry breaking the electrical conductivity remains infinite, even along the direction of symmetry breaking. This is easily understood from a field theoretic point of view.   Suppose that translation symmetry is spontaneously broken in such a way that the local density is $\rho = \rho_0 + \rho_1\cos(k(x-x_0))$.   This is a valid solution for any value of $x_0$.   So we expect phonon or `phason' modes -- analogous to the Goldstone bosons of a superfluid --  which globally translate $x_0$.    If we apply an electric field to this system, we will excite such zero wave number phonons, and the pattern of symmetry breaking will globally translate without any momentum relaxation, leading to an infinite conductivity.\footnote{In actual condensed matter systems, these phason modes are pinned by disorder so that, for example, Wigner crystals are insulators \cite{CDW}.}

There is a quick way to see how an infinite conductivity arises in holography.   Let us return to the derivation of \S \ref{sec:memmatrixderivation}.   We introduced disorder perturbatively via a sourced bulk scalar field $\varphi_0(k,r)\sim r^{d+1-\Delta}$ near the boundary.   For simplicity, take $\Delta>(d+1)/2$.   Since the source is fixed, the finite momentum perturbations in the bulk  $\delta \varphi(k,r) \sim r^\Delta$ near the boundary.   With these scalings one can check that $\delta \mathcal{P}_x\sim r^0$ near the boundary, which is consistent with our claim that we may impose the boundary condition $\delta \mathcal{P}_x=\rho/L^d$ at $r=0$.    However, suppose that $\varphi_0\sim r^\Delta$ as well near the boundary.   Then we find that $\delta \varphi/\varphi_0 = a_0r^0 + a_cr^c + \cdots $ near the boundary, with $c>0$ describing the next order correction, and $\delta \mathcal{P}_x \sim r^{2\Delta-d-1+c}$, implying that $\delta \mathcal{P}_x(r=0) =0$.   Hence, it is impossible to excite the bulk mode (\ref{eq:mmh3}).   Following the discussion in \S \ref{sec:memmatrixderivation}, the only bulk mode we can excite is the boost mode.   The perturbation of the gauge field in the bulk is hence \begin{equation}
\delta A_x \approx \left( \frac{Ts}{\rho} + p(r)\right)\mathrm{e}^{-\mathrm{i}\omega t},
\end{equation}
to leading order in $\omega$, away from the horizon.   But this is sufficient to compute the conductivity perturbatively, and we find \begin{equation}
\sigma(\omega) \approx \frac{\rho^2}{\epsilon+P} \frac{1}{-\mathrm{i}\omega}.
\end{equation}
At $\omega=0$, we recover a $\delta$ function as in (\ref{eq:Cleansigma2}).

\subsection{Zero temperature BKT transitions}
\label{sec:bkt}

We saw in \S\ref{sec:transition} above that nonzero temperature holographic phase transitions
were described by conventional Landau-Ginzburg physics. Fluctuations effects at the transition are suppressed
at large $N$, because they involve fluctuations of only a single mode, and hence the transitions are well captured
by mean field theory. In contrast, when holographic compressible phases are driven into ordered phases as a function of some source at zero temperature, even the large $N$ physics is strongly non-mean field. This subsection will show how
the emergent IR scaling regimes and the fact that the instability occurs when an operator is driven to have complex scaling exponents lead to certain `quantum BKT' infinite order quantum phase transitions.

What happens if, by varying some auxiliary sources in the theory, we continuously push the effective IR mass squared of an operator through the instability bound (\ref{eq:BFbound})? It was explained in \cite{Kaplan:2009kr} that this should be understood as the annihilation of two RG flow fixed points. In \S \ref{sec:multitrace} we noted that for the range of masses just above the instability (\ref{eq:BFbound})  -- here we generalize the discussion therein to general $z$, see \cite{Andrade:2012xy, Keeler:2012mb} --
\be
- \frac{D^2_\text{eff.}}{4} \leq m^2 L_\text{IR}^2 \leq - \frac{D^2_\text{eff.}}{4} + 1 \,,
\ee
two different normalizable boundary conditions were possible at the asymptotic boundary. These correspond to choosing the operator $\ocal$ dual to the bulk scalar field to have the different dimensions
\be
\Delta_\pm = \frac{1}{2} \left( D_\text{eff.} \pm \sqrt{D_\text{eff.} + 4 m^2 L^2_\text{IR}} \right) \,.
\ee
As noted in \S \ref{sec:multitrace}, the different boundary conditions then correspond to different scaling theories that
are related to each other by RG flow triggered by a double trace deformation $\int \mathrm{d}^{d+1}x \,  \ocal^2$. This deformation is irrelevant about the `standard' quantization with dimension $\Delta_+$ but relevant about the `alternate' quantization with dimension $\Delta_-$. These two fixed points merge as $m^2 L^2_\text{IR}$ is lowered to the onset of instability at $D_\text{eff.} + 4 m^2 L^2_\text{IR} = 0$. At the merger point the $\ocal^2$ coupling is marginal.

\begin{figure}
\centering
\includegraphics[width = 3.3in]{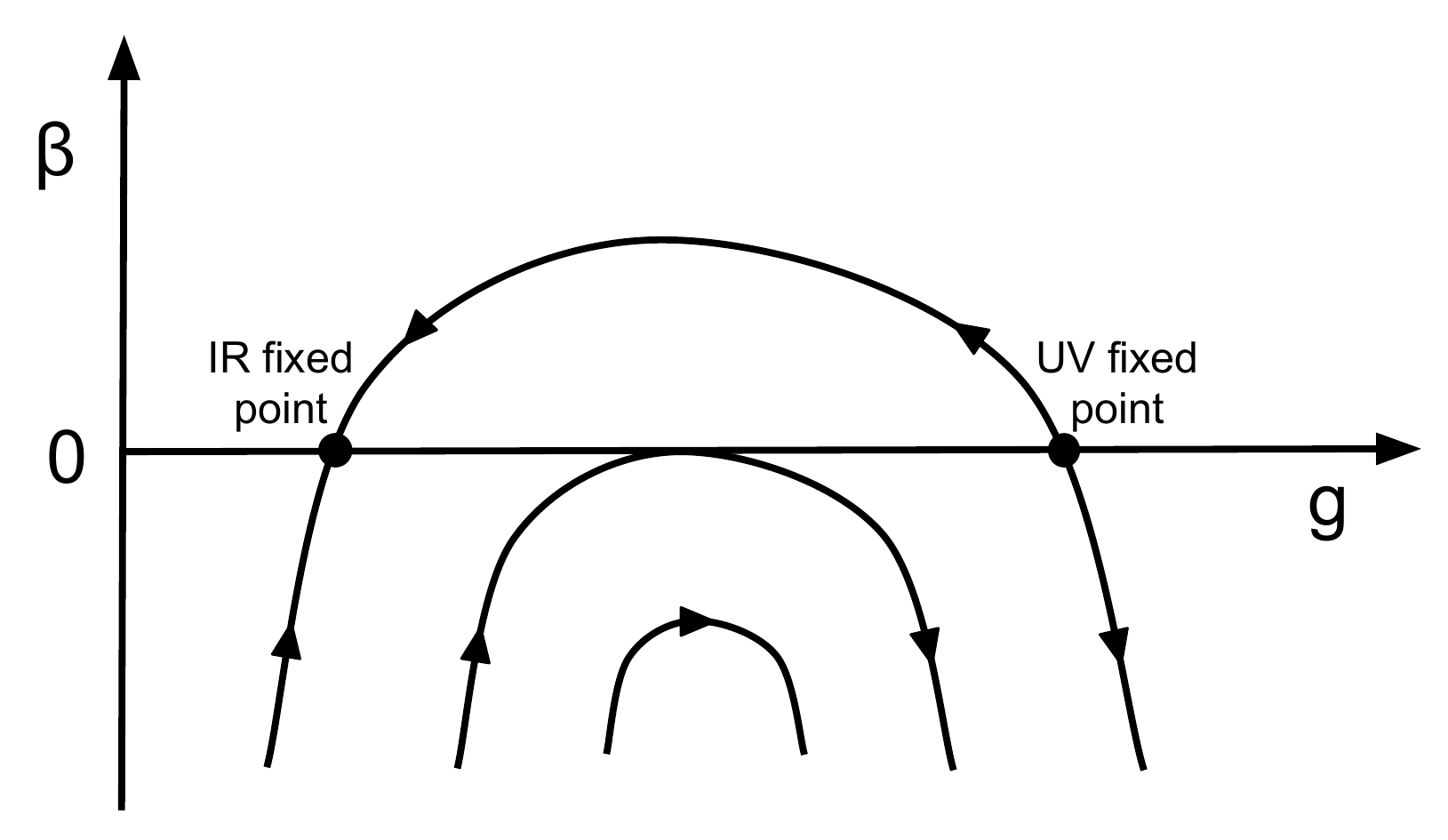}
\caption{\textbf{Annihilation of two RG fixed points.}  RG flows from (\ref{eq:bktrg}) with (from top to bottom) $\a > \a_\star$, $\a = \a_\star$ and $\a < \a_\star$.}
\label{fig:bkt}
\end{figure}

The merger of fixed points is described by the beta function \cite{Kaplan:2009kr}
\be\label{eq:bktrg}
\beta(g) = \mu \frac{\mathrm{d}g}{\mathrm{d}\mu} = (\a - \a_\star) - (g - g_\star)^2 \,,
\ee
where $g$ is the coupling of the $\int \mathrm{d}^{d+1}x \,  \ocal^2$ interaction and $\alpha$ is the parameter that tunes
the theory across the transition. For $\a > \a_\star$ the $\beta = 0$ fixed points are at $g_\pm = g_\star \mp \sqrt{\a - \a_\star}$. The fixed points merge at $\a = \a_\star$ and cease to exist at $\a < \a_\star$. This is illustrated in figure \ref{fig:bkt}. The lack of a fixed point indicates an IR instability. Such unstable flows typically result in the condensation of an operator.

The renormalization group flow equation (\ref{eq:bktrg}) is familiar from the classical BKT transition. In particular, this means that as the transition point is approached, a low energy scale is generated that is non-analytic in $\a$. For instance, just below the critical $\a_\star$, integrating the flow equation gives
\be\label{eq:BKTscaling}
\Lambda_\text{IR} \approx \Lambda_\text{UV} \mathrm{e}^{- \pi/\sqrt{\a_\star - \a}} \,.
\ee
For example, the scale $\Lambda_\text{IR}$ will set the critical temperature $T_{\mathrm{c}}$ at which a condensation instability is expected to occur for $\a < \a_\star$. This scale can be seen directly from the scaling geometry as follows \cite{Kaplan:2009kr}: Solve the wave equation for a field with $m^2 < m^2_\text{BF} = - D_\text{eff.}/(4 L_\text{IR}^2)$. Cut off the geometry in the UV and IR by imposing Dirichlet boundary conditions at $r=r_\text{UV}$ and $r = r_\text{IR}$, respectively. Then an unstable, growing in time solution to the wave equation appears when
\be
r_\text{IR}^z \approx r_\text{UV}^z \, \mathrm{e}^{\pi/\left(L_\text{IR}\sqrt{m^2_\text{BF} - m^2}\right)} \,.
\ee
This shows that the scale $r_\text{IR}$ is where the geometry will need to be modified by a condensate in order to remove the instability.

Explicit holographic models realizing these quantum BKT phase transitions were first studied in \cite{Jensen:2010ga, Iqbal:2010eh}. The tuning parameter in both cases is a magnetic field. In \cite{Iqbal:2010eh} the zero temperature destruction of holographic superconductivity by a magnetic field was shown to be in this class.

The quantum BKT transitions can be understood semi-holographically
\cite{Jensen:2011af, Iqbal:2011aj}. In fact we have already used this perspective in the formula (\ref{eq:GRrelate2}) above for the Green's function of the scalar field, obtained by matching with the IR geometry. The dynamics of quantum BKT transitions are characterized by coupling the gapless Goldstone mode (\ref{eq:deltaphi}), that is supported at intermediate radial scales in the spacetime and hence is not geometrized, as we noted in \S \ref{sec:coleman} above, to the large $N$ quantum critical bath that undergoes a merger with another critical point as we have just described above.

We noted above that the operator $\ocal^2$ that drives the RG flow between the two fixed points is necessarily marginal at the quantum critical coupling where the two fixed points merge. This has some interesting phenomenological consequences, especially at large $N$ where the single trace operator $\ocal$ then has dimension $\Delta = D_\text{eff.}/2$ at the critical point (more generally operator dimensions do not add). Such an operator can be coupled to fermions to obtain marginal Fermi liquid fermionic Green's functions \cite{Iqbal:2011aj, Vegh:2011aa} or can be coupled to currents to directly obtain $T$-linear resistivity \cite{Donos:2012ra}.

\section{Further topics}

\subsection{Probe branes}\label{sec:probebrane}

\subsubsection{Microscopics and effective bulk action}
\label{sec:microbrane}

In this section we discuss a class of holographic models that
are obtained from microscopically consistent string theoretic backgrounds in the bulk. They have an extra ingredient relative to the Kaluza-Klein compactifications discussed in \S \ref{sec:consistent}: the probe branes. The original motivation for considering probe branes came from the desire to include dynamical degrees of freedom in the fundamental representation (quarks) in holographic models for QCD \cite{Karch:2002sh}.
The types of construction considered in \S \ref{sec:consistent}, including for example the string theory background $\mathrm{AdS}_5\times \mathrm{S}^5$, are dual to quantum field theories such as $\mathcal{N}=4$ SYM that only contain adjoint degrees of freedom (gluons). We will see that probe branes are also interesting objects in the context of quantum matter.   For now, we digress from condensed matter physics temporarily. The upshot of the following motivational paragraphs will be the action (\ref{eq:DBI}) and (\ref{eq:Xdet}) below that we shall be adding to the bulk theory. A microscopic motivation is necessary in this case to justify the particular nonlinear form of the action. For a more detailed microscopic description, see \cite{CasalderreySolana:2011us}.

Recall from \S \ref{sec:malda} that $\mathcal{N}=4$ SYM in $d=3$ with gauge group $\mathrm{SU}(N)$ is the effective theory describing the low energy excitations of a coincident stack of $N$ D3-branes, in the regime where gravitational backreaction can be neglected. The massless degrees of freedom come from open strings stretching between pairs of the branes.   Such open strings carry two indices (one for each brane they end on) and hence are in the adjoint representation of $\mathrm{SU}(N)$.   If we want to construct quarks, which transform in (anti-)fundamental representations of the gauge group,  we need an additional object on which an open string can end, leaving a single endpoint on the D3 brane stack.  In particular, we can place another D$p$ brane somewhere -- wherever the new D$p$ brane intersects the D3 brane stack, massless ``quarks" exist due to strings stretching between D3 and D$p$ branes.

The crucial simplification of the probe brane limit is to consider many fewer than $N$ of the D$p$ branes. In general one considers $N_\text{f} \ll N$ D$p$ branes. The subscript f stands for `flavor', as each D$p$ brane corresponds to a distinct flavor of quark. To avoid clutter we will set $N_\text{f} = 1$. Increasing the string coupling as described in \S \ref{sec:malda} causes the $N$ D3 branes to backreact gravitationally on the spacetime and generate the $\mathrm{AdS}_5\times \mathrm{S}^5$ near horizon geometry. However, in the limit in which $N_\text{f} \ll N$, the D$p$ branes are not themselves heavy enough to backreact and hence they remain present in the backreacted spacetime. That is to say, to add $N_\text{f}$ quarks to $\mathcal{N}=4$ SYM, one must add $N_\text{f}$ D$p$ branes into the dual $\mathrm{AdS}_5\times \mathrm{S}^5$ spacetime. We will now describe what this means in practice.

Certain placements of the D$p$ brane in $\mathrm{AdS}_5\times \mathrm{S}^5$ preserve $\mathcal{N}=2$ supersymmetry, as discussed in \cite{Karch:2002sh}.
It is helpful to preserve supersymmetry because otherwise the probe branes typically have an instability in which they `slide off' the internal cycles of $\mathrm{S}^5$ on which they are placed. Two commonly studied configurations are: 

\begin{enumerate}
\item To obtain quarks propagating in all $d=3$ spatial dimensions of the QFT, we place a  D7 brane on $\mathrm{AdS}_5\times \mathrm{S}^3$.  The choice of $\mathrm{S}^3 \subset \mathrm{S}^5$ plays a physical role that we will elucidate below.
\item To obtain quarks propagating on a $d=2$ dimensional defect in the $d=3$ QFT, we place a D5 brane on $\mathrm{AdS}_4 \subset \mathrm{AdS}_5$,  wrapping two dimensions along an $\mathrm{S}^2\subset \mathrm{S}^5$. The degrees of freedom on the defect can exchange energy and momentum with the higher dimensional excitations of $\mathcal{N}=4$ SYM \cite{Karch:2000gx,DeWolfe:2001pq,Erdmenger:2002ex}.
\end{enumerate}

To describe the D$p$ brane, we must add additional bulk degrees of freedom that propagate on the $p+1$ dimensional worldvolume of the brane. There are two types of fields on the worldvolume. Firstly there are scalar fields that describe how the brane is embedded into the spacetime. Secondly there is a $\mathrm{U}(1)$ gauge field. As always in holography, this gauged symmetry corresponds to a global $\mathrm{U}(1)$ symmetry of the boundary theory. The global symmetry is sometimes referred to as `baryon number', as it counts the density of quarks in the theory. In the following we will show how the dynamics of the $\mathrm{U}(1)$ gauge field on the brane leads to novel transport effects that are not captured by the Einstein-Maxwell-dilaton class of theories with action (\ref{eq:EMDaction}) that we have considered so far. For brevity, we shall mostly ignore the scalar field dynamics. These can be consistently neglected in considering the Maxwell field dynamics, in the setups we consider, once the embedding of the D$p$ brane is given.

The new feature of the brane action is that it describes a nonlinear theory for the Maxwell field strength $F_{\alpha\beta}$ on the brane. Here $\alpha\beta$ denote worldvolume indices on the D$p$ brane. D branes can be shown to be described by the Dirac-Born-Infeld (DBI) action \cite{Polchinski:1998rq,Polchinski:1998rr}: \begin{equation}
S_{\mathrm{D}p} = -T_{\mathrm{D}p}\int \mathrm{d}^{p+1}x \sqrt{X} \,, \label{eq:DBI}
\end{equation}
where \begin{equation}
X \equiv - \det \left(g_{\alpha\beta} + 2\pi \alpha^\prime F_{\alpha\beta}\right) \,. \label{eq:Xdet}
\end{equation}
Here $2\pi \alpha^\prime$ is the fundamental string tension and $g_{\alpha\beta}$ is the induced metric on the brane (from its embedding into the spacetime).  Also recall (\ref{eq:strong}) and (\ref{eq:stringy}), and note that the brane tension $T_{\mathrm{D}p} \sim (2\pi  \alpha^\prime)^{-(p+1)/2} g_{\mathrm{s}}^{-1}$. The action we have written neglects various possible Wess-Zumino couplings to background form fields and to the dilaton \cite{Polchinski:1998rq,Polchinski:1998rr} that will not be important for our discussion. Those couplings can generate Chern-Simons terms for the worldvolume Maxwell field that can lead to spatially modulated instabilities analogous to those discussed in \S\ref{sec:inhomogS} above \cite{Bergman:2011rf,Jokela:2014dba}.
The DBI action is exact to all orders in the field strength (in units of $2\pi \alpha^\prime$), but is not valid for large gradients of the field strength. The action captures a certain class of string theoretic effects, and as such is a microscopically consistent action for nonlinear electrodynamics. This action has a long history \cite{Born:1934gh}. Our discussion will focus on qualitatively new effects due to the nonlinear interactions. Expanding the action for small $F$ leads to the usual quadratic Maxwell theory. With the action (\ref{eq:DBI}) at hand, computations are done using the standard holographic dictionary.

A single probe brane will negligibly backreact on the geometry, hence the name `probe'. This is consistent with the fact that the free energy of the D3 branes scales as $N^2$, compared to $N$ for the fundamental quarks dually described by the probe brane.  We also note for future reference that other thermodynamic and hydrodynamic quantities for the quark matter will also scale $\sim N$. The ability to neglect backreaction at large $N$ makes this class of models very tractable -- allowing a wider range of observables to be computed -- but also means that the large $N$ limit will miss a lot of physics that is relevant for transport.

Hence, we may place the D$p$ brane on top of the $\mathrm{AdS}_5$ (Schwarzschild) geometry, treating $g_{\alpha\beta}$ as fixed in (\ref{eq:Xdet}).   The induced metric $g_{\alpha\beta}$ is immediately found by restricting the $\mathrm{AdS}_5 \times \mathrm{S}^5$ metric  (\ref{eq:ads5s5}) to the probe brane worldvolume: 
\begin{align}
\frac{g_{\alpha\beta}\mathrm{d}x^\alpha \mathrm{d}x^\beta}{L^2} &= \left(\frac{1}{r^2f(r)} + \theta^\prime(r)^2\right)\mathrm{d}r^2 - \frac{f(r)}{r^2} \mathrm{d}t^2 \notag \\ &\;\;\; + \frac{\mathrm{d}\vec x^2_{p-k-1}}{r^2} + \cos^k \theta(r) \mathrm{d}\Omega_k^2.  \label{eq:gprobe}
\end{align}
where $k$ is defined as the dimension of the sphere which the brane wraps, and $L \cos\theta$ is the radius of the $\mathrm{S}^k$. Here we have allowed the size of the wrapped sphere to be radially dependent, although in several of the examples we consider below we will simply have $\theta = 0$. The effective description of the Maxwell field dynamics on the extended $p-k+1$ dimensions of the brane is given by integrating the full DBI action over the internal space to obtain
\begin{equation}
S = -T_{\mathrm{D}p} V_{\mathrm{S}^k}  \int \mathrm{d}^{p-k+1}x (L\cos\theta)^k \sqrt{-\det(\bar g_{ab}+2\pi\alpha^\prime F_{ab})} \,. \label{eq:reducedaction}
\end{equation}
Here $V_{\mathrm{S}^k}$ is the volume of a unit-radius $\mathrm{S}^k$, and $\bar g$ contains the first three terms of (\ref{eq:gprobe}). 

Upon Taylor expanding (\ref{eq:DBI}) in the limit of small $F_{ab}$, we readily recover the Maxwell action with electromagnetic coupling
\begin{equation}\label{eq:thetaofr}
\frac{1}{e^2} = T_{\mathrm{D}p}V_{\mathrm{S}^k} (L\cos\theta)^k (2\pi \alpha^\prime)^2,
\end{equation} 
When the size of the wrapped sphere is radially dependent, then (\ref{eq:thetaofr}) shows that the effective coupling $e$ is also radially dependent, somewhat analogously to the $Z(\Phi)$ factor in an EMD model such as (\ref{eq:EMDaction}).

\subsubsection{Backgrounds}
\label{sec:DBIback}

With the induced metric fixed to be (\ref{eq:gprobe}) we must find the background field $\theta(r)$ describing the brane embedding, as well as the electrostatic potential $A_t(r)$ that will determine the dual charge density in the usual way. Evaluated on such configurations, the reduced action (\ref{eq:reducedaction}) becomes
\begin{align}\label{eq:dbi2}
S &= -\mathcal{N} \int \mathrm{d}r \frac{\cos^k\theta(r)}{r^d} \times \notag \\
&\sqrt{\left(\frac{1}{r^2 f} + \theta'(r)^2 \right)\frac{f}{r^2} - \frac{(2 \pi \a')^2}{L^4} A_t'(r)^2} \,.
\end{align}
We have restricted to the case of `space-filling' branes with $p-k+1 = d+2$ for concreteness. The dimensionless prefactor of the $A_t'(r)^2$ term is just the 't Hooft coupling because $L^4 = \lambda \a'^2$ (this is the same equation as (\ref{eq:stringy}) in this case). The overall normalization of the action will not be important for us.

The bulk excitation described by $\theta(r)$ is dual to an operator that gives a mass to the quarks in the dual field theory (see e.g. \cite{Kobayashi:2006sb}). For massless quarks we set $\theta = 0$. At low temperatures and zero charge density, massive quarks means that the fundamental degrees of freedom described by the brane are gapped (bound into mesons). Analogously to our discussion in \S\ref{sec:gapped} above, we should expect the brane not to reach down into the far IR geometry. Indeed, in these cases the D$p$ brane  ``caps off" at some critical radius where $\theta(r_{\mathrm{crit}}) = \pi/2$, and the brane does not extend below this radius \cite{Mateos:2006nu, Albash:2006ew, Karch:2006bv}. With a nonzero charge density, this capping off does not occur even at low temperatures \cite{Kobayashi:2006sb}. Thus there are gapless charged degrees of freedom in this case, even with a mass for the quarks. This is analogous to massive free fermions with a chemical potential larger than the mass. We shall focus on the massless case with $\theta =0$, and we will comment on the effects of a nonzero mass where relevant.

With $\theta = 0$, the background profile for $A_t$ is easily seen to be
\be\label{eq:atsol}
A_t'(r) = - \frac{e^2}{L^{d-2}} \, \frac{\rho \, r^{d-2}}{\sqrt{1 + \hat \rho^2 r^{2d}}} \,,
\ee
here $\rho$ is the charge density, using (\ref{eq:atnearb}), while $\hat \rho = \frac{2 \pi \a'}{L^2} \rho$.
Note that wherever the nonlinearities in the DBI action are small -- which includes near the boundary $r \approx 0$ -- then the square root factor in the denominator of (\ref{eq:atsol}) is trivial and the behavior of a linear Maxwell field is recovered. Thus $e$ here is the effective Maxwell coupling (\ref{eq:thetaofr}). In the far IR at low temperatures, however, as $r \to \infty$ the nonlinearities are increasingly important and $A_t' \sim 1/r^2$. Eventually in this limit one must worry about backreaction of the brane on the background and, potentially more tractably, the possibility that stringy $\a'$ effects will become important \cite{Hartnoll:2009ns}.

The chemical potential corresponding to (\ref{eq:atsol}) is given by
$\mu = -\int_0^{r_+} A_t'(r) \mathrm{d}r$. In particular, at zero temperature where $r_+ \to \infty$, we have from (\ref{eq:atsol}) that
\be\label{eq:muzero0}
\hat \mu_0 \equiv \frac{L^{d-2}}{e^2} \frac{2\pi \a'}{L^2} \mu_{T=0} = C \hat \rho^{1/d}  \,,
\ee
where the constant $C = \Gamma\left(1 + \frac{1}{2d} \right) \Gamma \left(\frac{1}{2} - \frac{1}{2d} \right)/\sqrt{\pi}$ is relatively unimportant. The relationship $\mu \sim \rho^{1/d}$ follows from $z=1$ dimensional analysis at $T=0$. The quantity (\ref{eq:muzero0}) will appear in a couple of places shortly.

The thermodynamics of probe branes is easily computed using standard techniques. Somewhat unconventional powers of temperature arise, see e.g. \cite{Karch:2009eb}, although these should be understood as corrections to the order $N^2$ thermodynamics of the adjoint sector degrees of freedom. One relevant fact is that probe branes at nonzero charge density have a nonvanishing zero temperature entropy density, see e.g. \cite{Karch:2007br,Karch:2008fa,Karch:2009eb},
\be
s_{T=0} \propto \rho \,.  \label{eq:sT0probe}
\ee
Note that $\rho \sim N$, while the finite temperature entropy density scales as $N^2$, so (\ref{eq:sT0probe}) is formally subleading in $1/N$.  This is immediately reminiscent of the zero temperature entropy density (\ref{eq:sss}) of extermal AdS-RN black holes. The way the entropy arises technically in the probe brane case is that the free energy of the quarks is essentially given by the length that the brane extends from the boundary down to the horizon. This length grows linearly with low temperatures as the horizon recedes from the boundary. This result also holds with a quark mass. In the following \S\ref{sec:zeros} we will see another instance in which the low energy DBI dynamics resembles that of a $z=\infty$ scaling theory, even without backreaction onto the spacetime metric (which has $z=1$ in the present case).

\subsubsection{Spectral weight at nonzero momentum and `zero sound'}
\label{sec:zeros}

As previously, charge and current correlation functions are obtained from time- and space-dependent perturbations of the background described in the previous \S\ref{sec:DBIback}. Because there is no coupling to metric perturbations in the probe limit, these fluctuations obey equations similar to the those discussed in \S\ref{sec:maxw} for the charge dynamics of a zero density critical point. The novel effect is the nonlinearity of the DBI action that couples the perturbations to the background in a new way. Specifically, equations (\ref{eq:ax}) and (\ref{eq:ay}), with $z=1$, are replaced by \begin{widetext}\begin{subequations}\label{eq:dbiaeq}
\begin{align}
r^{d-2} \left(\frac{(1 + \hat \rho^2 r^{2d}) f}{(1 + \hat \rho^2 r^{2d}) \omega^2 - f k^2} \frac{1}{r^{d-2}} a_\parallel' \right)' + \frac{d \hat \rho^2 r^{2d-1} f}{(1 + \hat \rho^2 r^{2d})\omega^2-f k^2} a_\parallel'&= - \frac{a_\parallel}{f} \,, \label{eq:longa} \\
r^{d-2} \left(\frac{\left(1 + \hat \rho^2 r^{2d}\right) f}{r^{d-2}} a_\perp' \right)' - d \hat \rho^2 r^{2d-1}  f a_\perp' &= k^2 a_\perp - \left(1 + \hat \rho^2 r^{2d} \right)\frac{\omega^2}{f} a_\perp \,. \label{eq:transa}
\end{align}
\end{subequations}\end{widetext}
These equations reduce to (\ref{eq:ax}) and (\ref{eq:ay}) upon setting $\hat \rho = 0$, removing the nonlinearities.
Versions of these equations have been considered in several papers including \cite{Karch:2008fa,Kulaxizi:2008kv,Kulaxizi:2008jx,Davison:2011ek,Goykhman:2012vy,Anantua:2012nj}.

A key effect of the DBI nonlinearities is seen as follows. At zero temperature $f=1$ and the near horizon limit corresponds to $r \to \infty$. In both of the above equations we then see that $\hat \rho r^{2d}$ terms always dominate terms with $k^2$. This indicates that the momentum $k$ drops out of the near horizon equations, which strongly suggests that low energy dissipation can occur at nonzero momenta in these systems \cite{Hartnoll:2009ns,Goykhman:2012vy}. We will now show that this is indeed the case with a WKB analysis of the equations (\ref{eq:longa}) and (\ref{eq:transa}) that is valid at large $k$, following \cite{Anantua:2012nj}.

A WKB computation of the spectral density at large momenta proceeds as outlined around equation (\ref{eq:varWKB}) above. The computation can be done at any temperature. In both the longitudinal and transverse channels the low energy spectral weight (\ref{eq:rhok}) following from (\ref{eq:dbiaeq}) is found to be
\be\label{eq:probeWKB}
\rho(k) \sim \exp\left(- 2 k \int\limits_0^{r_+}\frac{\mathrm{d}r}{\sqrt{(1 + \hat \rho r^{2d})f} }\right)\,. 
\ee
Recal that $f$ is a function of $r/r_+$. Rescaling $r = r_+ \hat r$, it is clear that the exponent is a function of $\hat \rho/T^d$. In the high temperature limit (\ref{eq:probeWKB}) will simply become the exponential thermal Boltzmann suppression of nonzero momentum excitation in a $z=1$ theory (c.f. \cite{Son:2002sd}). At low temperatures we can set $f=1$ and $r_+ = \infty$. The integral in (\ref{eq:probeWKB}) then gives \cite{Anantua:2012nj}
\be\label{eq:kstarDBI}
\rho(k) \sim \mathrm{e}^{-k/k_\star} \,, \qquad k_\star = \frac{\hat \rho^{2/d}}{2 \hat \mu_0} \,.
\ee
The zero temperature chemical potential was found in (\ref{eq:muzero0}) above. This computation shows that zero temperature, zero frequency spectral weight is present for all momenta $k < k_\star$. The scale of $k_\star$ is set by the charge density.
The nonzero momentum spectral weight is more dramatic than that of extermal RN found in (\ref{eq:lowIm}) and (\ref{eq:Trhonu}), which vanished as a power law at low temperatures or frequencies. The spectral weight found in (\ref{eq:kstarDBI}) is precisely what might have been expected of a $2 k_{\mathrm{F}}$ singularity from a Fermi surface theory that has been smeared out by strong interactions.

Probe branes show a further feature reminiscent of Fermi surfaces: linearly dispersing collective modes at zero temperature. In an ordinary Fermi liquid, these modes essentially correspond to the ``sloshing" of the Fermi surface at fixed density. In contrast, in an ordinary sound mode, the charge density (along with the pressure) will oscillate at finite density. The holographic mode is found by solving the perturbation equations in the longitudinal channel (\ref{eq:longa}) at small momenta and frequencies and zero temperature. This can be done using, for instance, the method described in \S\ref{sec:diffusive} above. One finds a low frequency quasinormal mode with the dispersion \cite{Karch:2008fa}
\begin{equation}
\omega = \pm \frac{k}{\sqrt{d}} - \frac{\mathrm{i}k^2}{2d \hat \mu_0} \,.
\end{equation}
Recall that the zero temperature chemical potential was found in (\ref{eq:muzero0}) above. These zero temperature sound modes have been called holographic `zero sound' by analogy with the modes that exist in a Fermi surface. It is noteworthy that, as we noted in \S \ref{sec:sound} for analogous $T=0$ sound modes in backreacted EMD models, these modes coexist with low energy spectral weight at nonzero momentum. Perhaps a direct connection can be established in holographic models.

It is instructive to track the quasinormal modes in the complex plane upon increasing the temperature \cite{Bergman:2011rf, Davison:2011ek}. Unlike the backreacted models discussed in \S \ref{sec:sound}, the probe brane modes do not couple to metric fluctuations and hence are not involved in conventional $T>0$ sound modes (as occurs in the backreacted models). Therefore the `zero sound' mode of probe branes cannot cross over to the hydrodynamic nonzero temperature sound mode. Instead what one finds is that the pair of `zero sound' modes has a decreasing propagation speed and increasing attenuation constant.  At a critical temperature, the propagation speed vanishes, and the two poles begin moving in opposite directions along the imaginary frequency axis.   At high temperatures, the pole that moves towards the real axis becomes the ordinary hydrodynamic charge diffusion mode. The other pole is non-hydrodynamic. However, the presence of a non-hydrodynamic pole on the real axis will have a consequence that we discuss in the following section: a Drude peak in the optical conductivity (despite the fact that the probe brane does not conserve momentum, which can be lost to the bath).

The `zero sound' modes are seen in other probe brane models, including ones with a mass for the fundamental matter \cite{Kulaxizi:2008kv}. An interesting effect is seen if the background bath of adjoint matter (the geometry in which the probe brane is embedded) has $z \neq 1$. There the `zero sound' pole in the Green's function takes the form \cite{HoyosBadajoz:2010kd}
\be\label{eq:JJzero}
G^R_{J^x J^x}(\omega,k) \sim \frac{\omega^2}{k^2 - \omega^2/v^2 - c \, \omega^{2/z+1}} \,,
\ee
where $c$ is a complex constant. This shows that if $z > 2$ the mode is overdamped and does not exist as a sharp excitation. This condition is generalized in the presence of a hyperscaling violating background to $z > 2(1 - \theta/d)$ \cite{Pang:2013ypa, Dey:2013vja}.

\subsubsection{Linear and nonlinear conductivity}

Here we compute the electrical conductivity in these probe brane models, following \cite{Karch:2007pd}.  Before presenting the computation, however, we can think about how we expect the $N$ scaling of various quantities to affect the result. Recall the general hydrodynamic result (\ref{eq:Cleansigma2}) and the $N$ scalings discussed in \S\ref{sec:microbrane}:
\begin{equation}
\sigma = \sigma_{\textsc{q}} + \frac{\rho ^2}{\epsilon+P} \left(\frac{\mathrm{i}}{\omega}+\pi \delta(\omega)\right) \sim N + \frac{N^2}{N^2} \left(\frac{\mathrm{i}}{\omega}+\pi \delta(\omega)\right).\label{eq:sigmaNscaling}
\end{equation}
Here we used the fact that the charge dynamics is determined by the probe sector (which is order $N$) whereas the energy and pressure are dominated by the adjoint bath (which is order $N^2$). Hence, at leading order in the large $N$ limit, we will only be able to compute $\sigma_{\textsc{q}}$ and we will not see the effects of `momentum drag'. This is naturally understood from the fact that in the probe brane approximation, the backreaction of the geometry is neglected.   We hence cannot recover the ``boost" perturbation in the bulk corresponding to the $\delta$ function in (\ref{eq:Cleansigma2}), as we did in \S \ref{sec:memmatrixderivation}.   This lack of backreaction on the geometry does come with an interesting ``advantage" -- we may compute $\sigma$ beyond linear response without worrying about the heating of the black hole at nonlinear orders.    So below, we will compute the conductivity $\sigma \equiv J/E$,  but with $J$ a non-linear function of the electric field $E$.

The method we will use to compute the dc conductivity was developed by \cite{Karch:2007pd} and is similar in spirit to the method used in \S\ref{sec:dccond} above. The DBI equations of motion imply radially conserved currents, assuming that all bulk fields are spatially homogeneous:
\begin{widetext}
\begin{equation}
J^\mu = \frac{1}{2e^2(2\pi\alpha^\prime)} \left(\sqrt{X}(g+2\pi\alpha^\prime F)^{\mu r} - \sqrt{X}(g+2\pi\alpha^\prime F)^{r\mu}\right).
\end{equation}
\end{widetext}
Recall that $X$ was defined in (\ref{eq:Xdet}). Here the raised indices mean that the whole matrix $g+2\pi\alpha^\prime F$ is to be inverted, not that the indices are raised with the inverse metric. These quantities are, in fact, equal to the expectation value of the charge current operator in the boundary theory. As above we will focus on the case of massless quarks and with the background geometry being the $\mathrm{AdS}$-Schwarzchild geometry given in (\ref{eq:BH2}) and (\ref{eq:emblackfactor})  (in real time, with $z=1$ and $\theta=0$).  Upon applying the electric field $A_x = -Et + a_x(r)$, and assuming  a background gauge field $A_t(r)$, we obtain the set of equations \begin{subequations}\begin{align}
J^t &= \frac{L^{d-2}}{e^2}\frac{-r^{2-d}A_t^\prime}{\sqrt{1-\frac{(2\pi\alpha^\prime)^2}{L^4} r^4 \left(A_t^{\prime2} + f^{-1}(E^2-f^2a_x^{\prime2})\right)}} , \\
J^x &= \frac{L^{d-2}}{e^2}\frac{r^{2-d}fa_x^\prime}{\sqrt{1-\frac{(2\pi\alpha^\prime)^2}{L^4} r^4\left(A_t^{\prime2} + f^{-1}(E^2-f^2a_x^{\prime2})\right)}} .
\end{align}\end{subequations}

After a bit of algebra, we may rearrange the previous two equations into the following useful forms: \begin{subequations}\begin{align}
& \left(\frac{J^t}{J^x}\right)^2 = \left(\frac{A_t^\prime}{fa_x^\prime}\right)^2, \\
& \left(\frac{J^xe^2}{L^{d-2}}\right)^2 \left[1-{\textstyle{\frac{(2\pi\alpha^\prime)^2}{L^4}}} \frac{r^4}{f} E^2\right]  \\
&= \left[r^{2(2-d)} + {\textstyle{\left(\frac{e^2}{L^{d-2}}\frac{2\pi \alpha^\prime}{L^2} \right)^2}} r^4 \left(\left(J^t\right)^2 - \frac{(J^x)^2}{f}\right)\right](fa_x^\prime)^2. \notag
\end{align}\label{eq:stationaryE}\end{subequations}
We focus on the second equation.   As $f\rightarrow 0$ at the black hole horizon, the objects in square brackets on both sides are negative.  However, as $r\rightarrow0$ near the boundary, both brackets are positive.   Since they multiply manifestly positive quantities, these square brackets must both vanish at the same radius $r=r_\star$.  Setting the left hand side to zero implies
\begin{equation}
f(r_\star)  =  \textstyle{\frac{(2\pi\alpha^\prime)^2}{L^4}} r_\star^4 E^2 \,.
\label{eq:DBInewhor}
\end{equation}
Setting the right hand side to zero, we obtain a simple equation for the (nonlinear) conductivity $\sigma = J^x/E$: \begin{equation}
\sigma = \frac{L^{d-2}}{e^2}\sqrt{r_\star^{2(2-d)} + {\textstyle \left(\frac{e^2}{L^{d-2}}\frac{2\pi \alpha^\prime}{L^2} \right)^2} \rho^2 r_\star^4}.   \label{eq:sigmaDBI}
\end{equation}
In this last equation we have set $J^t = \rho$ for consistency with our earlier notation throughout.

At a quantum critical point, the nonlinear conductivity is subtle because the various limits $T \to 0, \omega \to 0$ and $E \to 0$ need not commute. These issues are explored for probe brane theories in \cite{Karch:2010kt}. If we take the linear response, $E \to 0$, limit of (\ref{eq:DBInewhor}) at fixed temperature then clearly $r_\star \to r_+$, the horizon. The dc conductivity then looks intriguingly like a nonlinear version of the `mean field' dc conductivity obtained previously in (\ref{eq:axionsigma}). In particular the first term in the square root is suggestive of a `pair creation' term while the second term in the square root depends explicitly on the charge density. This interpretation of the first term can be made more rigorous here: if a mass is given to the quarks then the first term picks up an additional factor that depends on the profile $\theta(r_\star)$. This factor becomes small if the mass of the quarks is large compared to the charge density \cite{Karch:2007pd}, consistent with the fact pair production is now costly. 
However, as we have seen in (\ref{eq:sigmaNscaling}) above, the probe brane conductivities do not know about momentum conservation and so the second term cannot be interpreted as a momentum drag or `Drude-like' term, despite appearances.

When the second term in the square root in (\ref{eq:sigmaDBI}) dominates (for instance with a mass for the quarks, as just described, or at temperatures low compared to the charge density) then the conductivity is proportional to the charge density. While this is natural in a limit of dilute charge carriers, the fact that it is true in the opposite limit of large densities is a nontrivial property of the DBI action in the bulk. Generalizing (\ref{eq:sigmaDBI}) to cases where the background metric has a general $z$ \cite{Hartnoll:2009ns}, and using (\ref{eq:Trp}) to relate $r_\star = r_+$ to the temperature, one obtains in this large density limit
\be
\sigma \propto \rho \, T^{-2/z} \,.
\ee
In particular, $z=2$ leads to $T$-linear resistivity.

Formulae similar to (\ref{eq:sigmaDBI}) can be obtained for probe brane models in a background magnetic field \cite{O'Bannon:2007in}.  

The equations in (\ref{eq:stationaryE}) describe a stationary state in the presence of a nonlinear electric field. Perturbation about this state allows the computation of thermal fluctuations of the current operator or `current noise' \cite{Sonner:2012if}.  For simplicity considering $d=2$ and $J^t=0$ -- i.e. at a zero density quantum critical point --  an explicit calculation \cite{Sonner:2012if} reveals that this out of equilibrium thermal noise obeys an `equilibrium' fluctuation-dissipation relation at an emergent temperature $T_\star$, related to $r_\star$ in (\ref{eq:DBInewhor}) by $T_\star = 1/\pi r_\star$. That is $r_\star$ rather than the horizon radius sets the effective temperature, c.f. (\ref{eq:Tgo}).

In models with backreaction, nonlinear reponse necessarily induces a time dependence due to Joule heating. The growth of the black hole horizon in time due to entropy production in a nonlinear electric field can be described exactly in certain simple cases \cite{Horowitz:2013mia}.

Finally, let us comment on the optical conductivity $\sigma(\omega)$. This can be obtained from the linearized perturbation equations given in the previous \S \ref{sec:zeros}. The conductivity exhibits a Drude-like peak at low frequencies \cite{Hartnoll:2009ns}. The peak becomes narrow (of width less than $T$) at low temperatures and, in fact, at zero temperature the peak becomes a delta function $\delta(\omega)$ if $z<2$. This can be seen from setting $k=0$ in the expression (\ref{eq:JJzero}) for the current-current Green's function. This delta function is not related to momentum conservation, as momentum does not couple to probe brane excitations. Instead, it suggests that probe branes have an additional conserved quantity at $T=0$ that is relaxed at any nonzero temperature. Perhaps the current operator itself. For a discussion of this open question, see \cite{Davison:2014lua}. We noted in the previous  \S \ref{sec:zeros} that the pole on the negative imaginary frequency axis responsible for this peak is a nonzero temperature remnant of the `zero sound' mode at zero temperature. The challenge of understanding the origin of the Drude peak in these models is therefore likely tied up with understanding the origin of the zero sound mode.

The divergence and Drude peak discussed in the previous paragraph depend
upon a nonzero charge density. In fact, at zero charge density,
it was shown in \cite{Karch:2010kt} that in $d=2$ boundary space dimensions the self-duality argument of \S \ref{sec:selfdual} extends to the nonlinear DBI action. This implies that the full nonlinear conductivity $\sigma (\omega, E,T)$ is independent of frequency in that case.

\subsubsection{Defects and impurities}

Probe D$p$ branes that are localized in the boundary spatial dimensions can be used to model defects or impurities that contain their own localized dynamical degrees of freedom (in the limit where these defects do not backreact on the ambient theory). A classic condensed matter model of such a system is the Kondo model, which couples a single impurity spin to a Fermi liquid. Holographic systems can generalize this class of models to describe a single impurity interacting with an ambient strongly coupled field theory. See \cite{Sachdev:2010uj} for a general discussion of defects coupled to higher dimensional CFTs and the relationship to probe branes.

Some aspects of Kondo physics were captured in a holographic probe brane model in \cite{Erdmenger:2013dpa}. In particular, an effect analogous to the screening of the impurity spin at low temperatures in the Kondo model was realized by a holographic superconductor-type instability on the defect. The condensation of a charged scalar reduces the electric flux in the IR, leading to the impurity transforming in a lower dimensional representation of $\mathrm{SU}(N)$ in the IR. This is the screening.

An earlier probe brane discussion of Kondo physics can be found in \cite{Harrison:2011fs}. In that construction the screening of the defect degrees of freedom at strong coupling was dually described by the breakdown of the probe limit and the backreaction of the defect branes on to the geometry. Such backreaction is described by a geometric transition \cite{Lin:2004nb, D'Hoker:2007fq} in which the brane itself disappears (the screening) and is replaced by various higher-form fluxes that thread new cycles that emerge in the geometry.

An interesting new phenomenon occurs when probe brane and anti-brane defects are placed a finite distance $R$ apart \cite{Kachru:2009xf}.   This is analogous to two ``opposite" impurities spatially separated in the field theory description.  When $TR\gg 1$,  each brane extends into the horizon. However, when $TR \ll 1$, the minimal free energy solution corresponds to a single brane which begins and ends on the AdS boundary, only extending a finite distance into the bulk. A preliminary study of the physics of this ``dimerization" transition also considered the possibility of a lattice of defects and anti-defects whose low temperature dimerization can lead to random configurations \cite{Kachru:2009xf, Kachru:2010dk}, possibly leading to glassy physics.

Rather than treating the probe branes themselves as defects, one can also consider probe brane actions (such as DBI) with inhomogeneous boundary conditions.  The discussion turns out to be very similar to that in \S\ref{sec:analyticstrongII}, and similar phenomenology holds: see e.g. \cite{Ryu:2011vq, Ikeda:2016rqh}.

\subsection{Disordered fixed points}\label{sec:disfpt}

It is possible that upon adding marginal or relevant disorder to a quantum field theory, there is an RG flow to a ``disordered fixed point"  where the disorder strength is either finite or infinite  \cite{fisher1999}. A disordered fixed point is possible in principle because disorder breaks translation invariance at all length scales. In the case of disordered statistical field theories (in which all dimensions including `time' are disordered), disordered fixed points can be accessed via standard tools that combine the replica methods with perturbative field theory theory: see \cite{PhysRevB.27.413} and references therein. However in the quantum mechanical case in which only the spatial dimensions are disordered, these methods typically do not work. This is because upon writing down the RG equations one finds \cite{ssbook,PhysRevB.49.4043}
\begin{equation}
\frac{\mathrm{d}\bar V}{\mathrm{d}\ell} = (d_{\mathrm{u.c.}}-d)\bar V + c \bar V^2 \,,
\end{equation}
where $\bar V$ is a coupling constant related to disorder strength, defined in (\ref{eq:vbar}) below, and $c>0$.  Hence below the critical dimension,  $\bar V \rightarrow \infty$ towards long distance scales, taking us outside of the perturbative regime of validity in the IR. This is to be contrasted with situations like the Wilson-Fisher fixed point where the second term in the beta function above has the opposite sign and so can balance the classical term. Even if some disordered quantum theory with the right sign quadratic term in the beta function were found -- for instance by effectively reducing the role of the time dimension \cite{PhysRevB.26.154,PhysRevB.65.024202} -- perturbative field theory would not be able to access the strongly disordered physics that is necessary to discuss incoherent transport of \S\ref{sec:in1}, or a possible localization transition.  In this section we will outline how holography may give rise to strongly disordered fixed points that can be studied without the replica trick or perturbation theory.

Let us recall how the notion of relevant versus irrelevant operators is
generalized to treat disordered couplings. In field theory this is called the Harris criterion \cite{0022-3719-7-9-009}. We can give a holographic derivation of this criterion \cite{Adams:2011rj,lucas1401,Hartnoll:2015rza}, which essentially amounts to the same power counting argument one would apply directly in the field theory.  Consider Einstein gravity coupled to a massive scalar field.  Neglecting the backreaction of scalar fields on the geometry, a bulk scalar field corresponding to a disordered boundary source takes the form \begin{equation}
\phi(\mathbf{x},r) = \int \frac{\mathrm{d}^dk}{(2\pi)^d} h(k) \mathrm{e}^{\mathrm{i} k\cdot x} \mathcal{F}_\Delta(kr)  r^{d+1-\Delta},
\end{equation}
where $h(\mathbf{k})$ are Gaussian quenched random variables obeying \begin{subequations}\begin{align}
\overline{h(k)} &= 0, \\
\overline{h(k) h(q)} &=  \bar V^2  \delta^{(d)}(k+q) \,, \label{eq:vbar}
\end{align}\end{subequations}
and $\mathcal{F}_\Delta(kr)$ can be expressed in terms of modified Bessel functions.  Its precise form is tangential to our discussion.  Even before averaging over disorder realizations, we immediately know that the metric will feel the first perturbative corrections due to this scalar hair at $\mathcal{O}(\bar V^2)$.   Of course, it may be the case that the perturbation is large compared to the background, in which case perturbation theory has failed.   Let us diagnose whether or not such a perturbative treatment about AdS is consistent in the UV ($r\rightarrow 0$).   This can be done quite simply \cite{lucas1401}.  Einstein's equations read \begin{align}
G_{ab} &\sim \frac{1}{r^2} \sim \overline{T_{ab}[\phi]} \sim \frac{1}{r^2} \times \bar V^2 r^{d+2-2\Delta}.  \label{eq:harris1}
\end{align}
We have assumed that we may average over disorder realizations to extract the long wavelength behavior of the metric.   The final step follows from straightforward scaling arguments -- importantly, the power of $r$ multiplying $\bar V^2$ is not $2(d+1-\Delta)$ (as it would be for a single cosine) but is $d+2-2\Delta$.   This is a consequence of the fact that $\bar V^2$ has a reduced dimension.   Evidently, as $r\rightarrow 0$, the right hand side of (\ref{eq:harris1}) is small compared to the left hand side -- and hence perturbation theory is sensible -- so long as \begin{equation}
2\Delta < d+2. \label{eq:harris2}
\end{equation}
(\ref{eq:harris2}) is known as the Harris criterion for the relevance of disorder.   In the case of more general hyperscaling-violating theories, (\ref{eq:harris2}) generalizes to \cite{lucas1401} \begin{equation}
2\Delta < d-\theta+2z.
\end{equation}

What happens when $\Delta$ is (Harris) marginal?   In this case $\bar V^2$ is a dimensionless number and there is a logarithmic divergence of the backreaction (\ref{eq:harris1}) towards the interior of the geometry \cite{Adams:2012yi}.    The full backreaction in this system was studied numerically in \cite{Hartnoll:2014cua}, where an emergent Lifshitz scaling in the IR was discovered. It was argued that the Lifshitz scaling amounted to a resummation of the logarithmic divergence, a fact that was checked analytically in perturbation theory to order $\bar V^4$.
Here we present a simpler argument for this resummation.     For simplicity, we focus on the case $d=1$.  Consider the following convenient ansatz for the metric (in units where $L=\kappa=1$):  \begin{equation}
\mathrm{d}s^2  = A(r) \left( \mathrm{d}r^2 + \mathrm{d}x^2\right) - B(r) \mathrm{d}t^2.
\end{equation}
The assumption that the metric is homogeneous is sensible -- averaging over disorder at leading order in perturbation theory, Einstein's equations now read \begin{subequations}\label{eq:disorderAB}\begin{align}
\frac{A}{\sqrt{B}}\partial_r \left(\frac{\partial_r B}{\sqrt{B}A}\right) &= \overline{(\partial_x \phi)^2 - (\partial_r \phi)^2}, \\
\frac{1}{\sqrt{B}}\partial_r \left(\frac{\partial_r B}{\sqrt{B}}\right) &= 4A + \frac{3A}{4} \overline{ \phi^2} .
\end{align}\end{subequations}
At leading order in perturbation theory, we may treat the right hand side of (\ref{eq:disorderAB}) as a small source, using $A=1/r^2$ as for AdS, which is manifestly homogeneous.  Employing scale invariance and the precise form of $\mathcal{F}_\Delta$ we find that at second order in perturbation theory: \begin{subequations}\begin{align}
\overline{\phi^2} &= \bar V^2 \int \frac{\mathrm{d}k}{2\pi} r \mathcal{F}_\Delta(kr)^2 = \bar V^2, \\
\overline{(\partial_x \phi)^2 - (\partial_r \phi)^2} &= \frac{\bar V^2}{4}.
\end{align}\end{subequations}
It is simple to check that a consistent solution to (\ref{eq:disorderAB}) is \begin{equation}
A = \frac{a_0}{r^2}, \;\;\;\; B = \frac{b_0}{r^{2+\bar V^2/4}},
\end{equation}
implying a non-trivial Lifshitz exponent $z=1+\bar V^2/8$.   Note that our normalization of $\bar V$ is different from \cite{Hartnoll:2014cua} due to different bulk scalar normalizations. A potentially surprising aspect of this results is that it implies the existence of a line of disordered fixed points parametrized by $\bar V$. A consistency check on the emergence of Lifshitz scaling was the construction of nonzero temperature disordered backgrounds, which were shown to have a low temperature entropy density scaling like $s \sim T^{1/z}$ with the anticiapted $\bar V$ dependent $z$ \cite{Hartnoll:2015faa}. Nonetheless, it would be desirable to have a more intrinsic IR understanding of the physics at work. This seems to require the development of more sophisticated techniques for solving the bulk Einstein equations with strong inhomogeneities.

With a view to understanding incoherent transport in strongly disordered gapless systems, the thermal conductivity $\kappa$ has been computed in these disordered fixed points, as well in new disordered fixed points constructed from relevant disorder \cite{Hartnoll:2015rza}. It was found that the thermal conductivity appears to exhibit log-periodic oscillations in temperature. Log-periodicity is symptomatic of discrete scale invariance and associated complex scaling exponents. It is possible, then, that these backgrounds then have instabilities in the general class studied in \S\ref{sec:holoS}. A possible endpoint of such putative instabilities are geometries with fragmented horizons, of the kind mentioned in \S\ref{sec:num510}. 

\subsection{Out of equilibrium I:  quenches}\label{sec:quench}

Our focus so far has been on the nature of quantum matter at zero and nonzero temperature and density, and on the consequences of perturbing states of quantum matter a little away from equilibrium, leading to transport coefficients like the electrical conductivity. The final two subsections will briefly consider quantum matter far from equilibrium.  This first section will focus on quenches in field theories in $d\ge 2$ spatial dimensions, where there are almost no techniques from field theory to apply.  As we will see, far from equilibrium dynamics will tend to drive quantum matter to finite temperature states.  By studying dynamics in black hole backgrounds, and/or black hole formation, holographic models make it possible to treat such thermal effects from first principles.  For this reason, the holographic approaches we derive below are a valuable tool, allowing for precise results beyond `dimensional analysis'. One caveat to keep in mind is that dissipation in holographic models is into a strongly interacting large $N$ `bath'. This may or may not be the dissipative mechanism of interest in other circumstances.

A simple way to drive a system far from equilibrium is through a quantum quench.   The idea is as follows:  consider a time dependent Hamiltonian \begin{equation}
H(t) = H_0 + H^\prime f(t),
\end{equation}
with $f(t)$ a function which varies over time scales $\tau$ which are often quite fast.   For simplicity, most research focuses on the case where $f(t)$ either interpolates between $f(-\infty)=0$ and $f(\infty)=1$  (quench from one state to another),   or where $f(\pm \infty)=0$ but $f(0)=1$  (pulsed quench).   We will see examples of both in this section.

The case where $f(t)=\sin(\omega t)\Theta(t)$ is also interesting, though we will have little to say on it in this review.   Such a drive generically does work on a system, possibly until it reaches infinite temperature. See \cite{Auzzi:2013pca} for a holographic study. 
On the other hand, let us also note the existence of curious (non-driven) states in field theory which do not thermalize (at large $N$) despite oscillating in time with non-vanishing energy density.  This is manifested holographically in fully nonlinear solutions of Einstein's equations in asymptotically Anti-de Sitter spacetime which oscillate: for example, gravitational analogues of standing waves which are nonlinearly stabilized \cite{Dias:2012tq, Buchel:2013uba}.

\subsubsection{Uniform quenches}
Let us begin with the case where $H_0$ describes a conformal field theory, and \begin{equation}
H^\prime = \lambda \mathcal{O}, \;\;\; f(t) \sim \mathrm{sech}\frac{t}{\tau},
\end{equation}
where $\mathcal{O}$ is the zero momentum mode of a relevant operator of dimension $0<\Delta\le d+1$, and we consider the limit of small $\lambda$.   The exact behavior of $f(t)$ is not important.    This setup was considered in a series of papers \cite{Bhattacharyya:2009uu, Buchel:2013gba, Das:2014jna, Das:2014hqa}, and we summarize the main conclusions.    The most interesting observation is that the energy density added to the CFT scales as \begin{equation}
\epsilon = \frac{\lambda^2}{\tau^{2\Delta-d-1}} \mathcal{E}\left(\tau^{d+1-\Delta} \lambda\right),   \label{721eq}
\end{equation}with $\mathcal{E}(x)$ a function of its dimensionless parameter.     The limit of interest is \begin{equation}
\lambda \tau^{d+1-\Delta} \rightarrow 0,   \label{eq:taulambda}
\end{equation}
where up to an overall constant, (\ref{721eq}) implies the physics in this limit is universal.

The universality of this result follows from the fact that this quench is ``fast".   Let us begin with the case $\Delta<d+1$.  Holographically, the time scale of black hole formation (or any other gravitational backreaction) associated with the quench is set by $\lambda^{-1/(d+1-\Delta)} \equiv \tau_\lambda $.  From (\ref{eq:taulambda}) $\tau_\lambda \gg \tau$.    Therefore, the quench dynamics is characterized by the fast dynamics of a scalar field in AdS, and the resulting slow dynamics of black hole formation (but with properties like $\epsilon$ already fixed on the fast time scale).  In order to fix $\epsilon \sim \lambda^2$, we employ a ``hydrodynamic" Ward identity (c.f. (\ref{eq:scalarward})) 
\begin{equation}
\epsilon = -\int\limits_{-\infty}^\infty \mathrm{d}t \; \langle \mathcal{O}(t)\rangle \partial_t (\lambda f(t)) \,,
\end{equation}
and $\langle \mathcal{O}\rangle \sim \lambda$ to leading order in $\lambda$, since $\langle \mathcal{O}\rangle_{\mathrm{CFT}} = 0$ as it is relevant.   Once the $\lambda$ scaling of $\epsilon$ is fixed, the $\tau$ scaling follows from dimensional analysis.   (\ref{721eq}) is valid for both free theories and strongly interacting theories \cite{Das:2014jna}.  Of course, holography provides strongly-coupled models where $\mathcal{E}(x)$ can actually be computed.

\begin{figure*}
\centering
\includegraphics[width = \textwidth]{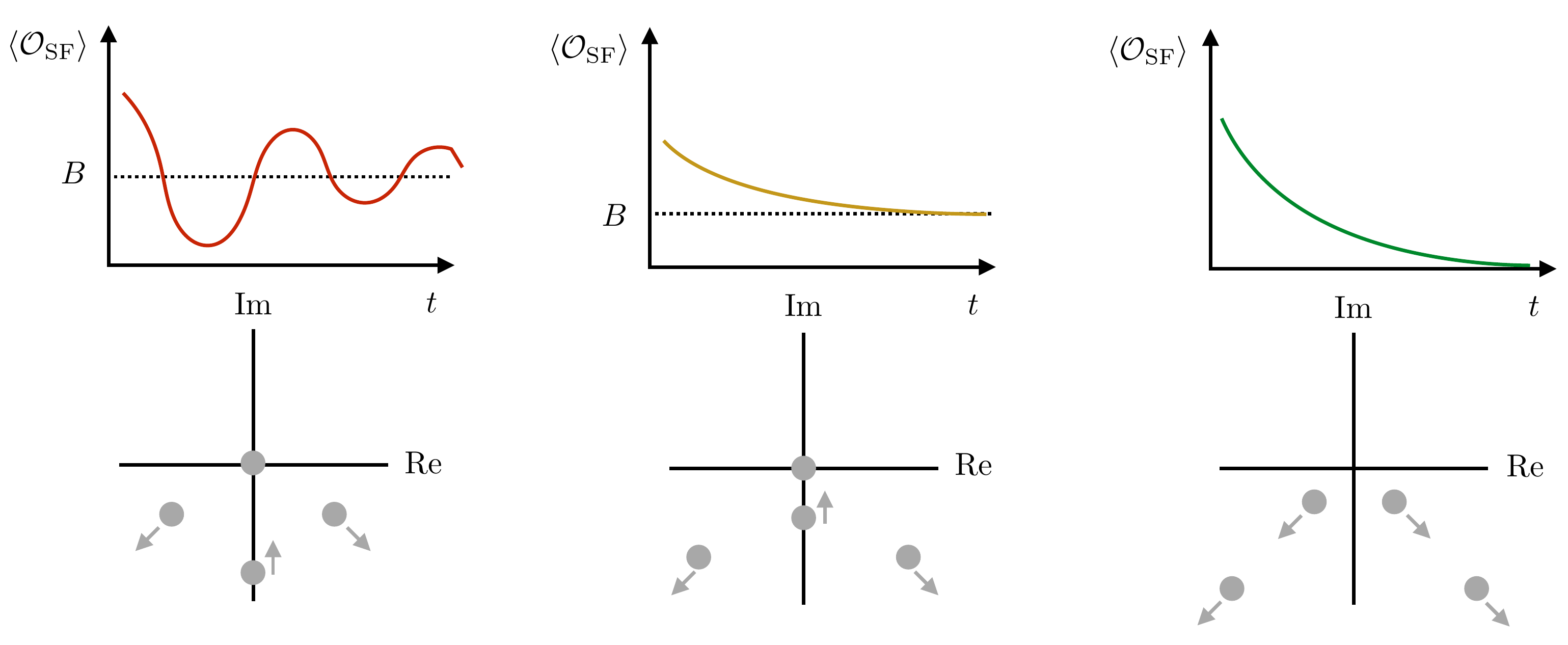}
\caption{\textbf{A sketch of the dynamics of the order parameter in a holographic superfluid quench.}   Top row: $\langle \mathcal{O}\rangle$ as a function of $t$.  Bottom row: location of the lowest lying QNMs, and their motion upon increasing $\lambda$.   Columns left to right:   $0<\lambda<\lambda_1$;  $\lambda_1<\lambda<\lambda_2$;  $\lambda>\lambda_2$.}
\label{fig:quenchfig1}
\end{figure*}

In the case of $\Delta=d+1$ (a marginal operator), \cite{Bhattacharyya:2009uu} was able to say much more about the resulting geometry.    Now $\lambda$ itself is a dimensionless parameter, and at leading order in the perturbative expansion in $\lambda$, they found that the metric could be well approximated by \begin{equation}
\mathrm{d}s^2 = \frac{L^2}{r^2}\left[-2\mathrm{d}r\mathrm{d}v - \left(1- \frac{M}{r^{d+1}} \mathcal{F}\left(\frac{v}{\tau}\right) \right) \mathrm{d}v^2 + \mathrm{d}x_i^2\right].  \label{eq:adsvaidya0}
\end{equation}
The coordinate $v$ is our ``time" coordinate, analogous to (\ref{eq:eddfink}).   The function $\mathcal{F}$ appearing above is an interpolating function between 0 at $v=-\infty$, and 1 at $v=\infty$.  The coefficient $M$ is given by $(4\pi T_*/d)^{d+1}$, where $T_*$ is the final temperature of the geometry.

When the force acts only over a short time, the function appearing in the metric (\ref{eq:adsvaidya0}) is effectively
\begin{equation}
F(v/\tau) = \Theta(v/\tau).  \label{eq:adsvaidya}
\end{equation}
This metric is called the AdS-Vaidya metric.   It is analogous to the original Vaidya metrics describing the geometry associated with infalling null dust \cite{vaidya}, and is an exact solution to Einstein's equations sourced by infalling pressureless dust.   The field theory interpretation of (\ref{eq:adsvaidya}) is clear, albeit quite surprising:  beyond time $t=0$ (in the boundary), the theory appears to thermalize instantaneously!   The reason is that we have added a very small amount of energy quickly.   If one adds an energy density $\epsilon \sim \tau^{-d-1}$ (by setting $\lambda=1$), then thermalization does not occur instantaneously.   This was observed numerically in \cite{Chesler:2008hg}, albeit in a slightly different model.   More recently, \cite{Anous:2016kss, Fitzpatrick:2016ive} have studied the dynamics of pure states in CFT2s of very high energy:  while they look thermal at $N=\infty$, non-perturbative corrections in $1/N$ restore unitarity (no thermalization).   It will be interesting to understand this physics more carefully from the bulk perspective, where such effects are non-perturbative in quantum gravity.

Another example of a spatially homogeneous quench is as follows.   Consider a holographic superfluid at an initial temperature $T<T_{\mathrm{c}}$ with a source for the superfluid order parameter which is pulsed analogously to above \cite{Bhaseen:2012gg}.    The pulse injects energy into the superfluid and the late time dynamics of the order parameter are described by \begin{equation}
\left|\langle \mathcal{O}_{\mathrm{SF}}(t)\rangle \right| \approx \left|  B+ A\mathrm{e}^{-\mathrm{i}\omega t - \gamma t}\right|.   \label{sfquencheq}
\end{equation}
Three possible behaviors were found, depending on the strength ``$\lambda$" of the quench, as shown in Figure \ref{fig:quenchfig1}.   When $0<\lambda<\lambda_1$,  both $\gamma$ and $\omega$ are positive, and $\langle \mathcal{O}_{\mathrm{SF}}(\infty)\rangle > 0$;  when $\lambda_1<\lambda<\lambda_2$, $\omega=0$ but $B > 0$;  when $\lambda>\lambda_2$, $\omega=0$ and $B = 0$.   Hence, for $\lambda<\lambda_2$, the final state is a superfluid,  while for $\lambda>\lambda_2$ it is a normal fluid.

As we have seen repeatedly in this review, this late time dynamics is naturally understood by studying the lowest lying QNMs of the bulk geometry.  In the superfluid phase, the existence of a Goldstone mode is manifested as a mode with $\gamma=\omega=0$, and so the relaxation in (\ref{sfquencheq}) is governed by the QNM with \emph{next} smallest imaginary part.   The ``motion" of these QNMs with increasing $\lambda$ is shown in Figure \ref{fig:quenchfig1}.   This reveals the holographic origin of the dynamical transition at $\lambda=\lambda_1$.   When the purely damped ($\omega=0$) mode collides with the Goldstone mode, these modes separate off the $\omega=0$ axis and $\gamma$ is increasing with increasing $\lambda$.   This corresponds to a quench which has injected enough energy so that the final state is in the normal phase.    These holographic dynamics reproduce the expectations of weakly interacting field theories \cite{barankov}, but it is instructive to see how the QNMs of ``superfluid" black holes recover this dynamics even at strong coupling.

\subsubsection{Spatial quenches}

\begin{figure*}
\centering
\includegraphics[width = \textwidth]{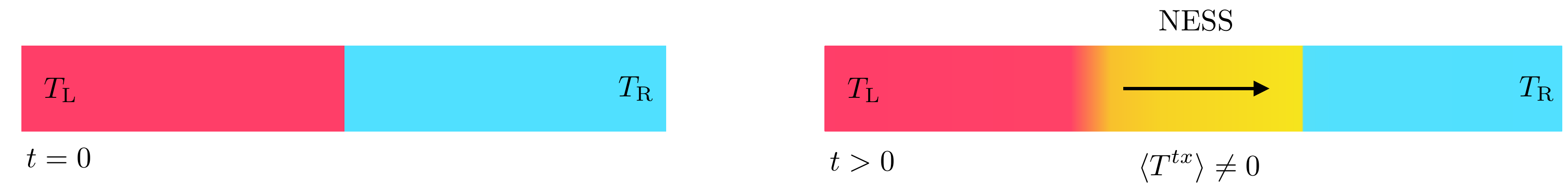}
\caption{\textbf{A quench connecting two heat baths.}  Left:  two CFTs placed next to each other at different temperature.   Right: a NESS forms as energy flows from the left to right bath.}
\label{figness}
\end{figure*}

Let us now consider a slightly more complicated set-up:  suppose we take two copies of a conformal field theory in $d$ spatial dimensions, one (L) defined for $x<0$ and the other (R) defined for $x>0$.  Now, suppose that we prepare L in a thermal state at temperature $T_{\mathrm{L}}$, and R in a thermal state at $T_{\mathrm{R}}$.    Without loss of generality,\footnote{So long as the theory is not chiral in $d=1$.}  we suppose that $T_{\mathrm{L}}>T_{\mathrm{R}}$.    At time $t=0$ we allow these theories to couple.  The set-up is depicted in Figure \ref{figness}.

What is often found in such quenches that the theory reaches a non-equilibrium steady state (NESS),  where the local density matrix near $x=0$ is given by $\exp[-\sum \lambda_i H_i]$, for all conserved quantities $H_i$ \cite{rigol}.   For the case of a conformal field theory, the answer is known exactly \cite{Bernard:2012je}.   For $|x|<t$,  the local density matrix takes the remarkably simple form \begin{equation}
\rho_{\mathrm{NESS}} =  \mathrm{e}^{-\beta(\cosh\theta H - \sinh\theta P)}, \label{rhoness}
\end{equation}
with \begin{equation}
\beta = (T_{\mathrm{L}}T_{\mathrm{R}})^{-1/2},\;\;\; \mathrm{e}^{2\theta} = \frac{T_{\mathrm{L}}}{T_{\mathrm{R}}}.
\end{equation}
$\rho_{\mathrm{NESS}}$ only contains two conserved quantities, despite the existence of an infinite number.  There is a very simple explanation: CFT2s consist of decoupled left and right moving theories, and so at the speed of light the left movers at $T_{\mathrm{R}}$ from the right bath and right movers at $T_{\mathrm{L}}$ from the left bath ``mix" near the interface.  This argument is rigorous and can be shown using CFT technology \cite{Bernard:2012je}; see \cite{bhaseenNP} for a holographic derivation.  

It was pointed out using hydrodynamic arguments in \cite{bhaseenNP,  Chang:2013gba} that the NESS could exist in any dimension.   Let us consider a CFT for simplicity, but the argument generalizes naturally, and focus on the case where $T_{\mathrm{L}}\approx T_{\mathrm{R}}$ to simplify the mathematics.   In this limit the system is almost in equilibrium and we can linearize the equations of hydrodynamics.   These equations lead to elementary sound waves and we conclude  that a NESS forms -- with density matrix analogous to (\ref{rhoness}) -- for $|x|<t/\sqrt{d}$ between a pair of sound waves propagating away from the interface.  The basic dynamics is depicted in Figure \ref{figness}.   The nonlinear calculation is more subtle but a NESS can be shown to form within hydrodynamics \cite{Lucas:2015hnv, Spillane:2015daa}.

At early times, the dynamics is complicated and possibly non-universal near the interface.   Holographic descriptions \cite{Amado:2015uza} allow us to access both the early and late time dynamics in a unified approach.   One immediate consideration is that \emph{if} a NESS exists, it is thermal (c.f. (\ref{rhoness})) in a holographic model:  this follows from black hole uniqueness theorems \cite{bhaseenNP}.  It is further proposed that knowledge of the heat current in the NESS allows one to determine \emph{fluctuations} of the heat current, giving the first solvable model of ``current noise" in higher dimensions \cite{bhaseenNP}.

\subsubsection{Kibble-Zurek mechanism and beyond}

An even more complicated quench set-up is as follows.    Consider a Hamiltonian $H$ with a thermal phase transition at temperature $T_{\mathrm{c}}$.   As an explicit example, we will keep in mind a normal-to-superfluid transition in $d=3$.   The effective action for the order parameter $\varphi$ dynamics is an O(2) model: \begin{equation}
\mathcal{L} = -|\partial \varphi|^2 + \epsilon |\varphi|^2 - \frac{\lambda}{2}|\varphi|^4.  \label{eq:kzml}
\end{equation}    
Suppose that we perform a thermal quench on the system in between times $-a_1\tau_{\textsc{q}} \le t \le a_2\tau_{\textsc{q}}$, for $a_{1,2}\sim \mathcal{O}(1)$, such that the ``reduced temperature" \begin{equation}
\epsilon(t) \equiv 1-\frac{T(t)}{T_{\mathrm{c}}} = \frac{t}{\tau_{\textsc{q}}}.
\end{equation}
A priori, a superfluid condensate forms, with $|\varphi|^2 = \epsilon/\lambda$, and nothing interesting happens.  But what is the phase of the complex number $\varphi$?    Some thermal fluctuation will lead to the local orientation of $\varphi$,  but these thermal fluctuations will kick $\mathrm{arg}(\varphi)$ to different values at different points.   We will now argue that this dynamics leads to the formation of vortex lines (or more general topological defects in more general models).

The canonical description of the resulting dynamics is due to Kibble \cite{Kibble:1980mv} and Zurek \cite{Zurek:1985qw}, and so is called the Kibble-Zurek mechanism (KZM).   In a theory of dynamical critical exponent $z$ and correlation length exponent $\nu$, the time scale associated with the dynamics is \begin{equation}
\tau(\epsilon) = \tau_0 |\epsilon|^{-z\nu}.
\end{equation}
If $\tau(\epsilon) \ll |t|$,  then we expect that the dynamics this far from the quench is adiabatic -- the system is in equilibrium (or as close to it as possible).   But when $\tau(\epsilon) \gg t$, then the system cannot reach equilibrium during the quench.   Hence, we define a freeze-out time when $\tau(\epsilon(t_{\mathrm{f}}))\sim t_{\mathrm{f}}$: \begin{equation}
t_{\mathrm{f}} \sim \tau_0 \left(\frac{\tau_{\textsc{q}}}{\tau_0}\right)^{z\nu/(z\nu+1)}.
\end{equation}
On this short time scale the system is far from equilibrium.   In particular, in the superfluid phase the early time dynamics ($0< t\ll t_{\mathrm{f}}$) will be in response to the thermal fluctuations imprinted on the system just above $T_{\mathrm{c}}$, which are of a characteristic size \begin{equation}
\xi_{\mathrm{f}} \sim \xi_0 \left(\frac{t_{\mathrm{f}}}{\tau_0}\right)^{1/z} \sim \xi_0 \left(\frac{\tau_{\textsc{q}}}{\tau_0}\right)^{\nu/(z\nu+1)}.
\end{equation}
The number density of vortices  in the superfluid at $T=0$ should be approximately set by this length scale: \begin{equation}
n_{\mathrm{vort}} \sim \frac{1}{\xi^{d-d_{\mathrm{top}}}_{\mathrm{f}}},   \label{eq:kzm}
\end{equation}
where $d_{\mathrm{top}}$ is the dimension of topological defects;  for a superfluid,  $d_{\mathrm{top}} = d-2$.   

Recent holographic studies have shown that this  argument holds for slow thermal quenches \cite{Sonner:2014tca, Chesler:2014gya}. The advantage of a holographic approach is that it incorporates both quantum and thermal processes, including dissipation, from first principles.  To recover this effect in holography is slightly subtle -- thermal fluctuations are suppressed by a factor of $1/N^2$ which vanishes in the classical gravity limit.  So \cite{Sonner:2014tca, Chesler:2014gya} simply added (small) random external forcing as a boundary condition to mimic these fluctuations -- the resulting dynamics will lead to defect formation across the phase transition.   
Holographic models recover the KZM in its regime of validity when $\tau_{\textsc{q}}$ is long.   As \cite{Chesler:2014gya} also emphasized,  holographic methods are well suited to model defect formation for rapid quenches, beyond the regime of validity of KZM.    The KZM argument fails once the condensate amplitude $|\varphi(t_{\mathrm{eq}})|^2 \sim \epsilon(t_{\mathrm{eq}})^{2\beta}$ for a critical exponent $\beta$;  in the example (\ref{eq:kzml}) above, $\beta=1/2$.   The amplitude of fluctuations, however, may be parametrically suppressed -- as in holography, where they are suppressed by $1/N^2$ corrections.   Hence the defects are frozen not at $t_{\mathrm{f}}$, but at $t_{\mathrm{eq}}$, which is generally much longer (since it takes longer for the exponentially growing $|\varphi(t_{\mathrm{eq}})|^2 $ to reach $\epsilon^{2\beta}$.   It is then $t_{\mathrm{eq}}$ that controls the density of defects, and this can lead to (parametrically, in the case of holography) small  prefactors in front of (\ref{eq:kzm}).  

When $t_{\mathrm{eq}} \gg \tau_{\textsc{q}} \gg t_{\mathrm{f}}$, the system may always be treated in linear response during the quench as the condensate is far from well-formed.   Hence, the final density of defects is governed only by the most unstable modes of the condensate at the final quench temperature.   This dynamics is also  associated with a characteristic length scale $\xi_{\mathrm{f}} \sim \epsilon_{\mathrm{f}}^{-\nu}$, which is independent of $\tau_{\textsc{q}}$.   Thus, the density of vortices is expected to scale as (using $\beta=\nu=1/2$, $z=2$ for the holographic model of \cite{Chesler:2014gya}) \begin{equation}
n \sim \left\lbrace\begin{array}{ll} \text{const.} &\ \tau_{\textsc{q}} \ll \tau_{\mathrm{f}} \\ \tau_{\textsc{q}}^{-1/2} &\ \tau_{\textsc{q}} \gg t_{\mathrm{f}}  \end{array}\right..
\end{equation}
This result was  recovered in holographic numerical simulations, although it should hold more generally.  

\subsection{Out of equilibrium II:  turbulence}\label{sec:turb}
One of the most fascinating open problems in all of physics is the nonlinear and chaotic dynamics of turbulent fluids, a problem which has important practical and theoretical applications for classical fluids like water and air \cite{davidson}.  It has been appreciated more recently that strongly interacting quantum fluids will share many of the same dynamical phenomena.   Indeed, given the fluid-gravity correspondence previously discussed, it is natural to expect that dynamical black holes can generically behave turbulently.   This has indeed been shown numerically \cite{Adams:2013vsa, Green:2013zba} for 3+1 dimensional asymptotically-AdS black holes.   The numerical methods necessary to study such dynamical problems are reviewed in \cite{Chesler:2013lia}.

As many of the phenomena associated with such turbulent black holes are familiar from classical turbulence, let us focus on interesting geometrical properties of a ``turbulent horizon".\footnote{We neglect subtleties with defining the black hole horizon in a dynamical spacetime -- such issues are less relevant in the fluid-gravity limit.}   A purely gravitational calculation \cite{Adams:2013vsa} shows that the horizon of such a black hole looks ``fractal" over the inertial range (the length scales over which turbulent phenomena appear self-similar).   This is reminiscent of the fact that turbulent fluids are more effective at dissipating energy than non-turbulent fluids.   Indeed, the rapidly growing horizon is a signature of the fast growth of the entropy in the dual theory.

Although holography has not taught us anything about classical turbulence,  it has proven to be more useful in the study of superfluid turbulence.   Superfluid turbulence is often called ``quantum" turbulence,  though this is a lousy name.    Although vortices are quantized in a superfluid, most turbulent phenomena are entirely classical in nature and insensitive to vortex quantization:  if we place $M\gg 1$ vortices with circulation $\Omega$ next to one another, they will behave much like a classical vortex of circulation $M\Omega$, and create a superfluid velocity field just like a large classical vortex.  The important difference between superfluid and classical turbulence actually arises at nonzero temperature. This is perhaps surprising, since most quantum phenomena are more pronounced at zero temperature.    At nonzero temperature, the normal fluid and superfluid can exchange energy and momentum, complicating the description of superfluid turbulence.

The major  difficulty with studying superfluid turbulence theoretically is the treatment of dissipation at nonzero temperature.   The standard approach is to use a damped Gross-Pitaevskii equation (see e.g. \cite{billam}),  but this approach is formally unjustified.    As we have seen repeatedly throughout this review, holographic methods are naturally adapted to study dissipative dynamics at nonzero temperature.   In \cite{Adams:2012pj},  the dynamics of superfluid vortices was studied in a probe limit.   In addition to seeing some signatures of turbulence including Kolmogorov scaling laws,  \cite{Adams:2012pj} was also able to give a ``microscopic" description of vortex pair annihilation, a process in which a vortex of winding number $+1$ collides with a vortex of winding number $-1$, releasing a nonlinear burst of sound which ultimately dissipates away.    By keeping track of the energy flux across the black hole horizon, they were able to show that dissipative processes during superfluid turbulence are essentially localized near vortex cores, over the ``healing length" defining the vortex core size.   This corresponds to the length scale over which superfluid hydrodynamics breaks down.  The simple holographic picture is that the vortices punch flux tubes from the boundary down through the bulk superconductor and into the horizon. This allows a localized channel through which energy can dissipate into the horizon. Finally, in classical turbulence in two spatial dimensions, vortices of like sign tend to clump together.   This signature was not strongly seen in holography, while other signatures of turbulence were observed, suggesting that the holographic computation may probe a new kind of finite temperature turbulence \cite{Adams:2012pj}.  It is not fully understood today whether this behavior is truly turbulent, or simply a signature of `overdamped' classical turbulence, as argued in \cite{billam2}.

The localized nature of dissipation in an manifestly nonzero temperature, dissipative theory of superfluid turbulence suggested the emergence of a simple effective description of superfluid turbulence, independent of holography.  \cite{Chesler:2014pka} emphasized this description, which is remarkably simple, as well as its practical consequences.   Let $X_n(t)$ denote the spatial position of vortex $n$ at time $t$;  one can argue on very general principles that the dynamics of a dilute mixture of vortices is essentially given by\begin{equation}
\rho_{\mathrm{s}}\Omega W_n \varepsilon^{ij} \left(\dot{X}^j_n - V^j_n(X_n-X_m)\right) = -\eta \dot{X}^i_n \,,
\end{equation}
with $V_n^i = W_n \epsilon_{ij}\partial_j  \log |X_n - X_m|$ the superfluid velocity flowing through the core of vortex $n$ and
$W_n=\pm1 $ the winding number of each vortex (higher winding number vortices are unstable and will rapidly break apart -- see e.g. \cite{Adams:2012pj}), $\rho_{\mathrm{s}}$ the superfluid density, $\Omega$ the quantum of circulation, and $\eta$ a dissipative coefficient whose determination requires a microscopic calculation, such as holography.   The dynamics is described by a single dimensionless parameter \begin{equation}
\hat \eta = \frac{\eta}{\rho_{\mathrm{s}}\Omega}.
\end{equation}
Classical turbulence is recovered in the limit $\hat\eta \lesssim 10^{-2}$;  in contrast, holographic simulations appear consistent with vortex dynamics with $\hat\eta \gtrsim 10^{-2}$.   This provides a simple explanation for the new behavior observed in \cite{Adams:2012pj}.    Furthermore, the vortex pair annihilation process, which plays a non-trivial role in holographic vortex dynamics, is described by \begin{equation}
\dot{N} \sim -\hat\eta N^2,  \label{dndt}
\end{equation}with $N(t)$ the number of vortices (of both $W_n$) at time $t$.  A different exponent (5/3) is proposed for classically turbulent flows as $\hat\eta\rightarrow 0$ \cite{Chesler:2014pka}.   

Equation (\ref{dndt}) is a remarkably simple prediction with ramifications for holographic and non-holographic theories, as well as experiments.   Two-dimensional cold atomic gases have allowed us to study superfluid turbulence in two dimensions experimentally \cite{neely, kwon}.  In fact, the effective description advocated in \cite{Chesler:2014pka} -- inspired by the holographic description of vortex annihilation -- may help lead to the first experimental detection of a turbulent superfluid flow in two spatial dimensions.   While it is easy to experimentally measure the vortex annihilation rate, no other experimentally accessible probe of turbulence is known.  It is likely that this effective description will also help to resolve the question of whether the holographic vortex dynamics of \cite{Adams:2012pj} is turbulent, albeit with substantial numerical effort.

Current holographic simulations of vortex annihilation ``experiments" \cite{Du:2014lwa} may not be at low enough $T/T_{\mathrm{c}}$ to be relevant for cold atomic gases.   A holographic computation of $\hat\eta(T)$, including low temperatures, is an interesting open problem.

\section{Connections to experiments}

As we noted in the introduction, \S \ref{sec:qmwqp}, modern quantum materials provide many examples of `strange metal' states
without quasiparticle excitations. Traditional field theoretic condensed matter studies of such states employ a variety of expansion methods
which extrapolate to strong-coupling from a weak-coupling starting point with quasiparticles. The advantage of the holographic
methods described here is that they directly yield a solvable framework for metallic states without quasiparticle excitations,
and this has many consequences for a complete description of the observable properties of such states. 

\subsection{Probing non-quasiparticle physics}

Before turning to specific materials below, the most important `practical' output of holography is that it suggests a shifted emphasis for what kind of observables reveal the nature of non-quasiparticle dynamics. This is helpful independently of the system of interest. 

\subsubsection{Parametrizing hydrodynamics}
\label{sec:hydroparam}

Holographic examples explicitly demonstrate that conducting quantum fluids exist in strongly interacting systems, and that one of their defining properties is rapid local equilbration in a time of order $\hbar/(k_{\mathrm{B}} T)$.  Classical hydrodynamics is the effective theory of such fluids beyond this time scale.  Indeed, solutions of holographic models point the way to direct hydrodynamic analyses of transport, without the need to extrapolate from a
quasiparticle framework; as we noted in \S \ref{sec:transcondmat}, the latter can often lead to misleading results.

A very basic distinction between quasiparticles and collective hydrodynamic transport is captured by the Lorenz ratio $L \equiv \kappa/(T \sigma)$. Quasiparticles transport both charge and heat and hence tie electronic and thermal conductivities together into the Wiedemann-Franz (WF) law (\ref{eq:WF}), as long as interactions can be neglected. Measured violations of the WF law have long been considered a tell-tale sign of non-quasiparticle or at the very least exotic physics, e.g. \cite{PhysRevB.72.214511,PhysRevLett.97.207001,PhysRevB.80.104510,Tanatar1320}. Hydrodynamic transport in particular, however, offers clean ways of thinking about violations of the WF law with distinctive signatures. In a hydrodynamic regime charge and heat are independent hydrodynamic variables. In an incoherent hydrodynamic metal they are essentially decoupled into distinct diffusive modes, see \S\ref{sec:in1} above and \cite{Hartnoll:2014lpa}. In hydrodynamic metals with a long-lived momentum they are related  by the relative efficiency with which sound propagation drags heat and charge as in e.g. equation (\ref{eq:alltrans}) above and \cite{Mahajan:2013cja}. A further effect in this regime (with large enough charge density) is a dramatic distinction between open and closed circuit thermal conductivity, $\kappa$ and $\bar \kappa$, discussed towards the end of \S\ref{sec:transcoeff}.

Recent experimental violations of the WF law have been usefully interpreted in terms of both coherent \cite{Crossno1058} and incoherent \cite{VO2} hydrodynamics. In a hydrodynamic regime the Lorenz ratio is also an interesting observable at higher temperatures. At higher temperatures one must either find a compelling way to subtract out the phonon contribution to the thermal conductivity \cite{VO2}, look at the Hall Lorenz ratio \cite{PhysRevLett.84.2219}, to which the neutral phonons do not directly couple, or alternatively understand the phonons themselves as an intrinsic part of the metallic system \cite{Zhang:2016ofh}.

A major open question from this standpoint is whether the many families of strange and bad metals with $T$-linear resistivity extending to high temperatures are in a non-quasiparticle hydrodynamic regime. If so, one would like to know whether the appropriate hydrodynamic framework is purely diffusive \cite{Hartnoll:2014lpa} or based around a further long-lived mode that could be momentum \cite{Davison:2013txa} or a Goldstone-related excitation such as a phase-disordered density wave \cite{CDW}. These long-lived modes directly couple to the currents and so their lifetime should be directly visible in the optical conductivity.

Direct evidence for hydrodynamic flow can be found in an unconventional sensitivity to the geometry of the current flow. Recent pioneering experiments have initiated progress in this direction \cite{Bandurin1055, Moll16}, as we discuss in more detail in \S\ref{sec:graphene} below. It will be very exciting if these and similar direct probes of hydrodynamic flow can be extended to other materials.


\subsubsection{Parametrizing low energy spectral weight}
\label{sec:spectralD}

The low energy spectral weight as a function of energy and momentum is a basic characteristic of any system. Conventional metallic phase of matter contain a Fermi surface of weakly interacting quasiparticles and this strongly determines the structure of the low energy spectral weight. Without quasiparticles the universally defined spectral weight to consider is that of the conserved charges and currents, such as $\text{Im}\, G^{\mathrm{R}}_{\rho\rho}(\omega,k)$. Direct, high resolution information about this observable would give invaluable insights into strange metal regimes.

A phenomenologically interesting possibility that emerges in some of the simplest holographic models is that of a $z=\infty$, semi-locally critical sector as in \S \ref{sec:ads2rd}. Such a sector retains some features of a Fermi surface without reference to single particle concepts. As described in \S \ref{sec:zeroTspectral} and \S \ref{sec:SWT} this leads to zero temperature spectral weight of the form $\omega^{2 \nu(k)}$ and a low temperature dependence of $T^{2 \nu(k)}$. The $k$ dependence of the exponent is generically present once $z=\infty$ and momentum is dimensionless. Requiring strictly local criticality with no $k$ dependence whatsoever -- as has occasionally been considered in the condensed matter literature --  amounts to a fine tuning. In clean systems, singular $k$ dependence linked to the underlying Fermi surface is generally expected, but such effects are suppressed in holographic analyses
in the leading large $N$ limit \cite{Faulkner:2012gt,Polchinski:2012nh,Sachdev:2012tj}.

While the spectral weights of currents and charges are the most universally defined, any operator in a locally critical regime can be expected to show similar scaling. Thus we discussed the impact of semi-local criticality on fermionic operators in \S \ref{sec:fermionz} and on `Cooper pair' operators in \S\ref{sec:transition}. The case of semi-locally critical fermions has motivated analysis of ARPES data with a continuously varying exponent \cite{reber2015power}, as well as a way to fit the form of measured unconventional quantum oscillations \cite{Tan287}. Quantum oscillations -- discussed briefly in \S\ref{sec:qoqo}  --  are especially interesting here. The Lifshitz-Kosevich formula for the amplitude of quantum oscillations with temperature is a direct measurement of Fermi-Dirac quasiparticles, see e.g. \cite{PhysRevB.81.140505}. If non-quasiparticle physics underlies at least some strange metals, strong deviations from Lifshitz-Kosevich must arise. It is challenging to measure quantum oscillations in quantum critical regimes because the effective mass becomes large, as seen in e.g. \cite{Ramshaw317}.

Direct, unambiguous evidence for semi-local criticality in any observable would be exciting as it would (\emph{i}) signal that the simplest holographic model captures the correct low temperature kinematics and (\emph{ii}) would allow the rich phenomenology of $z=\infty$ fixed points to be realized. The latter includes the strong effects of umklapp scattering on dc transport, as in (\ref{eq:tauinfty}), strong scattering of fermions by the critical sector, as in (\ref{eq:AdSRNprop}), tendency towards density wave instabilities, as in \S\ref{sec:inhomogS} and the ability to incorporate strong spatial inhomogeneities into the critical dynamics, mentioned in e.g. \S\ref{sec:num510}.

\subsubsection{Parametrizing quantum criticality}

Scaling arguments give a powerful way to organize observables in the absence of quasiparticle `building blocks'. It is well known that quantum critical points or phases lead to dynamics that is characterized by a dynamic critical exponent $z$ as well as by the scaling dimensions of operators in the critical theory \cite{ssbook}.

Compressible phases of matter offer a challenge to scaling theory because there is an additional scale, the charge density, which plays a key role. Even in a weakly coupled Fermi liquid the appropriate scaling theory for the low energy physics is nontrivial for this reason, as we discussed in \S\ref{sec:nfl}. In the strongly coupled compressible phases described holographically in \S\ref{sec:EMDgeom} and \S\ref{sec:anomalouscharge} two additional exponents played a central role: the hyperscaling violation exponent $\theta$ and the anomalous dimension for the charge density $\Phi$.
These exponents determine the temperature dependence of thermodynamic and transport obervables.

Experimental determination of the exponents $\theta$ and especially $\Phi$ in strange metals is of great interest. It was emphasized in \cite{Hartnoll:2015sea} that the Lorenz ratio, already discussed in \S\ref{sec:hydroparam}, is a direct probe of the exponent $\Phi$ in a scaling theory. If the Lorenz ratio has a nontrivial scaling with temperature in a quantum critical regime, then $\Phi$ must be nonzero. It has also been proposed to measure $\Phi$ through the nonlocal charge reponse that it induces \cite{Limtragool:2016gnl}.

Attempts to fit observed scaling of quantities as a function of temperature and magnetic field in the cuprates with the three exponents $\{z,\theta,\Phi\}$ have been only partially successful \cite{Hartnoll:2015sea,Khveshchenko:2015xea}. One challenge is that transport observables are potentially sensitive to irrelevant operators that break translation invariance, as we described in e.g. \S\ref{sec:memdrude} above. Various assumptions that need to be made for a scaling theory to get off the ground are outlined in \cite{Hartnoll:2015sea}. Finally, some of the observed scalings are better established than others. More systematic measurements to high temperatures of the Hall Lorenz ratio in the strange metal regime across the cuprate family, so far limited to \cite{PhysRevLett.84.2219, PhysRevB.72.054508,Matusiak2006376, 0295-5075-88-4-47005}, with contradictory results, would be especially desirable.

Holographic studies have also lead to new parametrizations of observables in zero density critical systems described by CFTs.
At high frequencies and short time scales, we learned that the operator product expansion controls deviations from criticality in response functions (\S \ref{sec:OPE} and \S \ref{sec:acsigma2}). At intermediate time scales holography motivates searching for universal behavior in the leading order non-hydrodynamic decay to equilibrium, which is controlled by quasinormal modes \cite{Bantilan:2016qos}. Going beyond linear response, we have seen the similar imprints of critical operator dimensions in the behavior of quantum many-body systems undergoing a quench (\S \ref{sec:quench}).     We hope that some of this novel far-from-equilibrium dynamics can be observed in quantum critical cold atomic gases \cite{xibo, endres}, and/or in the strange metals discussed above.

\subsubsection{Ordered phases and insulators}

It is an experimental fact that a large number of strange metallic regimes are unstable to ordering at low temperatures. As we noted in \S\ref{sec:sccondmat}, from a weakly-interacting single-particle perspective one expects quantum critical bosonic fluctuations to both enhance and inhibit ordering. The quantum critical modes can provide a strong `superglue' for pairing, but also strongly reduce the density of states available for pairing at the Fermi surface. In \S \ref{sec:holoS} we described an alternate non-quasiparticle description of ordering in a critical theory. The instability is ubiquitous in holographic models and occurs as a symmetry-breaking operator acquires a complex scaling exponent. This mechanism has several distinctive signatures including BKT-like exponential scalings, as discussed around equation (\ref{eq:BKTscaling}), and unconventional Cooper pair fluctuations above $T_c$, described by equation (\ref{eq:relate3}). We noted in \S\ref{sec:sccondmat} that a weakly interacting cousin of this physics seems to arise when the RG flow of a `Yukawa' coupling between fermions and a quantum critical boson couples to a nontrivial flow for the BCS coupling \cite{Raghu:2015sna}.

Many strange metals also arise in close proximity to localized phases. In \S\ref{sec:MIT} we saw that holographic models realized a novel scenario in which insulating behavior arises from relevant translational symmetry breaking operators in a $z=\infty$ scaling theory (see \S\ref{sec:spectralD}). This is an intrinsically non single-particle mechanism for localization. The temperature dependence of transport quantities in this scenario will be controlled by the scaling dimension of the relevant operator breaking translation invariance as in e.g. \cite{Donos:2012js}.

Both ordered and localized holographic phases often exhibit a soft, power law gap in e.g. the optical conductivity. This is due to the topologically ordered nature of holographic states, with deconfined gauge fields. We will discuss the possible relevance of topological order in the cuprates in \S\ref{sec:cuprates} below. Such soft gaps are also characteristic of quantum spin liquid candidates \cite{PhysRevB.86.155150,PhysRevLett.111.127401}.

\subsubsection{Fundamental bounds on transport}

Quasiparticles lead to an infinite number of long-lived operators, $\delta n_k$, and hence transport can be studied in terms of the scattering that these many variables experience. Without quasiparticles we have fewer moving parts to work with. It is natural, then, to turn to possible results that can apply to a large number of systems, following from basic principles of quantum mechanics and statistical mechanics (two handles on the system that we certainly always have!).

Fundamental bounds on dephasing times were discussed in \S \ref{sec:qmwqp}, and transport bounds were discussed in \S\ref{sec:in1}. These were motivated from several angles. Firstly it connects naturally to bounds that apply to quasiparticle systems and that can be derived from the uncertainty relations of single-particle quantum mechanics. Secondly, holographic models that are at infinite coupling could potentially have led to infinite scattering and hence vanishing transport coefficients. Yet, instead, appropriate dimensionless transport data in holography is typically `order one' in some suitable sense (see e.g. \S\ref{sec:dccond}). Thirdly, conducting states in holography were found to survive even with arbitrarily strong disorder in \S \ref{sec:in2} and \S \ref{sec:in3}: in particular, in some holographic models \cite{Grozdanov:2015qia, Grozdanov2} we saw that conductivities can be exactly bounded. Fourthly, a recent bound (\ref{MSSbound}) on chaotic timescales has been proven. This bound may plausibly have consequences for transport.

Examining non-quasiparticle transport through the lens of potential universal bounds has proved useful in several recent experimental studies \cite{Bruin2013, Analytis16, Zhang:2016ofh,2016arXiv161200815L}. Conductivity and/or diffusion bounds seem natural in holographic models, and it is possible there is a deep connection to such experiments.

\subsection{Experimental realizations of strange metals}

We now turn to some of the best studied families of experimentally realized strange metals where the ideas we have discussed in this review are (in our view) most likely to be relevant.

\subsubsection{Graphene}
\label{sec:graphene}

As we noted in \S \ref{sec:sit}, graphene is described at low energies by electronic excitations with a massless Dirac spectrum in 2+1 spacetime dimensions with the $\sim 1/r$ repulsive Coulomb interaction. 
It was argued in \cite{Muller:2008qx,Fritz08,MFS08} that breakdown of screening near the Dirac point should lead to strange
metal behavior which can be described by relativistic hydrodynamics in the presence of weak disorder. The needed transport results
were first worked out in \cite{hkms}, initially using holographic inspiration, but also by direct hydrodynamic arguments;
here we can extract the results from \S \ref{sec:incformal}. 

Crossno {\em et al.} \cite{Crossno1058} measured the thermal and electric conductivities of graphene for a range of temperatures
and densities near the charge neutrality point. We focus here on the Wiedemann-Franz ratio, which from e.g. (\ref{eq:allsigma}) is
\beq
\frac{\kappa}{T \sigma} = \frac{v_{\mathrm{F}}^2 \mathcal{M} \tau_{\rm imp}}{T^2 \sigma_{\textsc{q}}} \left( 1 + 
\frac{e^2 v_{\mathrm{F}}^2 \rho^2 \tau_{\rm imp}}{\mathcal{M} \sigma_{\textsc{q}}} \right)^{-2} . \label{WFpred}
\eeq
We have re-instated factors of the effective speed of light $v_F$ and fundamental charge $e$. We have also used the relativistic relations (\ref{eq:385}) for $\alpha_{\textsc{q}}$ and $\bar \kappa_{\textsc{q}}$.
The result (\ref{WFpred}) has some remarkable features in the clean limit, $\tau_{\rm imp} \rightarrow \infty$. Away from particle hole symmetry, $\rho \neq 0$, the Wiedemann-Franz ratio vanishes as  $\tau_{\rm imp} \rightarrow \infty$: the conductivity diverges because of momentum drag effects, while the thermal conductivity is finite due to the open circuit boundary conditions discussed in  \S\ref{sec:hydroparam} and \S\ref{sec:transcondmat}. On the other hand, at the particle-hole symmetric density, $\rho=0$, the Wiedemann-Franz ratio diverges as  $\tau_{\rm imp} \rightarrow \infty$:
this is because the conductivity now takes the finite value $\sigma_{\textsc{q}}$, while the thermal conductivity diverges because when $\rho = 0$ the open circuit boundary conditions no longer remove the long-lived momentum mode. See the divergence in (\ref{eq:kappagen}) as $\rho \to 0$.
Consequently (\ref{WFpred}) predicts a strong dependence on density in clean graphene. 
These predictions compare very well with the observations \cite{Crossno1058}. Further tests of the separate $T$ and $\rho$
dependencies of $\kappa$ and $\sigma$ agreed better with a theory in which the disorder was assumed to have long-wavelength
``puddle'' character, and the whole fluid was described by inhomogeneous hydrodynamics \cite{Lucas:2015sya}. 
Other work has examined the influence of long relaxation times between the electron and hole subsystems \cite{2009PhRvB..79h5415F,Seo:2016vks}.

Another test of the hydrodynamic nature of electron flow in graphene appeared in the experiments 
of Bandurin {\em et al.} \cite{Bandurin1055}. In a finite geometry with boundaries, they observed a `negative nonlocal electrical resistance' inconsistent with Ohmic diffusive electron flow, while consistent with viscous hydrodynamic flow.   Such an effect requires the electron-electron interaction time to be shorter  
than the electron-impurity scattering time.  This is a more direct probe of  hydrodynamic coefficients such as viscosity, compared to global transport measurements \cite{Crossno1058}.    Unfortunately, both of these experiments rely on electrical measurements.   If we study relativistic hydrodynamics near charge neutrality carefully (see \S \ref{sec:relativhydro}), we observe that the charge sector decouples from the energy-momentum sector.   Hence, we expect the viscous `backflow' signature of \cite{Bandurin1055} to vanish near charge neutrality.     Indeed, the signal of \cite{Bandurin1055} was strongest in the Fermi liquid regime of graphene, when $\mu/ k_{\mathrm{B}}T \sim 10$.    So while these explorations of viscous electronic dynamics are in many instances inspired by the holographic developments we have discussed, conventional kinetic theory (and its hydrodynamic limit) is the correct approach to quantitative modeling of these systems.   The quantitative modeling of the ballistic-to-hydrodynamic crossover in Fermi liquids, as in \cite{Molenkamp95},  may be the simplest way to quantitatively connect theories of hydrodynamic electron flow to experiment. 

An important open question is how to directly measure the viscosity of the charge neutral plasma in graphene in experiment.  This will allow us to test the theoretical prediction \cite{muller2009} that this plasma is very strongly interacting, with $\eta/s$ comparable to the `universal' holographic result (\ref{eq:KSSbound}):  $\eta/s=\hbar/4\pi k_{\mathrm{B}}$.    \cite{Lucas:2015sya} suggested that $\eta/s \sim 10\hbar/k_{\mathrm{B}}$, but the transport data of \cite{Crossno1058} is not specific enough to reliably measure this number.   Unfortunately, the presence of disorder makes a direct measurement of $\eta$ rather delicate.   One proposal has been the detection of electronic sound waves \cite{lucas2016}, but this may not be feasible.

\subsubsection{Cuprates}
\label{sec:cuprates}

The hole-doped cuprate high temperature superconductors provide a prominent strange metal near optimal doping \cite{Sachdev:2011cs}. It is likely that this strange metal exhibits the $T>0$ physics
of a $T=0$ critical point, or phase, near optimal doping. A great deal of theoretical and experimental work has tried
to deduce the theory of this critical point by examining the nature of the adjacent phases at lower and higher hole
densities. Much has been learnt sbout symmetry breaking in the under-doped `pseudogap'
regime: charge and spin density wave orders, Ising-nematic order, time-reversal and/or inversion symmetry breaking are present in a complex
phase diagram as a function of density and $T$ \cite{keimernature}. But it appears that none of these
order parameters can explain the gap-like features observed in most spectrocopic probes over a wide range range
of temperature in the underdoped regime. An attractive possibility, especially given the recent observation of a pseudogap
metal at low $T$ \cite{2016Natur.531..210B}, is that the fundamental characteristic of the pseudogap metal is the presence of topological
order (see \S \ref{sec:emerge2}). Then the various conventional symmetry-breaking order parameters are 
proposed to be incidental features of the topologically-ordered state.
The optimal doping criticality controlling the strange metal is associated with the loss of topological order into a
conventional Fermi liquid state at high doping. The critical theory of such a transition invariably involves a Fermi surface coupled
to emergent gauge fields, and such theories were briefly noted in \S \ref{sec:emerge2}; 
a specific candidate for such a `deconfined' theory is in \cite{SSDCNambu},
and this candidate includes spectator conventional orders that vanish at the critical point. From the perspective of the present article, 
the transport properties of such
critical theories with emergent gauge fields are likely to be described well by holographic models.

In the electrical transport properties of the strange metal, a prominent deviation from quasiparticle physics is in the frequency
dependence of the optical conductivity \cite{Marel2003}, $\sigma \sim \omega^{-\alpha}$ with $\alpha \approx 2/3$.
Remarkably, the exponent $\alpha = 2/3$ has been argued to be a robust property of Fermi surfaces coupled
to emergent gauge fields \cite{Furusaki1994,Eberlein16}.

The d.c. conductivity is characterized by a well-known linear-in-$T$ resistivity, which is 
a strong indication of the absence of quasiparticles. A direct signature of the absence of quasiparticles appears in the
Hall resistivity measurements \cite{Chien1991} which cannot be fit to a quasiparticle model \cite{PWA91}. Hydrodynamic and holographic models can provide a natural fit to the data \cite{Blake:2014yla} 
(see \S \ref{sec:magtrans}), although the connection of these models to the
microscopic theory has not been established.  

Recent experiments have focused on the thermal diffusivity of the strange metal \cite{Zhang:2016ofh}. 
They provide striking evidence
of strong electron-phonon coupling, with the both the electrons and phonons excitations exhibiting a scattering time of order $\hbar/(k_{\mathrm{B}} T)$. For the future, it would be of great interest to extend the graphene experiments discussed in \S \ref{sec:graphene} to the cuprates: those could provide crucial information on the nature of hydrodynamic flow in the
electron and phonon subsystems.

\subsubsection{Pnictides}
\label{sec:pnictides}

Unlike the hole-doped cuprates, the pnictides typically have a spin density wave quantum critical point near the
optimal doping for superconductivity. There also appears to be an interesting interplay between the spin density 
wave order, and Ising-nematic order which breaks tetragonal crystalline symmetry down to orthorhombic.
Moreover, strange metal behavior is also dominant near optimal doping \cite{Matsudareview}, 
suggesting a direct connection between
spin density wave or Ising-nematic criticality and the non-quasiparticle transport.
Electrical transport for spin density wave criticality, in the presence of weak disorder coupling
to the order parameter, leads to linear-in-$T$ resistivity \cite{Patel:2014jfa}, as we noted in \S \ref{sec:transcondmat}:
this is a possible explanation for the strange metal behavior. However, it remains to be seen whether 
such models can reproduce the remarkable scaling with $B/T$ in measurements of the magnetoresistance \cite{Analytis16}. The pnictides furthermore exhibit resistivities above the MIR bound, suggestive of non-quasiparticle physics.

\subsubsection{Heavy fermions}
\label{sec:heavy}

The rare-earth intermetallic compounds provide realizations of Kondo lattice models, where localized spin moments
residing on the rare-earth sites interact with itinerant electrons from the other elements. These provide numerous examples of
quantum criticality and strange metal behavior \cite{QSI10,2016arXiv160802925C}. Although the Kondo lattice models
look quite distinct from the single-band Hubbard models applied to the cuprates, the same fundamental issues on the nature of the
quantum phases and the phase transitions apply to both models. In particular, both models support the same
class of metallic states with topological order \cite{2012PhRvB..85s5123P,Punk:2015fha}.

The phase transitions in the Kondo lattice model are traditionally interpreted to be of two types \cite{QSI10}: ({\em i\/}) transitions with the
onset of spin density wave order, which are described by the Landau-Ginzburg-Wilson framework of
\S \ref{sec:sdw}; and ({\em ii\/}) Kondo-breakdown transitions, in which the local moments decouple from the conduction electrons.
Using a more general language, the second class of transitions are more precisely viewed as phase transitions involving
the onset of topological order, in which the size of the Fermi surface can change \cite{TSMVSS04}.
Theories of the second class of transition therefore involve emergent gauge fields, as in \S \ref{sec:emerge2}, 
and are likely to be closely related to theories of the cuprate strange metal. A remarkable example of a possible topological 
phase transition was studied in $\beta$-YbAlB$_4$ \cite{2015Sci...349..506T}.

\addcontentsline{toc}{section}{Acknowledgements}
\section*{Acknowledgements}

We are greatly indebted to our collaborators and colleagues over many years, who have built the holographic edifice described in this review.
We would like to thank Richard Davison, Blaise Gout\'eraux, Andreas Karch, Jorge Santos and Jan Zaanen for discussions relevant to the writing of this review.
This research was supported by the NSF under Grant DMR-1360789 and the MURI grant W911NF-14-1-0003 from ARO.   AL also acknowledges support from the Gordon and Betty Moore Foundation. SAH is partially supported by a DOE Early Career Award.
Research at Perimeter Institute is supported by the Government of Canada through Industry Canada and by the Province of Ontario 
through the Ministry of Research and Innovation. SS also acknowledges support from Cenovus Energy at Perimeter Institute.

\def\baselinestretch{1}\selectfont

\addcontentsline{toc}{section}{References}
\bibliography{rmprefs}

\end{document}